\documentclass{aa}

\usepackage{graphicx}
\usepackage{lscape}            
\usepackage{txfonts}

\usepackage{natbib}
\usepackage[colorlinks=true,urlcolor=blue,citecolor=blue,linkcolor=blue,breaklinks=true]{hyperref}

\bibpunct{(}{)}{;}{a}{}{,}

\newcommand{\nodata}{...}

\begin{document}

\title{A 3\,mm molecular line survey toward IRC+10216}

\titlerunning{3\,mm line survey toward IRC+10216}
\authorrunning{Ao et al.}

\author{
  Yi-Na Ao\inst{1} \and
  Yong Zhang\inst{1,2,3}\thanks{zhangyong5@mail.sysu.edu.cn} \and
  Jian-Jie Qiu\inst{4} \and
  Hao-Min Sun\inst{1} \and
  Xiao-Hu Li\inst{2,5,6}
}

\institute{School of Physics and Astronomy, Sun Yat-sen University, 2 Daxue Road, Tangjia, Zhuhai 519082, Guangdong Province, China \email{zhangyong5@email.sysu.edu.cn}
      \and
          Xinjiang Astronomical Observatory, Chinese Academy of Sciences, 150 Science 1-Street, Urumqi 830011, China
      \and
          CSST Science Center for the Guangdong-Hongkong-Macau Greater Bay Area, Sun Yat-Sen University, Guangdong Province, China
      \and
          School of Mathematics and Physics, Jinggangshan University, 28 Xueyuan Road, Qingyuan District, Ji’an 343009, Jiangxi Province, China
        \and 
        Xinjiang Key Laboratory of Radio Astrophysics, 150 Science 1-Street, Urumqi, Xinjiang 830011, China
        \and
        Key Laboratory of Radio Astronomy and Technology, Chinese Academy of Sciences, A20 Datun Road, Chaoyang District, Beijing, 100101, China
          }

\date{\today}

\abstract 
{
IRC+10216 is the brightest infrared source in the northern sky, known for its rich chemical composition. 
It is often used as a standard reference for studying the circumstellar envelope (CSE) of carbon-rich stars. 
While pioneering 3\,mm spectral surveys have laid foundational datasets, their system temperature limitations rendered spectral line detection thresholds inadequate for probing the source's complex organic molecule inventory at this
 band, which made superseding observations necessary.
}
{
We aim to gain an unbiased view regarding circumstellar chemistry and investigate whether IRC+10216 is typical or anomalous in terms of its chemical composition.
}
{
We carried out an in-depth spectral line survey of the circumstellar envelope of IRC+10216 utilizing the Arizona Radio Observatory 12\,m telescope. We achieved complete spectral sampling across the 90--116 GHz atmospheric window ($\lambda=2.6$--3.3\,mm).
}
{
A total of 214 emission lines belonging to 43 molecular species are identified in the CSE of IRC+10216, among which 28 lines are newly detected in this object and four emission lines remain unidentified. The excitation temperatures and column densities of 16 molecules are determined through rotation diagrams. We estimate the isotopic ratios of carbon,  oxygen, and silicon elements. 
For the majority of the molecular species, the line intensity ratios between IRC+10216 and CIT\,6 are inversely proportional to the square of their distance, which suggests that the chemical processes occurring within them are similar. 
Nevertheless, there is evidence suggesting that the emission of C$_{4}$H and C$_{3}$N in IRC+10216 is unusually strong.}
{These observations stand as the most sensitive and unbiased line survey of IRC+10216 within the $\lambda=3$ window carried out by a single-dish telescope. 
They offer a valuable reference for the astronomical community. 
They can facilitate comparative studies of the circumstellar chemistry of carbon-rich evolved stars and can act as a guiding framework for sensitive interferometric molecular mappings.}


\maketitle\section{Introduction}

Low- and intermediate-mass stars in the later stage of the asymptotic giant branch (AGB) phase generate a substantial amount of stellar winds. These winds strip away the outer layers of the stars, thereby leading to significant mass loss. The ejected material gives rise to a circumstellar envelope (CSE), which is rich in molecules and dust and is considered a natural site for molecule synthesis. To date, over 100 gas-phase molecules and 15 solid-state species have been detected within CSEs \citep{2021ARA&A..59..337D}. The evolutionary processes of materials ejected from stellar atmospheres during the expansion of the CSE remain poorly understood and require further investigation. To address that, observational data from a statistically significant sample of CSEs spanning various evolutionary stages across a wide frequency range are essential. Such comprehensive spectral line surveys provide an unbiased perspective on the chemical processes occurring in CSEs \citep[e.g., ][]{2011IAUS..280..237C}.

Our research group has initiated a long-term project of searching for circumstellar molecules utilizing single-dish telescopes \citep{2008ApJ...678..328Z,2009ApJ...691.1660Z,2009ApJ...700.1262Z,2012ApJ...760...66C,2013ApJ...773...71Z,2020ApJ...898..151Z,2020PASJ...72...46Z,2022ApJS..259...56Q,2023PASJ...75..853Y}. This study represents a continuation of our ongoing series as it presents a comprehensive 3\,mm spectral line survey of the archetypal carbon-rich AGB star IRC+10216. The observational data obtained through this survey enable a detailed comparative analysis of the chemical composition in CSEs across different evolutionary stages, providing critical insights into the molecular evolution processes during late stellar evolution.

IRC+10216, the nearest carbon-rich AGB star \cite[120\,pc,][]{1998MNRAS.293...18G},
has served as a cornerstone of circumstellar chemistry studies since its serendipitous discovery by \citet{1969ApJ...158L.133B}.  
Millimeter-wave breakthroughs commenced with pioneering molecular line detections by \citet{1971ApJ...167L..97W} and \citet{1971ApJ...163L..53S}, which established its status as a molecular factory. 
Subsequently, \cite{1971ApJ...167L..97W}, \cite{1971ApJ...163L..53S}, and others initiated the detection of specific molecules toward this source at the millimeter-wave band. 
\citet{1974ApJ...188..545T} detected SiC dust in IRC+10216 through identification of its characteristic 11.3\,$\mu$m  emission band, with gas-phase molecules SiCSi and SiC$_{2}$ probably being the key precursors for the formation of SiC dust grains. 
Modern observations reveal complex morphodynamics. Near-infrared interferometry uncovers a bipolar peanut-shaped core \citep{2007MNRAS.376L...6M}, while mid-IR imaging exposes nested nonconcentric dust shells with significant density fluctuations \citep{2011A&A...534A...1D}. This points to episodic mass loss. 
The extreme mass-loss rate \cite[$2.4\times10^{-5}$\,$M_{\odot}{\rm yr}^{-1}$,][]{2022ApJ...927L..33F} generates an envelope heavily enshrouded by dust. 
This dense dusty outflow creates a stratified chemical structure: ultraviolet photons from the interstellar radiation field are effectively attenuated in the outer envelope \citep{1982ApJ...252..201H}, while the inner regions sustain active chemistry with enhanced molecular abundances \citep{1982ApJ...252..201H,1993ApJ...419L..41C}. 
IRC+10216 hosts an exceptionally rich molecular inventory \citep[106 species,][hereafter T24]{2024ApJS..271...45T}, including not only carbon chains but also exotic metal-bearing species such as SiC$_{4}$, MgC$_{4}$H, and MgC$_{5}$N. These characteristics establish it as the archetypal laboratory for studying AGB circumstellar chemistry. 
However, its apparent chemical uniqueness raises a critical question: Can IRC+10216 truly represent the broader population of carbon-rich AGB stars? 
Resolving this requires systematic multifrequency observations to disentangle source-specific phenomena from universal chemical processes.
\defcitealias{2024ApJS..271...45T}{T24}

IRC+10216 has been established as a benchmark system for millimeter and submillimeter molecular line surveys; \citet{1984A&A...130..227J} detected 45 emission lines within the frequency range of 72.2--91.1\,GHz using the Onsala Space Observatory (OSO). 
Subsequent observations conducted by \citet {1995PASJ...47..853K} utilized the 45-meter radio telescope at the Nobeyama Radio Observatory (NRO) to systematically survey the 28–50 GHz frequency band. 
This spectroscopic campaign successfully detected 188 molecular lines belonging to 22 distinct chemical species, with notable detections including the long carbon-chain molecules HC$_{7}$N and HC$_{9}$N. 
\citet{2000A&AS..142..181C} conducted comprehensive observations of this source within the frequency range 129.0--172.5\,GHz using the Institut de Radioastronomie Millim{\'e}trique (IRAM) 30\,m telescope and discovered new species, including those containing rare isotopes of C, Mg, Si, S, and Cl. 
Employing the 10-meter Submillimeter Telescope at the Arizona Radio Observatory (ARO), \citet {2008ApJS..177..275H} conducted spectroscopic observations across the 131.2--160.3 GHz band. 
Concurrent observations using the Submillimeter Telescope (SMT) targeted the 219.5--245.5\,GHz and 251.5--267.5\,GHz ranges. 
This comprehensive study detected 377 molecular transitions, including the first detection of $^{13}$CCH and HN$^{13}$C.
\citet{2010ApJS..190..348T} conducted a line survey of the 214.5–-285.5\,GHz band using the 10-meter Submillimeter Telescope at the ARO. 
This study uncovered a significant number of unidentified spectral features, highlighting the critical need for complementary laboratory spectroscopic measurements. 
Using the Effelsberg 100-meter telescope, \citet{2015A&A...574A..56G} performed a spectroscopic survey across the 17.8--26.3 GHz frequency range. 
This observation campaign resulted in the detection of 78 spectral lines, 20 of which represent first-time detections in this source. 
\citet{2017A&A...606A..74Z} utilized the Tian Ma Radio Telescope to conduct a spectral line survey of the 13.3--18.5\,GHz band. 
This effort successfully identified 35 molecular transitions from 12 distinct chemical species, contributing to the characterization of molecular abundances in the observed region. 
\citet{2020PASJ...72...46Z} carried out line surveys of a representative sample
of CSEs, including IRC+10216, within the 20--25\,GHz band using the NRO 45\,m telescope and explored the circumstellar chemistry during different stages. 
These observations were designed to investigate the circumstellar chemistry across different evolutionary stages. 
Early molecular line surveys exhibited a notable spectral coverage deficit between 90--116\,GHz. To fill this observational gap,  \citetalias{2024ApJS..271...45T} conducted an 84.5--115.8\,GHz survey using the Purple Mountain Observatory's 13.7\,m radio telescope. 
These observations have been adopted as a fundamental benchmark to validate gas-phase chemical models \citep{2024A&A...682A.109M}. 
Recent observations from the Atacama Large Millimeter Array (ALMA) have raised higher requirements for the sensitivity of line surveys. 
Using ALMA, \citet{2024A&A...684A...4U} conducted high-sensitivity observations of three carbon-rich AGB stars, IRAS\,15194-5115, IRAS\,15082-4808, and IRAS\,07454-7112, in the frequency range of 85 to 116 GHz. 
While the authors compared molecular abundances derived from their observations with those documented for IRC+10216 in prior studies, the absence of a systematic spectral line census of IRC+10216 at equivalent sensitivity thresholds within this specific frequency range may compromise the robustness of their comparative analysis.

We present a comprehensive 3\,mm spectral line survey of IRC+10216 conducted with the upgraded ARO 12\,m telescope at Kitt Peak, now equipped with the ALMA prototype receivers that doubled sensitivity compared to pre-2013 configurations. 
The remainder of this paper is structured as follows. 
In Section~\ref{obs}, we provide a concise description of the observations and data reduction procedures. 
Section~\ref{Results} presents the outcomes of the spectral line identification and rotation diagram analysis. 
In Section~\ref{dis}, we compare our observational results with those reported by \citetalias{2024ApJS..271...45T} and discuss the implications of our findings for circumstellar chemistry. 
Finally, Section~\ref{conclusion} offers a summary of our work.

\begin{figure*}[!htbp]
\centering
\includegraphics[width = 0.9 \textwidth]{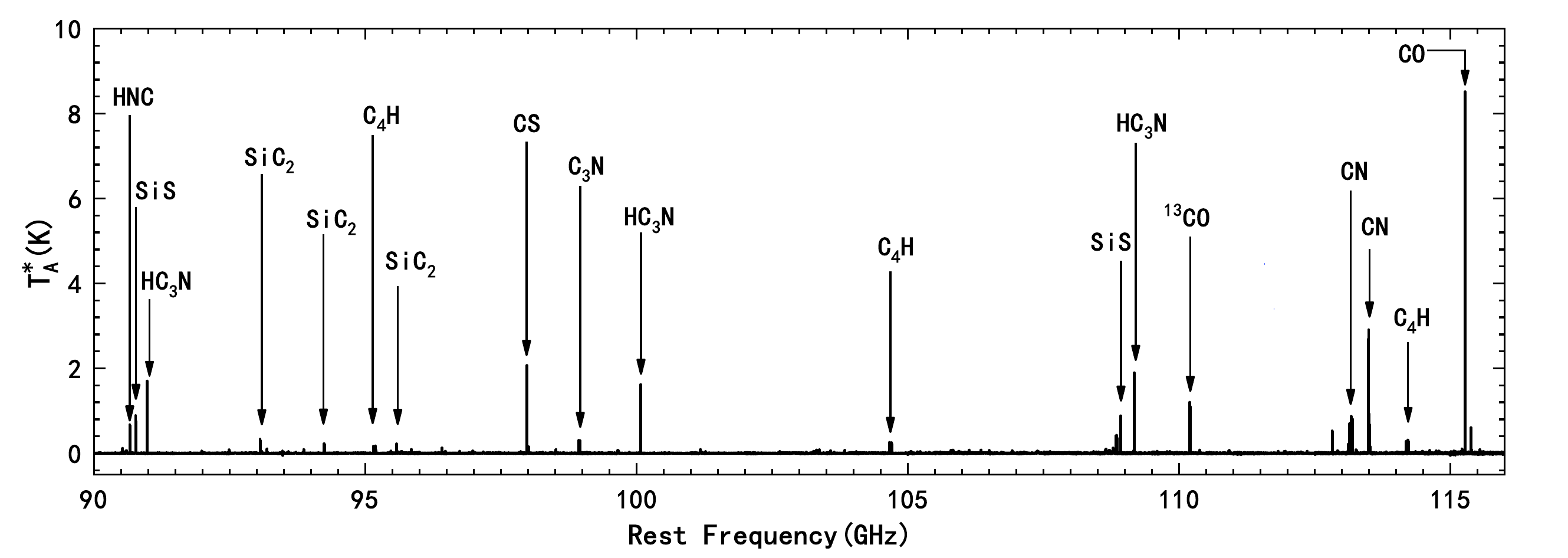}
\caption{{Full spectrum of IRC+10216 from 90 to 116 GHz with the strong features marked.}\label{Fig:zoomv1}}
\end{figure*}

\section{Observations and data reduction}\label{obs}
The observations were carried out during May--June 2015.
Our target sample comprises two archetypal carbon-rich evolved stars: IRC+10216 (this work's primary focus) and CIT\,6. 
The $\lambda=3$\,mm line survey  of CIT\,6 has been reported by \citet{2023PASJ...75..853Y}, and serves as a key reference for our comparative analysis with IRC+10216. 
The receivers were set to operate in the single sideband (SSB) dual polarization mode. 
 For the lower sideband, the typical image rejection ratio was approximately 20\,dB. In contrast, the upper sideband exhibited a lower image rejection ratio. We carried out a comprehensive examination of the image frequencies of strong lines in the image sidebands and identified several image features. The most prominent among them is located at 112820 MHz, which has been misclassified as a "U" line in \citet{2023PASJ...75..853Y}. In reality, this feature is a result of sideband leakage from the HC$_{3}$N $J=11 \to 10$ line in the image sideband.
The spectrometer back ends of the ARO 12\,m telescope were two 500\,MHz 256 channels filter banks (FBs) connected to a millimeter autocorrelator (MAC) with 3072 channels at 195\,kHz per channel. 
The typical system temperatures ($T_{\rm sys}$) range from 100
to 250\,K, with the value depending on the observed frequency band. 
Observations were conducted in position switching mode with an azimuth beam throw of $\pm2\arcsec$. 
The schedule of observations was primarily determined by the rise and fall times of IRC+10216 and CIT\,6. 
Efforts were made to complete the observations of both sources within a specific frequency band before switching to another band, to reduce the calibration time and minimize comparison errors with the work of \citet{2023PASJ...75..853Y}. 
The spectra cover a frequency range from 90 to 116\,GHz, and the corresponding half-power beam width (HPBW) ranges from 41$\arcsec$ to 71$\arcsec$.  
The HPBW is sufficiently large to accommodate the entire emission region of the objects.  The typical on-source time is about 40 minutes for each band. 
The spectra were calibrated to the $T_{\rm A}^{*}$ scale with a correction for atmospheric attenuation, radiative loss, and rearward and forward scattering. 
The main beam temperature, $T_{\rm mb}$ , is derived through the expression of $T_{\rm mb} = T_{\rm R}^{*}/\eta_{\rm m}^{*}$, where the corrected beam efficiency, $\eta_{\rm m}^{*}$ , varies with frequency and is about 0.95 in the 3\,mm window.

Data reduction was performed using the CLASS software package in GILDAS\footnote[1]{GILDAS is developed and distributed by the Observatory de Grenoble and IRAM.}. 
We identified and removed the bad scans that are seriously affected by bandpass irregularities and then combined the spectra from individual scans. 
In addition, a low-order polynomial fit was used to subtract the baseline. 
To improve the signal-to-noise ratio, the spectra were smoothed (combined channels) by a factor of 4. 
The full spectra are shown in Figure.~\ref{Fig:zoomv1}.

\section{Results}\label{Results}
\subsection{Line identifications}\label{line iden}
\begin{figure*}[!htbp]
\centering
\includegraphics[width = 0.8\textwidth]{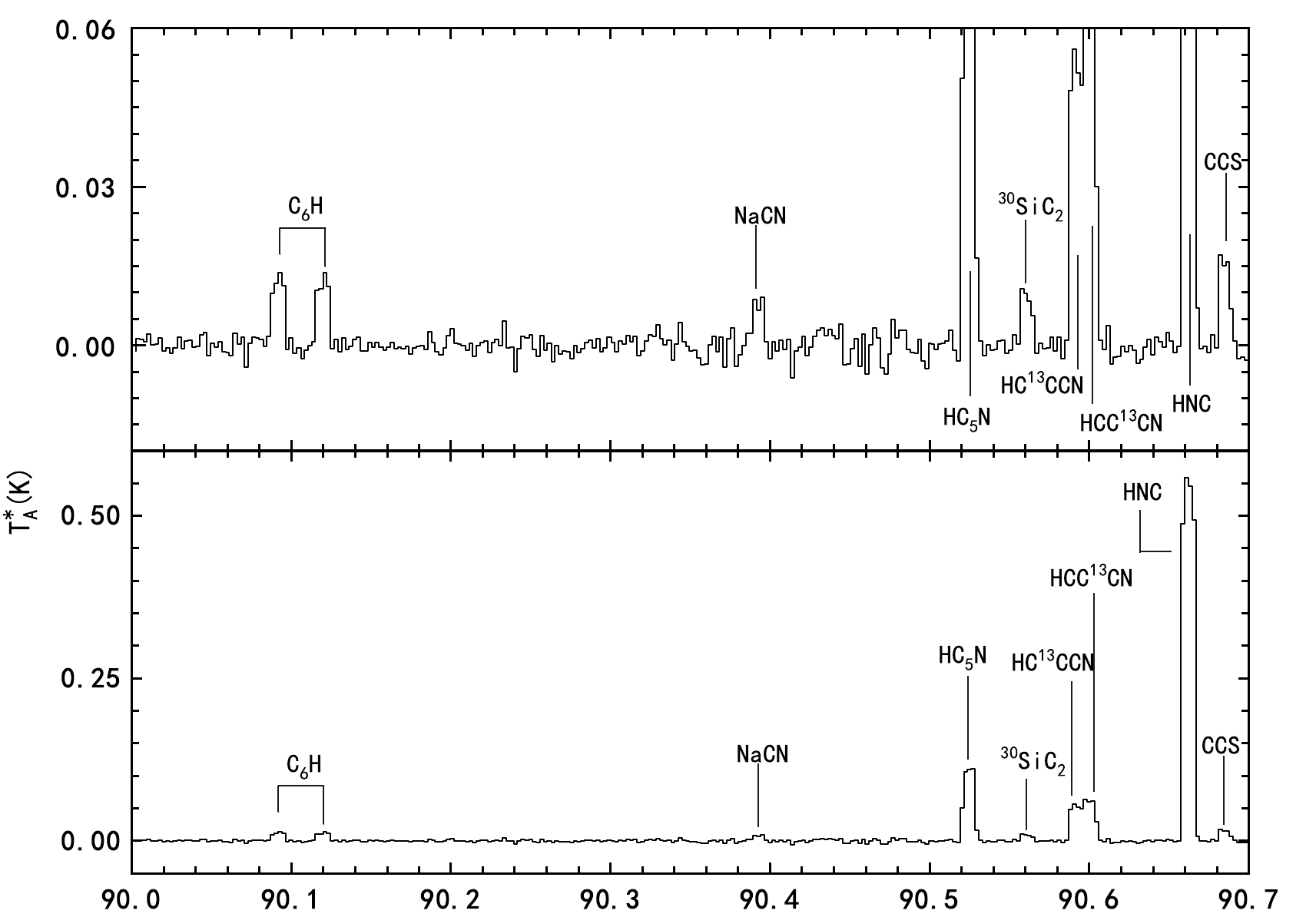}
\caption{{Zoom-in spectra of IRC+10216 with all detected lines labeled. To provide a clearer visualization, the spectra have been smoothed to a resolution of 3\,MHz. "SL" denotes that the spectral feature arises from sideband leakage rather than a genuine detection. The complete set of spectra is shown in Figure.~\ref{Fig:irc_complete}. } \label{Fig:irc}}
\end{figure*}

Generally, we consider an emission line to be a real detection when its signal-to-noise ratio exceeds 3. 
However, certain lines are relatively weak, having a signal-to-noise ratio below 3. 
We consider these lines as valid detections when other transitions of the same molecule have been detected in IRC+10216 or are commonly observed in other evolved stars. 
The identification of emission lines is based on the Spalatalogue\footnote[2]{http://splatalogue.online.} database, which incorporates the catalogs from the Cologne Database for Molecular Spectroscopy catalogs (CDMS\footnote[3]{http://www.astro.uni-koeln.de/cdms/catalog.}) \citep{2005JMoSt.742..215M}, the Jet Propulsion Laboratory Catalog\footnote[4]{http://spec.jpl.nasa.gov.} (JPL) \citep{1998JQSRT..60..883P}, and Lovas\footnote[5]{http://www.nist.gov/pml/data/micro/index.cfm.} \citep{2004JPCRD..33..177L}. 
In this work, we detect a total of 214 emission lines, including four unidentified lines. 
The lines that have been identified belong to 43 molecules. 
With the exceptions of SiS and NaCl, all of these molecules contain carbon. 
Although we discovered no new molecules, we detected 28 molecular lines for the first time in this source. 
As shown in Figure.~\ref{Fig:zoomv1}, the CO $J=1 \to 0$ transition exhibits the highest intensity, followed by the CN $J=1 \to 0$ transition. 
The zoomed-in spectra, smoothed using a four-channel boxcar filter, are presented in Figure.~\ref{Fig:irc}.
The line profiles of most detected lines are fitted using a stellar-shell model. 
The fitting results for each line, including velocity-integrated intensity, line width, and line center temperature, 
are listed in Table~\ref{Tab:irclines}.  
For lines with blended hyperfine structures, only the total velocity-integrated intensity and the intrinsic strength ratio of each line are presented. In what follows, a brief description of the detected molecules is provided.

\begin{figure*}[!htbp]
\centering
\includegraphics[width = 0.45 \textwidth]{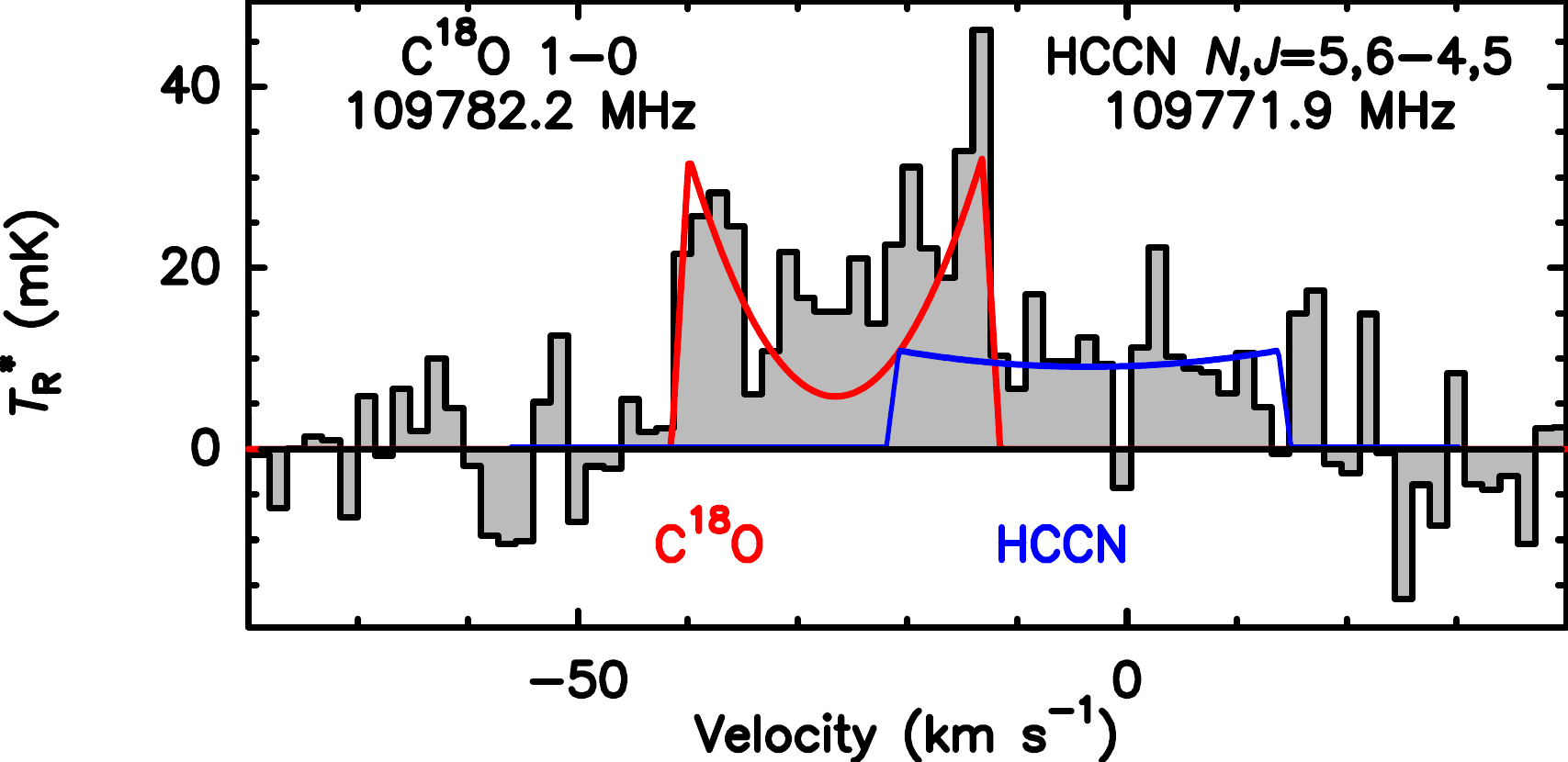}
\hspace{0.05\textwidth}
\includegraphics[width = 0.45 \textwidth]{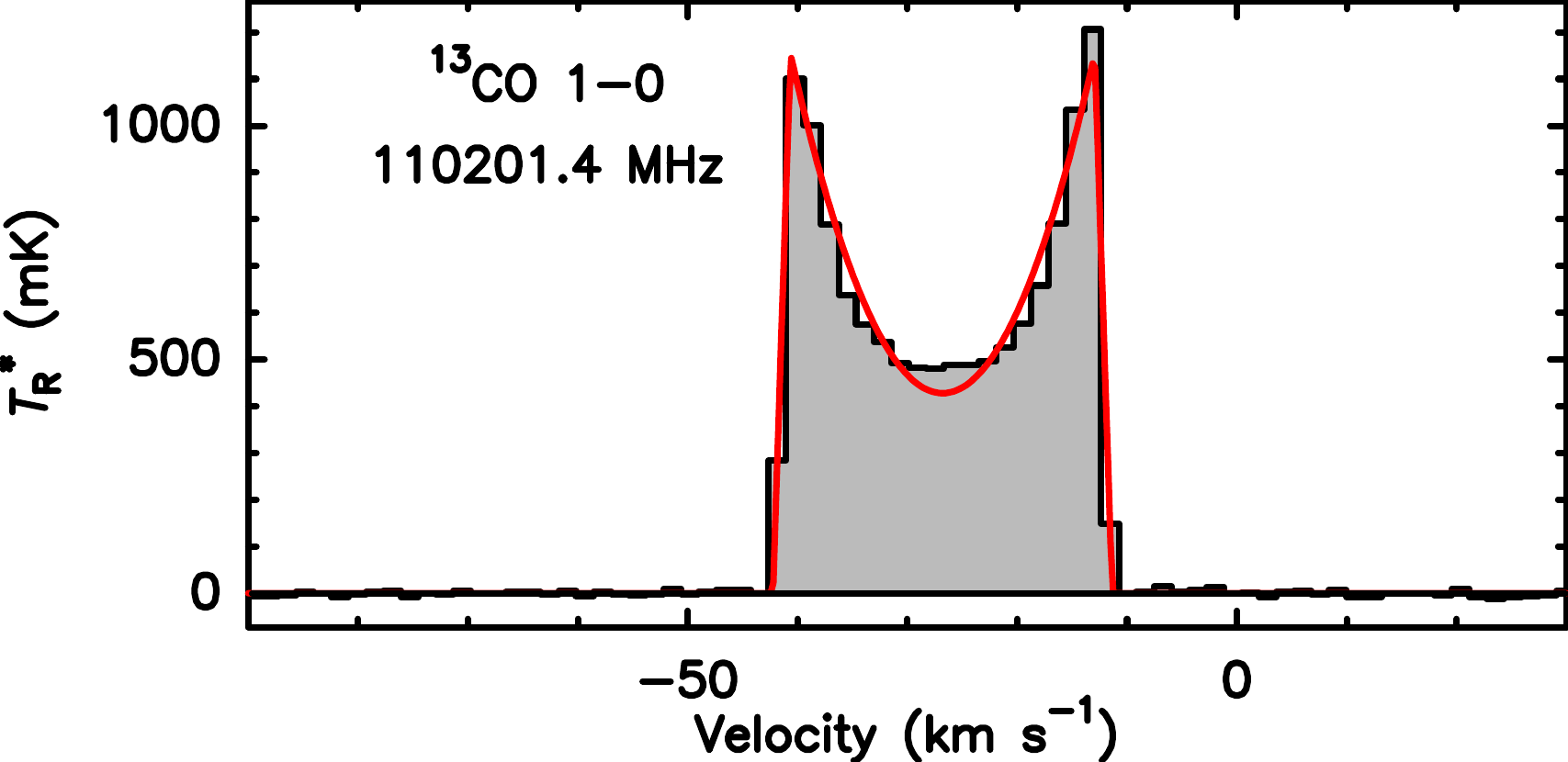}
\vspace{0.1cm}
\includegraphics[width = 0.45 \textwidth]{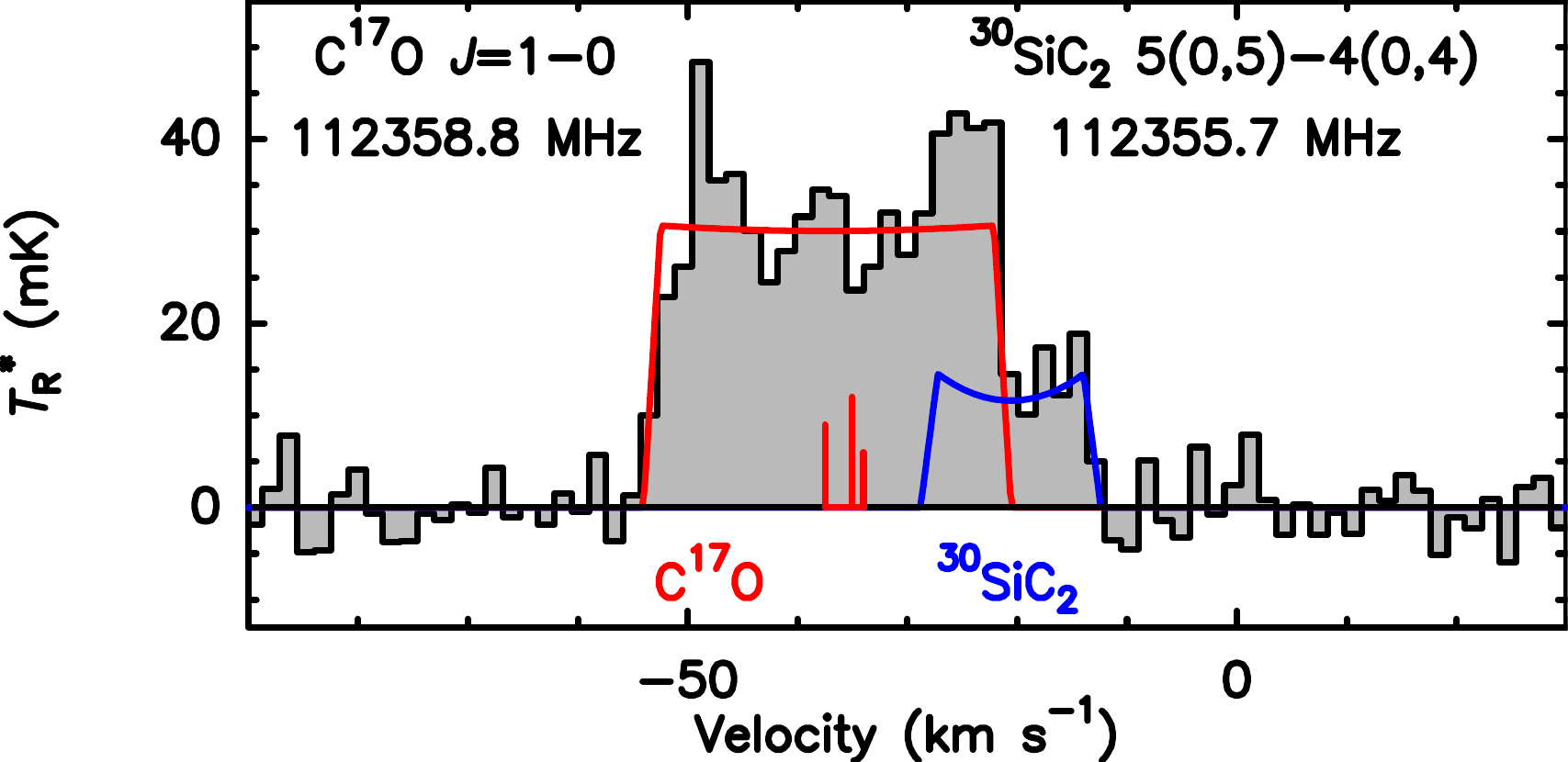}
\hspace{0.05\textwidth}
\includegraphics[width = 0.45 \textwidth]{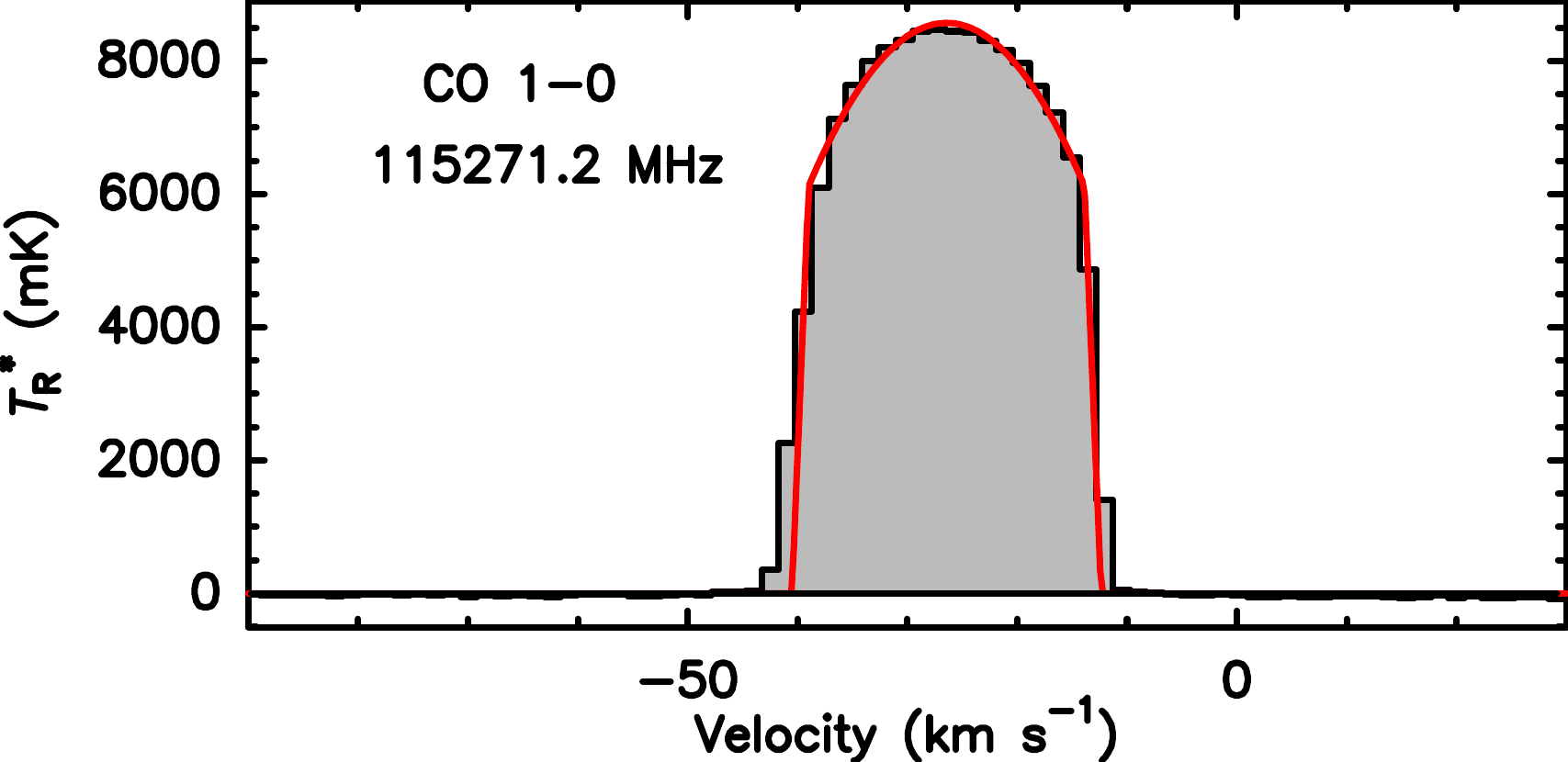}
\vspace{0.1cm}
\includegraphics[width = 0.45 \textwidth]{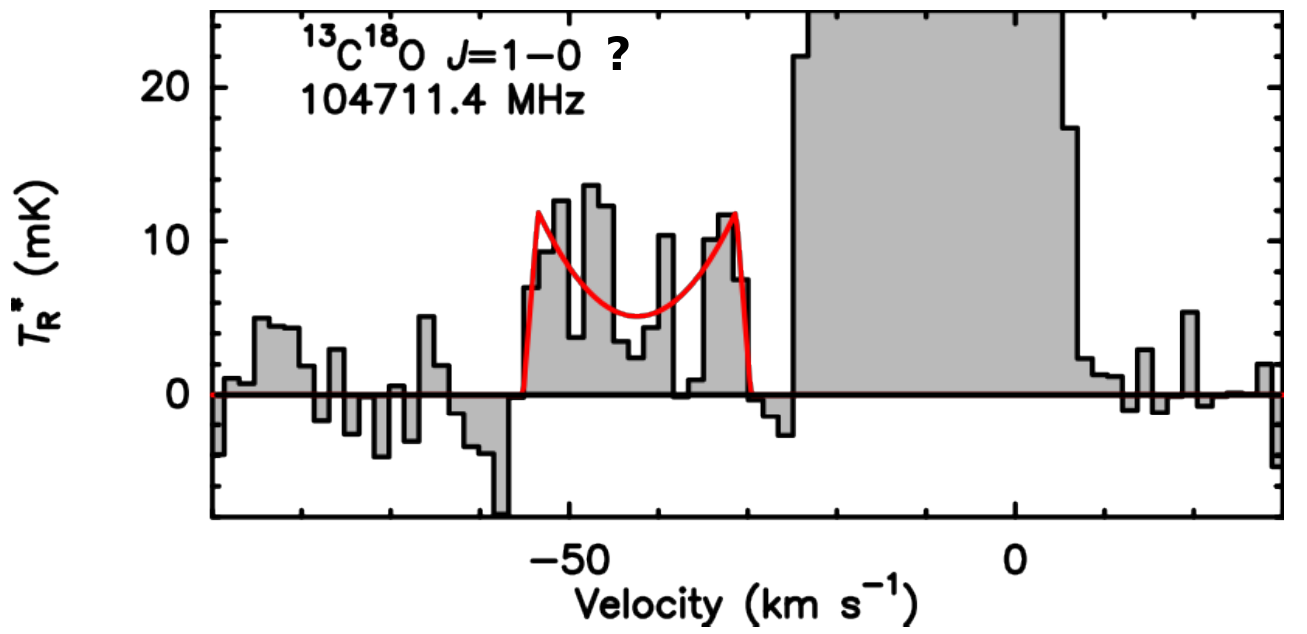}
\caption{{Line profiles of CO and its isotopologues.
The red and green curves represent the multiple fitting profiles obtained through the application of the stellar-shell model. The vertical solid lines indicate the wavelengths and relative intensities of the hyperfine components. Note that the identification
of the $\rm ^{13}C^{18}O$ $J = 1 \to 0$ line is tentative. The complete figure set is shown in Figure.~\ref{Fig:fitting_2}--Figure.~\ref{Fig:fitting_27}.
}\label{Fig:fitting_1}}
\end{figure*}

\paragraph{CO.} 
The $J=1 \to 0$ transitions of CO, $^{13}$CO, C$^{17}$O, and C$^{18}$O are detected, with the CO $J=1 \to 0$ line being the strongest. 
These observations probe gas components with varying optical depths, thereby uncovering distinct kinematic characteristics. 
The first detection of the CO $J=1 \to 0$ line in IRC+10216 was made by \citet{1971ApJ...163L..53S} using the 36-foot antenna of the National Radio Astronomy Observatory. 
As shown in Figure.~\ref{Fig:fitting_1}, the line profiles of the $J=1 \to 0$ transitions of CO and $^{13}$CO are parabolic and double-peaked respectively, 
which correspond to the optically thick and thin scenarios. 
For the $J=1 \to 0$ transitions of C$^{17}$O and C$^{18}$O, their profiles cannot be unambiguously determined because of blending with other lines. 
Nevertheless, it appears that they possess multiple spectral components, which is consistent with limb-brightened outflows. At 104.71\,GHz, a weak spectral feature is observed, which precisely coincides with the frequency of the $\rm ^{13}C^{18}O$ $J = 1 \to 0$ transition line. Nevertheless, if we assume that the intensity ratio of the $\rm C^{18}O$ $J = 1 \to 0$ to the $\rm ^{13}C^{18}O$ $J = 1 \to 0$ lines follows the $\rm ^{12}C/^{13}C$ abundance ratio \citep[45,][]{2000A&AS..142..181C}, the integrated intensity of the $\rm ^{13}C^{18}O$ $J = 1 \to 0$ line is expected to be 13\,mK\,km\,s$^{-1}$, which is approximately an order of magnitude lower than the value obtained from our measurements. Given this significant discrepancy, we conclude that the identification of this feature as the $\rm ^{13}C^{18}O$ $J = 1 \to 0$ should be considered tentative, pending further confirmation.

\paragraph{C$_{\rm n}$S.}
We detected the $J=2 \to 1$ transition of CS, six rotational transitions of C$_{2}$S, five rotational transitions of C$_{3}$S, and a weak vibrationally excited transition in the $\nu=1$ state of CS $J=2 \to 1$ .
We also detected the isotopologues of CS, including $^{13}$CS, C$^{33}$S, and C$^{34}$S. 
As shown in Figure.~\ref{Fig:fitting_2}, the line profile of CS $J=2 \to 1$ shows a parabolic shape with a slightly asymmetric wing, where the red-shifted side is brighter. 
The line profile of C$^{34}$S $J=2 \to 1$ shows a double-peaked shape. 
The C$_{2}$S $(N,J)=(7,8) \to (6,7)$ line is partly blended with a C$_{4}$H line, as shown in Figure.~\ref{Fig:fitting_4}. 
The remaining lines present slight double-peaked profiles. 
High spatial resolution maps of IRC+10216 
reveal that the CS emission is distinctly clumpy, with its substructures presenting in a pattern resembling concentric shells or circular arcs \citep{2019A&A...629A.146V}.

\paragraph{SiS.}
We detected a total of five emission lines of SiS and its isotopologues. 
The SiS $J=5 \to 4$ transition  
exhibits a distinct double-peaked morphology characterized by an asymmetrical velocity wing demonstrating enhanced emission on the red-shifted side, as clearly depicted in Figure.~\ref{Fig:fitting_5}.  In contrast, the
 $J=6 \to 5$ line displays a smooth parabolic-shaped profile. This contrasting behavior between adjacent rotational transitions aligns with similar spectral discrepancies previously reported by \citet{2019A&A...629A.146V}

\paragraph{SiC$_{2}$.}
We detected seventeen emission lines of SiC$_{2}$ and its $^{13}$C isotopologues. 
Among these, we report on three lines for the first time in IRC+10216, while the rest were previously documented in earlier observations \citep{1986A&A...167L...9C,2018A&A...618A...4C}. 
As depicted in Figure.~\ref{Fig:fitting_7}, the line profiles of SiC$_{2}$ closely resemble those of SiS, characterized by a double-peaked structure featuring a brighter red wing. These lines exhibit a relatively narrow linewidth, with the local standard rest velocity  ($v_{\rm LSR}$)
distributed within the range of $-40$ to $-10$ km s$^{-1}$.
Interferometric observations have revealed that the spatial distribution of SiC$_{2}$ takes the form of a shell encircling 
the central star. However, this shell is far from uniform; rather, it is clumpy and nonspherical, as previously reported in the literature \citep{1992ApJ...388L..31G,1992PASJ...44..469T,1995ApJ...439..445G}.
The nonuniform nature of the SiC$_{2}$ shell implies complex physical processes at play in the circumstellar environment, potentially involving inhomogeneous mass loss from the central star, or interactions between the circumstellar gas and the surrounding interstellar medium.

\paragraph{SiC$_{4}$.}
There are eight SiC$_{4}$ transitions within the frequency range of the current observations.
Six of these, from $J=30 \to 29$ to $35 \to 34$, are detected with a peak temperature of about 10\,mK. 
As shown in Figure.~\ref{Fig:fitting_8}, the $J=30 \to 29$ and $33 \to 32$ are unresolved and blended with the emission lines of $^{26}$MgNC and C$_{6}$H, respectively. 
The SiC$_{4}$ $J=36 \to 35$ and $37 \to 36$ lines are too weak to be detected.

\paragraph{H$_{2}$C$_{3}$ and H$_{2}$C$_{4}$.}
We detected three emission lines of H$_{2}$C$_{3}$ 
and, as depicted in Figure.~\ref{Fig:fitting_10}, two of these lines are blended with one another.
The H$_{2}$C$_{3}$ $5(1,4) \to 4(1,3)$ line exhibits a double-peaked profile. H$_{2}$C$_{3}$ is an isomer of $c$-H$_{2}$C$_{3}$, a molecule that has been widely detected in the interstellar medium. 
The first potential detection of $c$-H$_{2}$C$_{3}$ was  reported by \citet{1991ApJ...368L..39C}. 
However, in our study, we did not observe any emission lines associated with $c$-H$_{2}$C$_{3}$. 
On the other hand, we detected nine spectral lines of 
 H$_{2}$C$_{4}$. Among these, five had been previously reported in earlier observations \citep{1991ApJ...368L..43C,2018A&A...615L...4P}. As depicted in Figure.~\ref{Fig:fitting_11}, the line profiles of 
 H$_{2}$C$_{4}$ exhibit both double-peaked and flat-topped shapes.

\paragraph{$l$-C$_{3}$H.}
The spectral analysis reveals complex blending in the $l$-C$_{3}$H transitions due to unresolved adjacent hyperfine components. We detected eight hyperfine lines of $l$-C$_{3}$H, two of which blended together and remained unresolved as evidenced by asymmetric profiles in Figure.~\ref{Fig:fitting_12}. The double-peaked line profiles indicate optically thin emission from an expanding circumstellar envelope. Observations of these lines 
have been previously reported \citep{1985ApJ...294L..49T,1986A&A...164L...1C}. Comparative analysis with its cyclic isomer $c$-C$_{3}$H  reveals significant chemical differentiation processes \citep{2008JChPh.128c4301S}. Although many transition lines of $c$-C$_{3}$H lie within our observed frequency range, no lines are detected. Therefore, we conclude that the abundance of $c$-C$_{3}$H in IRC+10216 is substantially lower than that of its linear isomer.
This may arise from the fact that linear chains preferentially form through radical-radical recombination in the outer CSE, while cyclic isomers require a higher density environment that is unavailable in IRC+10216's extended envelope.

\paragraph{C$_{4}$H.}
Our observations revealed 27 rotational transitions of C$_{4}$H.
Each rotational transition resolves into two hyperfine components with a similar intensity.
Ground-state transitions display double-peaked profiles
with slight red-wing enhancement,  as shown in Figure.~\ref{Fig:fitting_13}.   
The C$_{4}$H $v_{7}$=2$^{0}$, $N = 10 \to 9, J=19/2 \to 17/2$ line is badly blended with the C$_{6}$H $^{2}\Pi_{3/2} J=69/2 \to 67/2 f$ line.
Spatially resolved ALMA observations \citep{2017A&A...601A...4A} constrain the C$_{4}$H distribution to a hollow spherical shell.

\paragraph{C$_{5}$H.}
Our  observations detected 20 rotational transitions of the C$_{5}$H radical, including two newly detected transitions. The remaining lines are consistent with previous detections \citep{1986A&A...164L...1C,1995ApJ...445L..47Z,2018A&A...615L...4P}. 
As shown in Figure.~\ref{Fig:fitting_14}, each line splits into two unresolved blended hyperfine structures.

\paragraph{C$_{6}$H.}
Among all identified molecules, C$_{6}$H exhibits the highest number of spectral lines, totaling 33.
The C$_{6}$H $^{2}\Pi_{1/2} J=75/2-73/2 e,f$ transition, which falls within our observed frequency range but is blended with C$_{4}$H lines, could not be conclusively detected. 
The $^{2}\Pi_{1/2}, J=81/2 \to 79/2 e$ and $^{2}\Pi_{1/2}, J=81/2 \to 79/2 f$ 
transitions represent new detections, while the remaining lines have been previously reported \citep{1987A&A...175L...5G,1987A&A...181L...1C,1995ApJ...445L..47Z,2010A&A...517L...2A}. 
Similar to C$_{4}$H, each rotational transition of C$_{6}$H undergoes splitting into two $\Lambda$ doubling components. 
As illustrated in Figure \ref{Fig:fitting_15}, these lines display distinct double-peaked profiles, which suggests that C$_{6}$H is likely to be optically thin.

\paragraph{CN.}
The significant abundance of CN observed in CSE and the interstellar medium arises from its efficient formation via HCN photodissociation and resilience in UV-irradiated environments. 
Its complex hyperfine structure arises from the spin-rotation interaction and the hyperfine structure due to the nuclear spin of nitrogen ($I=1$), which produces nine measurable hyperfine structure components in the $N=1 \to 0$ transition. 
Our observations detected all nine components, as shown in Figure.~\ref{Fig:fitting_16}.
The rarer isotopologue $^{13}$CN shows more intricate splitting, with 14 hyperfine components in its $N=1 \to 0$ transition (113.488\,GHz), which are categorized into three distinct groups. 
This substructure emerges from combined $^{13}$C nuclear spin ($I=1/2$) and $^{14}$N quadrupole moment ($I=1$) interactions, creating additional splitting compared to the main isotopologue. 
Previous observations by \citetalias{2024ApJS..271...45T} detected all the CN lines but failed to observe $^{13}$CN.

\paragraph{C$_{3}$N and C$_{5}$N.}
Our observations revealed four ground-state rotational transitions and one $\nu_5=1$ vibrationally excited state transition of C$_3$N. 
The ground-state transitions exhibit characteristic double-peaked profiles with enhanced red-wing emission (Figure.~\ref{Fig:fitting_18}), which is consistent with optically thin radiation  from an expanding circumstellar envelope. 
While these ground-state lines were previously observed by \citetalias{2024ApJS..271...45T}, we report the first detection of the $\nu_5=1$ excited state transition, which indicates non-local thermodynamic equilibrium (non-LTE) excitation in the inner envelope. 
C$_5$N was first detected in IRC+10216 by \citet{1998A&A...335L...1G}. 
Subsequent observations by \citet{2022A&A...658A..39P} identified the $N=14 \to 13$ and $15 \to 14$ transitions. 
In this work, we detected five new hyperfine components of the $N=40 \to 39$, $J=79/2 \to 77/2$ transition. 
These hyperfine structures show severe blending that prevents reliable profile fitting.

\paragraph{HC$_{3}$N and HC$_{5}$N.}
We detected nine emission lines of HC$_{3}$N, consisting of three ground-state lines and six $v_{7}=1$ vibrationally excited lines. 
As depicted in Figure.~\ref{Fig:fitting_21}, the parabolic profiles of the ground-state lines indicate that the HC$_{3}$N lines are optically thick. 
For the lines of HCC$^{13}$CN and HC$^{13}$CCN, which are from the same energy level, a small frequency difference leads to partial blending. 
As shown in Figure.~\ref{Fig:fitting_22}, we detected ten ground-state transitions of HC$_{5}$N, ranging from $J = 34 \to 33$ to $J = 43 \to 42$. 
Their flat-topped line profiles suggest an optically thin and spatially unresolved scenario.

\paragraph{MgNC and MgCN.}
We detected four MgNC lines, one $^{25}$MgNC emission line, and two $^{26}$MgNC lines. 
All of these lines had been previously reported in studies \citep{1995A&A...297..183G}. As depicted in Figure.~\ref{Fig:fitting_23}, the line profiles of MgNC $(N,J)=(8,15/2) \to (7,13/2)$ and $(8,17/2) \to (7,15/2)$ exhibit a double-peaked shape. The $(9,17/2) \to (8,15/2)$ and $(9,19/2) \to (8,17/2)$ lines are blended with each other, making it impossible to determine their individual profiles. 
The rotational transitions of MgCN ($N=9 \to 8, 10 \to 9$, and $11 \to 10$) at 91.7, 101.9, and 112.07\,GHz, respectively, fall within our observational frequency range. Among these, the $N=10 \to 9$ is distinctly detected in the blue wing of a C$_6$H line. However, the $N=9 \to 8$ and $11 \to 10$
transitions are heavily obscured by noise, rendering them undetectable in our data.
This molecular species was first identified in IRC+10216 through 3\,mm wavelength observations by \citet{1995ApJ...445L..47Z}. A direct comparison of our spectra with their published results demonstrates remarkable consistency in the detected line profiles and intensities, thereby corroborating the presence of MgCN in this object.

\paragraph{NaCN and NaCl.}
We detected two sodium-containing species, NaCN and NaCl, toward IRC+10216. 
The NaCl $7 \to 6$ and $8 \to 7$ lines fall within our frequency coverage and are both detected, as depicted in Figure.~\ref{Fig:fitting_25}. 
In total, eight NaCN lines are detected, as shown in Figure.~\ref{Fig:fitting_26}, and all of them have been previously reported \citep{1994ApJ...426L..97T,2012A&A...543A..48A,2019A&A...630L...2C}.

\paragraph{CH$_{3}$CN.}
We detected seven rotational transitions of CH$_{3}$CN, all of which had been previously reported \citep{2014A&A...570A..45A,2017A&A...607L...5Q}. 
Each $J+1 \to J$ rotational transition undergoes centrifugal distortion, splitting into $J$ hyperfine components that become spectrally blended (Figure.~\ref{Fig:fitting_27}), which prevents individual line profile analysis. 
ALMA mapping observations toward IRC+10216 revealed that the majority of CH$_{3}$CN is distributed in a hollow shell approximately \(2\arcsec\) from the star, with a central void that has a radius of up to \(1\arcsec\) \citep{2015ApJ...814..143A}. 
This distribution contradicts the standard chemical model of IRC+10216 and significantly differs from that of other molecules. 
The observed discrepancy may be attributed to the clumpy structure of the envelope, which enables UV photons to penetrate the inner regions.

\paragraph{Unidentified lines.}
We detected four emission lines that could not be correlated with any potentially detectable transitions listed in the spectroscopic catalog. In Table~\ref{Tab:irclines}, these lines are marked with "U".
The U line at 98.05\,GHz coincides with CH$_{3}^{13}$CH$_{2}$CN $11_{5,7} \to 10_{5,6}$ observed in Orion~KL \citep{2007A&A...466..255D}. 
However, we exclude the molecule as the carriers based on the non-detection of it's numerous predicted transitions within our spectral coverage. 
The U line at 112.04 GHz was also detected by \citet{1995ApJ...445L..47Z}.
These U lines likely originate from novel circumstellar molecules,  known moelcules with complex internal degrees of freedom (e.g., internal rotation in HC$_{2n+1}$N polymers), among other possibilities. 
Future experimental investigations are crucial to characterize the nature of these unassigned U lines.

\subsection{Rotational temperatures and column densities}\label{T and density}
The rotational diagram method is used to calculate the excitation temperatures ($T_{\rm ex}$) and column densities ($N$) of the molecules detected in our observations. 
We assume that the emission lines are optically thin and that the level populations of the molecules are in LTE. 
The expression of rotation diagram is 
\begin{equation}\label{f1}
{\rm ln}(\frac{3k \int T_{\rm s} \, {\rm d v}}{8\pi^{3}\nu\mu^{2}S})~=~{\rm ln}(\frac{N_{\rm u}}{g_{\rm u}}) ~=~{\rm ln}(\frac{N}{Q})-\frac{E_{\rm u}}{kT_{\rm ex}},
\end{equation}
where $\int T_{s}\, {\rm dv}$ represents the integration of the source brightness temperature ($T_{\rm s}$) over the velocity, 
$\nu$ is the rest frequency of each transitions, 
$\mathbf{\mu}$ is the permanent dipole moment, $\mathbf{S}$ is the transition moment, 
$g_{\rm u}$ is the degeneracy of the upper state, 
$E_{\rm u}$ is the upper level energy of the transition, and
$Q$ is the partition function that varies as a function of excitation temperature. 
If multiple distinct transitions of a given molecule are detected, a straight line is utilized to fit them by means of Equation \ref{f1}, where $\ln\frac{N_{\rm u}}{g_{\rm u}}$ represents the ordinate and $\frac{E_{\rm u}}{k}$ serves as the abscissa.  We can determine 
$T_{\rm ex}$ and $N$  from the slope and intercept of the fitting line. 
To account for beam dilution effects in our single-dish observations, we applied the standard correction formula $T_{\rm s} = T_{R}(\theta_{\rm s}^{2}+\theta_{\rm b}^{2})/\theta_{\rm s}^{2}$, where $\theta_{\rm b}$  (55$\arcsec$--70$\arcsec$) represents the frequency-dependent HPBW.
Although molecular emission regions exhibit $\theta_{s}$ variations \citepalias{2024ApJS..271...45T}, we adopted a uniform value, $\theta_{s}=30\arcsec$.  
For optically thin transitions, rotational diagrams constructed via Equation~\ref{f1} show enhanced reliability with increasing numbers of detected transitions. 
Deviations from linearity may arise from non-LTE excitation processes (e.g., radiative pumping in vibrationally excited states), moderate optical depth effects, line blending, or molecular misassignments.

Figure.~\ref{Fig:RD} illustrates the rotational diagrams of 16 molecules detected in our observations. 
While rotational diagrams for most species in our sample are well fitted by a single straight line, C$_{6}$H, HC$_{3}$N, and HC$_{5}$N exhibit significant deviations requiring dual excitation temperature components. 
The spatial thermal structures of these molecules are likely to be stratified. 
Among the detected molecules, the rotational diagram of HC$_{5}$N shows the most distinct thermal stratification. 
As shown in Figure.~\ref{Fig:RD}, based on our observations, the rotational temperature of HC$_{5}$N is 18.7\,K, with higher-$J$ lines tracing hotter regions \citep{2000A&AS..142..181C}, while lower-$J$ transitions probe colder regions \citep{2022A&A...658A..39P}. 
The derived $T_{\rm ex}$ and $N$ of 16 molecules are listed in Table~\ref{Tab:irc_rd}.

CN ranks among the most abundant molecules in IRC+10216, with an abundance approximately five times lower than that of HCN \citep{1995ApJ...439..996D}. Our observations reveal two fine-structure components of the CN $(N=1-0)$ transition at approximately 113.1 and 113.5\,GHz, which allows us to derive their optical depths. 
The intrinsic intensity ratio of these two components is two. 
However, a statistical survey of 42 carbon-rich stars by \citet{1997A&A...319..235B} reported an average ratio of 1.3, which suggests significant optical thickness effects. 
By defining the intensity ratio between the stronger and weaker components as $R,$ the optical depth of the stronger component ($\tau_{s}$) can be determined using the relation 
$R = \frac{1-\rm exp (-\tau_{s})}{1-\rm exp (-\tau_{s}/2)}$.
For IRC+10216, our analysis yields an optical depth of $\tau_{s}=0.55$ for the stronger component.

\subsection{\texorpdfstring{Fractional abundances relative to H$_{2}$}{Fractional abundances relative to H2}}\label{abundance}
Assuming that the CSE is a spherical shell formed by a constant mass-loss rate, the molecular emission is optically thin, and a uniform $T_{\rm ex}$,
we calculated the fractional abundances of the molecules relative to H$_{2}$ ($f_{\rm}$) using the formula given by \citet{1997IAUS..178..457O}, 
\begin{equation}\label{f2}
f_{\rm X}=1.7\times10^{-28}\frac{v_{e}\theta_{\rm b}D}{\dot{M}_{\rm H_{2}}}\frac{Q(T_{\rm ex})\nu^{2}}{g_{\rm u}A_{\rm ul}} \frac{e^{E_{\rm l}/kT_{\rm ex}}\int T_{R}\,{\rm dv}}{\int_{x_{i}}^{x_{e}}e^{-4x^{2}ln2}\,{\rm dx}},
\end{equation}
where $v_{e}$ is the expansion velocity of the CSE, with the measured line width taken as the value, in unit of km\,s$^{-1}$, 
$D$ is the distance in pc \cite[120\,pc;][]{1998MNRAS.293...18G}, 
$\dot{M}_{\rm H_{2}}$ is the mass-loss rate in $M_\odot$ yr$^{-1}$, 
$g_{\rm u}$ is the statistical weight of the upper level, 
$A_{\rm ul}$ is the Einstein coefficient for the transition, and $x_{i,e}$ is calculated by using the formula $x_{i,e}=R_{i,e}/(D\theta_{b})$, 
where $R_{i}$ and $R_{e}$ are the inner and outer radii of the shell obtained by \citet{2003A&A...402..617W}. 
The mass-loss rate is obtained via the CO $J=1 \to 0$ line using the prescription from \citet{2002A&A...388..609W}, with a CO abundance assumption of $f_{\rm CO} = 1 \times 10^{-3}$. 
We obtain $\dot{M}_{\rm H_{2}}=6.7\times10^{-6}$. 
The calculated molecular abundances are presented in Table~\ref{Tab:irc_rd}. 
We compared our results with those of previous studies by \citet{2008ApJS..177..275H}, \citet{2015A&A...574A..56G}, \citet{2017A&A...606A..74Z}, and \citet{2022A&A...658A..39P}. 
Our findings show a good agreement with the earlier results.

\section{Discussion}\label{dis}

\subsection{\texorpdfstring{Comparison with \citetalias{2024ApJS..271...45T}}{Comparison with 2024ApJS..271...45T}}\label{com tuo}
\begin{figure*}[!htbp]
\centering
\includegraphics[width = 0.47\textwidth]{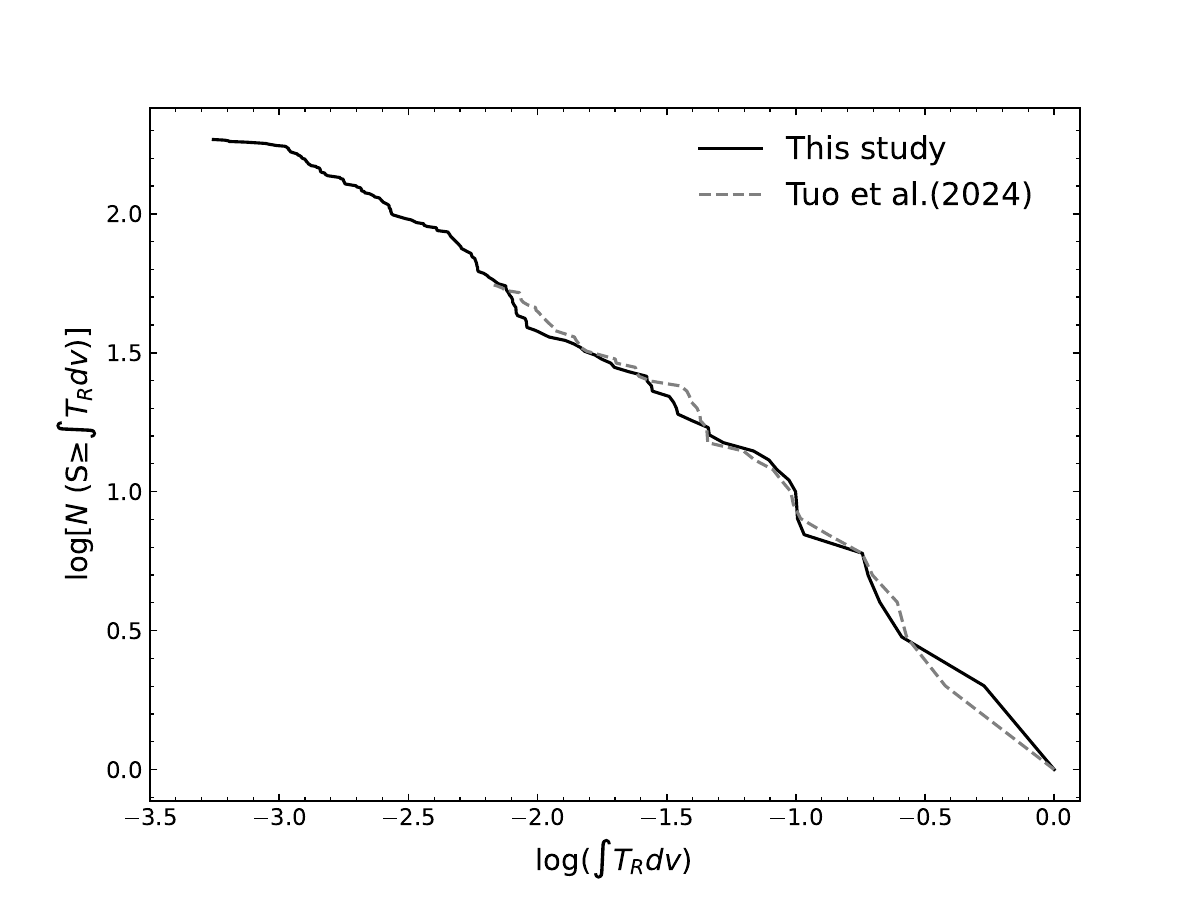}
\includegraphics[width = 0.47\textwidth]{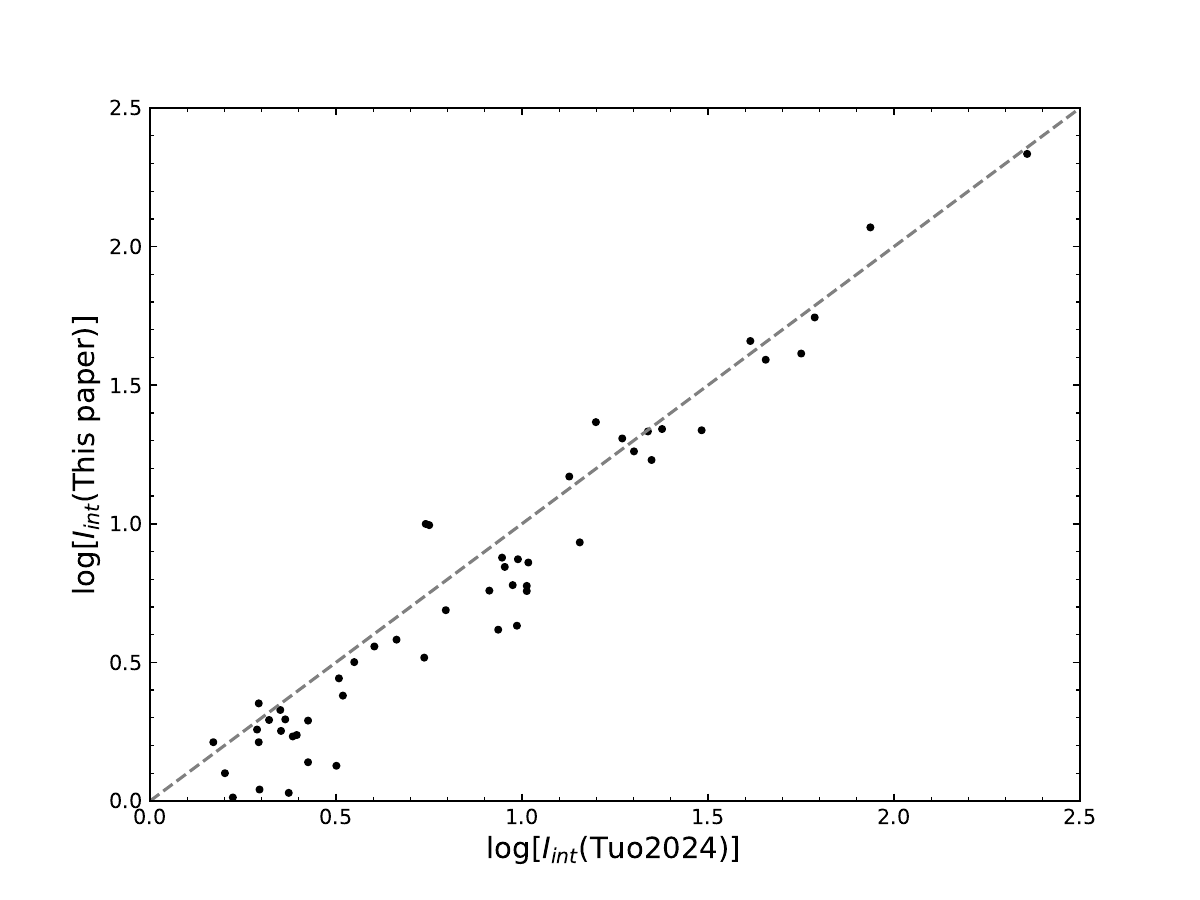}
\caption{{Number of detected spectral lines above a given $\int T_{R}dv$ value. The maximum $\int T_{R}dv$ value has been normalized to 1 (left panel). Comparison of the line intensities of molecules detected in this work and \citetalias{2024ApJS..271...45T}. The dashed line represents $x=y$ (right panel).}
\label{Fig:compare number with Tuo and our}}
\end{figure*}

The comparative analysis between our 3 mm spectral survey and the prior work of \citetalias{2024ApJS..271...45T} (Figure.~\ref{Fig:compare number with Tuo and our}) reveals both consistency and advancement in probing IRC+10216's molecular inventory. 
Within the overlapping 90--115.8\,GHz spectral range, both surveys demonstrate robust agreement for bright emission lines. 
We compared the integrated intensities of the common emission lines detected in our observations and those in \citetalias{2024ApJS..271...45T} , as depicted in Figure \ref{Fig:compare number with Tuo and our}. 
The data have been corrected for the impact of beam dilution. 
Overall, the ratios of line intensities between our results and those of \citetalias{2024ApJS..271...45T} are close to unity. 
This indicates a high degree of consistency in the intensities of the emission lines between the two datasets. 
This mutual validation strengthens confidence in line identifications for 57 previously cataloged transitions. 
Our observations achieve a seven-fold sensitivity improvement.
This enables the detection of 157 additional weak lines.

However, a more detailed analysis reveals variations in the beam-dilution corrected intensity ratios of individual molecules between \citetalias{2024ApJS..271...45T} and this work [$I_{\rm int}({\rm Tuo2024})$ versus $I_{\rm int}$({\rm this paper})], as shown in Figure.~\ref{Fig:compare intensity with Tuo and our}. 
During beam dilution corrections, an identical source size was assumed for both datasets. 
Given the larger telescope beam of our observations, our measurements preferentially sample more extended spatial regions compared to \citetalias{2024ApJS..271...45T}. 
Consequently, if species exhibit differing spatial distributions, the $I_{\rm int}({\rm Tuo2024})/I_{\rm int}$({\rm this paper}) ratios may deviate from unity.

\subsection{Isotopic ratio}\label{Isotopic Ratio}
The central star of IRC+10216 is thought to have evolved to the late AGB phase \citep{2000A&AS..142..181C}, where the third dredge-up processes have transported nucleosynthetic products from the stellar interior to the surface through convective mixing, subsequently expelling them into the CSE via stellar winds. 
Isotopic abundance ratios of molecular species in the CSE serve as critical tracers of these internal nucleosynthetic and mixing processes \citep{2005ARA&A..43..435H}. 
Based on observed molecular transition lines, we derived the carbon and silicon isotopic ratios in IRC+10216. 
The results are listed in Table \ref{Tab:isotopic} alongside values for the Sun, CIT\,6, and CRL\,3068.

\begin{figure}[!htbp]
\centering
\includegraphics[width = 0.4 \textwidth]{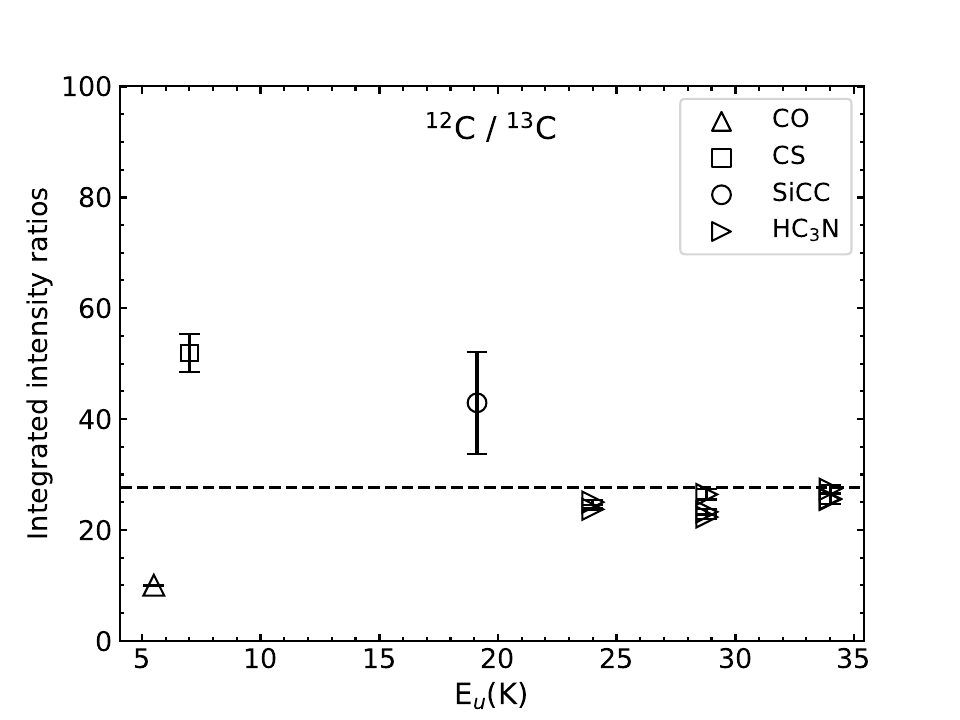}
\caption{{$^{12}$C$/^{13}$C isotopic ratios  derived from CO, CS, SiCC, and HC$_{3}$N. The ordinate represents the line intensity ratio derived from two distinct isotopologues. The abscissa denotes the upper-level energy of the transitions.}
\label{Fig:isotopic ratio CO}}
\end{figure}

Carbon isotopic ratios prove particularly sensitive to stellar processing. 
The third dredge-up typically enhances surface $^{12}${C}/$^{13}${C} ratios, while cool bottom processing (CBP) in low-mass red giants ($M < 2\,M_\odot$) counteracts this trend through extra mixing \citep{1995ApJ...453L..41C,1998A&A...332..204C}. 
We derived the $^{12}${C}/$^{13}${C} ratio through five $^{13}$C-bearing species ($^{13}$CO, $^{13}$CS, Si$^{13}$CC, H$^{13}$CCCN, HC$^{13}$CCN), employing line intensity ratios as shown in Figure.~\ref{Fig:isotopic ratio CO}. 
Due to optical thickness in CO and CS lines, the derived $^{12}${CO}/$^{13}${CO} and $^{12}${CS}/$^{13}${CS} ratios represent lower limits. 
The derived $^{12}${C}/$^{13}${C} ratio is consistent with that reported by
\citet{2000A&AS..142..181C}, while showing substantial depletion relative to solar ($\sim$89) and Orion Bar ($\sim$67) values \citep{1998ApJ...494L.107K}. 
This depletion supports CBP-induced mixing in IRC+10216's progenitor. 

\begin{table*}[!hbt]
\caption{Isotopic ratios.}\label{Tab:isotopic}             
\normalsize
\centering                                     
\begin{tabular}{cccccccc}        
\hline\hline                      
     Isotopic ratio     & \multicolumn{3}{c}{IRC+10216}      & CIT\,6 & CRL 3068$^{d}$   &Solar$^{e}$  \\
\cline{2-4} 
                 & mean  & Cernicharo$^{a}$  &      &      &   &   \\
\hline                                                                                
$^{12}$C$/$$^{13}$C     &                               27.6            &               45              &               & $14.3^{b}$      &       9.2     &       89      &       \\
$^{28}$Si$/$$^{29}$Si   &               19.6            &       15.4    &               &   $4.3^{b}$     &       11.5    &               19.6    &       \\
$^{28}$Si$/$$^{30}$Si   &               28.2            &         20.3  &               & $8.8^{c}$       &       28.8    &       29.9    &       \\
$^{29}$Si$/$$^{30}$Si   &               1.5             &       1.5     &               &       $1.0^{c}$       &       2.5     &       1.5     &       \\
\hline
\end{tabular}
\tablefoot{\\$^{a}$ From \citet{2000A&AS..142..181C}. \\$^{b}$ From \citet{2023PASJ...75..853Y}. \\$^{c}$ From \citet{2009ApJ...691.1660Z}. \\$^{d}$ From \citet{2009ApJ...700.1262Z}. \\$^{e}$ From \citet{2003ApJ...591.1220L}.}
\end{table*}

The  $^{17}\mathrm{O}/^{18}\mathrm{O}$ ratio in AGB
envelopes can be significantly enhanced through hot bottom burning (HBB) nucleosynthesis. This process occurs when the base of the convective envelope reaches temperatures exceeding $10^7\,\mathrm{K}$, which enables proton-capture reactions that preferentially produce $^{17}\mathrm{O}$ via $^{16}\mathrm{O}(p,\gamma)^{17}\mathrm{F}(\beta^+\nu)^{17}\mathrm{O}$ while depleting $^{18}\mathrm{O}$ through $^{18}\mathrm{O}(p,\alpha)^{15}\mathrm{N}$.
Using the integrated intensity ratio of $\mathrm{C}^{17}\mathrm{O}$ $J = 1 \to 0$ to $\mathrm{C}^{18}\mathrm{O}$ $J = 1 \to 0$ rotational transitions, combined with the isotopic abundance calculation framework established by \citet{2017A&A...600A..71D}, we derive $^{17}\mathrm{O}/^{18}\mathrm{O} = 1.51 \pm 0.16$ for  IRC\,+10216. This measurement shows reasonable consistency with the value of $1.19 \pm 0.05$ reported in \citet{2017A&A...600A..71D}, yet demonstrates a striking contrast to the solar value of $0.19 \pm 0.02$. The observed $^{17}\mathrm{O}/^{18}\mathrm{O}$ enhancement  provides direct observational evidence for the operation of HBB in this object. 

Silicon isotopic ratios remain largely unaffected by AGB nucleosynthesis. 
Our detection of four silicon isotopologues ($^{29}$SiS, $^{30}$SiS, $^{29}$SiCC, $^{30}$SiCC; Figure.~\ref{Fig:isotopic ratio 28Si29Si}) reveals silicon isotopic ratios consistent with solar system abundances. 
The low $^{29,30}$Si abundances minimize optical depth effects, ensuring robust determination of primordial silicon isotope ratios inherited from the interstellar medium.

\begin{figure*}[!htbp]
\centering
\includegraphics[width = 1 \textwidth]{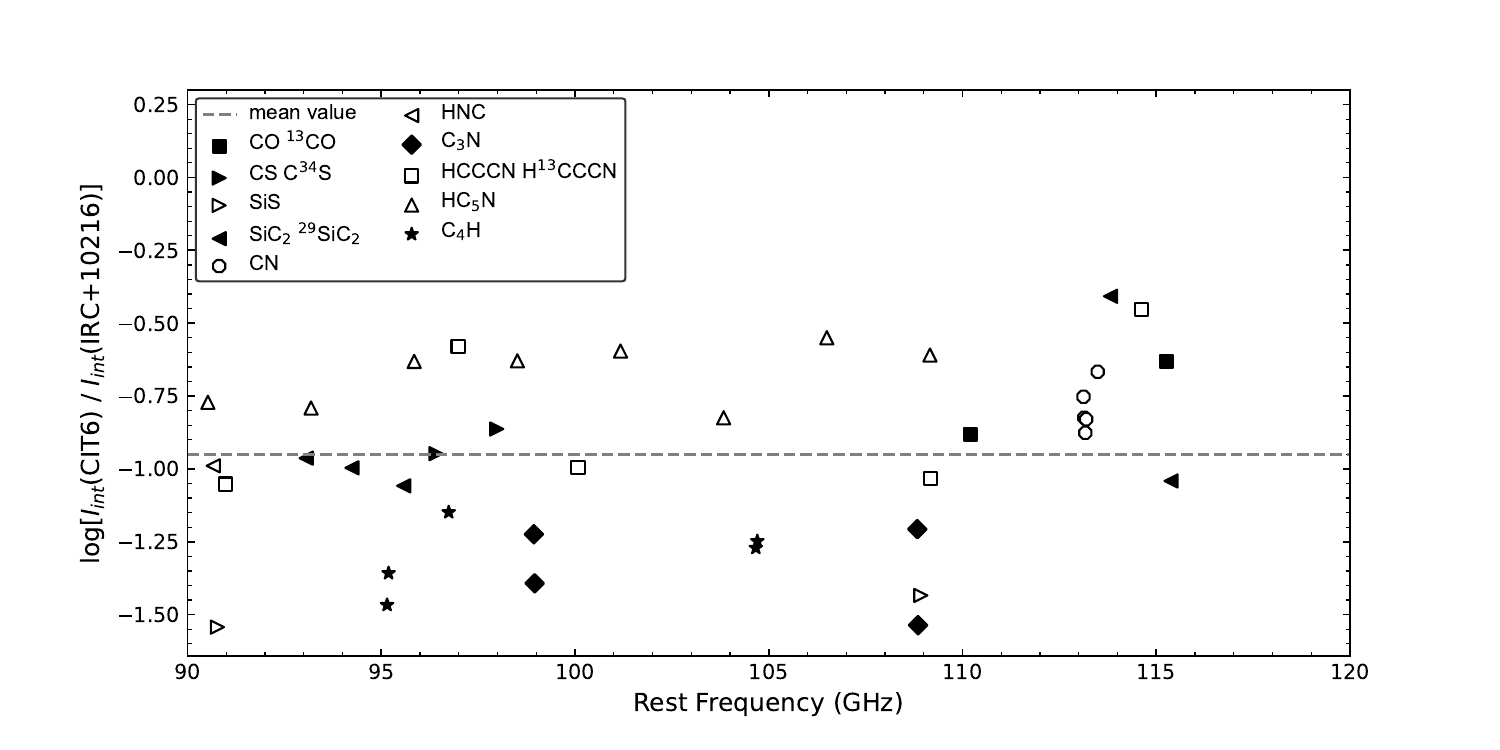}
\caption{{Integrated intensity ratios of the $\lambda=3$\,mm  molecular lines between IRC+10216 and CIT\,6. The dashed line represents the mean value.}\label{Fig:compare with 10216 and CIT6}}
\end{figure*}

\subsection{Comparison with CIT\,6}\label{compare_cit6}
CIT\,6 is the second most infrared-bright carbon-rich star, immediately following IRC+10216. 
To investigate whether IRC+10216 represents a typical carbon-rich AGB star, we compared its 3\,mm spectrum with that observed for CIT\,6, which was acquired under identical instrumental setup \citep{2023PASJ...75..853Y}. 
Every emission line detected in CIT 6 was also observed in IRC+10216, which suggests a  similarity in their underlying physical and chemical processes. 
The beam-dilution corrected integrated line intensity ratios between the two sources, $I_{\rm int}$(CIT\,6)/$I_{\rm int}$(IRC+10216), are presented in 
Figure.~\ref{Fig:compare with 10216 and CIT6}, which can provide insights into differences in molecular abundances, excitation conditions, or the physical structure of their circumstellar environments. 
The mean value of log[$I_{\rm int}$(CIT\,6)/$I_{\rm int}$(IRC+10216)] is about -0.95, with all emission lines displaying values distributed around this mean.
The ratio of distances between CIT\,6 and IRC+10216 is approximately 3. 
Given the inverse-square relationship between intensity and distance, the observed differences in line intensities can be attributed to their varying distances, with IRC+10216 being located at a closer proximity, rather than intrinsic source properties.

Although the principal variations in line intensities are mainly attributed to the difference in distance between these sources, a careful scrutiny of Figure.~\ref{Fig:compare with 10216 and CIT6} discloses subtle disparities between the two. 
The intensity ratios of the SiS, C$_{4}$H, and C$_{3}$N lines are beneath the mean value, while those of SiC$_{2}$, CN, and HC$_{5}$N are higher. 
These discrepancies could serve as tracers of CSE evolution, specifically probing the chemical processing occurring within these environments. SiC$_{2}$, as a daughter molecule of SiS \citep{2023FrASS..1015642F}, could experience an increasing SiC$_{2}$/SiS abundance ratio over the course of its evolution.

Cyanopolyynes are hypothesized to be synthesized through a bottom-up process similar to that of  polyynes in circumstellar environments. 
Consider the following potential reactions for the formation of $\mathrm{HC_{5}N}$:
\begin{equation}\label{HC3N_1}
\mathrm{C_{2}H + HC_{3}N} \longrightarrow \mathrm{HC_{5}N + H},
\end{equation}
\begin{equation}\label{HC3N_2}
\mathrm{CN + C_{4}H_{2}} \longrightarrow \mathrm{HC_{5}N + H},
\end{equation}
and
\begin{equation}\label{HC3N_3}
\mathrm{C_{3}N + C_{2}H_{2}} \longrightarrow \mathrm{HC_{5}N + H}.
\end{equation}
The activation barrier of the reaction described in Equation~\ref{HC3N_1} is approximately 800\,K. 
As a result, this reaction has a relatively low rate in the environment of IRC+10216 \citep{2017A&A...601A...4A}. 
Although chemical equilibrium calculations indicate that both Equations~\ref{HC3N_2} and \ref{HC3N_3} proceed rapidly at low temperatures and are considered equally significant for the formation of $\mathrm{HC_{5}N}$, a comparison between theoretical calculations and observational data reveals that the reaction in Equation~\ref{HC3N_3} predominantly occurs in the inner regions of the CSE, where the abundance of $\mathrm{C_{3}N}$ is relatively low. 
Consequently, the formation of $\mathrm{HC_{5}N}$ is mainly attributed to the reaction described in Equation~\ref{HC3N_2}. 
Moreover, the formation of $\mathrm{HC_{3}N}$ takes place closer to the inner regions of the CSE and at an earlier stage compared to $\mathrm{HC_{5}N}$. 
As depicted in Figure.~\ref{Fig:compare with 10216 and CIT6}, we observe that the abundance of $\mathrm{HC_{5}N}$ in CIT\,6 is obviously higher than that in IRC+10216, while the abundances of $\mathrm{HC_{3}N}$ in both objects are nearly identical.

Therefore, we conclude that the higher abundances of SiC$_{2}$ and HC$_{5}$N in CIT\,6 imply a more advanced evolutionary stage of CIT\,6 relative to IRC+10216. 
This finding is in accordance with the results of previous studies \citep{2009ApJ...691.1660Z,2023PASJ...75..853Y}. 
Additionally, we note that the line intensities of C$_{4}$H and C$_3$N in IRC+10216 are anomalously high compared to those in CIT\,6. 
As proposed by \citet{2017A&A...601A...4A}, HC$_{3}$N is considered a first-generation daughter molecule in the C$_{2}$H$_{2}$ reaction chain, while C$_{4}$H and C$_{3}$N are second-generation daughter molecules. 
Consequently, HC$_{3}$N is expected to form earlier. 
Nevertheless, the results presented in Figure.~\ref{Fig:compare with 10216 and CIT6} seem to show the opposite trend. 
This discrepancy indicates that other factors likely influence the abundances of C$_{4}$H and C$_{3}$N. 
Given that the formation of both C$_{4}$H and C$_{3}$N is driven by UV photons, we infer that the UV-induced reactions are more pronounced in IRC+10216. 
Some studies \citep[e.g.,][]{2015A&A...575A..91C,2022ApJ...941...90S} have suggested the presence of a companion star within IRC+10216. 
The UV radiation from the hot companion star would enhance the overall UV radiation field in the envelope, which would potentially explain the abundances patterns of C$_{4}$H and HC$_{3}$N. 
However, this conclusion remains tentative and requires further observational confirmation in the future.

\section{Summary}\label{conclusion}
This paper presents a high-sensitivity spectral line survey of the carbon-rich AGB star IRC+10216, carried out within the frequency range of 90--116 GHz. 
In total, 214 emission lines were detected, among which 28 are newly reported. 
We identified 211 spectral lines corresponding to 43 distinct molecules. However, four "U" lines could not be assigned to any known molecule. 
With the exception of SiS and NaCl, all detected molecules are carbon-bearing species. 
C$_{6}$H exhibits the largest number of emission lines, with 32 lines detected, followed by  C$_{4}$H with 27 lines. 
For molecules with optically thin emission lines, we employed the rotational diagram method to calculate their excitation temperatures, column densities, and fractional abundances. 
Based on the line intensity ratios, we determined the isotopic ratios of carbon, oxygen, and silicon elements. 
Our findings indicate that the $^{12}$C$/^{13}$C ratio in IRC+10216 is characteristic of evolved stars and is significantly lower than the solar value.

Through a comparison of the overlapping spectra between our observations and those of \citetalias{2024ApJS..271...45T}, we determined a high degree of consistency in the detection of stronger lines. 
Specifically, every line identified by \citetalias{2024ApJS..271...45T} was also detected in our study. 
Moreover, our observations revealed a multitude of weaker lines that evaded detection in the work of \citetalias{2024ApJS..271...45T}, underscoring the enhanced sensitivity of our observational setup. 
Regarding the line intensity ratios between the two observational datasets, they are found to be close to unity. 
The moderate deviations from unity can be ascribed to variations in the spatial extents of different molecular species. 
Our high-sensitivity spectra establish a robust spectral reference framework for comparative analysis of ALMA observations toward carbon-rich evolved stars.

Our systematic comparison of IRC+10216 and CIT\,6 reveals critical insights into chemical evolution during the AGB phase. 
While both sources exhibit remarkably similar molecular inventories, with line intensity differences primarily attributable to their different distances, we have identified crucial evolutionary indicators. 
The relatively weaker SiC$_{2}$ and HC$_{5}$N line intensities in IRC+10216 suggest advanced evolutionary processing. 
Conversely, the enhanced C$_{4}$H and C$_{3}$N intensities in IRC+10216 defy standard chemical evolution models. 
Their photodissociation-driven formation pathways implicate a stronger UV radiation field potentially generated by a binary system.

\begin{acknowledgements}
\label{acknowledgments}
We sincerely express our gratitude to an anonymous reviewer for his/her valuable comments, which have significantly enhanced the quality of this paper.
The financial supports of this work are from 
the National Natural Science Foundation of China (NSFC, No. 12473027,  12333005, 11973099, and 12463006),
the Guangdong Basic and Applied Basic Research Funding (No.\,2024A1515010798),
the Greater Bay Area Branch of the National Astronomical Data Center (No.\,2024B1212080003),
and the science research grants from the China Manned Space Project (NO. CMS-CSST-2021-A09, CMS-CSST-2021-A10, etc).
Y.Z. thanks the Xinjiang Tianchi Talent Program (2023). 
X.H.L. acknowledges support from the Natural Science Foundation of Xinjiang Uygur Autonomous Region (No. 2024D01E37) and the National Science Foundation of China (12473025).
\end{acknowledgements}

\bibliography{ref}

\begin{thebibliography}{75}
\expandafter\ifx\csname natexlab\endcsname\relax\def\natexlab#1{#1}\fi

\bibitem[{{Ag{\'u}ndez} {et~al.}(2014){Ag{\'u}ndez}, {Cernicharo}, \& {Gu{\'e}lin}}]{2014A&A...570A..45A}
{Ag{\'u}ndez}, M., {Cernicharo}, J., \& {Gu{\'e}lin}, M. 2014, \aap, 570, A45

\bibitem[{{Ag{\'u}ndez} {et~al.}(2010){Ag{\'u}ndez}, {Cernicharo}, {Gu{\'e}lin}, {Kahane}, {Roueff}, {K{\l}os}, {Aoiz}, {Lique}, {Marcelino}, {Goicoechea}, {Gonz{\'a}lez Garc{\'\i}a}, {Gottlieb}, {McCarthy}, \& {Thaddeus}}]{2010A&A...517L...2A}
{Ag{\'u}ndez}, M., {Cernicharo}, J., {Gu{\'e}lin}, M., {et~al.} 2010, \aap, 517, L2

\bibitem[{{Ag{\'u}ndez} {et~al.}(2017){Ag{\'u}ndez}, {Cernicharo}, {Quintana-Lacaci}, {Castro-Carrizo}, {Velilla Prieto}, {Marcelino}, {Gu{\'e}lin}, {Joblin}, {Mart{\'\i}n-Gago}, {Gottlieb}, {Patel}, \& {McCarthy}}]{2017A&A...601A...4A}
{Ag{\'u}ndez}, M., {Cernicharo}, J., {Quintana-Lacaci}, G., {et~al.} 2017, \aap, 601, A4

\bibitem[{{Ag{\'u}ndez} {et~al.}(2015){Ag{\'u}ndez}, {Cernicharo}, {Quintana-Lacaci}, {Velilla Prieto}, {Castro-Carrizo}, {Marcelino}, \& {Gu{\'e}lin}}]{2015ApJ...814..143A}
{Ag{\'u}ndez}, M., {Cernicharo}, J., {Quintana-Lacaci}, G., {et~al.} 2015, \apj, 814, 143

\bibitem[{{Ag{\'u}ndez} {et~al.}(2012){Ag{\'u}ndez}, {Fonfr{\'\i}a}, {Cernicharo}, {Kahane}, {Daniel}, \& {Gu{\'e}lin}}]{2012A&A...543A..48A}
{Ag{\'u}ndez}, M., {Fonfr{\'\i}a}, J.~P., {Cernicharo}, J., {et~al.} 2012, \aap, 543, A48

\bibitem[{{Bachiller} {et~al.}(1997){Bachiller}, {Fuente}, {Bujarrabal}, {Colomer}, {Loup}, {Omont}, \& {de Jong}}]{1997A&A...319..235B}
{Bachiller}, R., {Fuente}, A., {Bujarrabal}, V., {et~al.} 1997, \aap, 319, 235

\bibitem[{{Becklin} {et~al.}(1969){Becklin}, {Frogel}, {Hyland}, {Kristian}, \& {Neugebauer}}]{1969ApJ...158L.133B}
{Becklin}, E.~E., {Frogel}, J.~A., {Hyland}, A.~R., {Kristian}, J., \& {Neugebauer}, G. 1969, \apjl, 158, L133

\bibitem[{{Cernicharo} {et~al.}(2011){Cernicharo}, {Ag{\'u}ndez}, \& {Gu{\'e}lin}}]{2011IAUS..280..237C}
{Cernicharo}, J., {Ag{\'u}ndez}, M., \& {Gu{\'e}lin}, M. 2011, in The Molecular Universe, ed. J.~{Cernicharo} \& R.~{Bachiller}, Vol. 280, 237--248

\bibitem[{{Cernicharo} {et~al.}(2019){Cernicharo}, {Cabezas}, {Pardo}, {Ag{\'u}ndez}, {Berm{\'u}dez}, {Velilla-Prieto}, {Tercero}, {L{\'o}pez-P{\'e}rez}, {Gallego}, {Fonfr{\'\i}a}, {Quintana-Lacaci}, {Gu{\'e}lin}, \& {Endo}}]{2019A&A...630L...2C}
{Cernicharo}, J., {Cabezas}, C., {Pardo}, J.~R., {et~al.} 2019, \aap, 630, L2

\bibitem[{{Cernicharo} {et~al.}(1991{\natexlab{a}}){Cernicharo}, {Gottlieb}, {Guelin}, {Killian}, {Paubert}, {Thaddeus}, \& {Vrtilek}}]{1991ApJ...368L..39C}
{Cernicharo}, J., {Gottlieb}, C.~A., {Guelin}, M., {et~al.} 1991{\natexlab{a}}, \apjl, 368, L39

\bibitem[{{Cernicharo} {et~al.}(1991{\natexlab{b}}){Cernicharo}, {Gottlieb}, {Guelin}, {Killian}, {Thaddeus}, \& {Vrtilek}}]{1991ApJ...368L..43C}
{Cernicharo}, J., {Gottlieb}, C.~A., {Guelin}, M., {et~al.} 1991{\natexlab{b}}, \apjl, 368, L43

\bibitem[{{Cernicharo} {et~al.}(2018){Cernicharo}, {Gu{\'e}lin}, {Ag{\'u}ndez}, {Pardo}, {Massalkhi}, {Fonfr{\'\i}a}, {Velilla Prieto}, {Quintana-Lacaci}, {Marcelino}, {Marka}, {Navarro}, \& {Kramer}}]{2018A&A...618A...4C}
{Cernicharo}, J., {Gu{\'e}lin}, M., {Ag{\'u}ndez}, M., {et~al.} 2018, \aap, 618, A4

\bibitem[{{Cernicharo} {et~al.}(2000){Cernicharo}, {Gu{\'e}lin}, \& {Kahane}}]{2000A&AS..142..181C}
{Cernicharo}, J., {Gu{\'e}lin}, M., \& {Kahane}, C. 2000, \aaps, 142, 181

\bibitem[{{Cernicharo} {et~al.}(1987){Cernicharo}, {Guelin}, {Menten}, \& {Walmsley}}]{1987A&A...181L...1C}
{Cernicharo}, J., {Guelin}, M., {Menten}, K.~M., \& {Walmsley}, C.~M. 1987, \aap, 181, L1

\bibitem[{{Cernicharo} {et~al.}(1986{\natexlab{a}}){Cernicharo}, {Kahane}, {Gomez-Gonzalez}, \& {Guelin}}]{1986A&A...167L...9C}
{Cernicharo}, J., {Kahane}, C., {Gomez-Gonzalez}, J., \& {Guelin}, M. 1986{\natexlab{a}}, \aap, 167, L9

\bibitem[{{Cernicharo} {et~al.}(1986{\natexlab{b}}){Cernicharo}, {Kahane}, {Gomez-Gonzalez}, \& {Guelin}}]{1986A&A...164L...1C}
{Cernicharo}, J., {Kahane}, C., {Gomez-Gonzalez}, J., \& {Guelin}, M. 1986{\natexlab{b}}, \aap, 164, L1

\bibitem[{{Cernicharo} {et~al.}(2015){Cernicharo}, {Marcelino}, {Ag{\'u}ndez}, \& {Gu{\'e}lin}}]{2015A&A...575A..91C}
{Cernicharo}, J., {Marcelino}, N., {Ag{\'u}ndez}, M., \& {Gu{\'e}lin}, M. 2015, \aap, 575, A91

\bibitem[{{Charbonnel}(1995)}]{1995ApJ...453L..41C}
{Charbonnel}, C. 1995, \apjl, 453, L41

\bibitem[{{Charbonnel} {et~al.}(1998){Charbonnel}, {Brown}, \& {Wallerstein}}]{1998A&A...332..204C}
{Charbonnel}, C., {Brown}, J.~A., \& {Wallerstein}, G. 1998, \aap, 332, 204

\bibitem[{{Chau} {et~al.}(2012){Chau}, {Zhang}, {Nakashima}, {Deguchi}, \& {Kwok}}]{2012ApJ...760...66C}
{Chau}, W., {Zhang}, Y., {Nakashima}, J.-i., {Deguchi}, S., \& {Kwok}, S. 2012, \apj, 760, 66

\bibitem[{{Cherchneff} \& {Glassgold}(1993)}]{1993ApJ...419L..41C}
{Cherchneff}, I. \& {Glassgold}, A.~E. 1993, \apjl, 419, L41

\bibitem[{{Dayal} \& {Bieging}(1995)}]{1995ApJ...439..996D}
{Dayal}, A. \& {Bieging}, J.~H. 1995, \apj, 439, 996

\bibitem[{{De Nutte} {et~al.}(2017){De Nutte}, {Decin}, {Olofsson}, {Lombaert}, {de Koter}, {Karakas}, {Milam}, {Ramstedt}, {Stancliffe}, {Homan}, \& {Van de Sande}}]{2017A&A...600A..71D}
{De Nutte}, R., {Decin}, L., {Olofsson}, H., {et~al.} 2017, \aap, 600, A71

\bibitem[{{Decin}(2021)}]{2021ARA&A..59..337D}
{Decin}, L. 2021, \araa, 59, 337

\bibitem[{{Decin} {et~al.}(2011){Decin}, {Royer}, {Cox}, {Vandenbussche}, {Ottensamer}, {Blommaert}, {Groenewegen}, {Barlow}, {Lim}, {Kerschbaum}, {Posch}, \& {Waelkens}}]{2011A&A...534A...1D}
{Decin}, L., {Royer}, P., {Cox}, N.~L.~J., {et~al.} 2011, \aap, 534, A1

\bibitem[{{Demyk} {et~al.}(2007){Demyk}, {M{\"a}der}, {Tercero}, {Cernicharo}, {Demaison}, {Margul{\`e}s}, {Wegner}, {Keipert}, \& {Sheng}}]{2007A&A...466..255D}
{Demyk}, K., {M{\"a}der}, H., {Tercero}, B., {et~al.} 2007, \aap, 466, 255

\bibitem[{{Feng} {et~al.}(2023){Feng}, {Li}, {Millar}, {Szczerba}, {Wang}, {Quan}, {Qin}, {Fang}, {Tuo}, {Miao}, {Ma}, {Xu}, {Sun}, {Jiang}, {Chang}, {Yang}, {Hou}, {Li}, \& {Zhang}}]{2023FrASS..1015642F}
{Feng}, Y., {Li}, X., {Millar}, T.~J., {et~al.} 2023, Frontiers in Astronomy and Space Sciences, 10, 1215642

\bibitem[{{Fonfr{\'\i}a} {et~al.}(2022){Fonfr{\'\i}a}, {DeWitt}, {Montiel}, {Cernicharo}, \& {Richter}}]{2022ApJ...927L..33F}
{Fonfr{\'\i}a}, J.~P., {DeWitt}, C.~N., {Montiel}, E.~J., {Cernicharo}, J., \& {Richter}, M.~J. 2022, \apjl, 927, L33

\bibitem[{{Gensheimer} {et~al.}(1992){Gensheimer}, {Likkel}, \& {Snyder}}]{1992ApJ...388L..31G}
{Gensheimer}, P.~D., {Likkel}, L., \& {Snyder}, L.~E. 1992, \apjl, 388, L31

\bibitem[{{Gensheimer} {et~al.}(1995){Gensheimer}, {Likkel}, \& {Snyder}}]{1995ApJ...439..445G}
{Gensheimer}, P.~D., {Likkel}, L., \& {Snyder}, L.~E. 1995, \apj, 439, 445

\bibitem[{{Gong} {et~al.}(2015){Gong}, {Henkel}, {Spezzano}, {Thorwirth}, {Menten}, {Wyrowski}, {Mao}, \& {Klein}}]{2015A&A...574A..56G}
{Gong}, Y., {Henkel}, C., {Spezzano}, S., {et~al.} 2015, \aap, 574, A56

\bibitem[{{Groenewegen} {et~al.}(1998){Groenewegen}, {Whitelock}, {Smith}, \& {Kerschbaum}}]{1998MNRAS.293...18G}
{Groenewegen}, M.~A.~T., {Whitelock}, P.~A., {Smith}, C.~H., \& {Kerschbaum}, F. 1998, \mnras, 293, 18

\bibitem[{{Guelin} {et~al.}(1987){Guelin}, {Cernicharo}, {Kahane}, {Gomez-Gonzalez}, \& {Walmsley}}]{1987A&A...175L...5G}
{Guelin}, M., {Cernicharo}, J., {Kahane}, C., {Gomez-Gonzalez}, J., \& {Walmsley}, C.~M. 1987, \aap, 175, L5

\bibitem[{{Guelin} {et~al.}(1995){Guelin}, {Forestini}, {Valiron}, {Ziurys}, {Anderson}, {Cernicharo}, \& {Kahane}}]{1995A&A...297..183G}
{Guelin}, M., {Forestini}, M., {Valiron}, P., {et~al.} 1995, \aap, 297, 183

\bibitem[{{Guelin} {et~al.}(1998){Guelin}, {Neininger}, \& {Cernicharo}}]{1998A&A...335L...1G}
{Guelin}, M., {Neininger}, N., \& {Cernicharo}, J. 1998, \aap, 335, L1

\bibitem[{{He} {et~al.}(2008){He}, {Dinh-V-Trung}, {Kwok}, {M{\"u}ller}, {Zhang}, {Hasegawa}, {Peng}, \& {Huang}}]{2008ApJS..177..275H}
{He}, J.~H., {Dinh-V-Trung}, {Kwok}, S., {et~al.} 2008, \apjs, 177, 275

\bibitem[{{Herwig}(2005)}]{2005ARA&A..43..435H}
{Herwig}, F. 2005, \araa, 43, 435

\bibitem[{{Huggins} \& {Glassgold}(1982)}]{1982ApJ...252..201H}
{Huggins}, P.~J. \& {Glassgold}, A.~E. 1982, \apj, 252, 201

\bibitem[{{Johansson} {et~al.}(1984){Johansson}, {Andersson}, {Ellder}, {Friberg}, {Hjalmarson}, {Hoglund}, {Irvine}, {Olofsson}, \& {Rydbeck}}]{1984A&A...130..227J}
{Johansson}, L.~E.~B., {Andersson}, C., {Ellder}, J., {et~al.} 1984, \aap, 130, 227

\bibitem[{{Kawaguchi} {et~al.}(1995){Kawaguchi}, {Kasai}, {Ishikawa}, \& {Kaifu}}]{1995PASJ...47..853K}
{Kawaguchi}, K., {Kasai}, Y., {Ishikawa}, S.-I., \& {Kaifu}, N. 1995, \pasj, 47, 853

\bibitem[{{Keene} {et~al.}(1998){Keene}, {Schilke}, {Kooi}, {Lis}, {Mehringer}, \& {Phillips}}]{1998ApJ...494L.107K}
{Keene}, J., {Schilke}, P., {Kooi}, J., {et~al.} 1998, \apjl, 494, L107

\bibitem[{{Lodders}(2003)}]{2003ApJ...591.1220L}
{Lodders}, K. 2003, \apj, 591, 1220

\bibitem[{{Lovas}(2004)}]{2004JPCRD..33..177L}
{Lovas}, F.~J. 2004, Journal of Physical and Chemical Reference Data, 33, 177

\bibitem[{{Menut} {et~al.}(2007){Menut}, {Gendron}, {Schartmann}, {Tuthill}, {Lopez}, {Danchi}, {Wolf}, {Lagrange}, {Flament}, {Rouan}, {Cl{\'e}net}, \& {Berruyer}}]{2007MNRAS.376L...6M}
{Menut}, J.~L., {Gendron}, E., {Schartmann}, M., {et~al.} 2007, \mnras, 376, L6

\bibitem[{{Millar} {et~al.}(2024){Millar}, {Walsh}, {Van de Sande}, \& {Markwick}}]{2024A&A...682A.109M}
{Millar}, T.~J., {Walsh}, C., {Van de Sande}, M., \& {Markwick}, A.~J. 2024, \aap, 682, A109

\bibitem[{{M{\"u}ller} {et~al.}(2005){M{\"u}ller}, {Schl{\"o}der}, {Stutzki}, \& {Winnewisser}}]{2005JMoSt.742..215M}
{M{\"u}ller}, H. S.~P., {Schl{\"o}der}, F., {Stutzki}, J., \& {Winnewisser}, G. 2005, Journal of Molecular Structure, 742, 215

\bibitem[{{Olofsson}(1997)}]{1997IAUS..178..457O}
{Olofsson}, H. 1997, IAU Symposium, 178, 457

\bibitem[{{Pardo} {et~al.}(2022){Pardo}, {Cernicharo}, {Tercero}, {Cabezas}, {Berm{\'u}dez}, {Ag{\'u}ndez}, {Gallego}, {Tercero}, {G{\'o}mez-Garrido}, {de Vicente}, \& {L{\'o}pez-P{\'e}rez}}]{2022A&A...658A..39P}
{Pardo}, J.~R., {Cernicharo}, J., {Tercero}, B., {et~al.} 2022, \aap, 658, A39

\bibitem[{{Pardo} {et~al.}(2018){Pardo}, {Cernicharo}, {Velilla Prieto}, {Fonfr{\'\i}a}, {Ag{\'u}ndez}, {Quintana-Lacaci}, {Massalkhi}, {Tercero}, {G{\'o}mez-Garrido}, {de Vicente}, {Gu{\'e}lin}, {Kramer}, {Marka}, {Teyssier}, \& {Neufeld}}]{2018A&A...615L...4P}
{Pardo}, J.~R., {Cernicharo}, J., {Velilla Prieto}, L., {et~al.} 2018, \aap, 615, L4

\bibitem[{{Pickett} {et~al.}(1998){Pickett}, {Poynter}, {Cohen}, {Delitsky}, {Pearson}, \& {M{\"u}ller}}]{1998JQSRT..60..883P}
{Pickett}, H.~M., {Poynter}, R.~L., {Cohen}, E.~A., {et~al.} 1998, \jqsrt, 60, 883

\bibitem[{{Qiu} {et~al.}(2022){Qiu}, {Zhang}, {Zhang}, \& {Nakashima}}]{2022ApJS..259...56Q}
{Qiu}, J.-J., {Zhang}, Y., {Zhang}, J.-S., \& {Nakashima}, J.-i. 2022, \apjs, 259, 56

\bibitem[{{Quintana-Lacaci} {et~al.}(2017){Quintana-Lacaci}, {Cernicharo}, {Velilla Prieto}, {Ag{\'u}ndez}, {Castro-Carrizo}, {Fonfr{\'\i}a}, {Massalkhi}, \& {Pardo}}]{2017A&A...607L...5Q}
{Quintana-Lacaci}, G., {Cernicharo}, J., {Velilla Prieto}, L., {et~al.} 2017, \aap, 607, L5

\bibitem[{{Sheehan} {et~al.}(2008){Sheehan}, {Parsons}, {Zhou}, {Garand}, {Yen}, {Moore}, \& {Neumark}}]{2008JChPh.128c4301S}
{Sheehan}, S.~M., {Parsons}, B.~F., {Zhou}, J., {et~al.} 2008, \jcp, 128, 034301

\bibitem[{{Siebert} {et~al.}(2022){Siebert}, {Van de Sande}, {Millar}, \& {Remijan}}]{2022ApJ...941...90S}
{Siebert}, M.~A., {Van de Sande}, M., {Millar}, T.~J., \& {Remijan}, A.~J. 2022, \apj, 941, 90

\bibitem[{{Solomon} {et~al.}(1971){Solomon}, {Jefferts}, {Penzias}, \& {Wilson}}]{1971ApJ...163L..53S}
{Solomon}, P., {Jefferts}, K.~B., {Penzias}, A.~A., \& {Wilson}, R.~W. 1971, \apjl, 163, L53

\bibitem[{{Takano} {et~al.}(1992){Takano}, {Saito}, \& {Tsuji}}]{1992PASJ...44..469T}
{Takano}, S., {Saito}, S., \& {Tsuji}, T. 1992, \pasj, 44, 469

\bibitem[{{Tenenbaum} {et~al.}(2010){Tenenbaum}, {Dodd}, {Milam}, {Woolf}, \& {Ziurys}}]{2010ApJS..190..348T}
{Tenenbaum}, E.~D., {Dodd}, J.~L., {Milam}, S.~N., {Woolf}, N.~J., \& {Ziurys}, L.~M. 2010, \apjs, 190, 348

\bibitem[{{Thaddeus} {et~al.}(1985){Thaddeus}, {Gottlieb}, {Hjalmarson}, {Johansson}, {Irvine}, {Friberg}, \& {Linke}}]{1985ApJ...294L..49T}
{Thaddeus}, P., {Gottlieb}, C.~A., {Hjalmarson}, A., {et~al.} 1985, \apjl, 294, L49

\bibitem[{{Treffers} \& {Cohen}(1974)}]{1974ApJ...188..545T}
{Treffers}, R. \& {Cohen}, M. 1974, \apj, 188, 545

\bibitem[{{Tuo} {et~al.}(2024){Tuo}, {Li}, {Sun}, {Millar}, {Zhang}, {Qiu}, {Quan}, {Esimbek}, {Zhou}, {Gao}, {Chang}, {Xiao}, {Feng}, {Miao}, {Ma}, {Szczerba}, \& {Fang}}]{2024ApJS..271...45T}
{Tuo}, J., {Li}, X., {Sun}, J., {et~al.} 2024, \apjs, 271, 45

\bibitem[{{Turner} {et~al.}(1994){Turner}, {Steimle}, \& {Meerts}}]{1994ApJ...426L..97T}
{Turner}, B.~E., {Steimle}, T.~C., \& {Meerts}, L. 1994, \apjl, 426, L97

\bibitem[{{Unnikrishnan} {et~al.}(2024){Unnikrishnan}, {De Beck}, {Nyman}, {Olofsson}, {Vlemmings}, {Tafoya}, {Maercker}, {Charnley}, {Cordiner}, {de Gregorio}, {Humphreys}, {Millar}, \& {Rawlings}}]{2024A&A...684A...4U}
{Unnikrishnan}, R., {De Beck}, E., {Nyman}, L.~{\r{A}}., {et~al.} 2024, \aap, 684, A4

\bibitem[{{Velilla-Prieto} {et~al.}(2019){Velilla-Prieto}, {Cernicharo}, {Ag{\'u}ndez}, {Fonfr{\'\i}a}, {Quintana-Lacaci}, {Marcelino}, \& {Castro-Carrizo}}]{2019A&A...629A.146V}
{Velilla-Prieto}, L., {Cernicharo}, J., {Ag{\'u}ndez}, M., {et~al.} 2019, \aap, 629, A146

\bibitem[{{Wilson} {et~al.}(1971){Wilson}, {Penzias}, {Jefferts}, {Kutner}, \& {Thaddeus}}]{1971ApJ...167L..97W}
{Wilson}, R.~W., {Penzias}, A.~A., {Jefferts}, K.~B., {Kutner}, M., \& {Thaddeus}, P. 1971, \apjl, 167, L97

\bibitem[{{Winters} {et~al.}(2002){Winters}, {Le Bertre}, {Nyman}, {Omont}, \& {Jeong}}]{2002A&A...388..609W}
{Winters}, J.~M., {Le Bertre}, T., {Nyman}, L.~{\r{A}}., {Omont}, A., \& {Jeong}, K.~S. 2002, \aap, 388, 609

\bibitem[{{Woods} {et~al.}(2003){Woods}, {Sch{\"o}ier}, {Nyman}, \& {Olofsson}}]{2003A&A...402..617W}
{Woods}, P.~M., {Sch{\"o}ier}, F.~L., {Nyman}, L.~{\r{A}}., \& {Olofsson}, H. 2003, \aap, 402, 617

\bibitem[{{Yang} {et~al.}(2023){Yang}, {Zhang}, {Qiu}, {Ao}, \& {Li}}]{2023PASJ...75..853Y}
{Yang}, K., {Zhang}, Y., {Qiu}, J., {Ao}, Y., \& {Li}, X. 2023, \pasj, 75, 853

\bibitem[{{Zhang} {et~al.}(2017){Zhang}, {Zhu}, {Li}, {Chen}, {Wang}, \& {Zhang}}]{2017A&A...606A..74Z}
{Zhang}, X.-Y., {Zhu}, Q.-F., {Li}, J., {et~al.} 2017, \aap, 606, A74

\bibitem[{{Zhang}(2020)}]{2020ApJ...898..151Z}
{Zhang}, Y. 2020, \apj, 898, 151

\bibitem[{{Zhang} {et~al.}(2020){Zhang}, {Chau}, {Nakashima}, \& {Kwok}}]{2020PASJ...72...46Z}
{Zhang}, Y., {Chau}, W., {Nakashima}, J.-i., \& {Kwok}, S. 2020, \pasj, 72, 46

\bibitem[{{Zhang} {et~al.}(2008){Zhang}, {Kwok}, \& {Dinh-V-Trung}}]{2008ApJ...678..328Z}
{Zhang}, Y., {Kwok}, S., \& {Dinh-V-Trung}. 2008, \apj, 678, 328

\bibitem[{{Zhang} {et~al.}(2009{\natexlab{a}}){Zhang}, {Kwok}, \& {Dinh-V-Trung}}]{2009ApJ...691.1660Z}
{Zhang}, Y., {Kwok}, S., \& {Dinh-V-Trung}. 2009{\natexlab{a}}, \apj, 691, 1660

\bibitem[{{Zhang} {et~al.}(2009{\natexlab{b}}){Zhang}, {Kwok}, \& {Nakashima}}]{2009ApJ...700.1262Z}
{Zhang}, Y., {Kwok}, S., \& {Nakashima}, J.-i. 2009{\natexlab{b}}, \apj, 700, 1262

\bibitem[{{Zhang} {et~al.}(2013){Zhang}, {Kwok}, {Nakashima}, {Chau}, \& {Dinh-V-Trung}}]{2013ApJ...773...71Z}
{Zhang}, Y., {Kwok}, S., {Nakashima}, J.-i., {Chau}, W., \& {Dinh-V-Trung}. 2013, \apj, 773, 71

\bibitem[{{Ziurys} {et~al.}(1995){Ziurys}, {Apponi}, {Guelin}, \& {Cernicharo}}]{1995ApJ...445L..47Z}
{Ziurys}, L.~M., {Apponi}, A.~J., {Guelin}, M., \& {Cernicharo}, J. 1995, \apjl, 445, L47

\end{thebibliography}
\bibliographystyle{aa}

\begin{appendix}
\onecolumn
\begin{landscape}
\section{Molecular lines detected in this survey}
\begin{longtable}{cccccccccc}
\caption{Lines detected in the spectral survey of IRC+10216.} \label{Tab:irclines}\\
\hline \hline
\begin{tabular}{c} Frequency\\(MHz) \end{tabular}     & Species      & \begin{tabular}{c} Transition \\(upper--lower) \end{tabular} & \begin{tabular}{c} RMS\\(mK)  \end{tabular}      & \begin{tabular}{c} $\int T_{\rm mb}dv$\\(mK km s$^{-1}$) \end{tabular} & \begin{tabular}{c}  $V_{\rm exp}$\\(km s$^{-1}$) \end{tabular} & \begin{tabular}{c} $T_{\rm mb}$ \\mK  \end{tabular}& \begin{tabular}{c} E$_{\rm u}/K$\\K  \end{tabular}& Notes \\
\hline \hline
\endfirsthead 
\caption{continued.}\\ 
\hline \hline
\begin{tabular}{c} Frequency\\(MHz) \end{tabular}     & Species      & \begin{tabular}{c} Transition \\(upper--lower) \end{tabular} & \begin{tabular}{c} rms\\(mK)  \end{tabular}      & \begin{tabular}{c} $\int T_{\rm mb}dv$\\(mK km s$^{-1}$) \end{tabular} & \begin{tabular}{c}  $V_{\rm exp}$\\(km s$^{-1}$) \end{tabular} & \begin{tabular}{c} $T_{\rm mb}$ \\mK  \end{tabular}& \begin{tabular}{c} E$_{u}/K$\\K  \end{tabular}& Notes \\
\hline \hline
\endhead 
\hline
\endfoot
115271.2        &       CO      &       $       J=1-0   $       &       23.6         &       215971.2$\pm$2397.1     &       13.33  $\pm$  0.81      &       9073.7         &       5.5     &               \\
110201.4        &       $^{13}$CO       &       $       J=1-0   $       &       5.8         &       21540.3$\pm$31.4        &       14.64  $\pm$  0.05      &       452.6         &       5.3     &               \\
112358.8        &       C$^{17}$O       &       $       J=1-0, F={3/2}-5/2      $       &       3.3         &       955.6$\pm$24.3  &       16.64  $\pm$  1.63      &       31.6         &       5.4     &       a,b     \\
112359.0        &       C$^{17}$O       &       $       J=1-0, F=7/2-5/2        $       &       \nodata &       \nodata &       \nodata &       \nodata &       5.4         &       a,b     \\
112360.0        &       C$^{17}$O       &       $       J=1-0, F=5/2-5/2        $       &       \nodata &       \nodata &       \nodata &       \nodata &       5.4         &       a,b     \\
109782.2        &       C$^{18}$O       &       $       J=1-0   $       &       6.9         &       590.1$\pm$43.2  &       14.25  $\pm$  2.82      &       14.7         &       5.3     &       b       \\
104711.4        &       $^{13}$C$^{18}$O        &       $J=1-0$ &       3.2     &       177.6$\pm$56.3  &       12.33  $\pm$  0.13    &       5.5     &       5.0     &       ?       \\
97271.0         &       CS      &       $       \nu=1,2-1       $       &       2.9         &       120.2$\pm$9.8   &       11.85  $\pm$  2.21      &       6.3         &       1837.4  &       c       \\
\end{longtable}

\tablefoot{
(a)--Unsolved hyperfine structures;
(b)--Blending with other transitions;
(c)--Marginal detection;
(N)--New detection in this source. 
(?)--Tentative identification.
}
\end{landscape}

\clearpage
\onecolumn
\begin{landscape}
\begin{longtable}{ccccccccccc}
\caption{Column densities and rotational temperatures of the molecules detected in IRC+10216.}
\label{Tab:irc_rd}\\
\hline\hline
& \multicolumn{4}{c}{This work} & & \multicolumn{4}{c}{Other study} &  \\
\cline{2-5} \cline{7-10}
&      &  \begin{tabular}{c} $T_{\rm ex}$ \\ (K) \end{tabular}      &  \begin{tabular}{c} $N$ \\ (cm$^{-2}$)  \end{tabular}     & f$_{X}$     &    & \begin{tabular}{c} $T_{\rm ex}$ \\ (K) \end{tabular}  & \begin{tabular}{c} $N$ \\ (cm$^{-2}$)  \end{tabular}   & f$_{X}$ & Ref.  \\
\hline\hline
\endfirsthead
\caption{continued.}\\
\hline\hline
& \multicolumn{4}{c}{This work} & & \multicolumn{4}{c}{Other study} &  \\
\cline{2-5} \cline{7-10}
      &      &  \begin{tabular}{c} $T_{\rm ex}$ \\ (K) \end{tabular}      &  \begin{tabular}{c} $N$ \\ (cm$^{-2}$)  \end{tabular}     & f$_{X}$    &    & \begin{tabular}{c} $T_{\rm ex}$ \\ (K) \end{tabular}  & \begin{tabular}{c} $N$ \\ (cm$^{-2}$)  \end{tabular}   & f$_{X}$ & Ref.  \\
\hline  
\endhead
\hline  
\endfoot
C$_{4}$H$(v=0)$ &               &       $       45.3\pm2.3      $       &       $       (1.9\pm0.3)\times 10^{15} $       &       $       3.1\times 10^{-6}       $       &               &       $       53.0\pm3.0      $       &       $       (8.1\pm1.1)\times 10^{15} $       &       $               $       &       a       &       \\
        &               &       $               $       &                               &                               &               &       $       4.9\pm0.6       $       &       $       (1.8\pm0.3)\times 10^{14} $       &       $               $       &       b       &       \\
        &               &       $               $       &                               &                               &               &       $       18.5\pm7.6      $       &       $       (2.4\pm0.2)\times 10^{15} $       &       $       1.1\times 10^{-6}       $       &       d       &       \\
C$_{4}$H$(^{2}\Pi_{1/2} v_{7}=1)$       &               &       $       52.8\pm3.4      $       &       \nodata &       \nodata &               &       $       53.0\pm4.0      $       &       $       (5.9\pm2.1)\times 10^{15} $       &       $               $       &       a       &       \\
        &               &       $               $       &                               &                               &               &       $       10.0\pm5.0      $       &       $       (2.2\pm0.4)\times 10^{13} $       &       $               $       &       b       &       \\
C$_{4}$H$(^{2}\Pi_{3/2} v_{7}=1)$       &               &       $       57.8\pm3.3      $       &       \nodata &       \nodata &               &       $       53.0\pm4.0      $       &       $       (5.9\pm2.1)\times 10^{15} $       &       $               $       &       a       &       \\
\hline
\end{longtable} 
\tablefoot{\\$^{a}$ From \citet{2008ApJS..177..275H}. \\$^{b}$ From \citet{2022A&A...658A..39P}. \\$^{c}$ From \citet{2017A&A...606A..74Z}. \\$^{d}$ From \citet{2015A&A...574A..56G}.}
\end{landscape}
\twocolumn

\clearpage
\begin{figure*}[!htbp]
\section{Spectra}
\centering
\includegraphics[width = 0.8\textwidth]{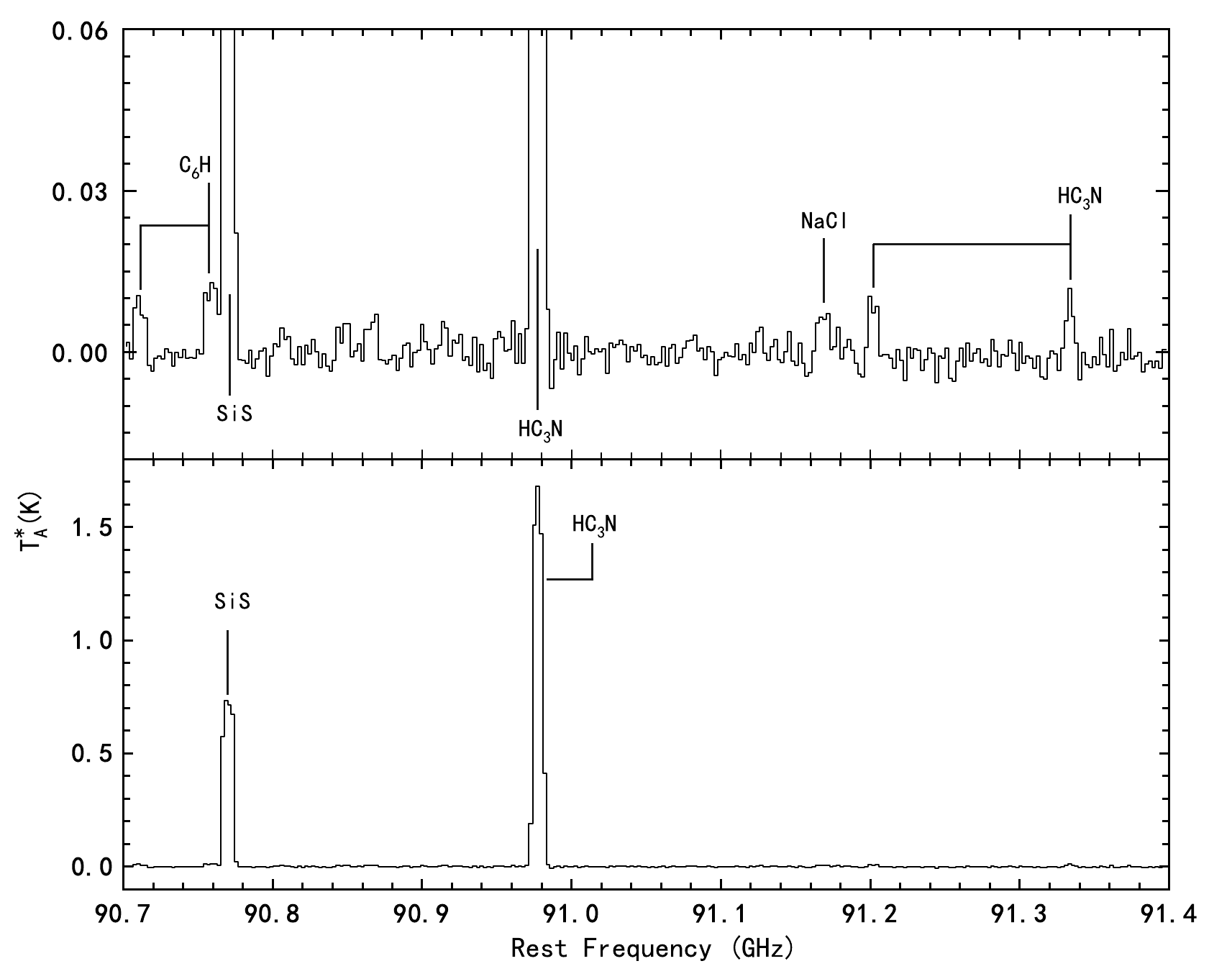}
\includegraphics[width = 0.8 \textwidth]{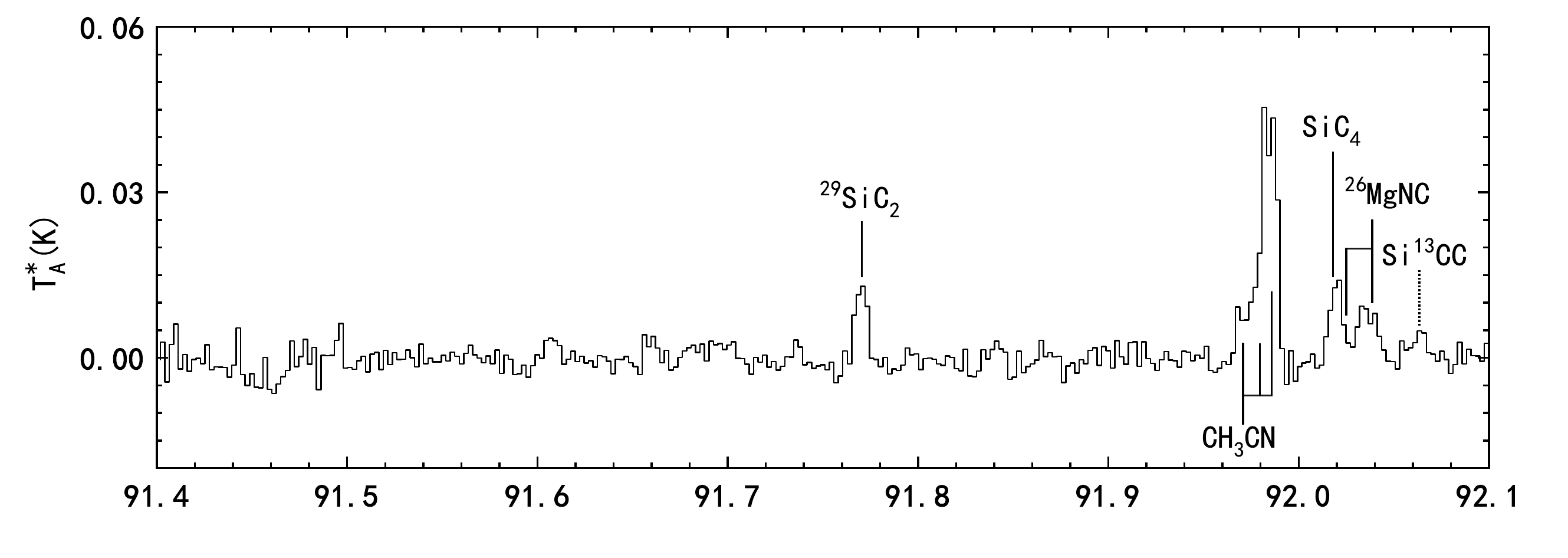}
\caption{{Same as Figure.~\ref{Fig:irc}, but for the frequency range 90.7--116\,GHz.}\label{Fig:irc_complete}}
\end{figure*}

\begin{figure*}[!htbp]
\centering
\includegraphics[width = 0.8 \textwidth]{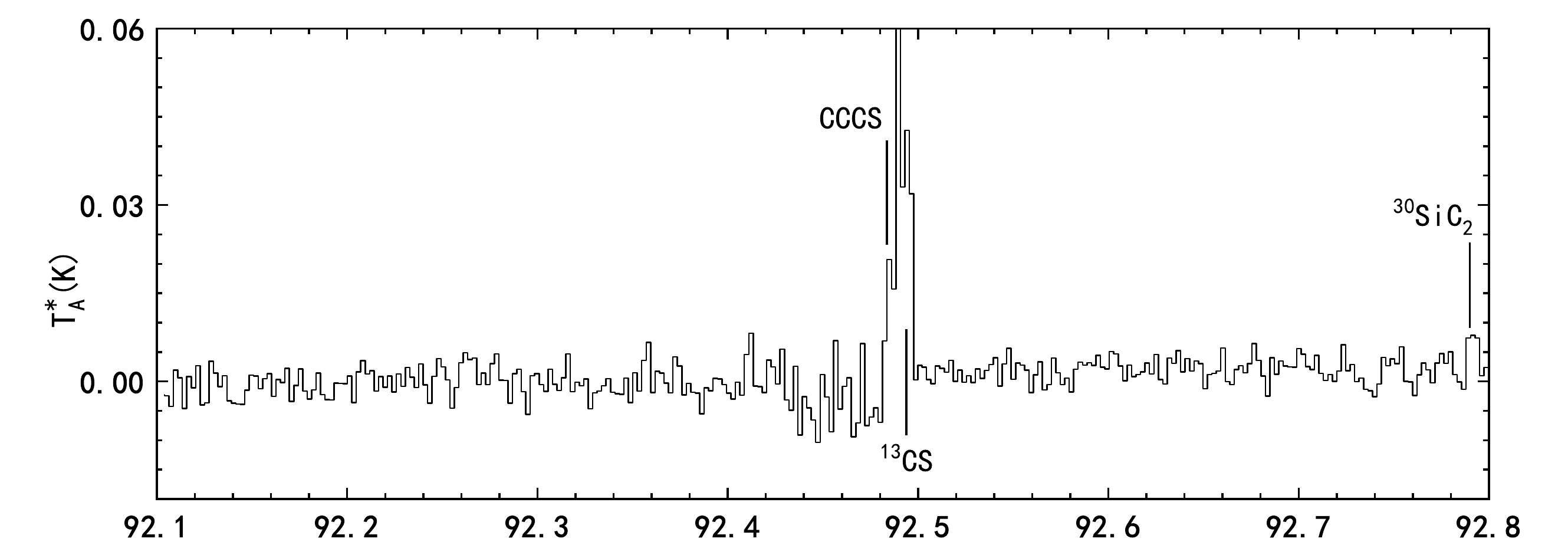}
\includegraphics[width = 0.8 \textwidth]{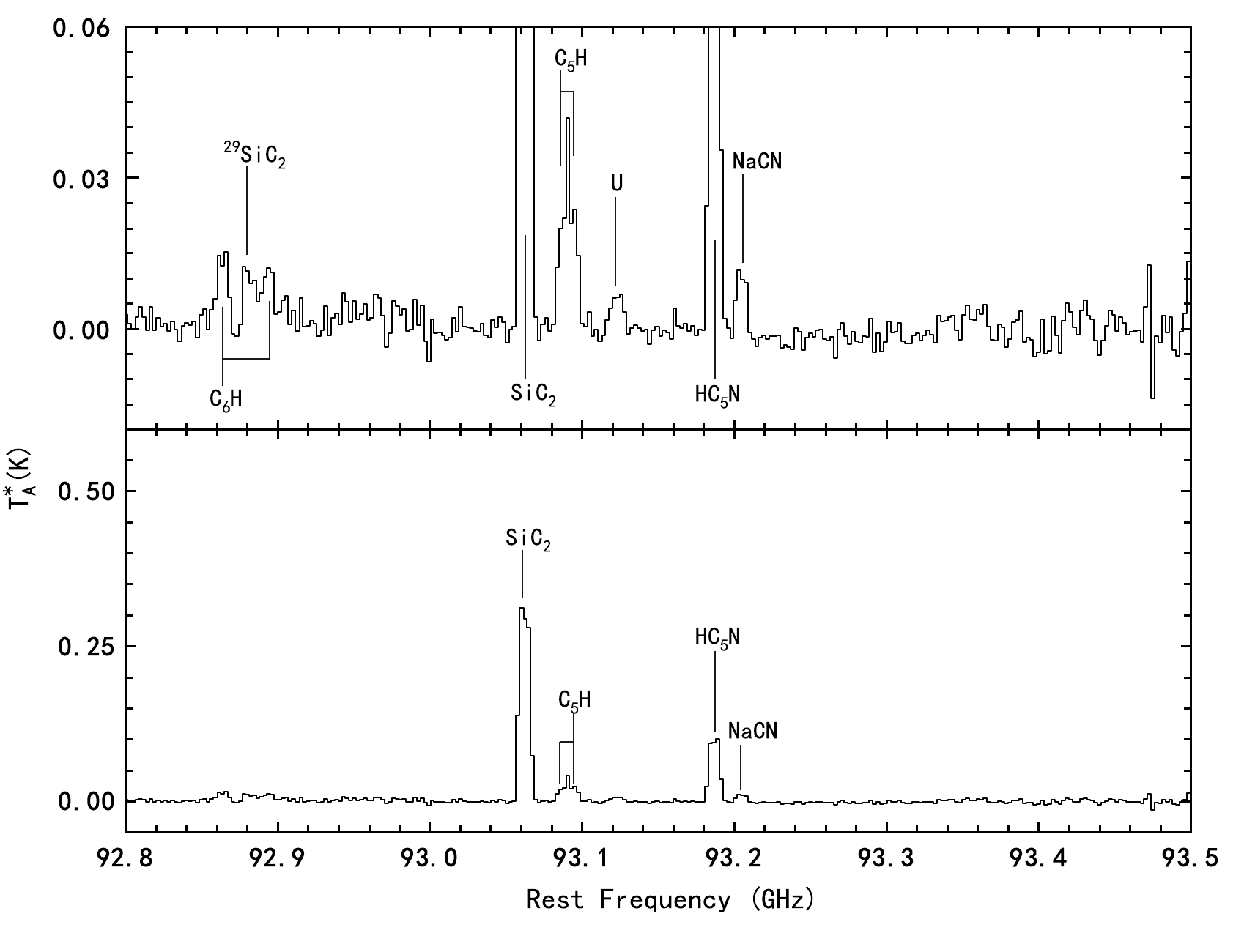}
\centerline{Figure.~\ref{Fig:irc_complete}. --- continued.}
\end{figure*}

\begin{figure*}[!htbp]
\centering
\includegraphics[width = 0.8 \textwidth]{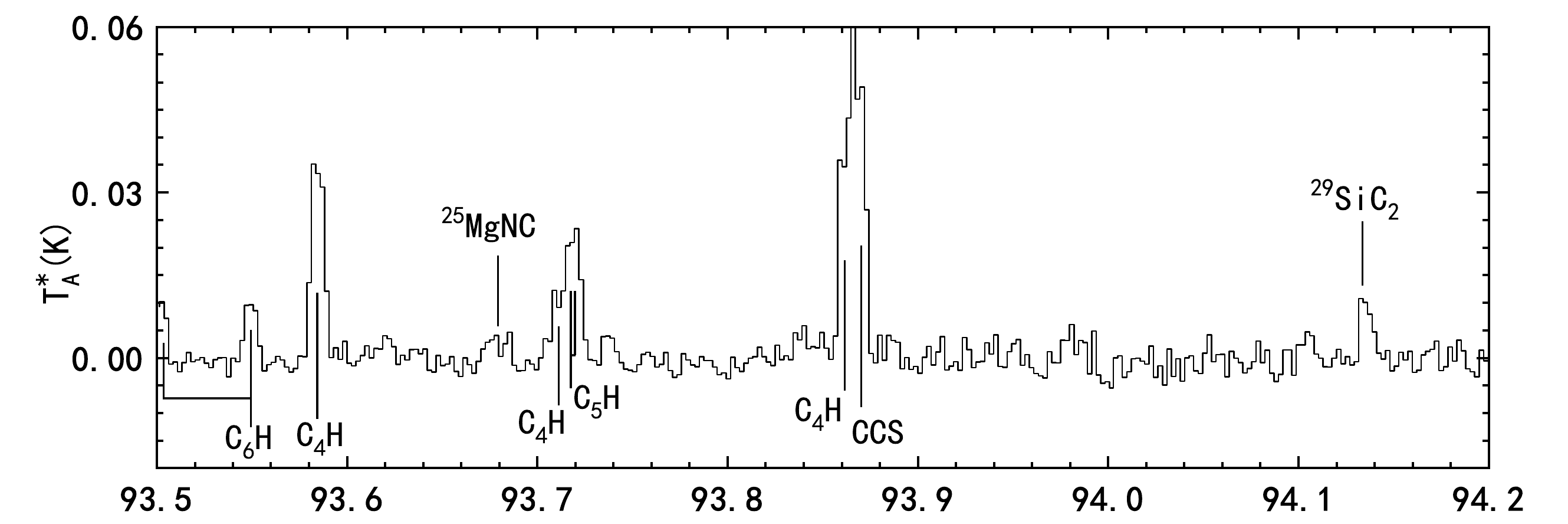}
\includegraphics[width = 0.8 \textwidth]{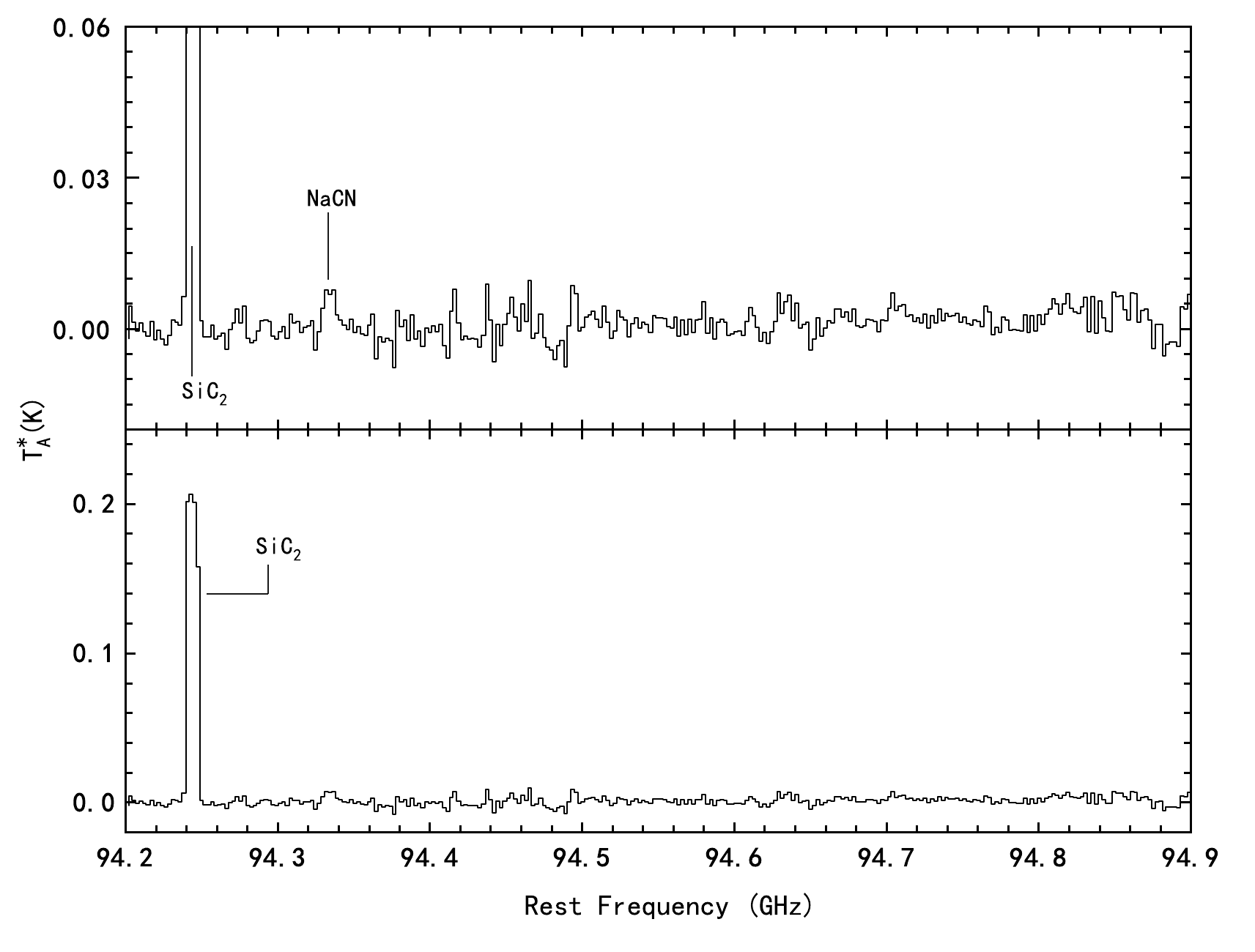}
\centerline{Figure.~\ref{Fig:irc_complete}. --- continued.}
\end{figure*}

\begin{figure*}[!htbp]
\centering
\includegraphics[width = 0.8 \textwidth]{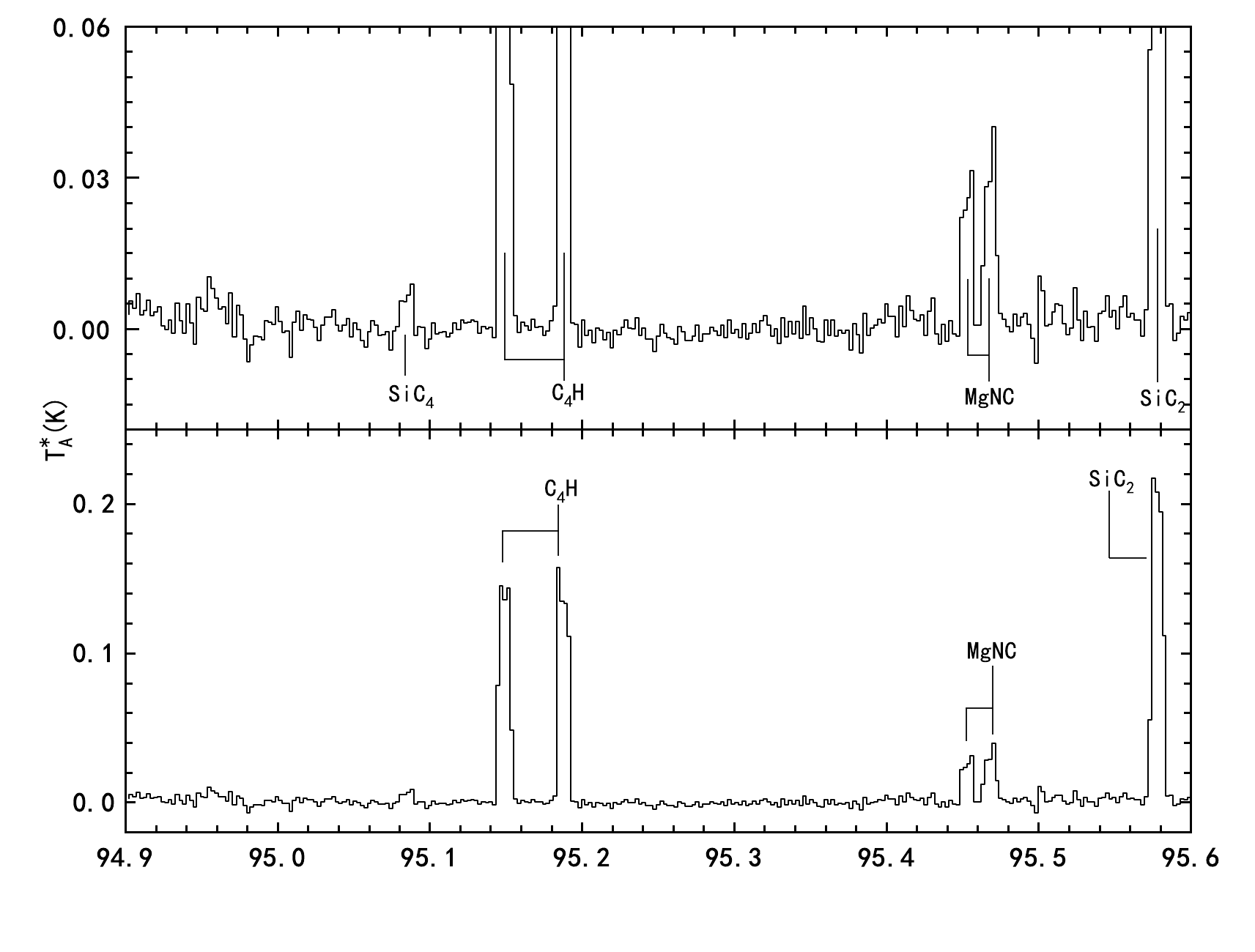}
\includegraphics[width = 0.8 \textwidth]{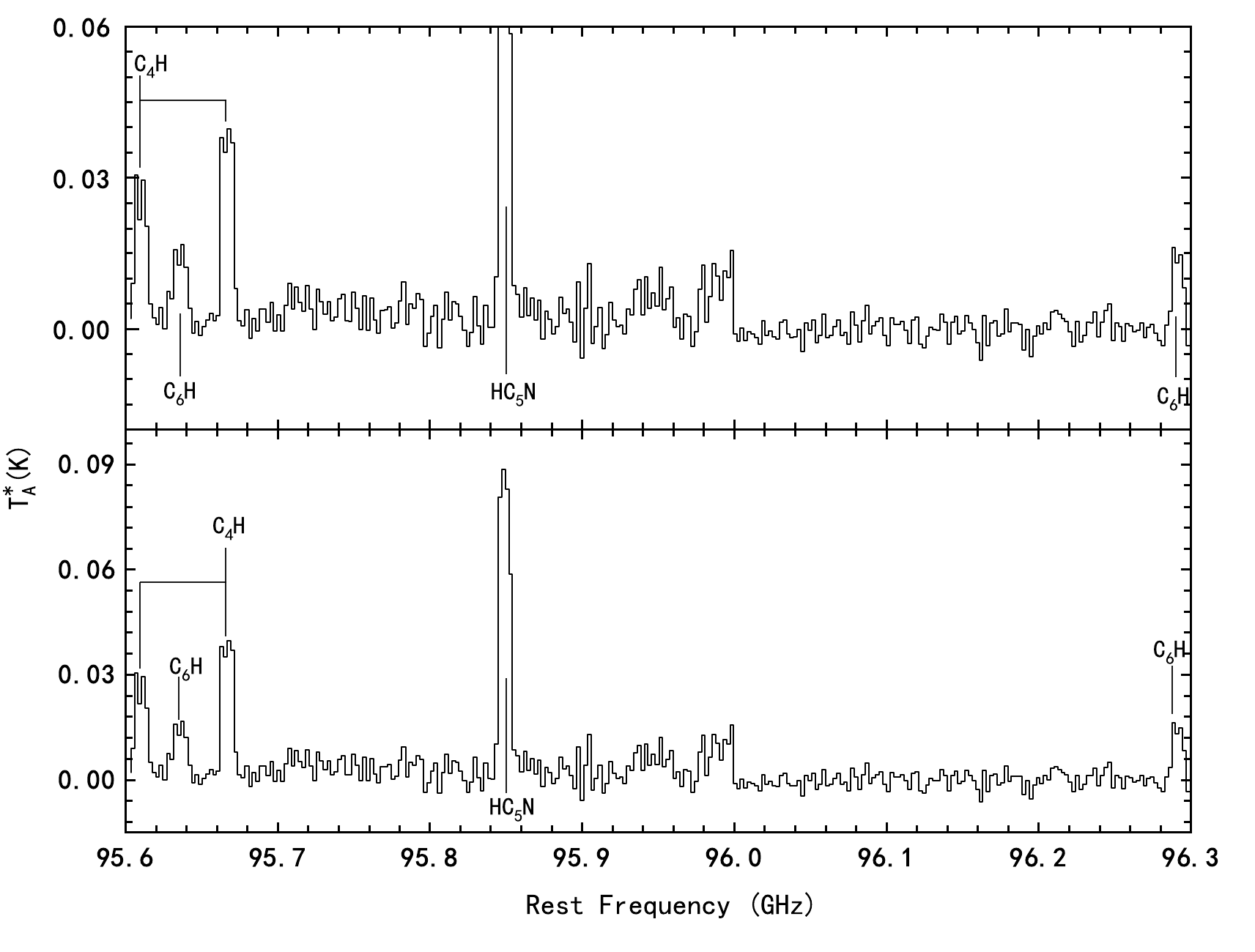}
\centerline{Figure.~\ref{Fig:irc_complete}. --- continued.}
\end{figure*}

\begin{figure*}[!htbp]
\centering
\includegraphics[width = 0.8 \textwidth]{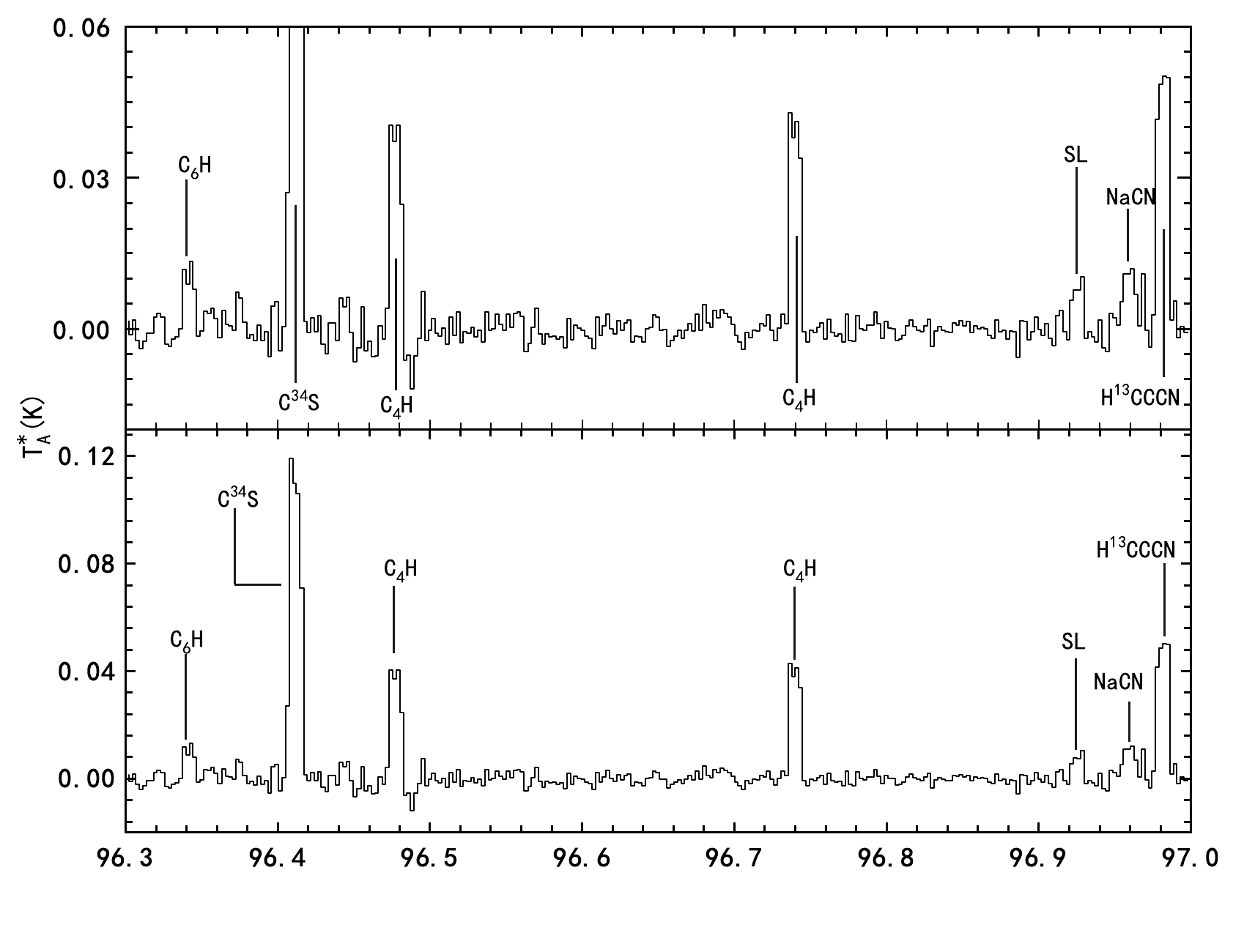}
\includegraphics[width = 0.8 \textwidth]{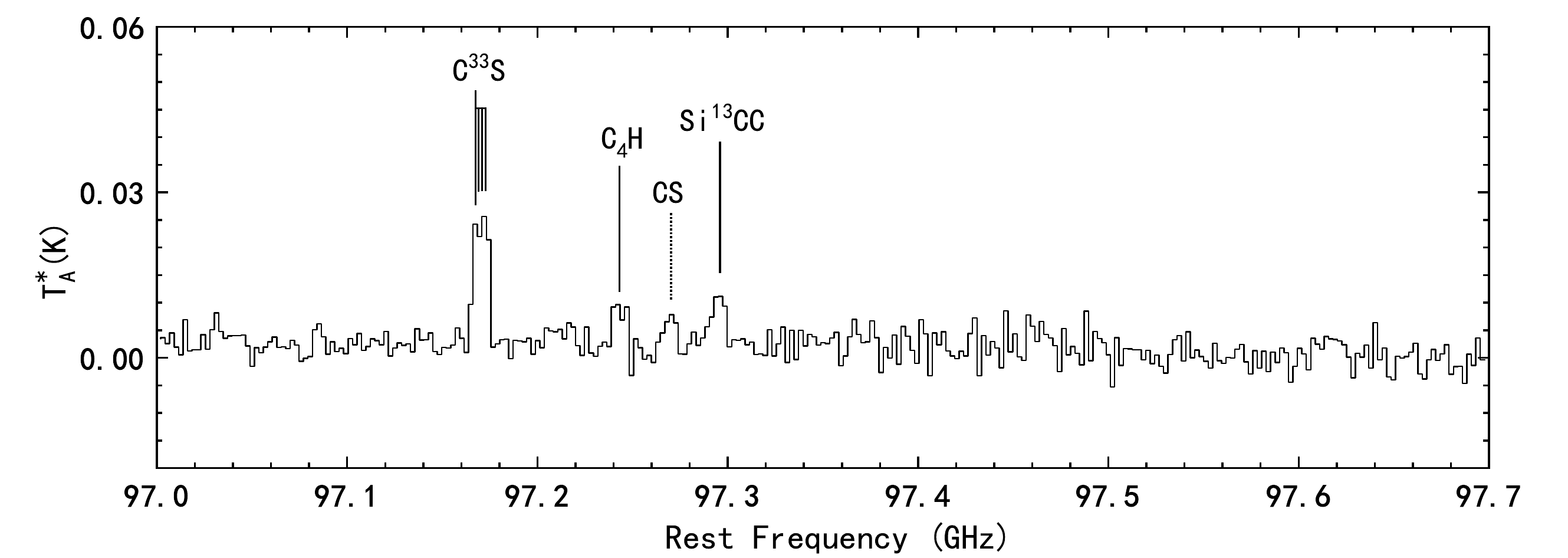}
\centerline{Figure.~\ref{Fig:irc_complete}. --- continued.}
\end{figure*}

\begin{figure*}[!htbp]
\centering
\includegraphics[width = 0.8 \textwidth]{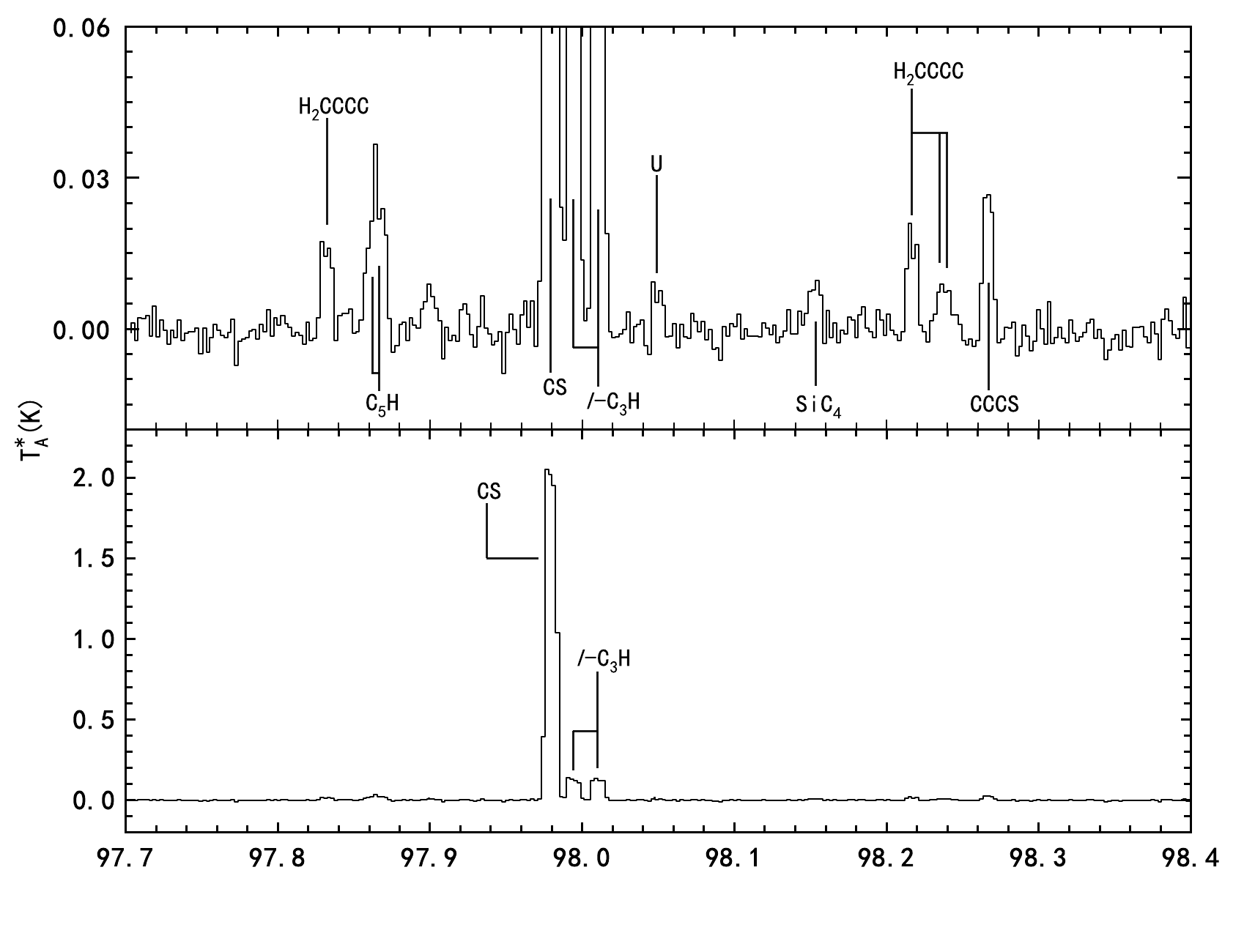}
\includegraphics[width = 0.8 \textwidth]{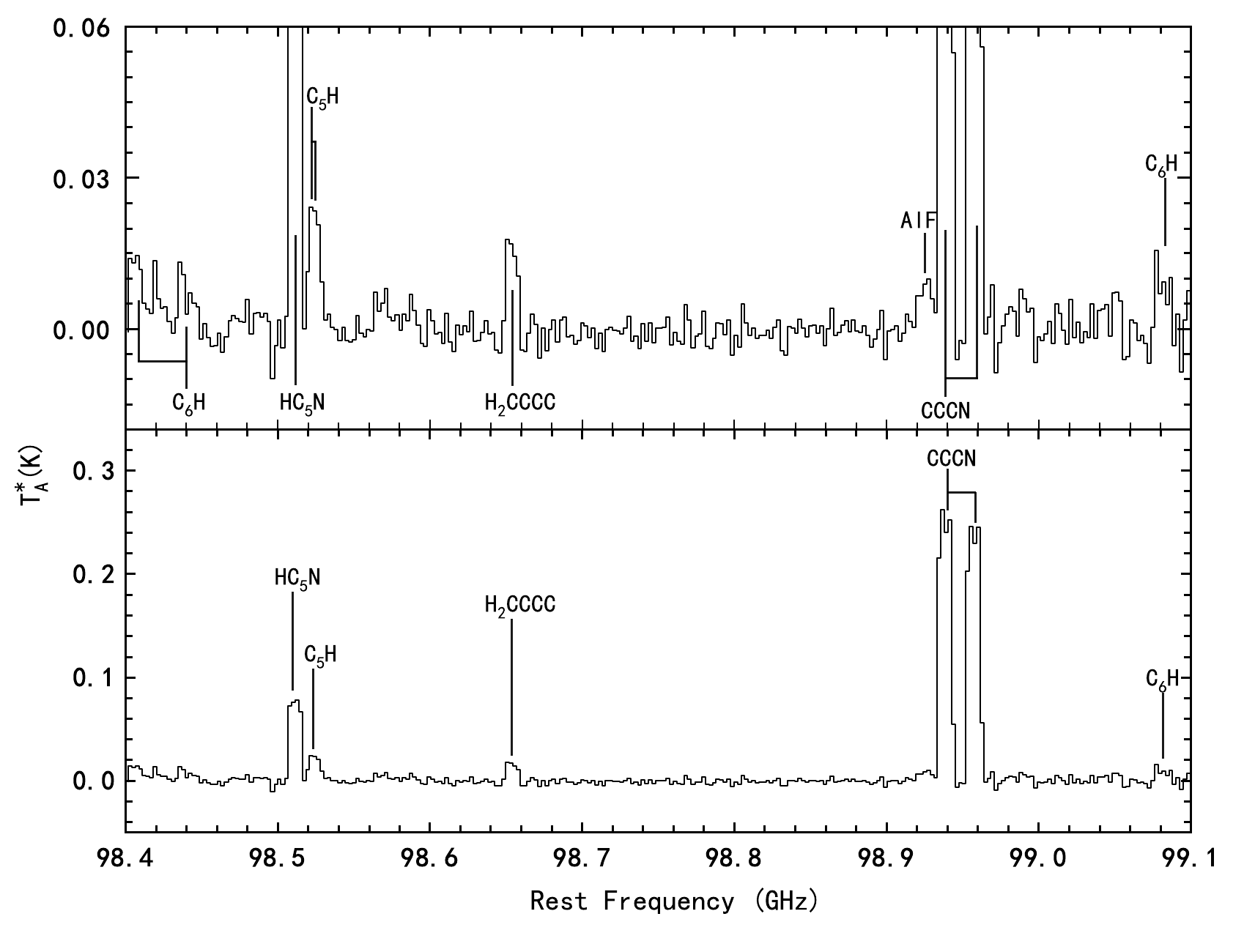}
\centerline{Figure.~\ref{Fig:irc_complete}. --- continued.}
\end{figure*}

\begin{figure*}[!htbp]
\centering
\includegraphics[width = 0.8 \textwidth]{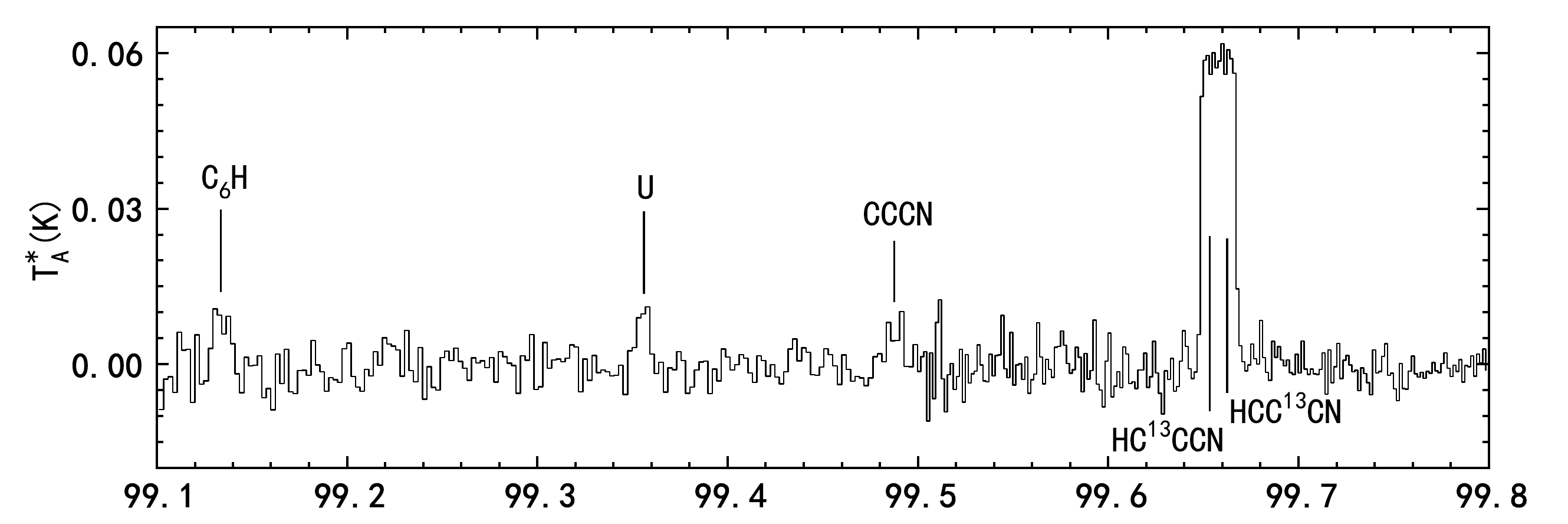}
\includegraphics[width = 0.8 \textwidth]{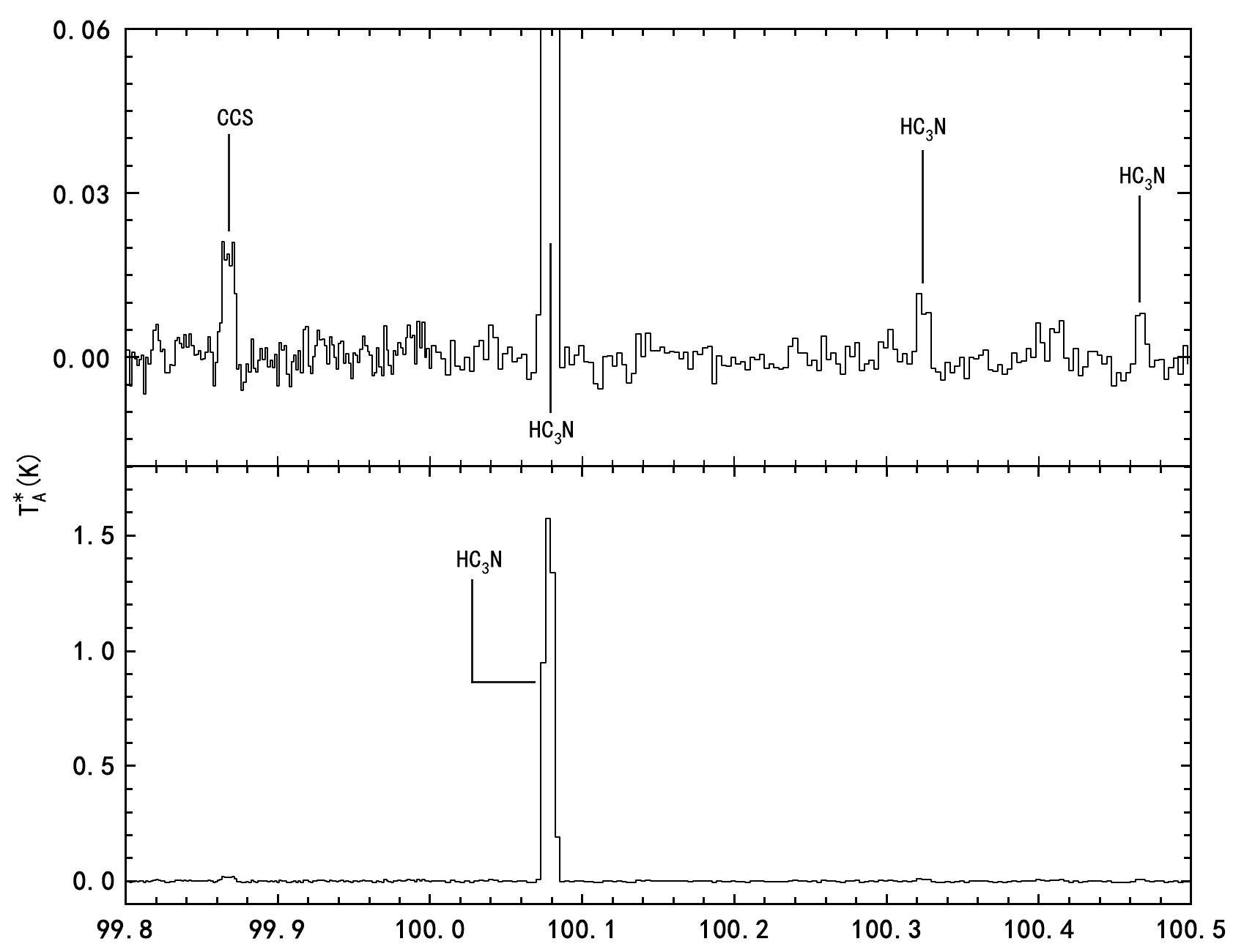}
\includegraphics[width = 0.8 \textwidth]{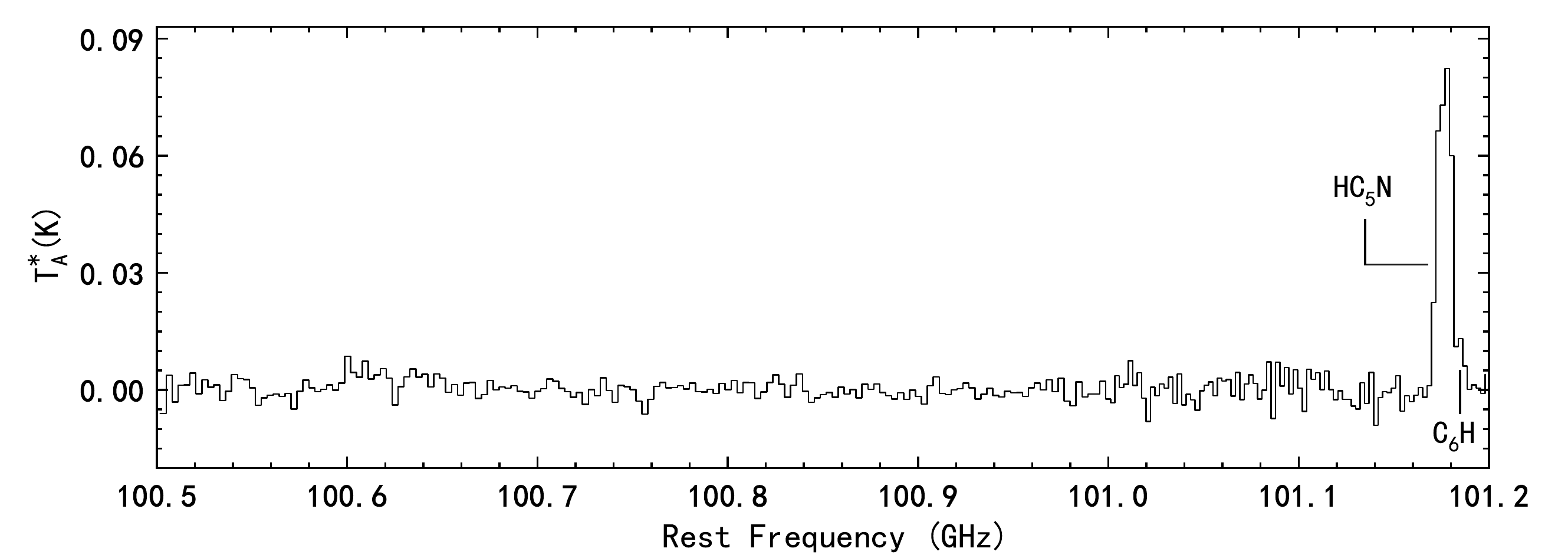}
\centerline{Figure.~\ref{Fig:irc_complete}. --- continued.}
\end{figure*}

\begin{figure*}[!htbp]
\centering
\includegraphics[width = 0.8 \textwidth]{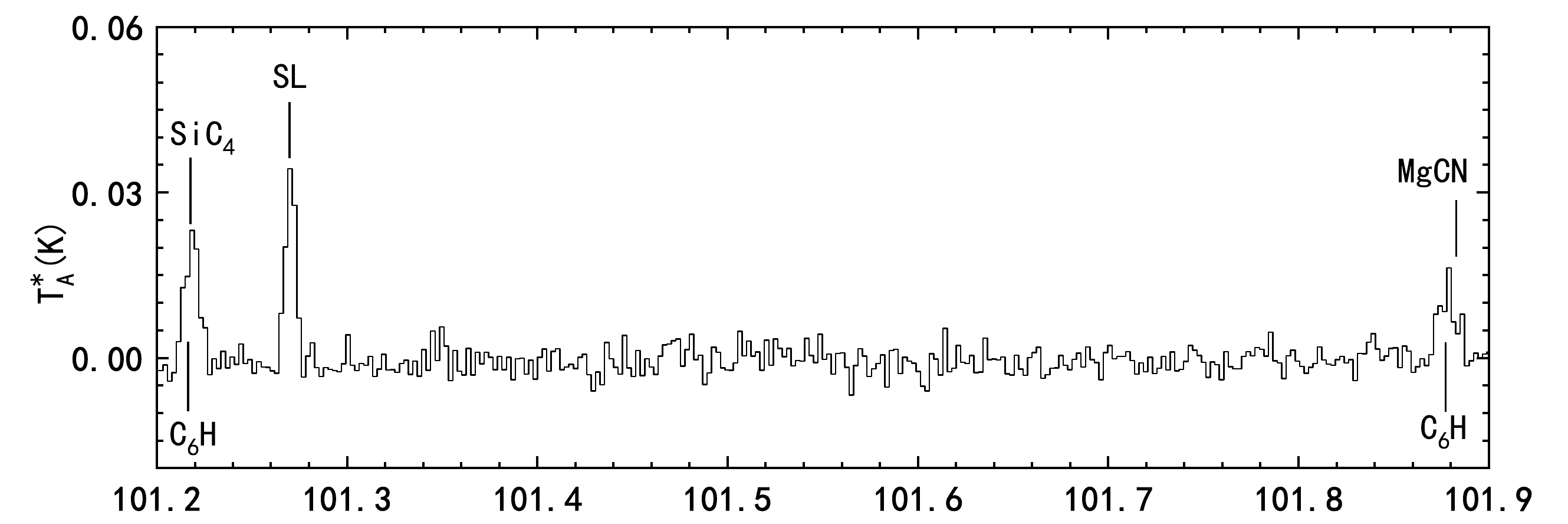}
\includegraphics[width = 0.8 \textwidth]{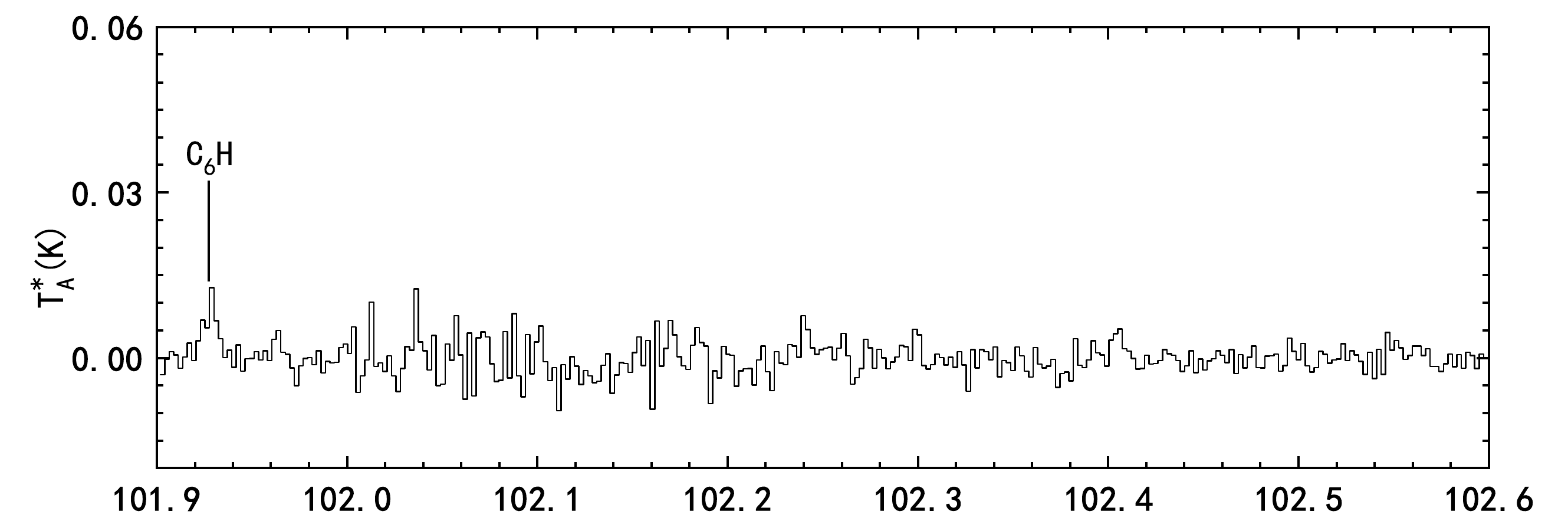}
\includegraphics[width = 0.8 \textwidth]{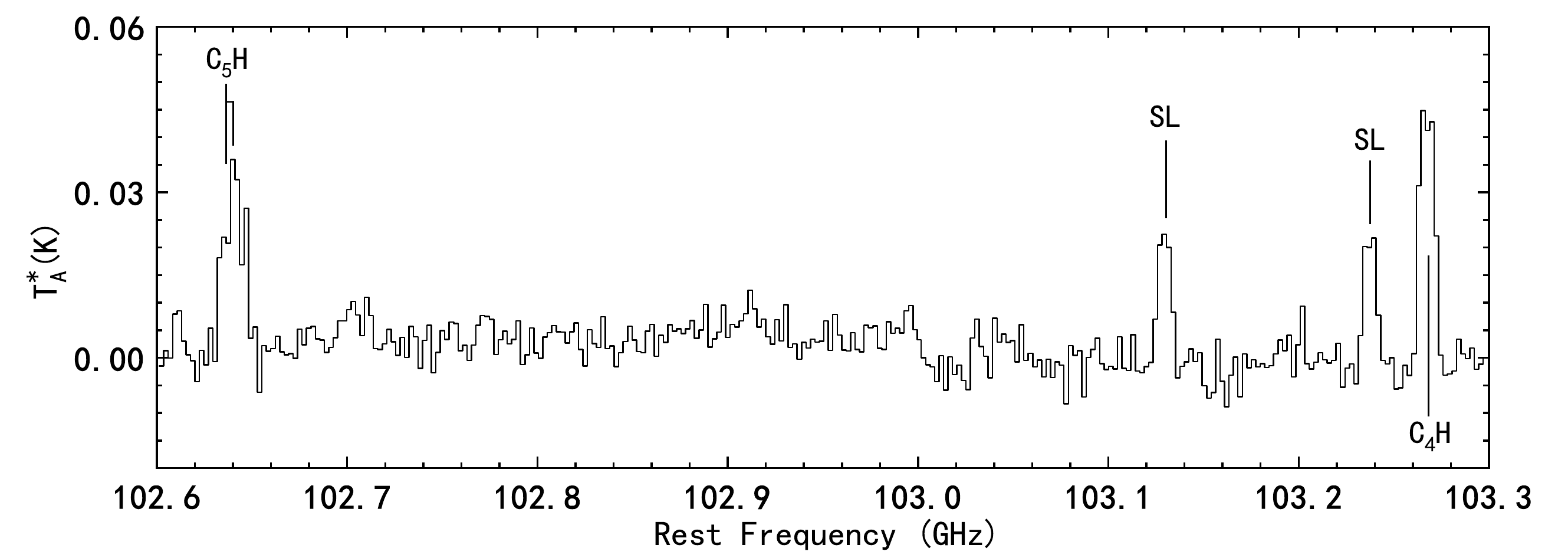}
\centerline{Figure.~\ref{Fig:irc_complete}. --- continued.}
\end{figure*}

\begin{figure*}[!htbp]
\centering 
\includegraphics[width = 0.8 \textwidth]{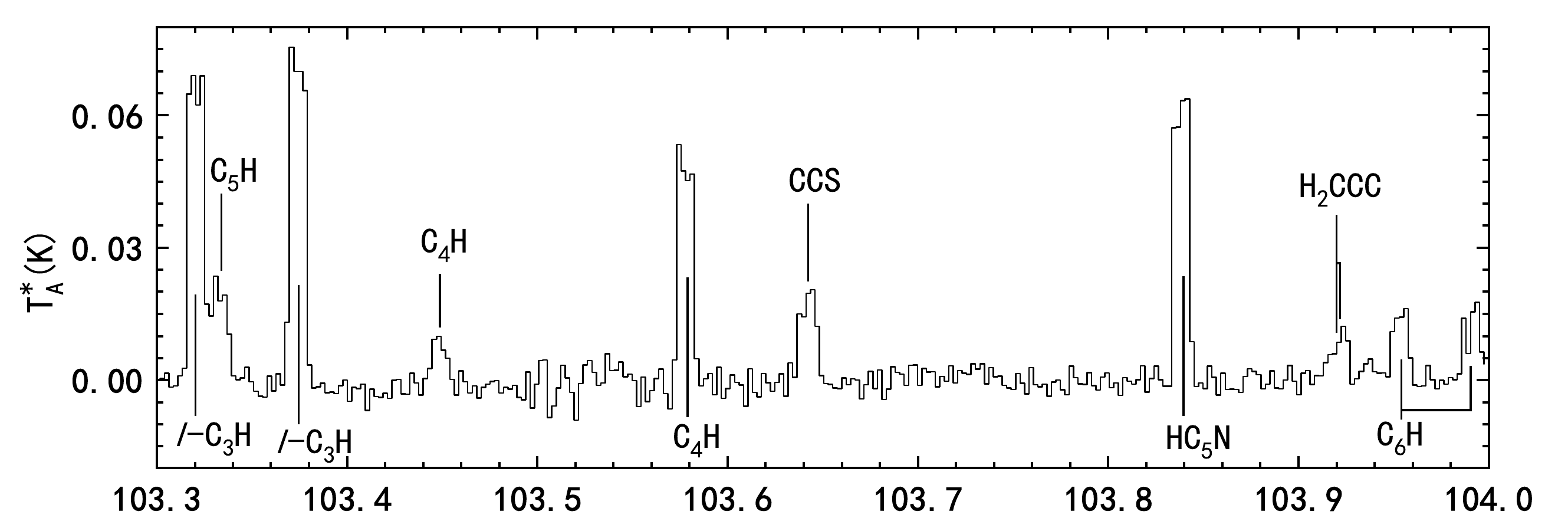}
\includegraphics[width = 0.8 \textwidth]{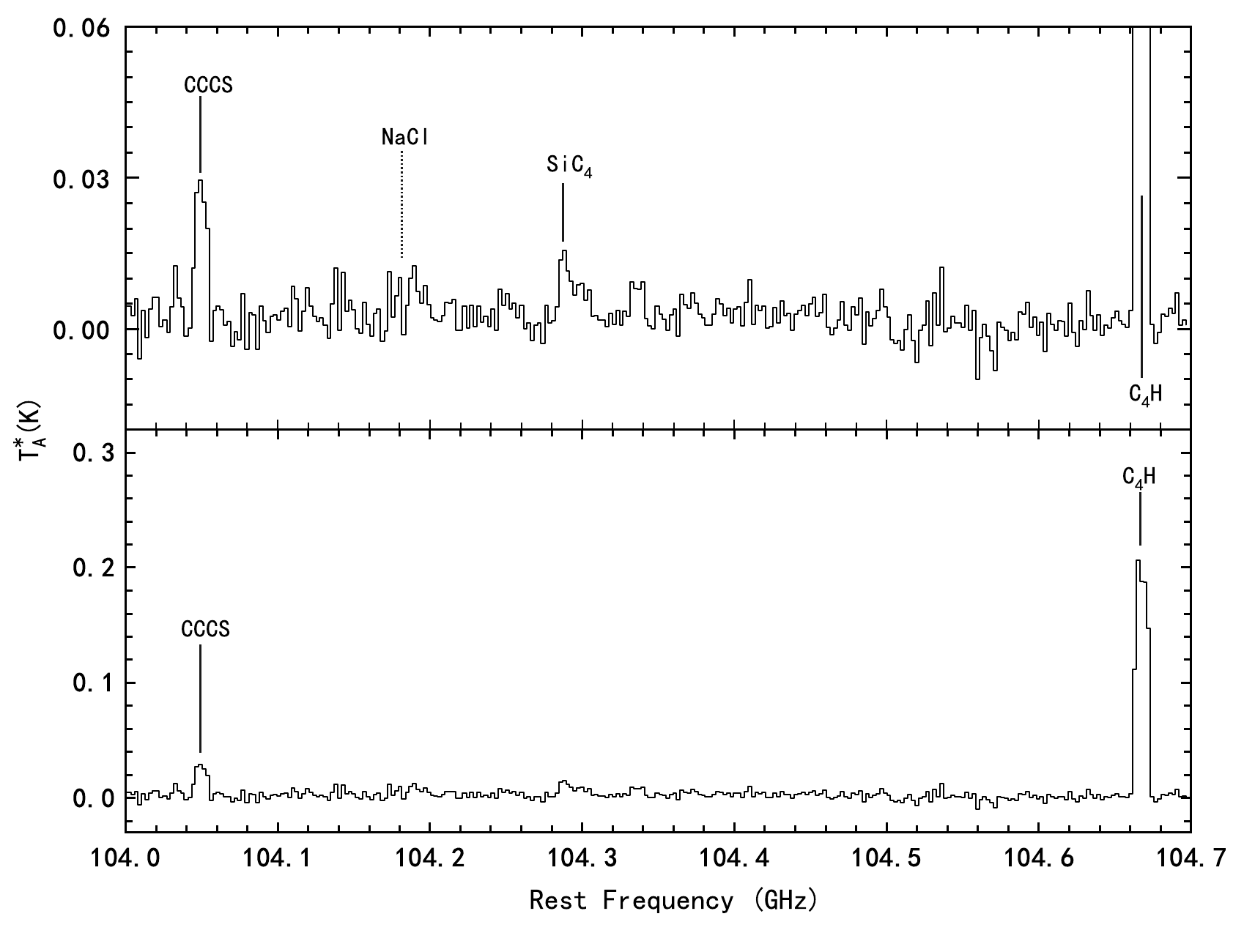}
\centerline{Figure.~\ref{Fig:irc_complete}. --- continued.}
\end{figure*}

\begin{figure*}[!htbp]
\centering 
\includegraphics[width = 0.8 \textwidth]{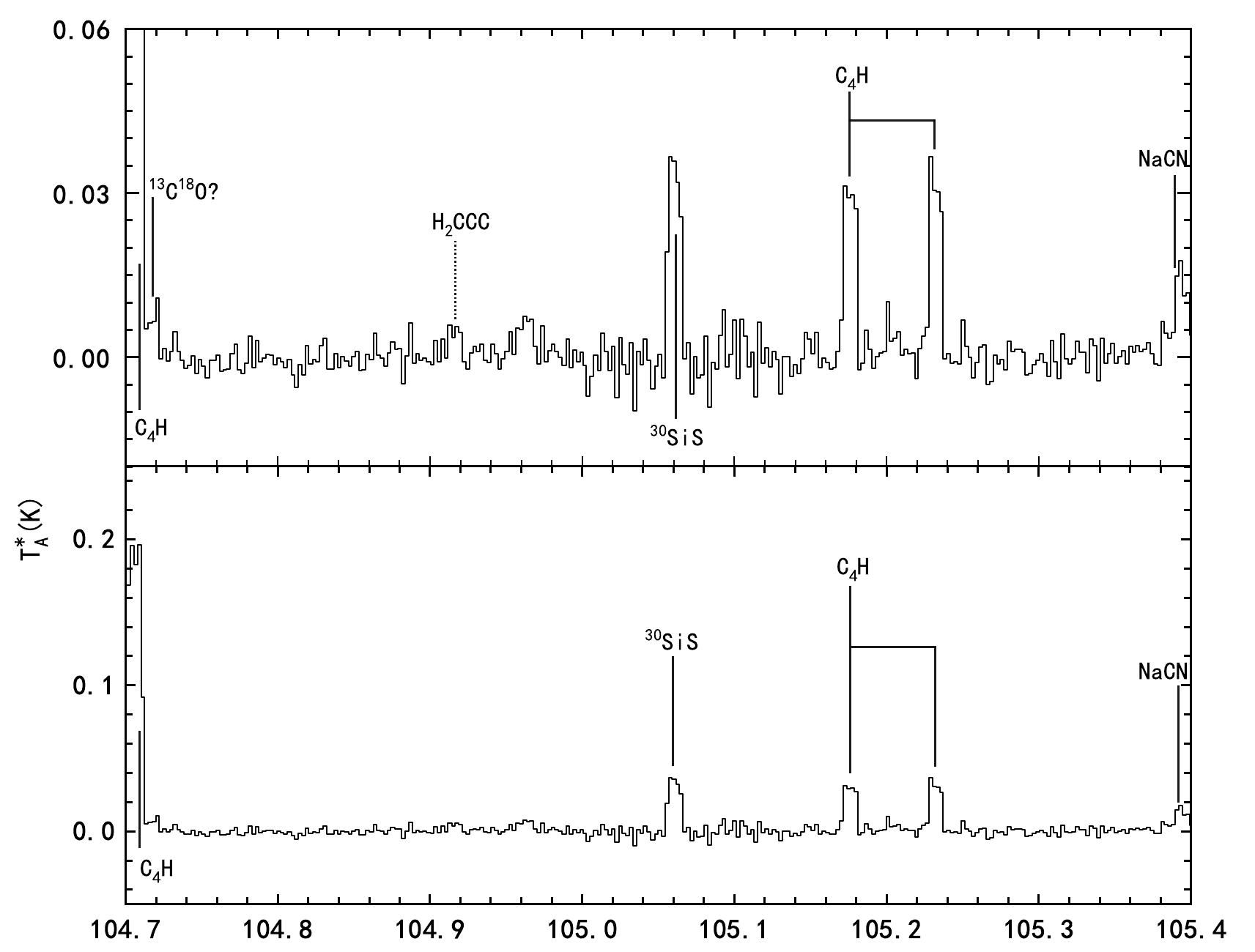}
\includegraphics[width = 0.8 \textwidth]{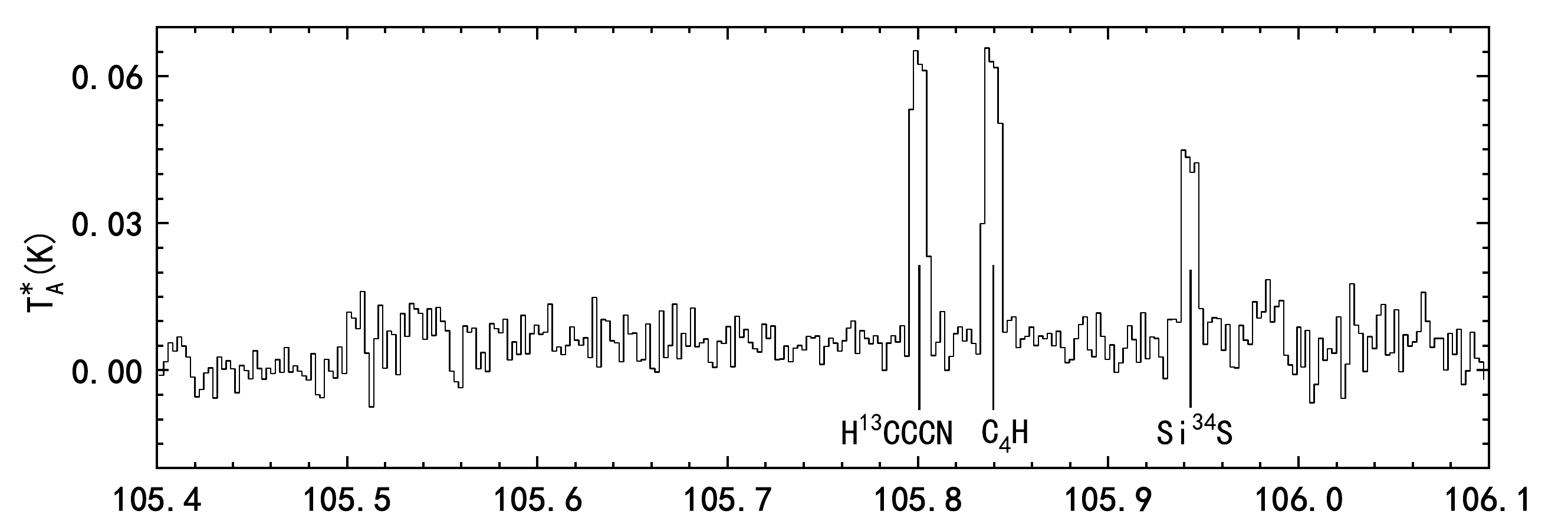}
\includegraphics[width = 0.8 \textwidth]{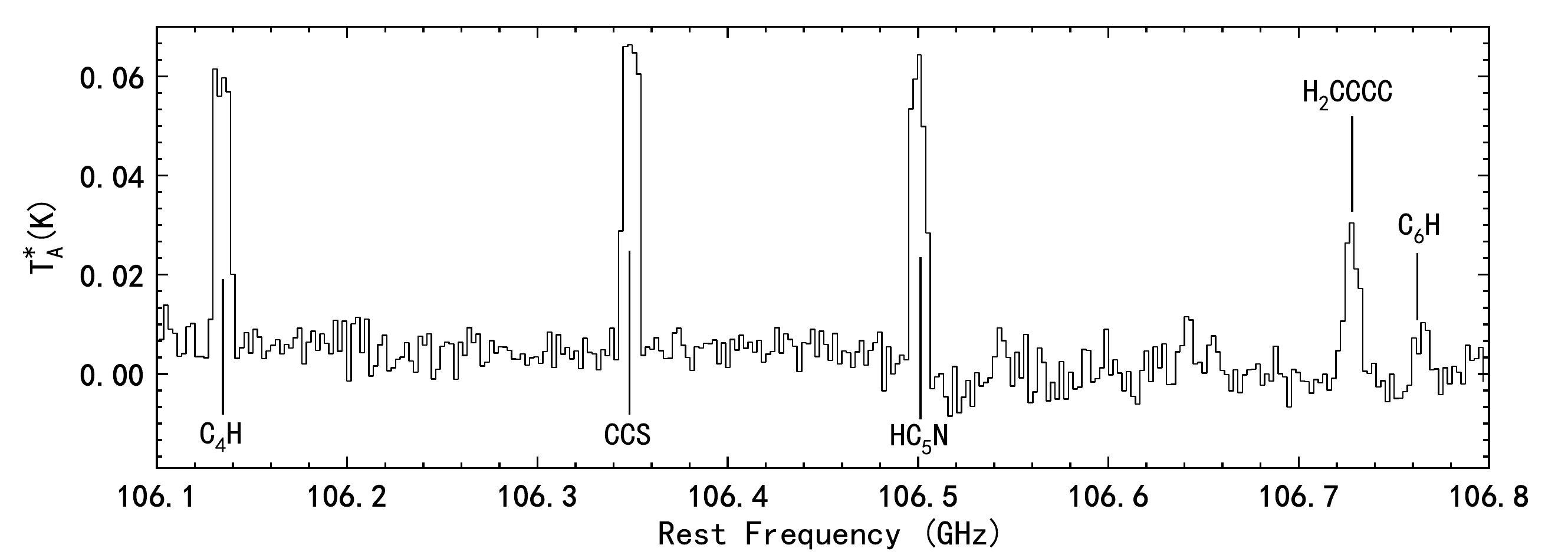}
\centerline{Figure.~\ref{Fig:irc_complete}. --- continued.}
\end{figure*}

\begin{figure*}[!htbp]
\centering 
\includegraphics[width = 0.8 \textwidth]{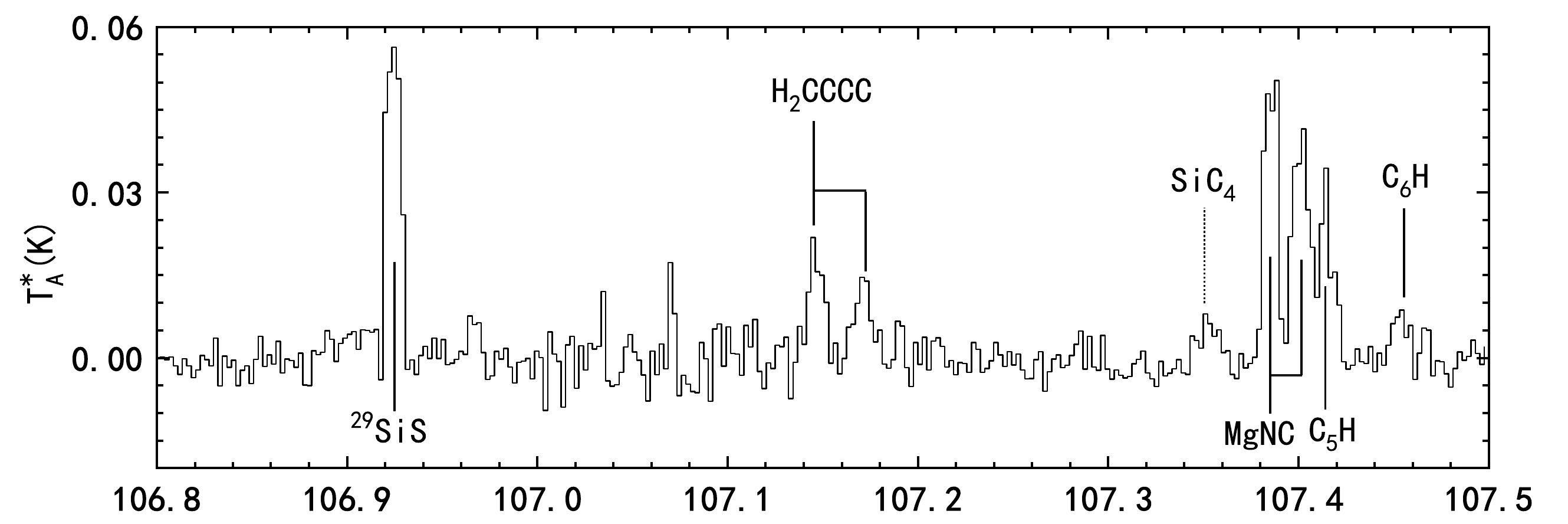}
\includegraphics[width = 0.8 \textwidth]{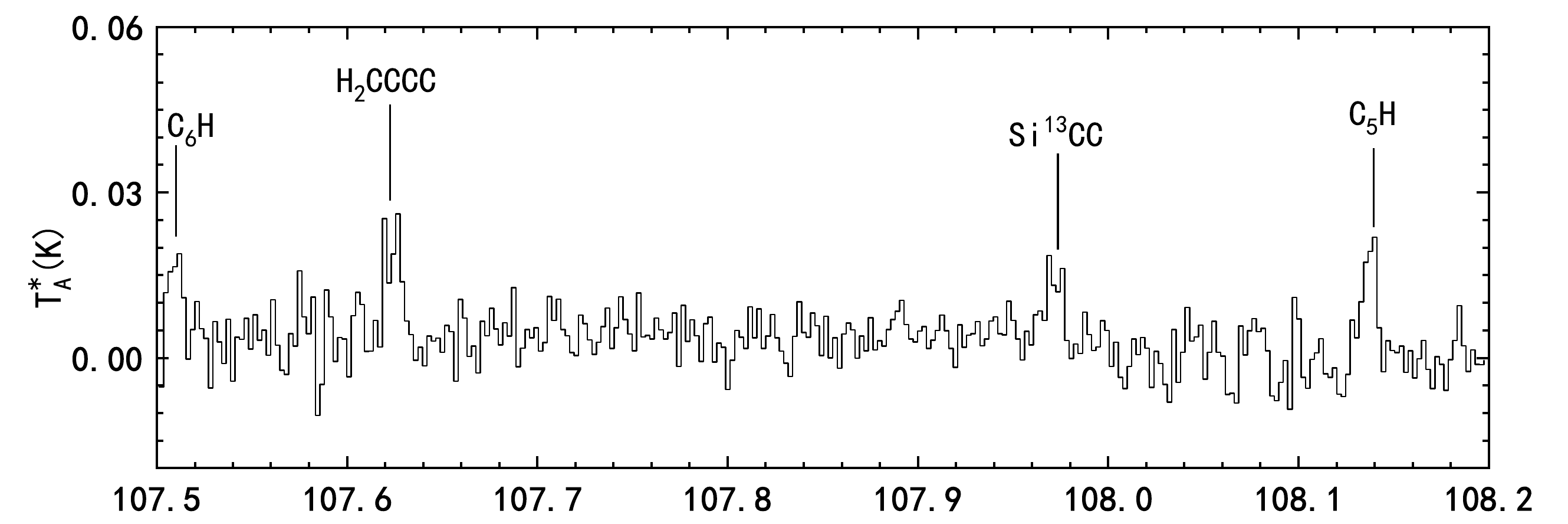}
\includegraphics[width = 0.8 \textwidth]{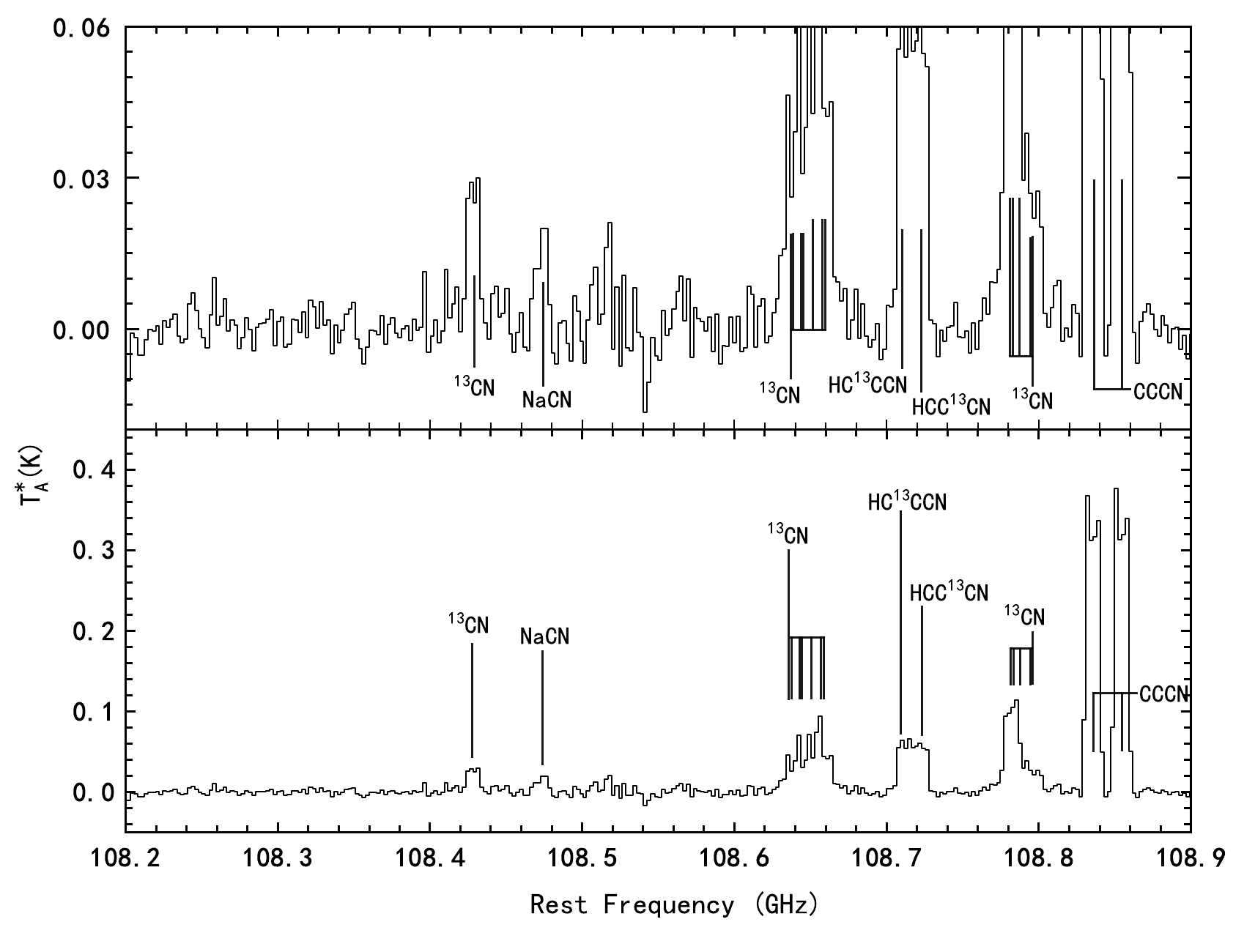}
\centerline{Figure.~\ref{Fig:irc_complete}. --- continued.}
\end{figure*}

\begin{figure*}[!htbp]
\centering
\includegraphics[width = 0.8 \textwidth]{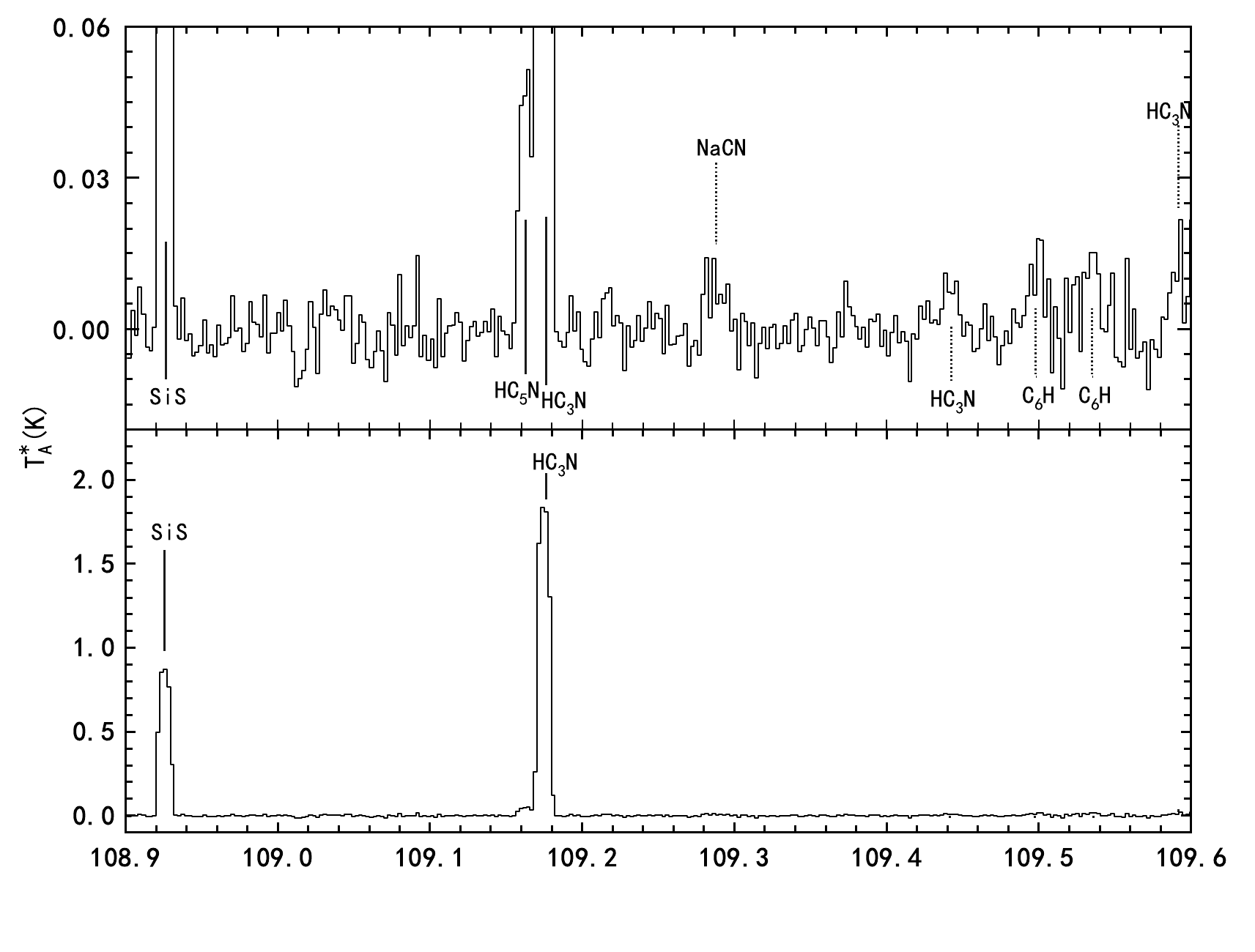}
\includegraphics[width = 0.8 \textwidth]{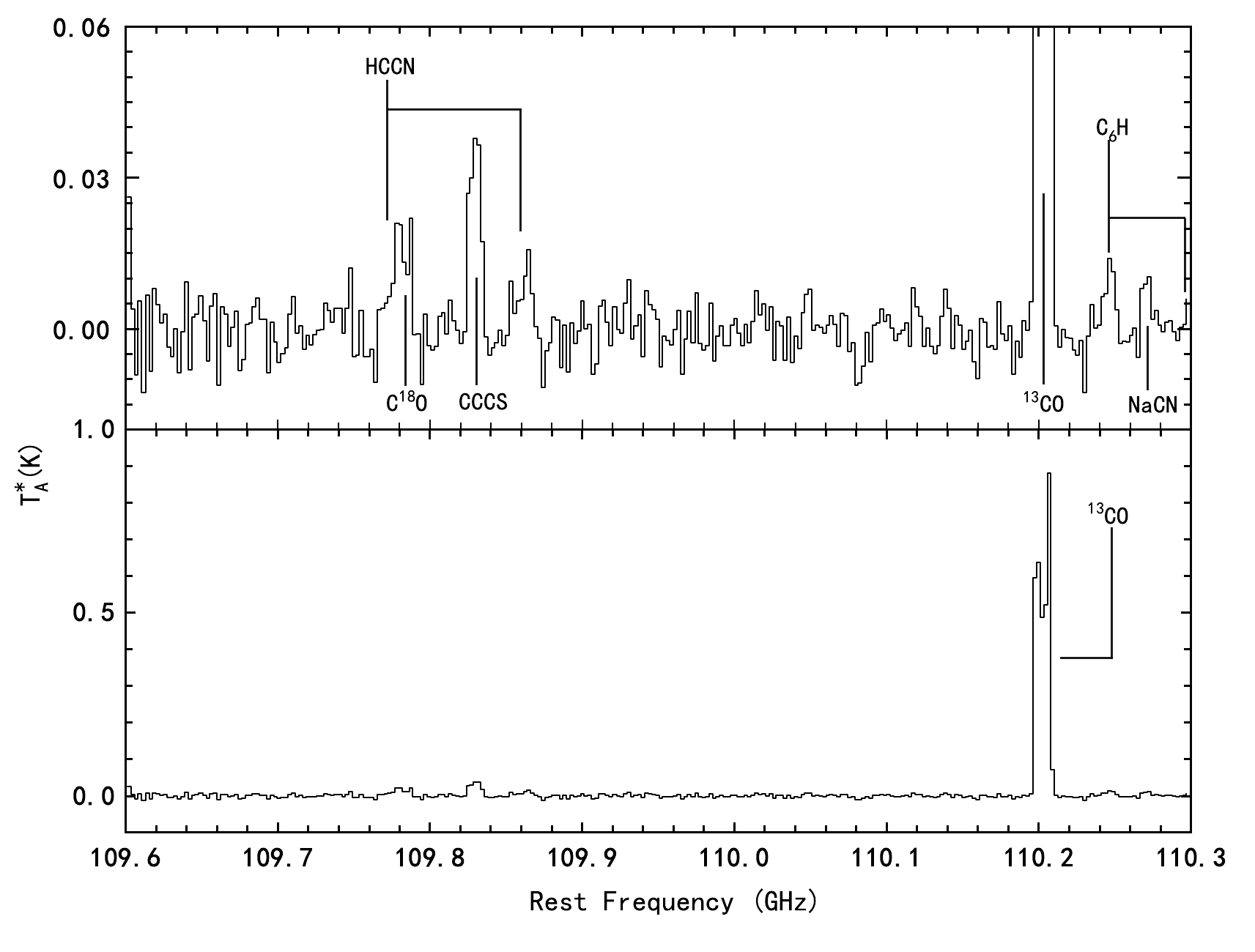}
\centerline{Figure.~\ref{Fig:irc_complete}. --- continued.}
\end{figure*}

\begin{figure*}[!htbp]
\centering
\includegraphics[width = 0.8 \textwidth]{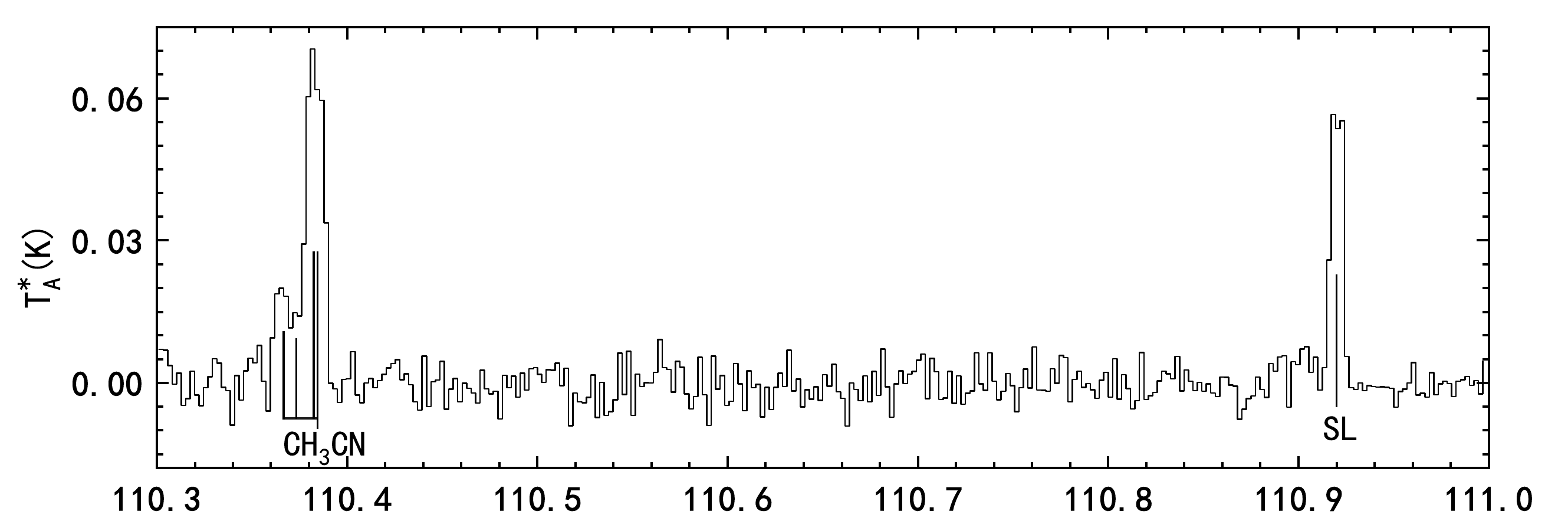}
\includegraphics[width = 0.8 \textwidth]{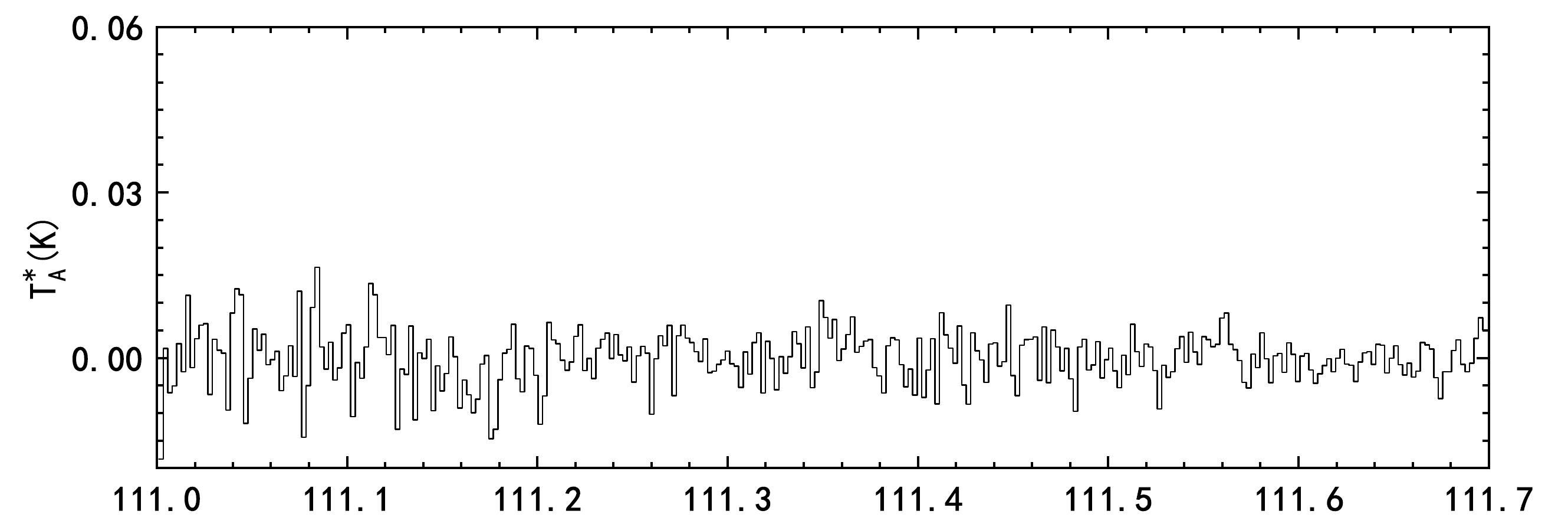}
\includegraphics[width = 0.8 \textwidth]{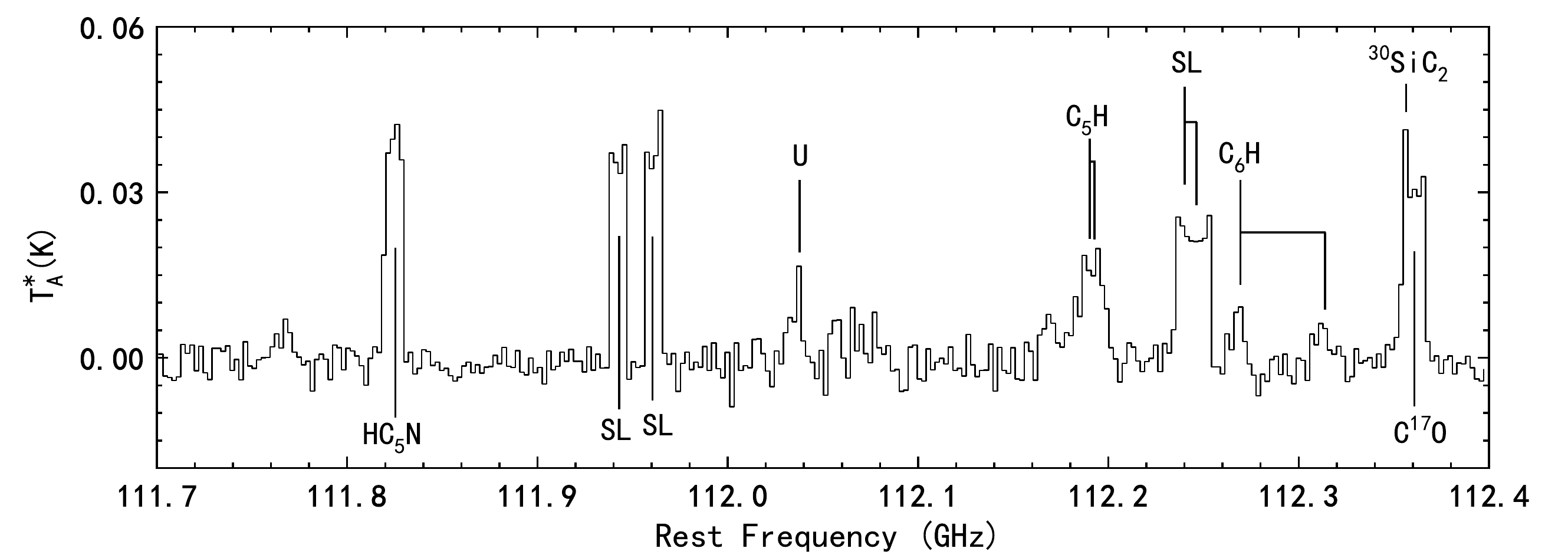}
\centerline{Figure.~\ref{Fig:irc_complete}. --- continued.}
\end{figure*}

\begin{figure*}[!htbp]
\centering
\includegraphics[width = 0.8 \textwidth]{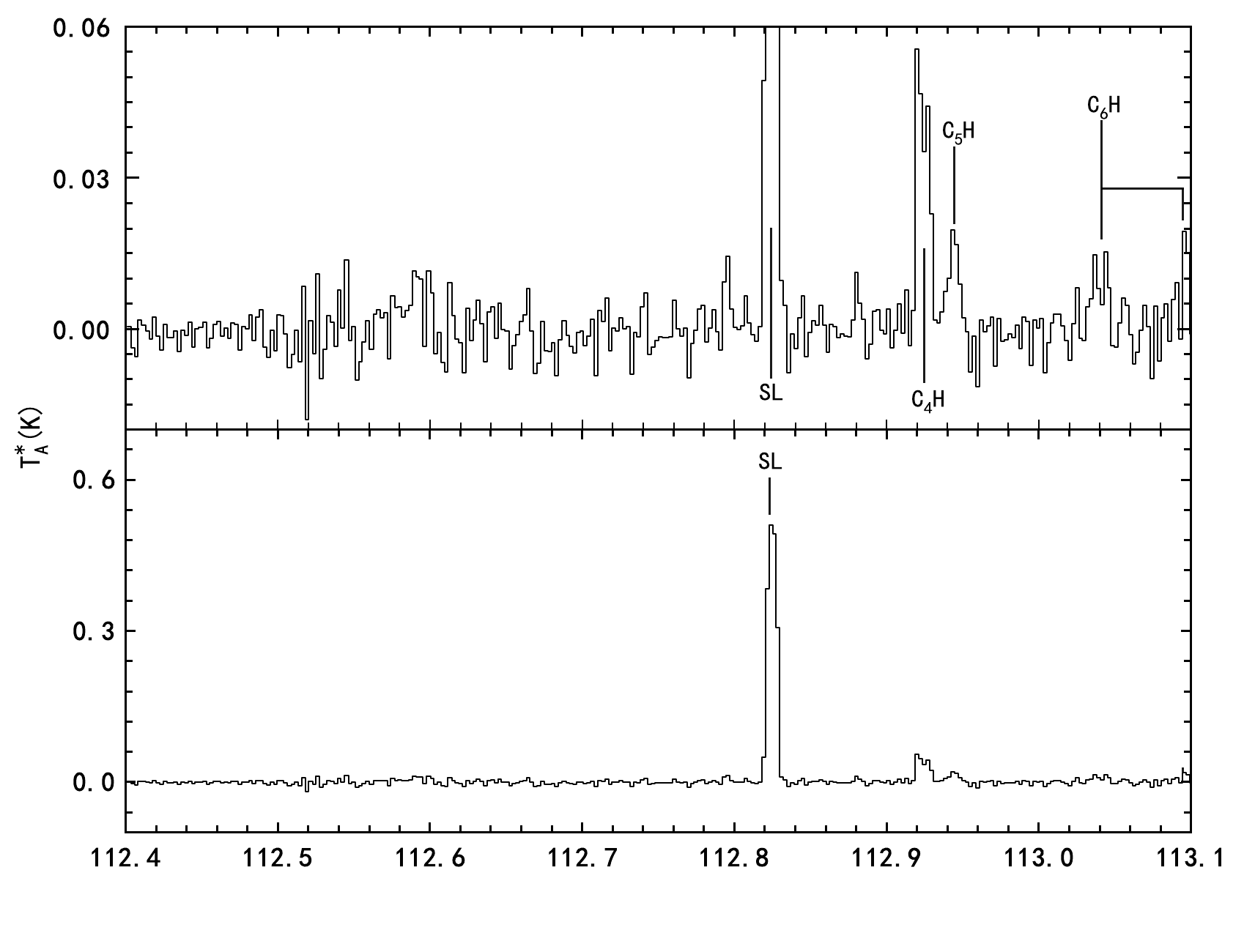}
\includegraphics[width = 0.8 \textwidth]{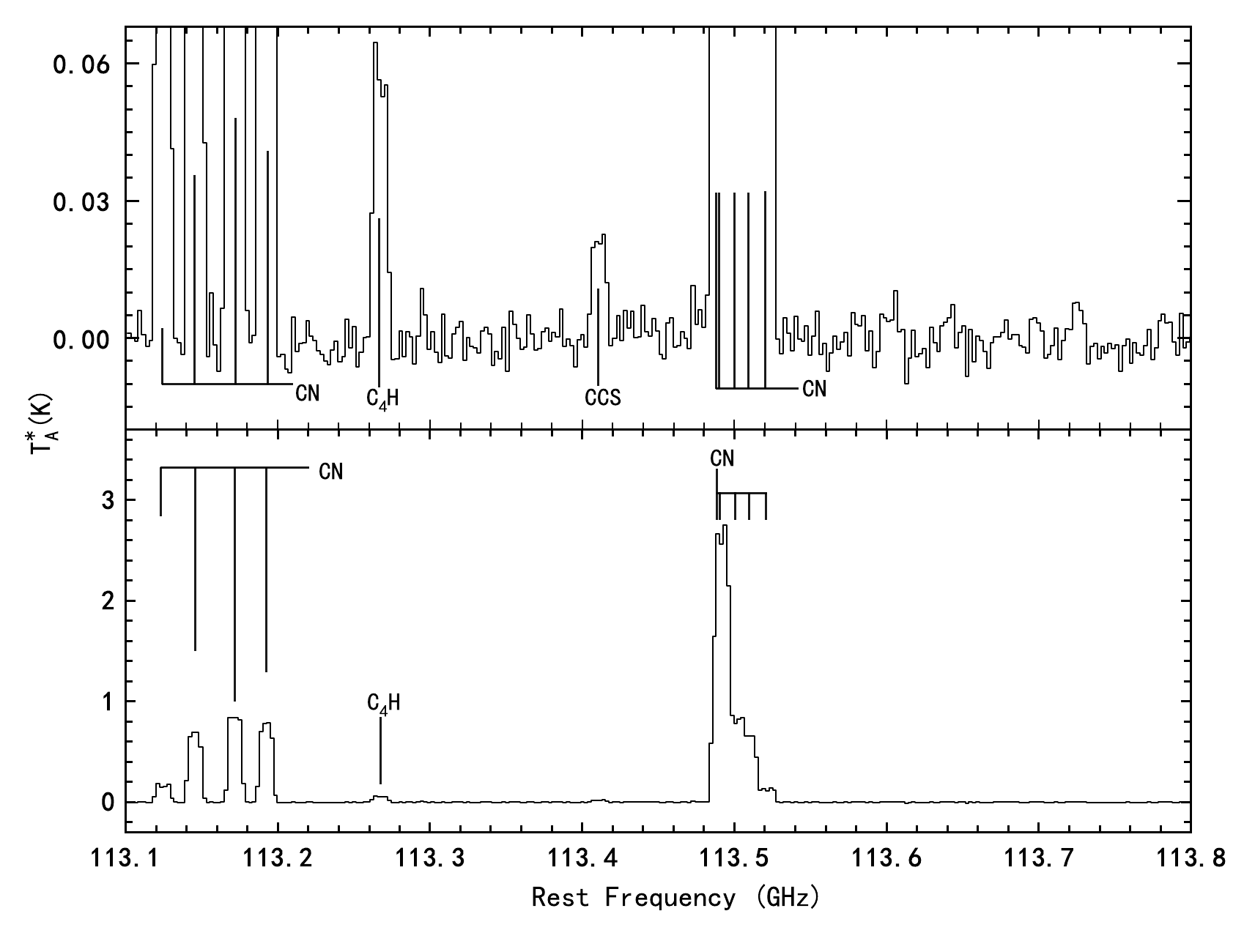}
\centerline{Figure.~\ref{Fig:irc_complete}. --- continued.}
\end{figure*}

\begin{figure*}[!htbp]
\centering
\includegraphics[width = 0.8 \textwidth]{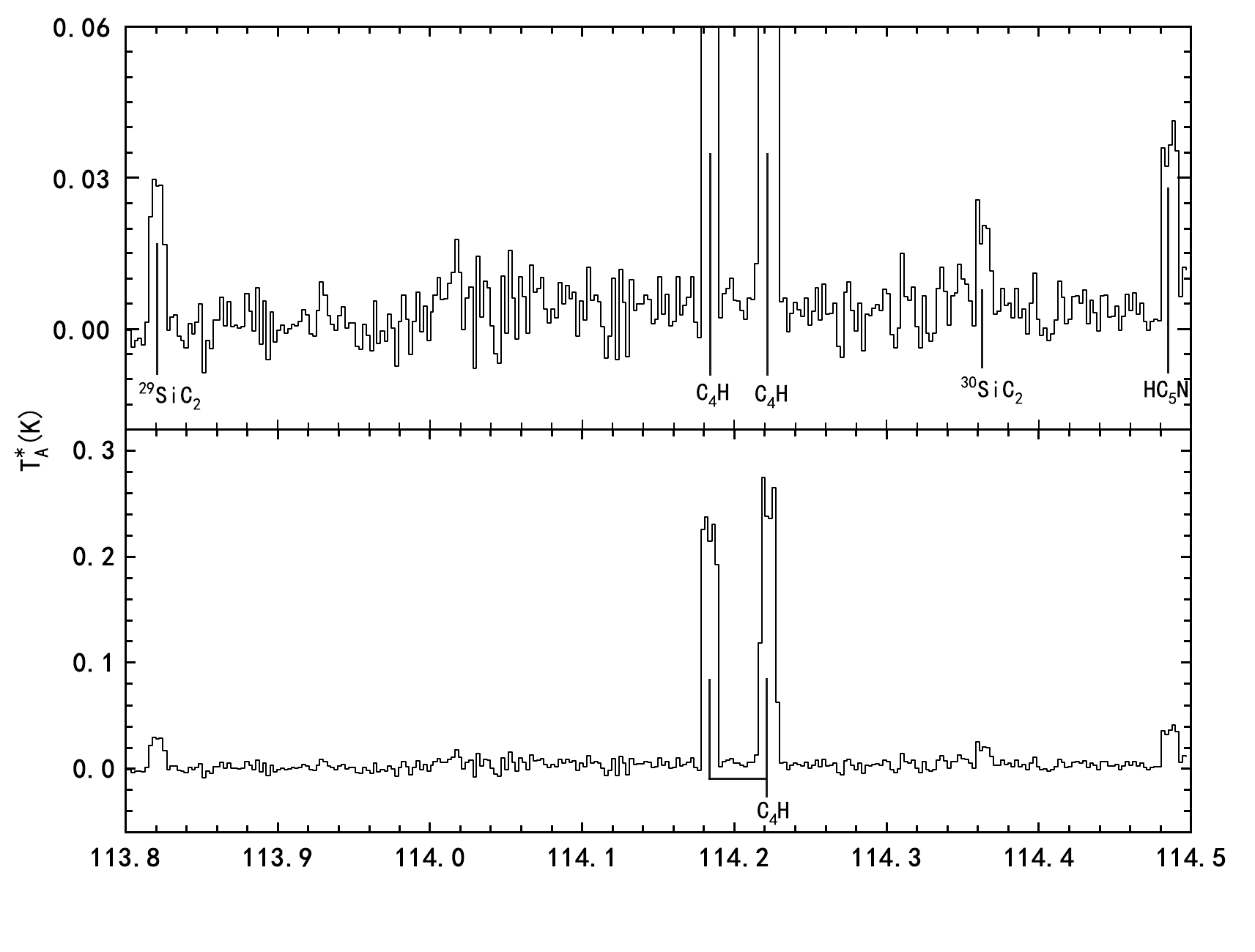}
\includegraphics[width = 0.8 \textwidth]{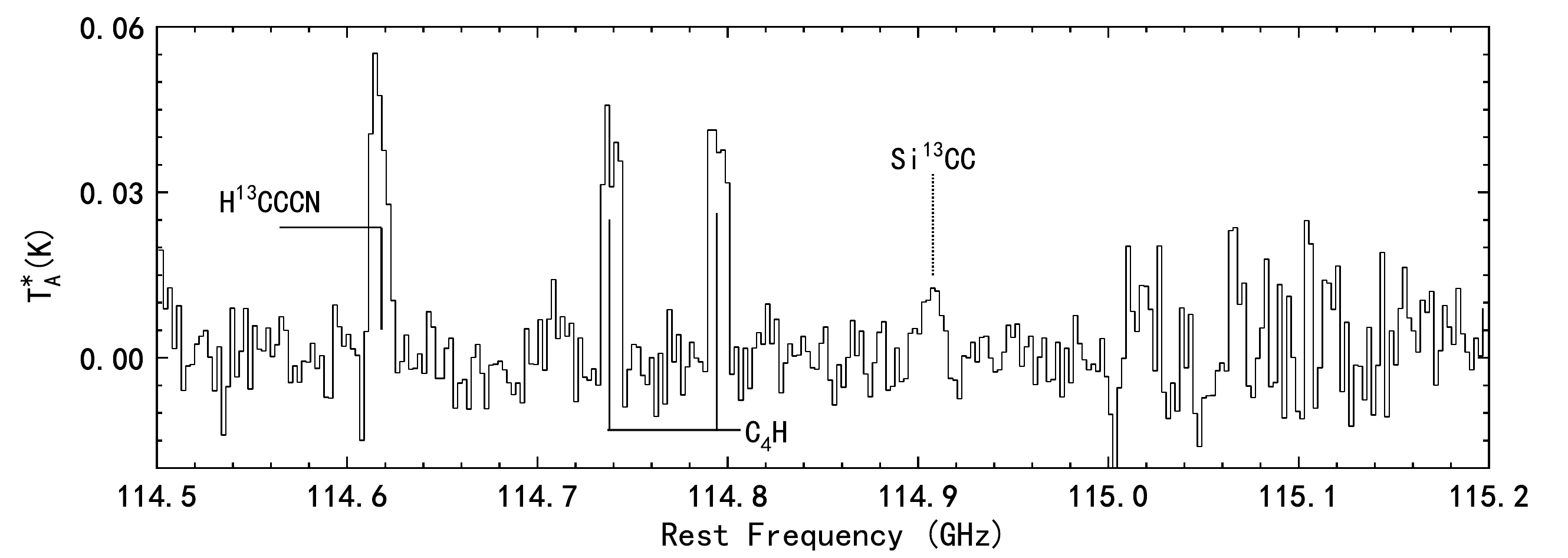}
\centerline{Figure.~\ref{Fig:irc_complete}. --- continued.}
\end{figure*}

\begin{figure*}[!htbp]
\centering
\includegraphics[width = 0.9 \textwidth]{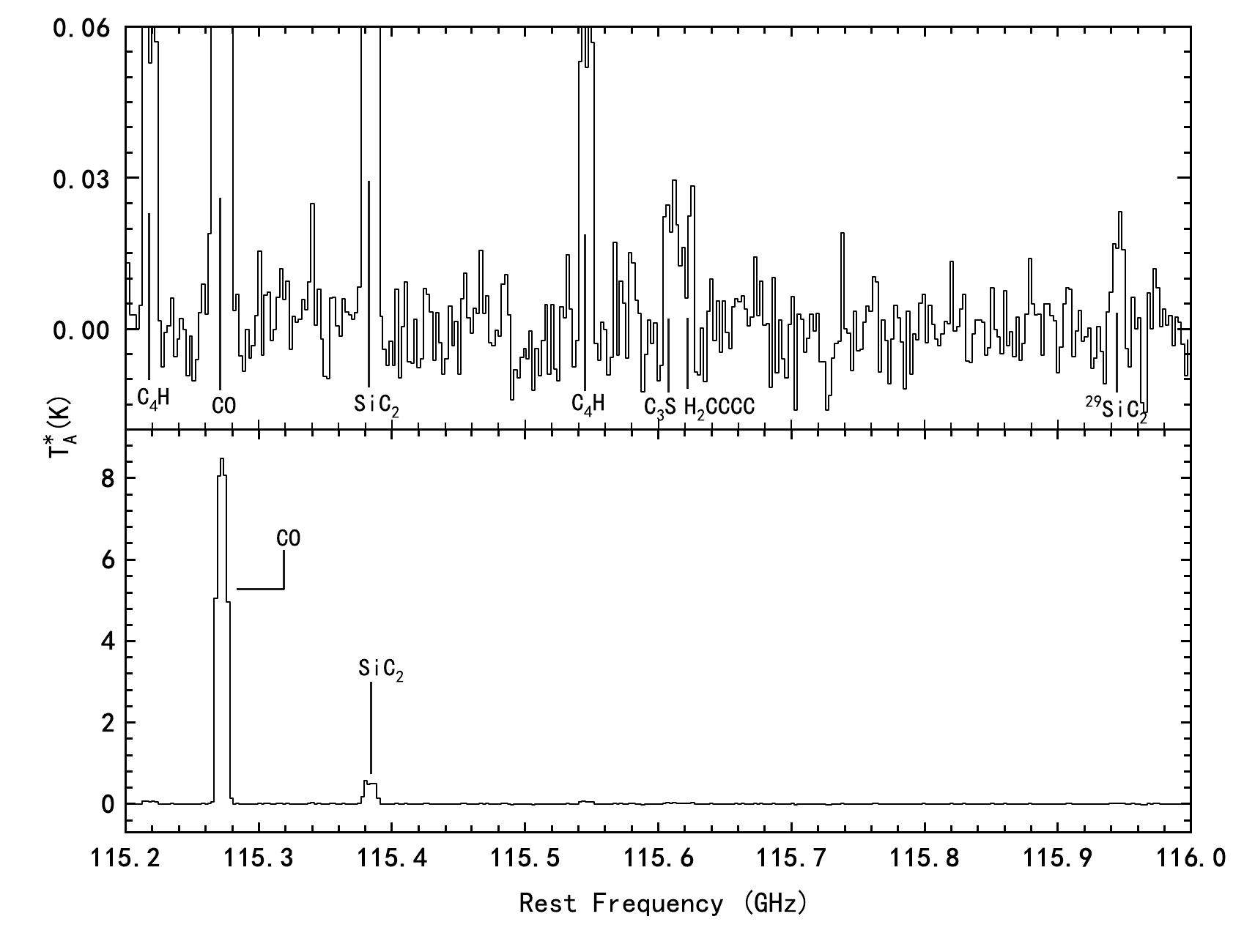}
\centerline{Figure.~\ref{Fig:irc_complete}. --- continued.}
\end{figure*}

\begin{figure*}[!htbp]
\section{Spectral line profiles}
\centering
\includegraphics[width = 0.45 \textwidth]{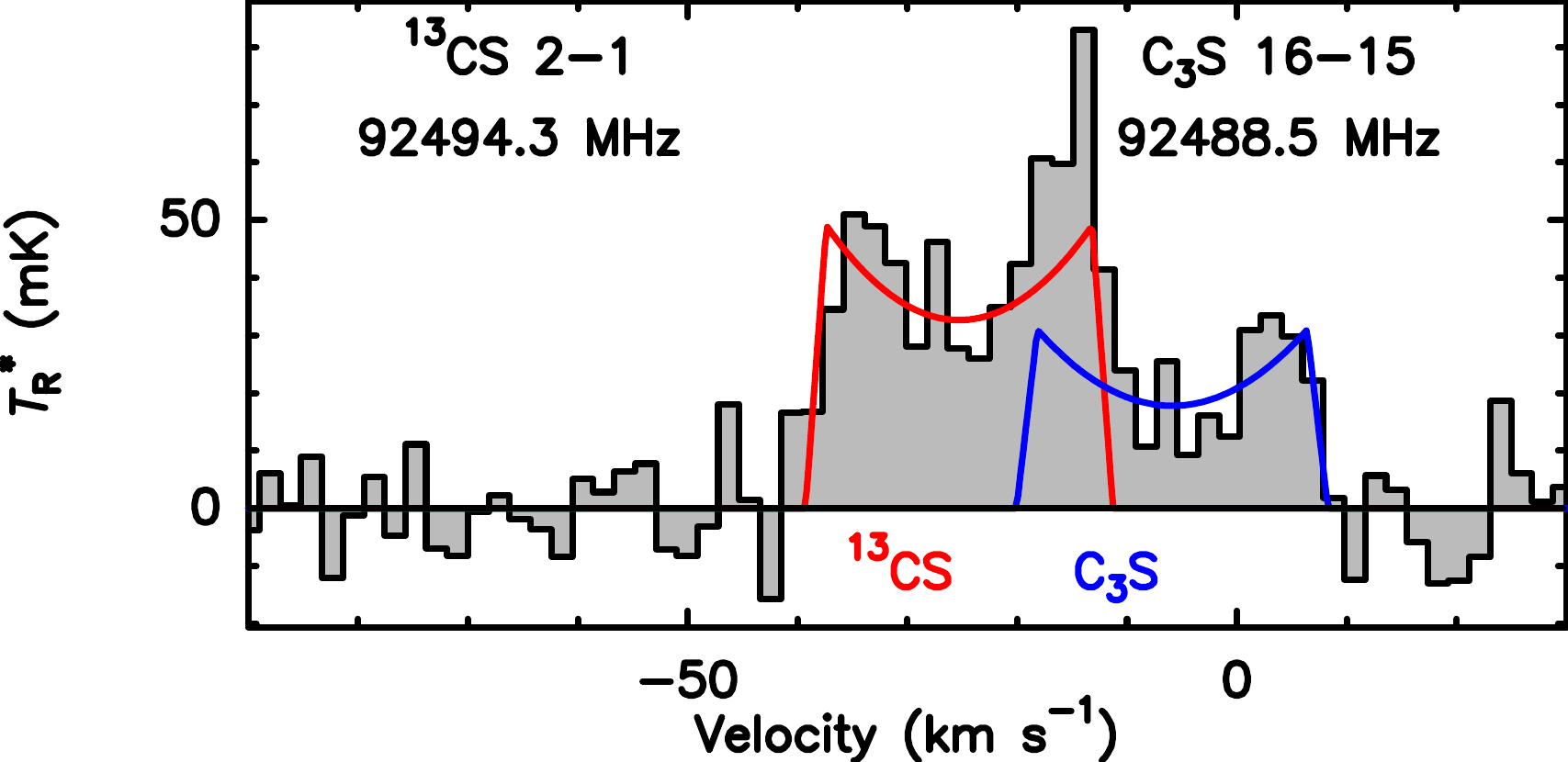}
\hspace{0.05\textwidth}
\includegraphics[width = 0.45 \textwidth]{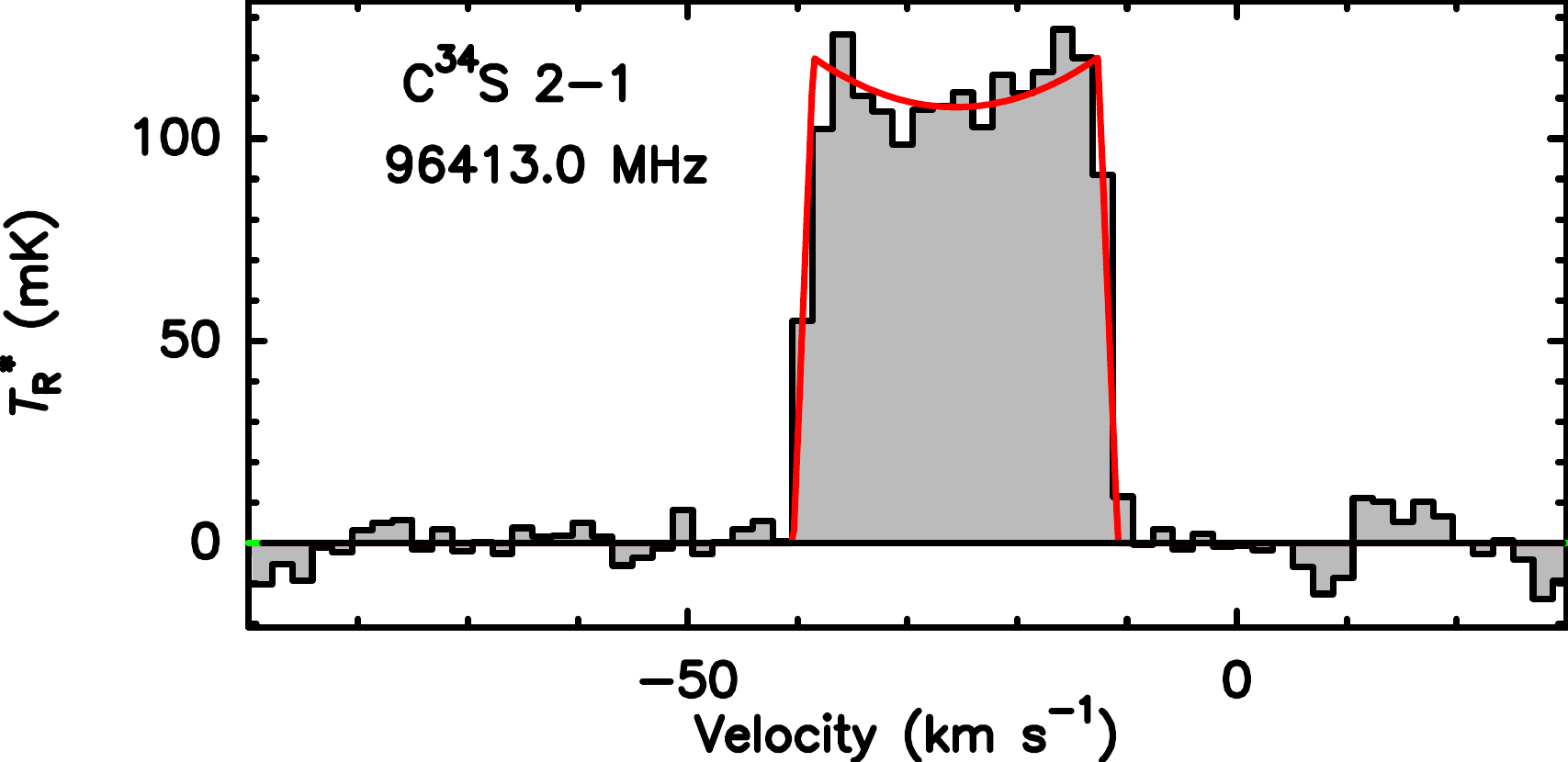}
\vspace{0.1cm}
\includegraphics[width = 0.45 \textwidth]{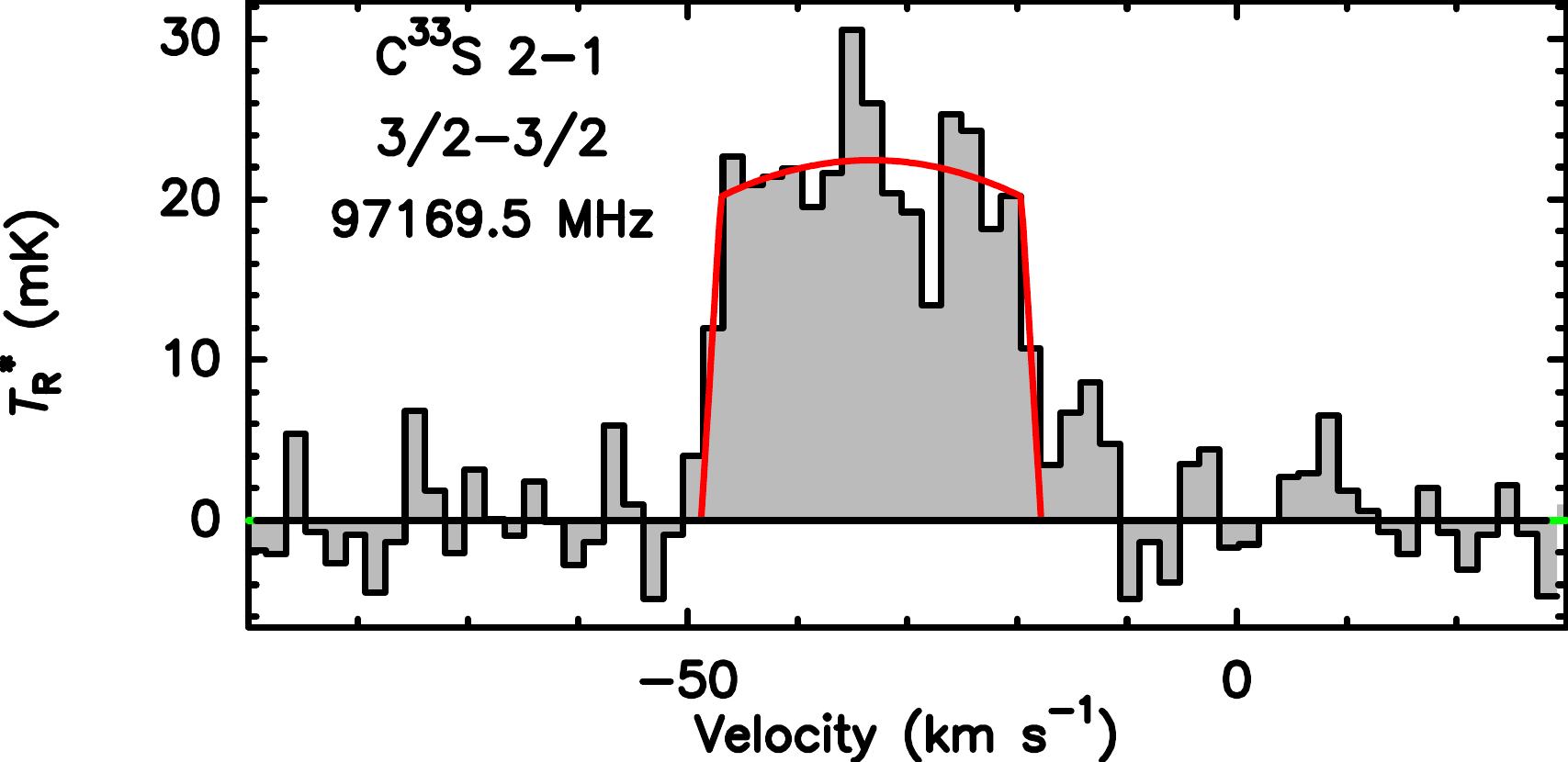}
\hspace{0.05\textwidth}
\includegraphics[width = 0.45 \textwidth]{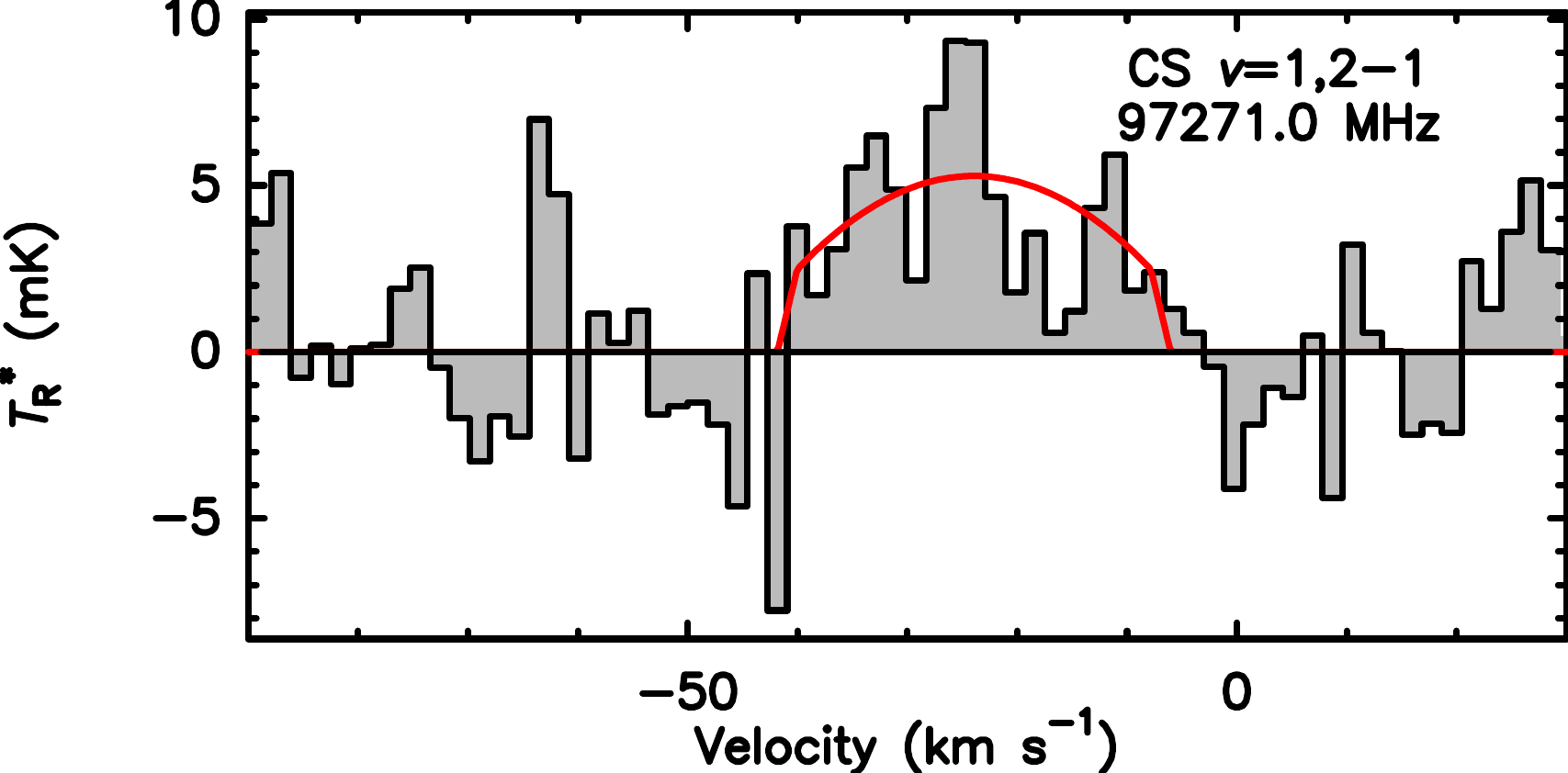}
\includegraphics[width = 0.45 \textwidth]{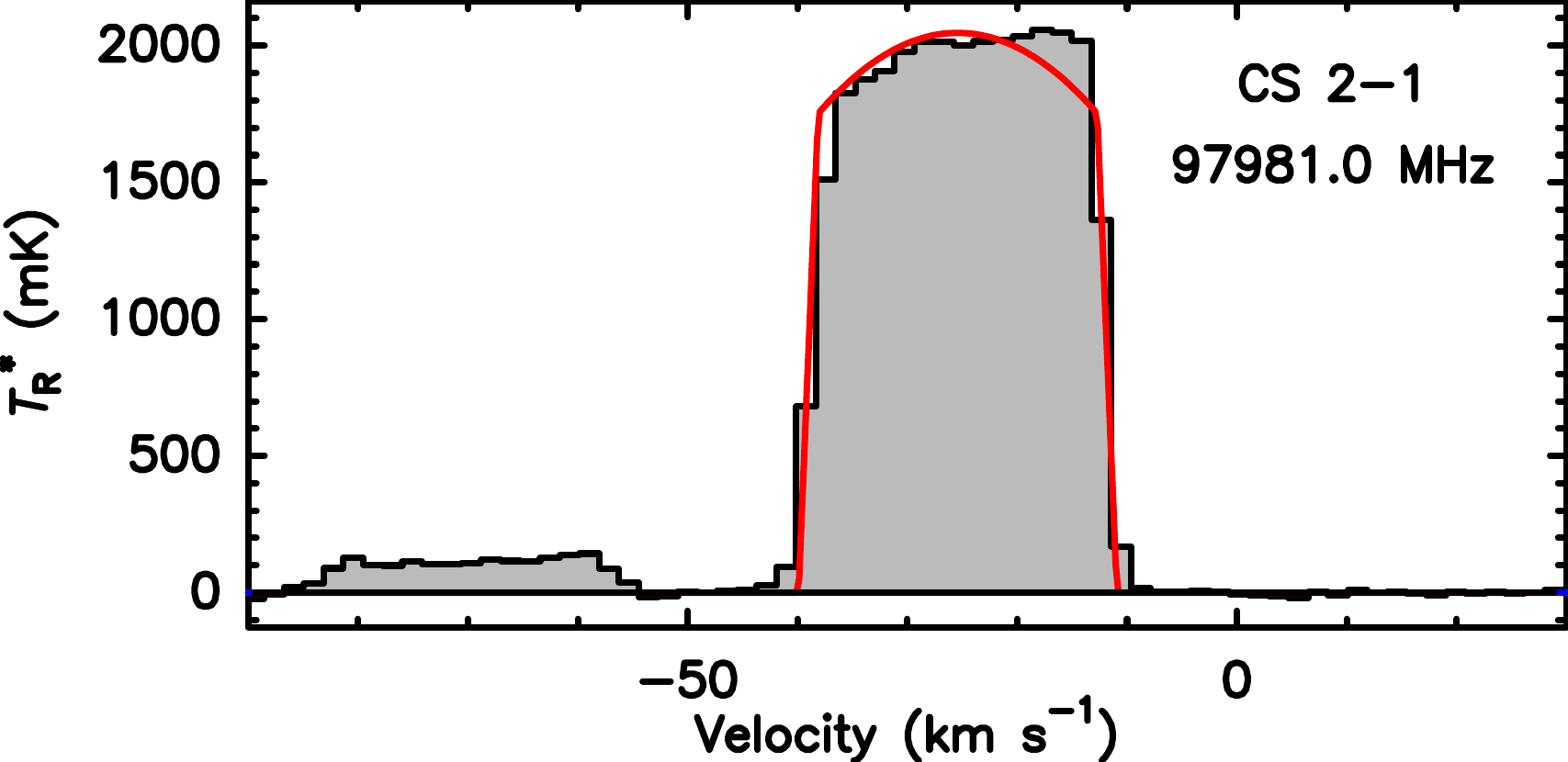}
\vspace{0.1cm}
\caption{{Same as Figure.~\ref{Fig:fitting_1}, but for CS and its isotopologues.}\label{Fig:fitting_2}}
\end{figure*}

\begin{figure*}[!htbp]
\centering
\includegraphics[width = 0.45 \textwidth]{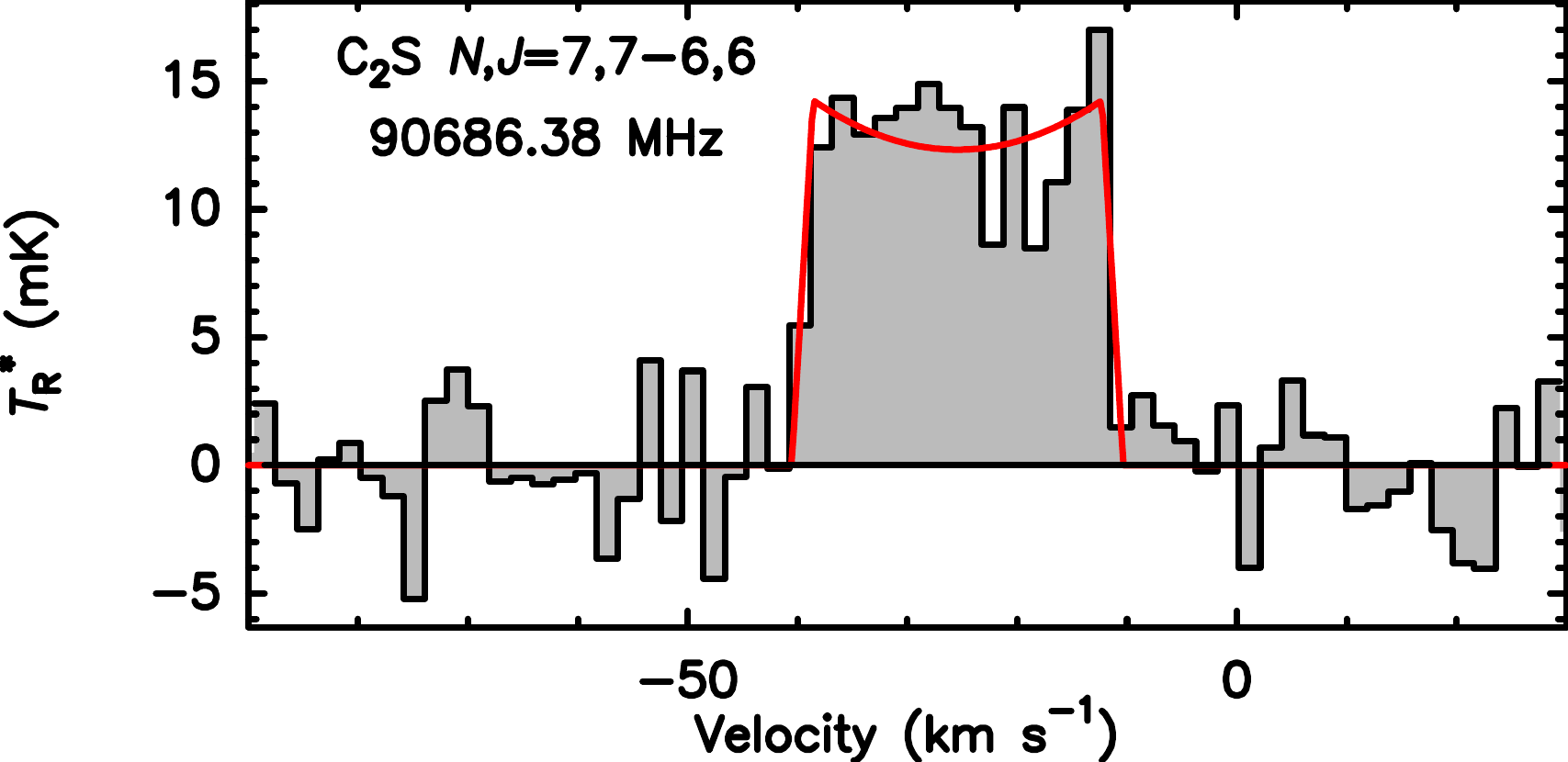}
\hspace{0.05\textwidth}
\includegraphics[width = 0.45 \textwidth]{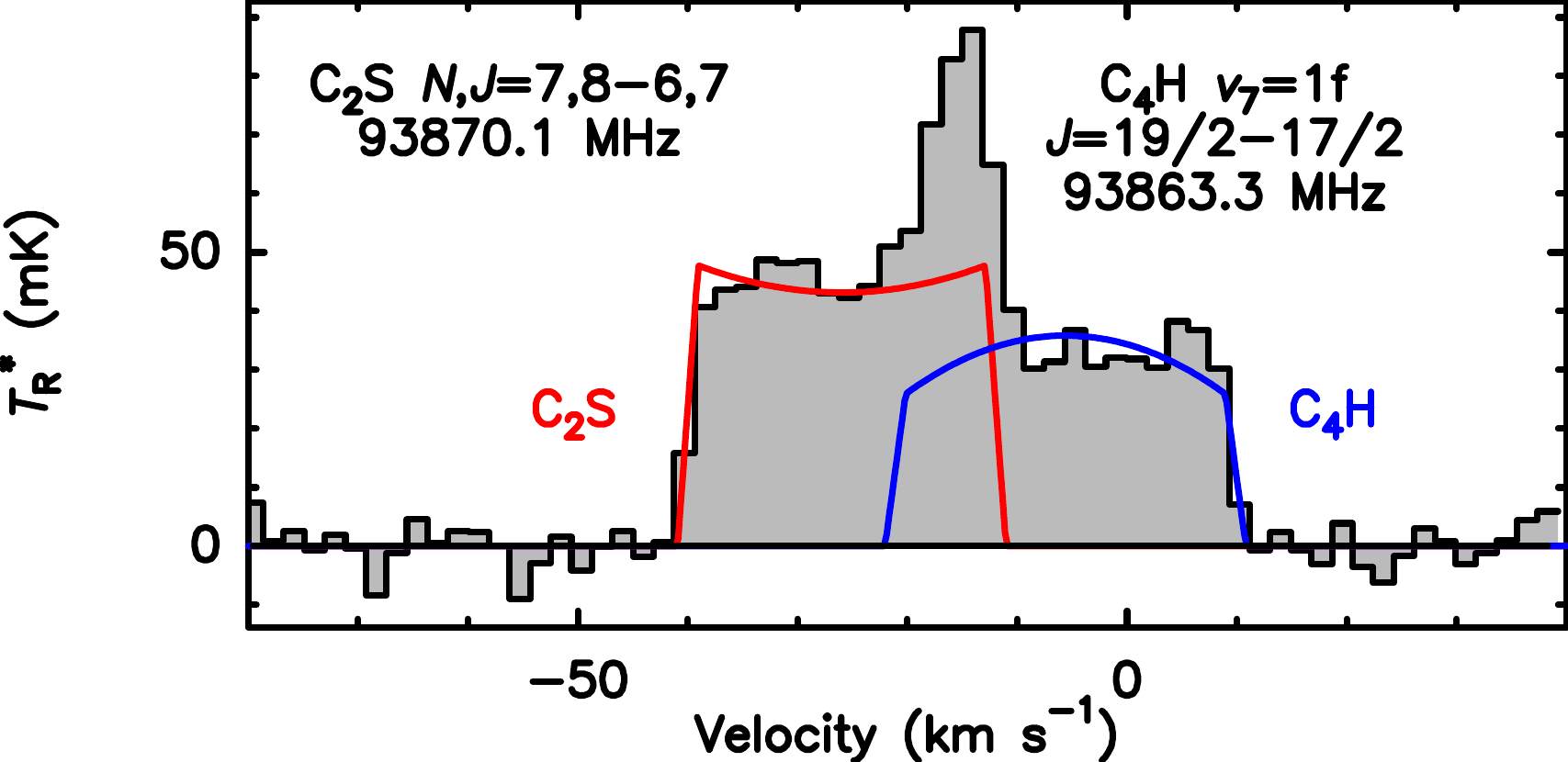}
\vspace{0.1cm}
\includegraphics[width = 0.45 \textwidth]{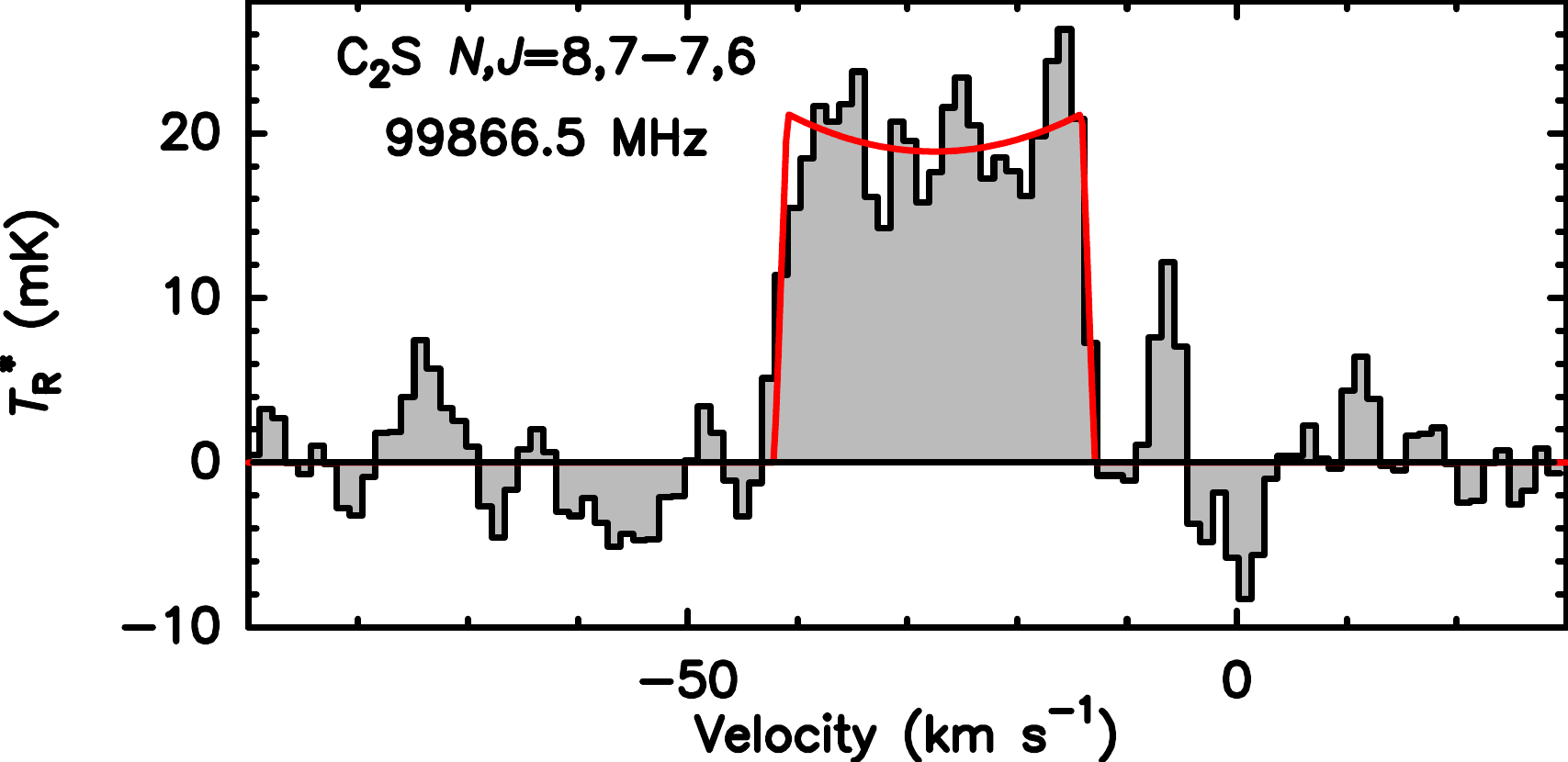}
\hspace{0.05\textwidth}
\includegraphics[width = 0.45 \textwidth]{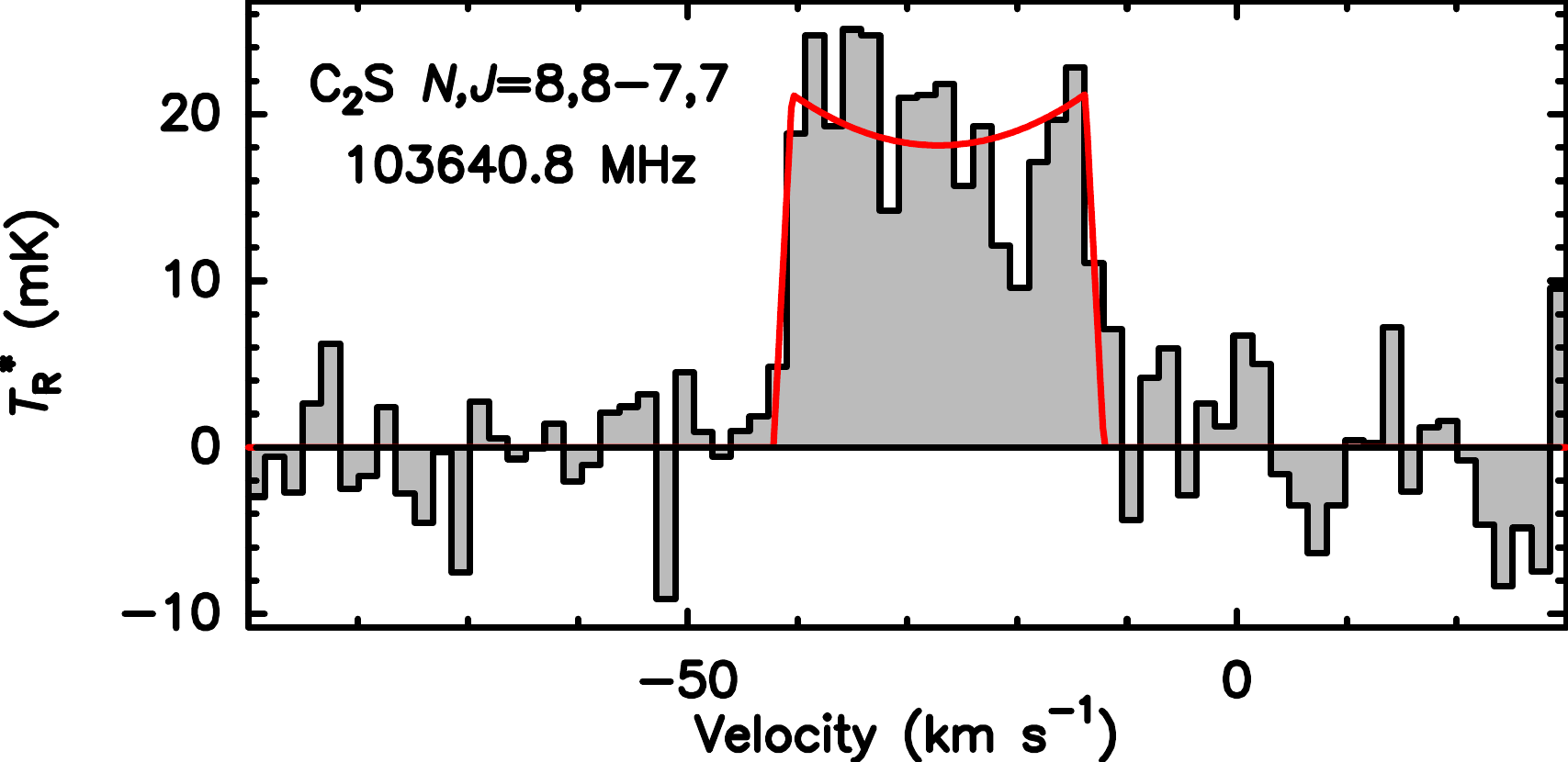}
\vspace{0.1cm}
\includegraphics[width = 0.45 \textwidth]{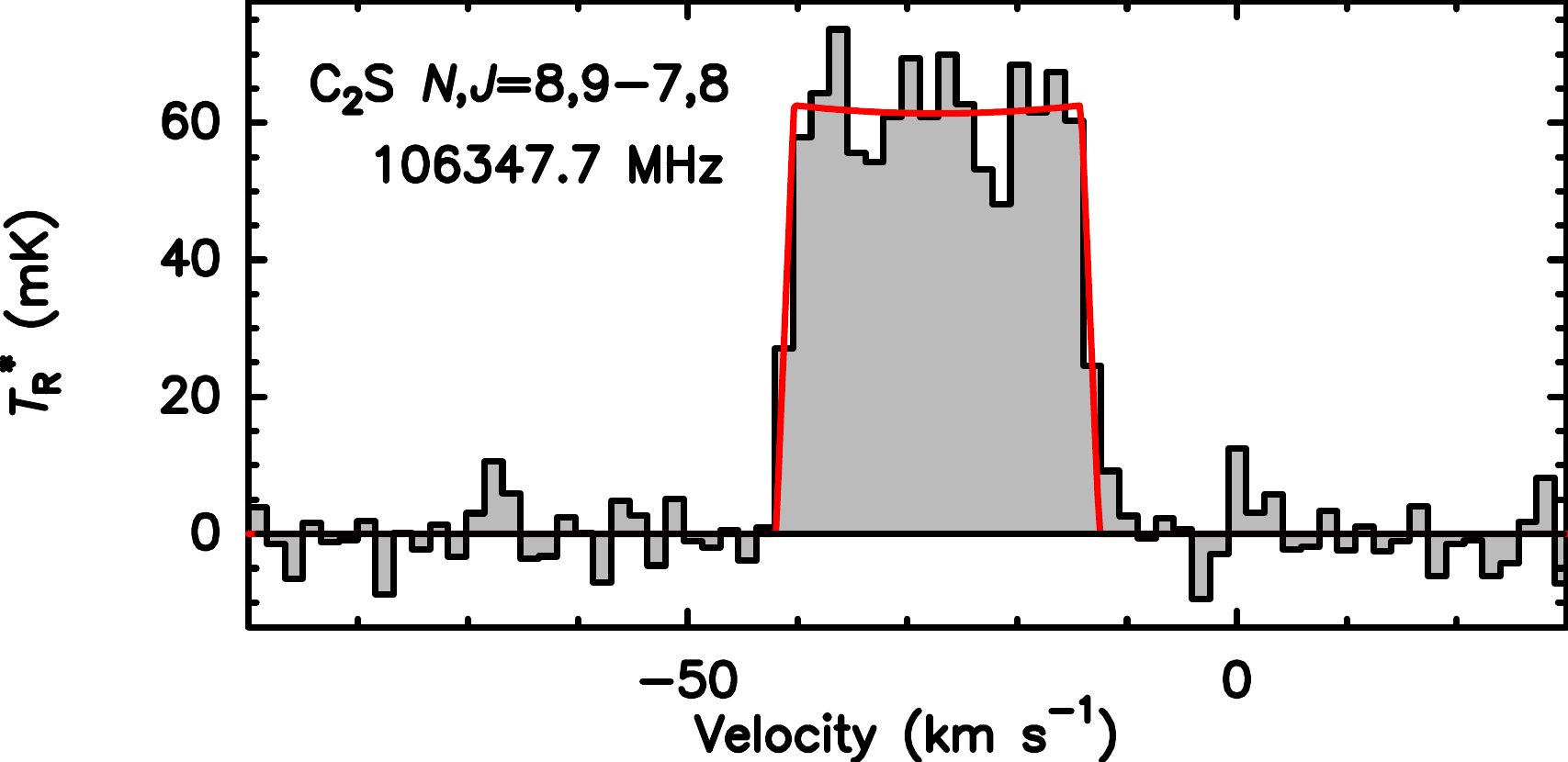}
\hspace{0.05\textwidth}
\includegraphics[width = 0.45 \textwidth]{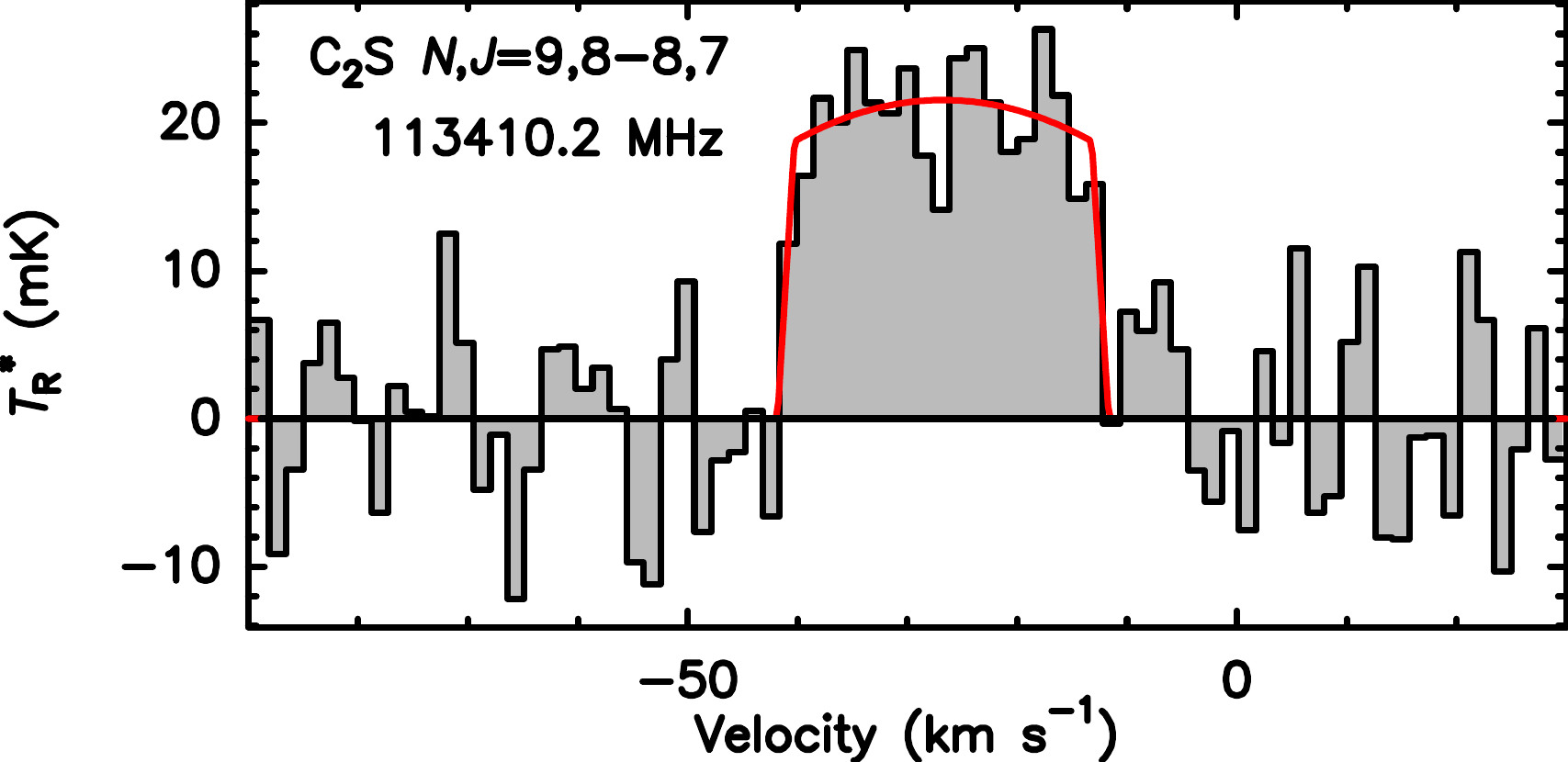}
\vspace{0.1cm}
\includegraphics[width = 0.45 \textwidth]{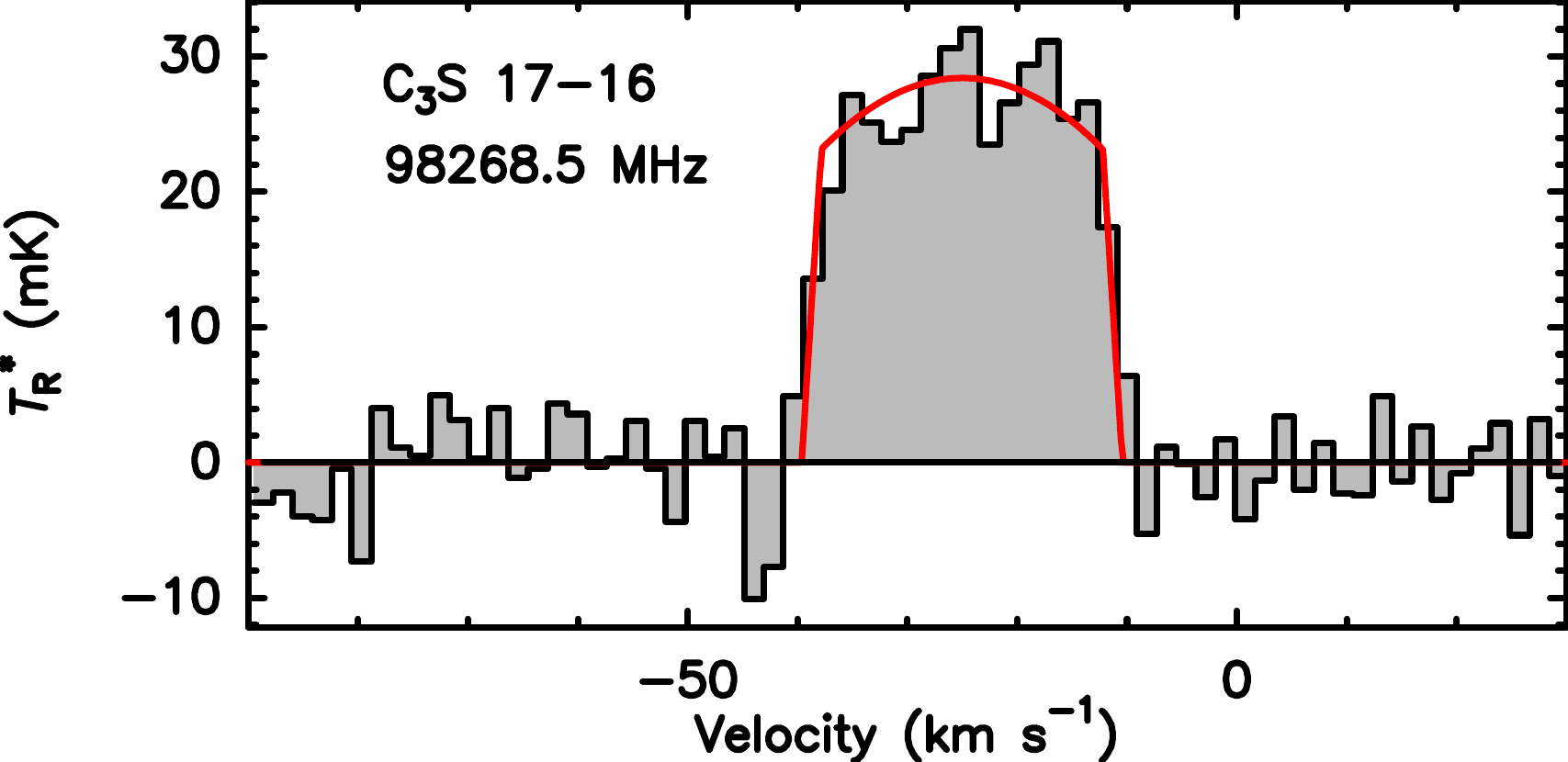}
\hspace{0.05\textwidth}
\includegraphics[width = 0.45 \textwidth]{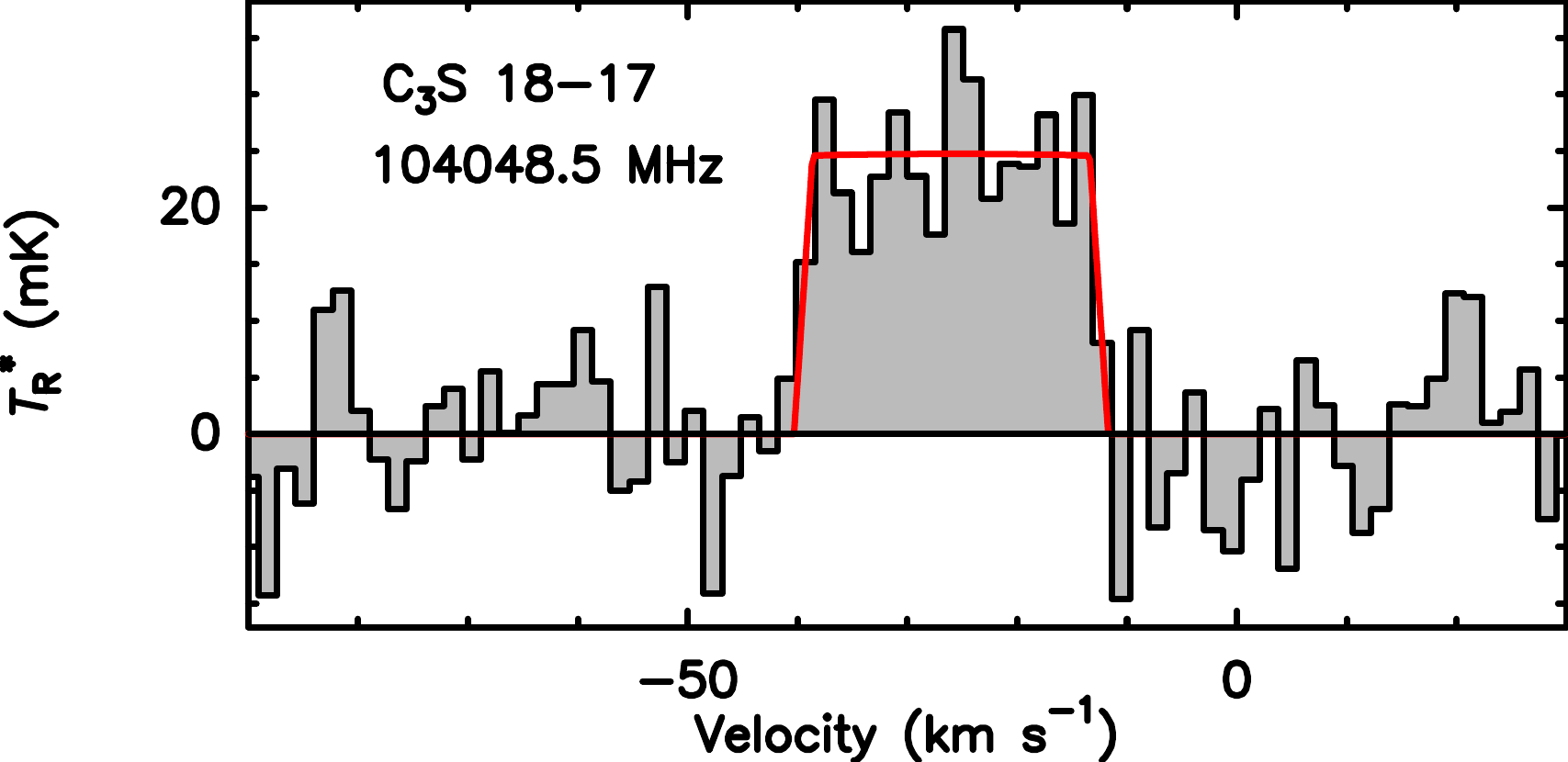}
\vspace{0.1cm}
\includegraphics[width = 0.45 \textwidth]{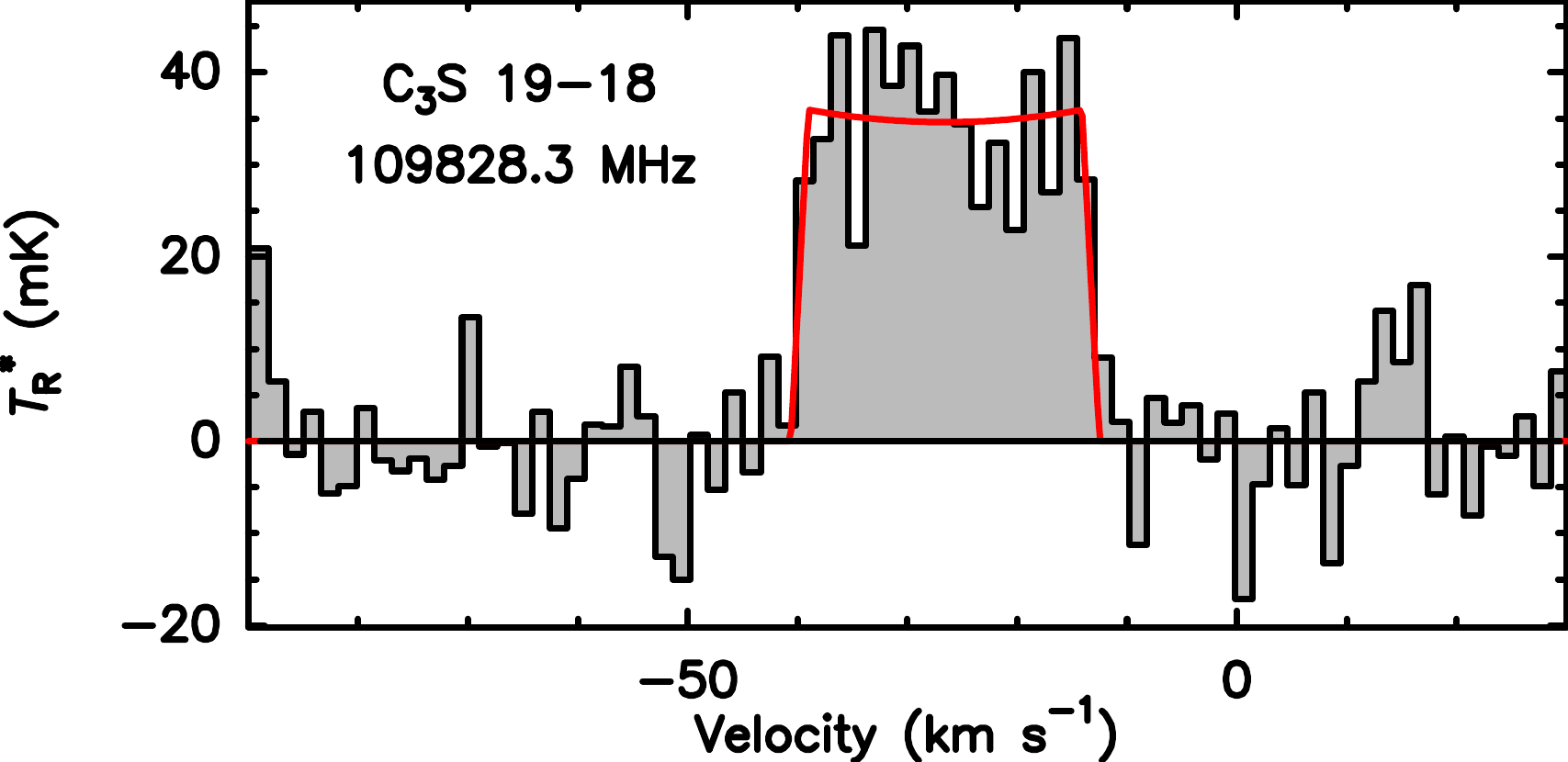}
\hspace{0.05\textwidth}
\includegraphics[width = 0.45 \textwidth]{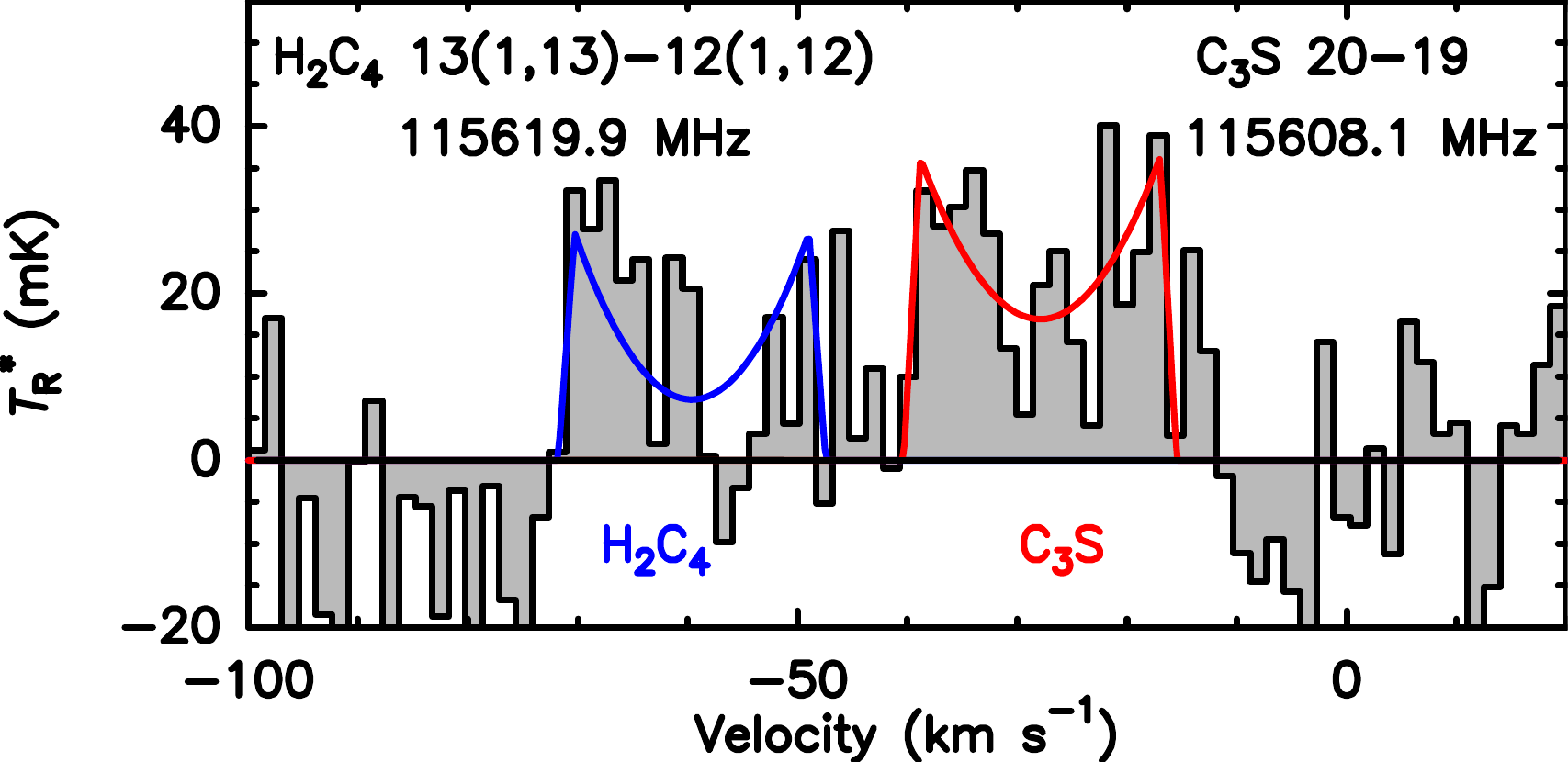}
\caption{{The same as Figure.~\ref{Fig:fitting_1}, but for C$_{2}$S and C$_{3}$S. }\label{Fig:fitting_4}}
\end{figure*}

\begin{figure*}[!htbp]
\centering
\includegraphics[width = 0.45 \textwidth]{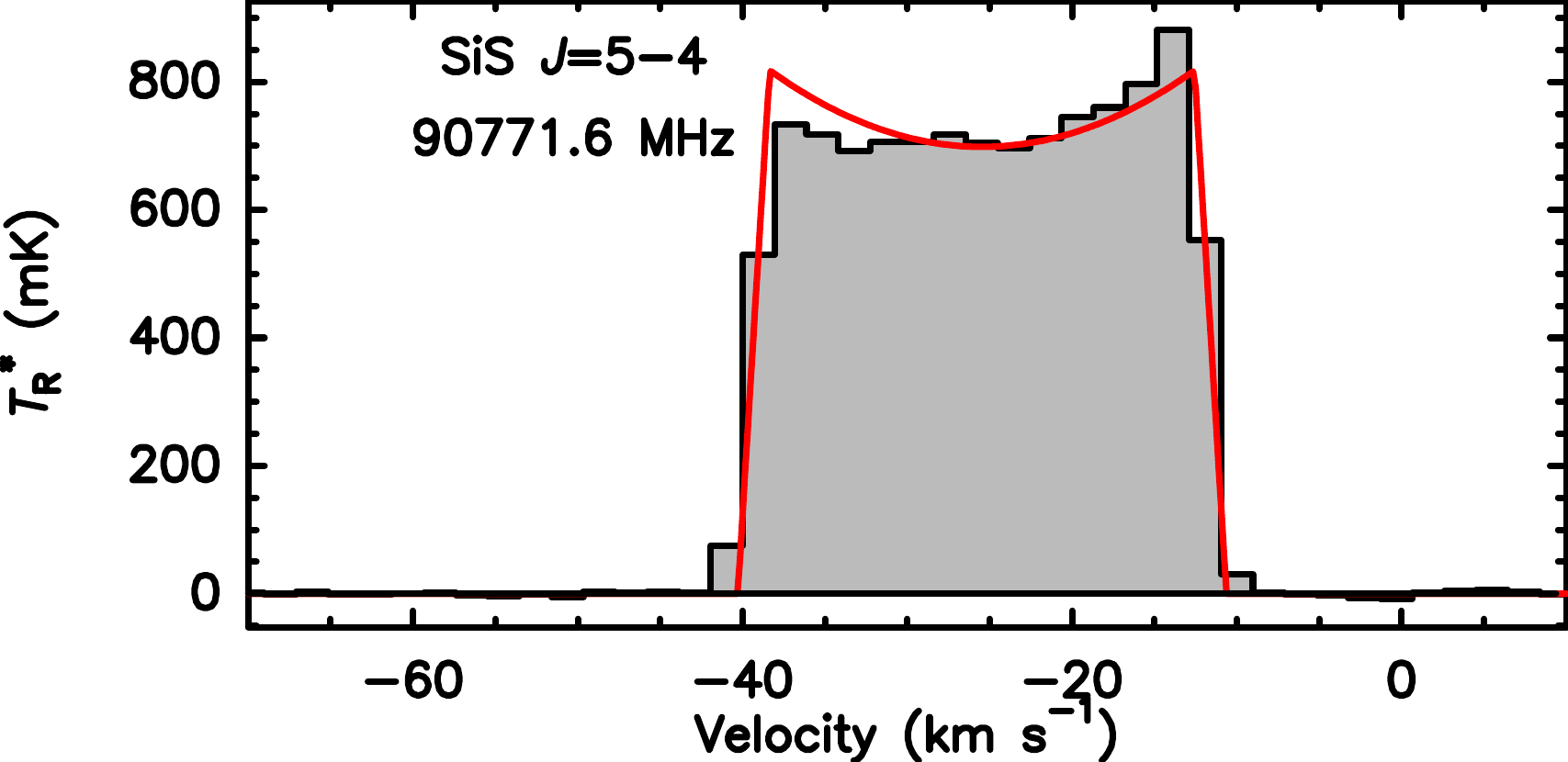}
\hspace{0.05\textwidth}
\includegraphics[width = 0.45 \textwidth]{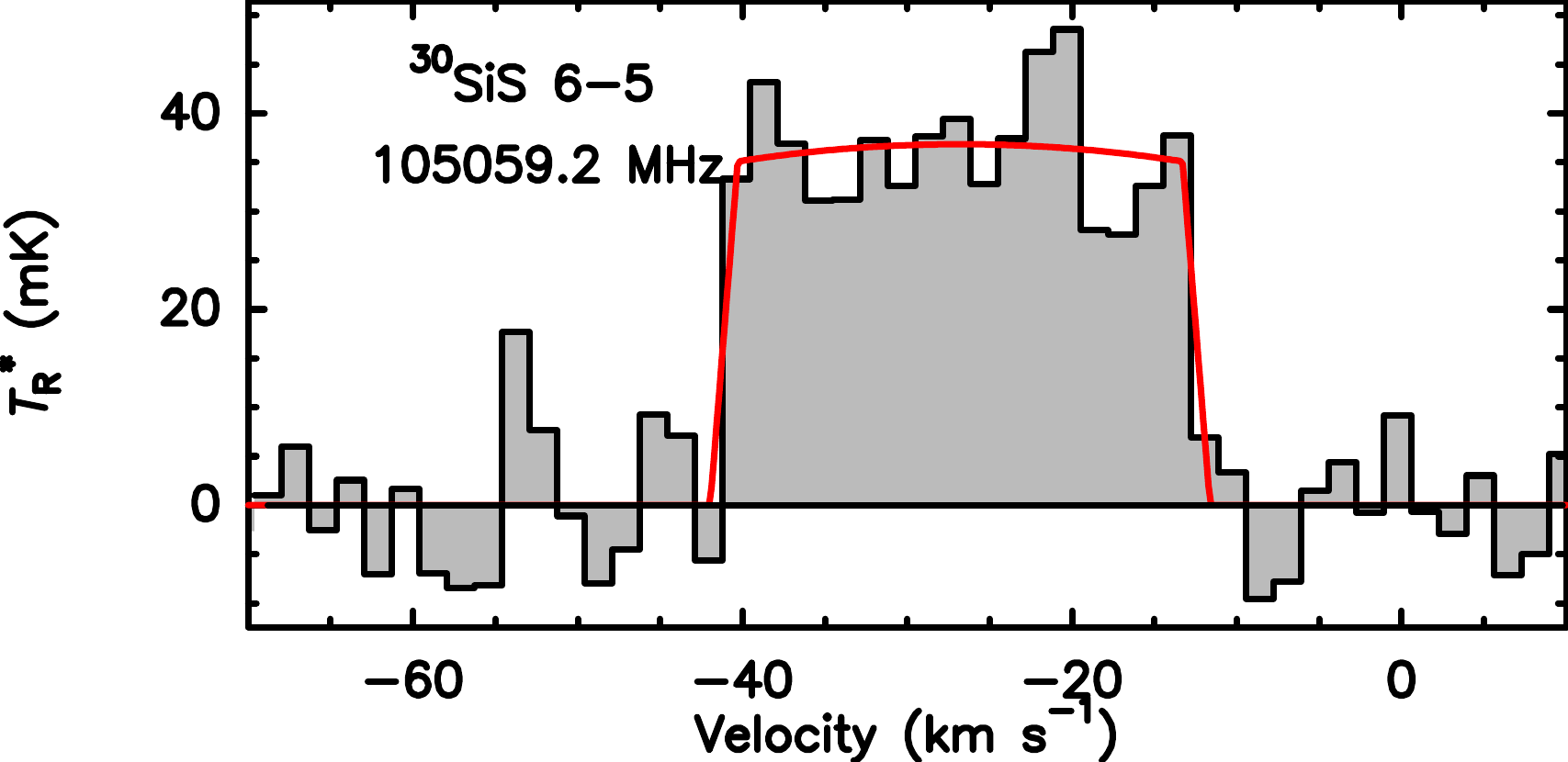}
\vspace{0.1cm}
\includegraphics[width = 0.45 \textwidth]{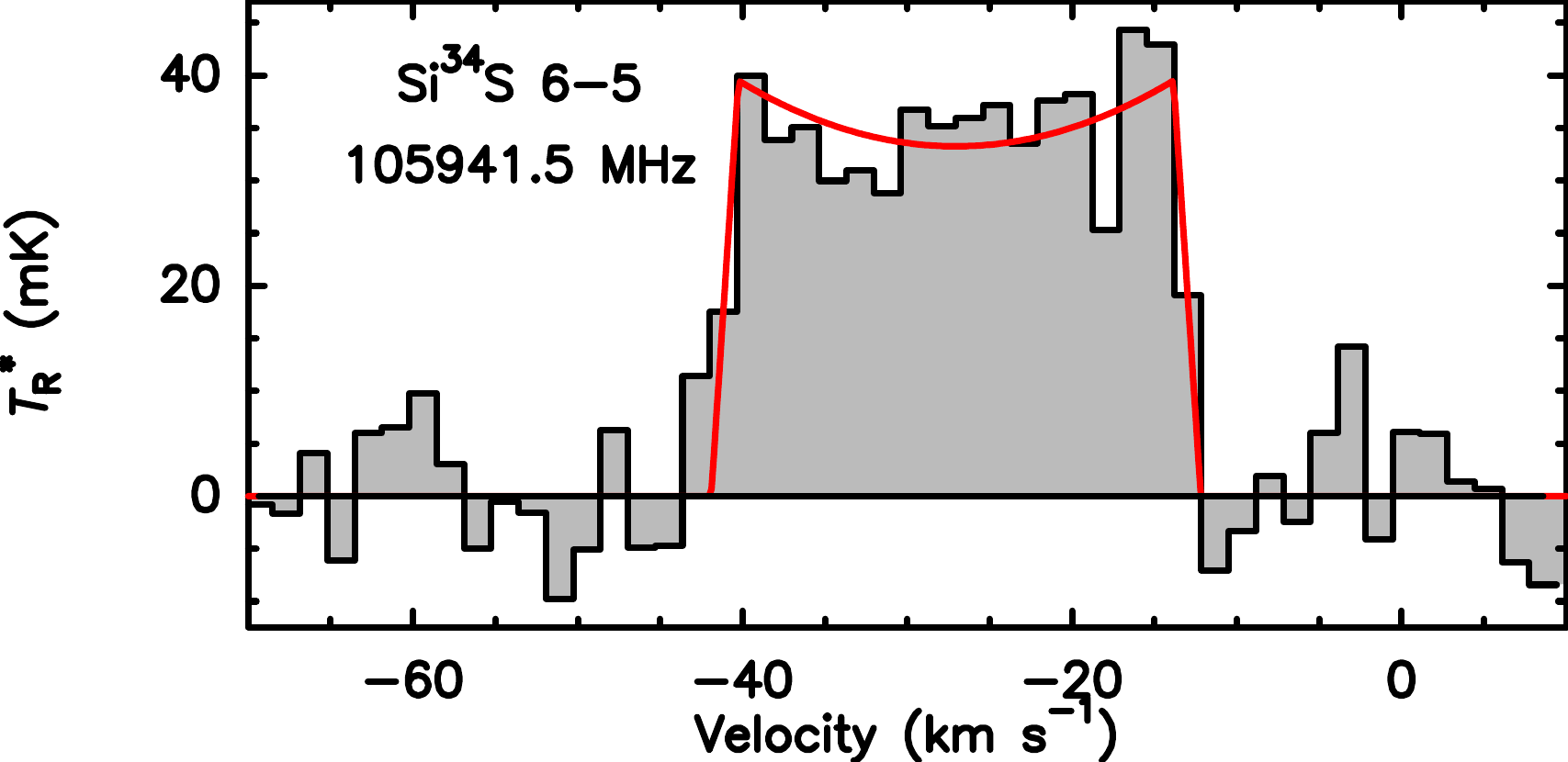}
\hspace{0.05\textwidth}
\includegraphics[width = 0.45 \textwidth]{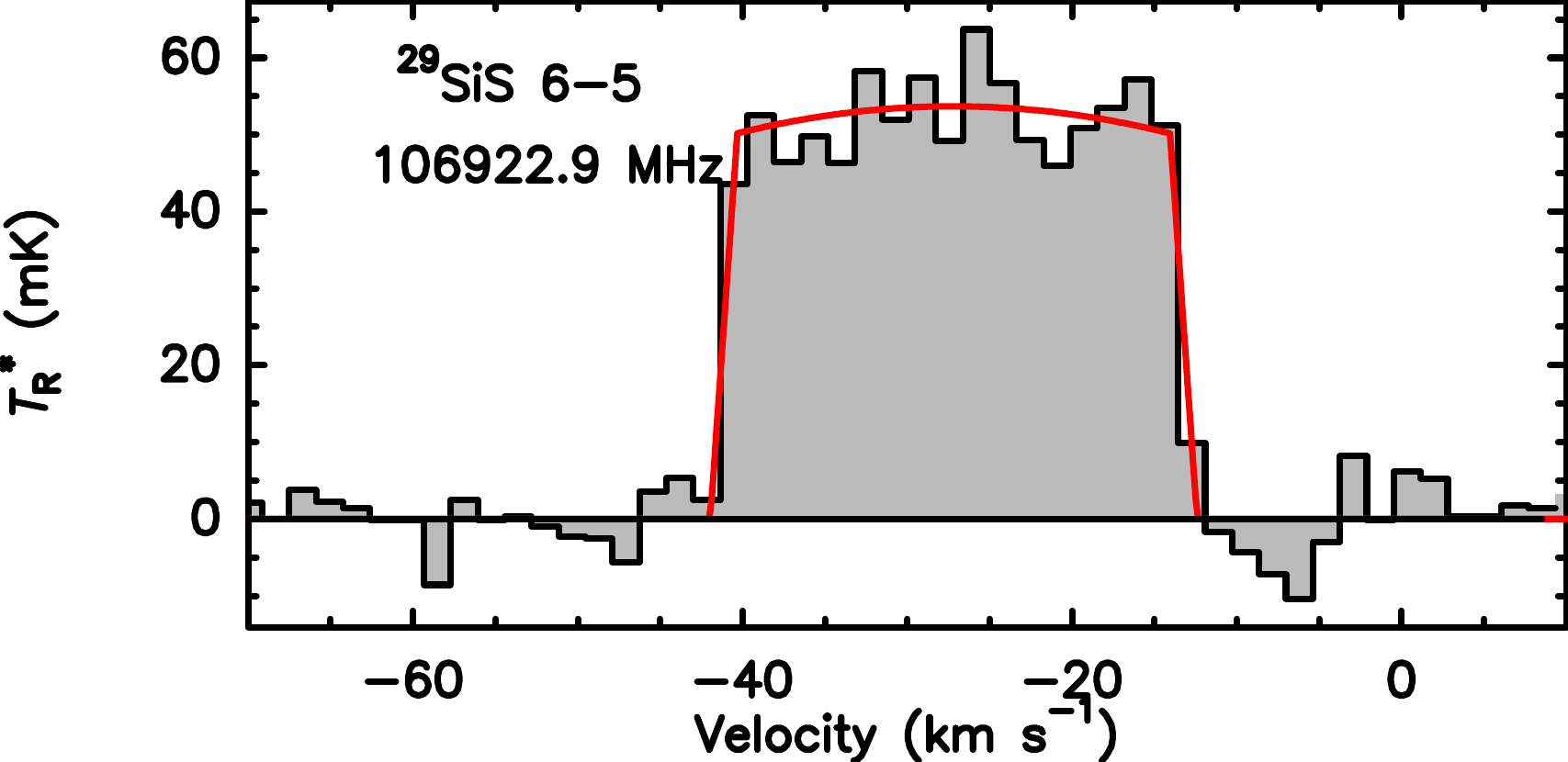}
\includegraphics[width = 0.45 \textwidth]{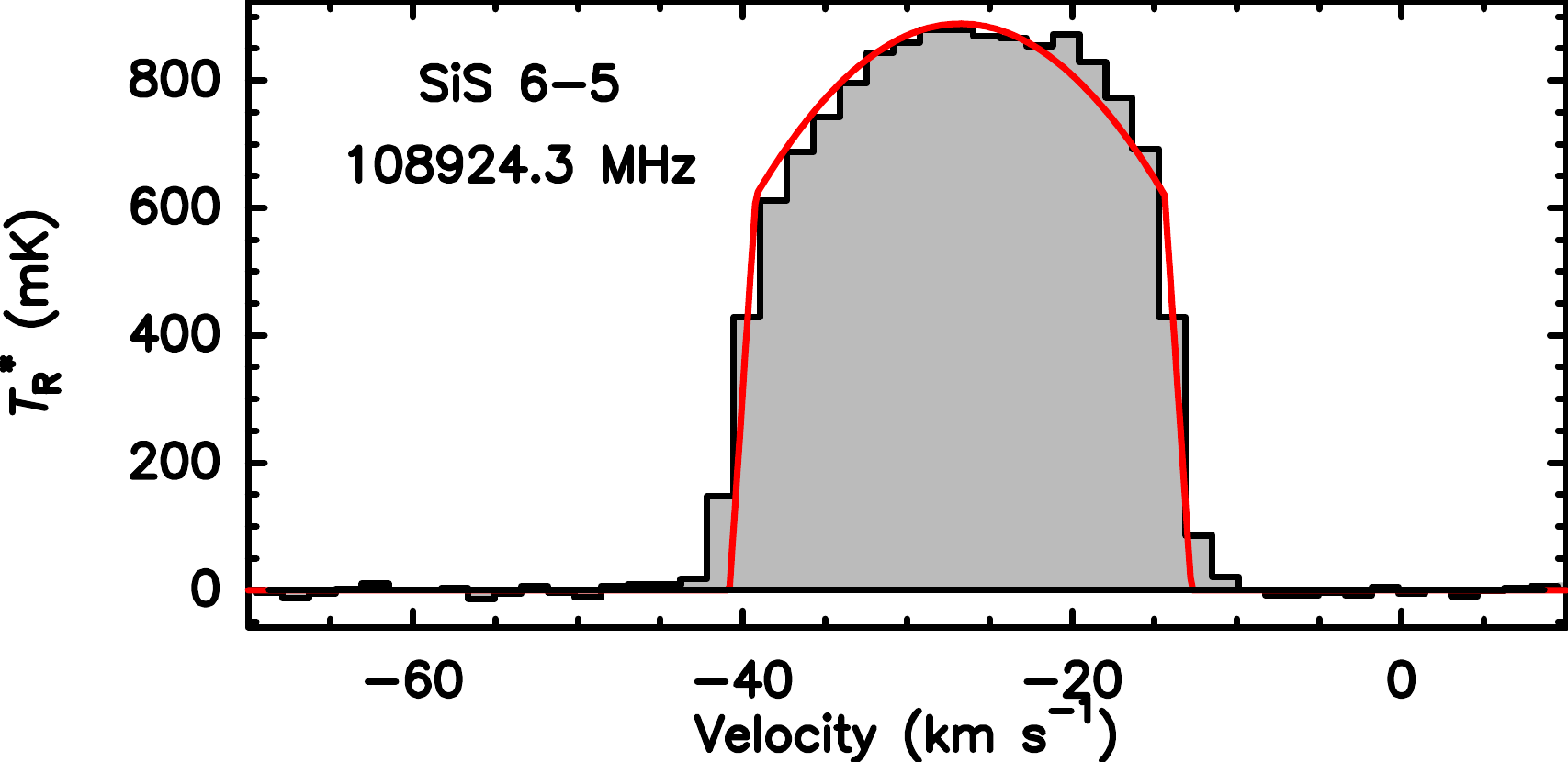}
\caption{{Same as Figure.~\ref{Fig:fitting_1}, but for SiS and its isotopologues.}\label{Fig:fitting_5}}
\end{figure*}

\begin{figure*}[!htbp]
\centering
\includegraphics[width = 0.45 \textwidth]{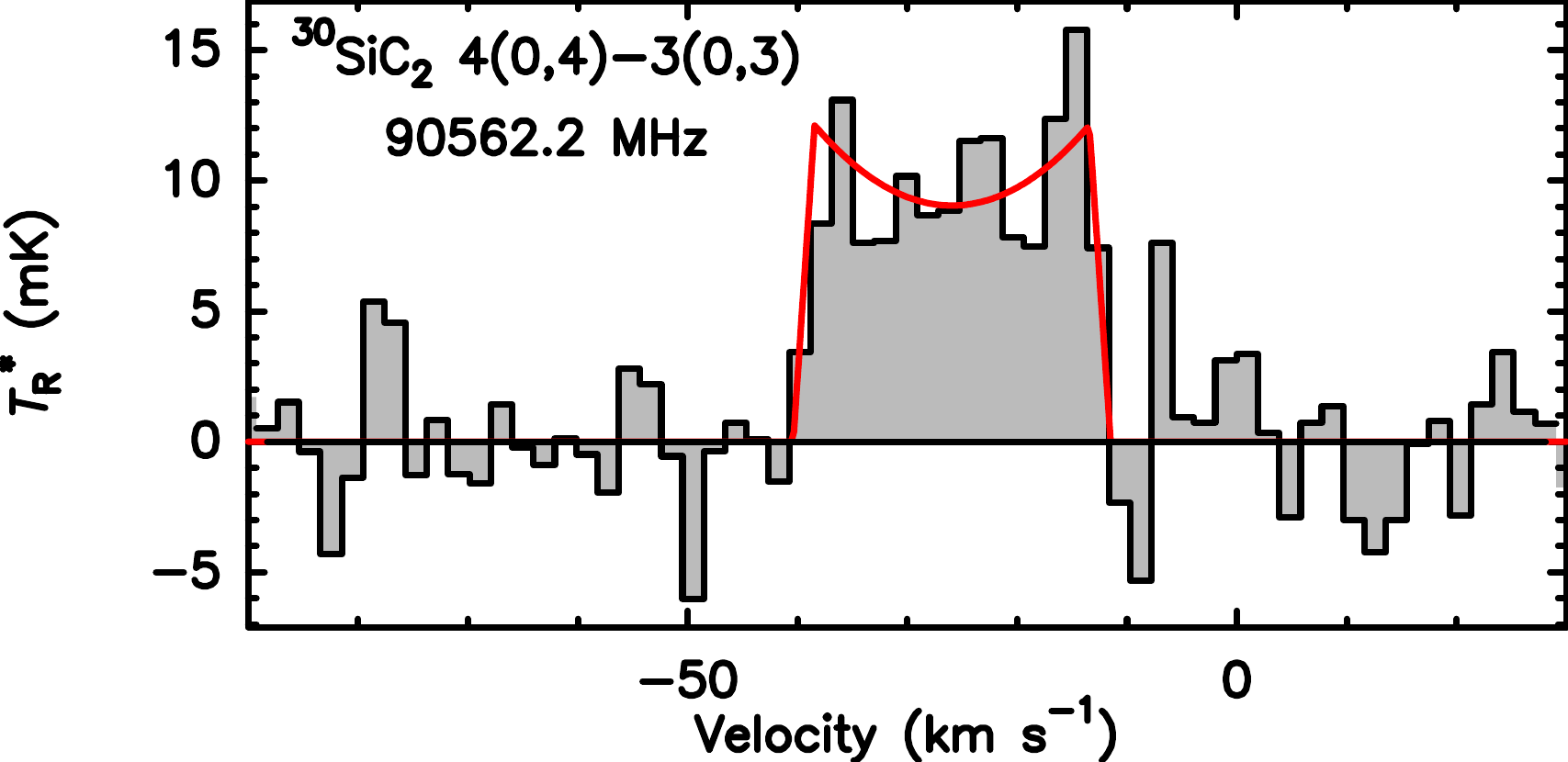}
\vspace{0.1cm}
\includegraphics[width = 0.45 \textwidth]{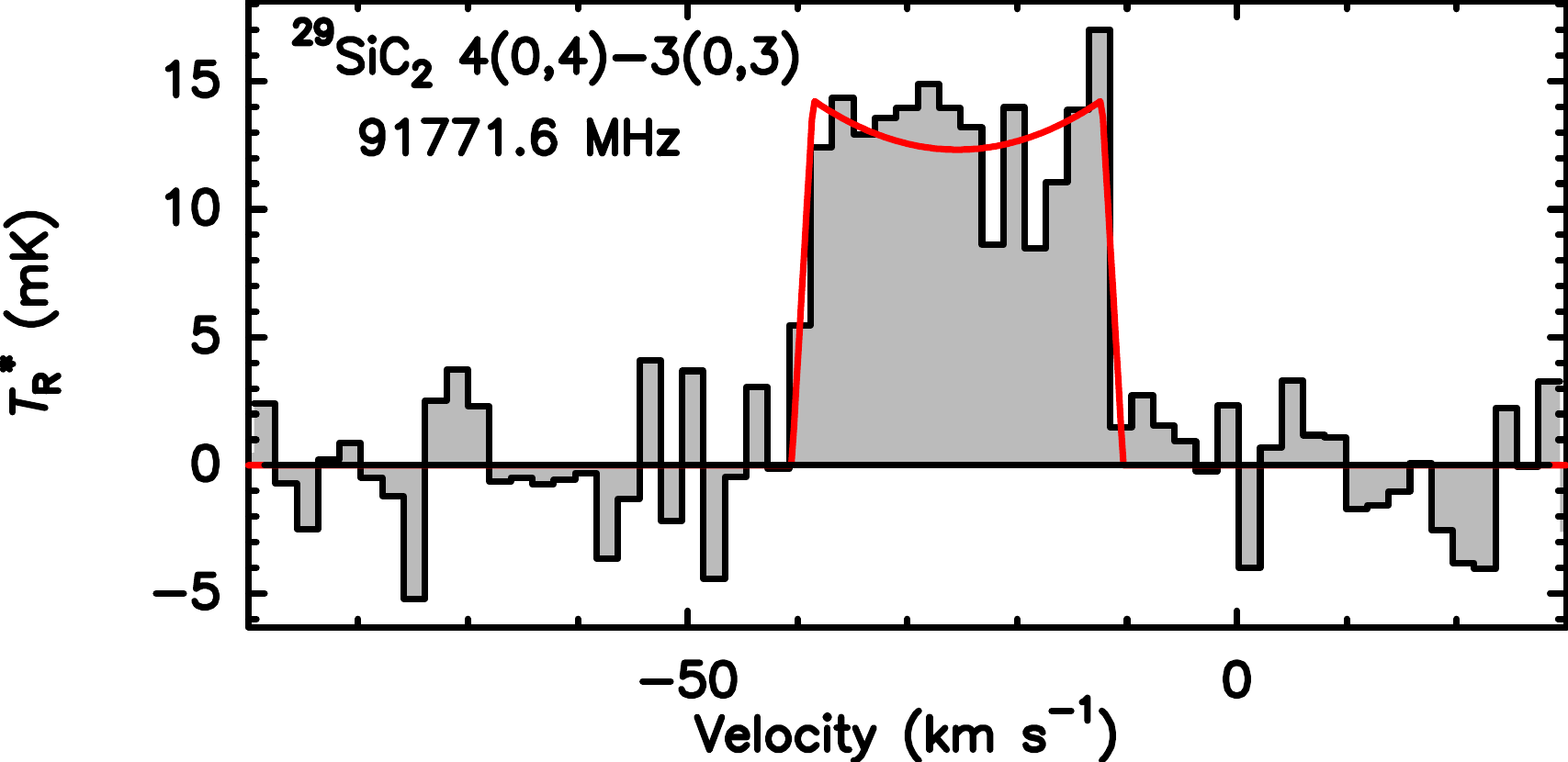}
\hspace{0.05\textwidth}
\includegraphics[width = 0.45 \textwidth]{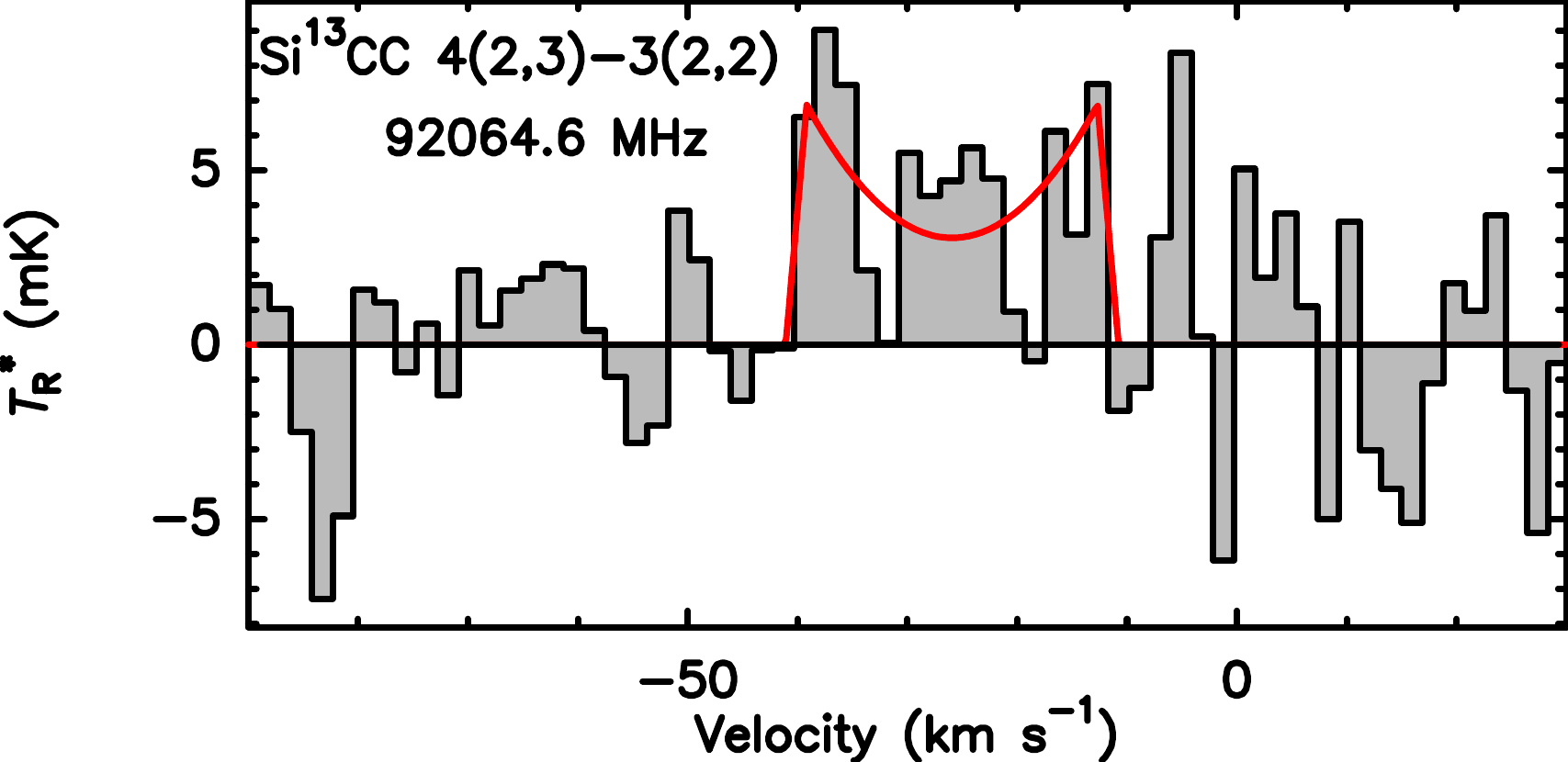}
\vspace{0.1cm}
\includegraphics[width = 0.45 \textwidth]{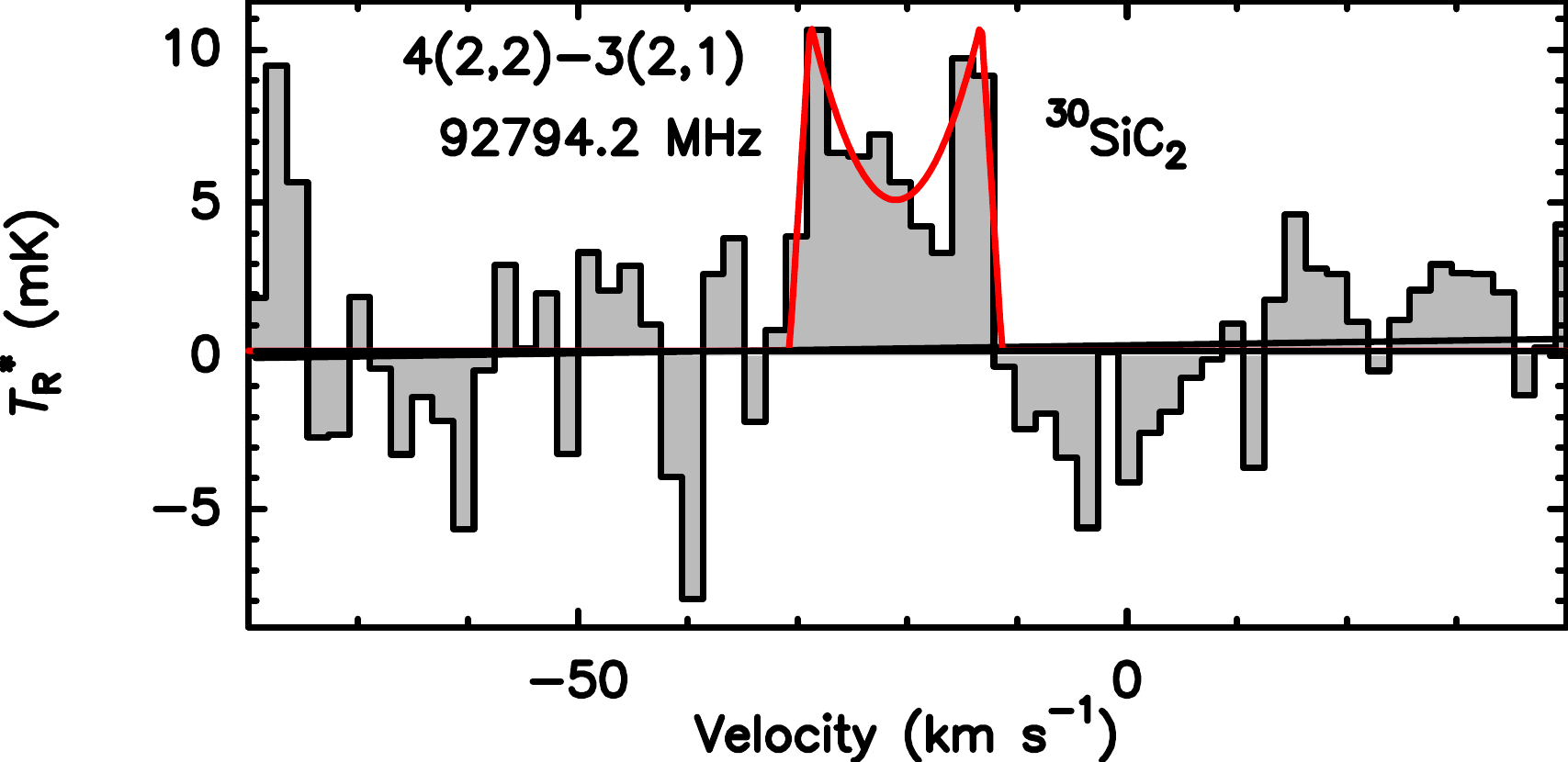}
\hspace{0.05\textwidth}
\includegraphics[width = 0.45 \textwidth]{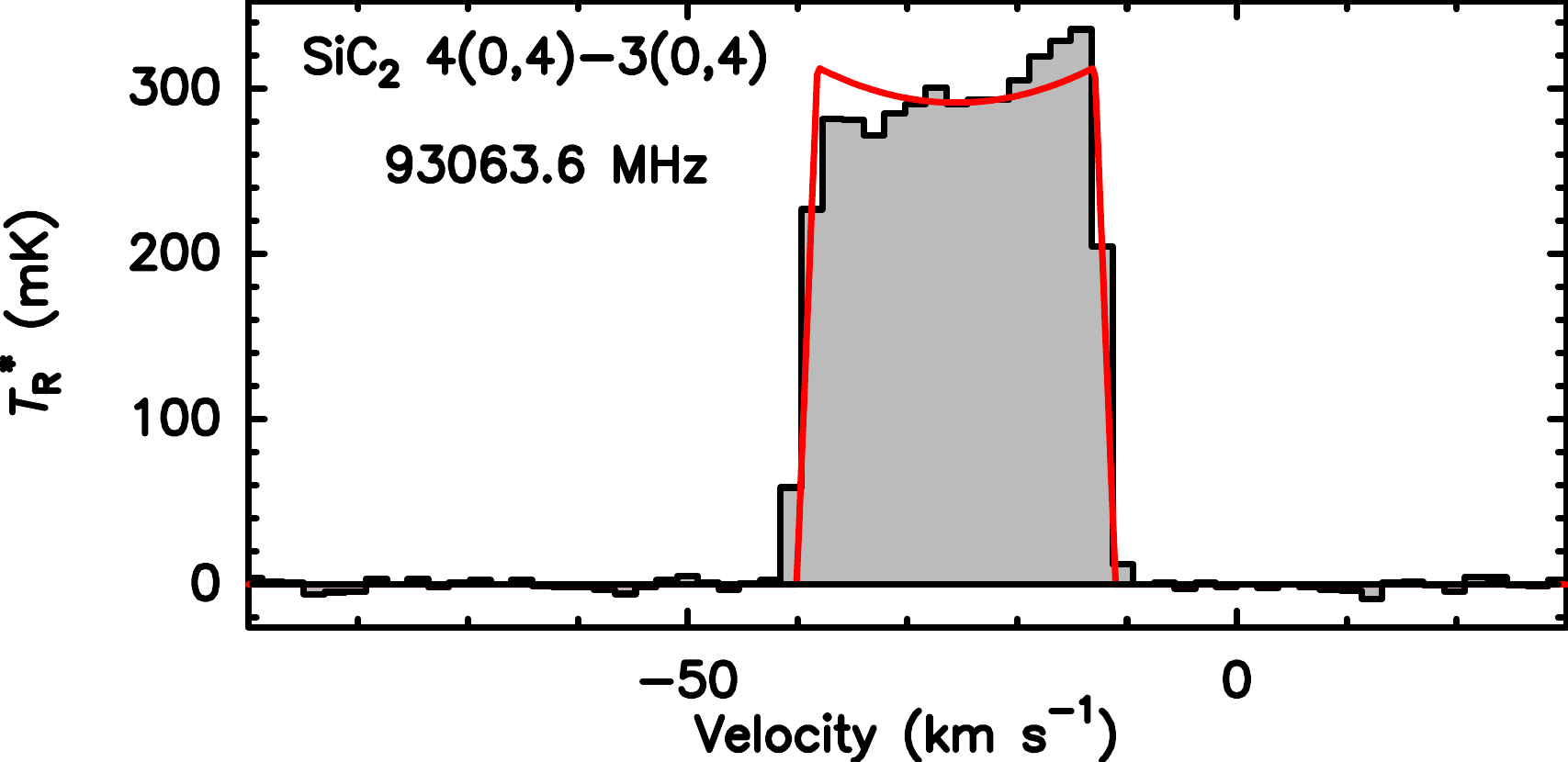}
\vspace{0.1cm}
\includegraphics[width = 0.45 \textwidth]{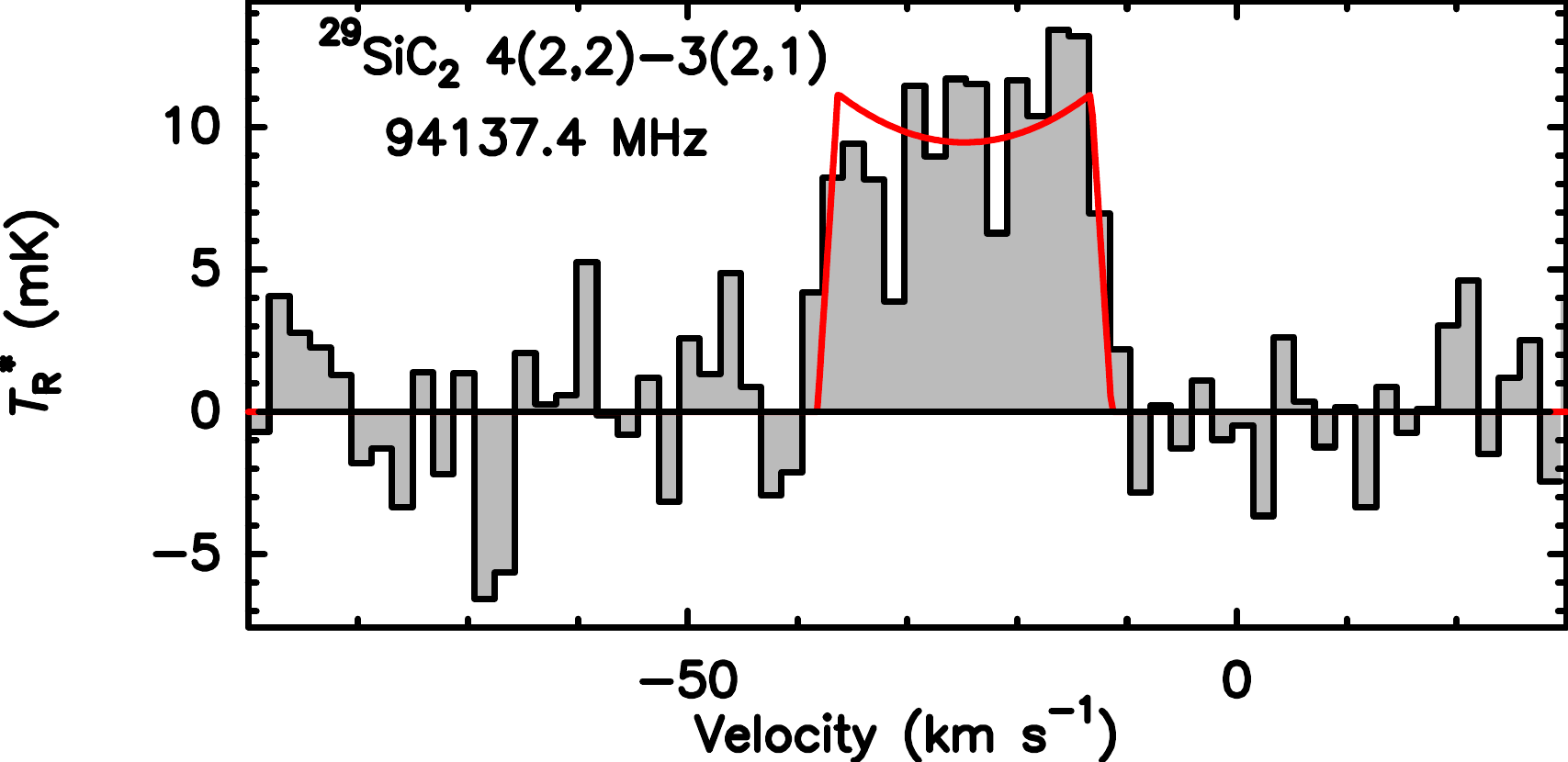}
\hspace{0.05\textwidth}
\includegraphics[width = 0.45 \textwidth]{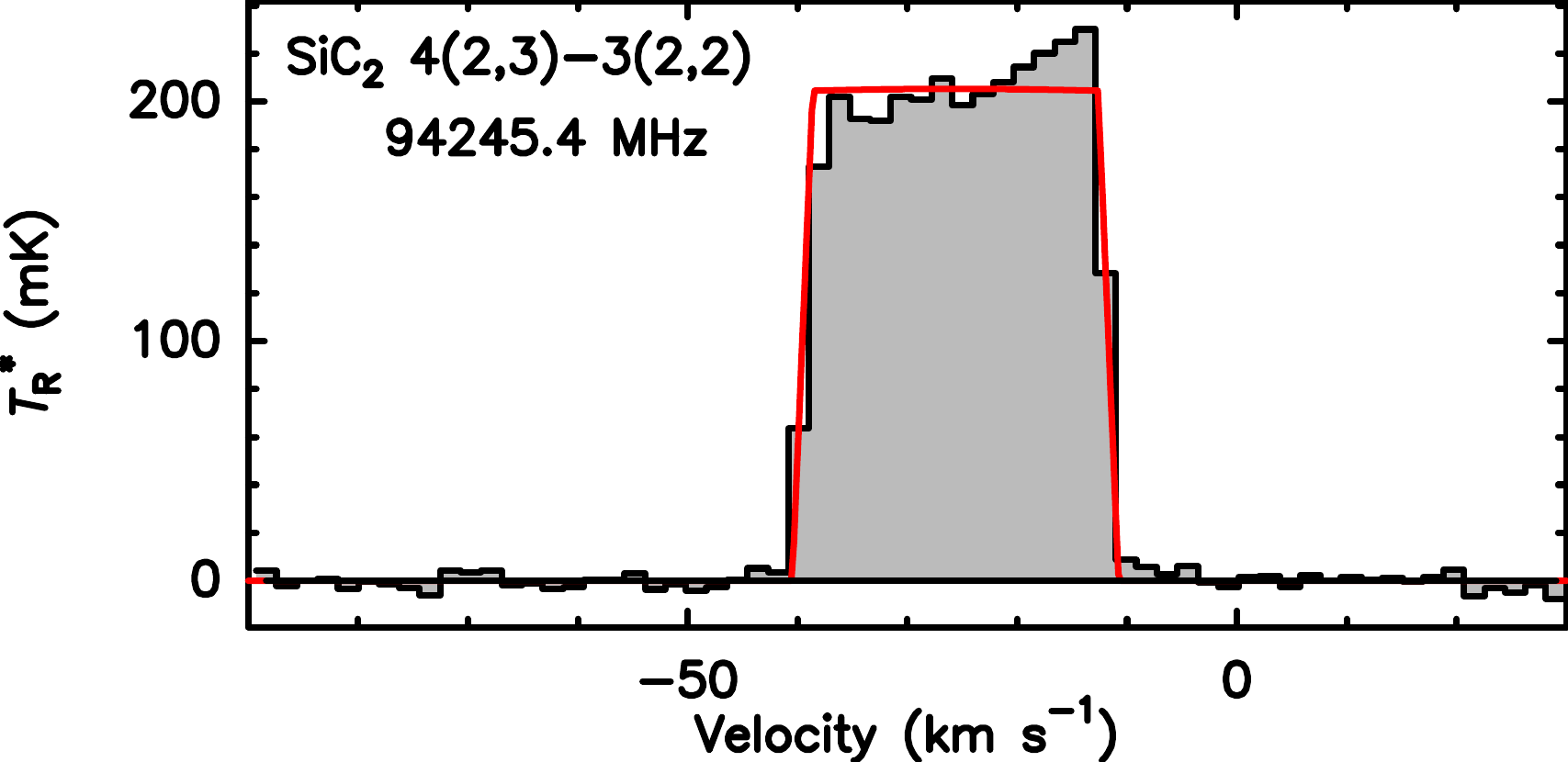}
\vspace{0.1cm}
\includegraphics[width = 0.45 \textwidth]{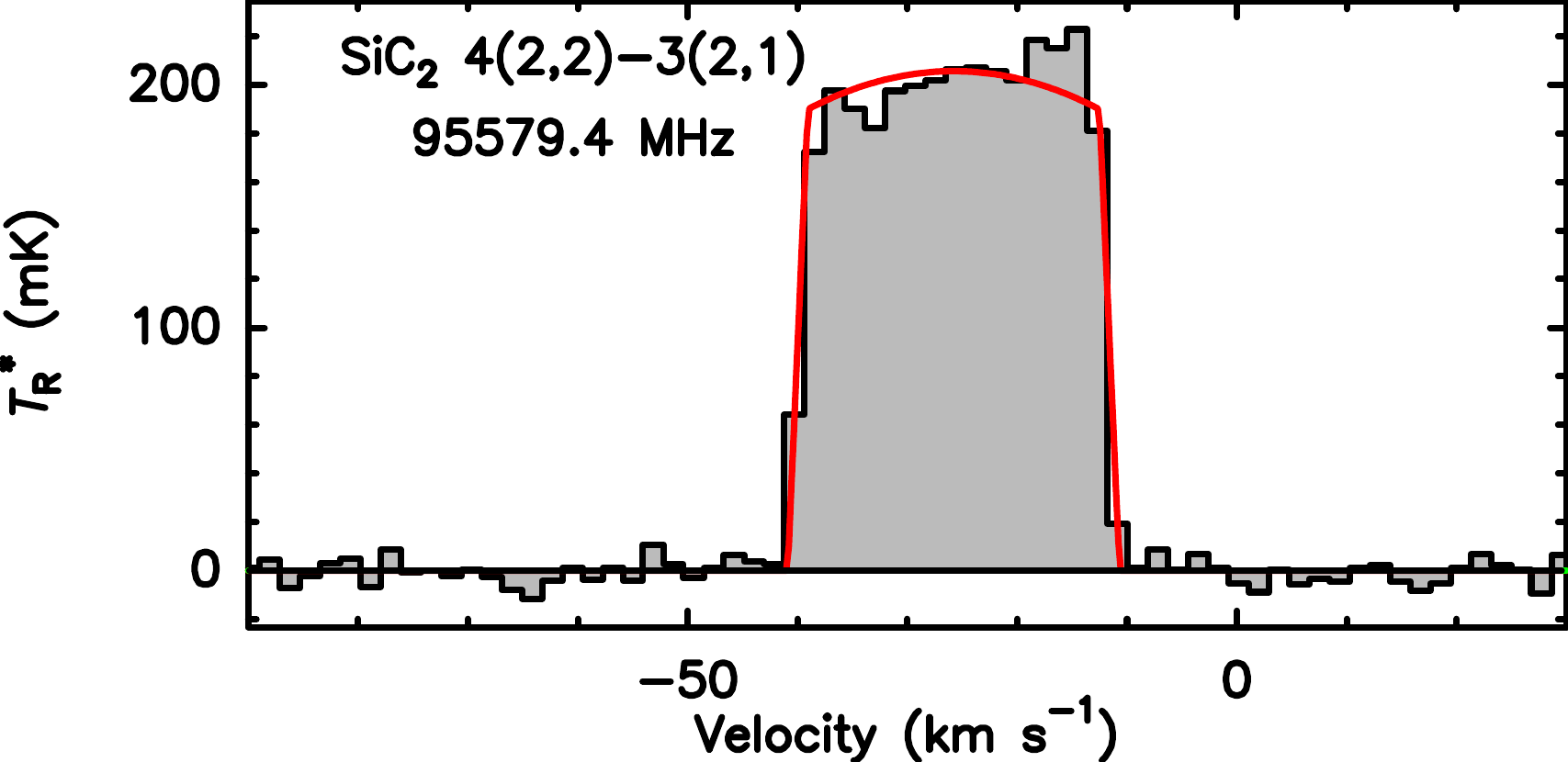}
\hspace{0.05\textwidth}
\includegraphics[width = 0.45 \textwidth]{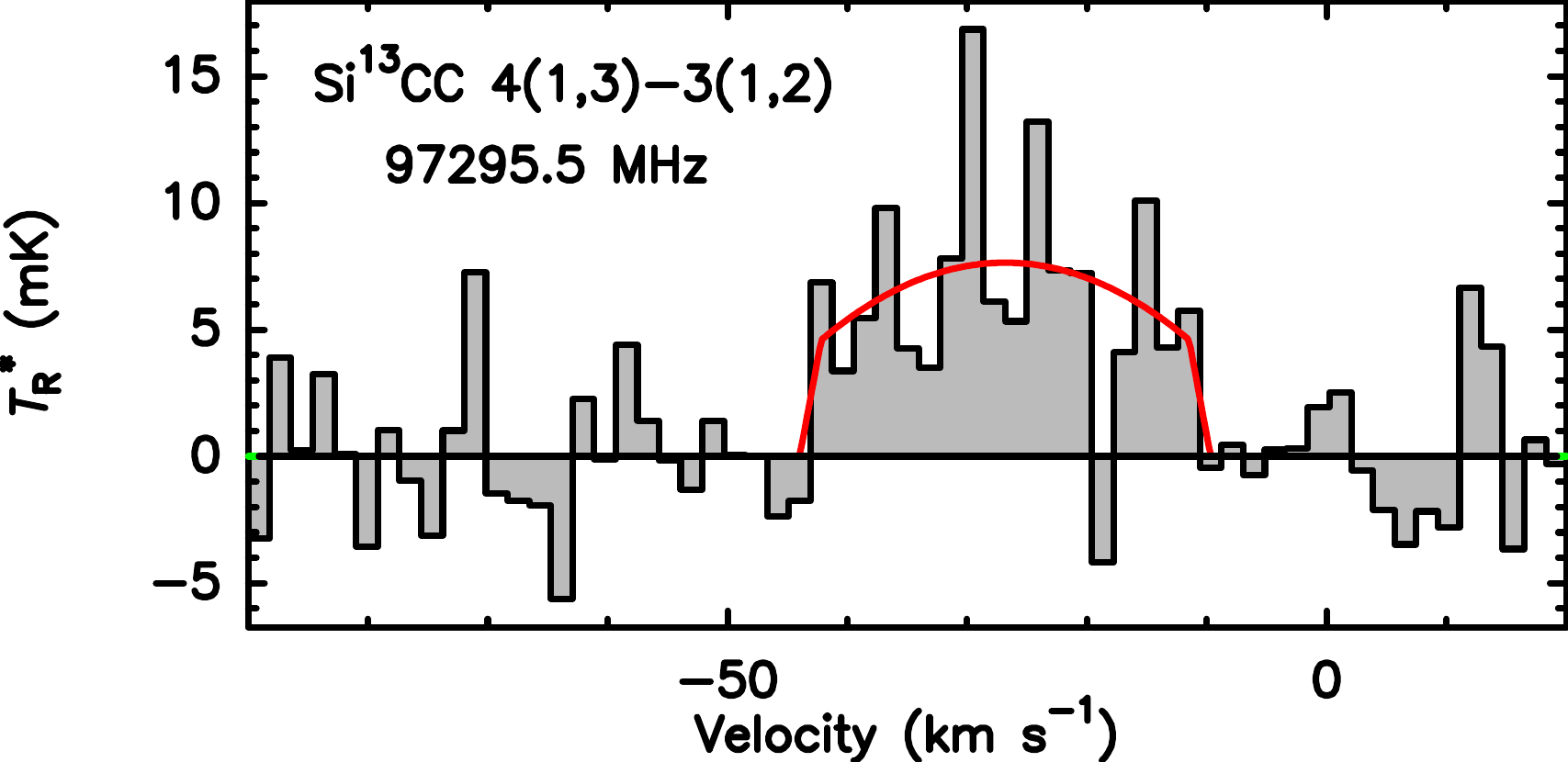}
\caption{{Same as Figure.~\ref{Fig:fitting_1}, but for SiC$_{2}$ and its isotopologues.}\label{Fig:fitting_7}}
\end{figure*}

\begin{figure*}[!htbp]
\centering
\includegraphics[width = 0.45 \textwidth]{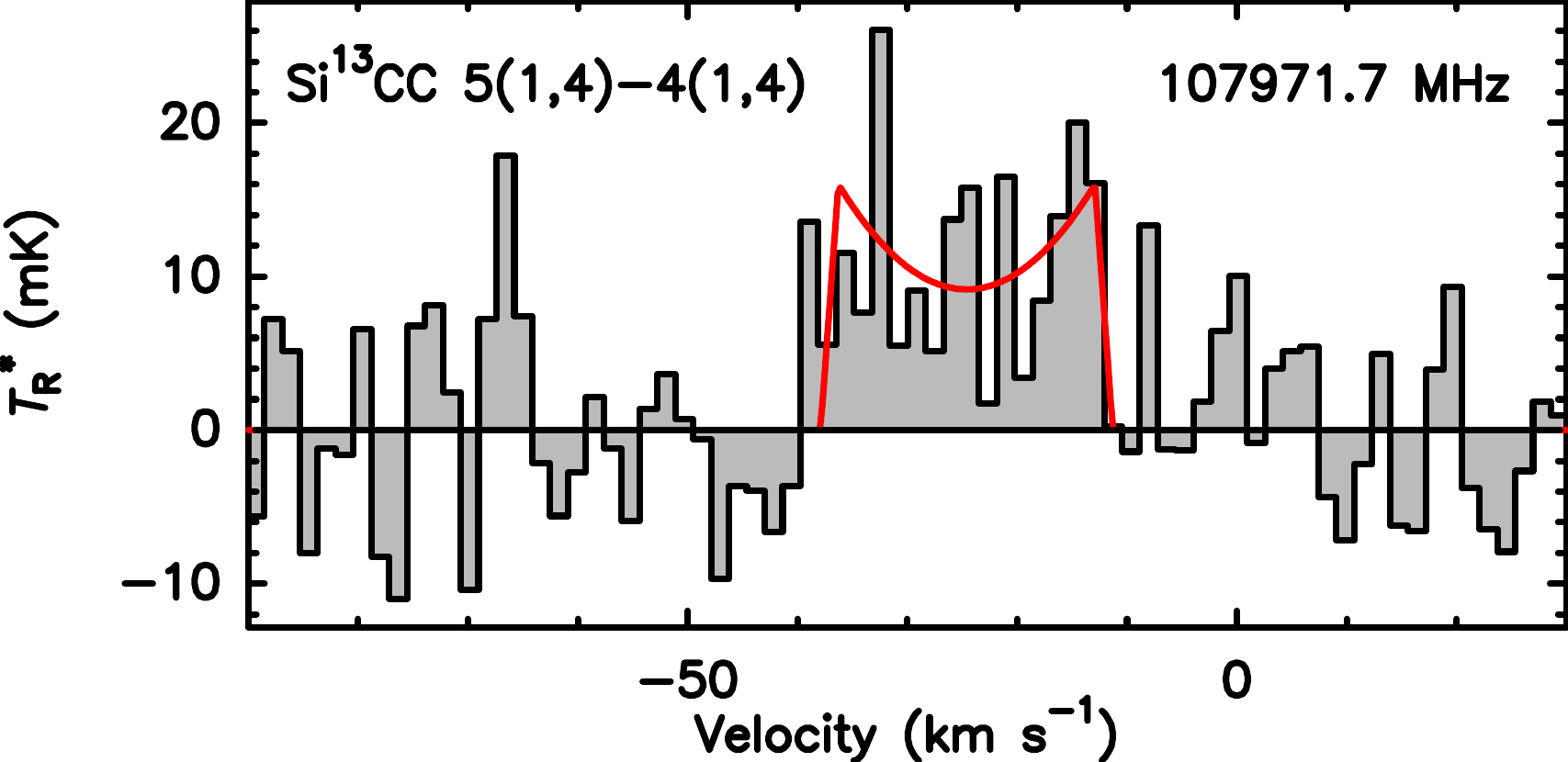}
\hspace{0.05\textwidth}
\includegraphics[width = 0.45 \textwidth]{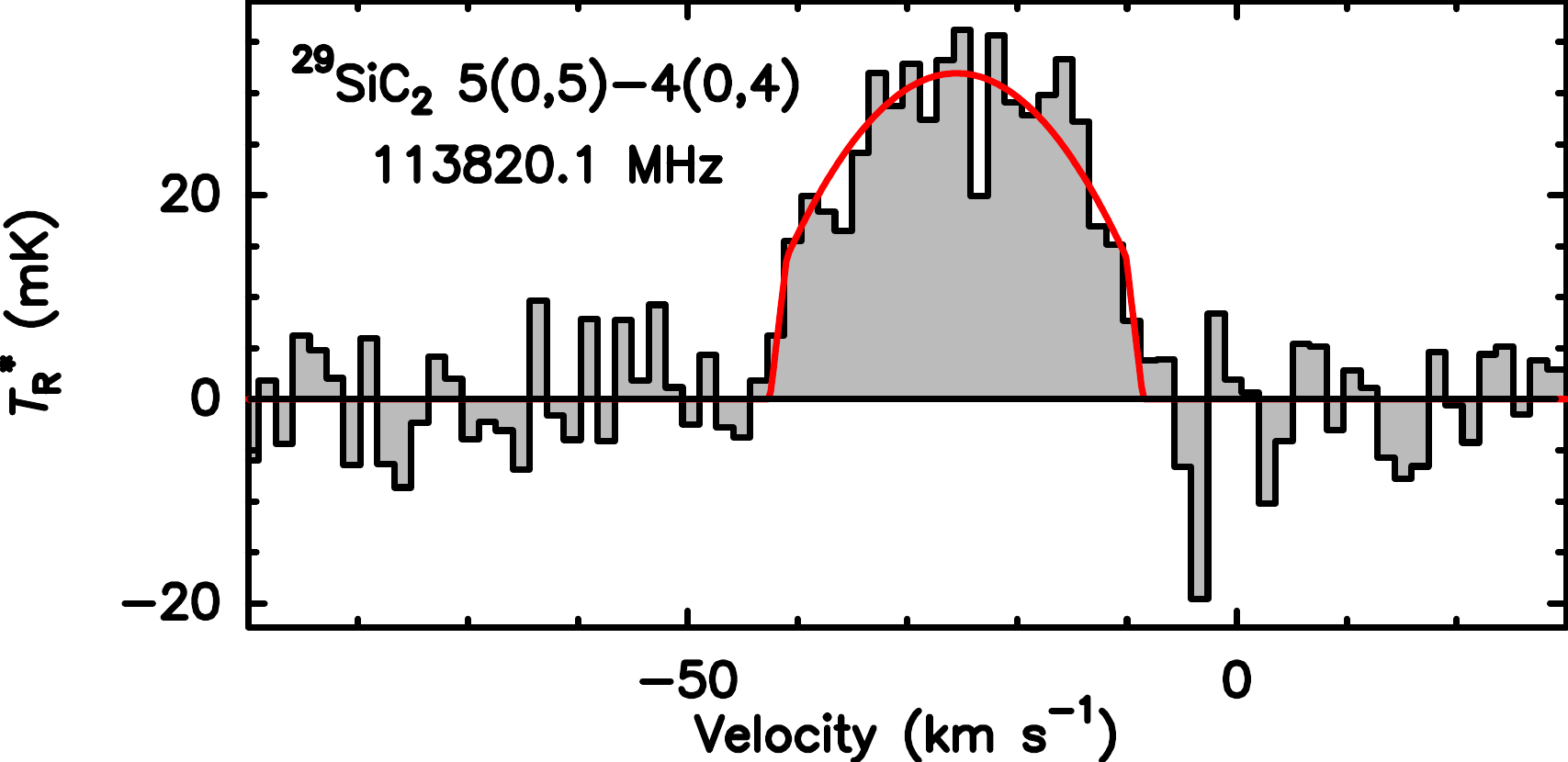}
\vspace{0.1cm}
\includegraphics[width = 0.45 \textwidth]{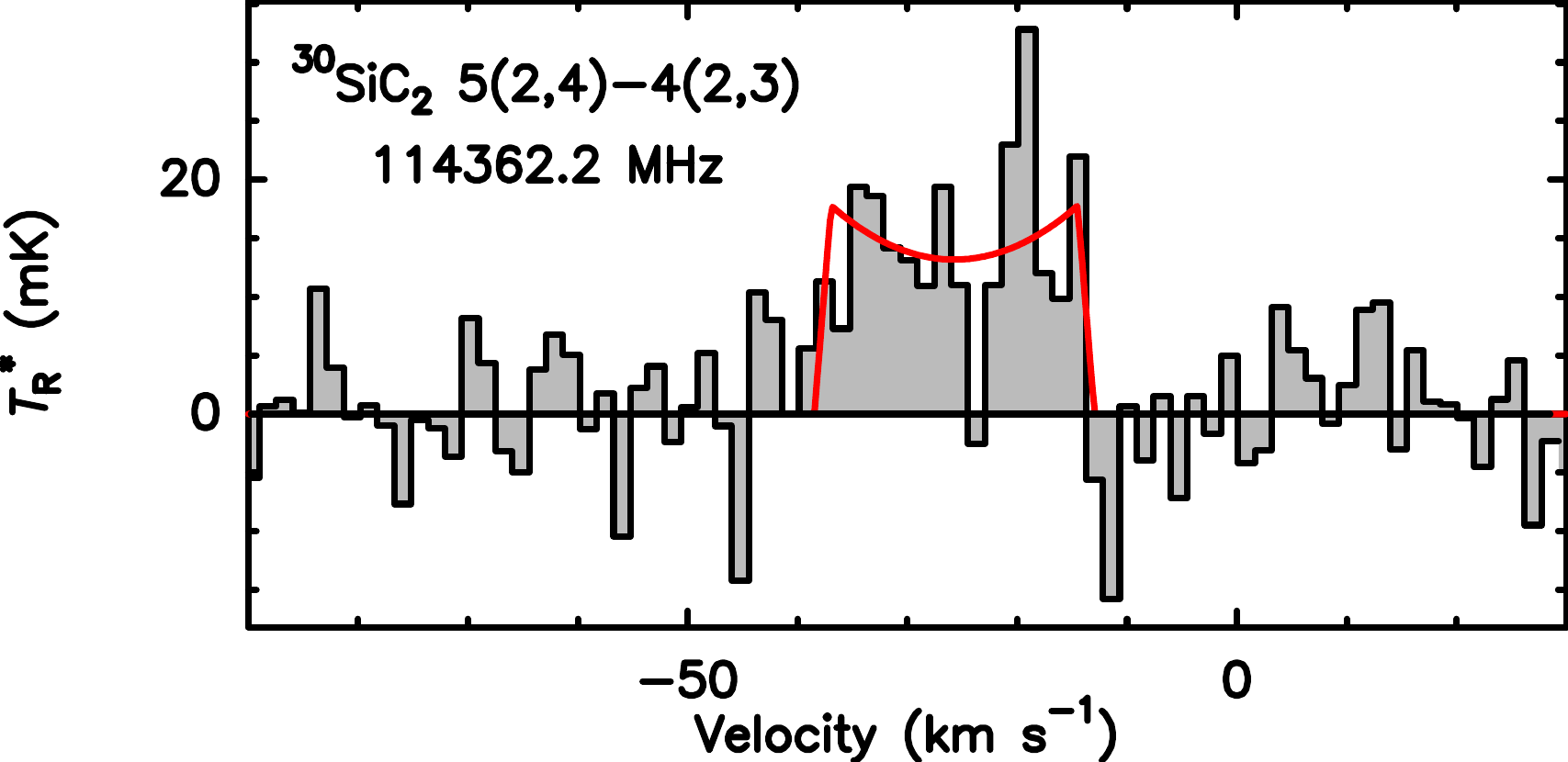}
\hspace{0.05\textwidth}
\includegraphics[width = 0.45 \textwidth]{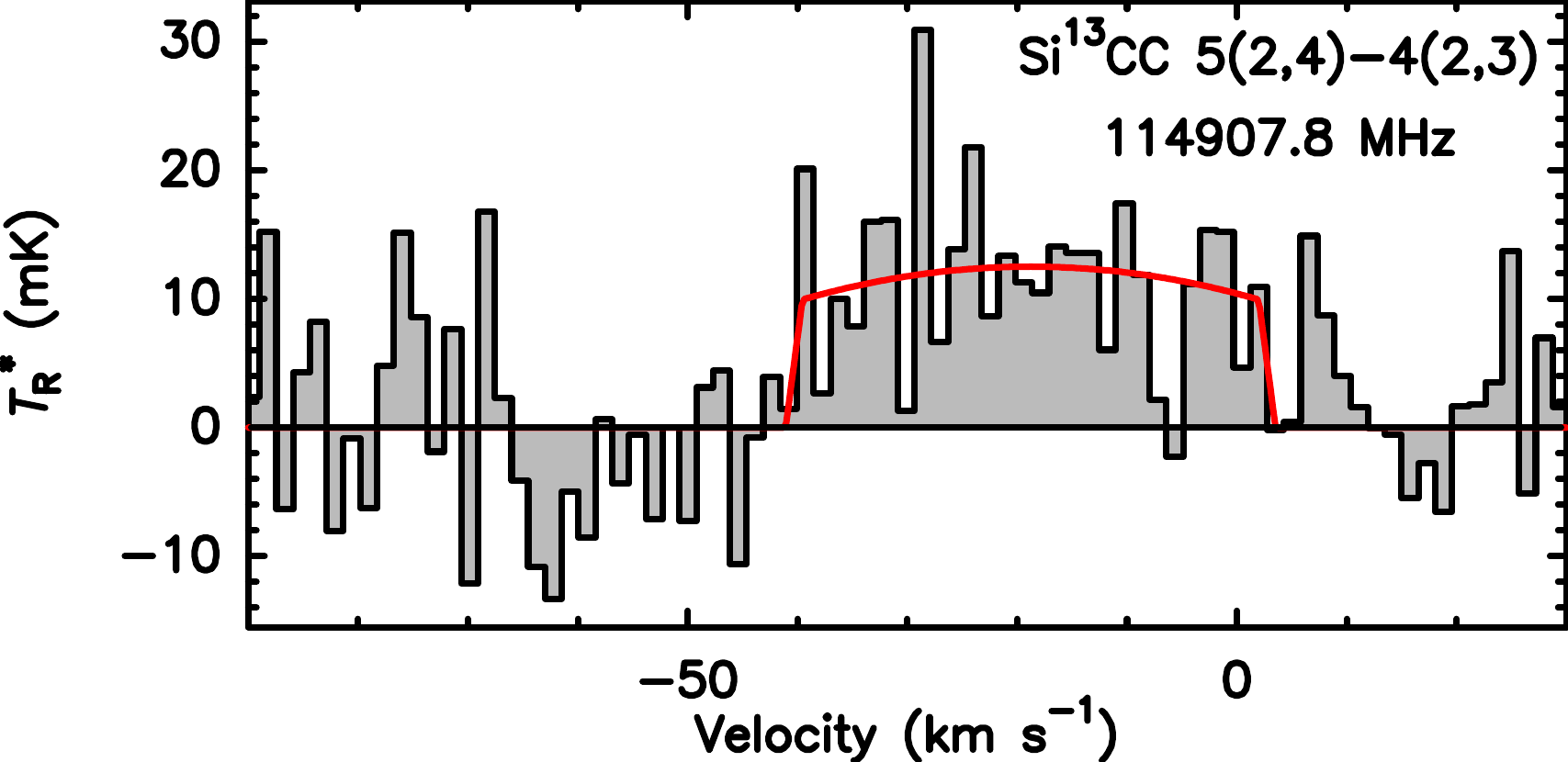}
\vspace{0.1cm}
\includegraphics[width = 0.45 \textwidth]{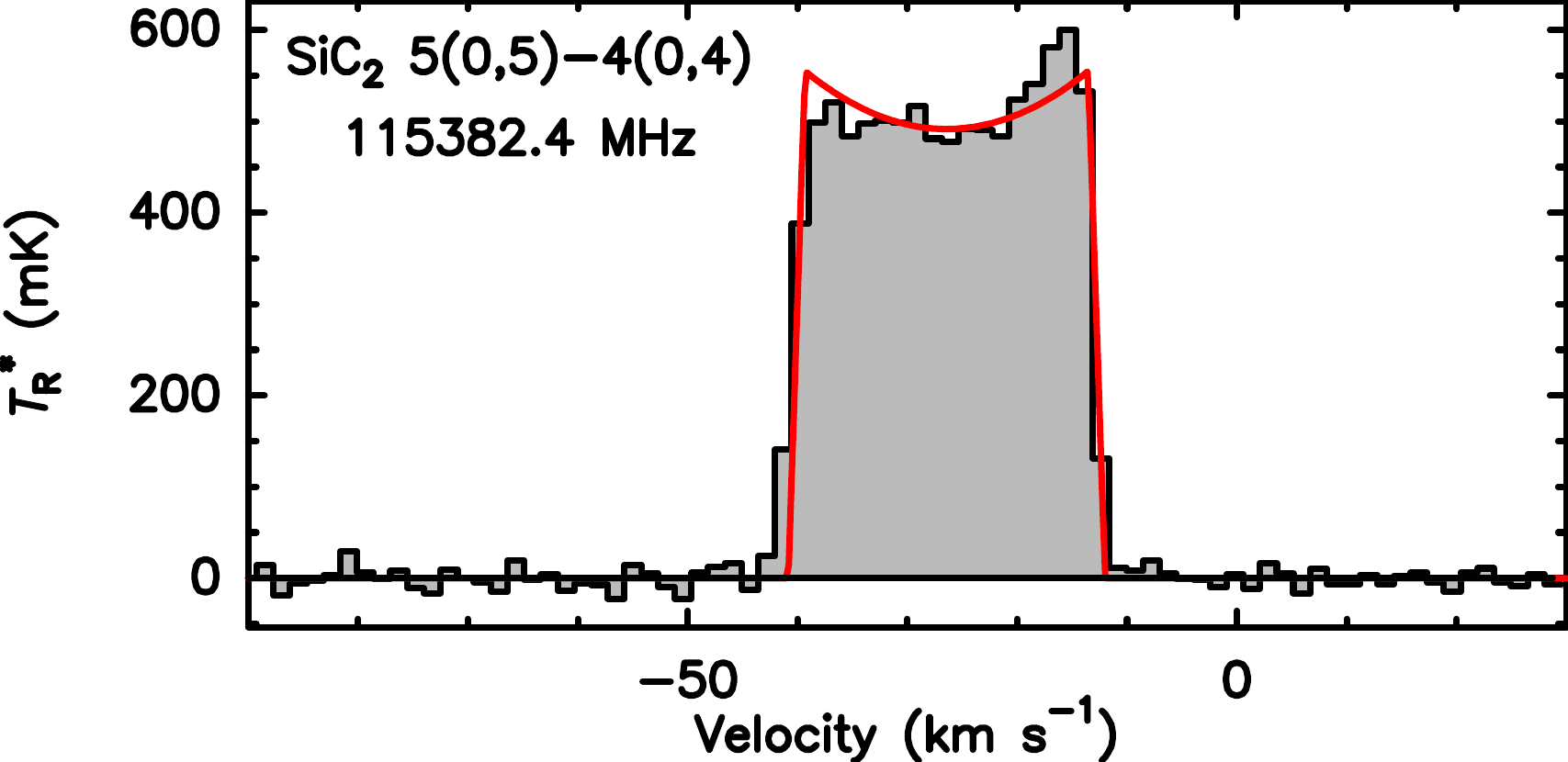}
\hspace{0.05\textwidth}
\includegraphics[width = 0.45 \textwidth]{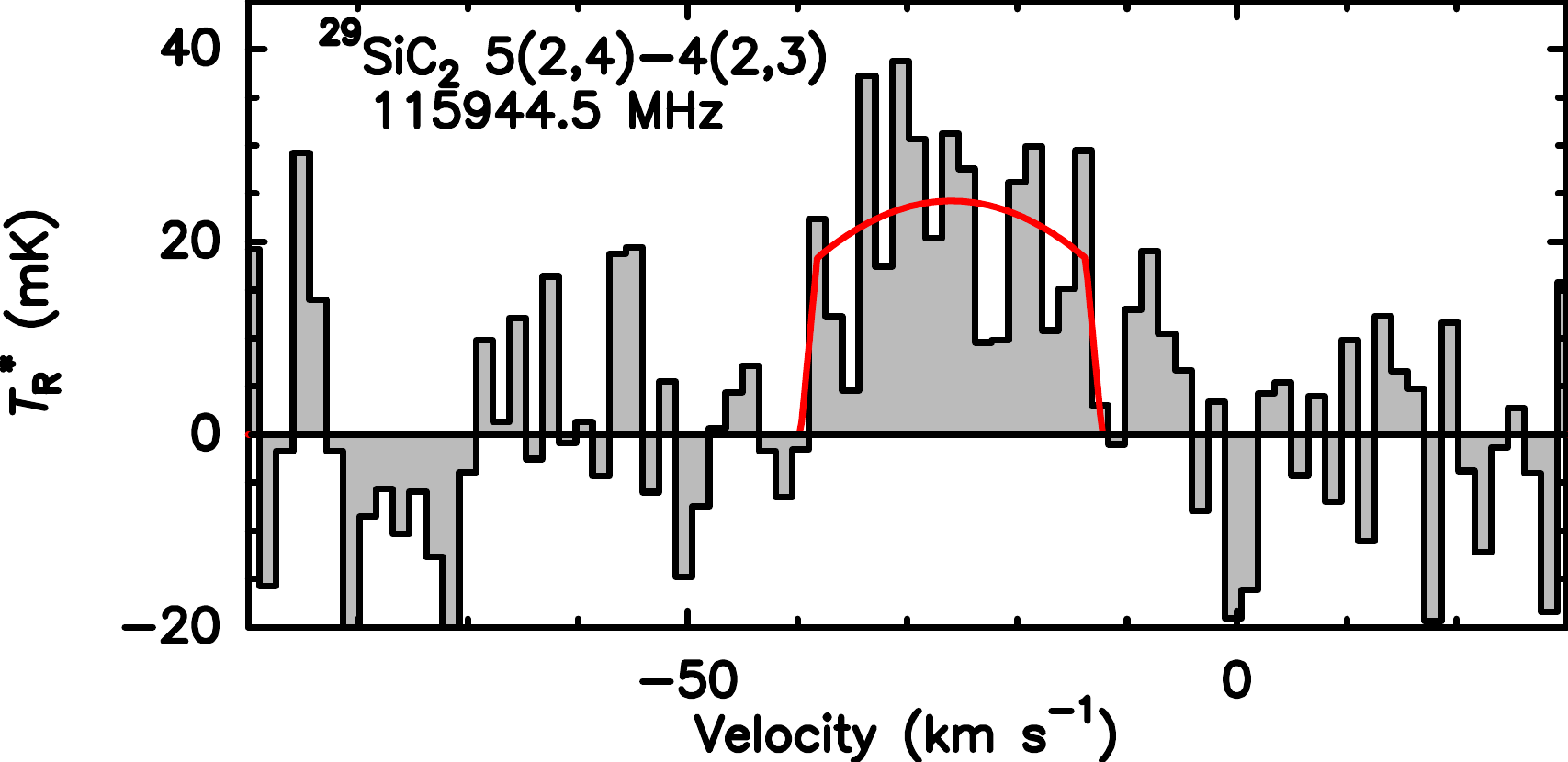}
\centerline{Figure \ref{Fig:fitting_7}. --- continued}
\end{figure*}

\begin{figure*}[!htbp]
\centering
\includegraphics[width = 0.45 \textwidth]{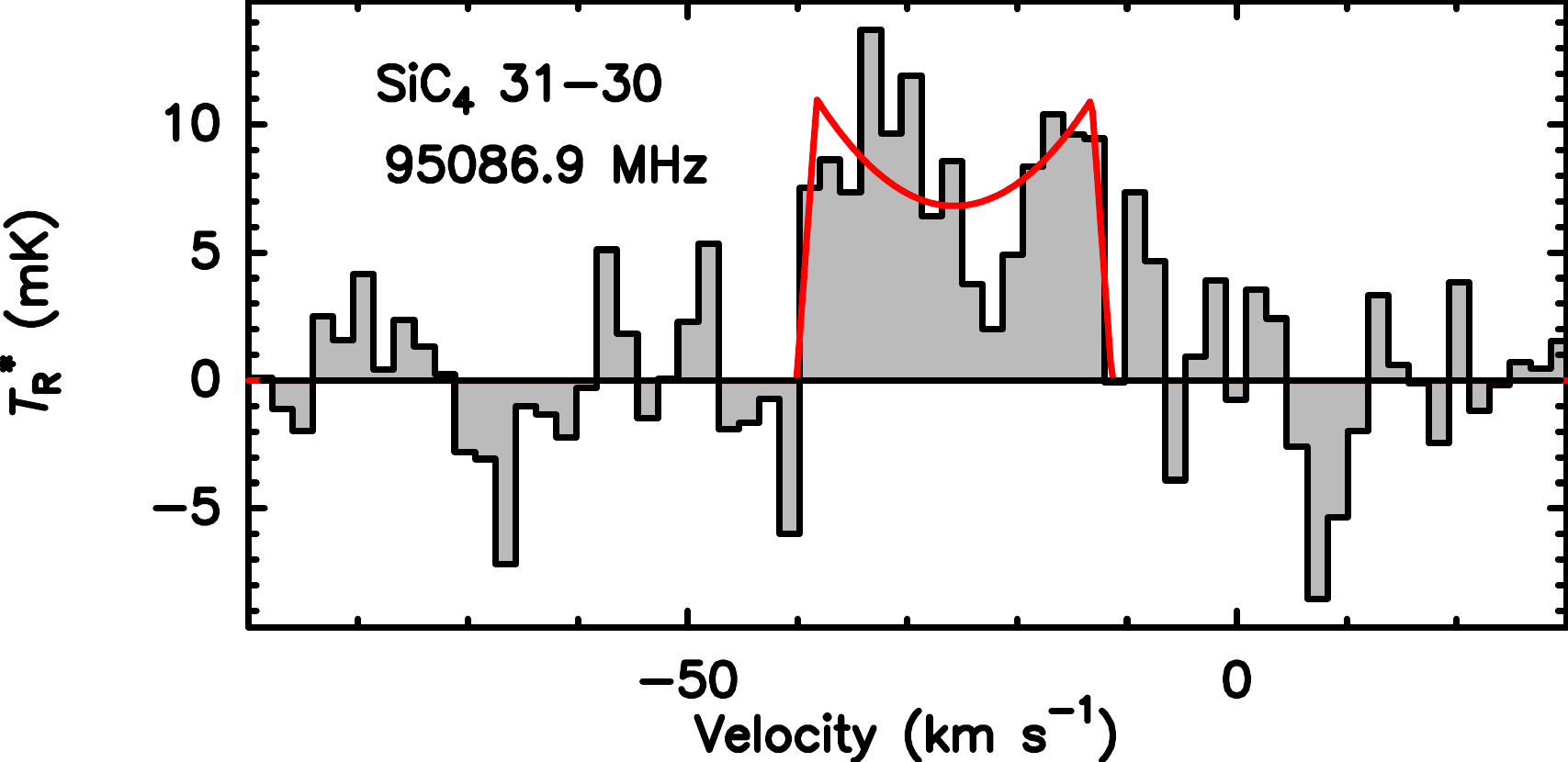}
\hspace{0.05\textwidth}
\includegraphics[width = 0.45 \textwidth]{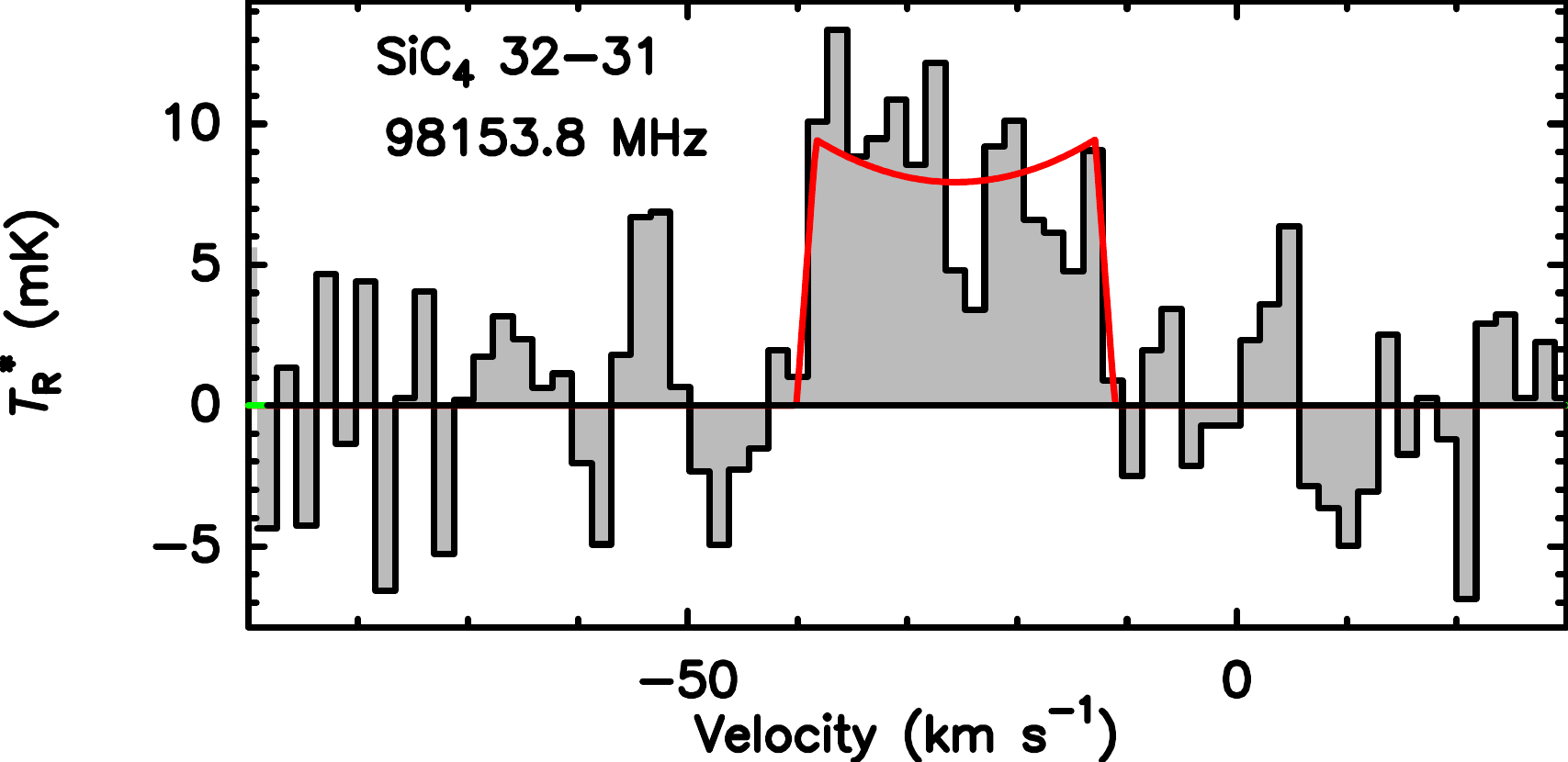}
\vspace{0.1cm}
\includegraphics[width = 0.45 \textwidth]{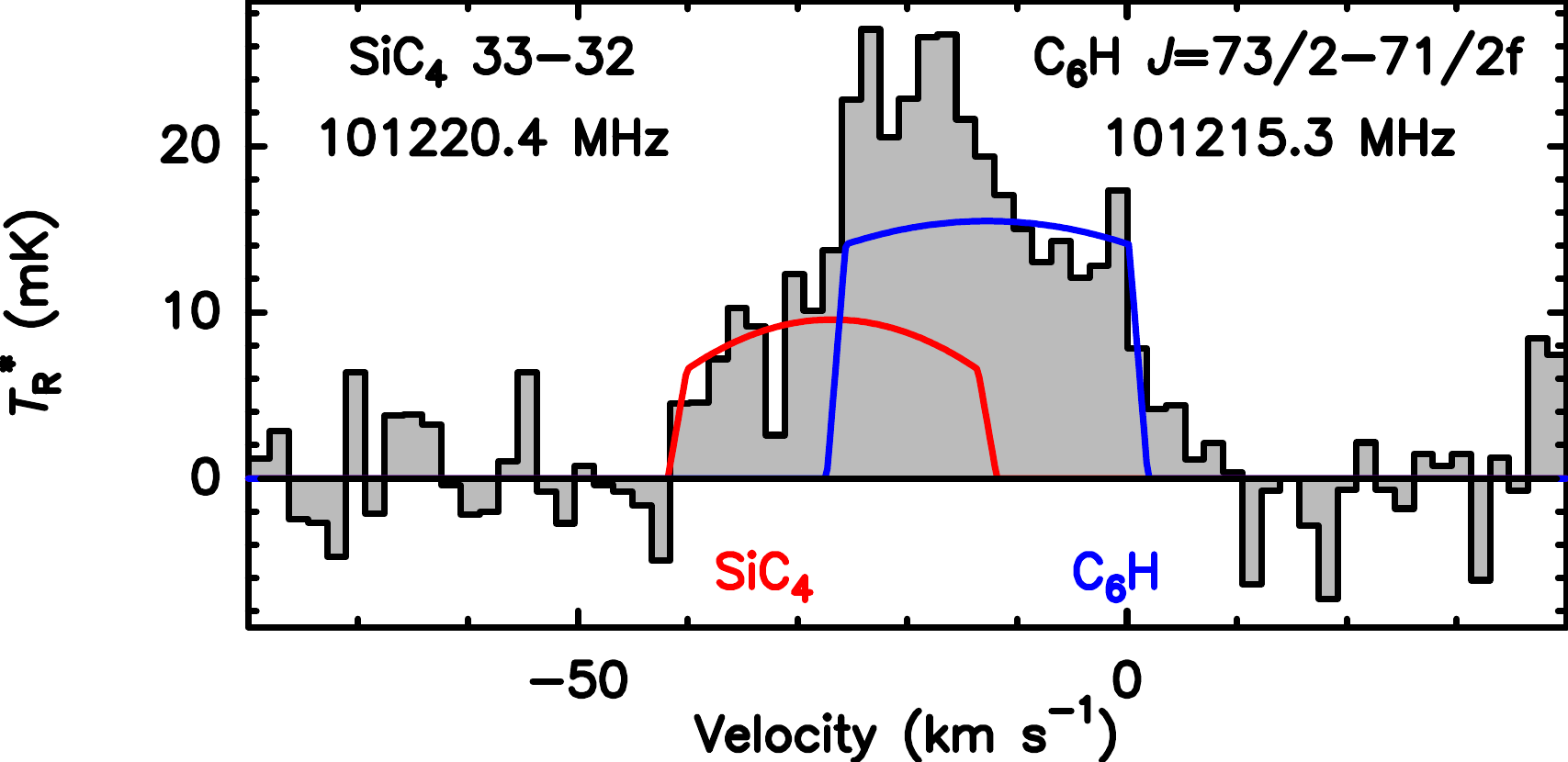}
\hspace{0.05\textwidth}
\includegraphics[width = 0.45 \textwidth]{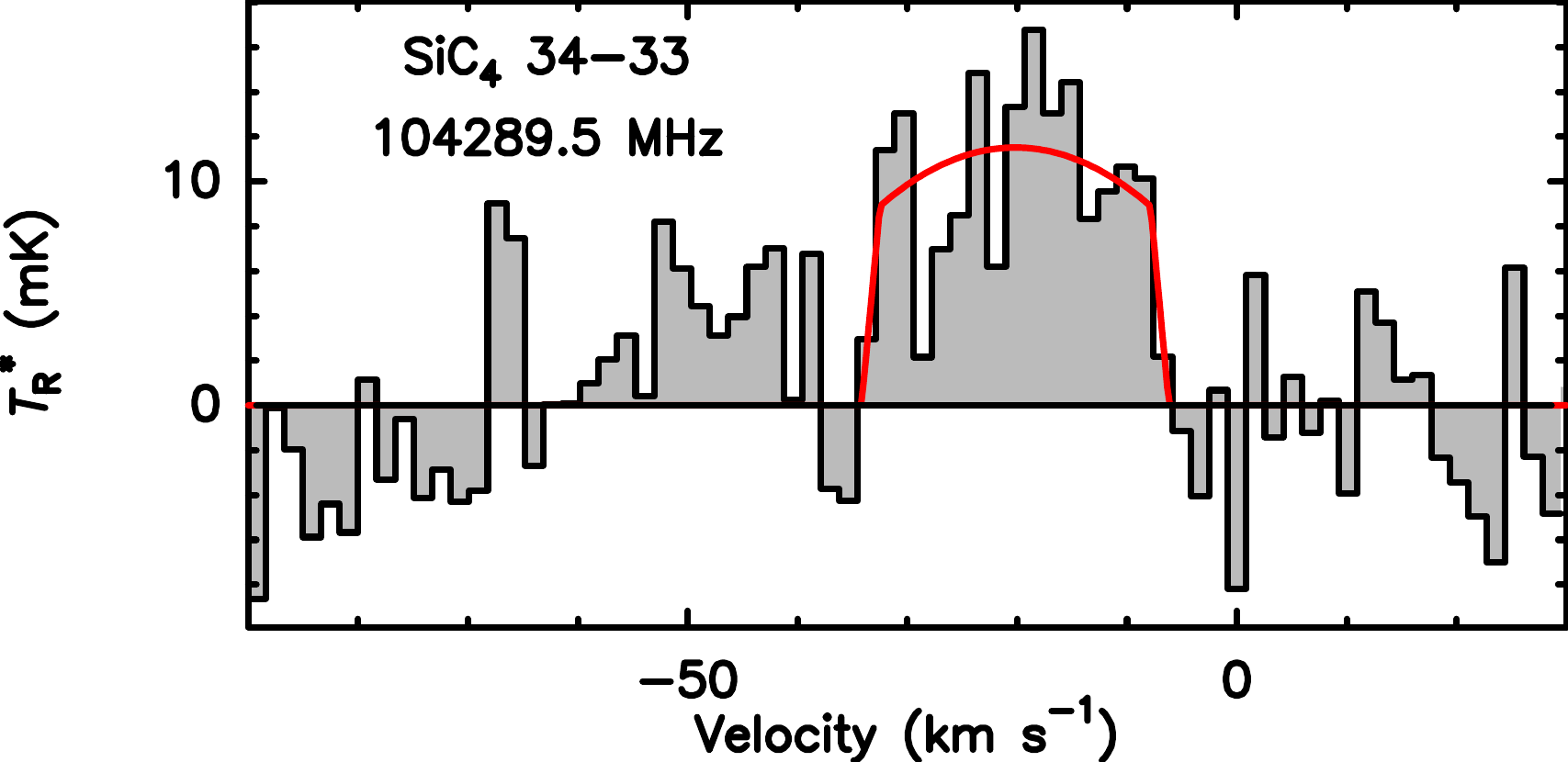}
\vspace{0.1cm}
\includegraphics[width = 0.45 \textwidth]{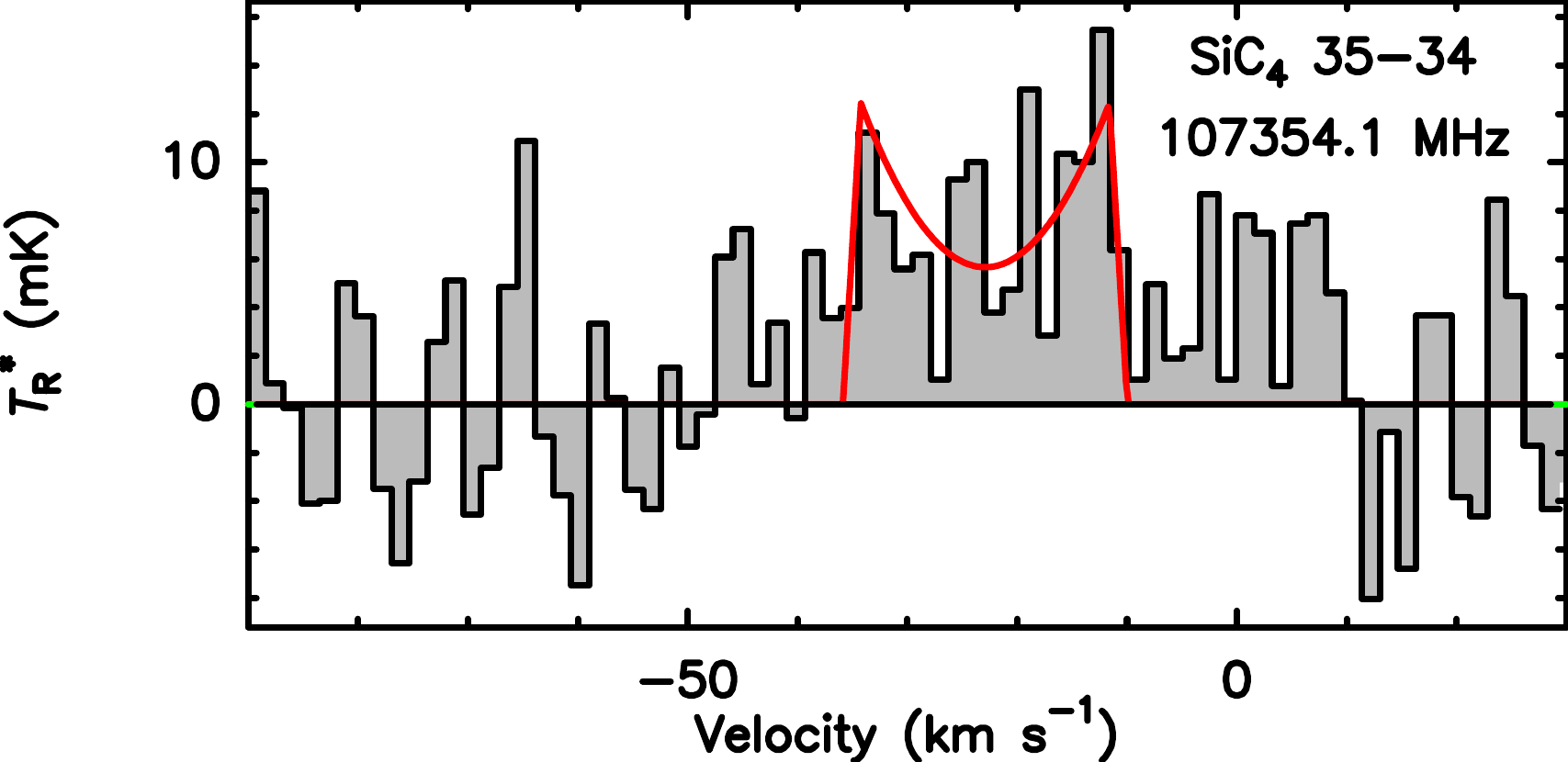}
\caption{{Same as Figure.~\ref{Fig:fitting_1}, but for SiC$_{4}$.}\label{Fig:fitting_8}}
\end{figure*}

\begin{figure*}[!htbp]
\centering
\includegraphics[width = 0.45 \textwidth]{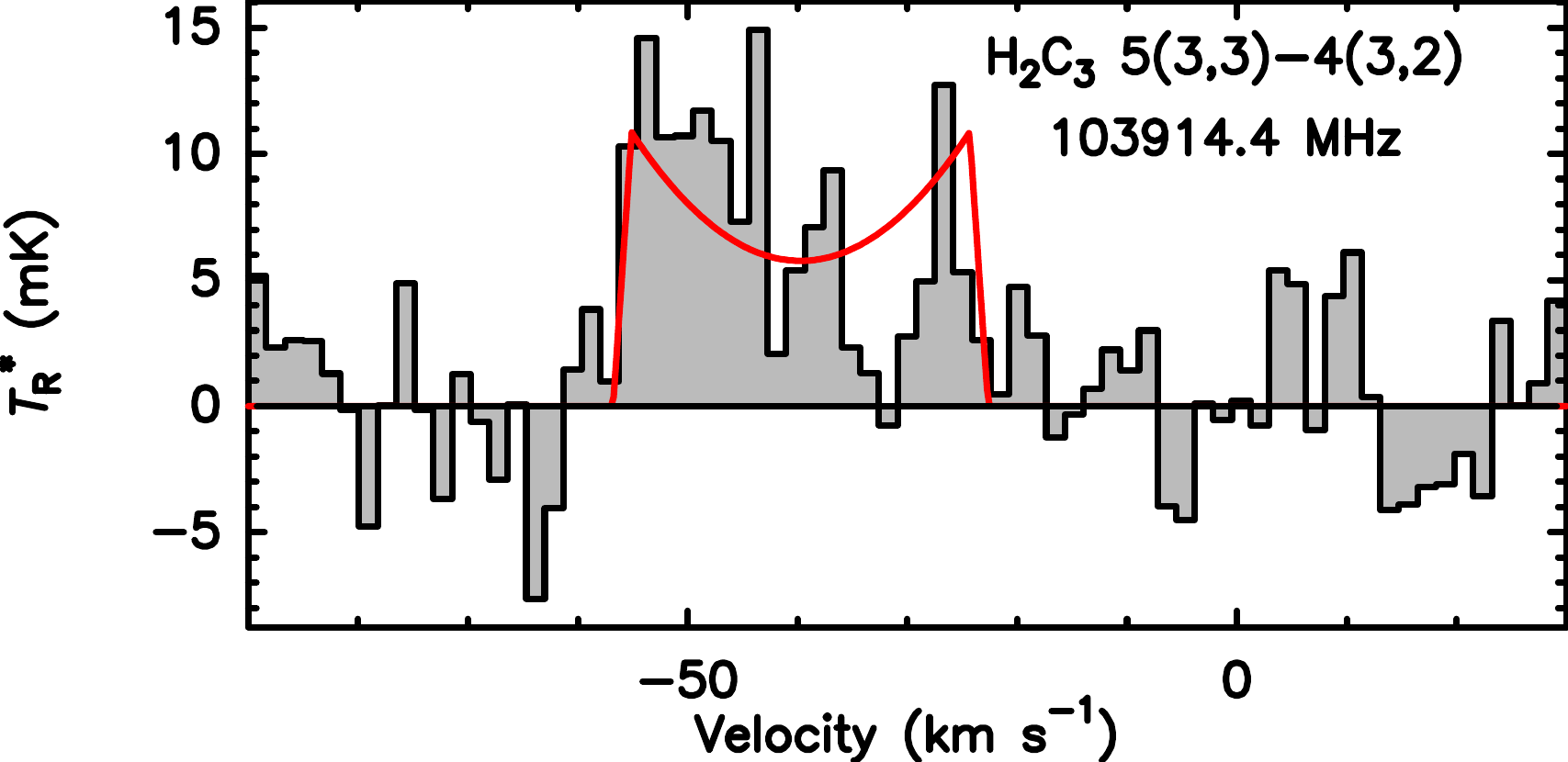}
\hspace{0.05\textwidth}
\includegraphics[width = 0.45 \textwidth]{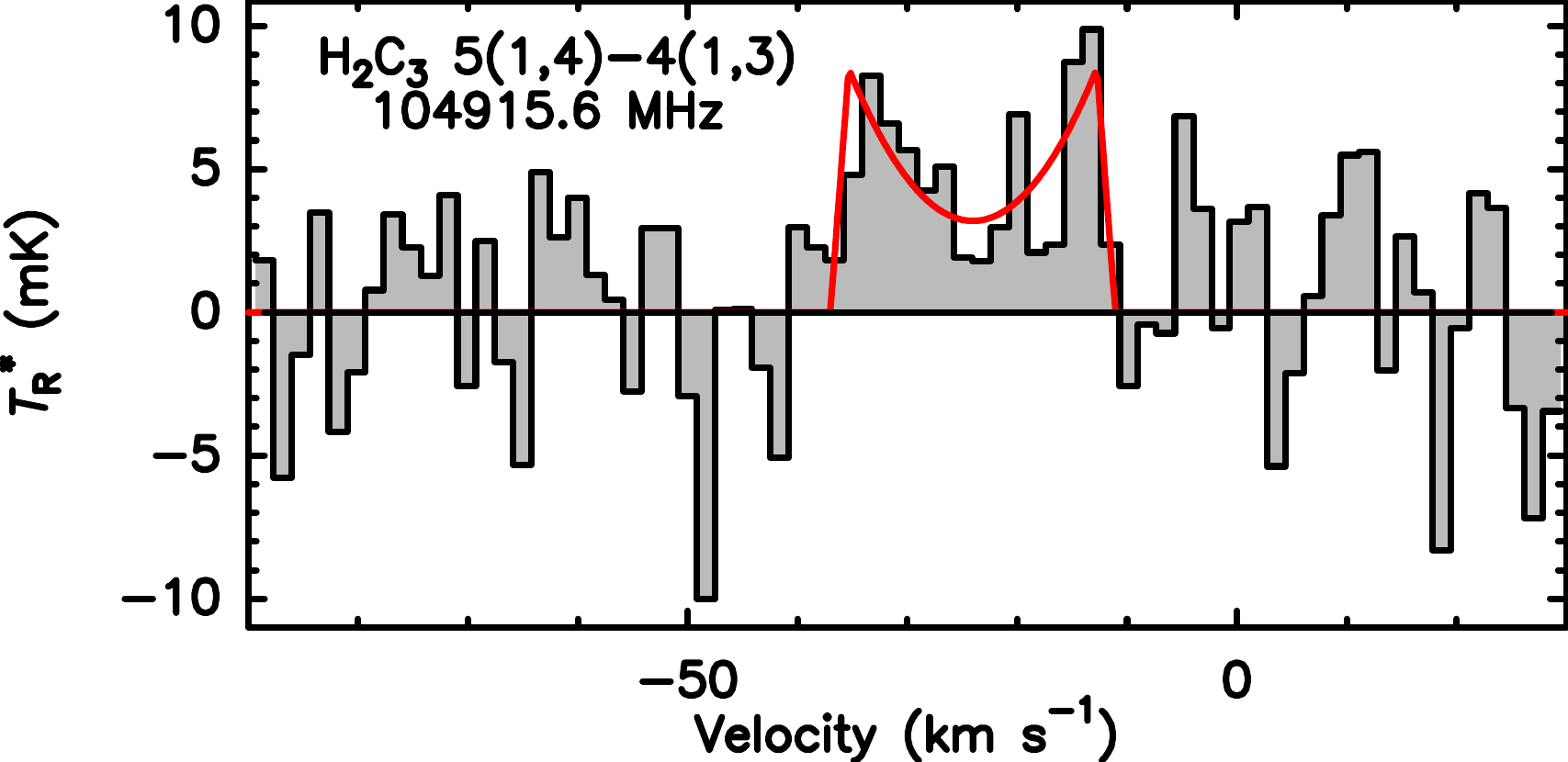}
\caption{{Same as Figure.~\ref{Fig:fitting_1}, but for H$_{2}$C$_{3}$.}\label{Fig:fitting_10}}
\end{figure*}

\begin{figure*}[!htbp]
\centering
\includegraphics[width = 0.45 \textwidth]{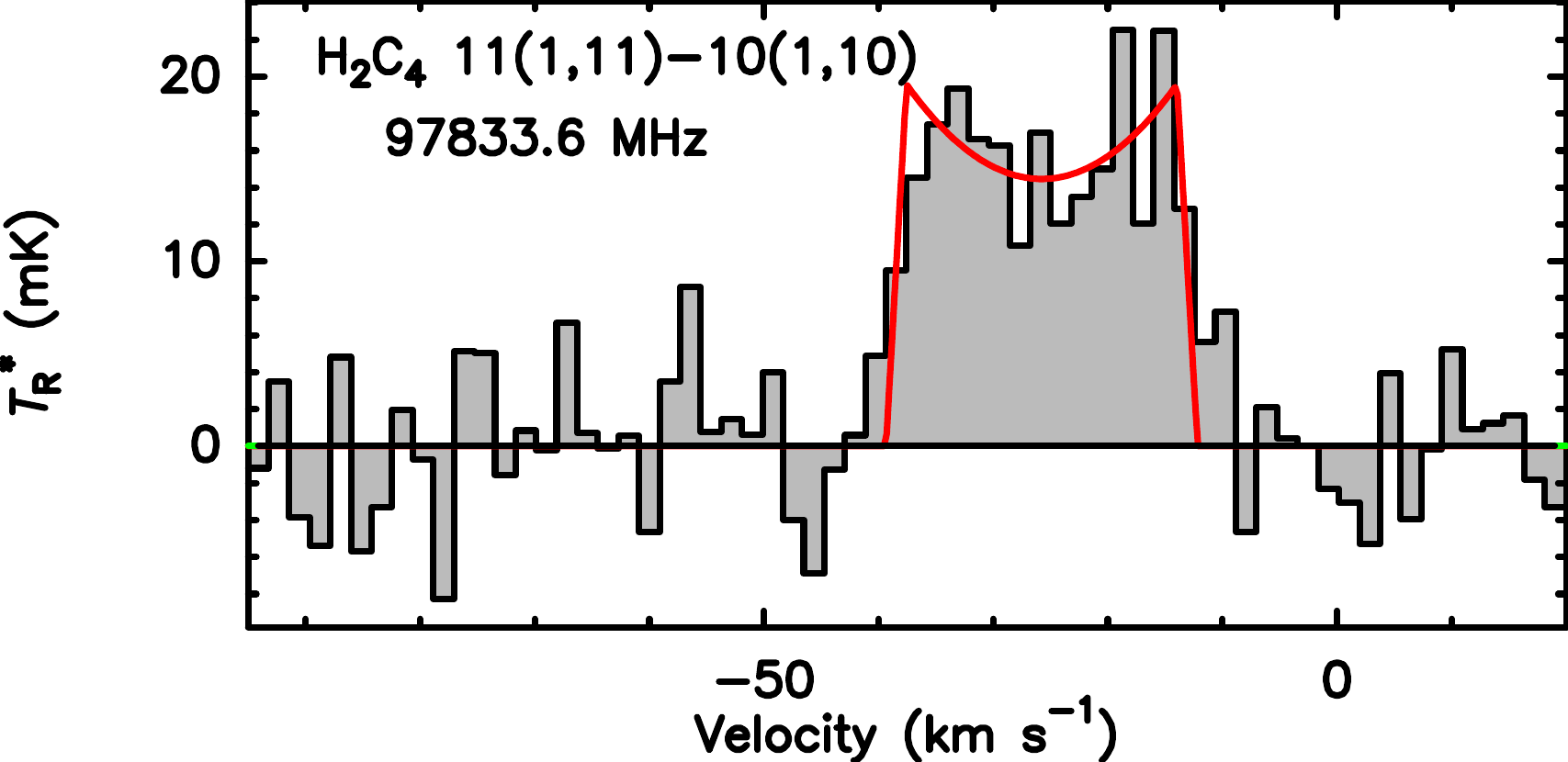}
\hspace{0.05\textwidth}
\includegraphics[width = 0.45 \textwidth]{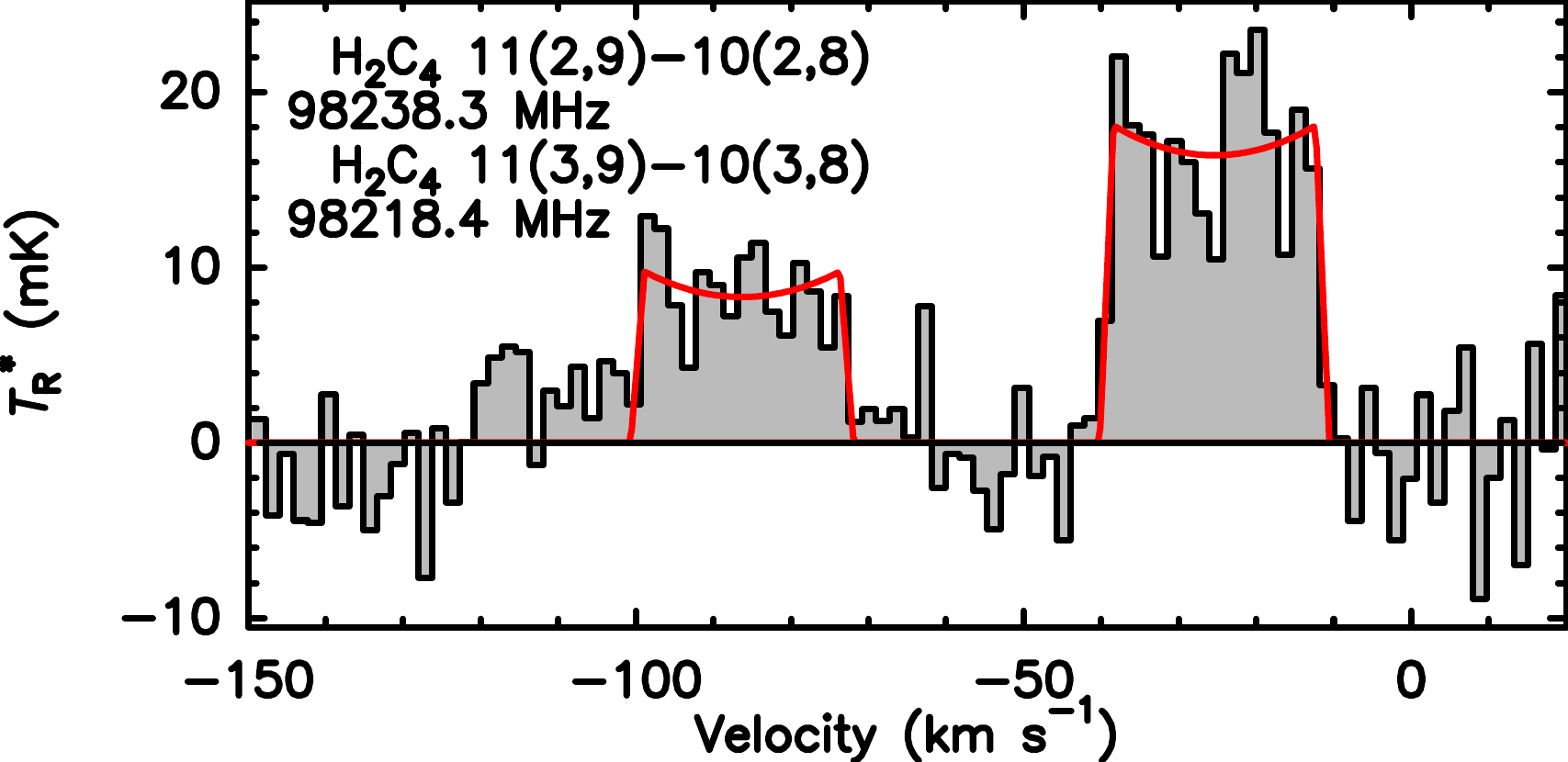}
\vspace{0.1cm}
\includegraphics[width = 0.45 \textwidth]{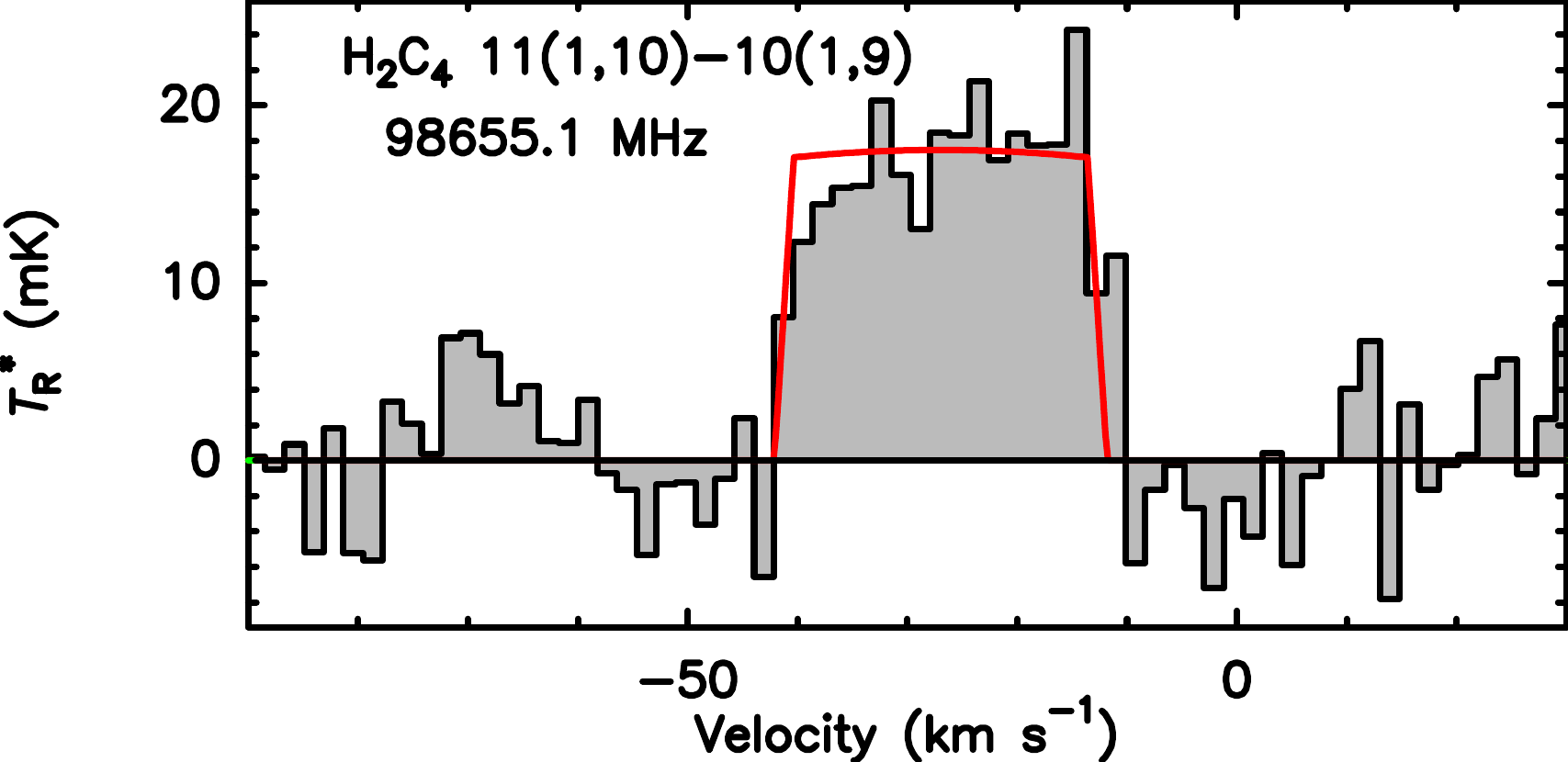}
\hspace{0.05\textwidth}
\includegraphics[width = 0.45 \textwidth]{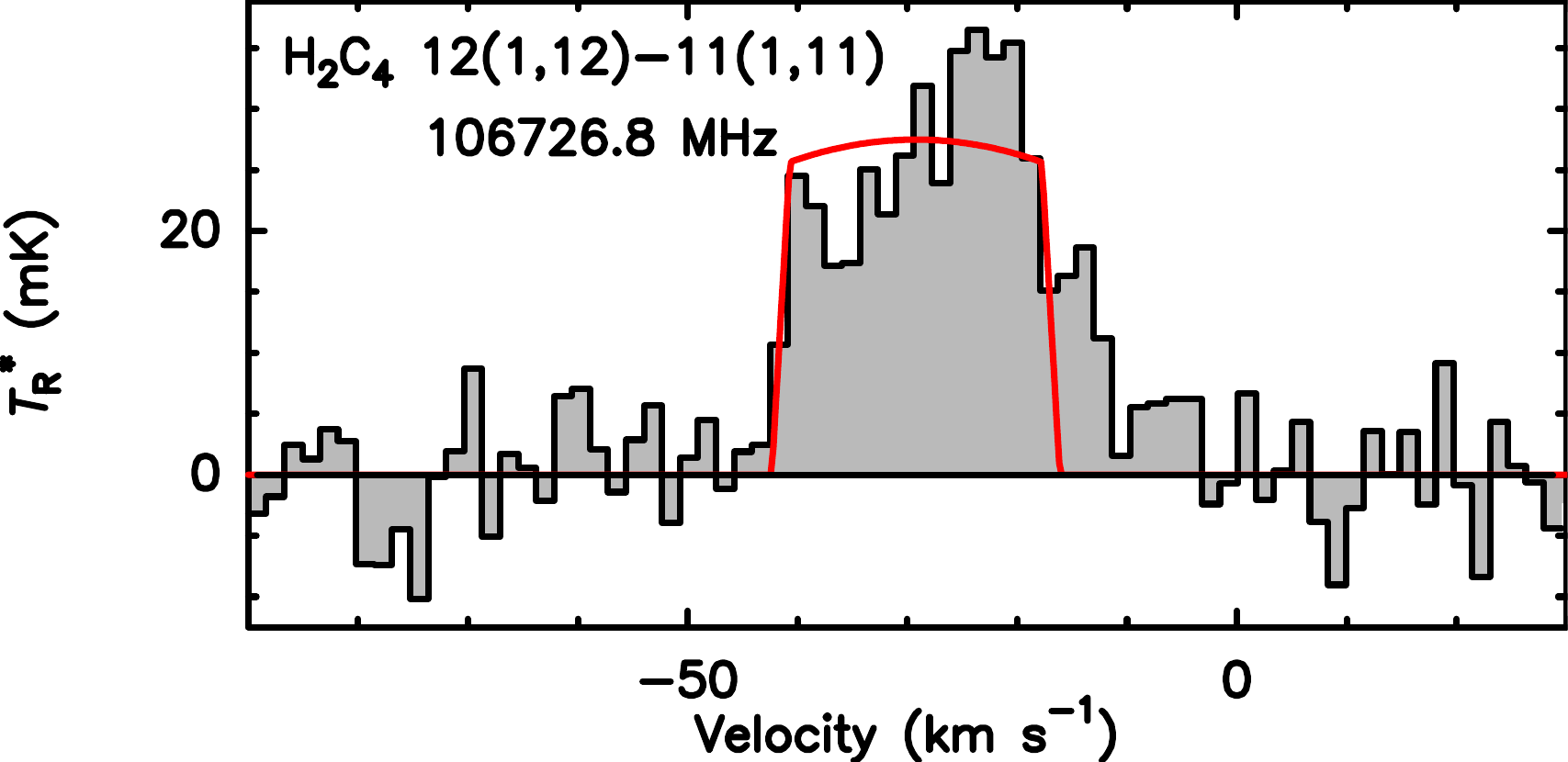}
\vspace{0.1cm}
\includegraphics[width = 0.45 \textwidth]{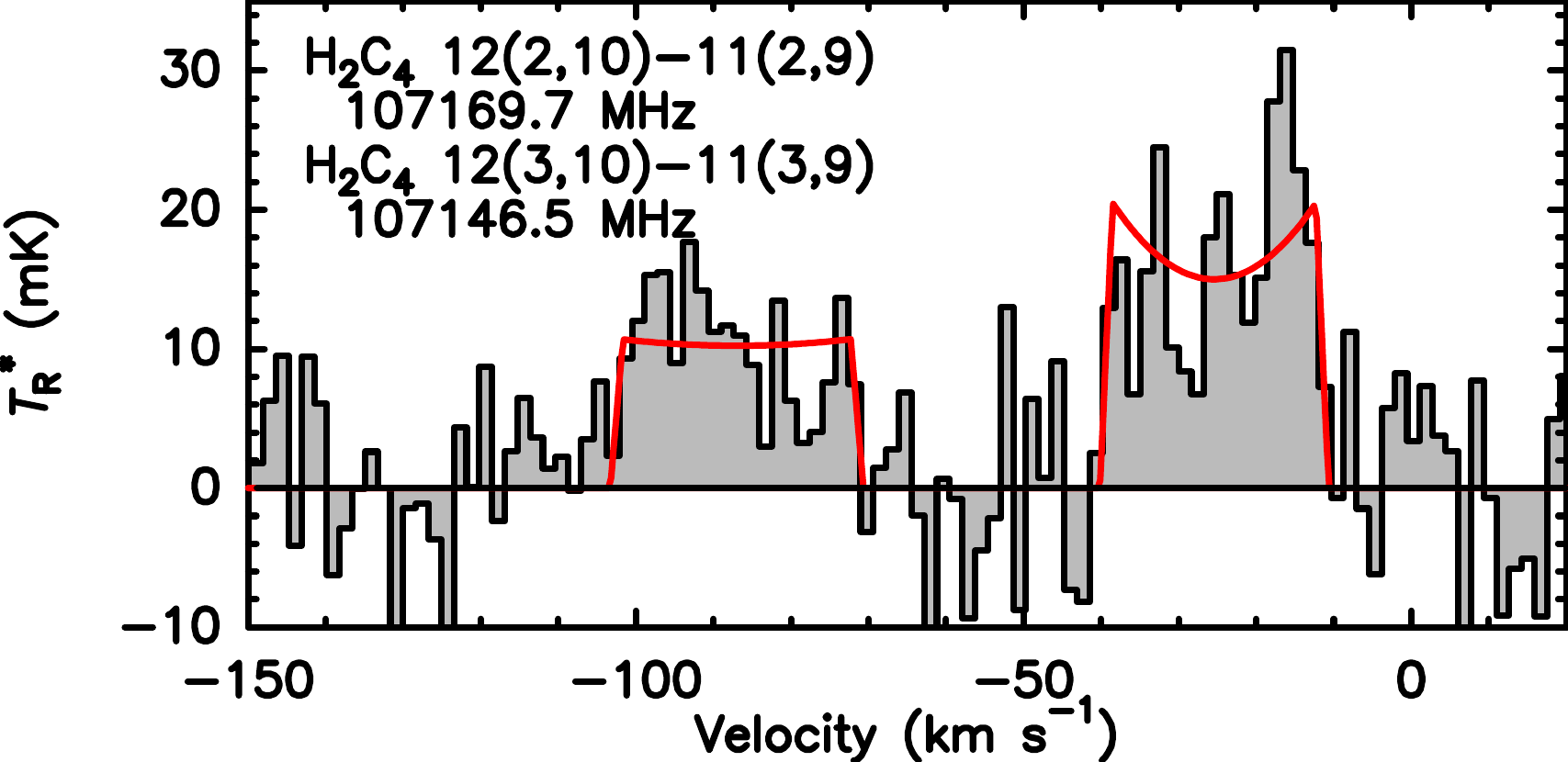}
\hspace{0.05\textwidth}
\includegraphics[width = 0.45 \textwidth]{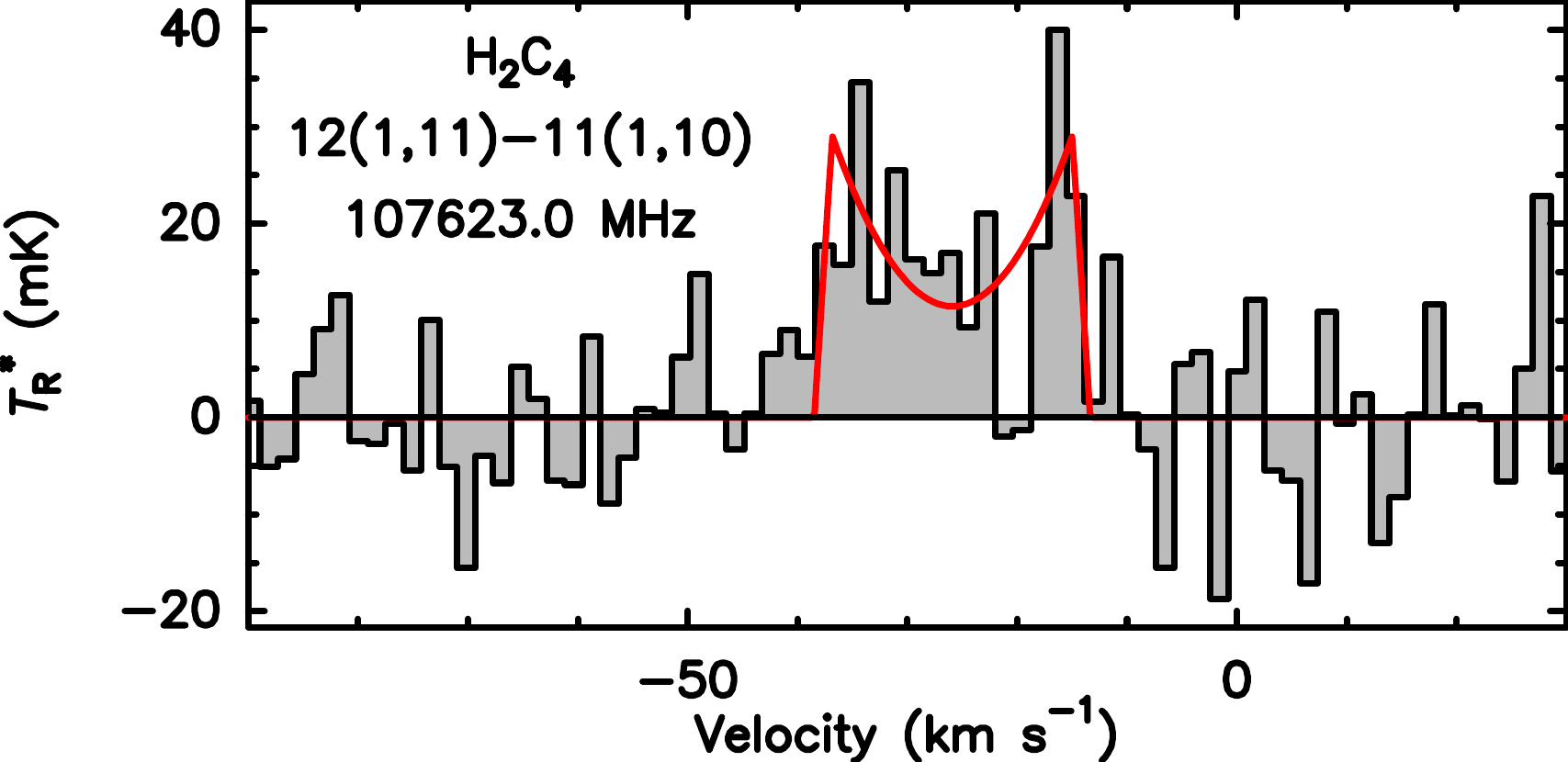}
\caption{{Same as Figure.~\ref{Fig:fitting_1}, but for H$_{2}$C$_{4}$. }\label{Fig:fitting_11}}
\end{figure*}

\begin{figure*}[!htbp]
\centering
\includegraphics[width = 0.45 \textwidth]{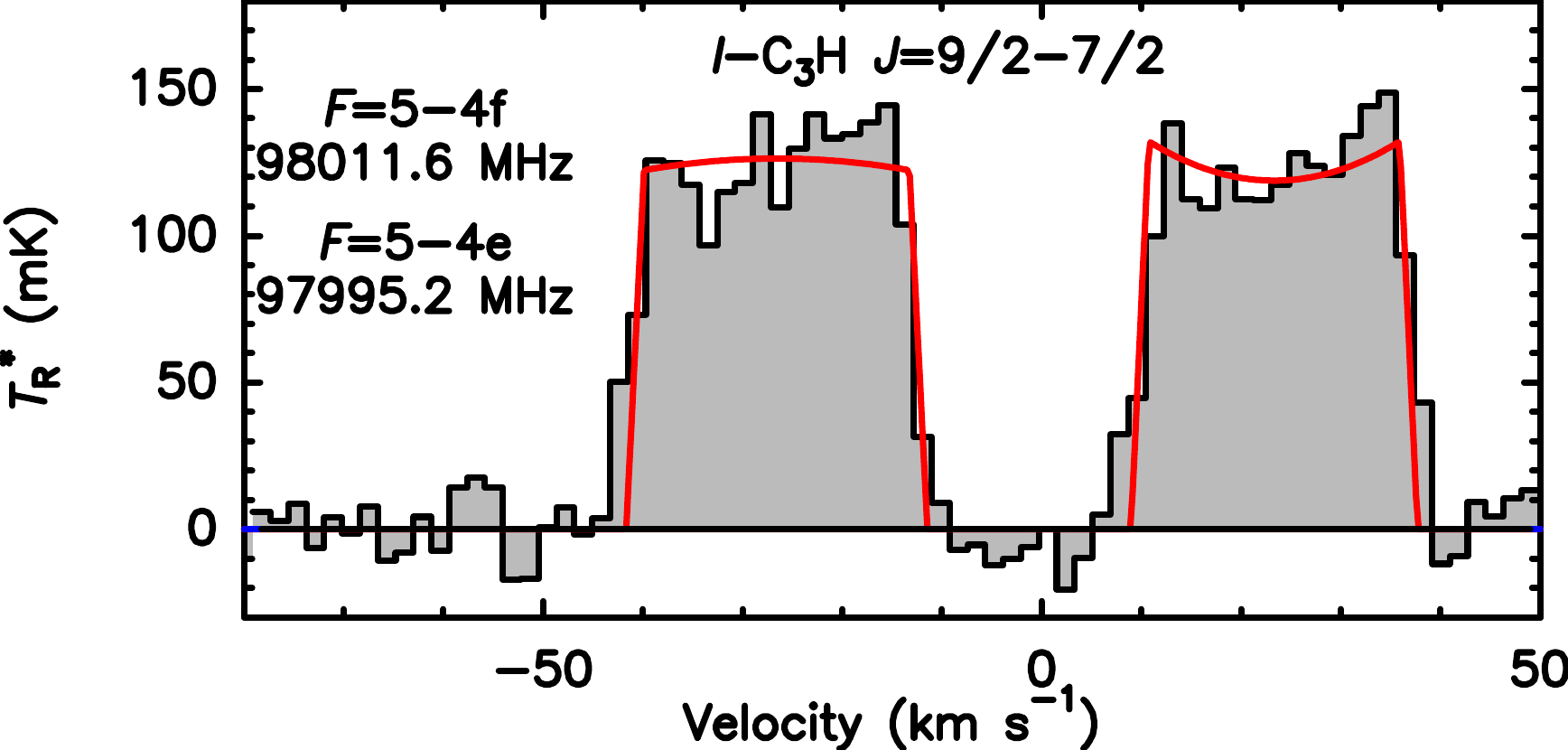}
\hspace{0.05\textwidth}
\includegraphics[width = 0.45 \textwidth]{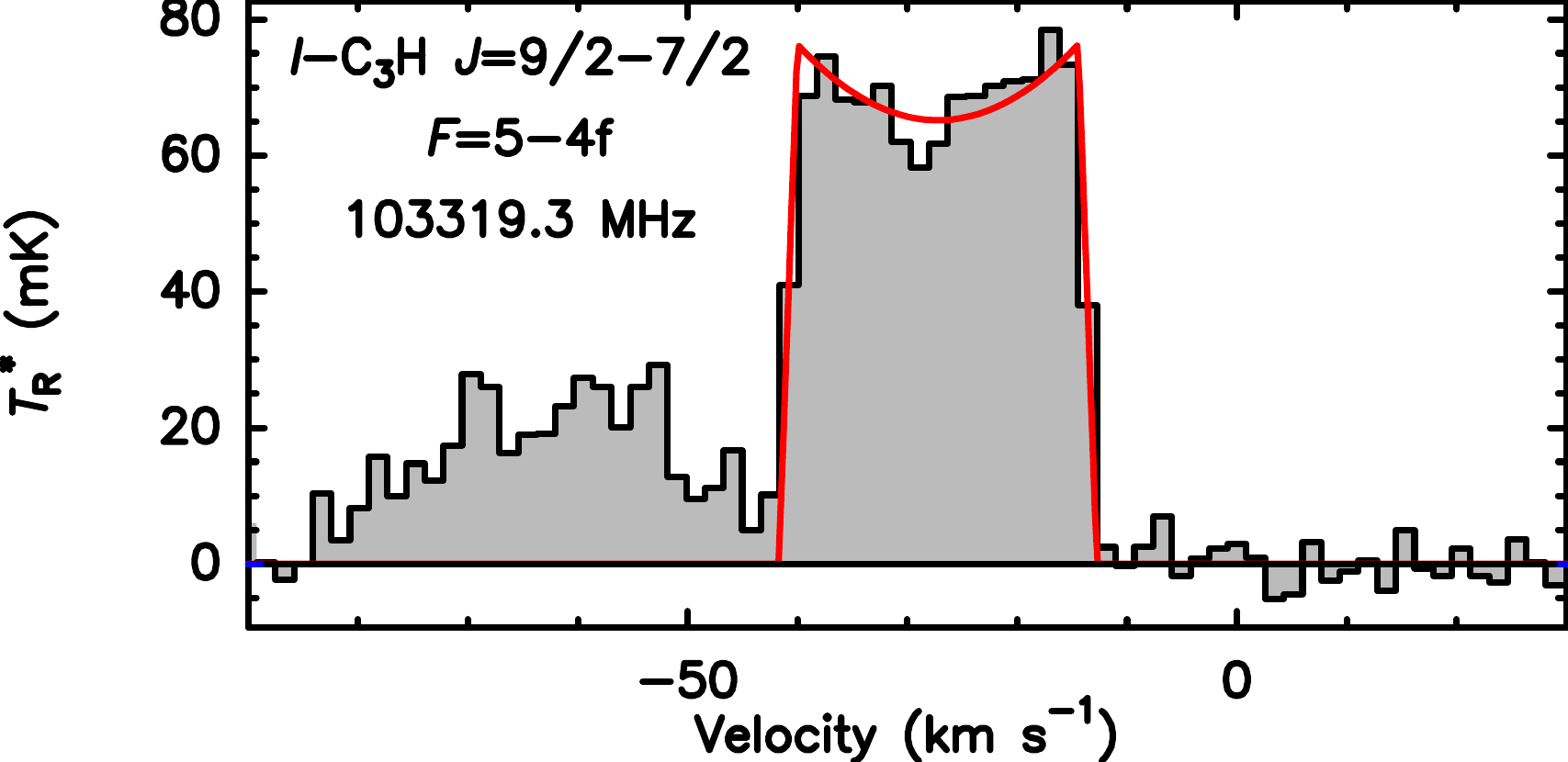}
\vspace{0.1cm}
\includegraphics[width = 0.45 \textwidth]{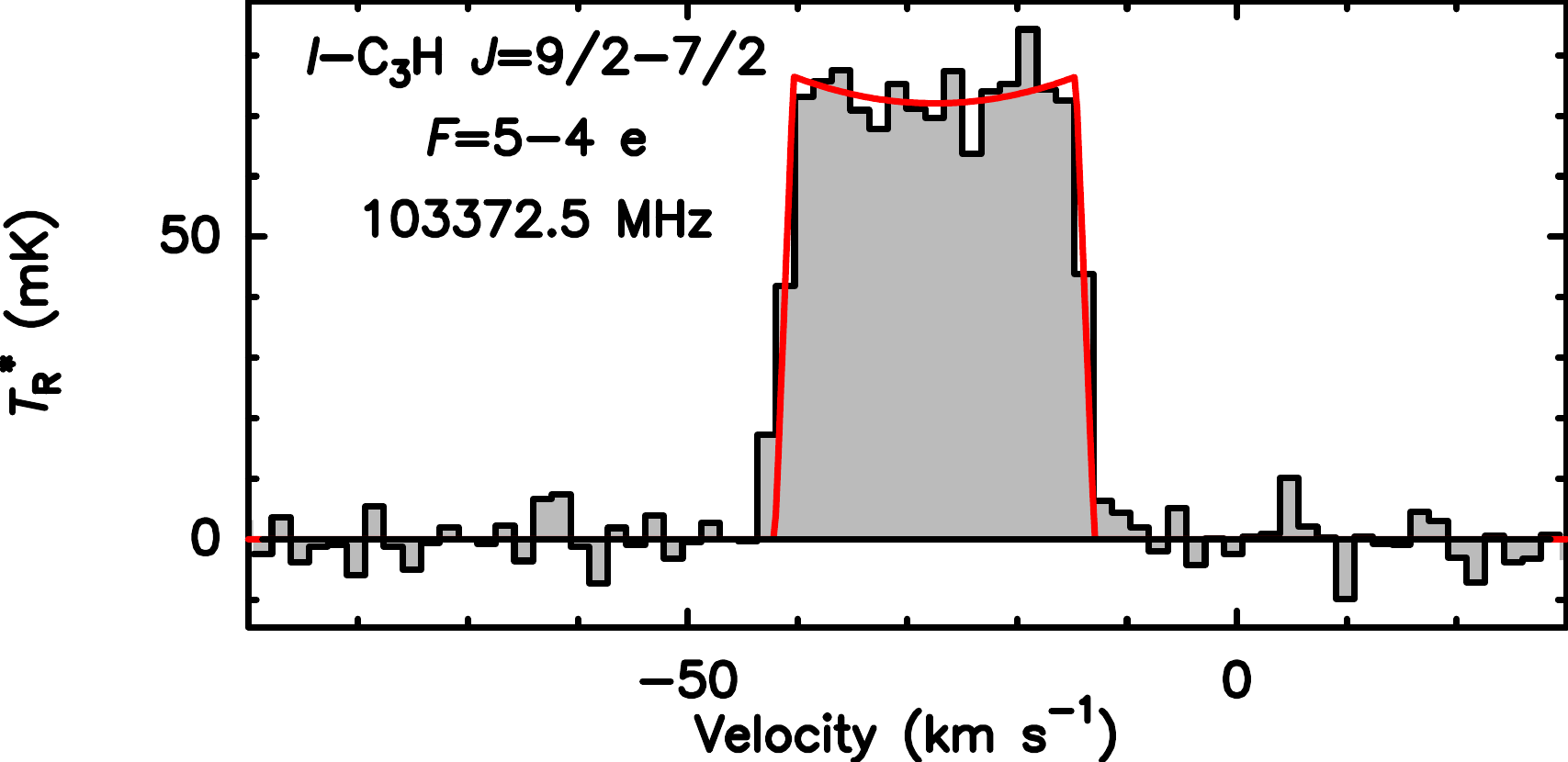}
\caption{{Same as Figure.~\ref{Fig:fitting_1}, but for $l$-C$_{3}H$. }\label{Fig:fitting_12}}
\end{figure*}

\begin{figure*}[!htbp]
\centering
\includegraphics[width = 0.45 \textwidth]{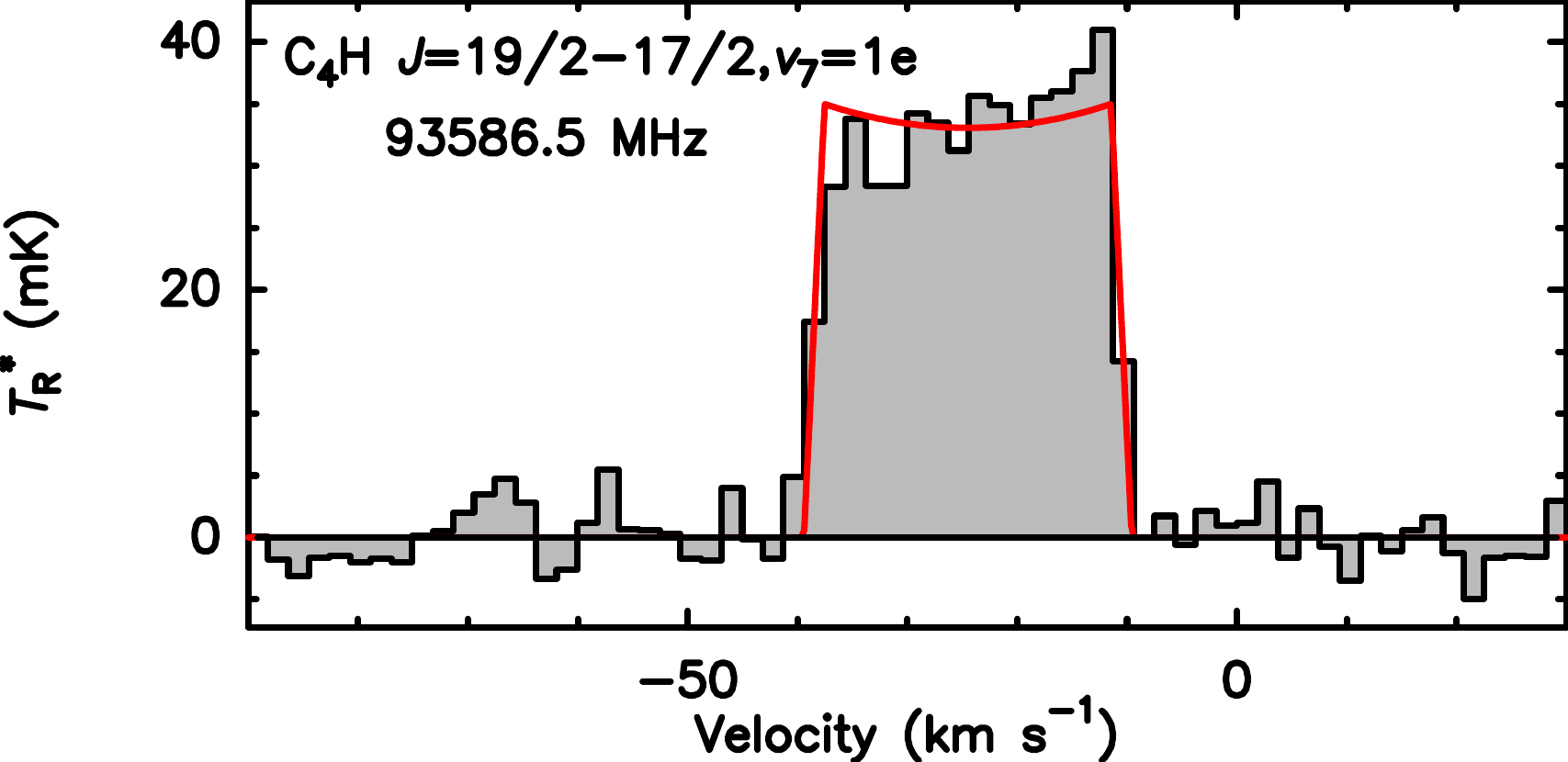}
\hspace{0.05\textwidth}
\includegraphics[width = 0.45 \textwidth]{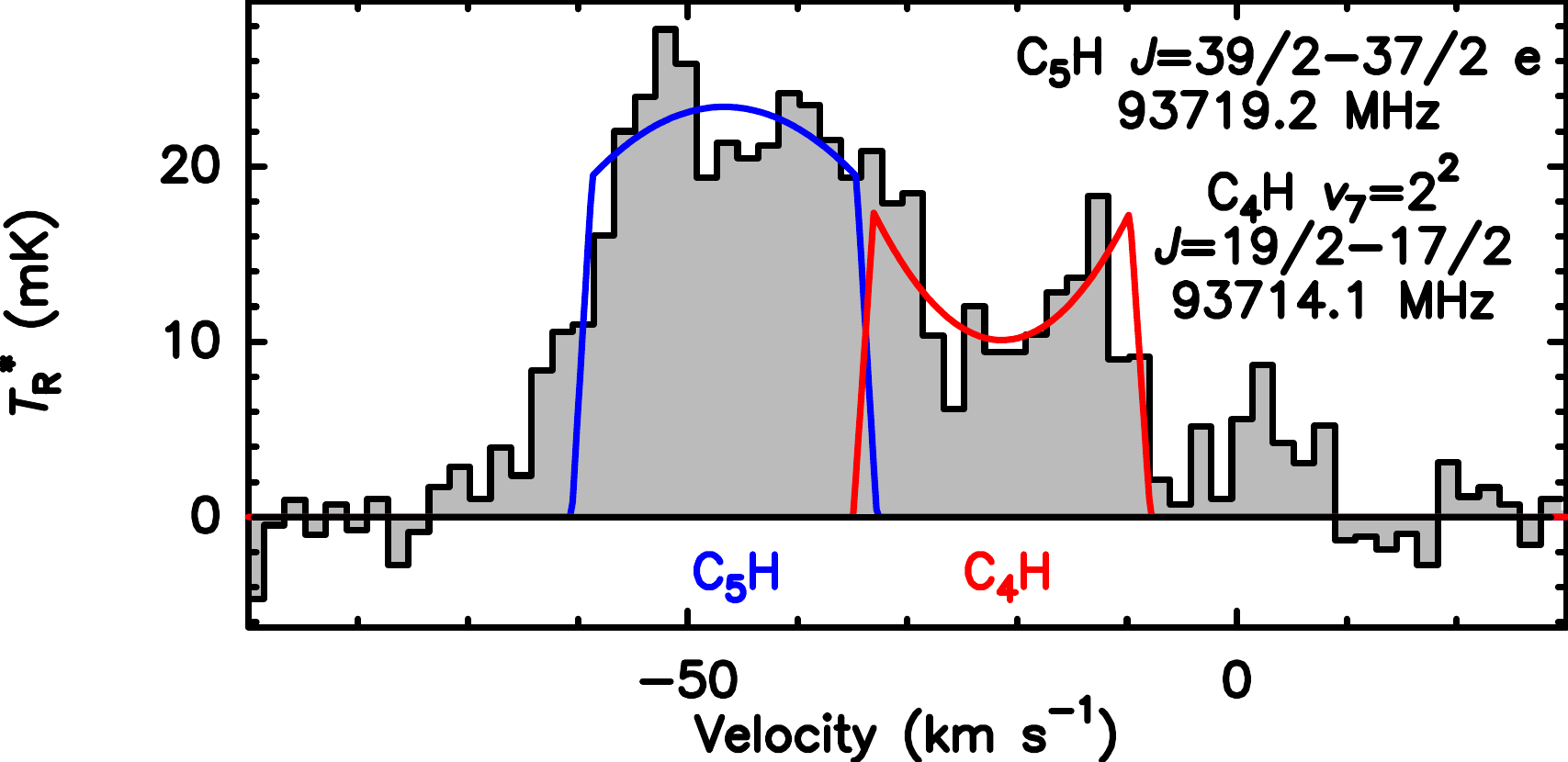}
\vspace{0.1cm}
\includegraphics[width = 0.45 \textwidth]{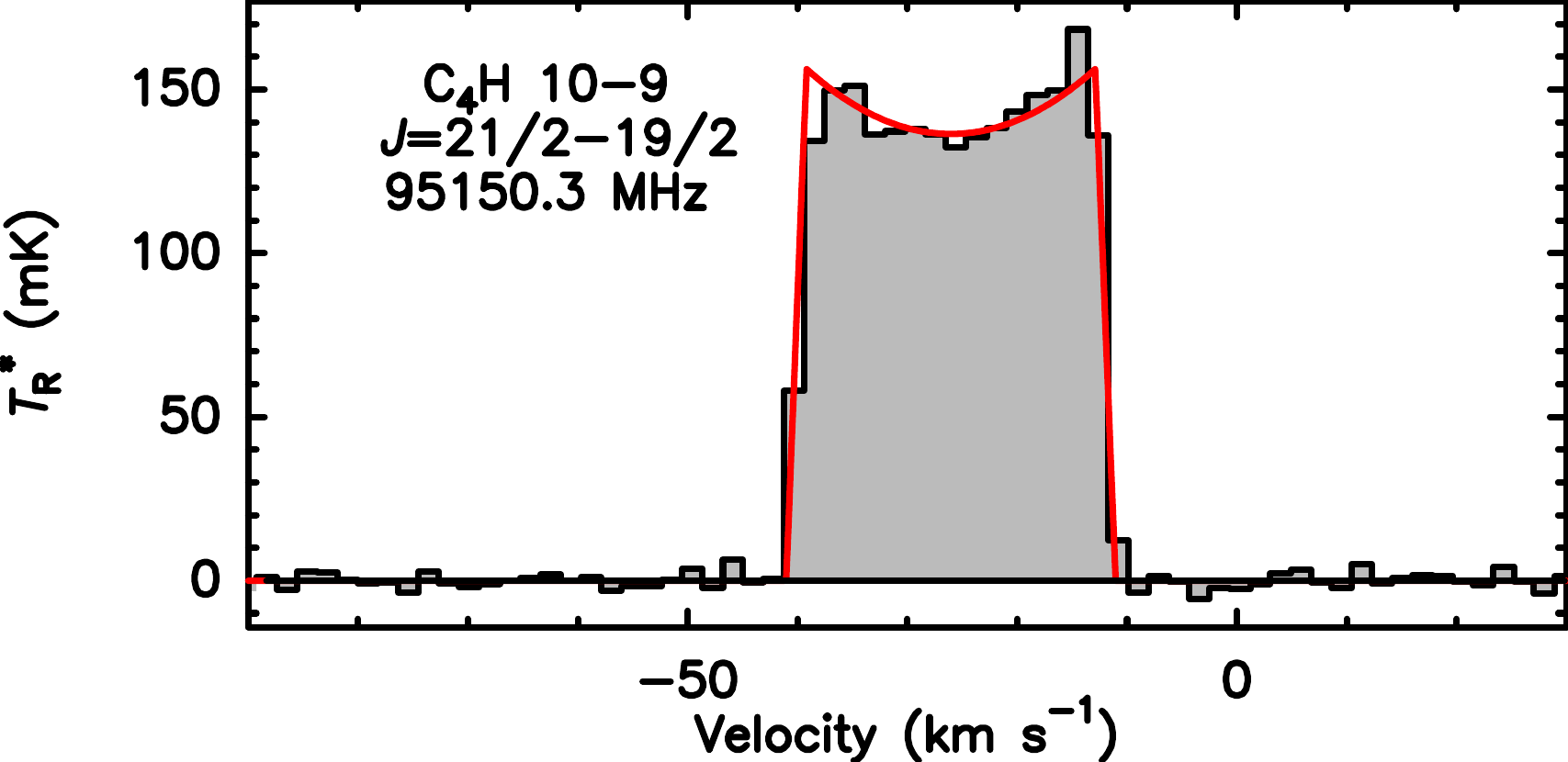}
\hspace{0.05\textwidth}
\includegraphics[width = 0.45 \textwidth]{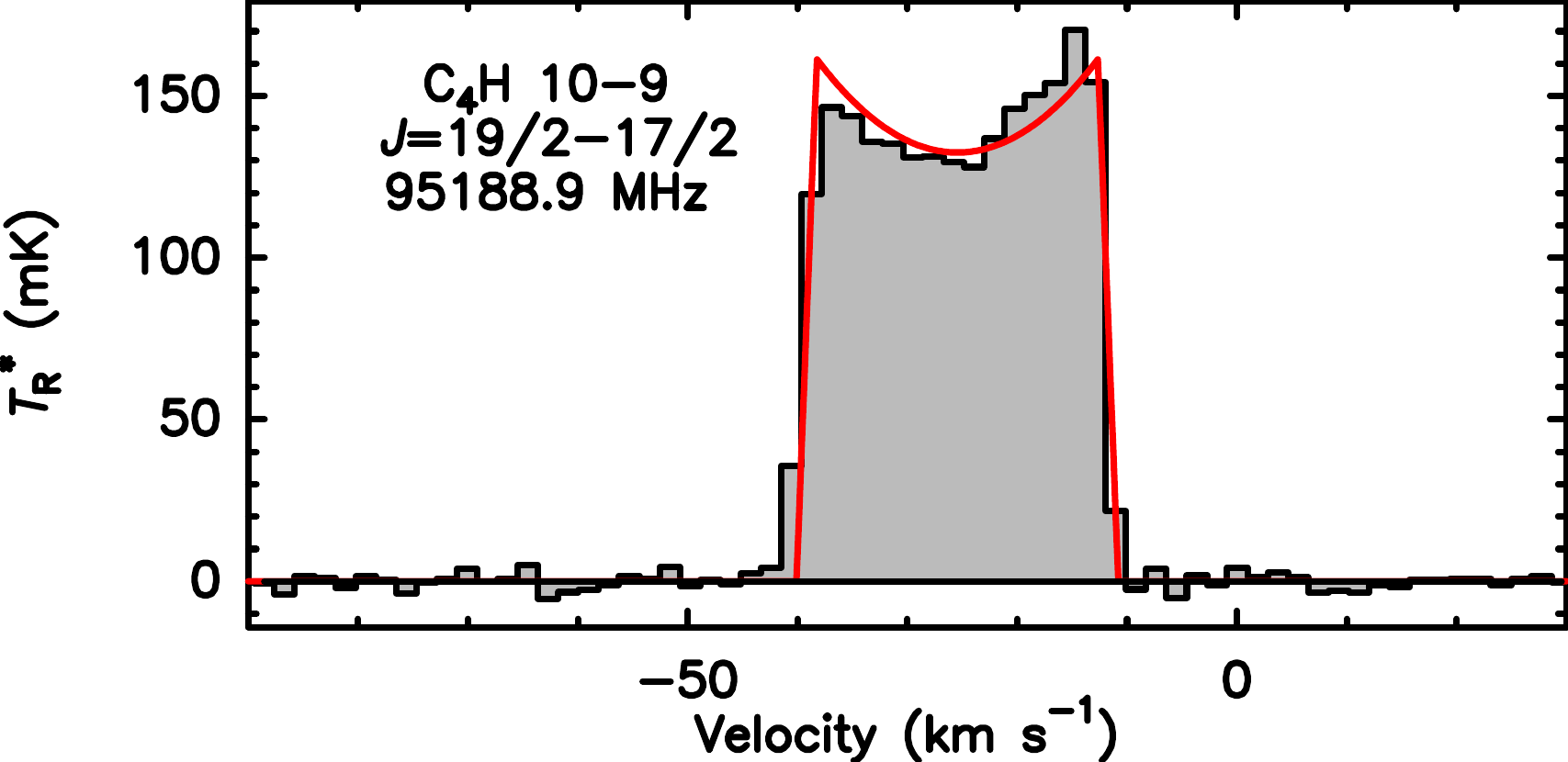}
\vspace{0.1cm}
\includegraphics[width = 0.45 \textwidth]{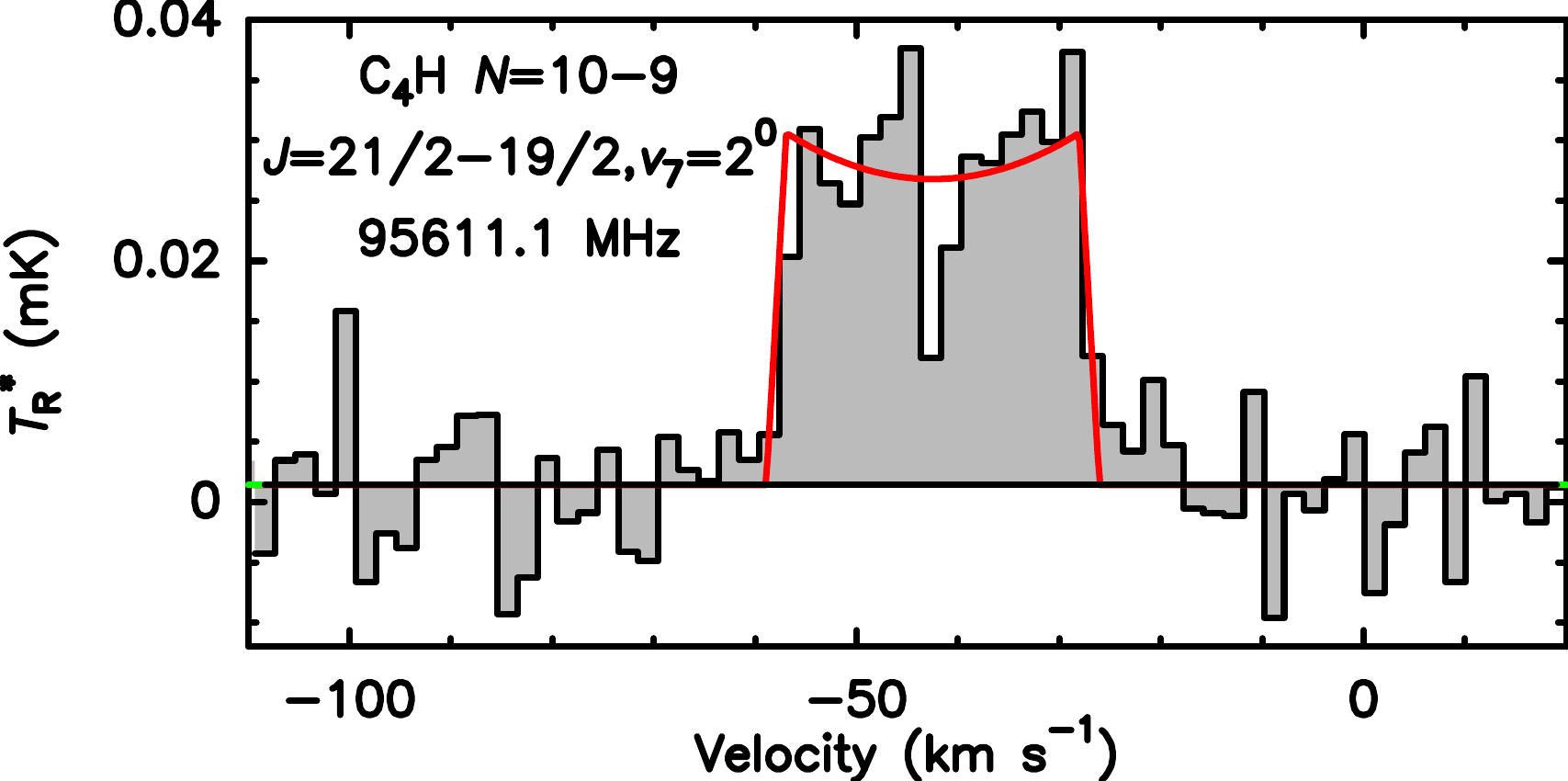}
\hspace{0.05\textwidth}
\includegraphics[width = 0.45 \textwidth]{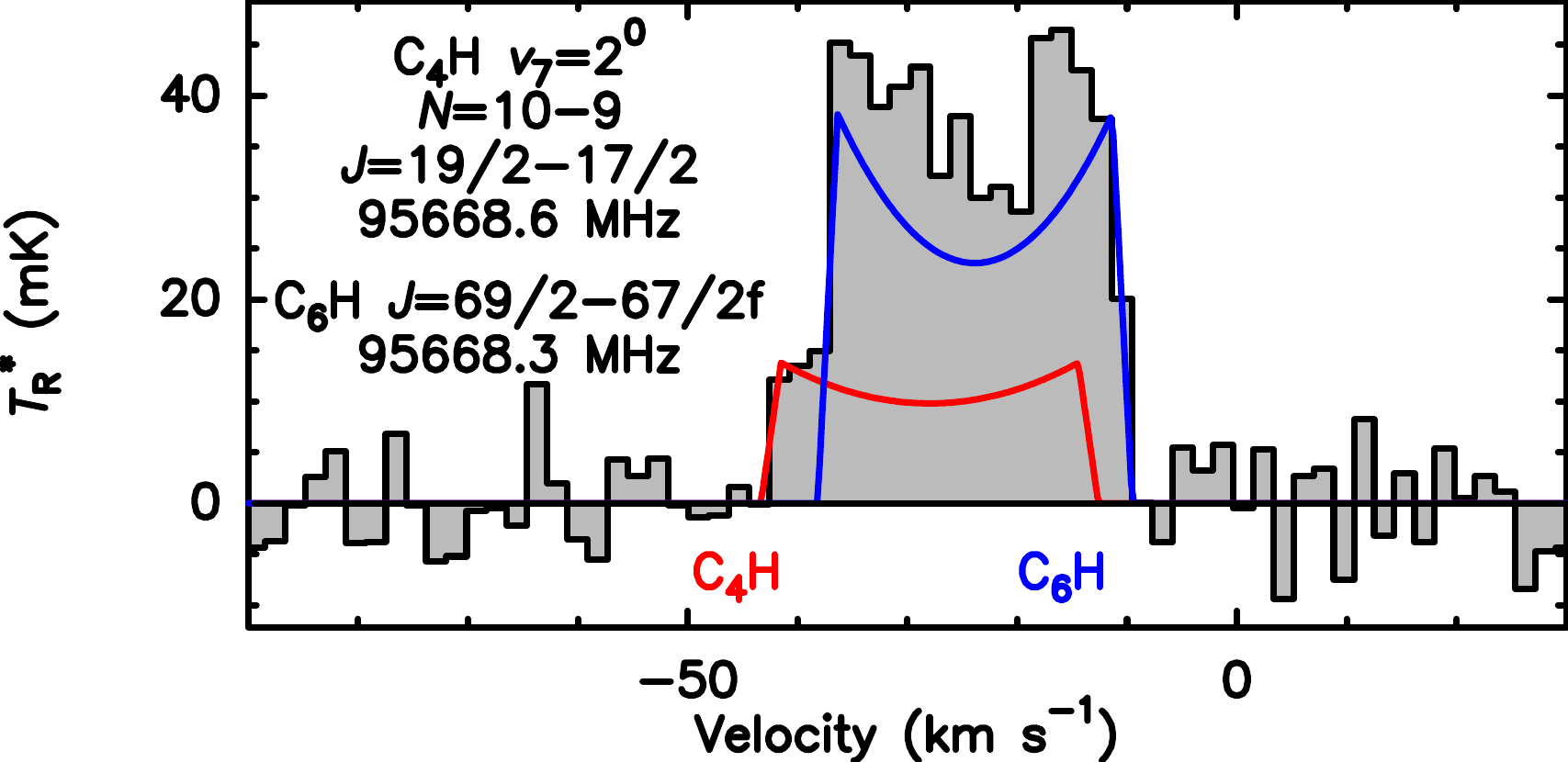}
\vspace{0.1cm}
\includegraphics[width = 0.45 \textwidth]{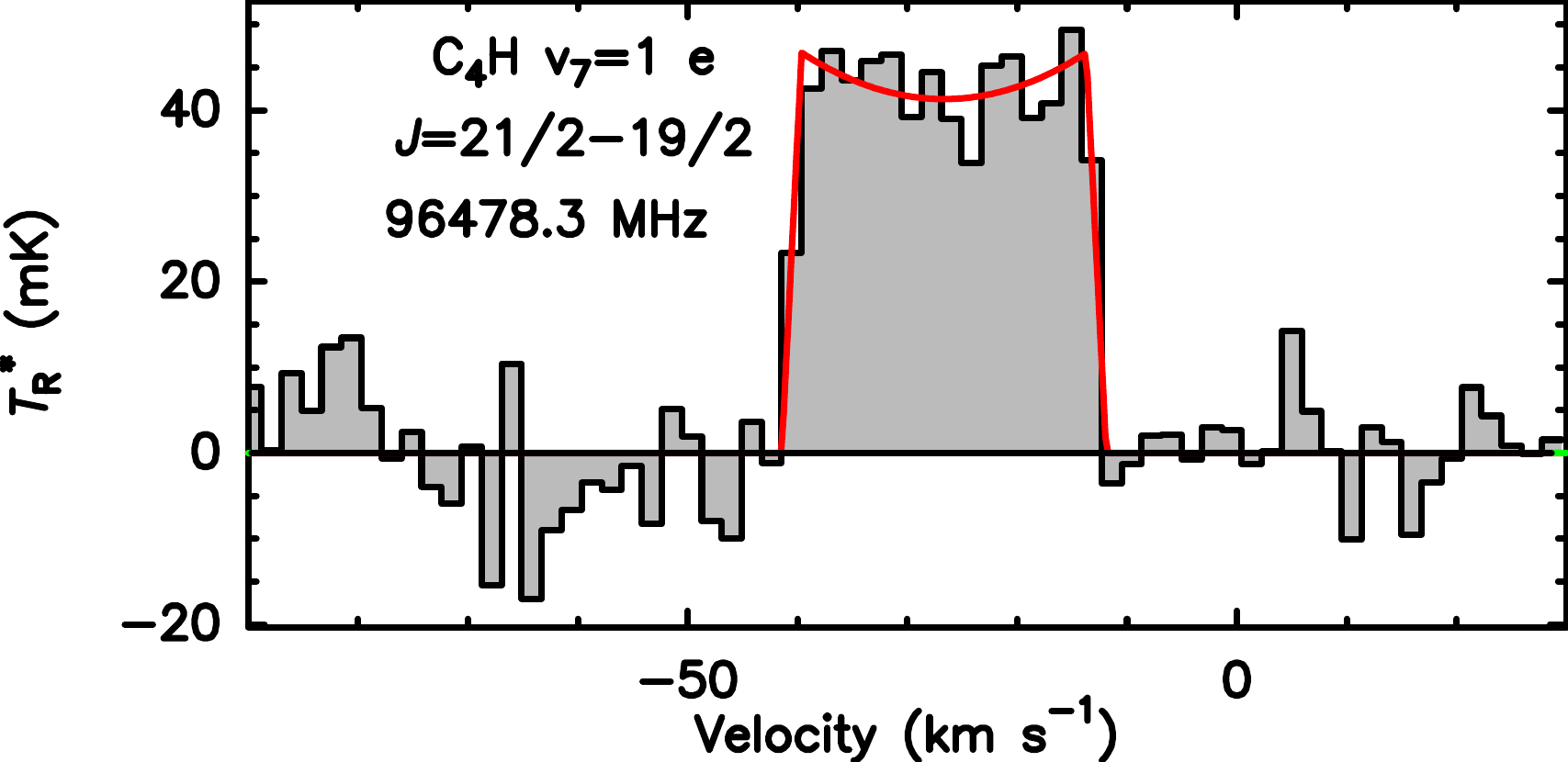}
\hspace{0.05\textwidth}
\includegraphics[width = 0.45 \textwidth]{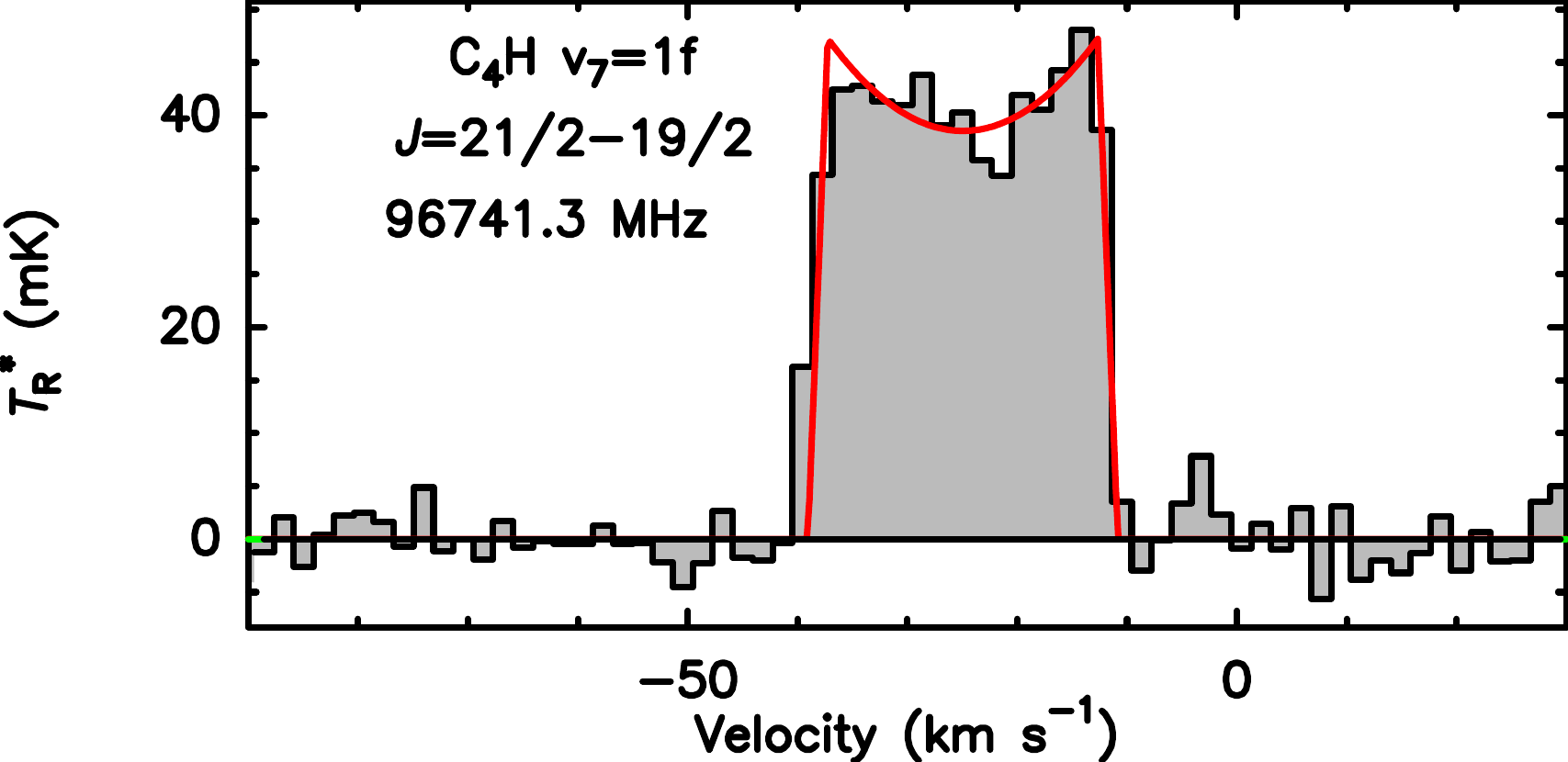}
\vspace{0.1cm}
\includegraphics[width = 0.45 \textwidth]{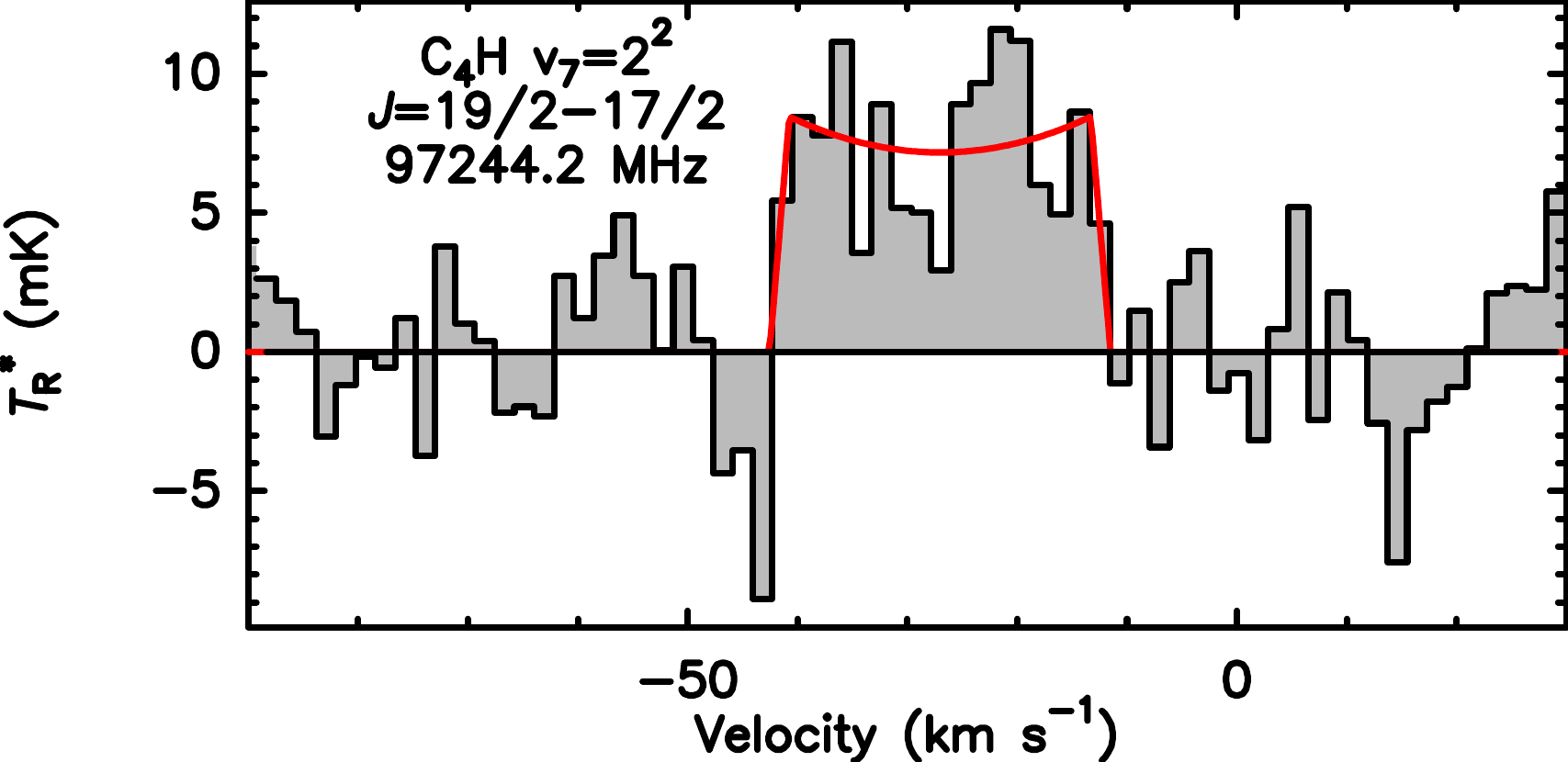}
\hspace{0.05\textwidth}
\includegraphics[width = 0.45 \textwidth]{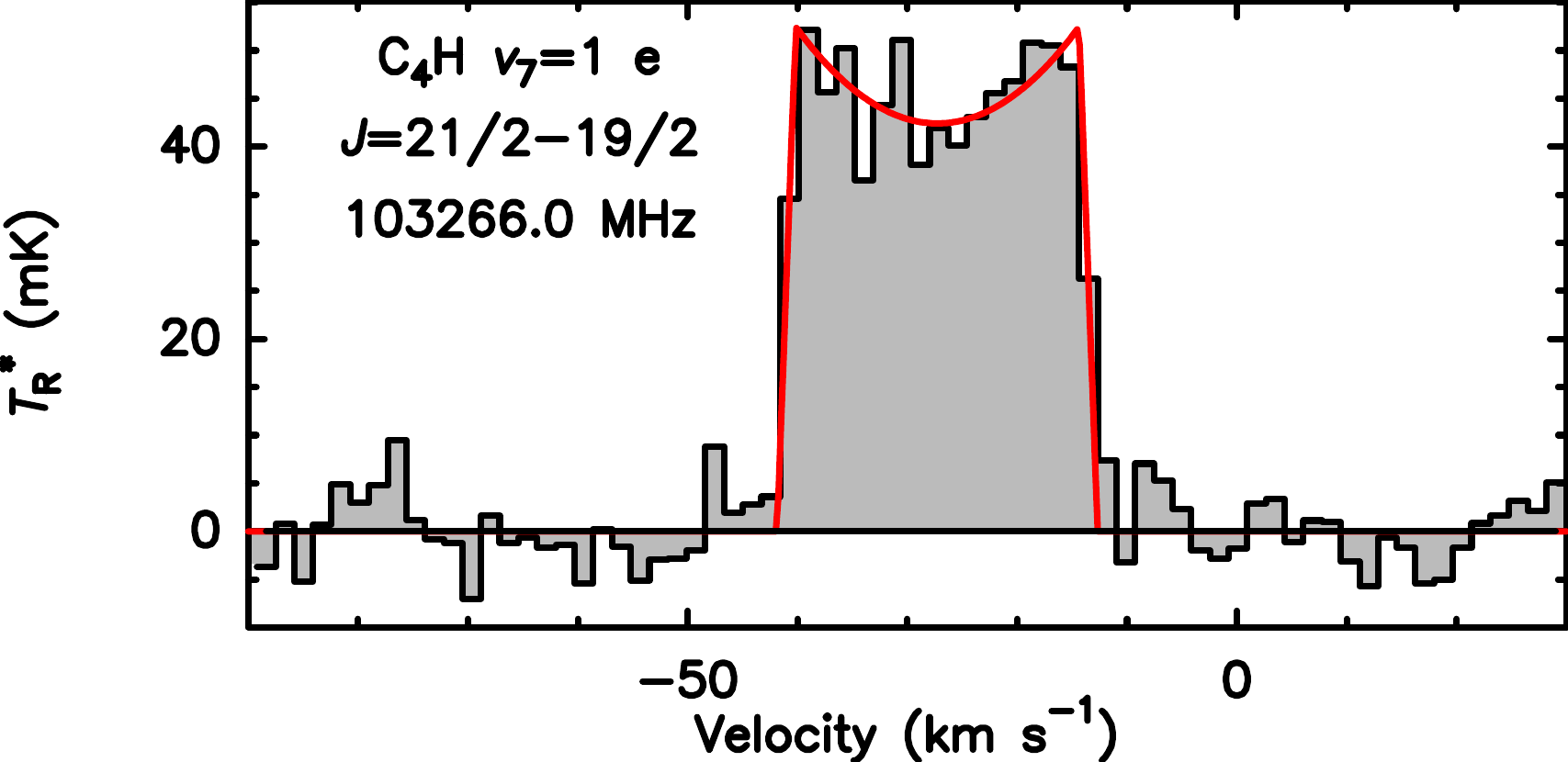}
\caption{{Same as Figure.~\ref{Fig:fitting_1}, but for C$_{4}$H. }\label{Fig:fitting_13}}
\end{figure*}

\begin{figure*}[!htbp]
\centering
\includegraphics[width = 0.45 \textwidth]{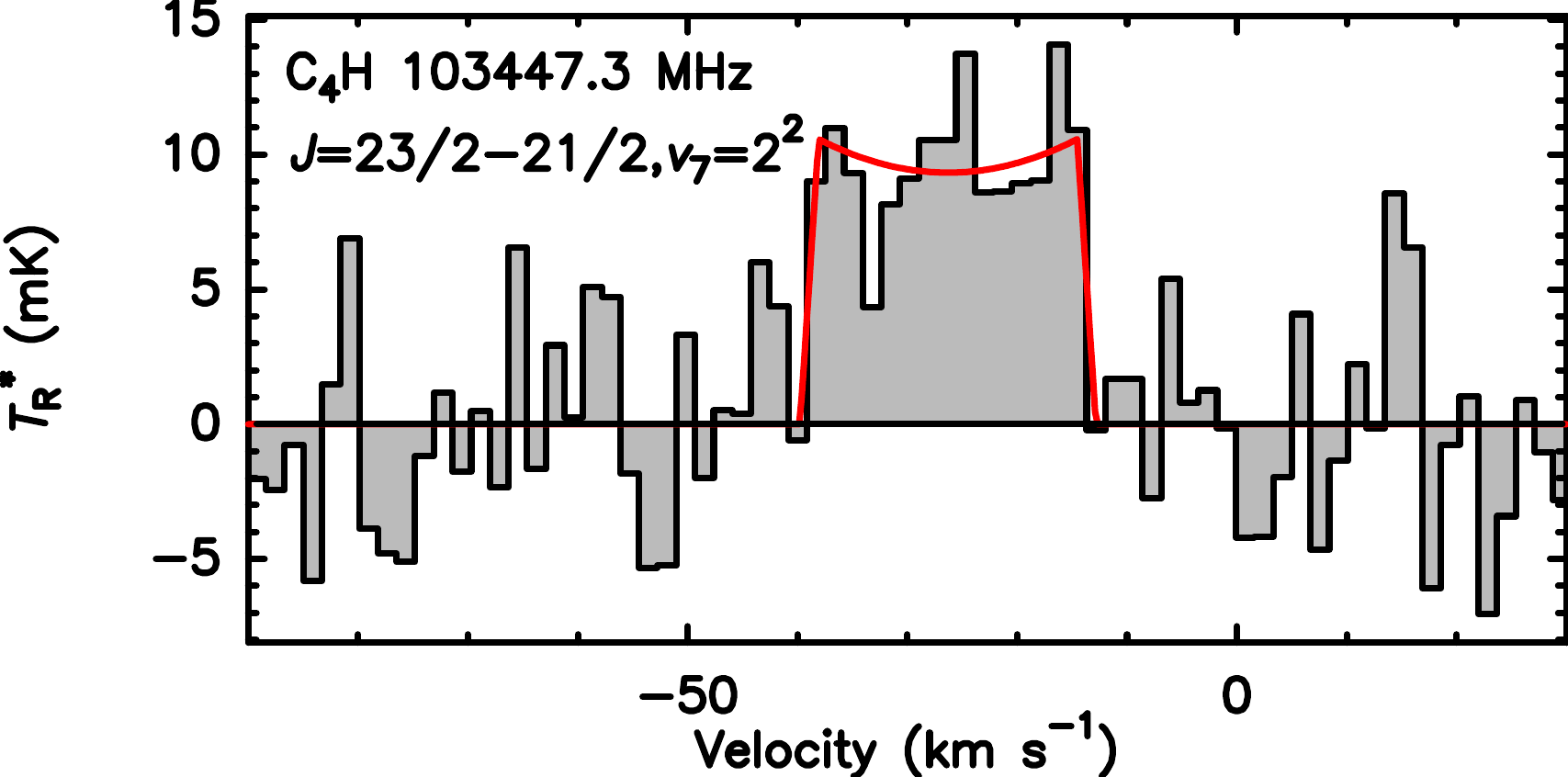}
\hspace{0.05\textwidth}
\includegraphics[width = 0.45 \textwidth]{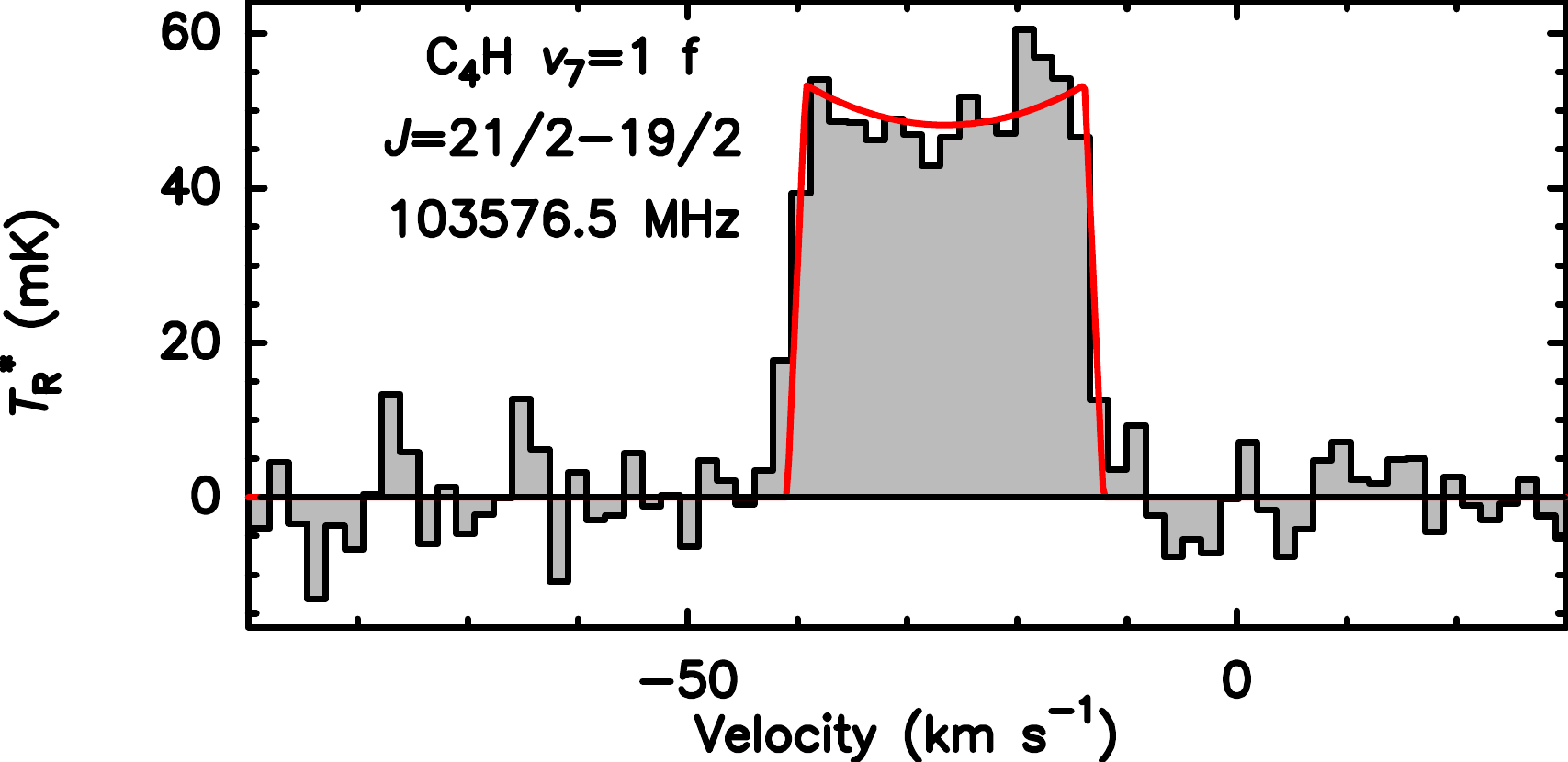}
\vspace{0.1cm}
\includegraphics[width = 0.45 \textwidth]{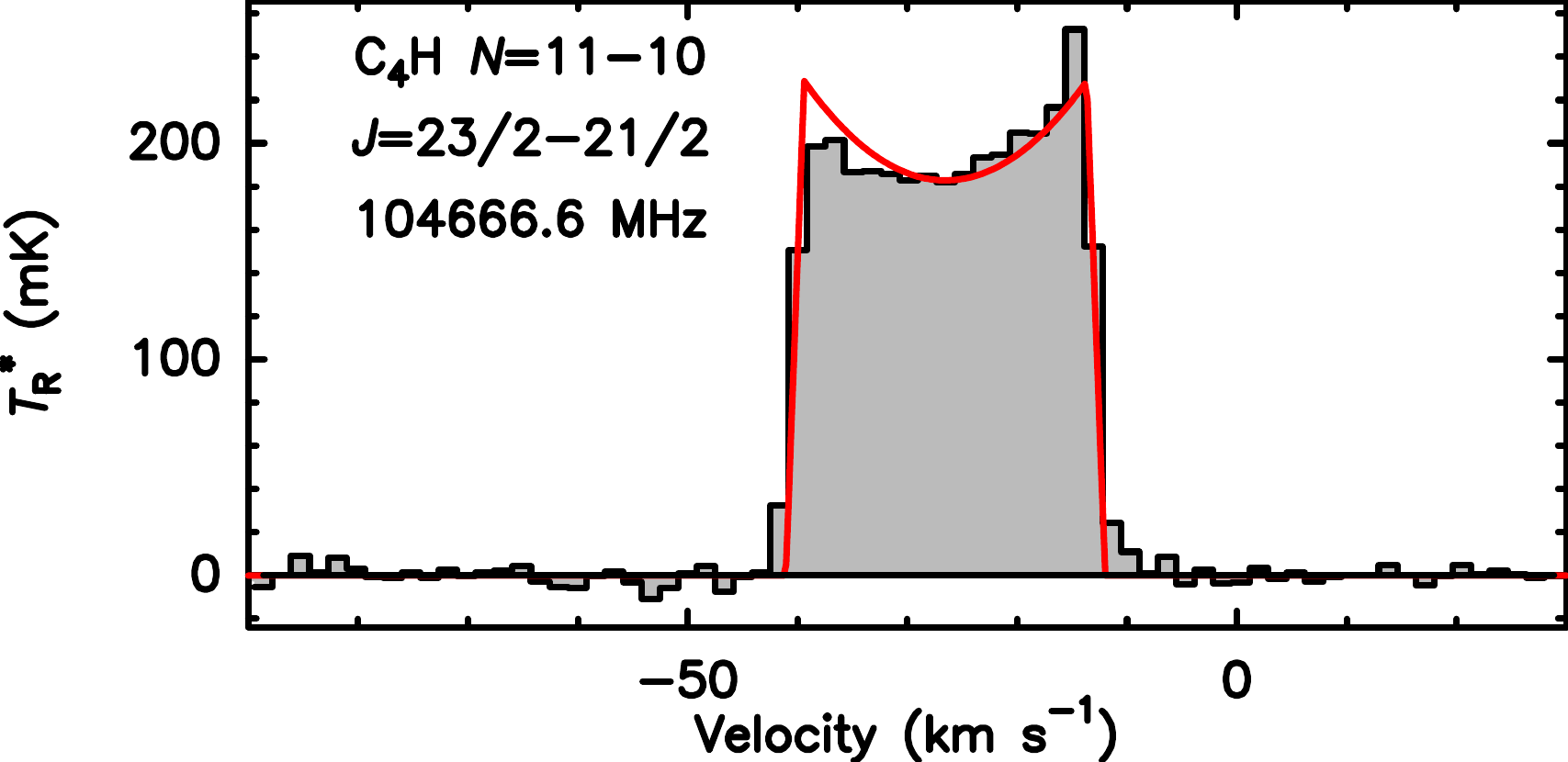}
\hspace{0.05\textwidth}
\includegraphics[width = 0.45 \textwidth]{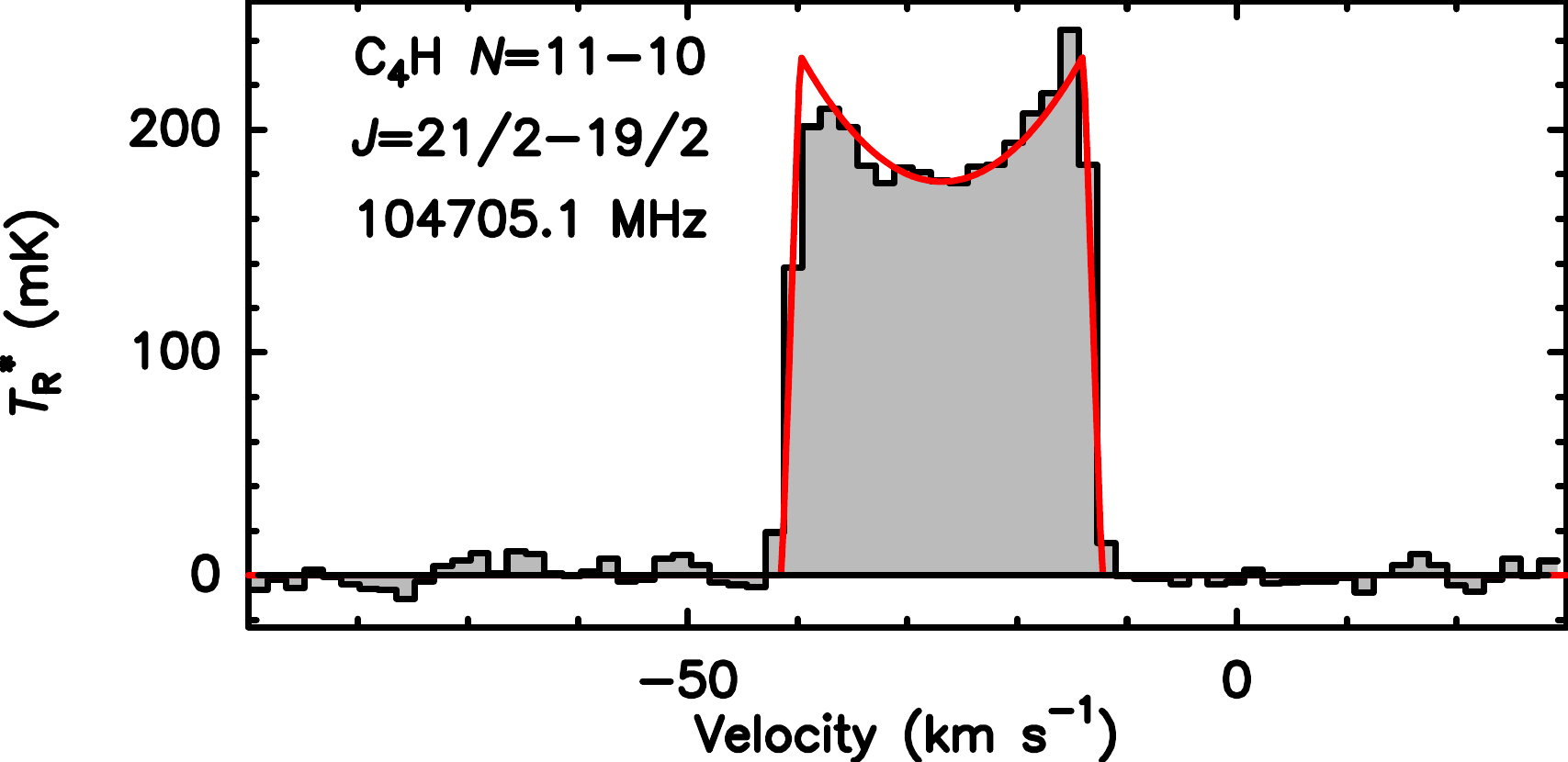}
\vspace{0.1cm}
\includegraphics[width = 0.45 \textwidth]{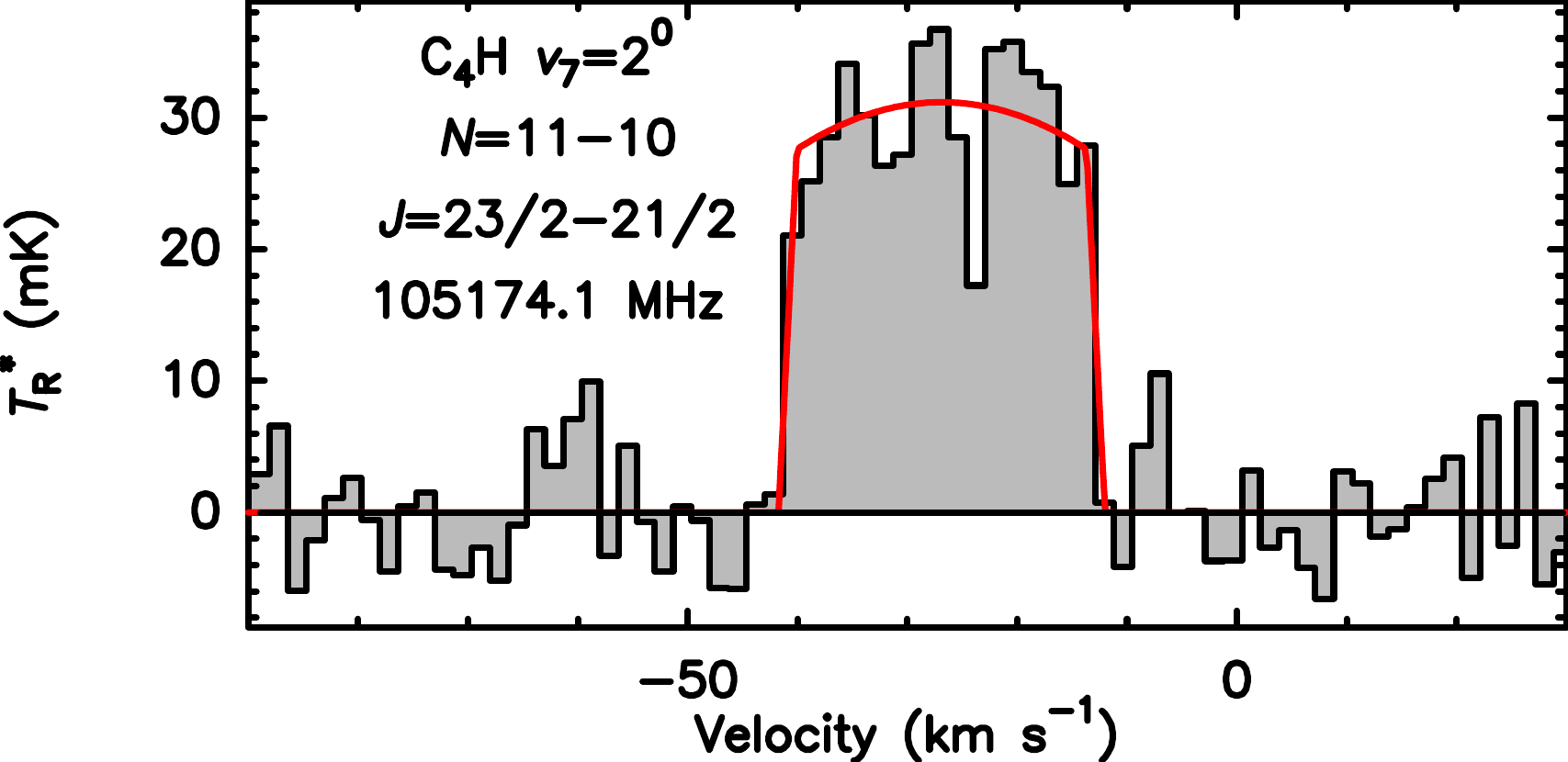}
\hspace{0.05\textwidth}
\includegraphics[width = 0.45 \textwidth]{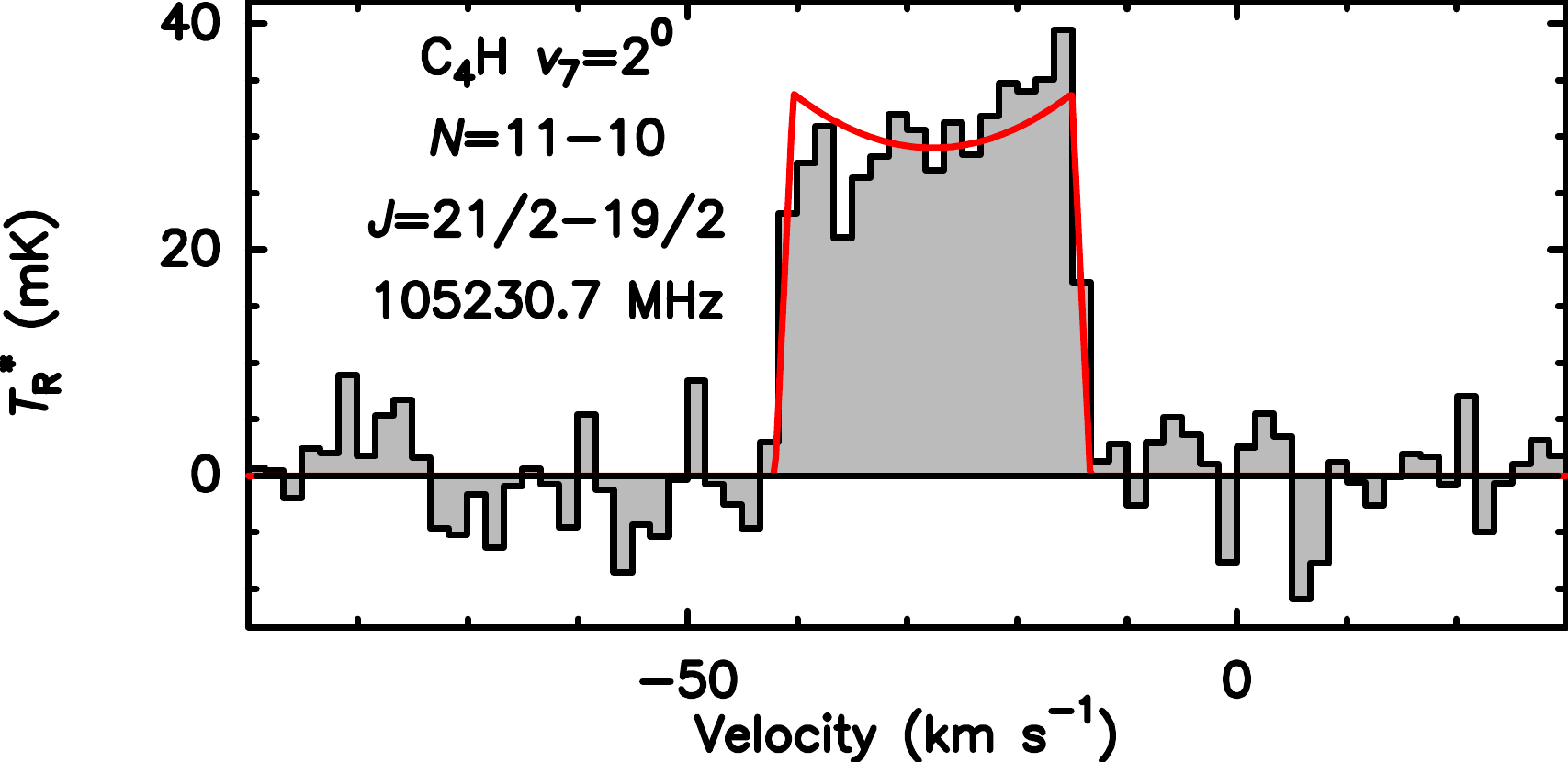}
\vspace{0.1cm}
\includegraphics[width = 0.45 \textwidth]{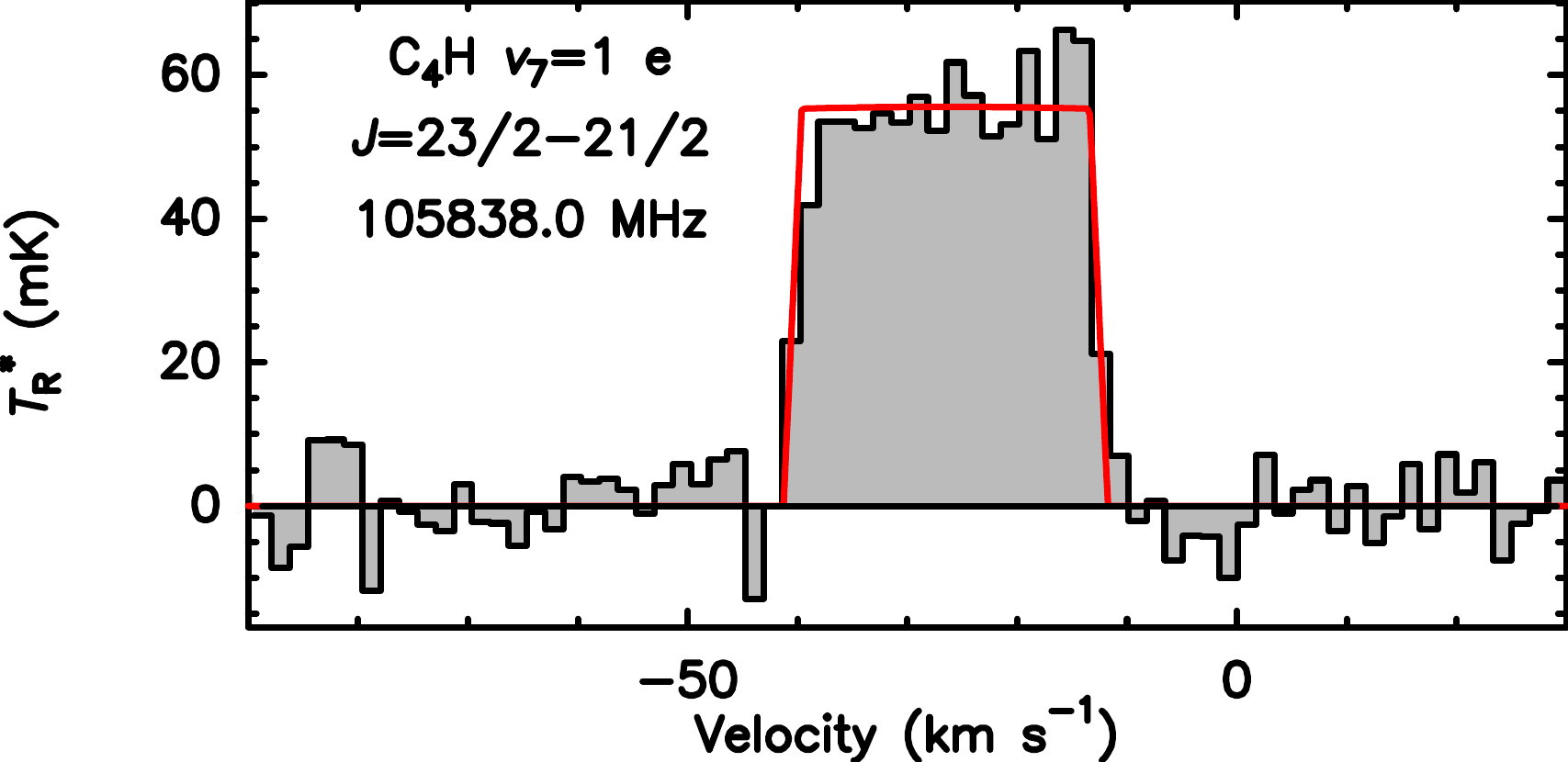}
\hspace{0.05\textwidth}
\includegraphics[width = 0.45 \textwidth]{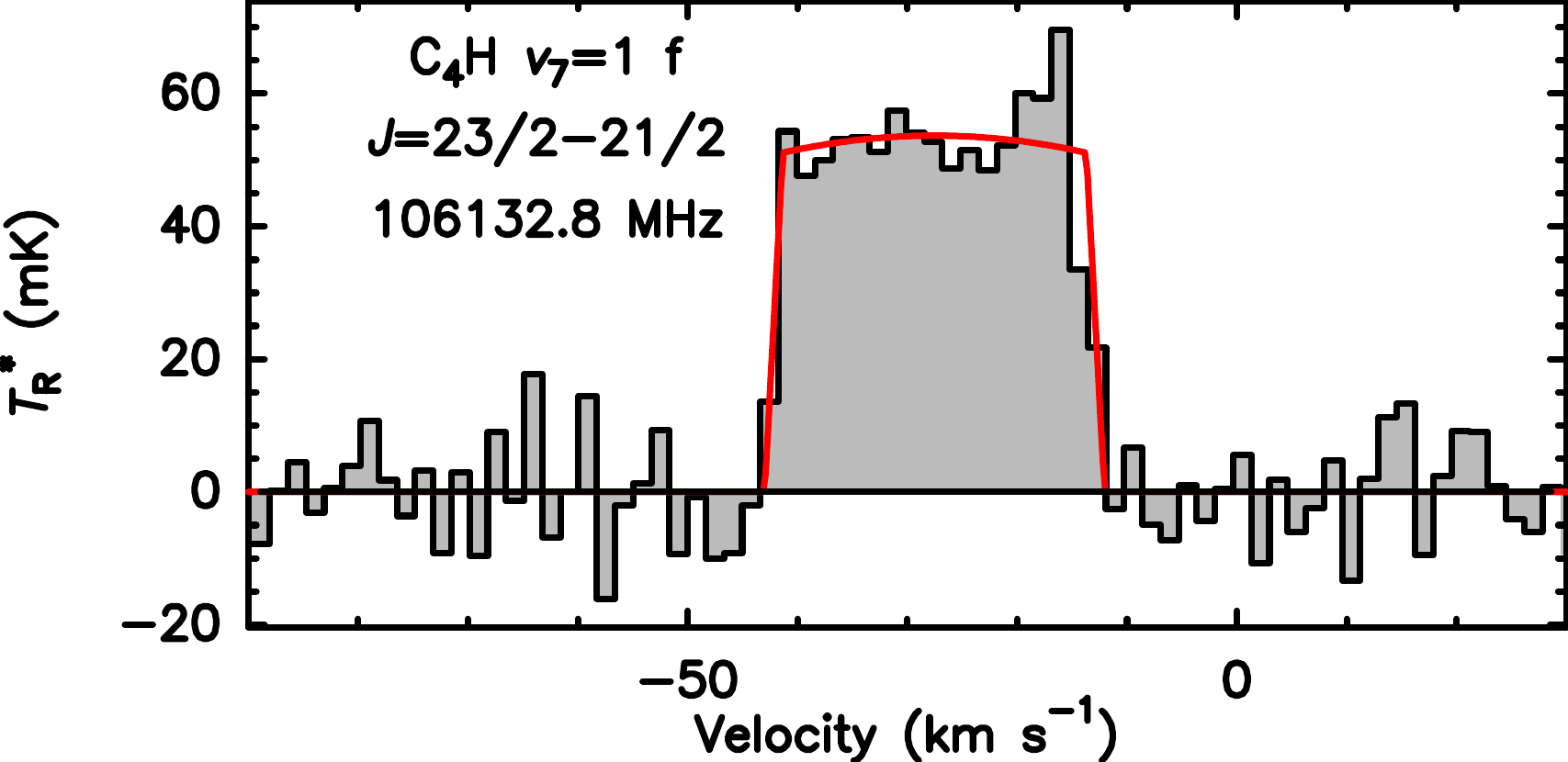}
\vspace{0.1cm}
\includegraphics[width = 0.45 \textwidth]{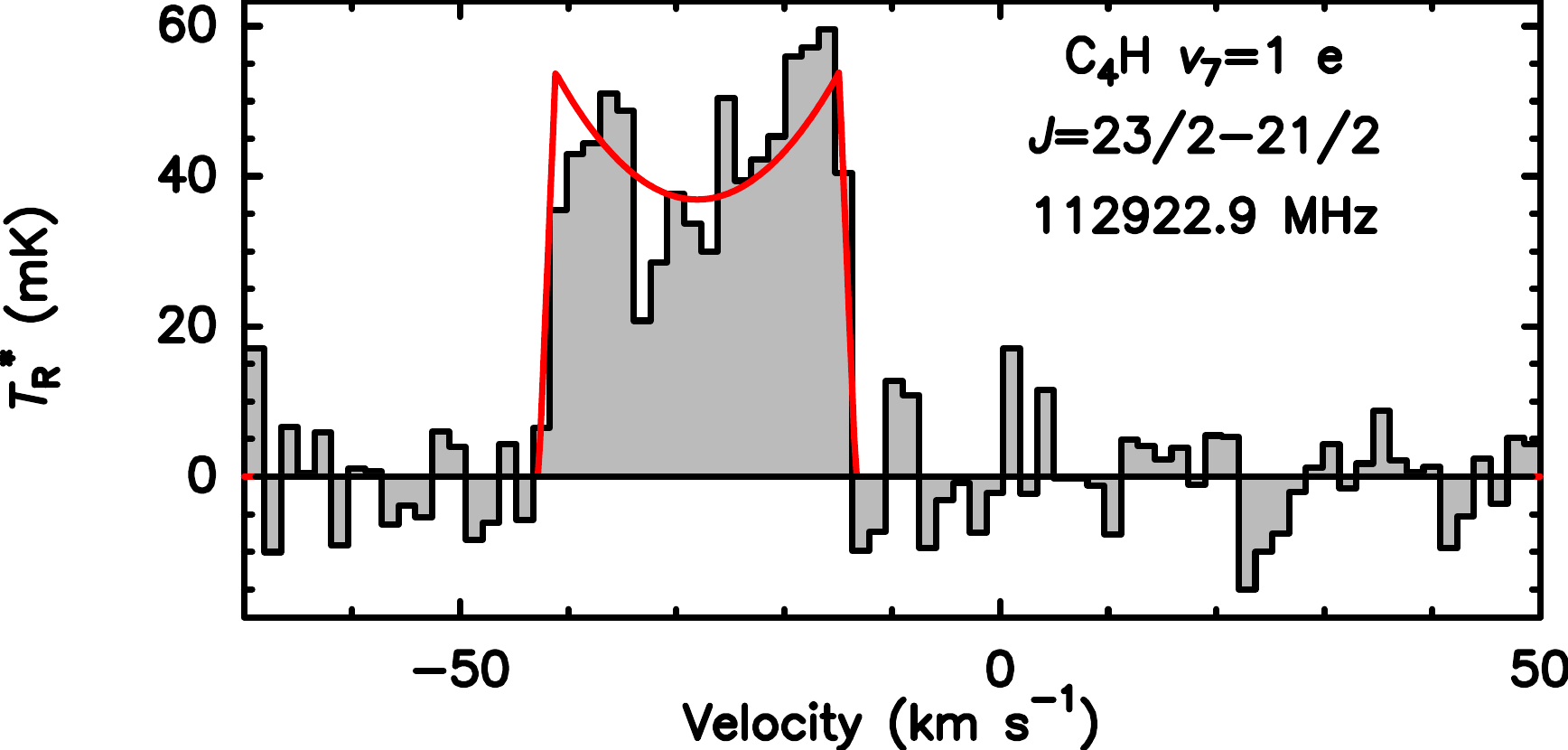}
\hspace{0.05\textwidth}
\includegraphics[width = 0.45 \textwidth]{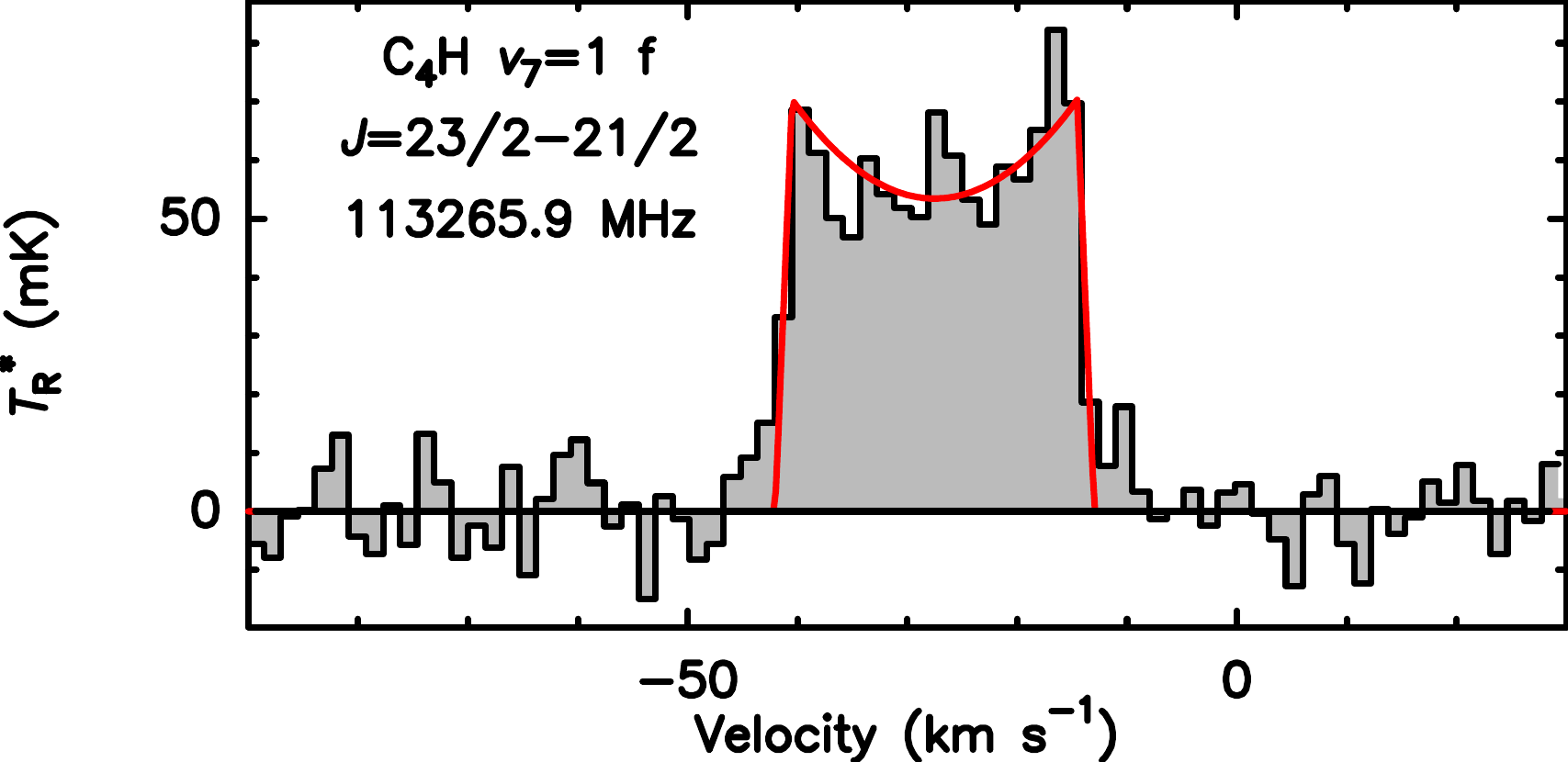}
\centerline{Figure \ref{Fig:fitting_13}. --- continued}
\end{figure*}

\begin{figure*}[!htbp]
\centering
\includegraphics[width = 0.45 \textwidth]{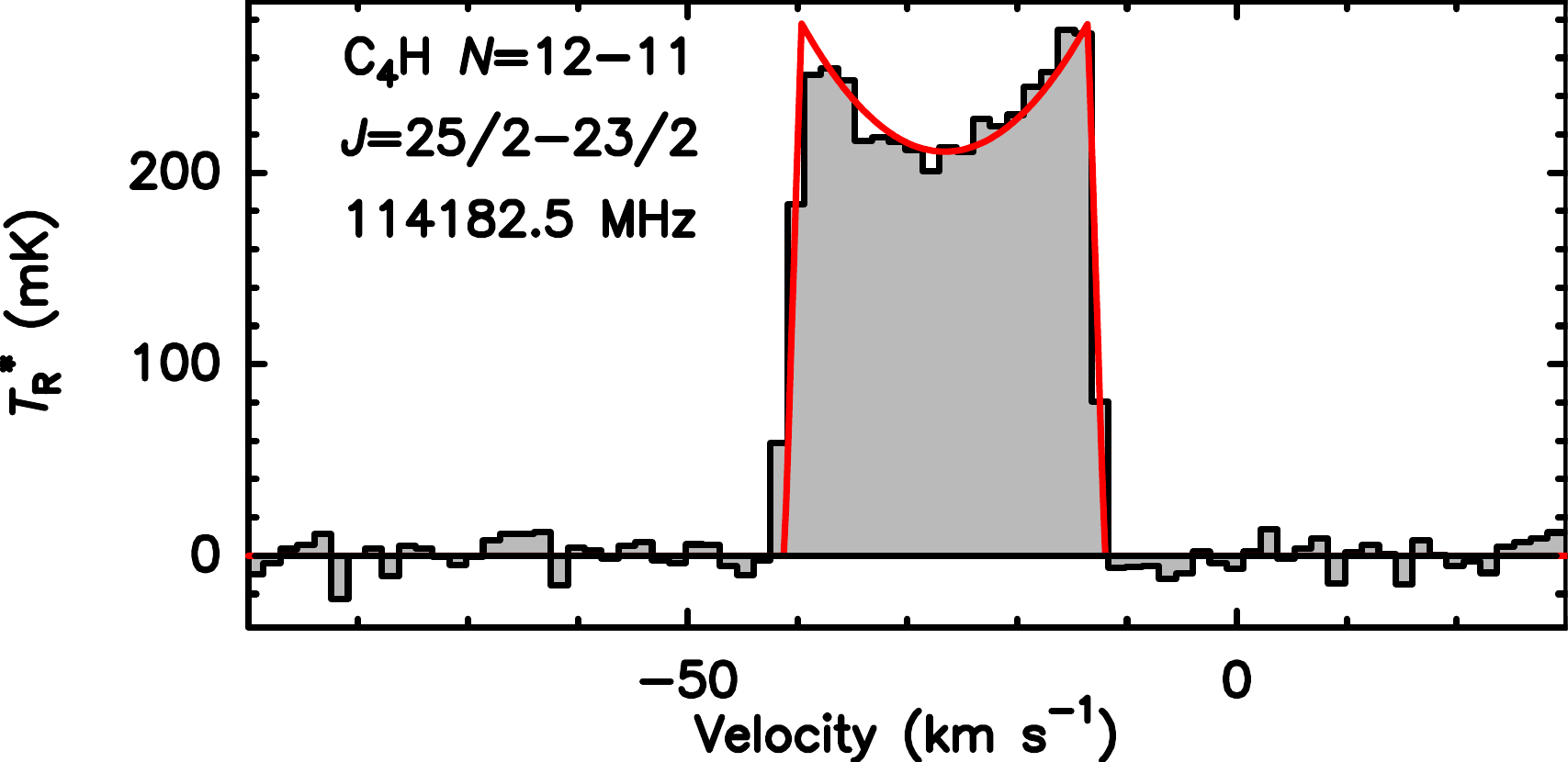}
\hspace{0.05\textwidth}
\includegraphics[width = 0.45 \textwidth]{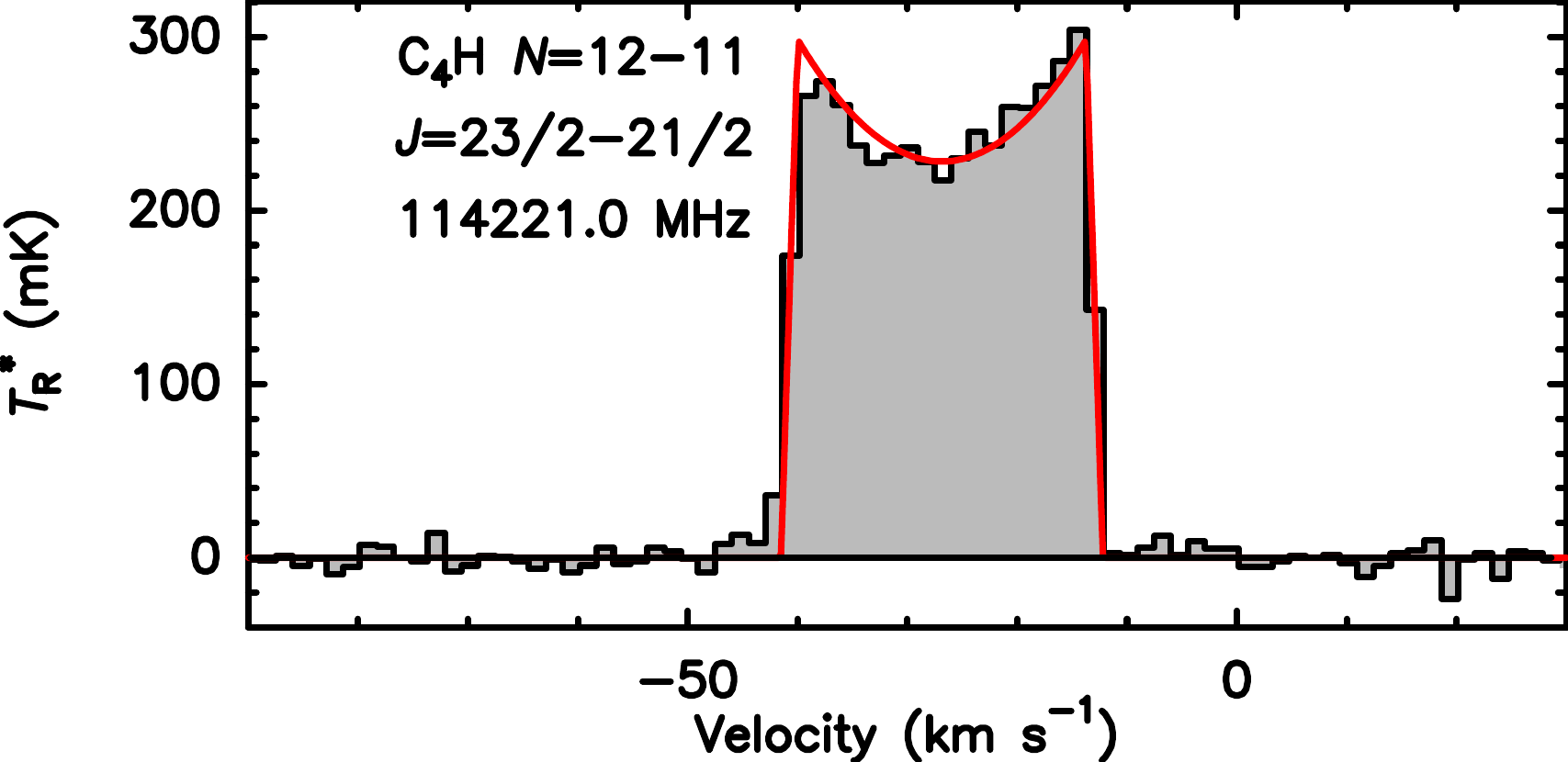}
\vspace{0.1cm}
\includegraphics[width = 0.45 \textwidth]{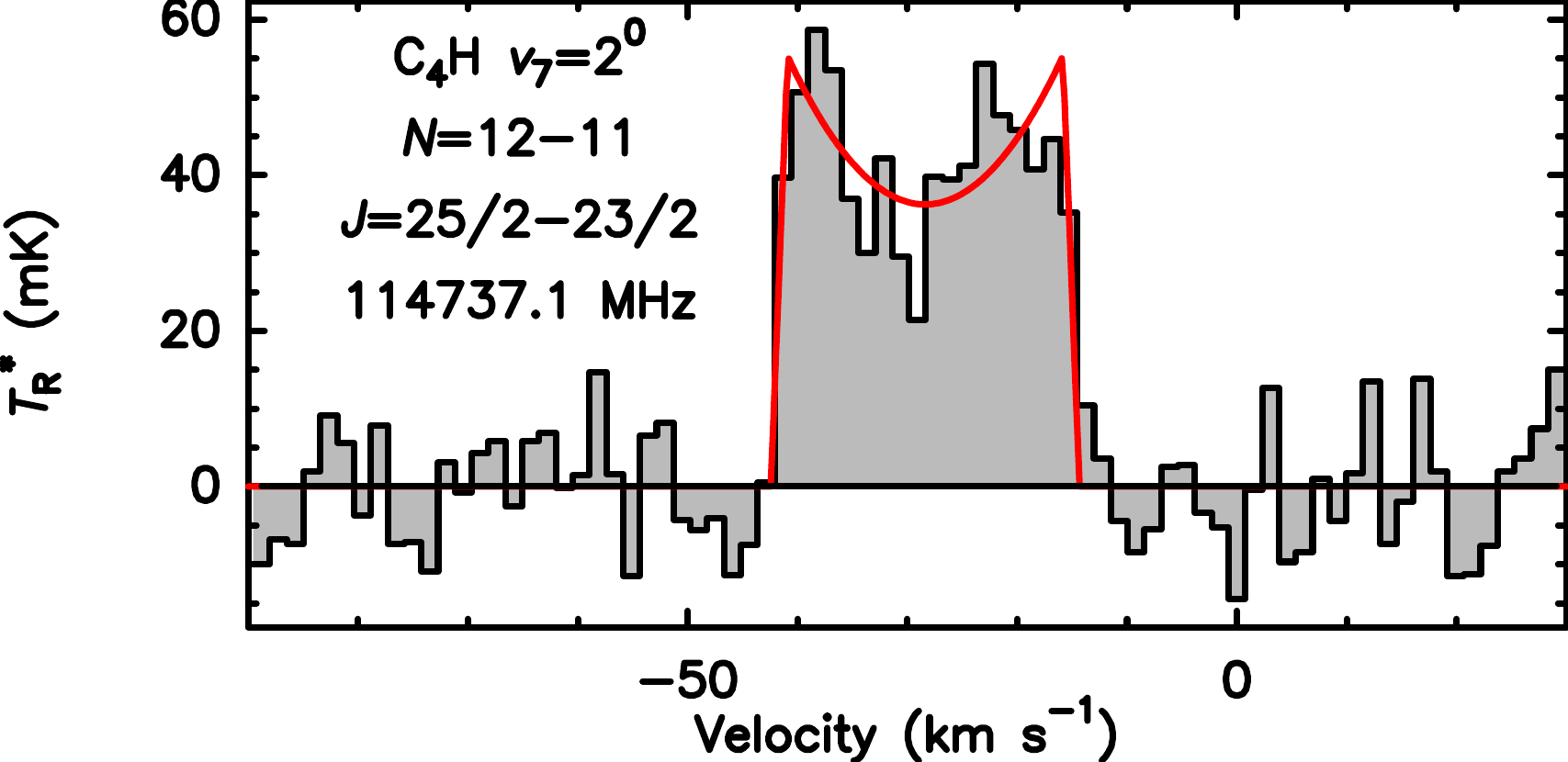}
\hspace{0.05\textwidth}
\includegraphics[width = 0.45 \textwidth]{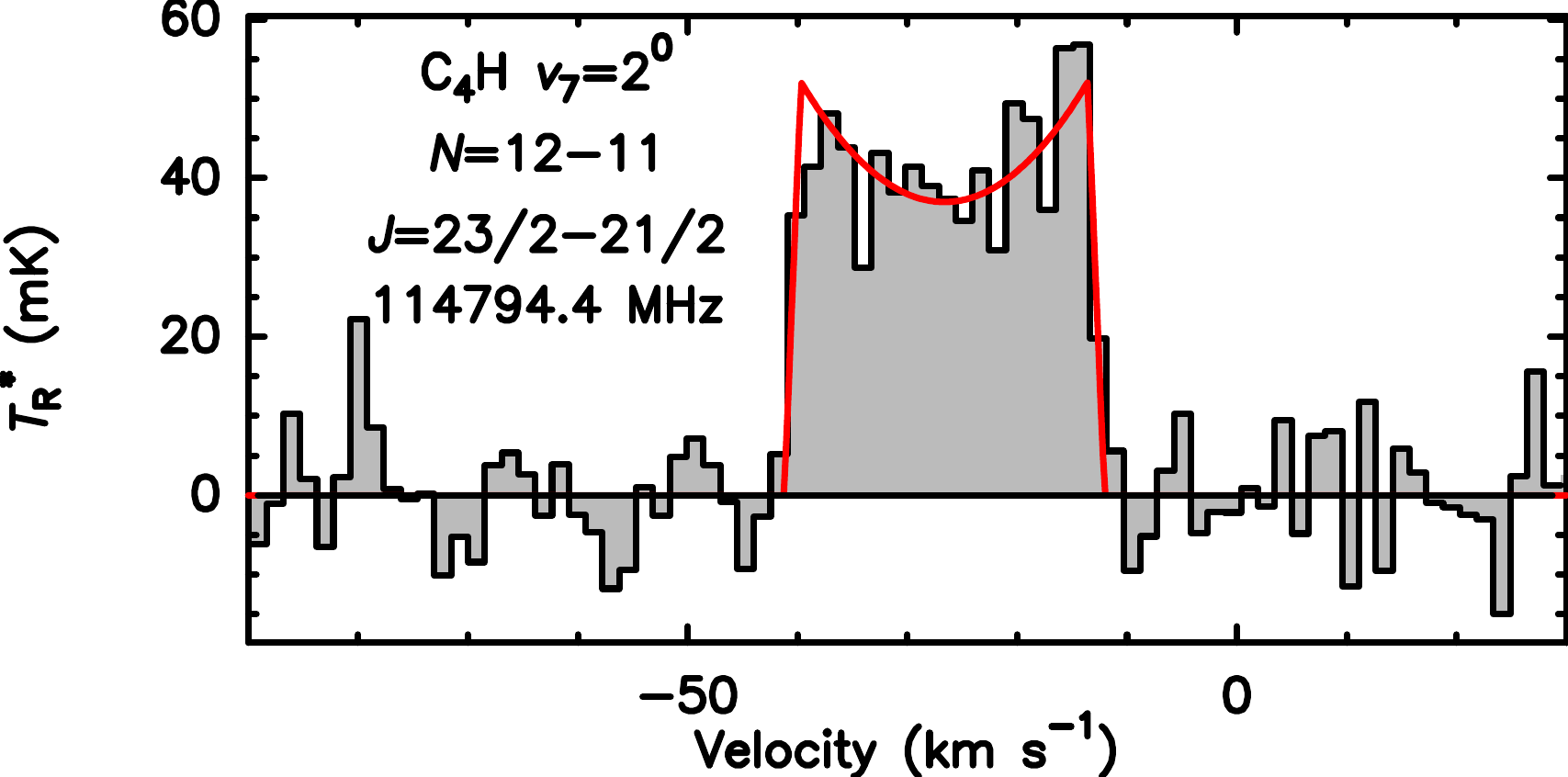}
\vspace{0.1cm}
\includegraphics[width = 0.45 \textwidth]{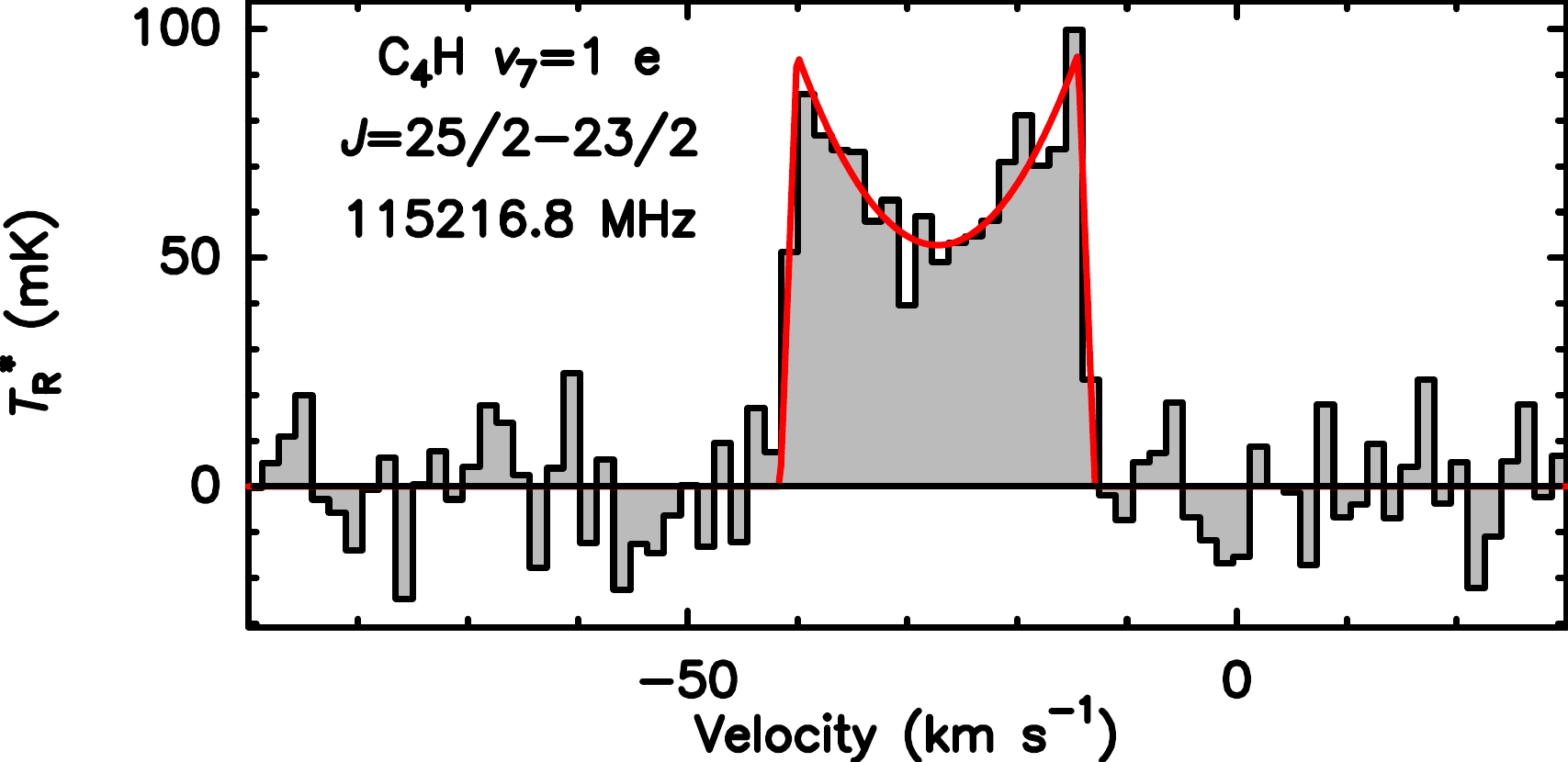}
\hspace{0.05\textwidth}
\includegraphics[width = 0.45 \textwidth]{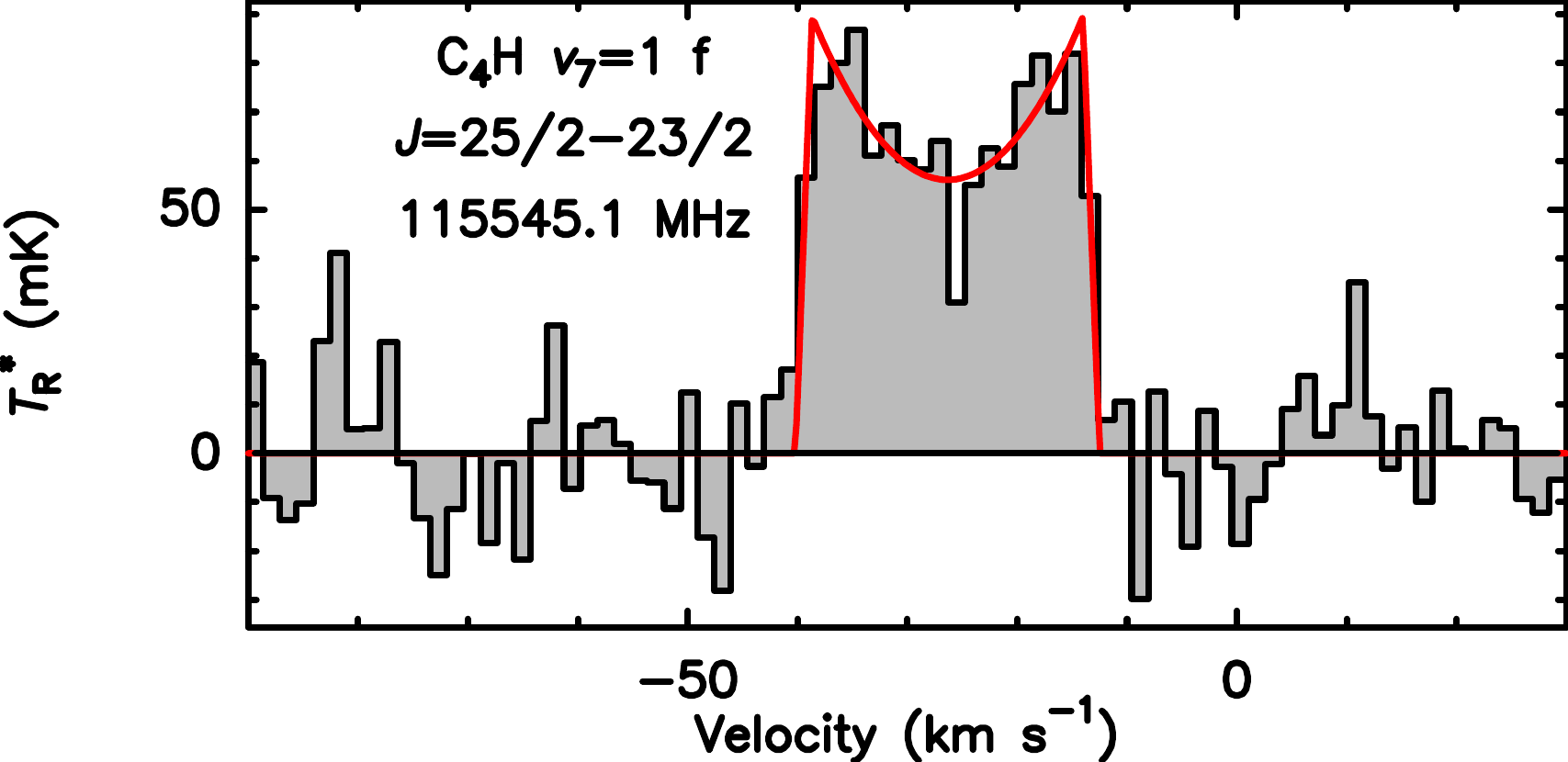}
\vspace{0.1cm}
\centerline{Figure \ref{Fig:fitting_13}. --- continued}
\end{figure*}

\begin{figure*}[!htbp]
\centering
\includegraphics[width = 0.45 \textwidth]{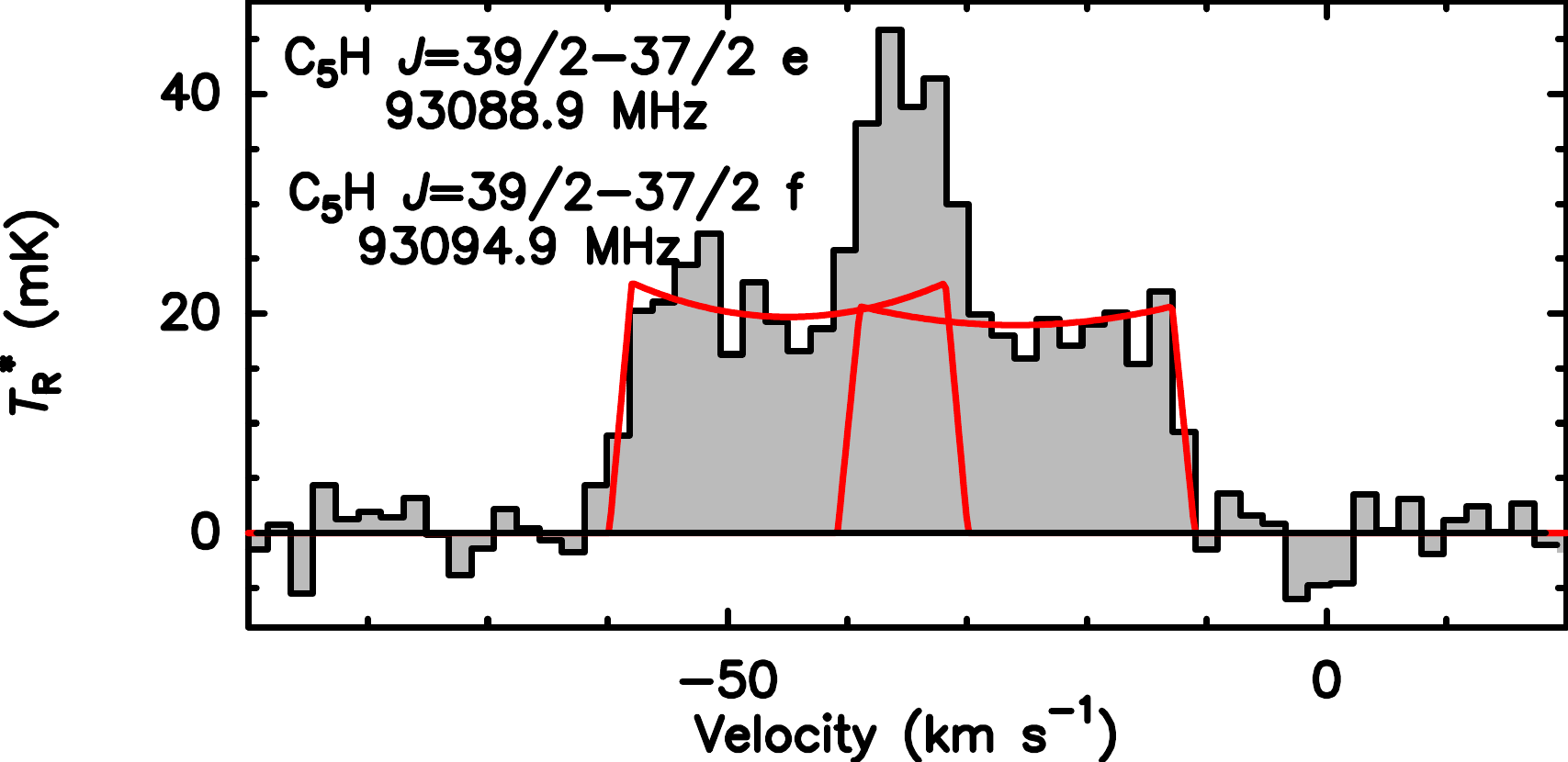}
\hspace{0.05\textwidth}
\includegraphics[width = 0.45 \textwidth]{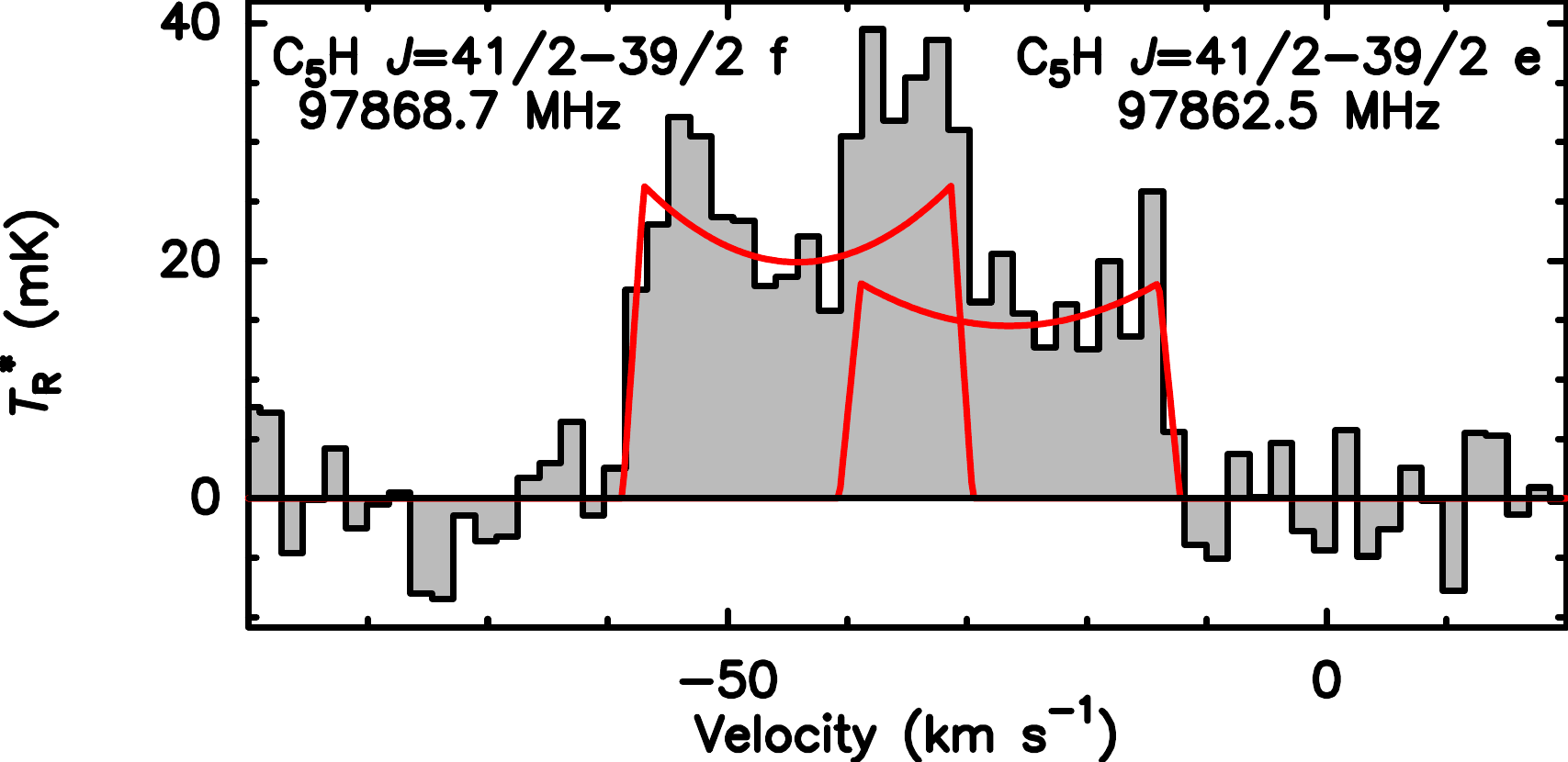}
\vspace{0.1cm}
\includegraphics[width = 0.45 \textwidth]{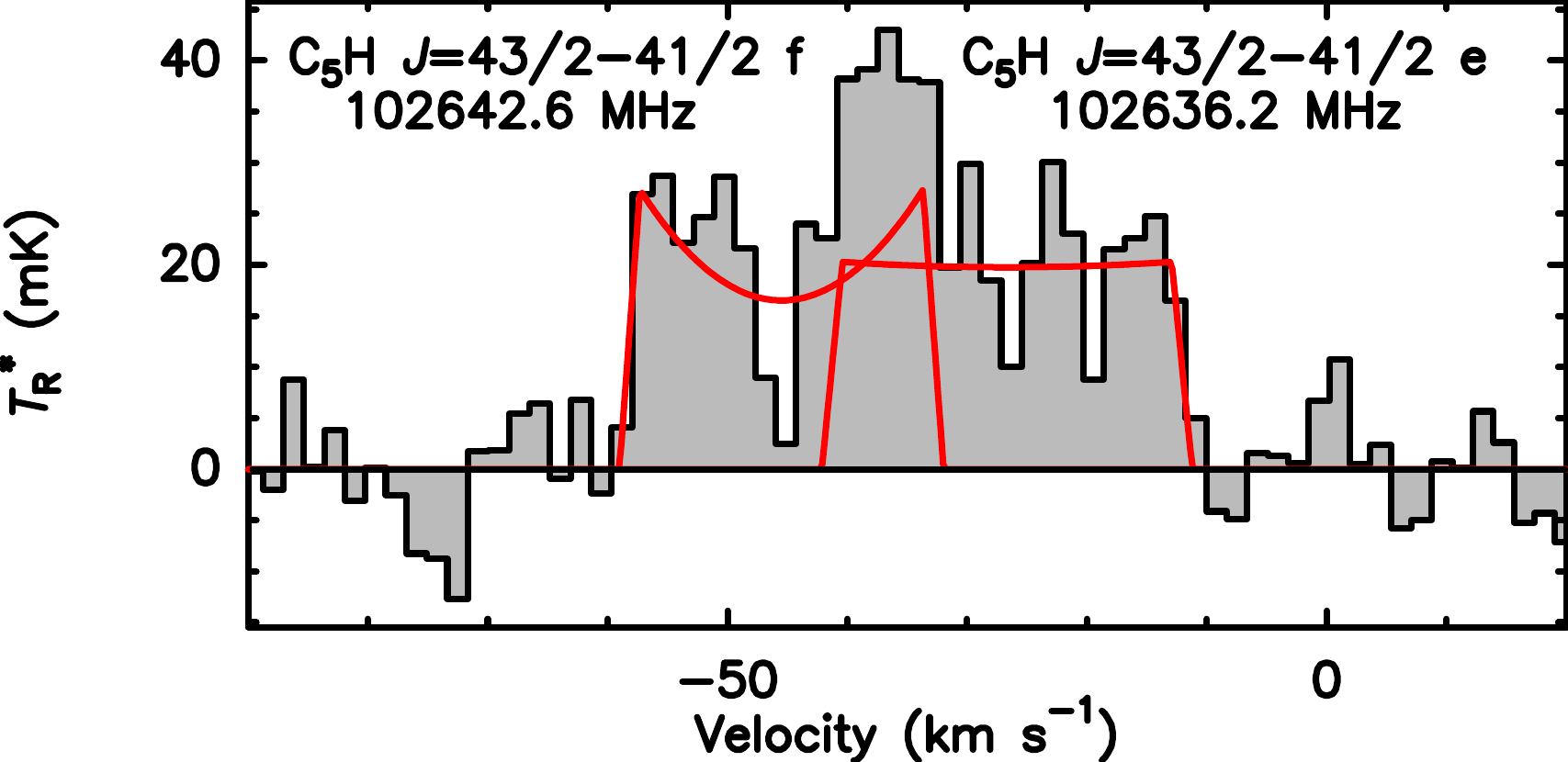}
\hspace{0.05\textwidth}
\includegraphics[width = 0.45 \textwidth]{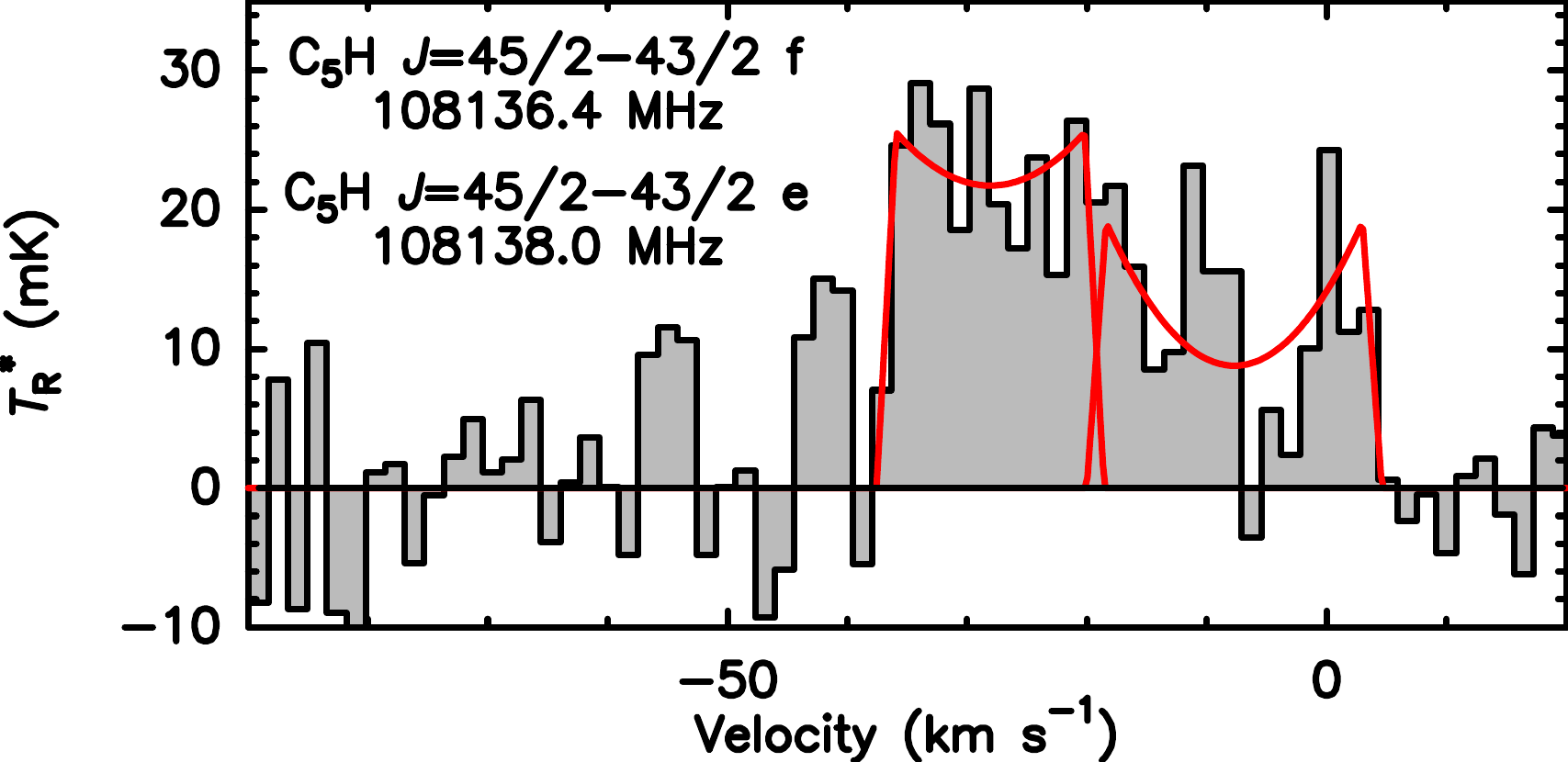}
\vspace{0.1cm}
\includegraphics[width = 0.45 \textwidth]{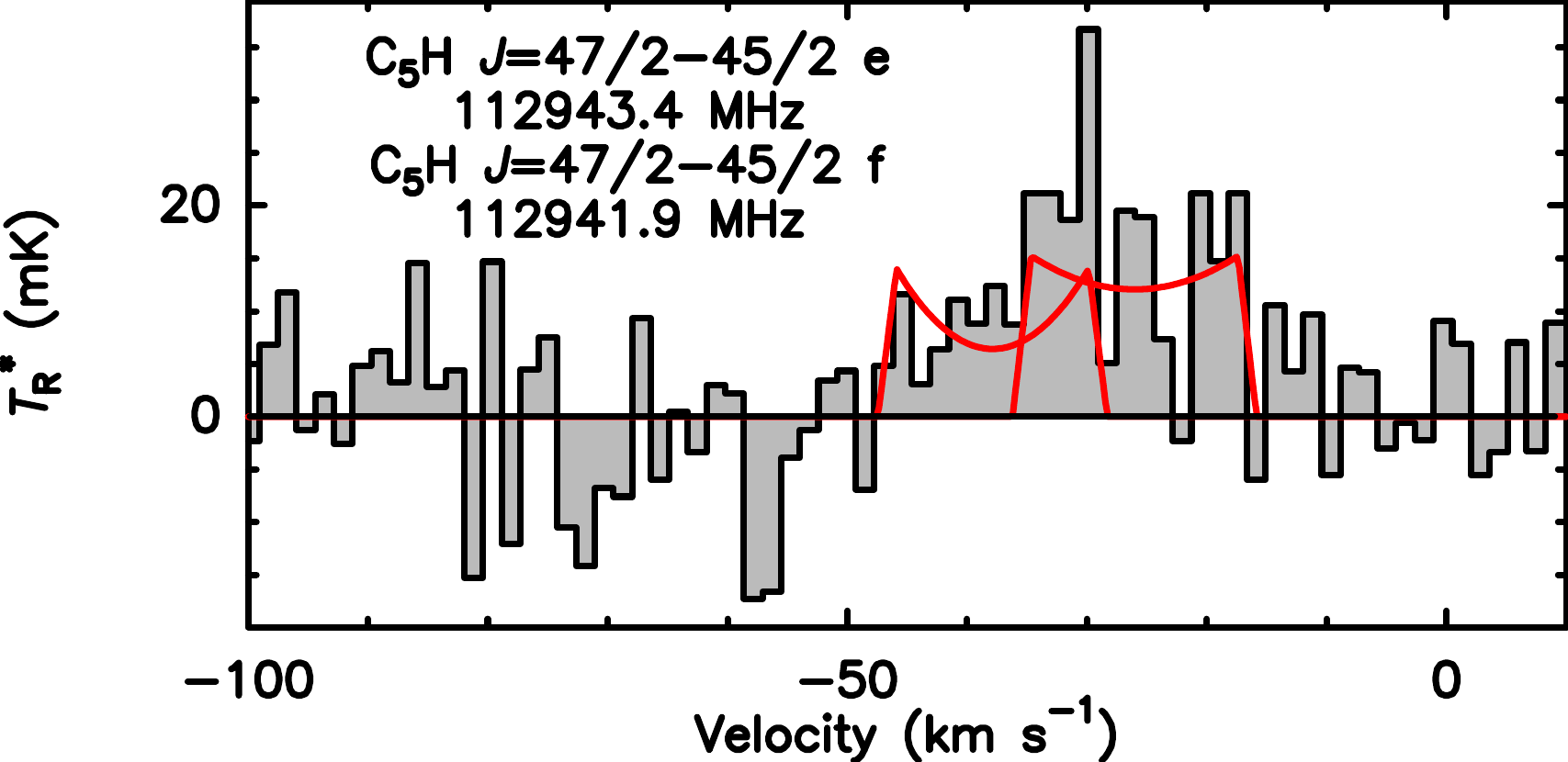}
\hspace{0.05\textwidth}
\includegraphics[width = 0.45 \textwidth]{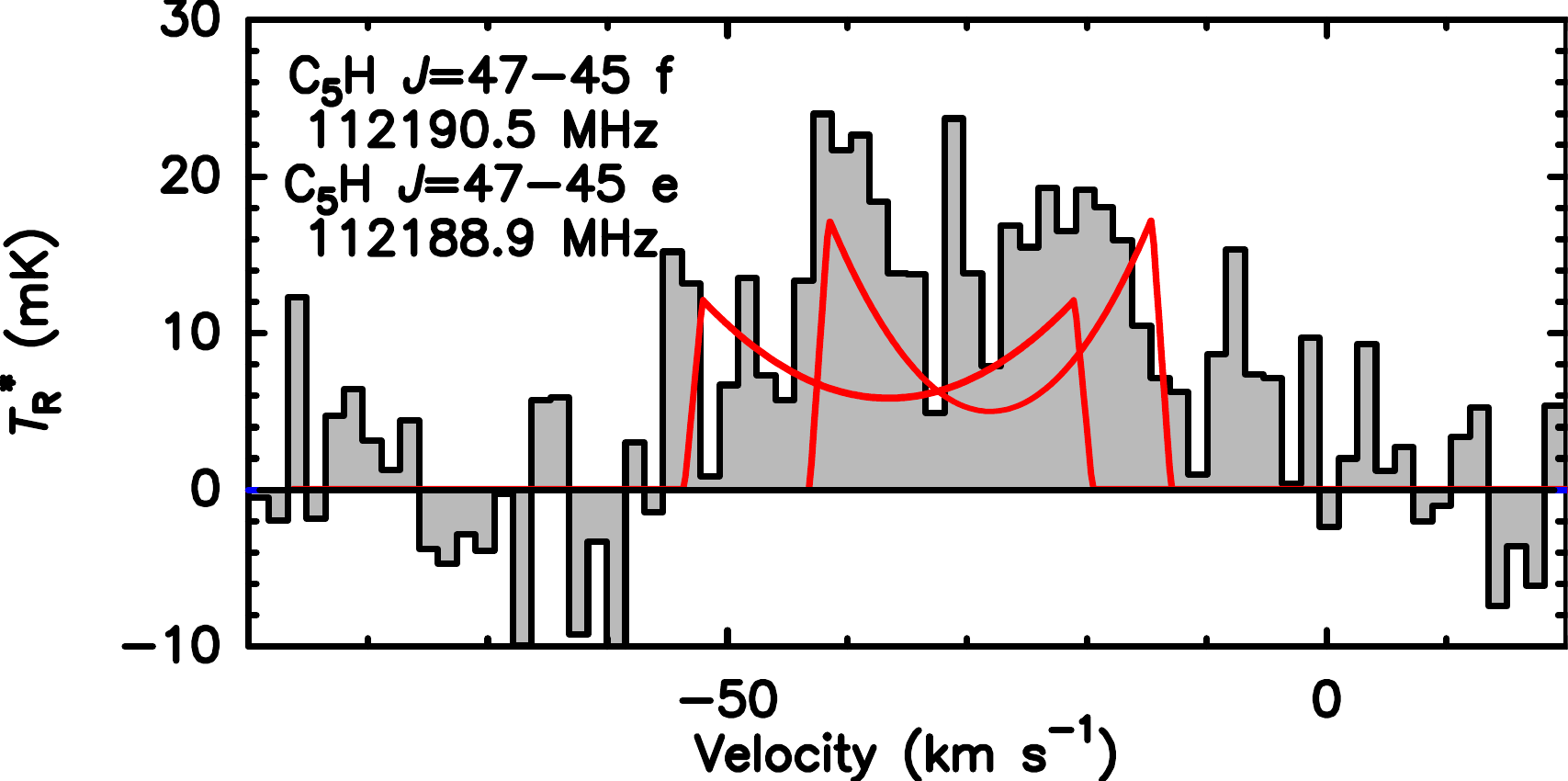}
\vspace{0.1cm}
\includegraphics[width = 0.45 \textwidth]{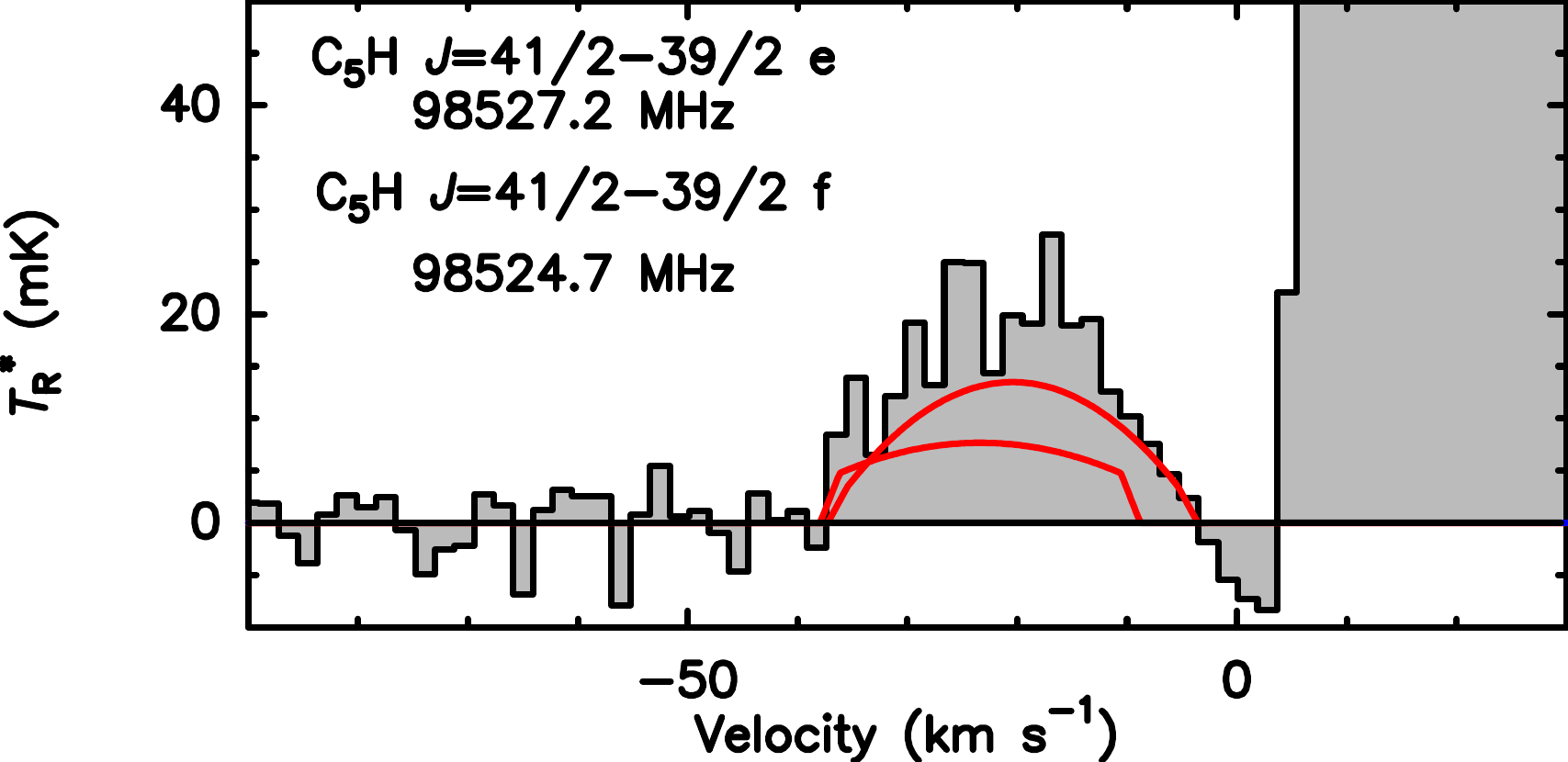}
\hspace{0.05\textwidth}
\includegraphics[width = 0.45 \textwidth]{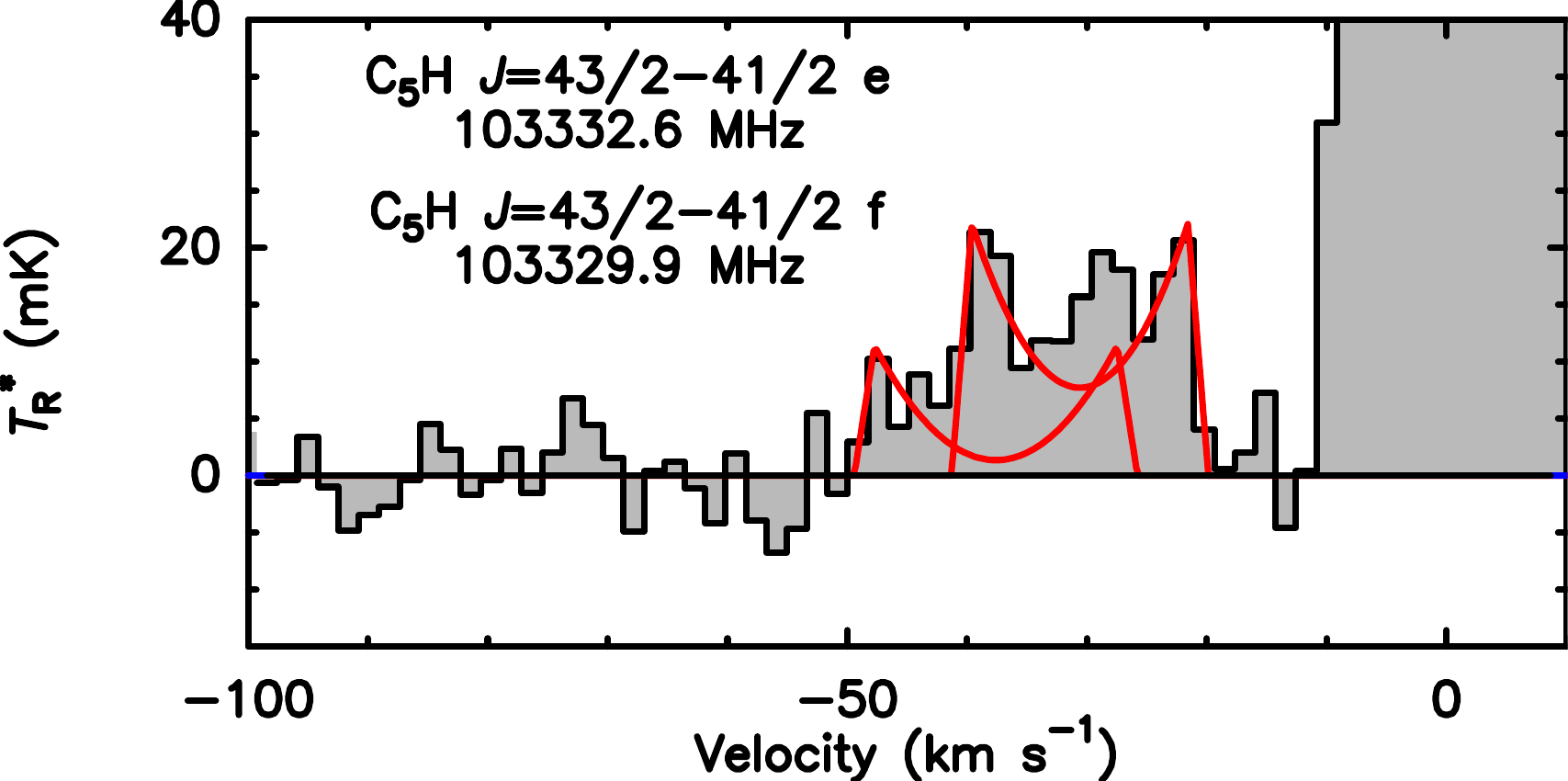}
\caption{{Same as Figure.~\ref{Fig:fitting_1}, but for C$_{5}$H. }\label{Fig:fitting_14}}
\end{figure*}

\begin{figure*}[!htbp]
\centering
\includegraphics[width = 0.45 \textwidth]{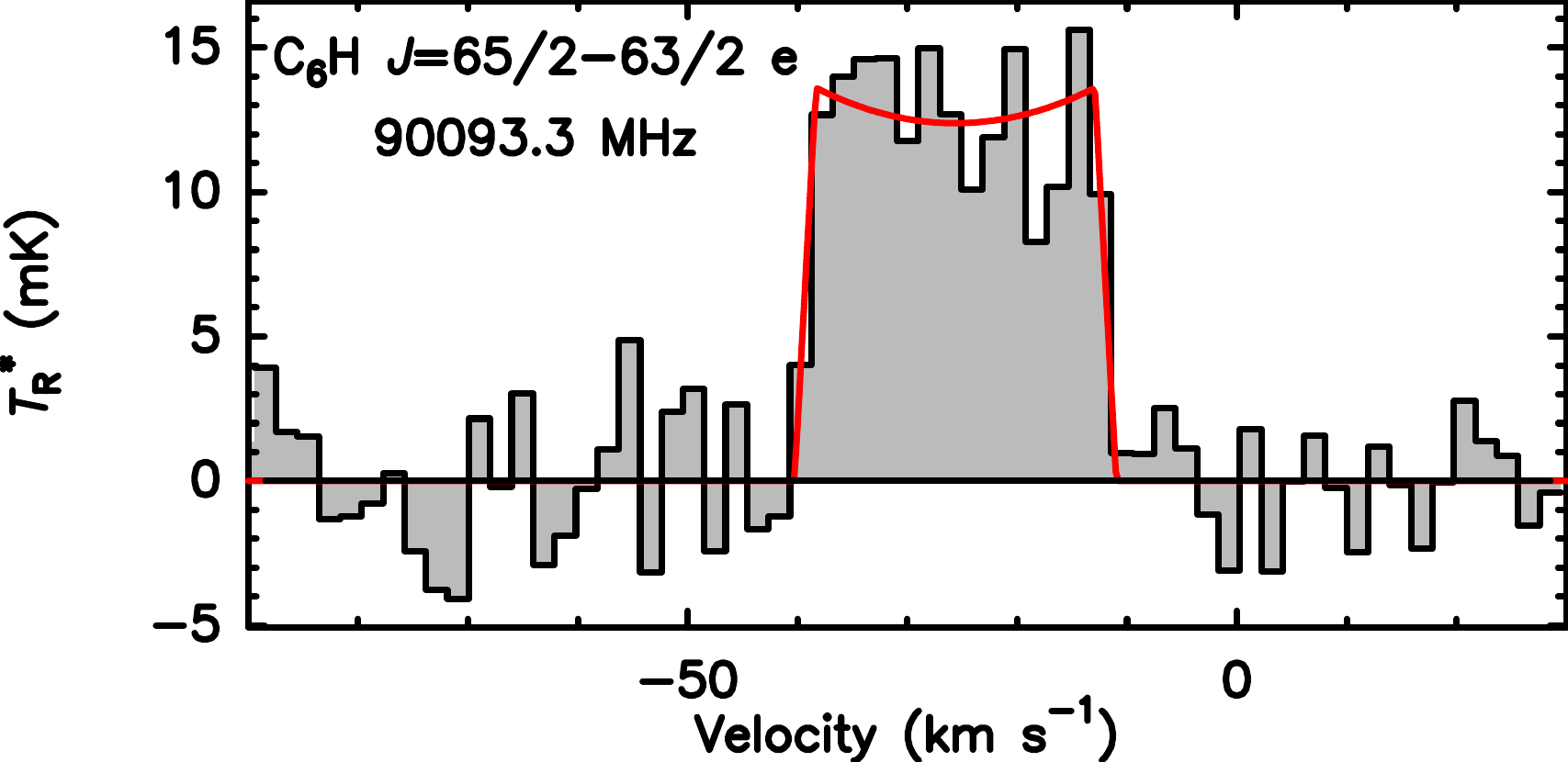}
\hspace{0.05\textwidth}
\includegraphics[width = 0.45 \textwidth]{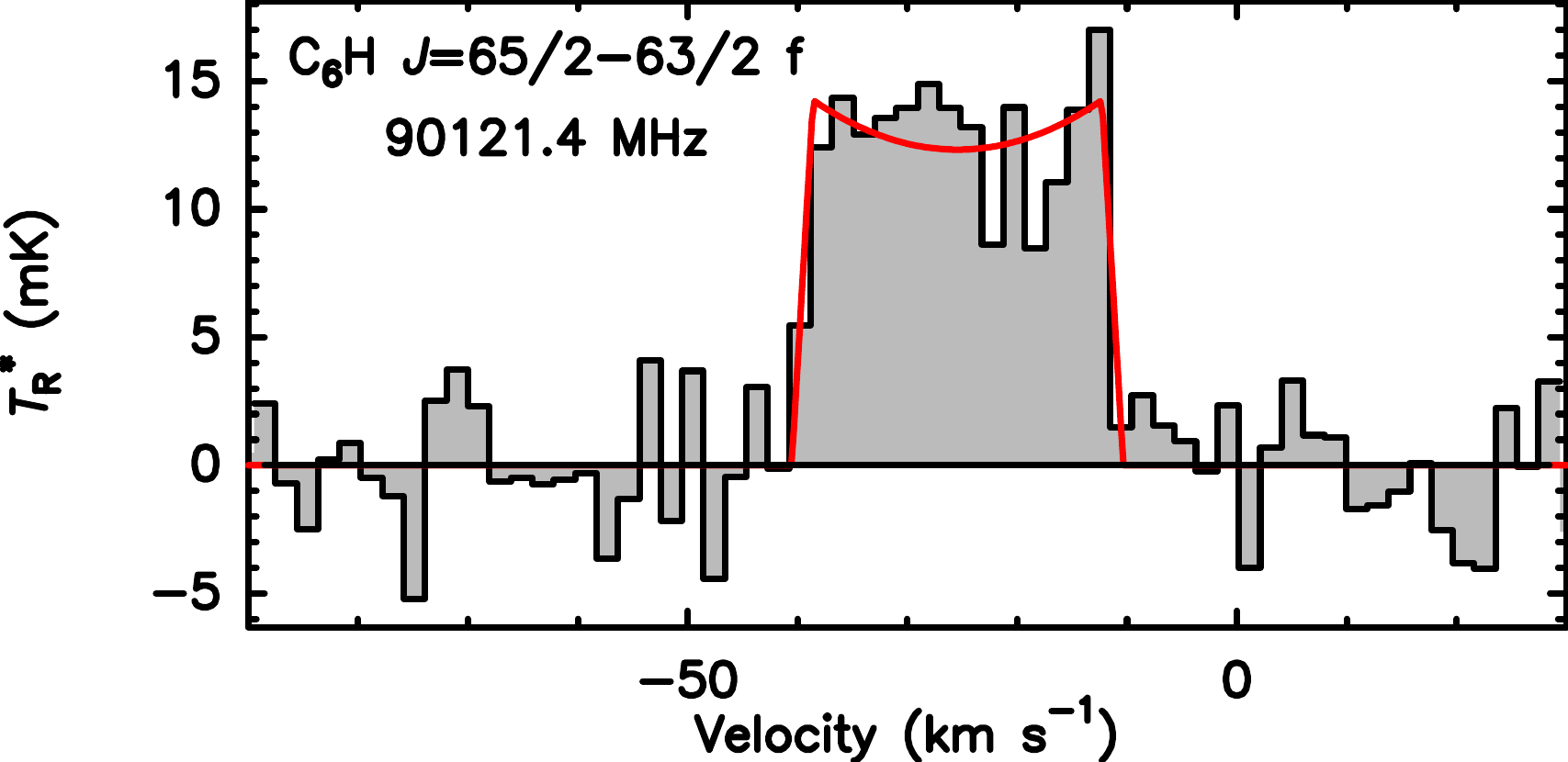}
\vspace{0.1cm}
\includegraphics[width = 0.45 \textwidth]{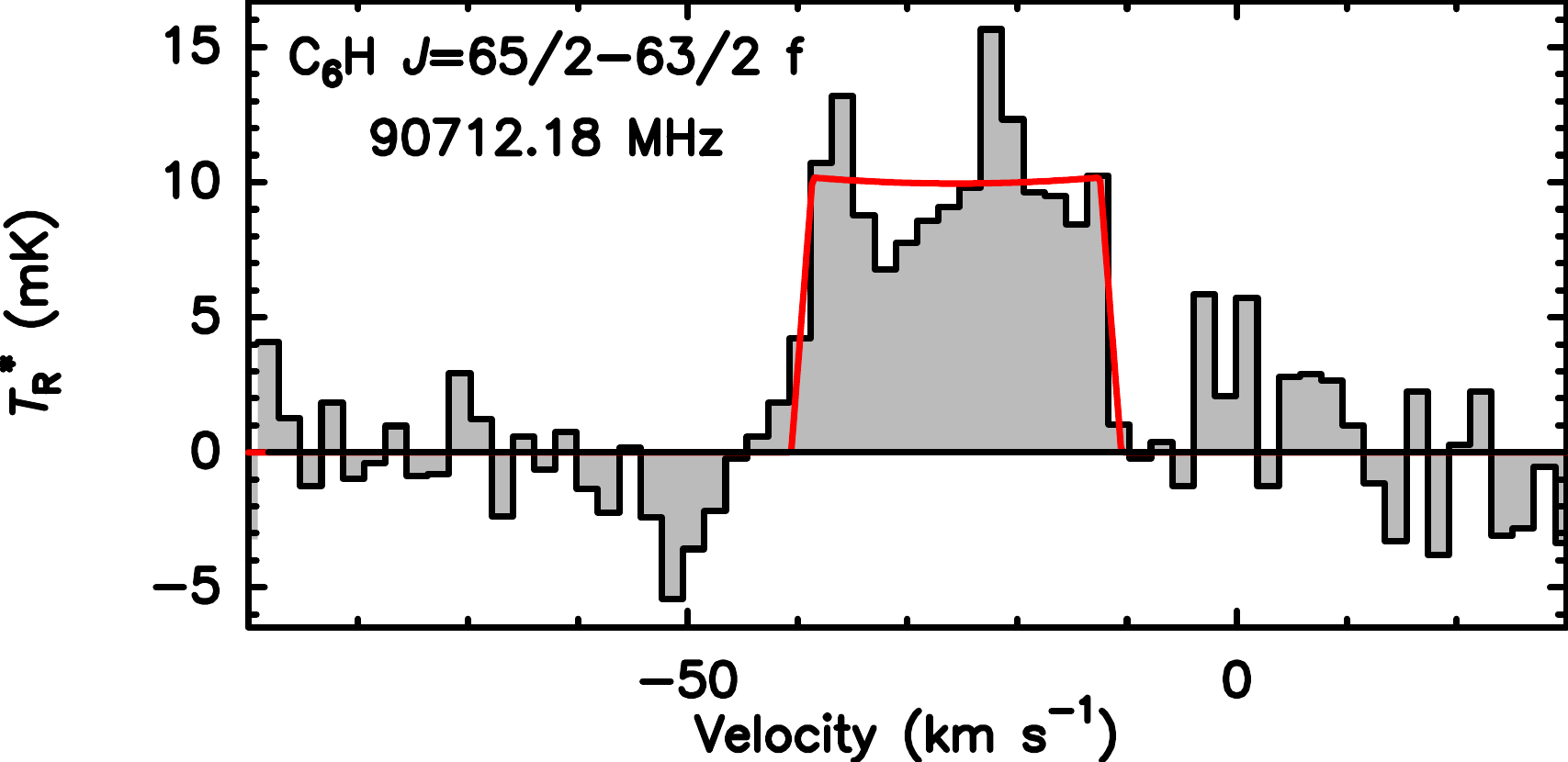}
\hspace{0.05\textwidth}
\includegraphics[width = 0.45 \textwidth]{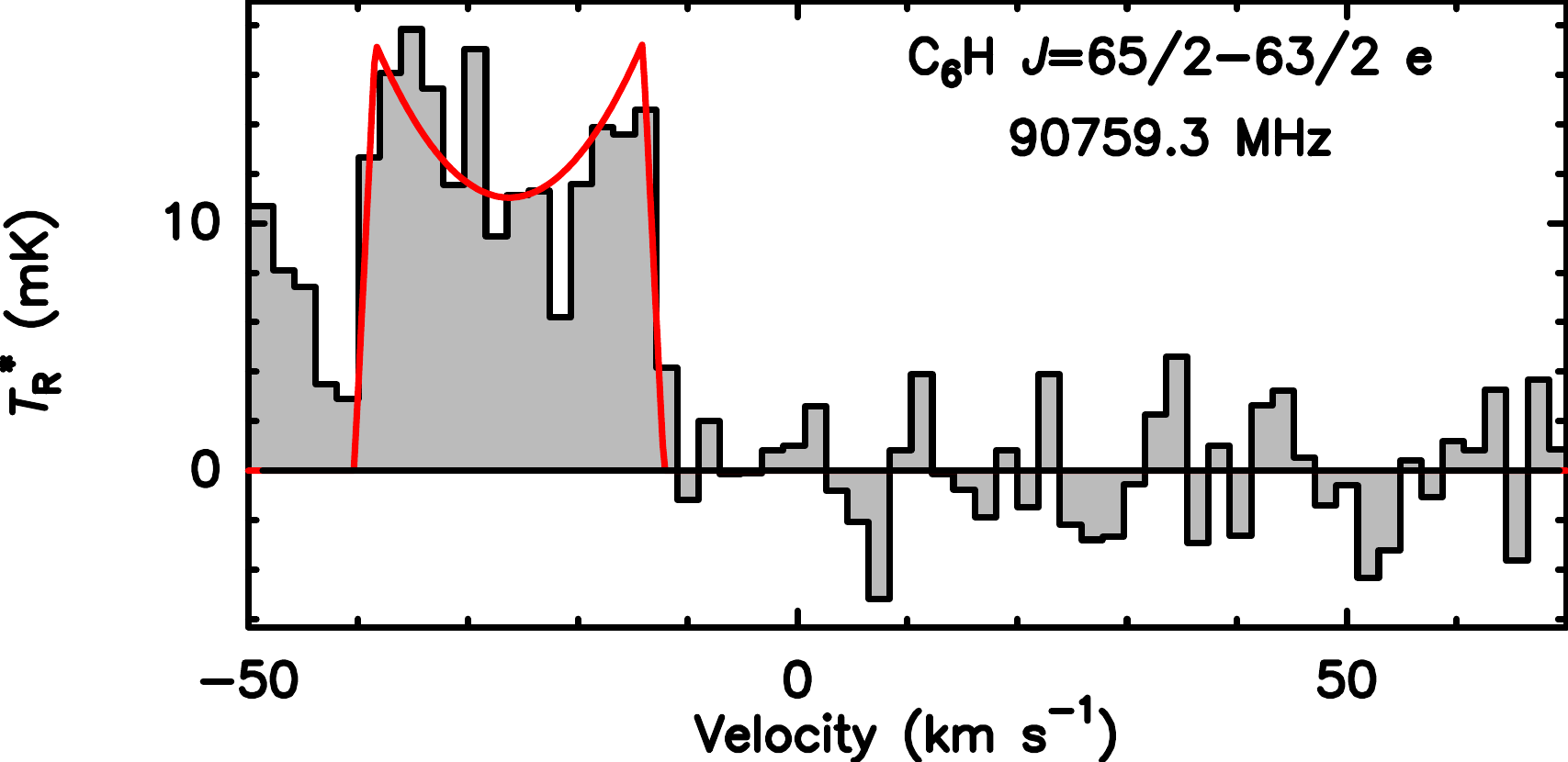}
\vspace{0.1cm}
\includegraphics[width = 0.45 \textwidth]{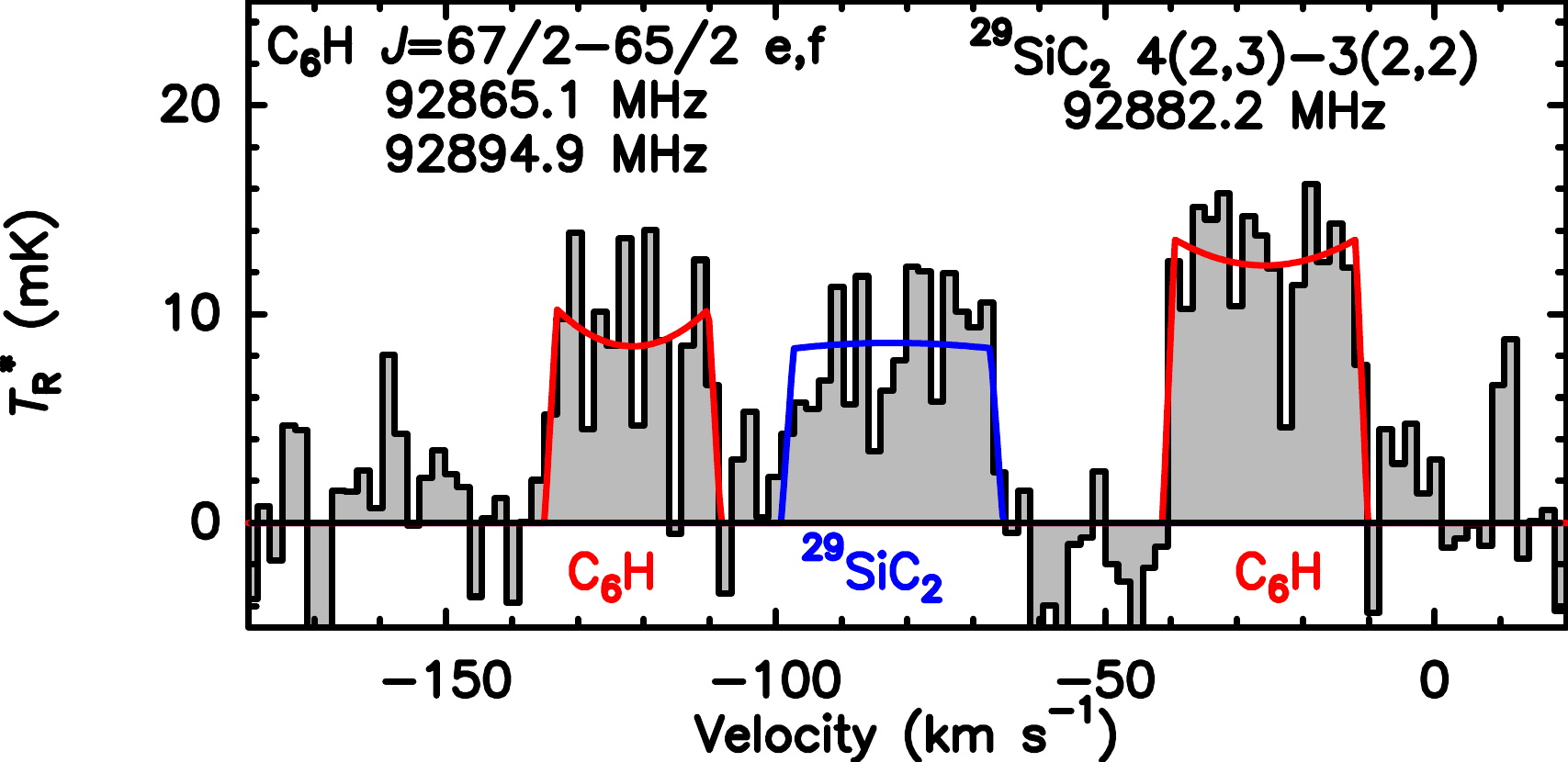}
\hspace{0.05\textwidth}
\includegraphics[width = 0.45 \textwidth]{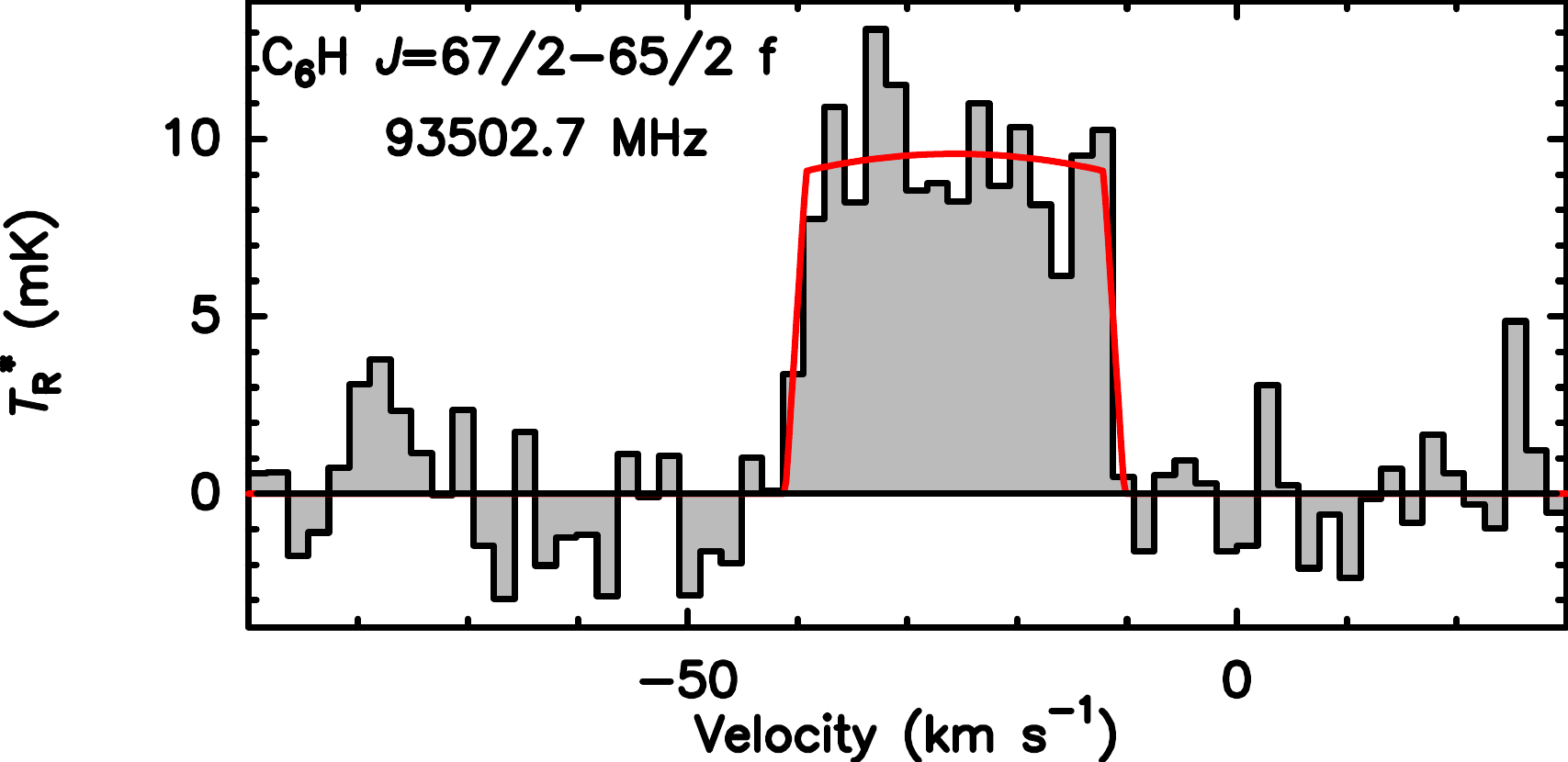}
\vspace{0.1cm}
\includegraphics[width = 0.45 \textwidth]{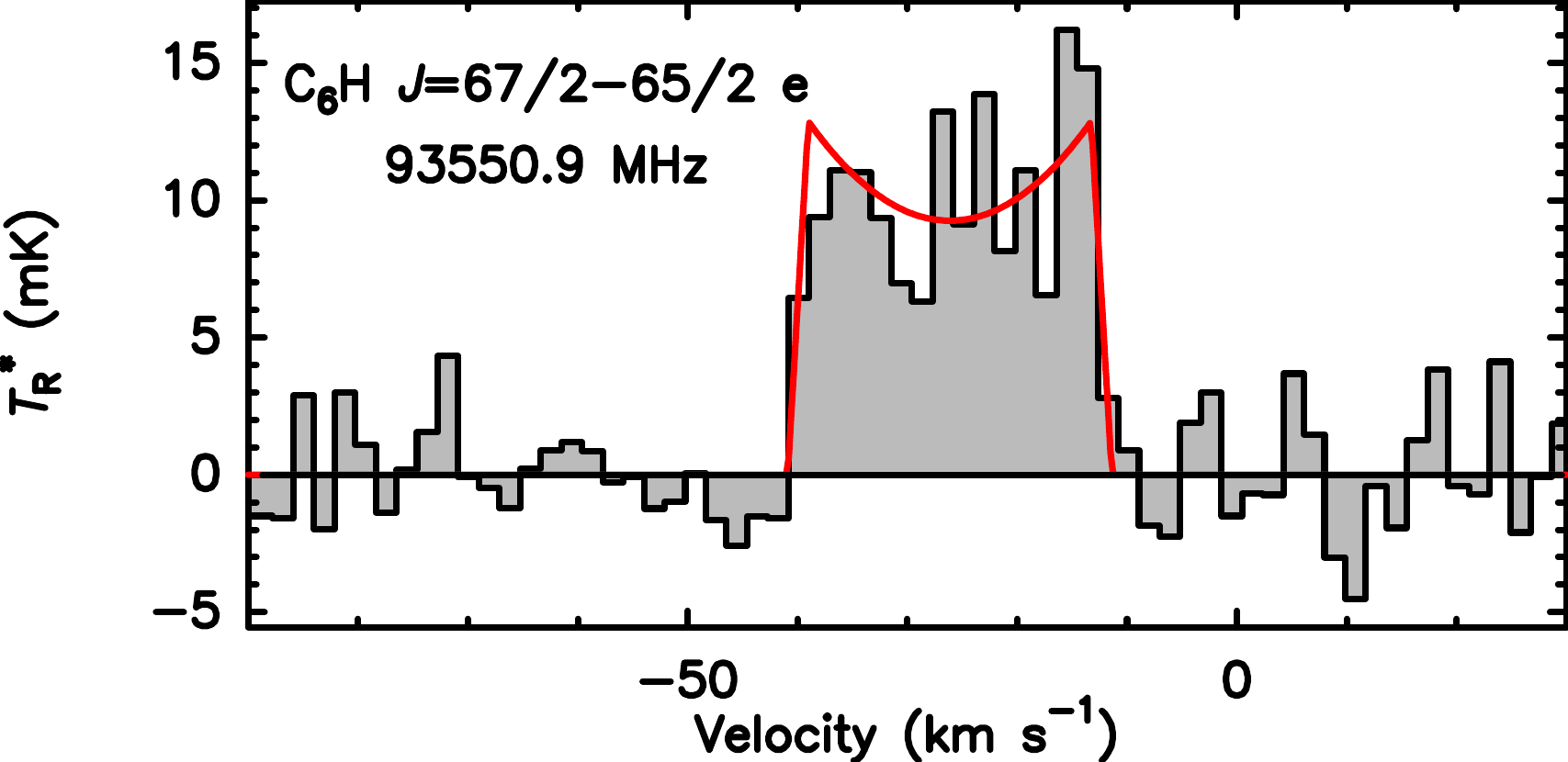}
\hspace{0.05\textwidth}
\includegraphics[width = 0.45 \textwidth]{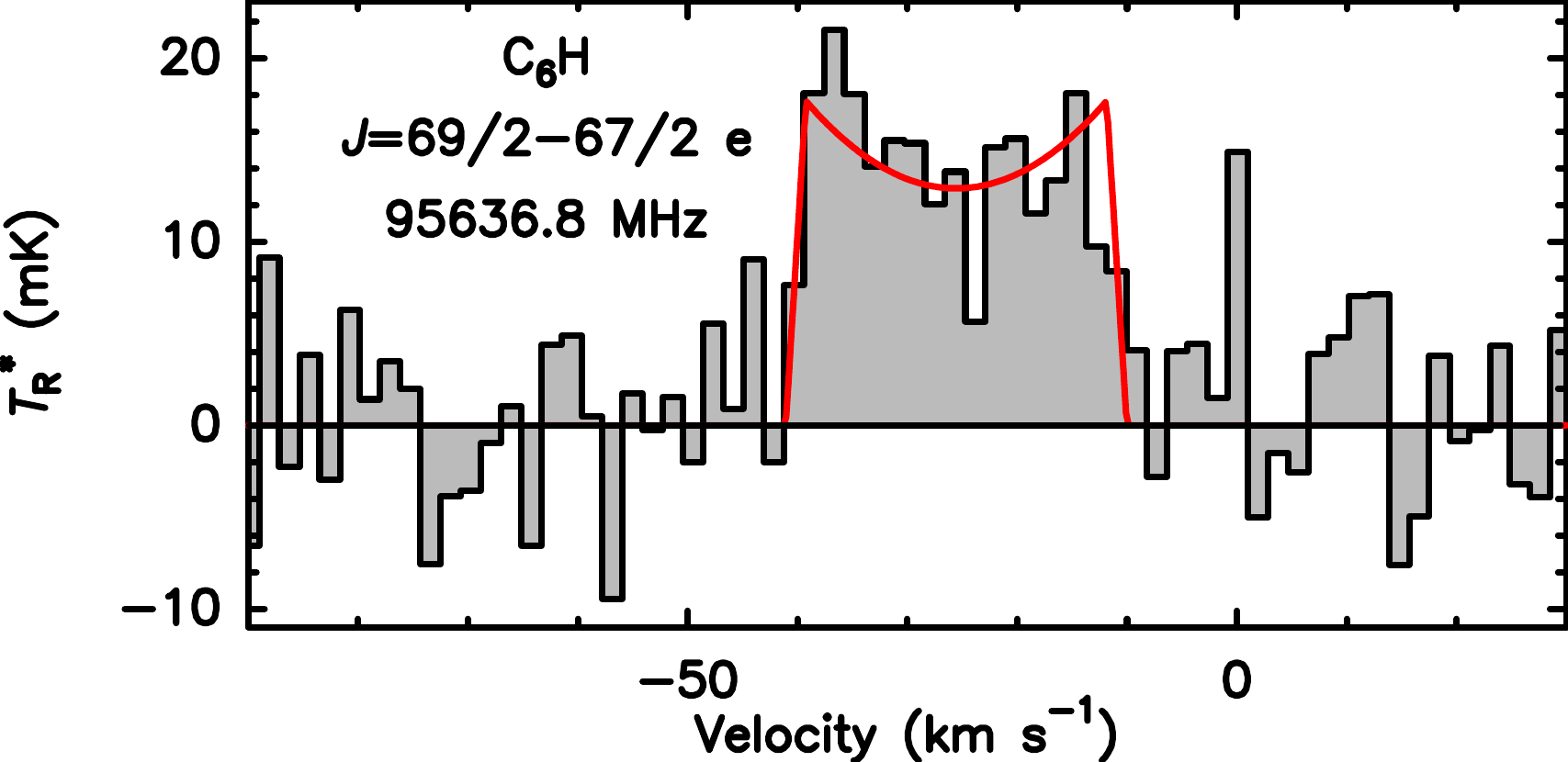}
\vspace{0.1cm}
\includegraphics[width = 0.45 \textwidth]{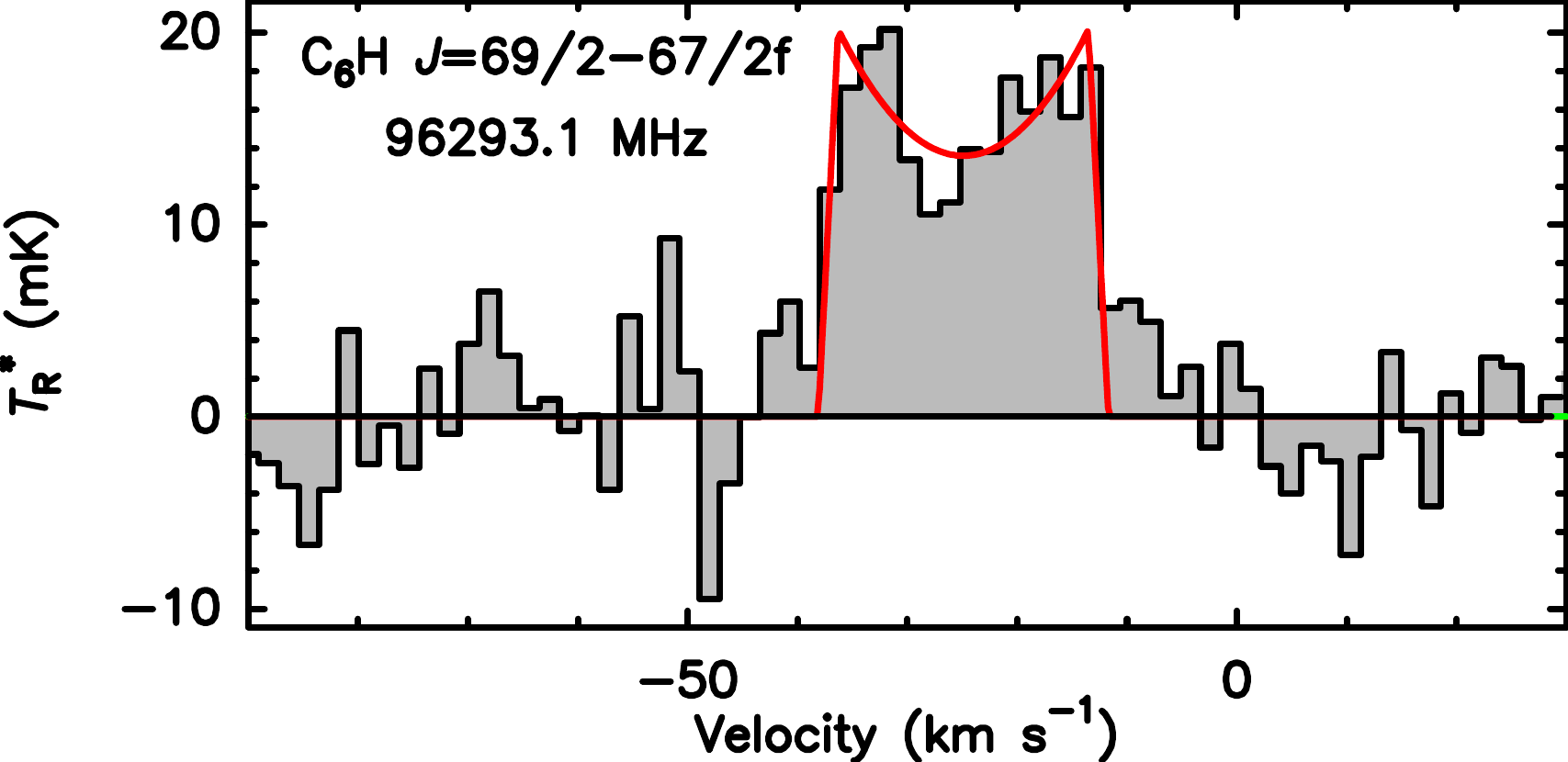}
\hspace{0.05\textwidth}
\includegraphics[width = 0.45 \textwidth]{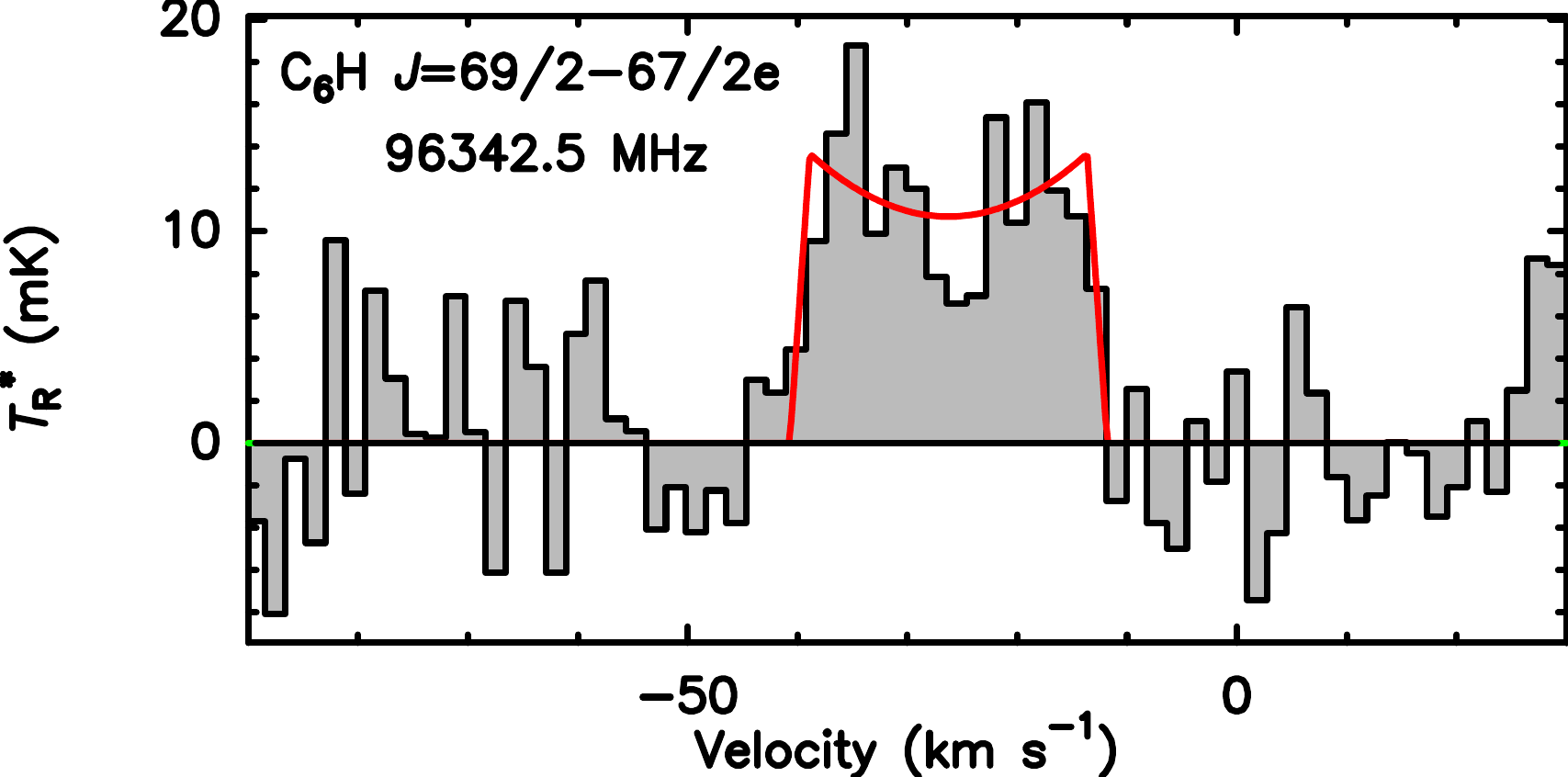}
\caption{{Same as Figure.~\ref{Fig:fitting_1}, but for C$_{6}$H. }\label{Fig:fitting_15}}
\end{figure*}

\begin{figure*}[!htbp]
\centering
\includegraphics[width = 0.45 \textwidth]{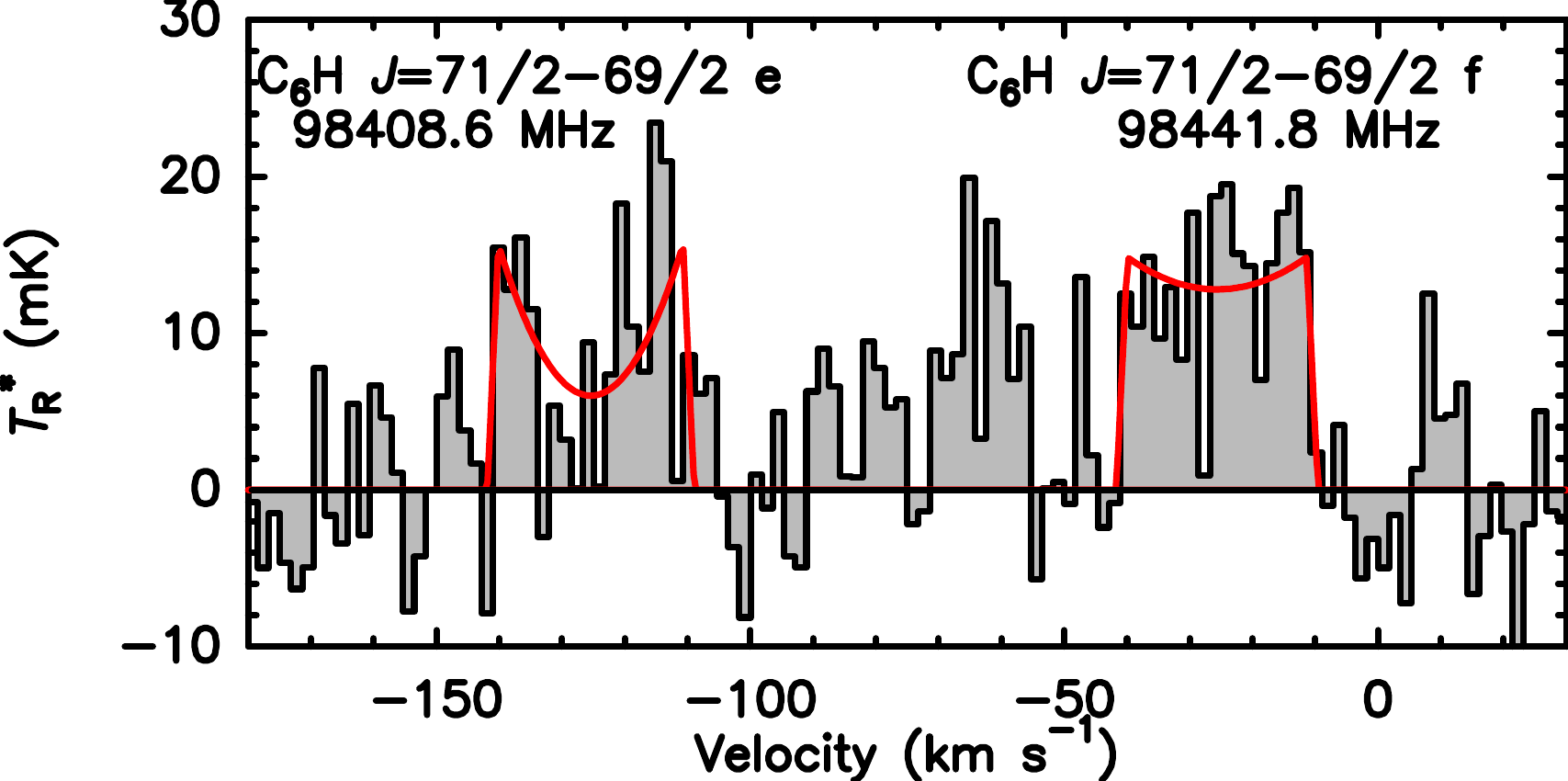}
\hspace{0.05\textwidth}
\includegraphics[width = 0.45 \textwidth]{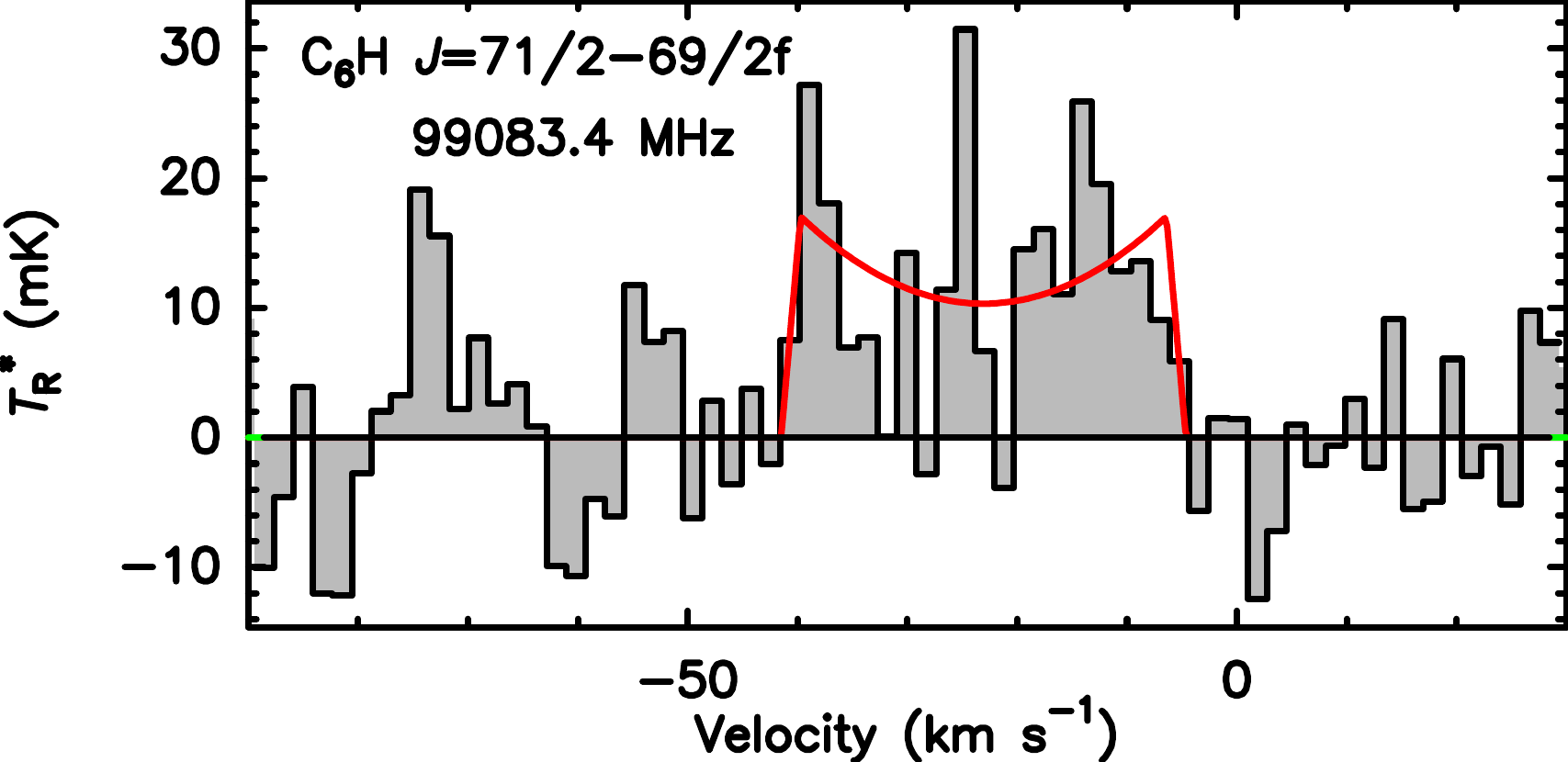}
\vspace{0.1cm}
\includegraphics[width = 0.45 \textwidth]{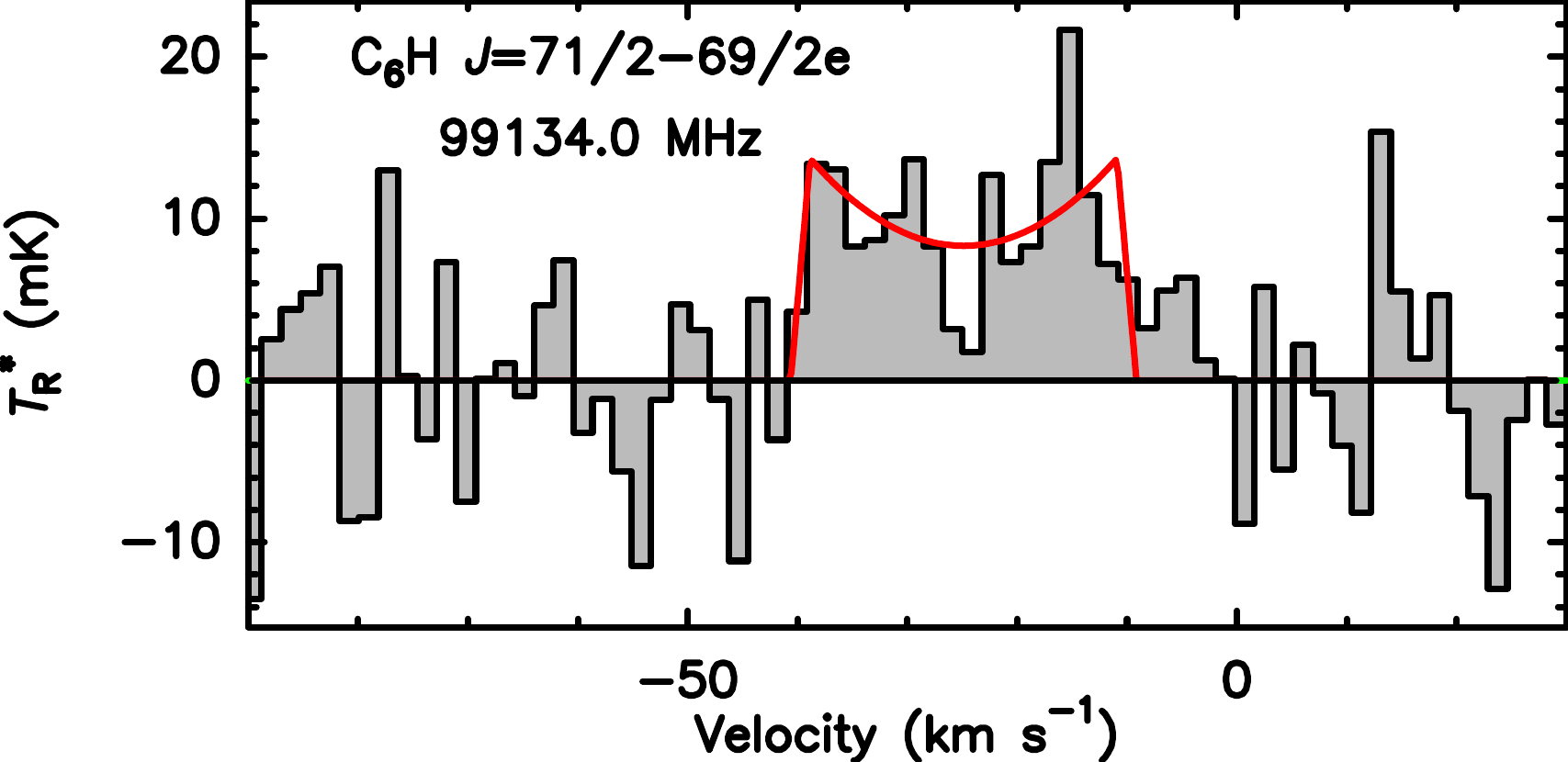}
\hspace{0.05\textwidth}
\includegraphics[width = 0.45 \textwidth]{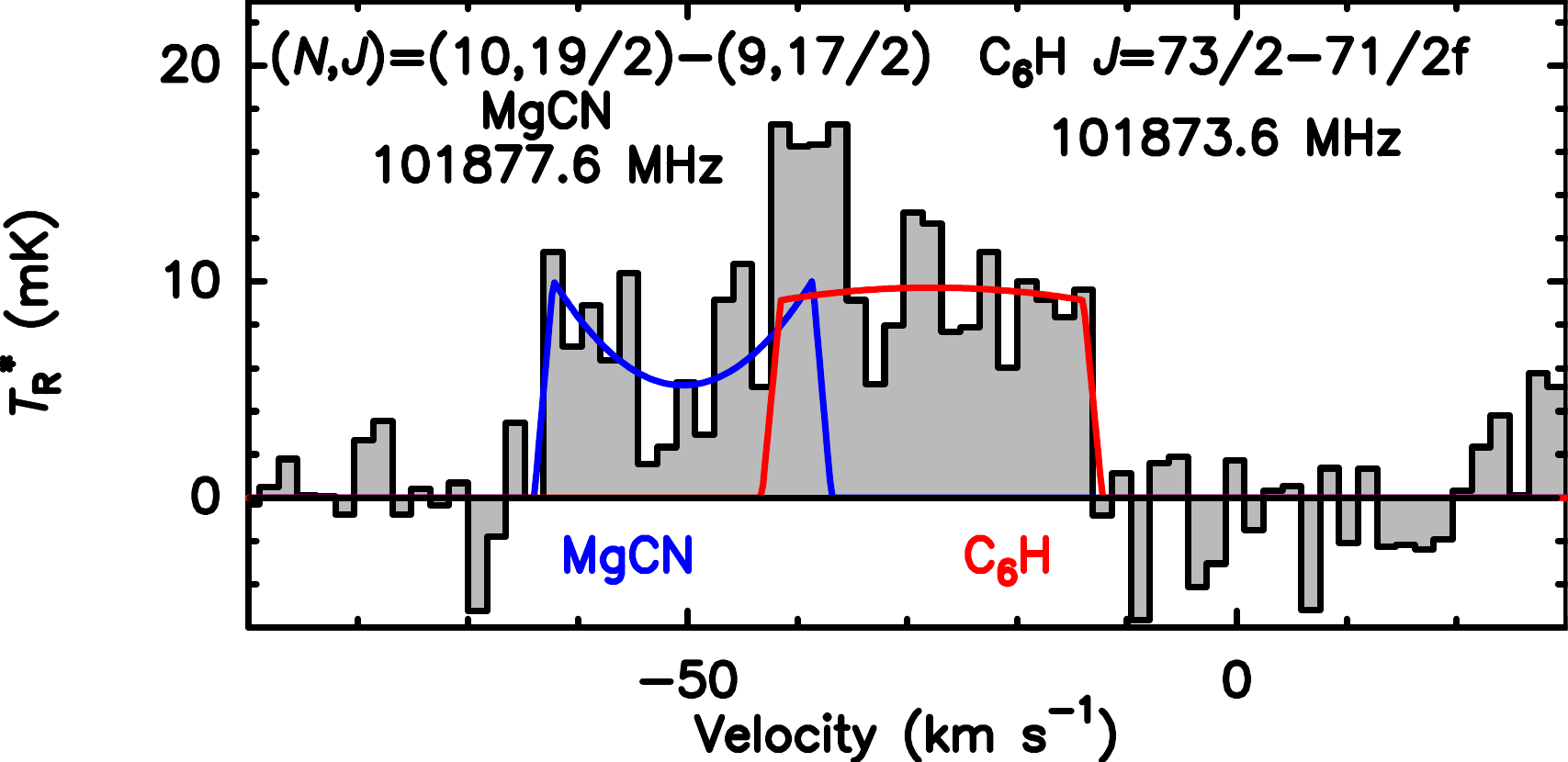}
\vspace{0.1cm}
\includegraphics[width = 0.45 \textwidth]{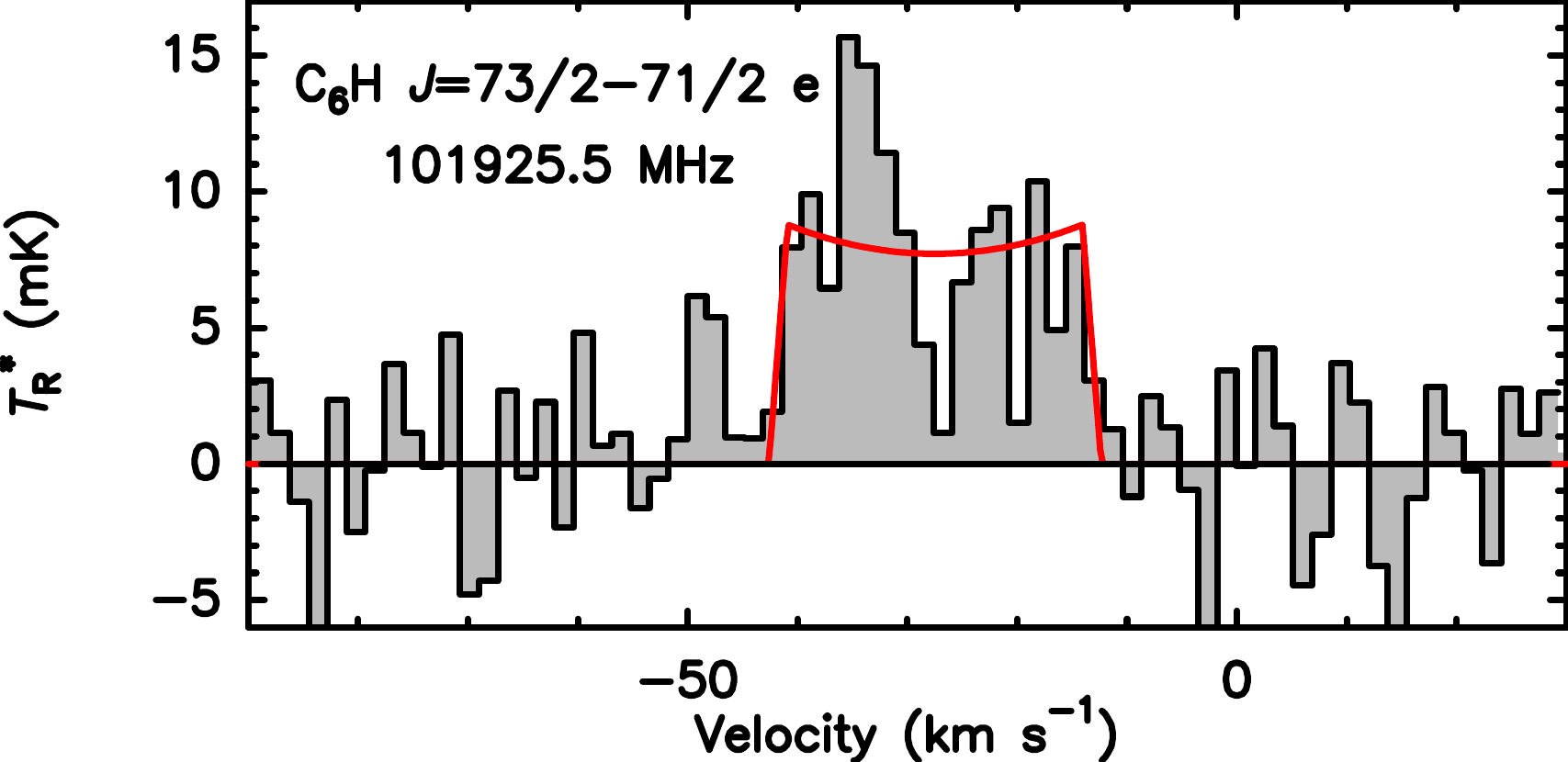}
\hspace{0.05\textwidth}
\includegraphics[width = 0.45 \textwidth]{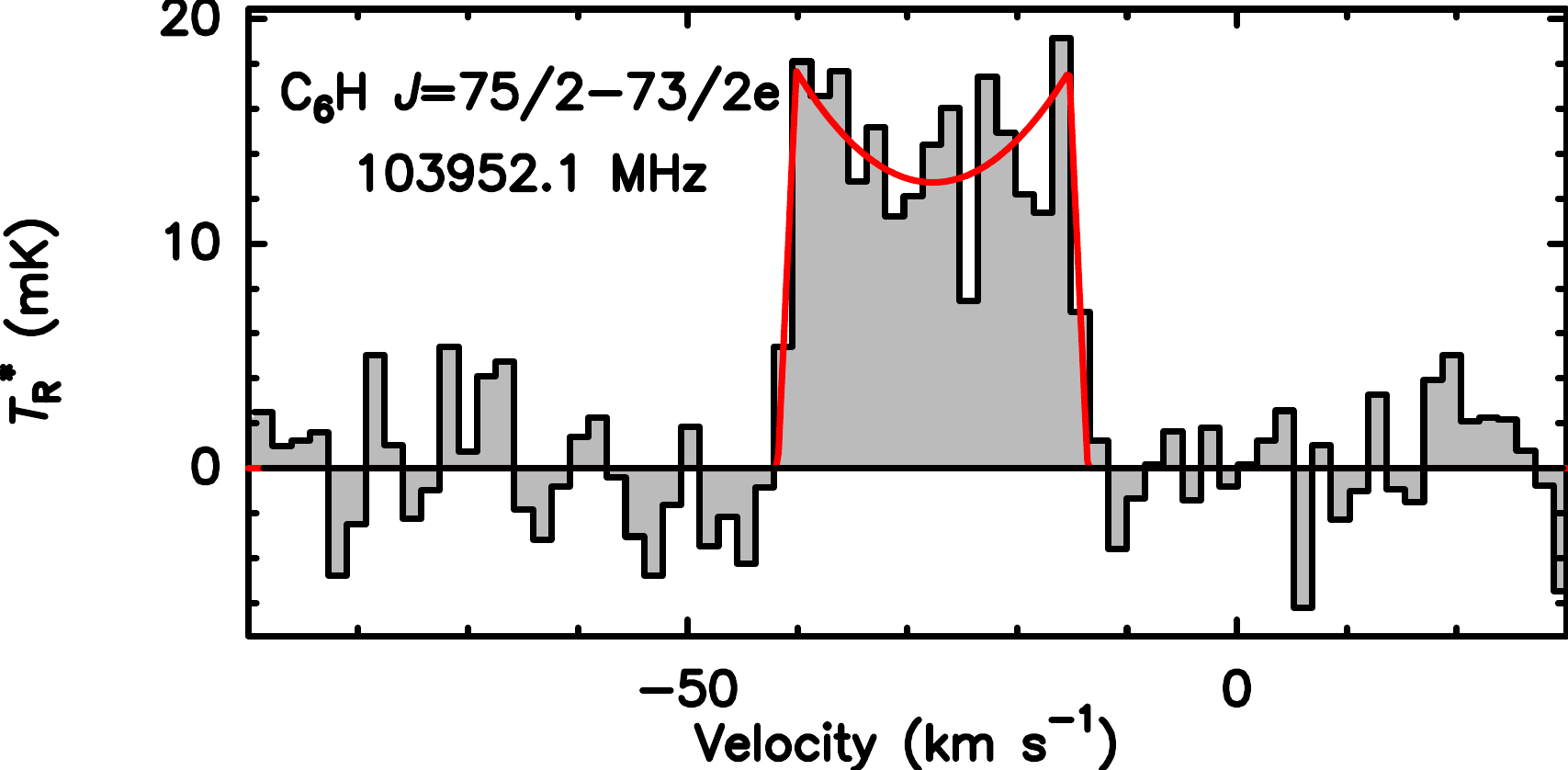}
\vspace{0.1cm}
\includegraphics[width = 0.45 \textwidth]{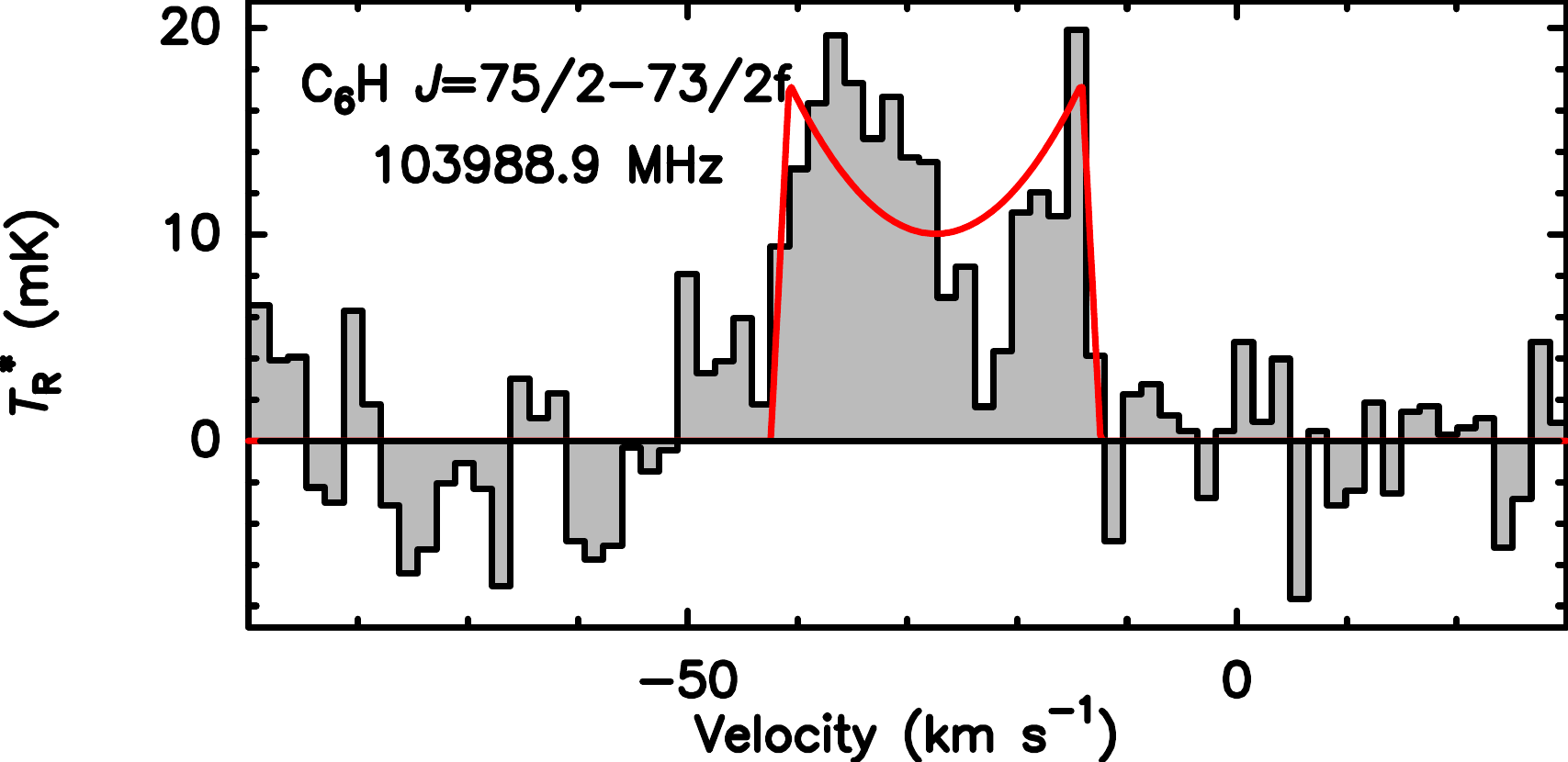}
\hspace{0.05\textwidth}
\includegraphics[width = 0.45 \textwidth]{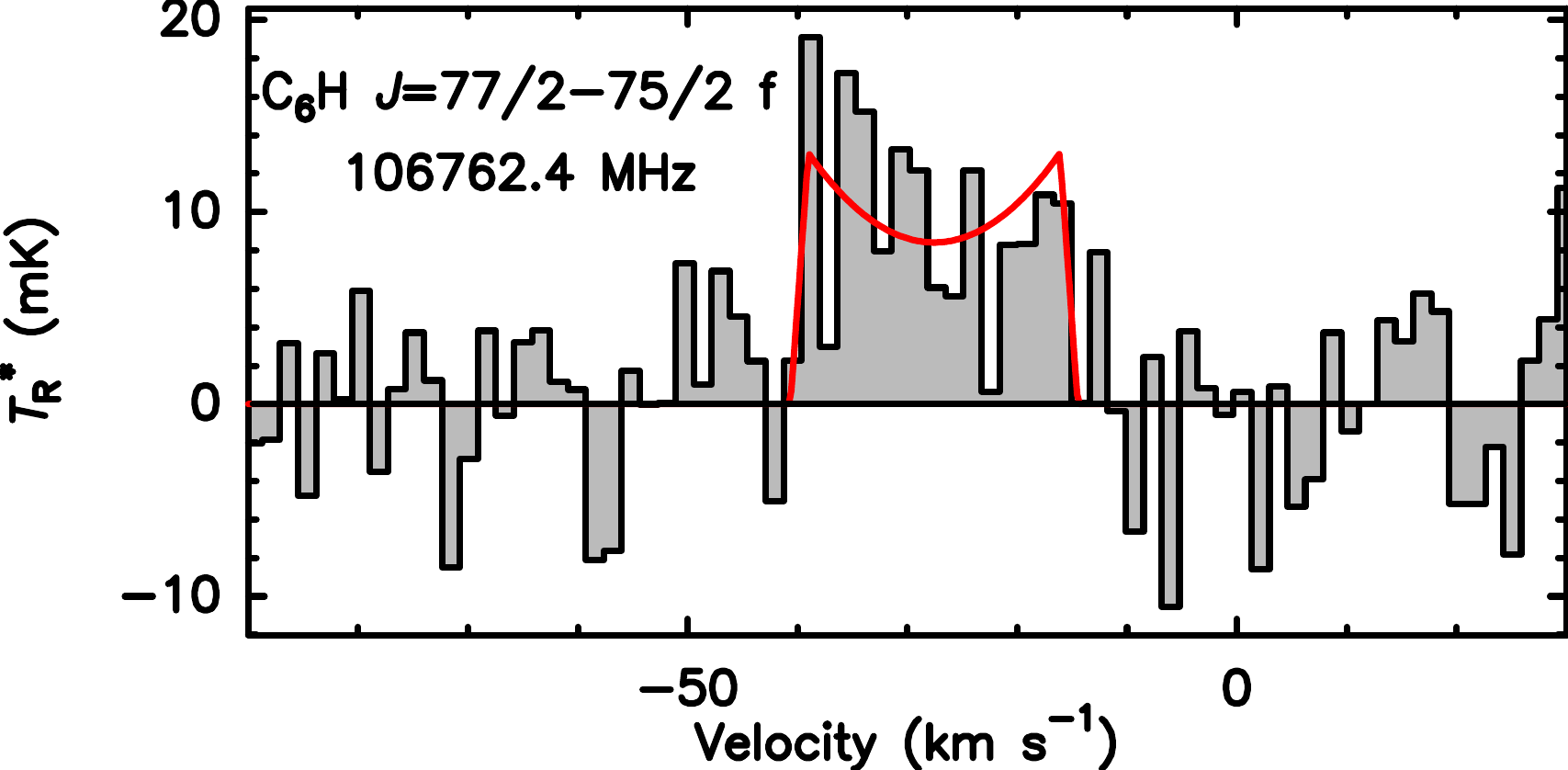}
\vspace{0.1cm}
\includegraphics[width = 0.45 \textwidth]{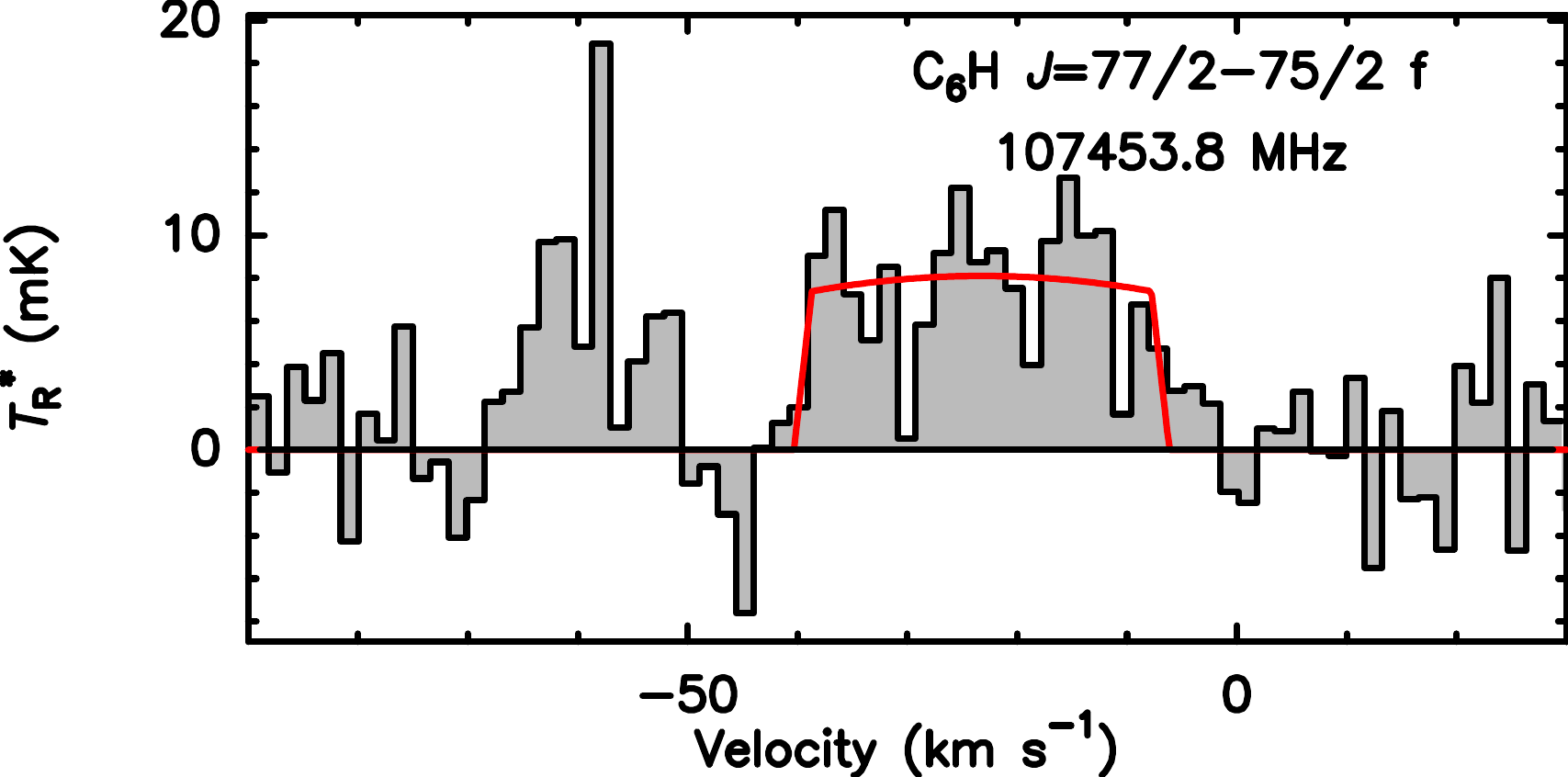}
\hspace{0.05\textwidth}
\includegraphics[width = 0.45 \textwidth]{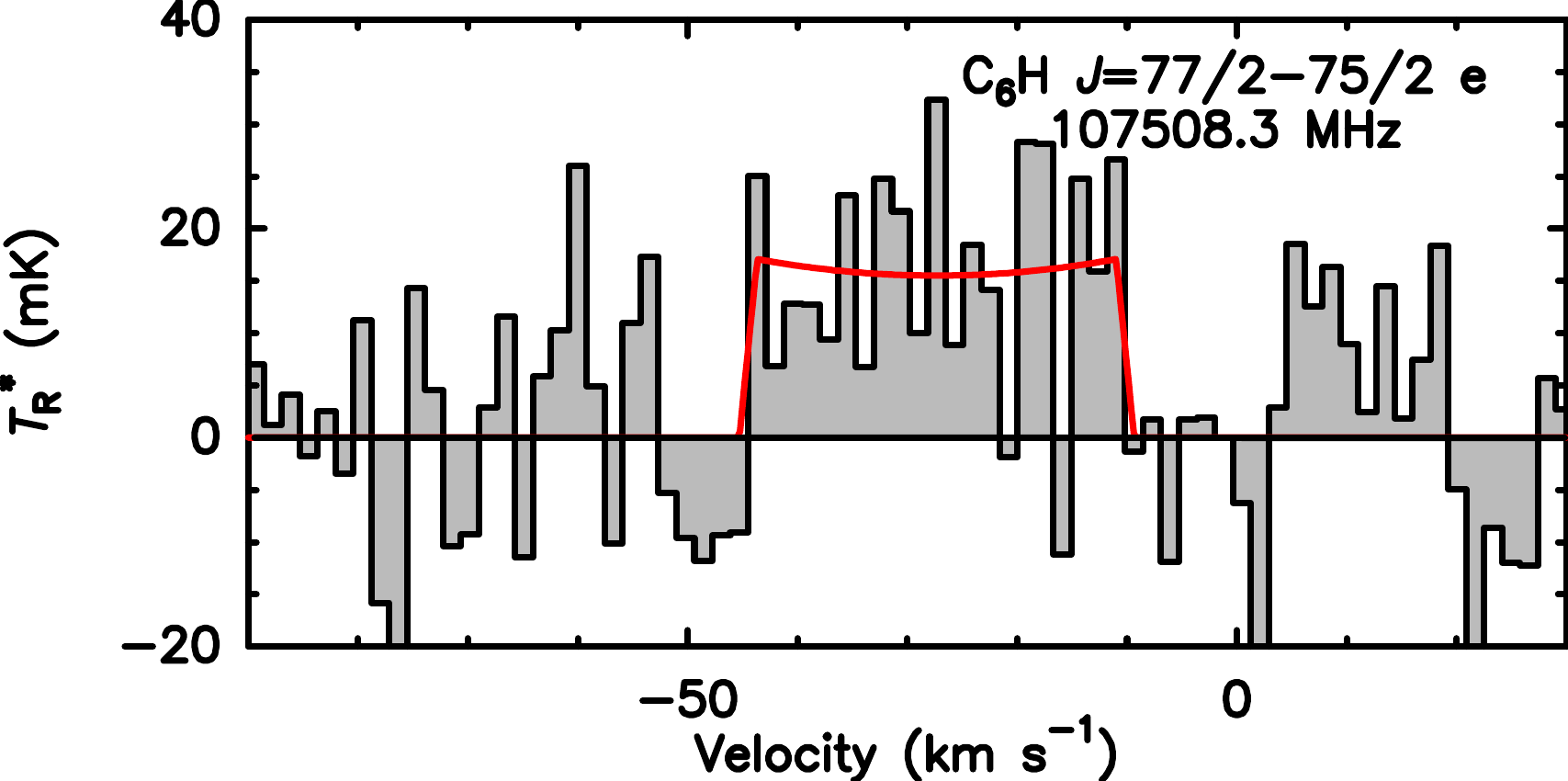}
\centerline{Figure \ref{Fig:fitting_15}. --- continued}
\end{figure*}

\begin{figure*}[!htbp]
\centering
\includegraphics[width = 0.45 \textwidth]{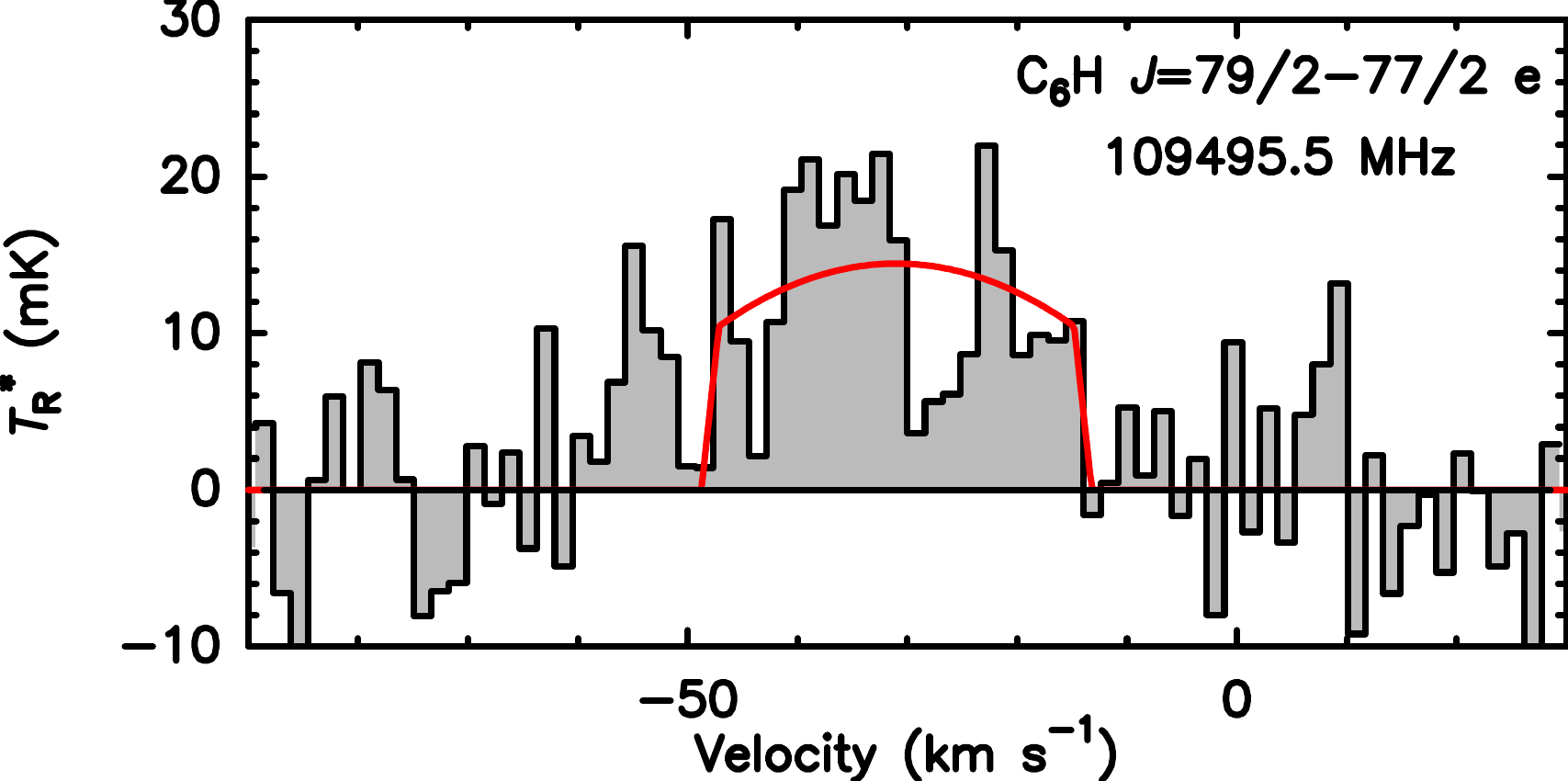}
\hspace{0.05\textwidth}
\includegraphics[width = 0.45 \textwidth]{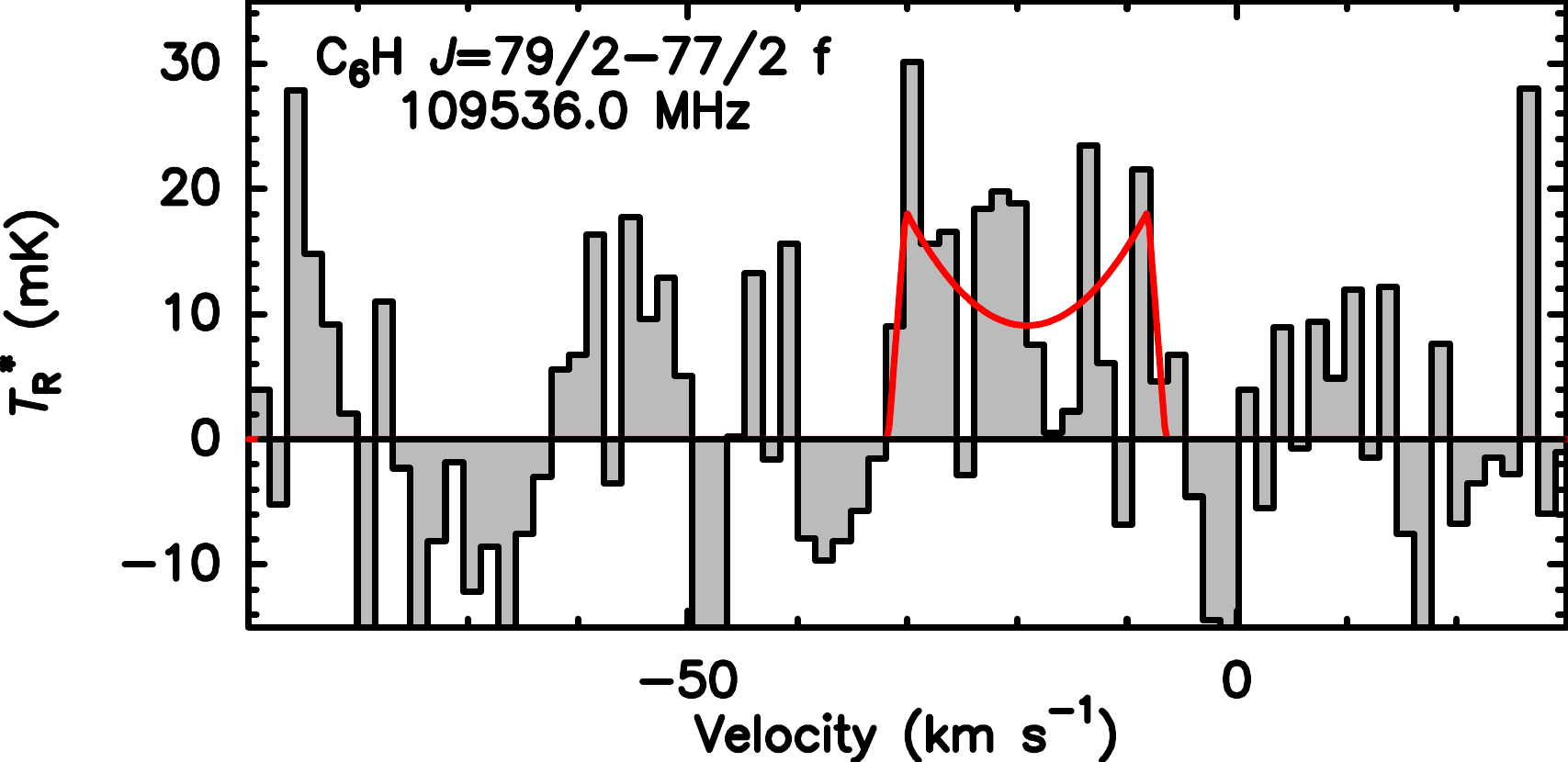}
\vspace{0.1cm}
\includegraphics[width = 0.45 \textwidth]{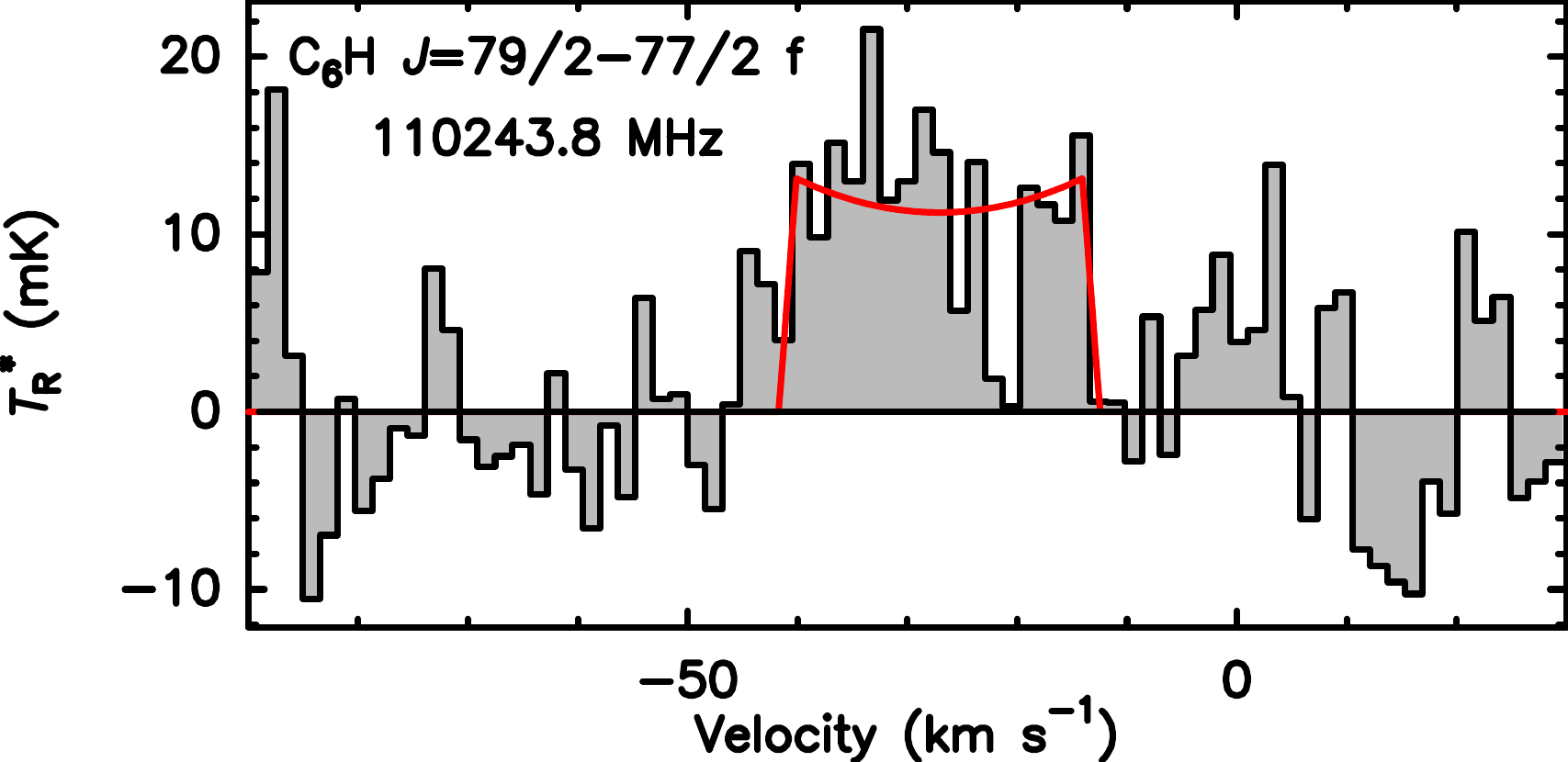}
\hspace{0.05\textwidth}
\includegraphics[width = 0.45 \textwidth]{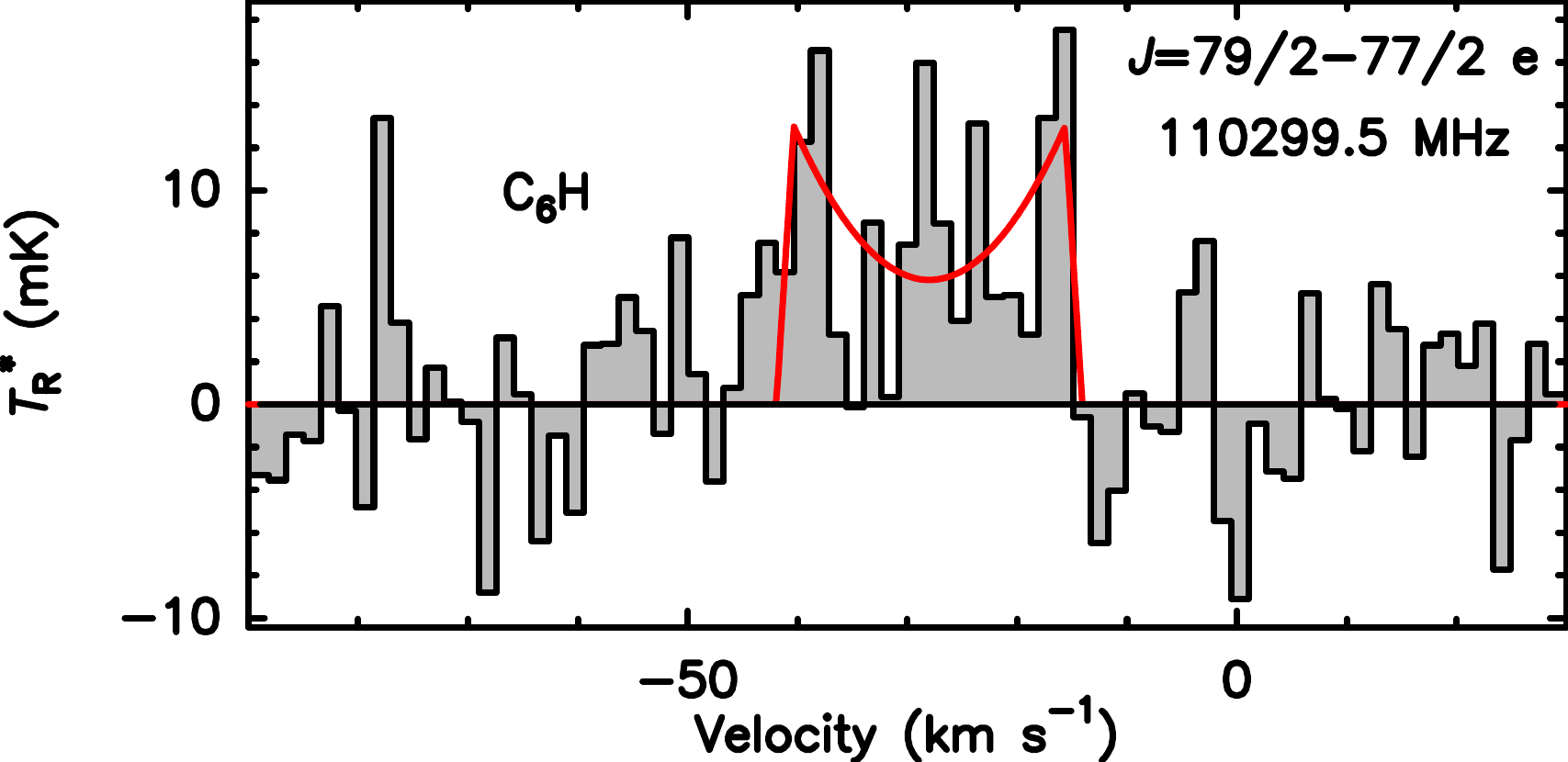}
\vspace{0.1cm}
\includegraphics[width = 0.45 \textwidth]{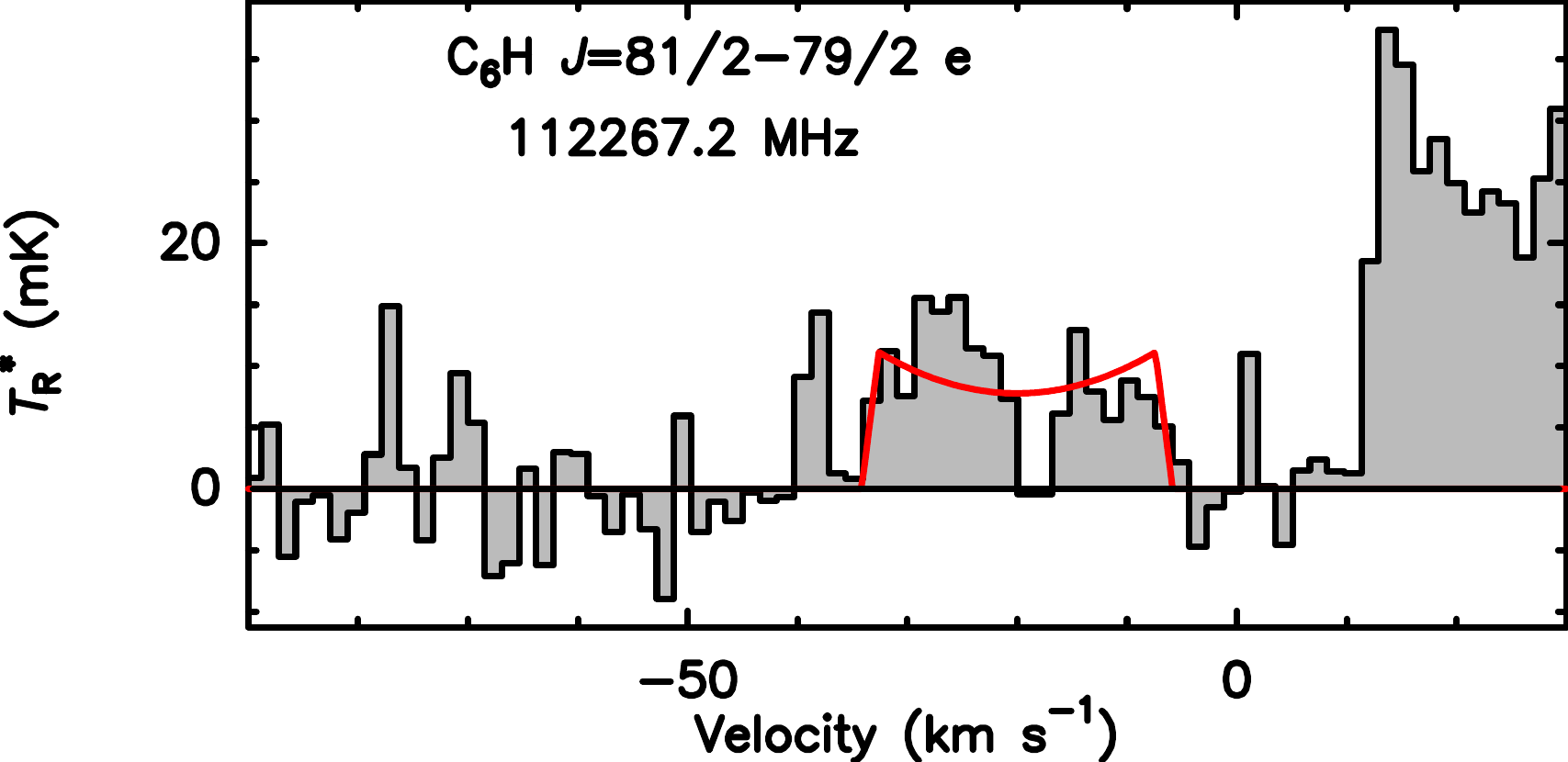}
\hspace{0.05\textwidth}
\includegraphics[width = 0.45 \textwidth]{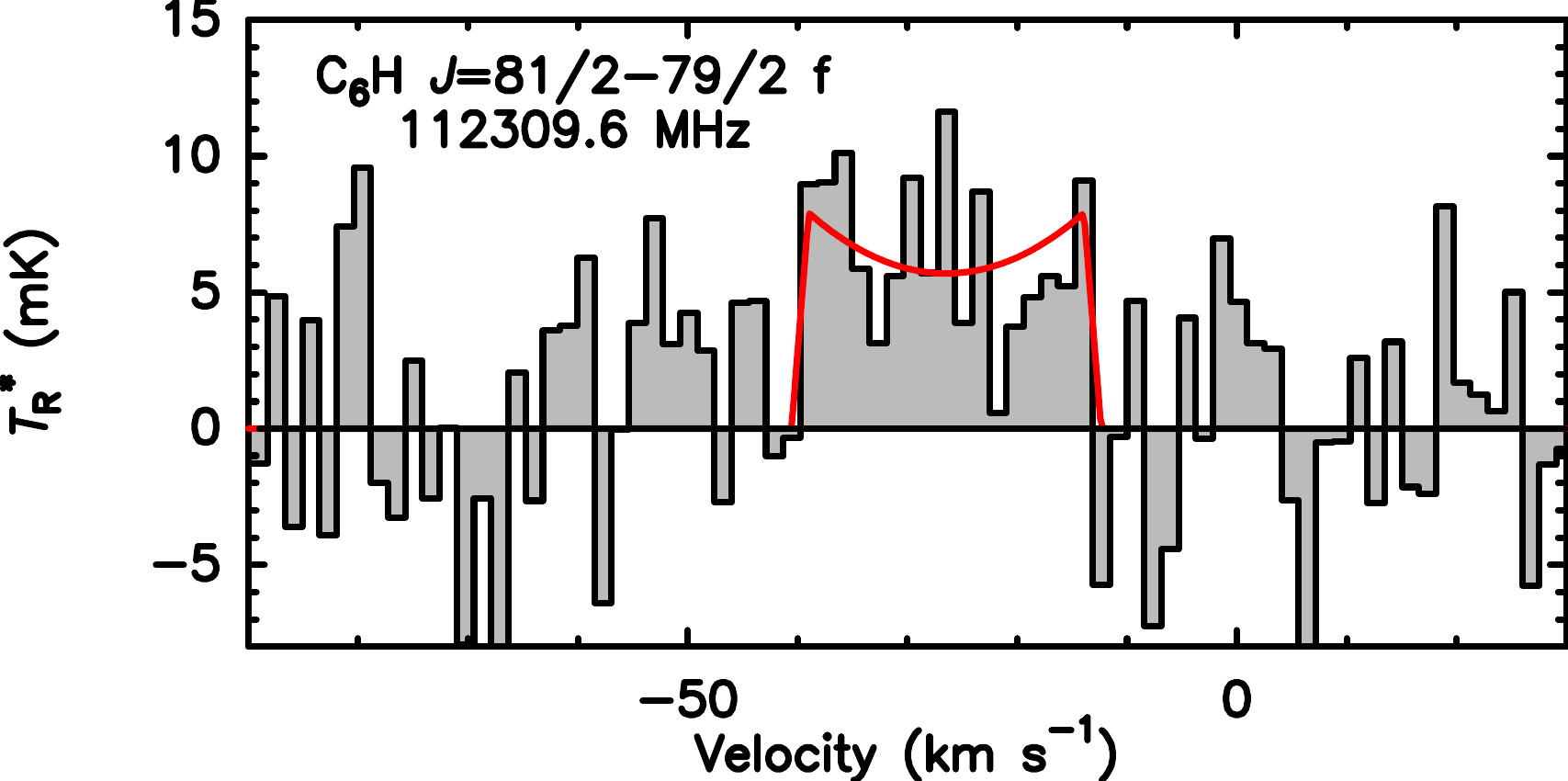}
\vspace{0.1cm}
\includegraphics[width = 0.45 \textwidth]{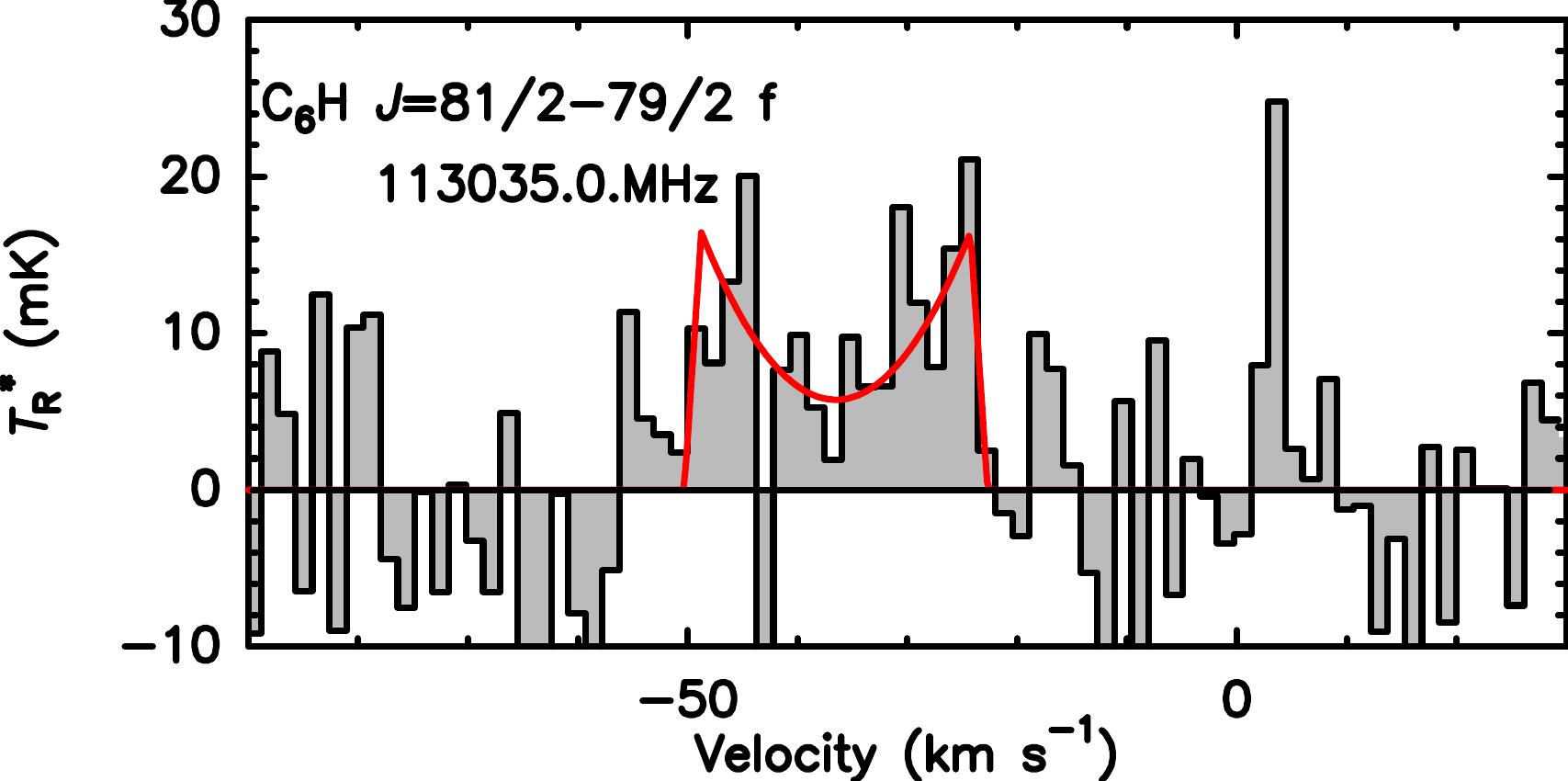}
\hspace{0.05\textwidth}
\includegraphics[width = 0.45 \textwidth]{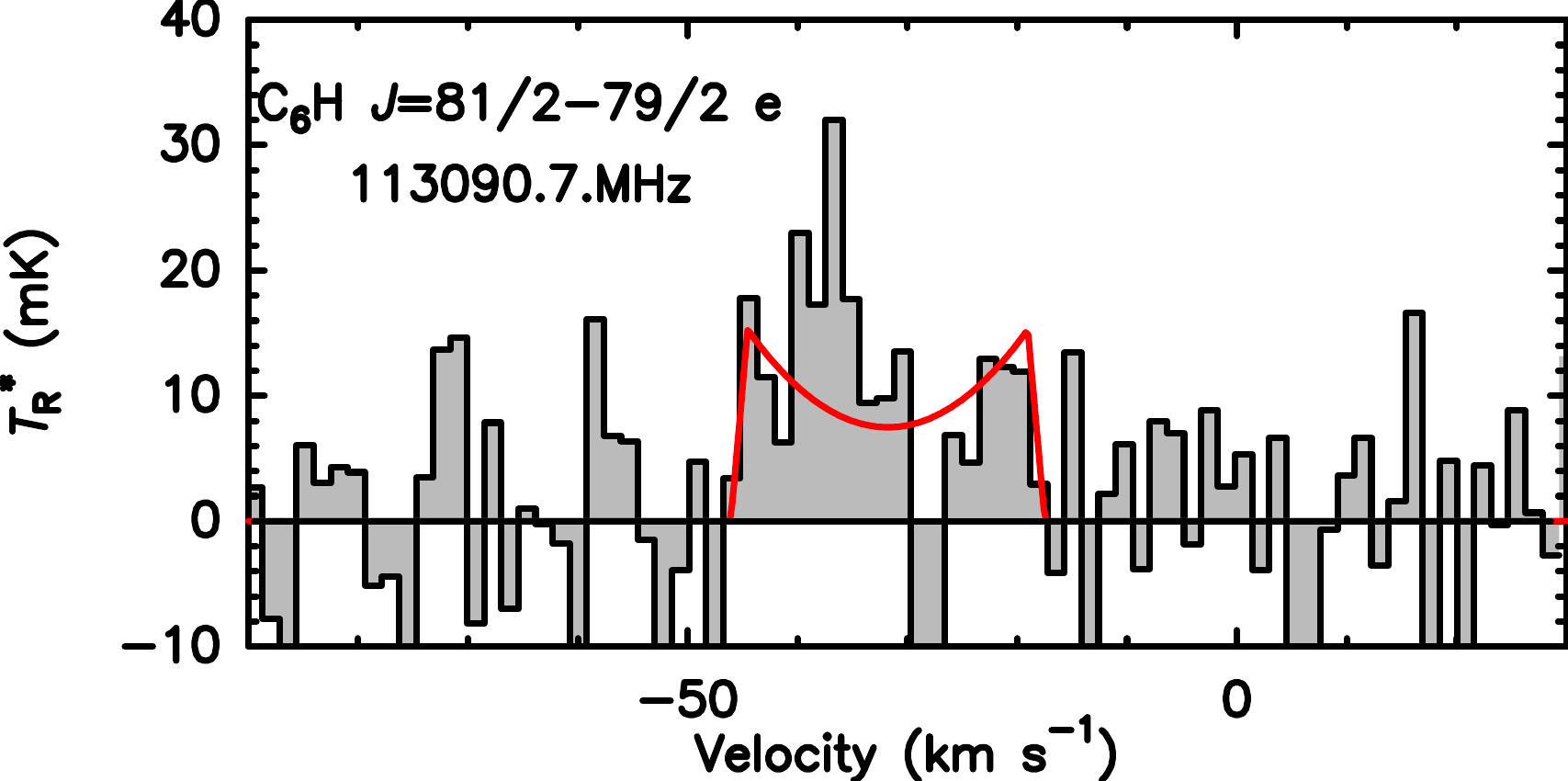}
\centerline{Figure \ref{Fig:fitting_15}. --- continued}
\end{figure*}

\begin{figure*}[!htbp]
\centering
\includegraphics[width = 0.45 \textwidth]{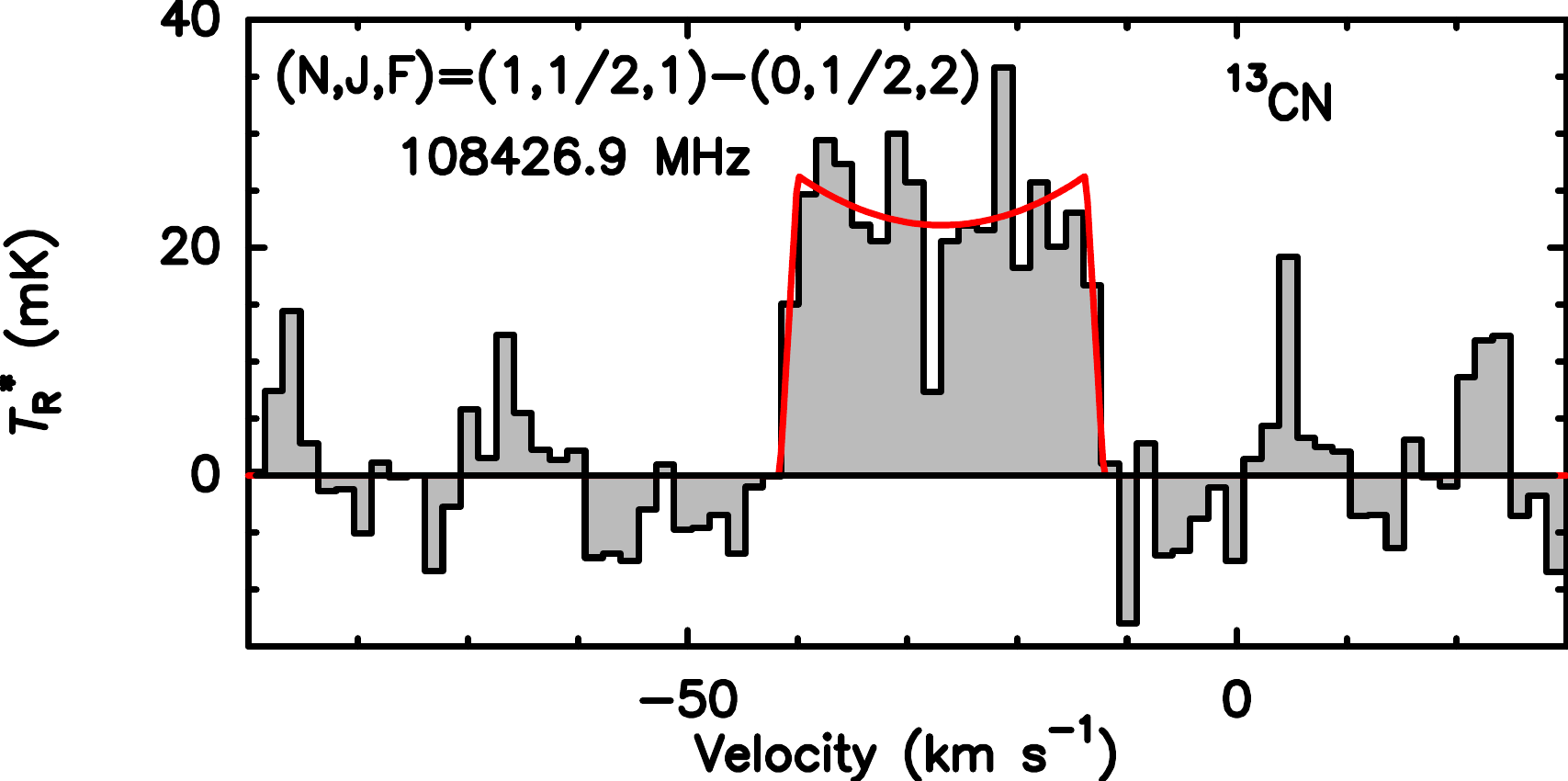}
\hspace{0.05\textwidth}
\includegraphics[width = 0.45 \textwidth]{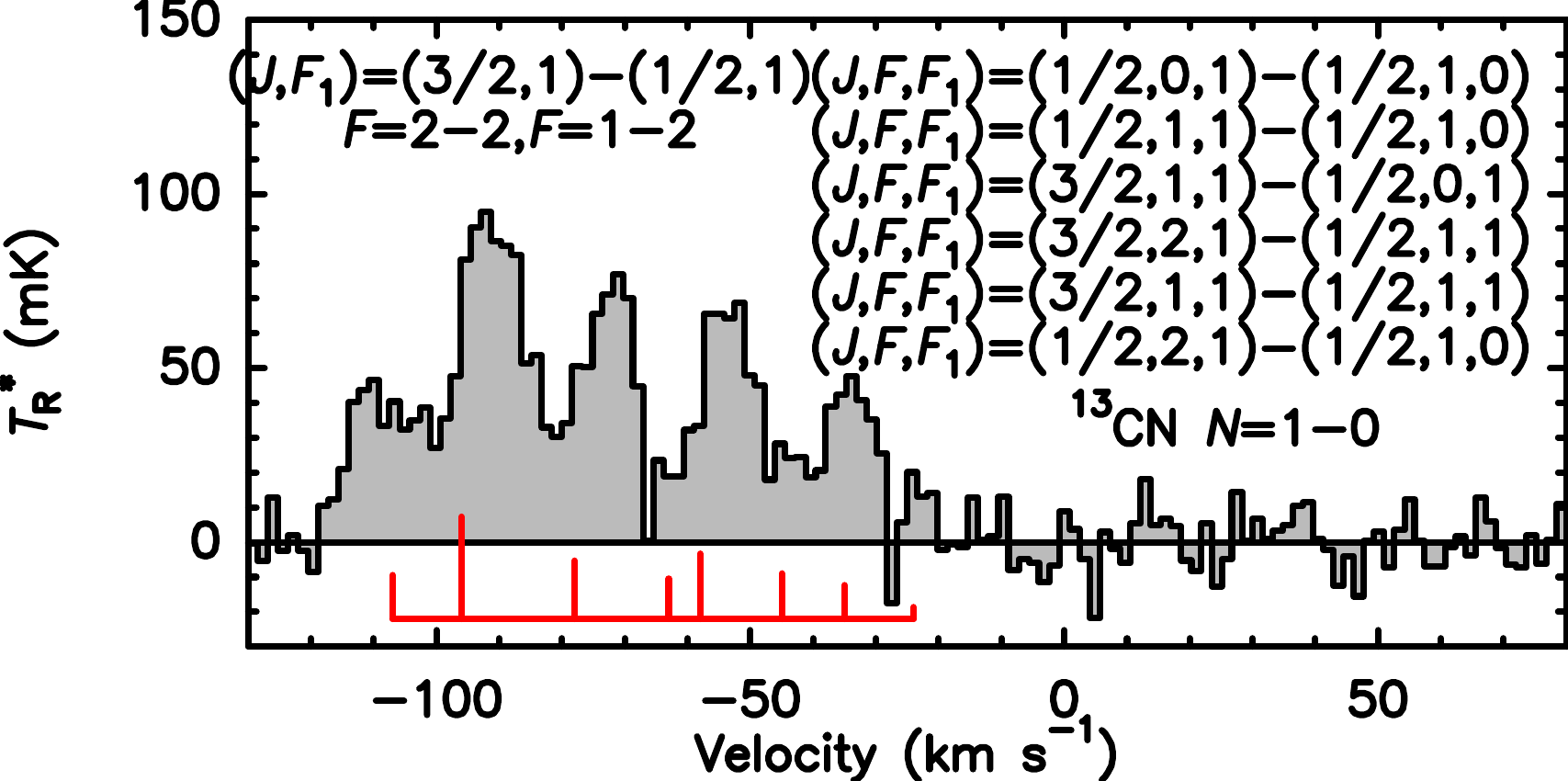}
\vspace{0.1cm}
\includegraphics[width = 0.45 \textwidth]{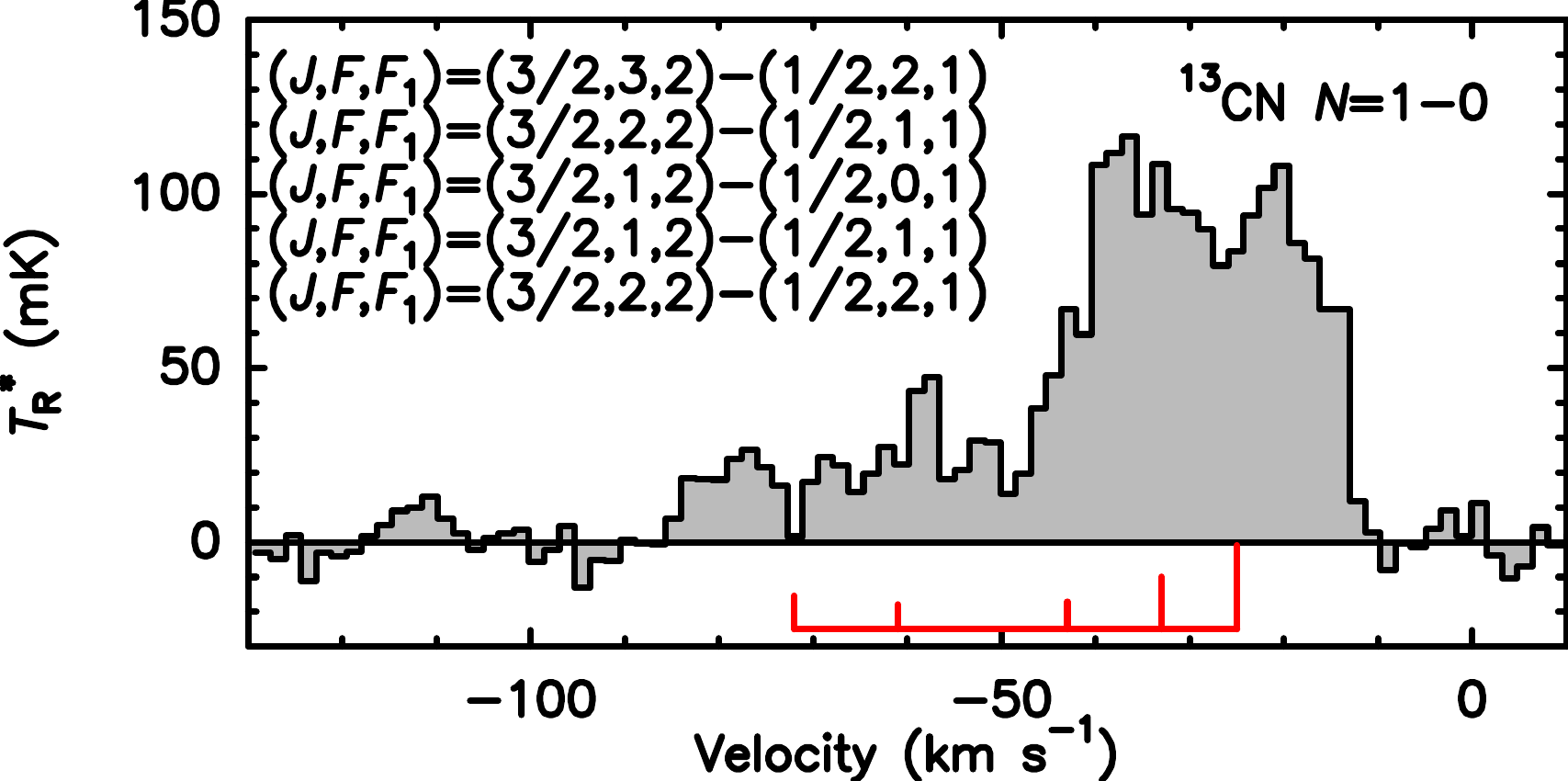}
\hspace{0.05\textwidth}
\includegraphics[width = 0.45 \textwidth]{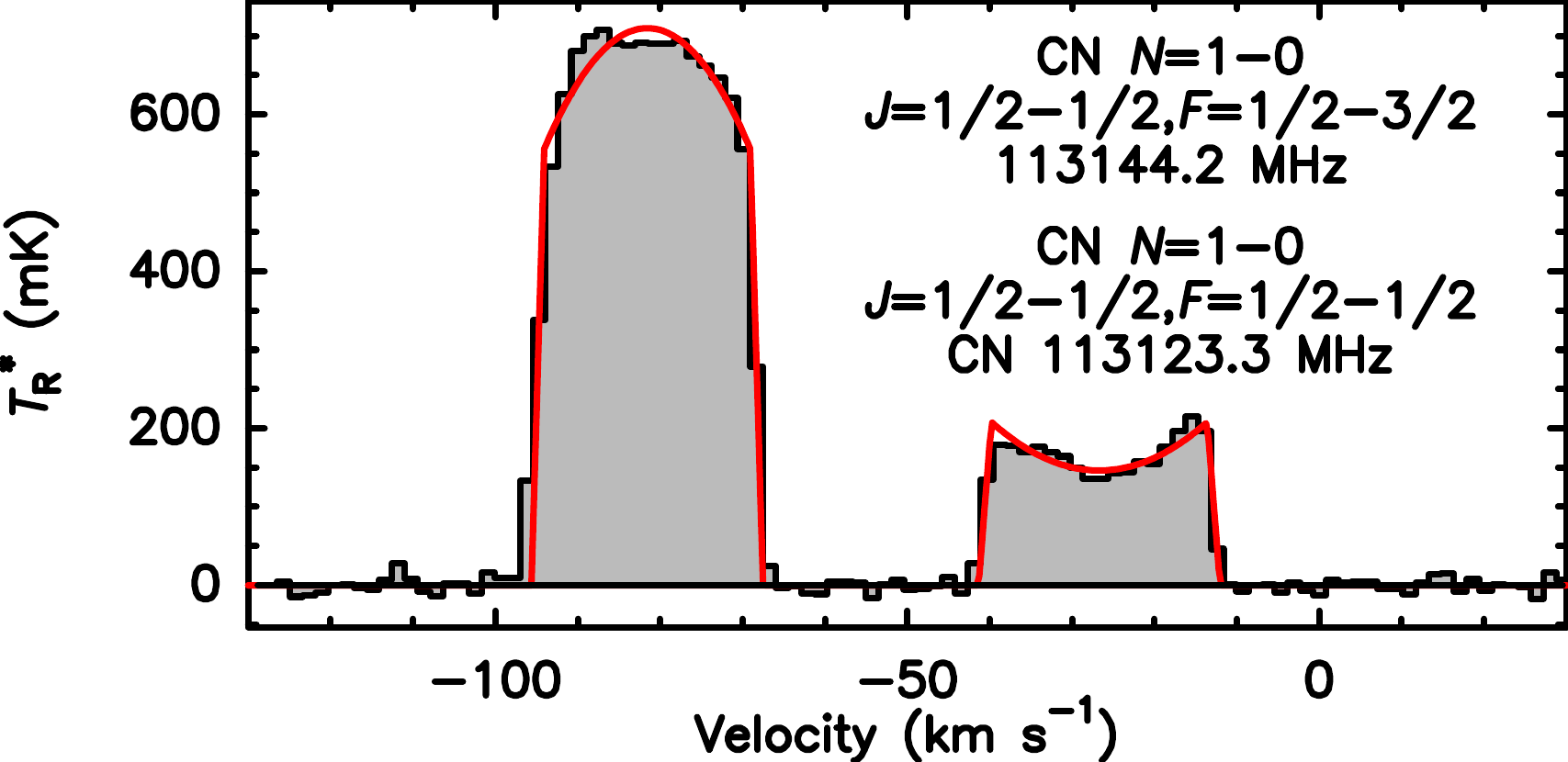}
\vspace{0.1cm}
\includegraphics[width = 0.45 \textwidth]{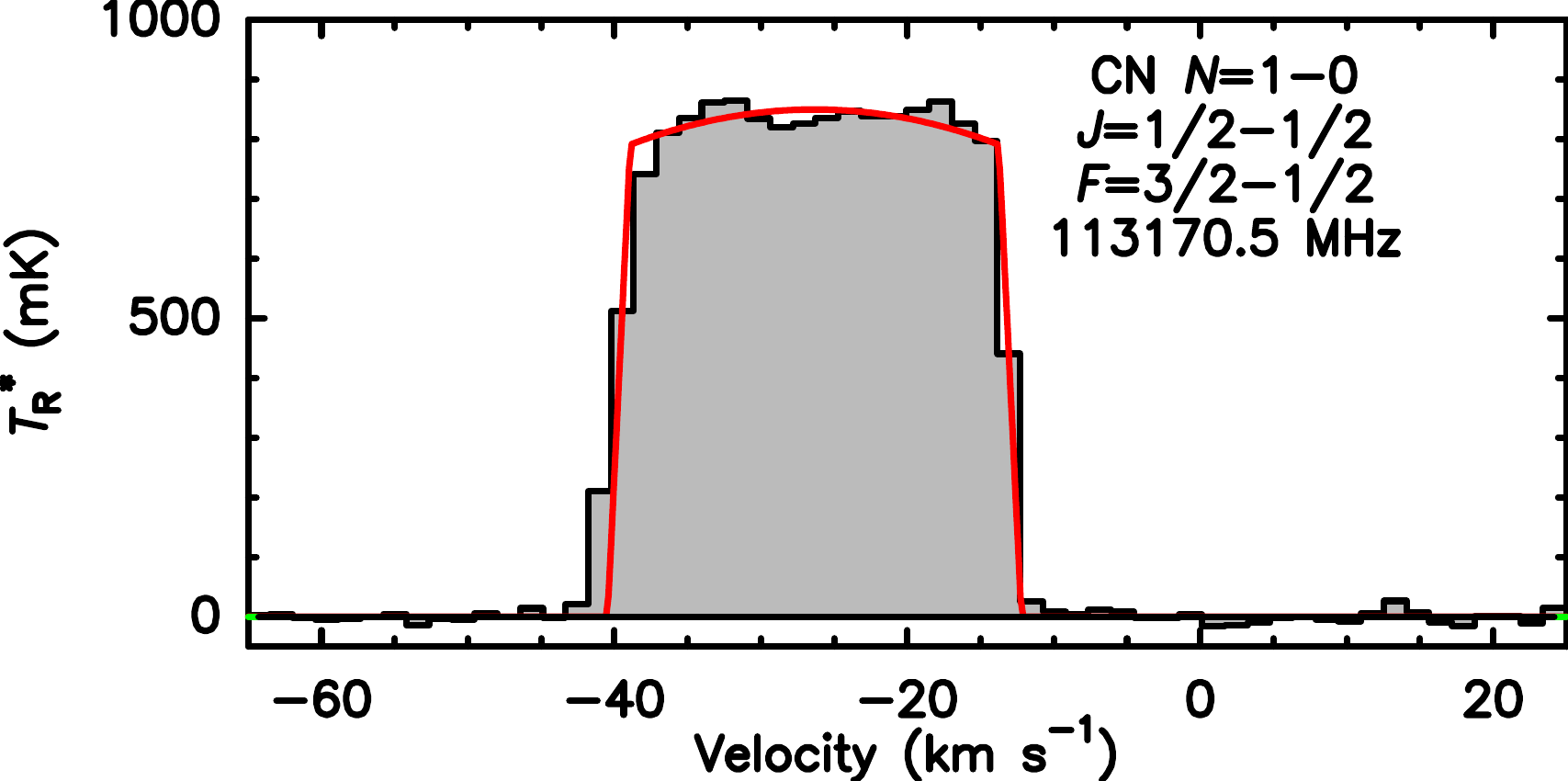}
\hspace{0.05\textwidth}
\includegraphics[width = 0.45 \textwidth]{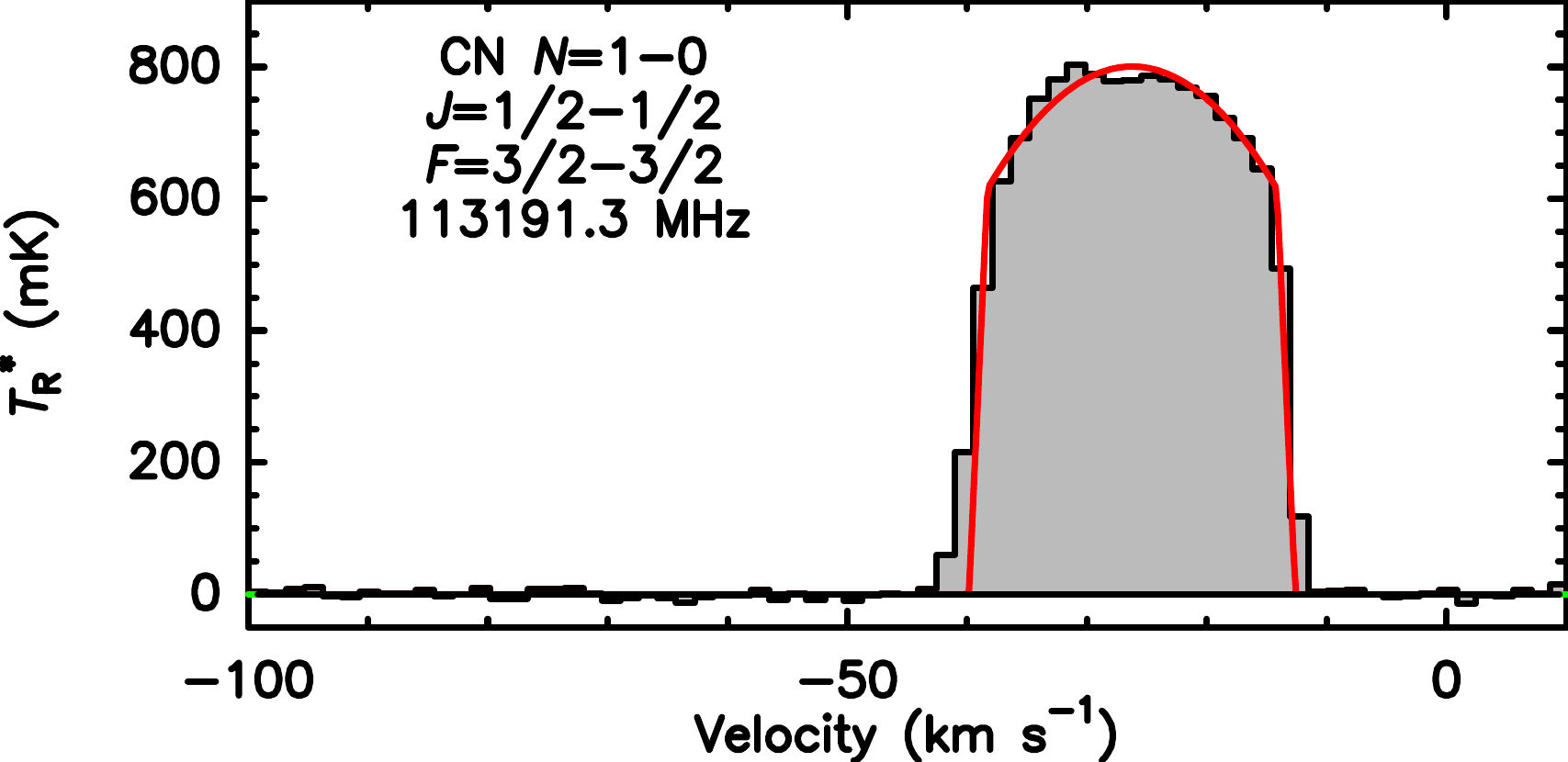}
\vspace{0.1cm}
\includegraphics[width = 0.45 \textwidth]{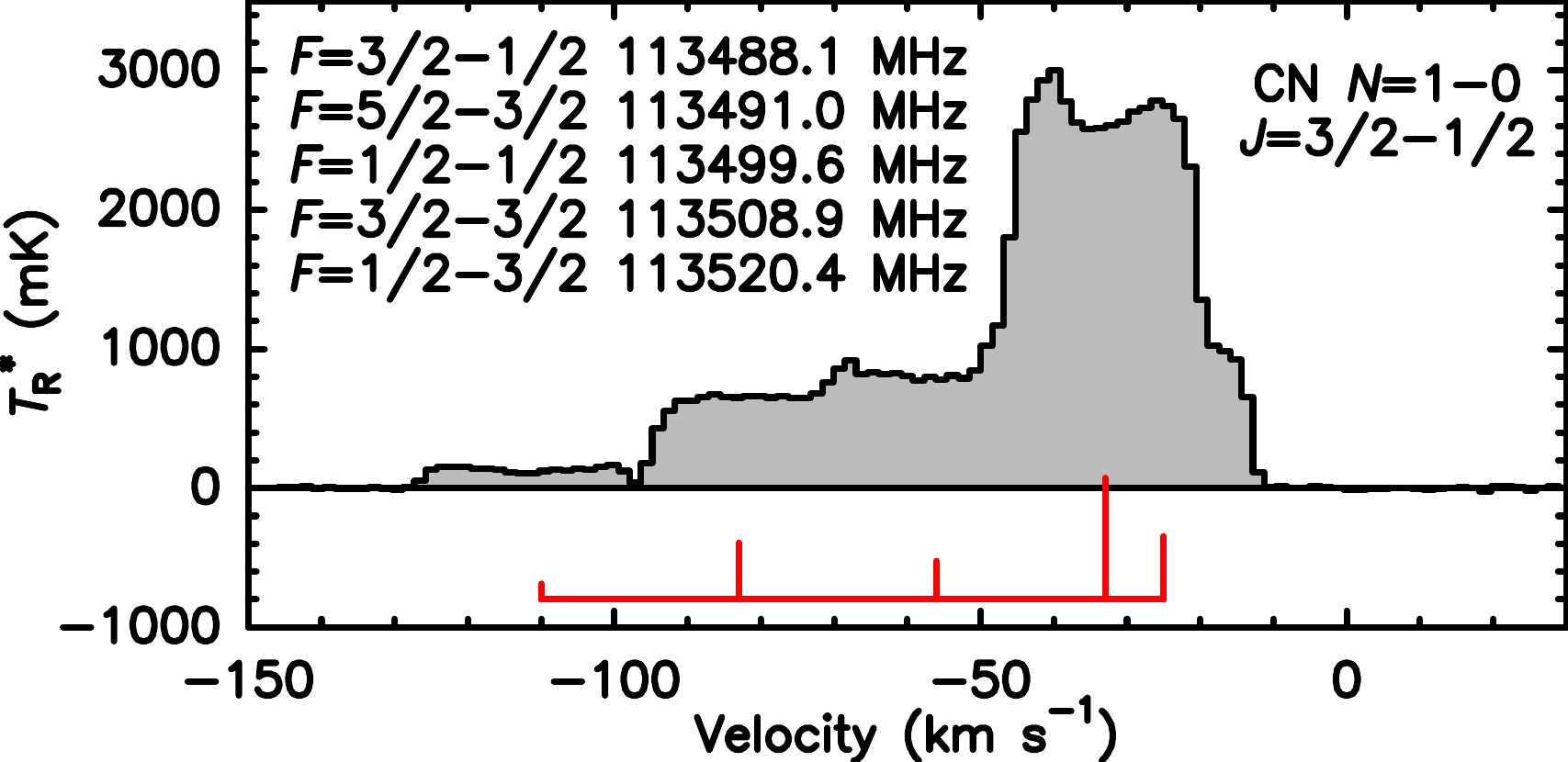}
\hspace{0.05\textwidth}
\includegraphics[width = 0.45 \textwidth]{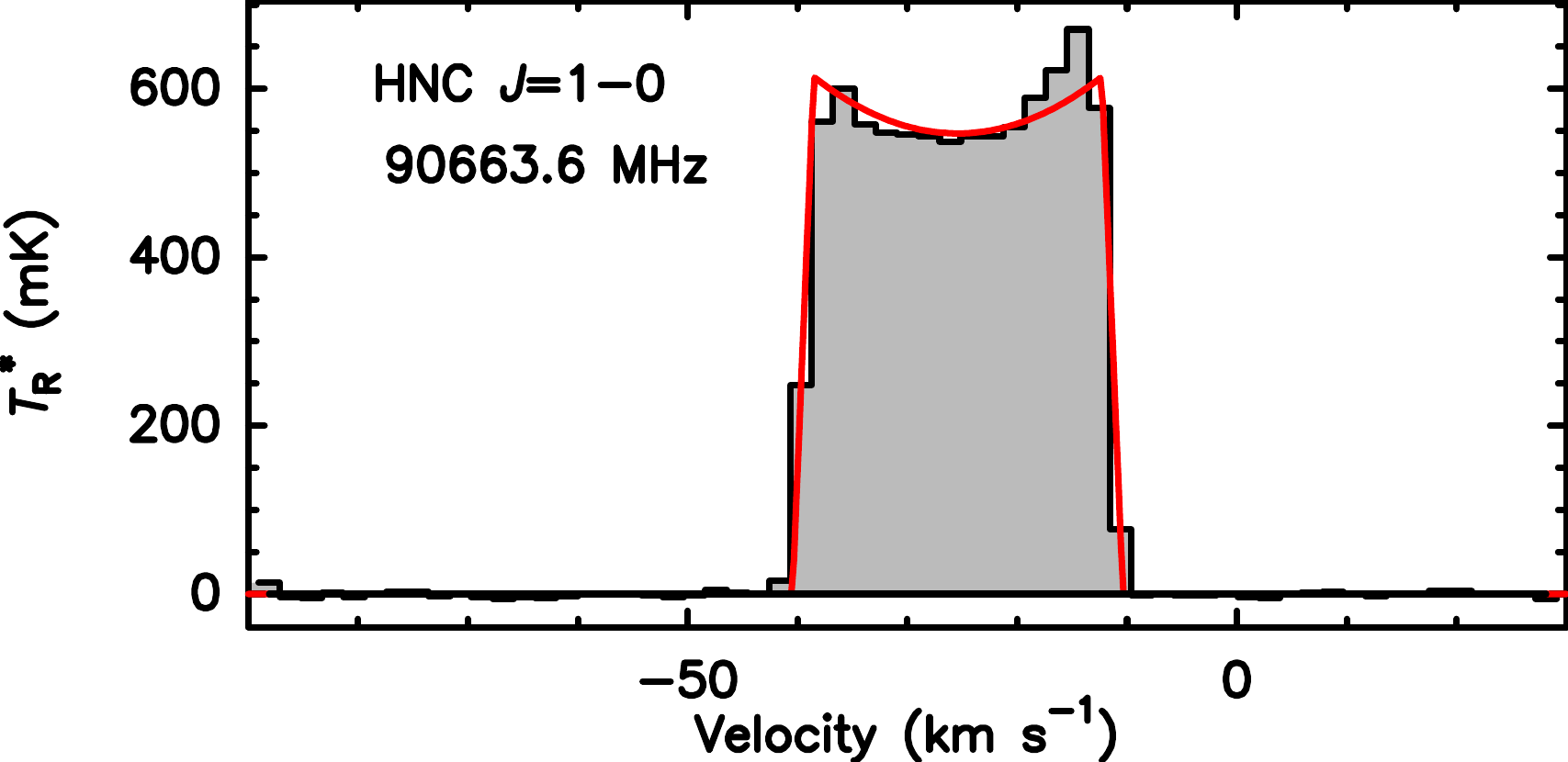}
\caption{{Same as Figure~\ref{Fig:fitting_1}, but for HNC, CN, and $^{13}$CN. }\label{Fig:fitting_16}}
\end{figure*}

\begin{figure*}[!htbp]
\centering
\includegraphics[width = 0.45 \textwidth]{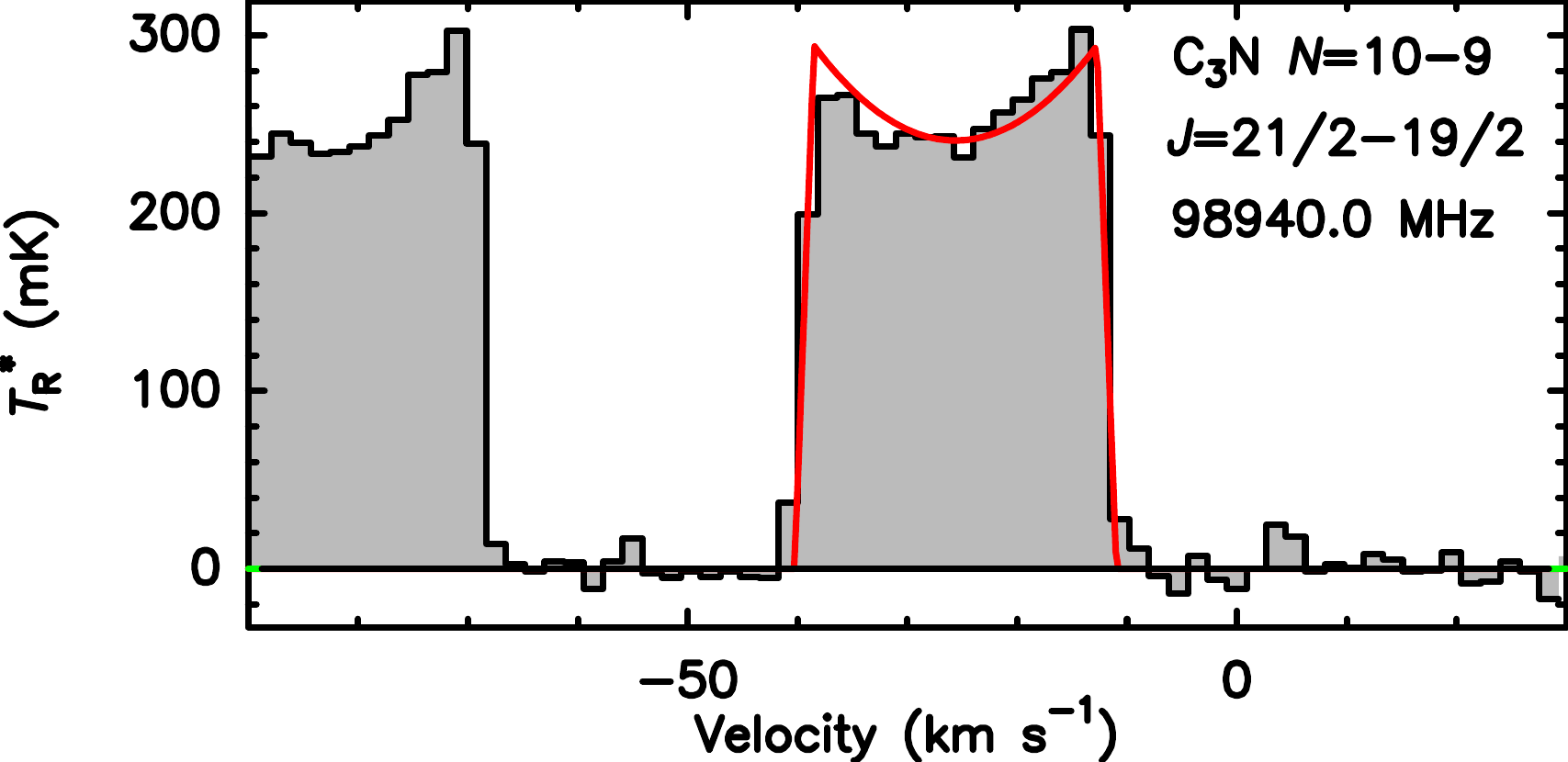}
\hspace{0.05\textwidth}
\includegraphics[width = 0.45 \textwidth]{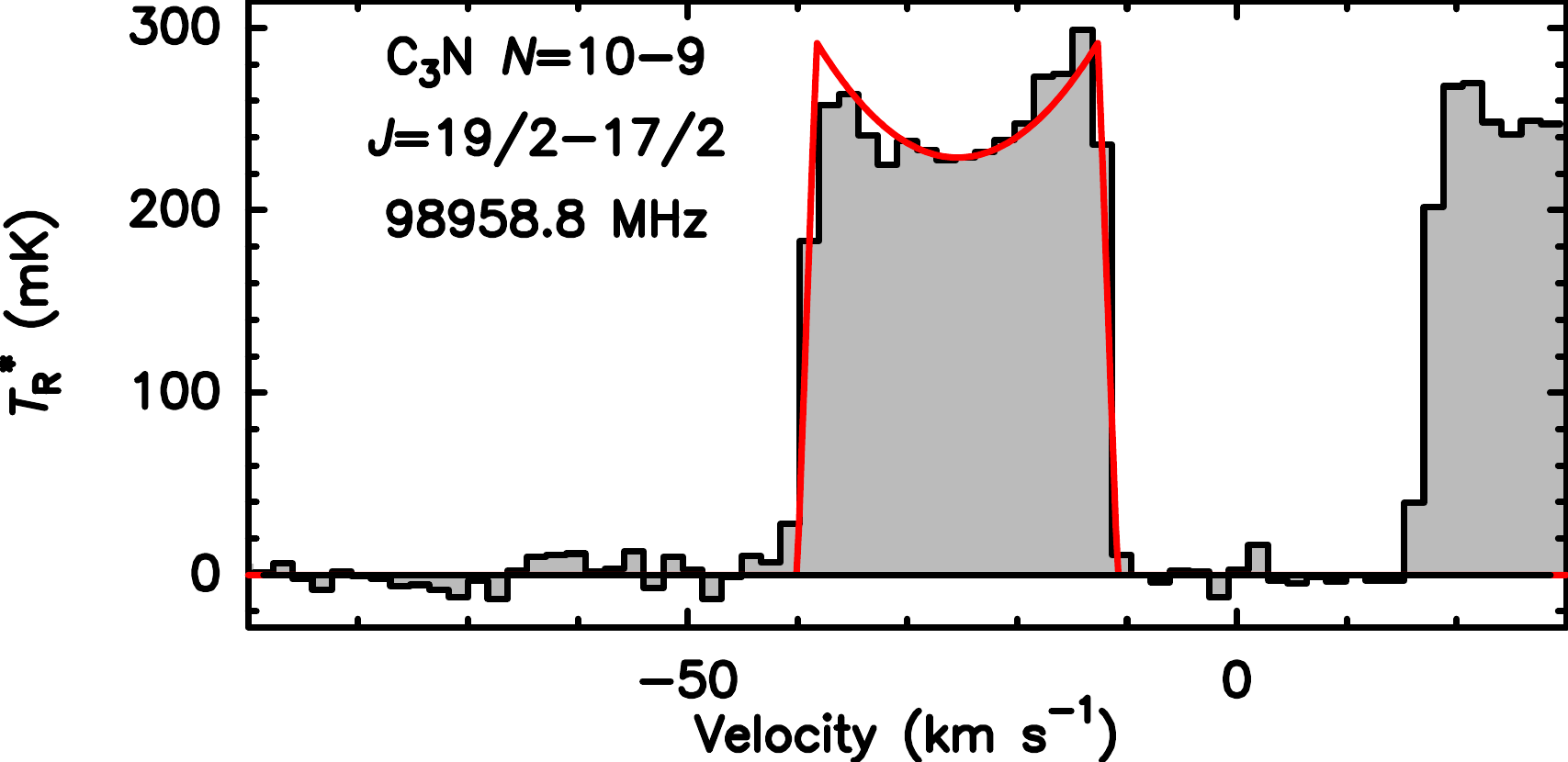}
\vspace{0.1cm}
\includegraphics[width = 0.45 \textwidth]{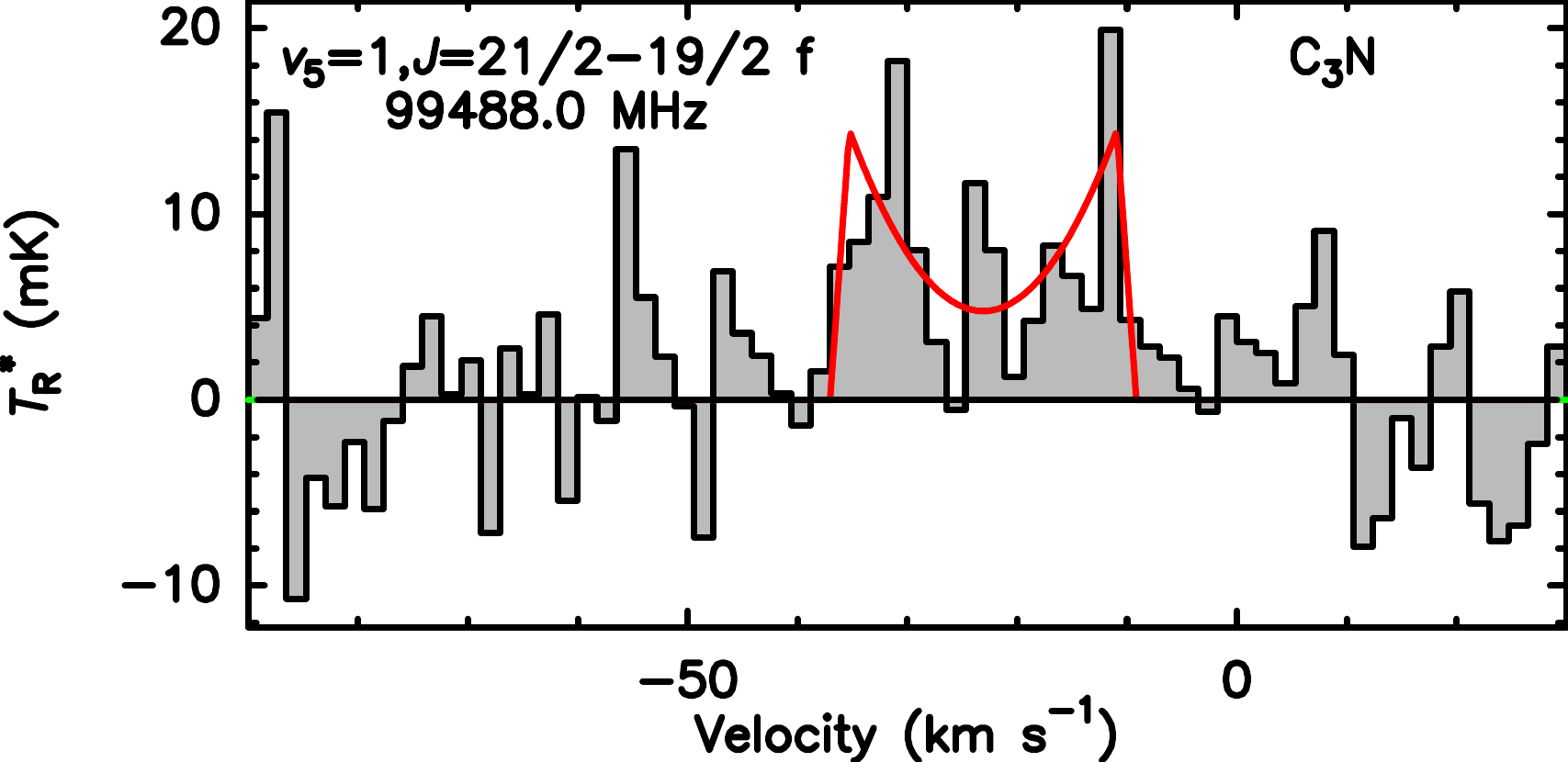}
\hspace{0.05\textwidth}
\includegraphics[width = 0.45 \textwidth]{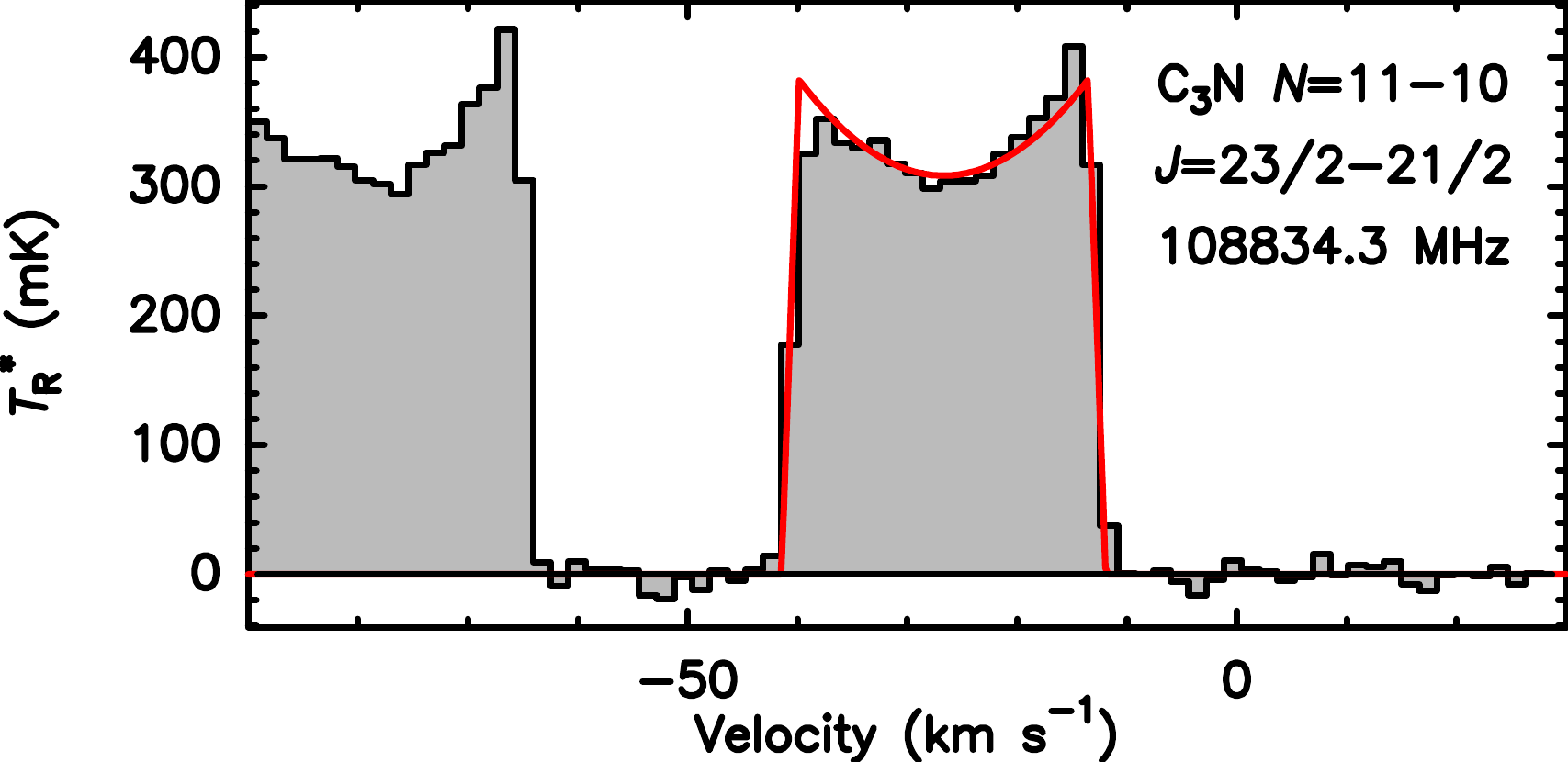}
\vspace{0.1cm}
\includegraphics[width = 0.45 \textwidth]{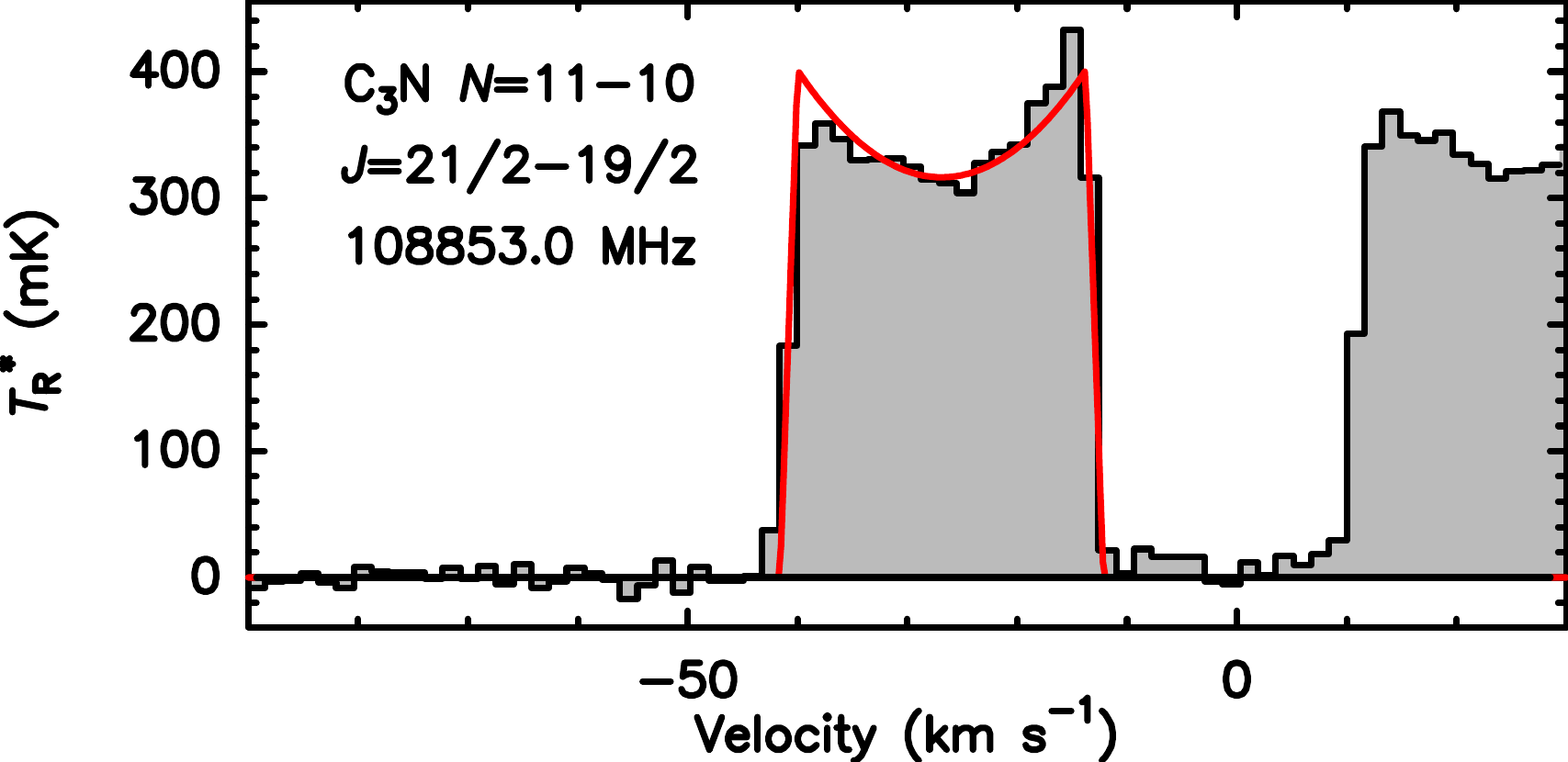}
\hspace{0.05\textwidth}
\includegraphics[width = 0.45 \textwidth]{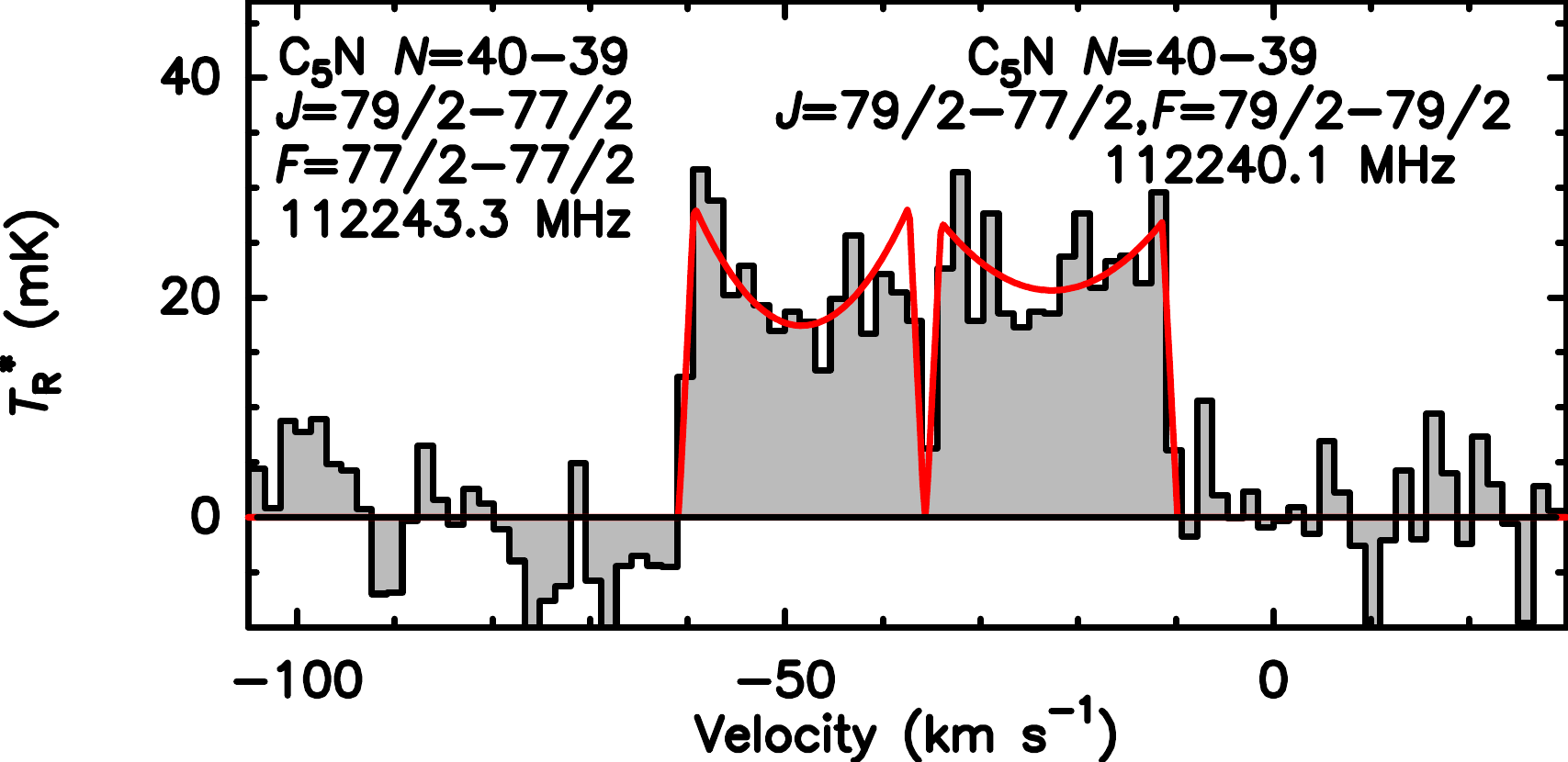}
\caption{{Same as Figure.~\ref{Fig:fitting_1}, but for C$_{3}$N and C$_{5}$N. }\label{Fig:fitting_18}}
\end{figure*}

\begin{figure*}[!htbp]
\centering
\includegraphics[width = 0.45 \textwidth]{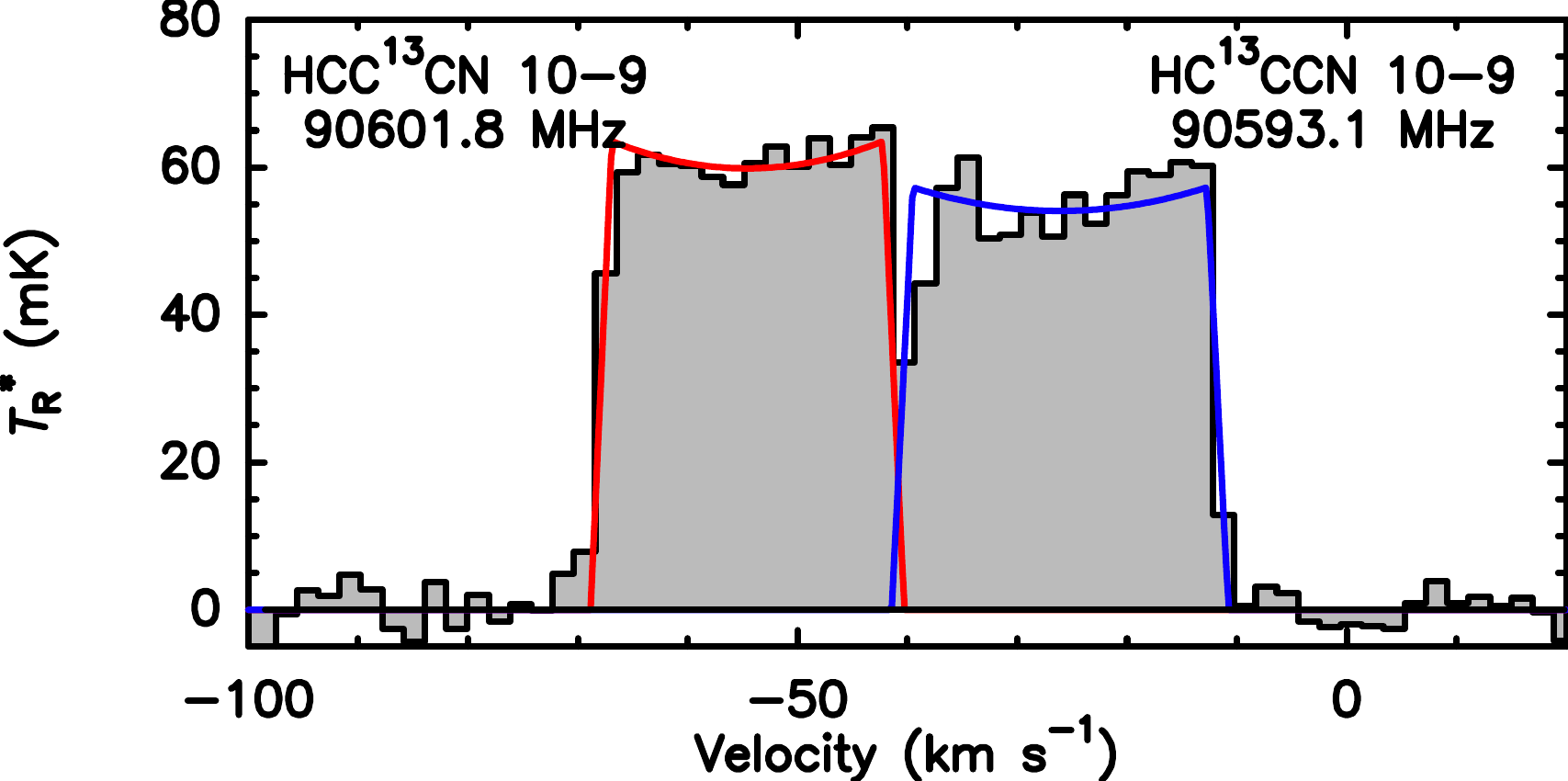}
\hspace{0.05\textwidth}
\includegraphics[width = 0.45 \textwidth]{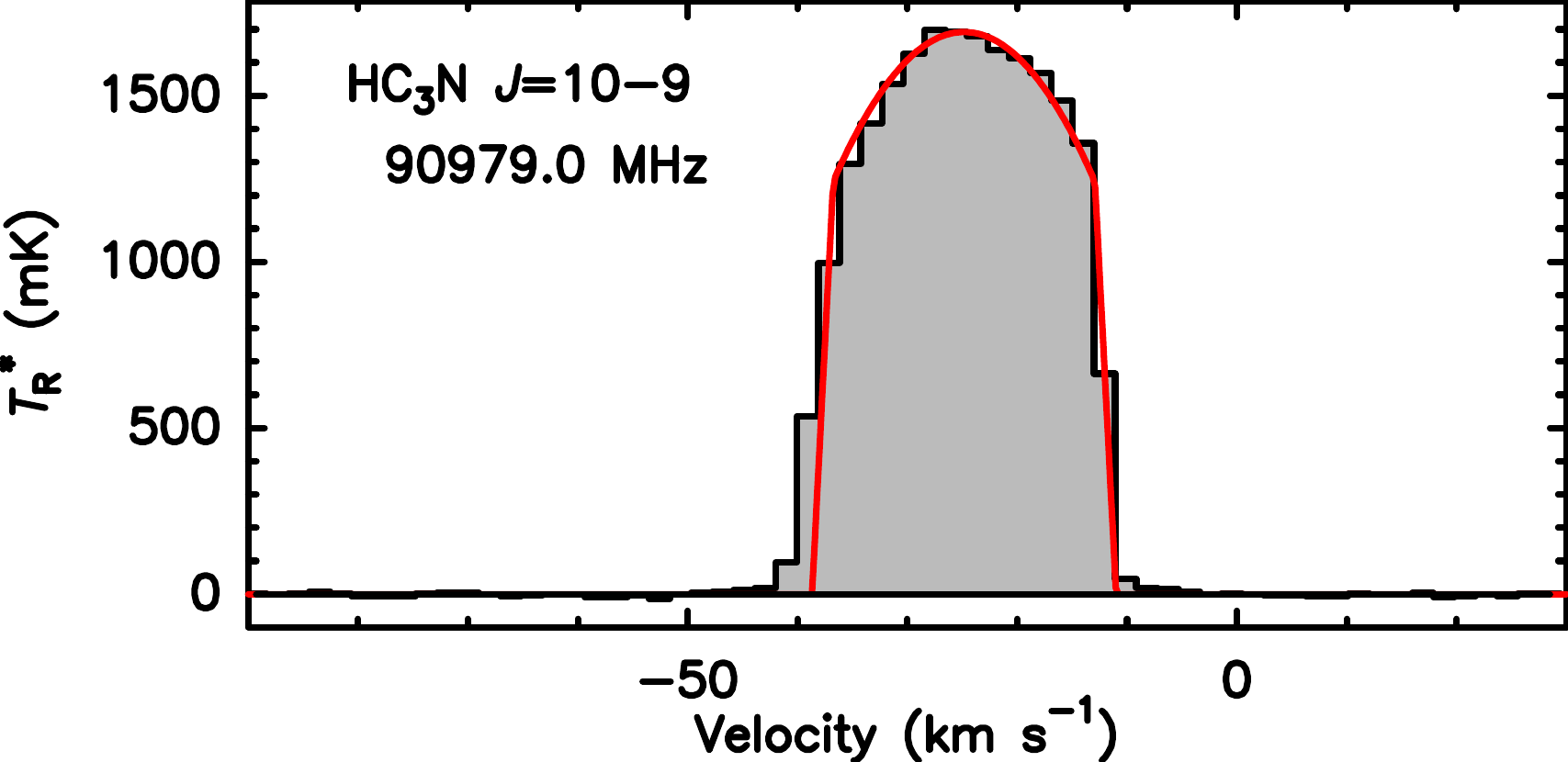}
\vspace{0.1cm}
\includegraphics[width = 0.45 \textwidth]{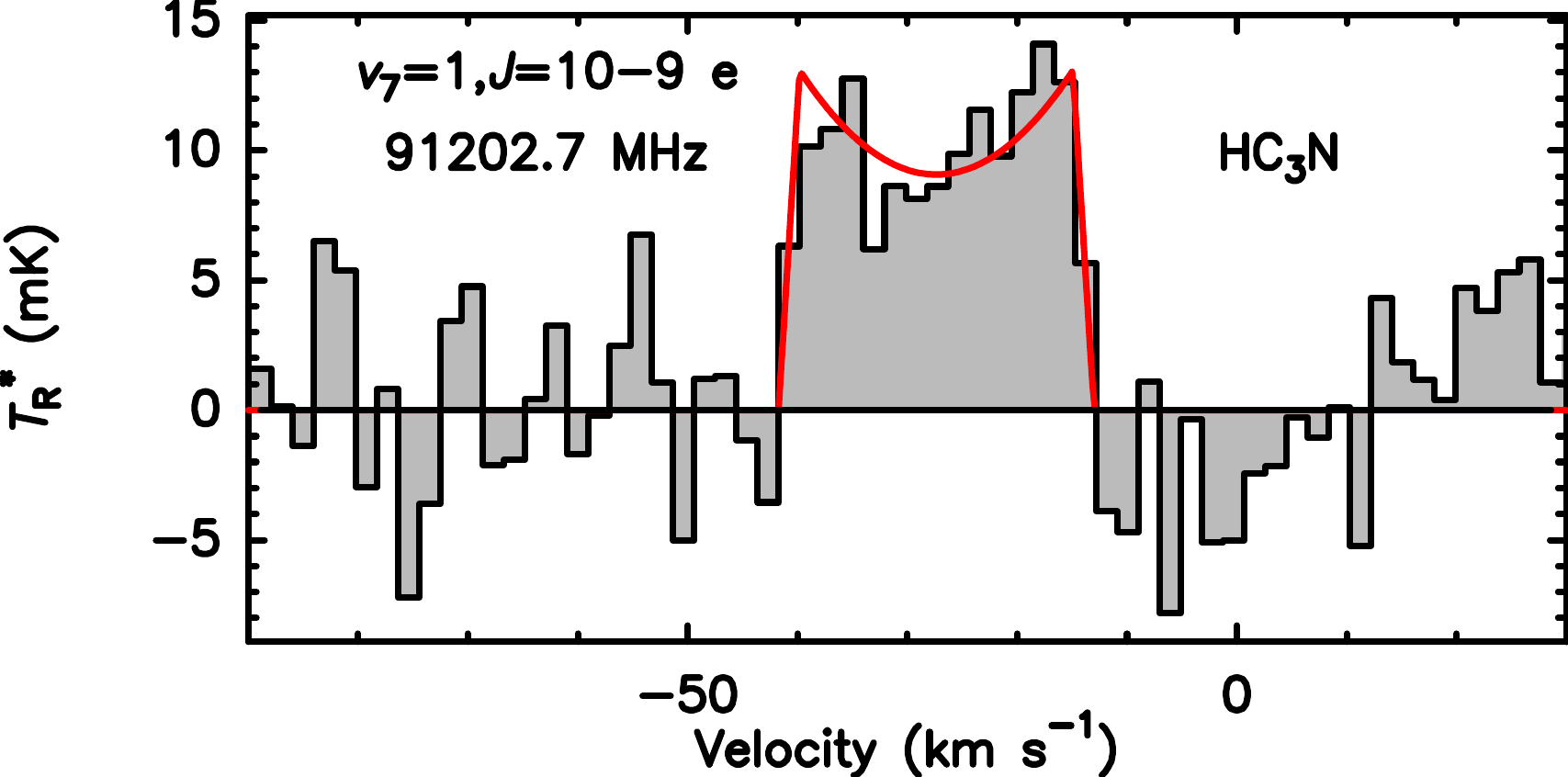}
\hspace{0.05\textwidth}
\includegraphics[width = 0.45 \textwidth]{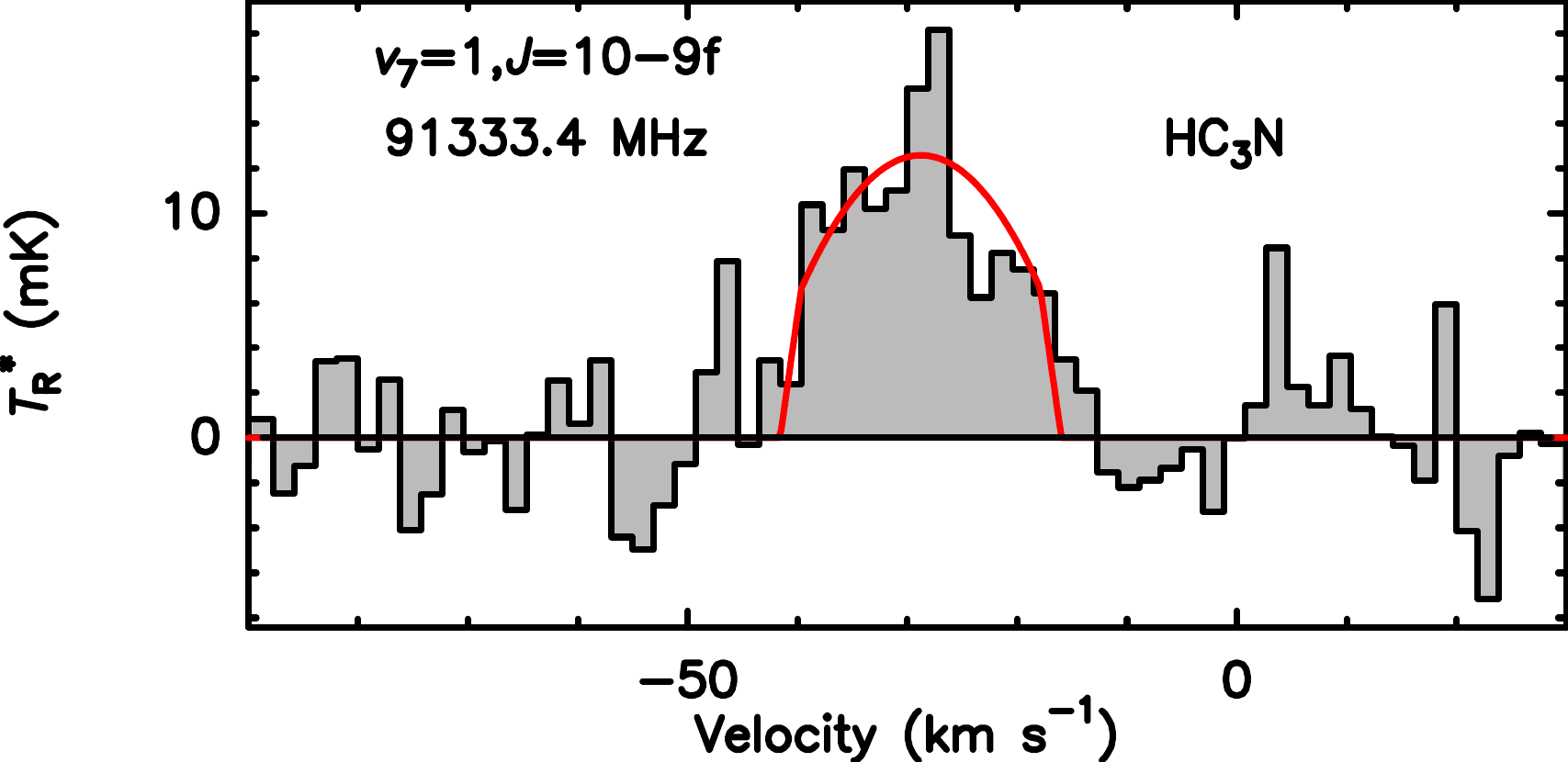}
\vspace{0.1cm}
\includegraphics[width = 0.45 \textwidth]{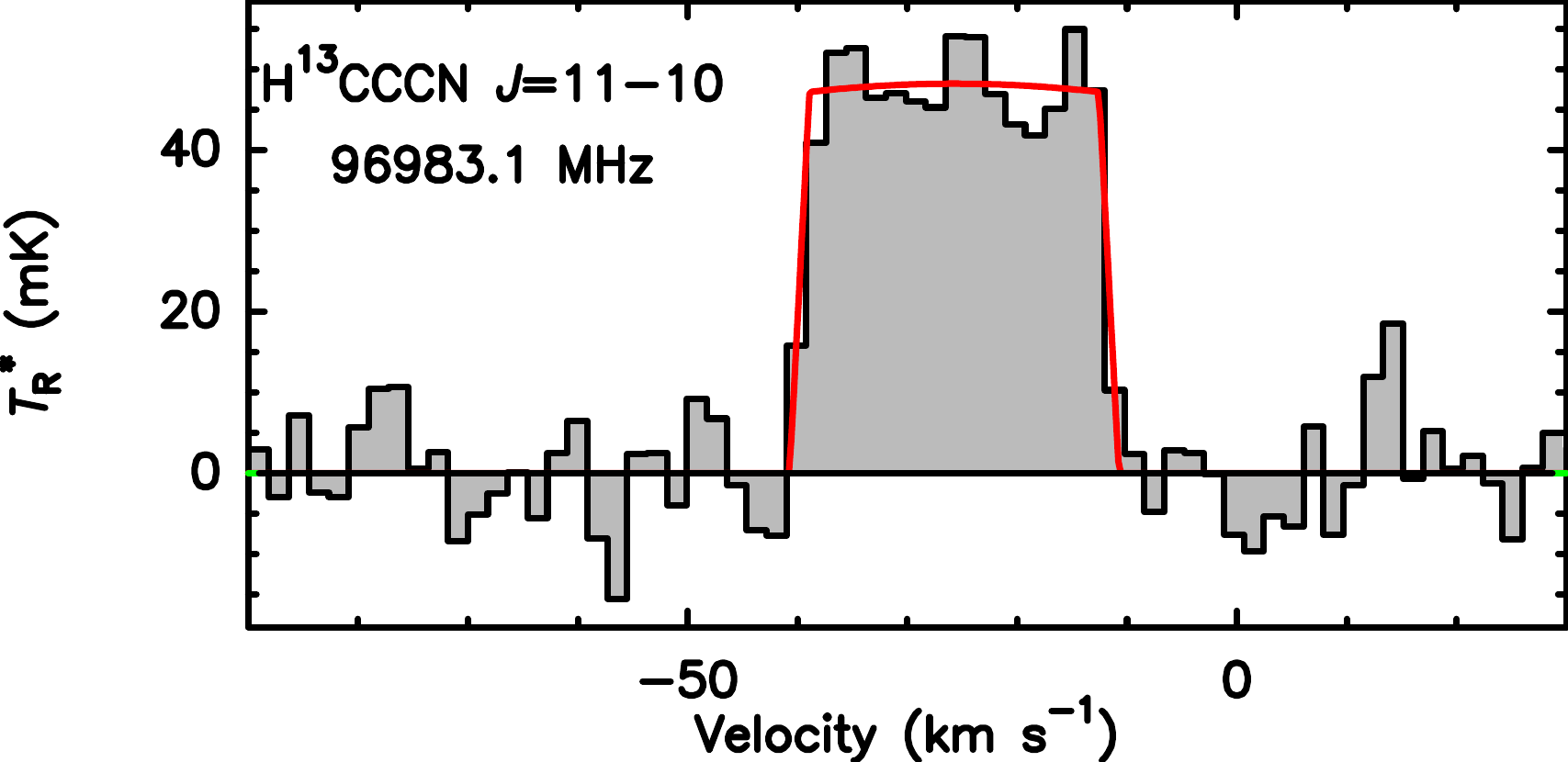}
\hspace{0.05\textwidth}
\includegraphics[width = 0.45 \textwidth]{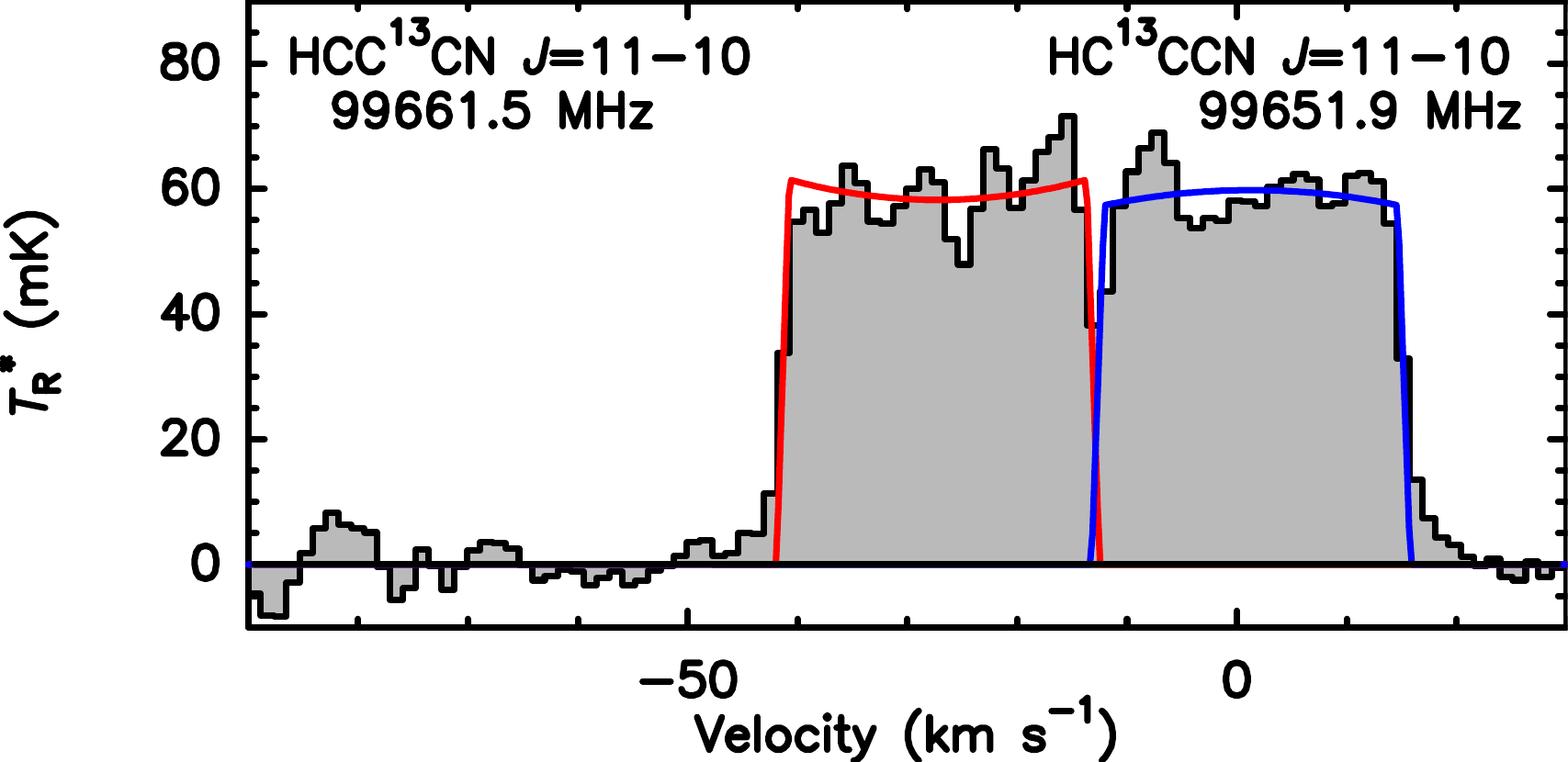}
\vspace{0.1cm}
\includegraphics[width = 0.45 \textwidth]{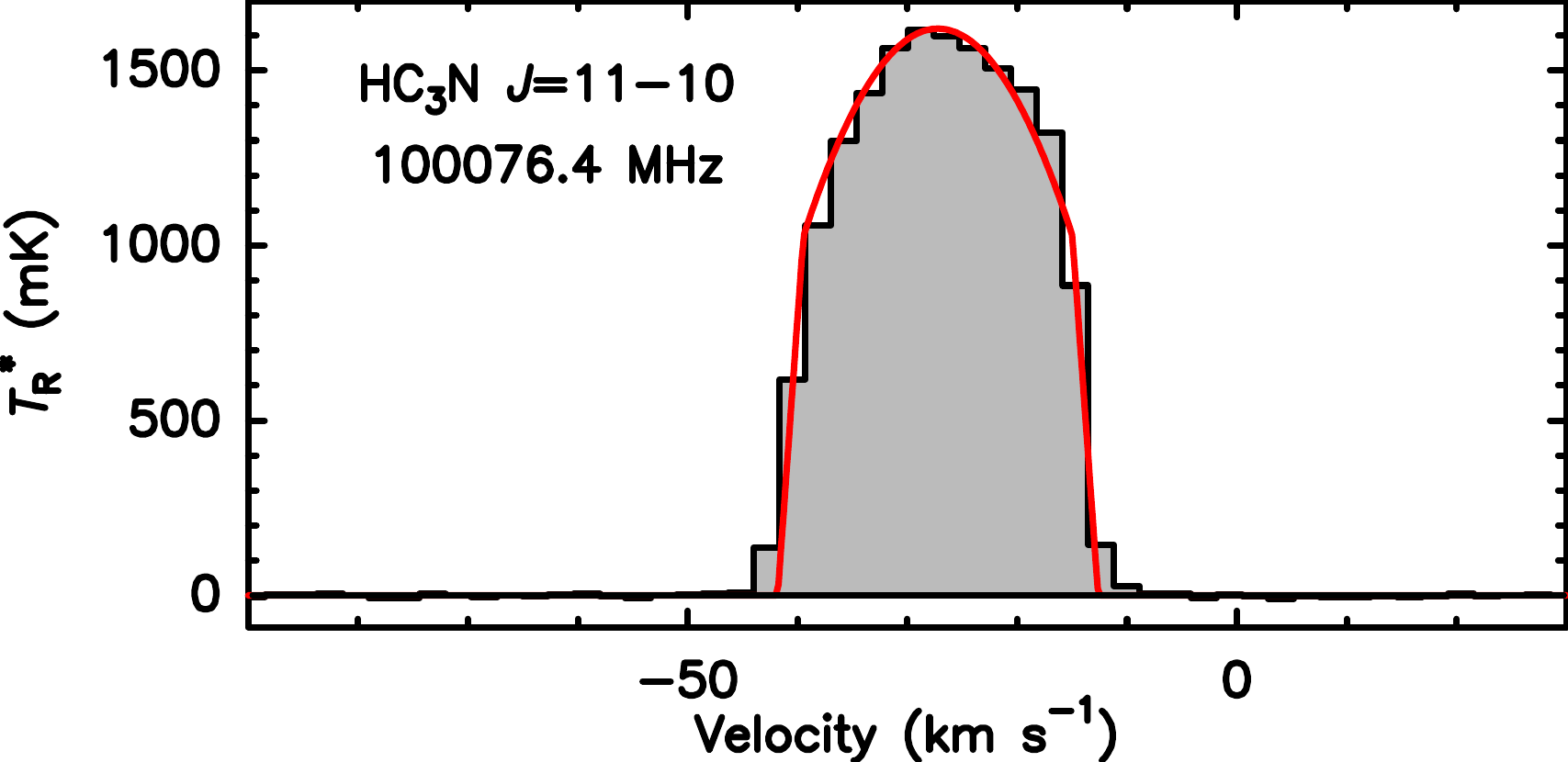}
\hspace{0.05\textwidth}
\includegraphics[width = 0.45 \textwidth]{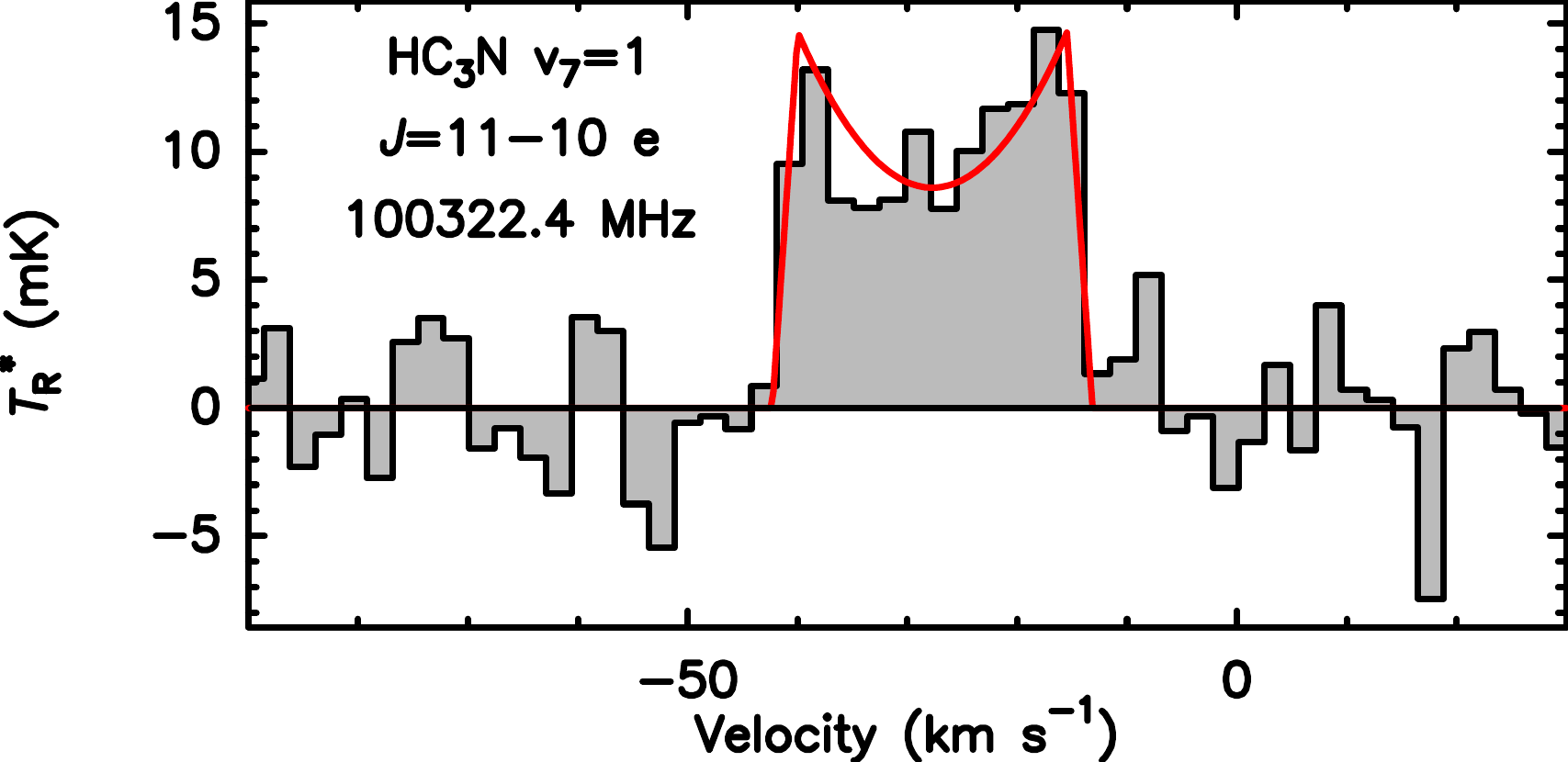}
\vspace{0.1cm}
\includegraphics[width = 0.45 \textwidth]{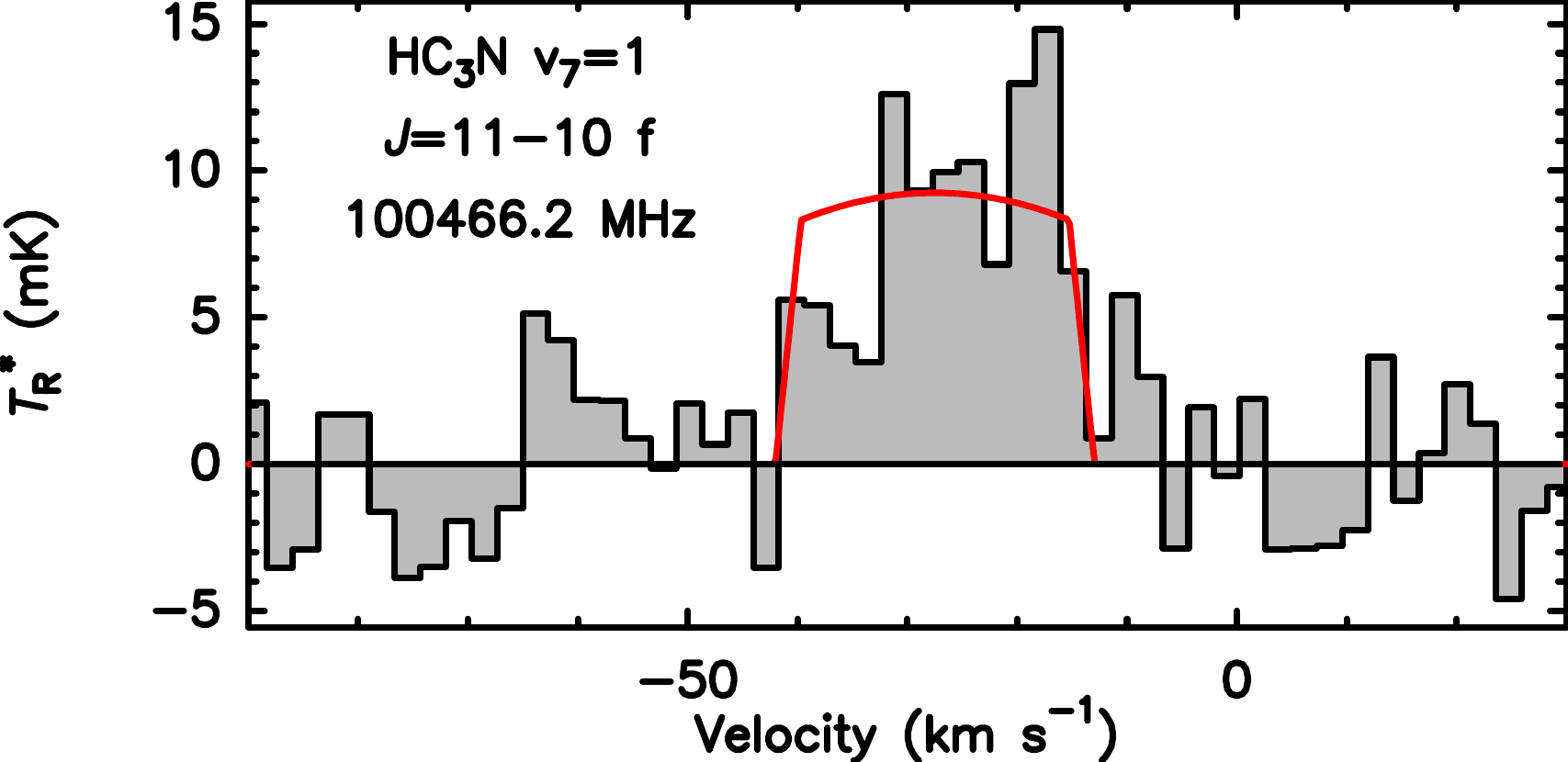}
\hspace{0.05\textwidth}
\includegraphics[width = 0.45 \textwidth]{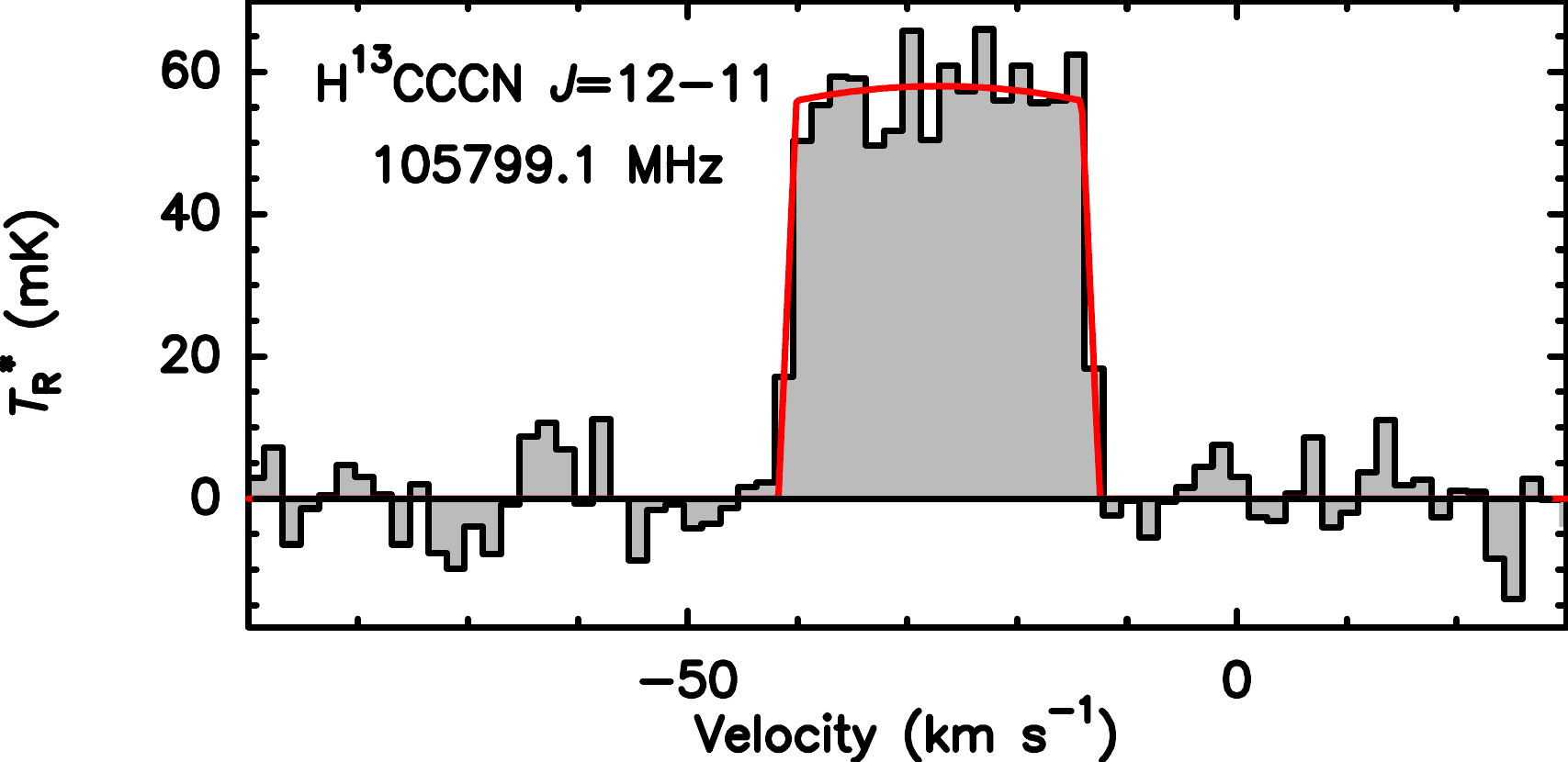}
\caption{{Same as Figure.~\ref{Fig:fitting_1}, but for  HC$_{3}$N, its isotopologues, and HC$_{2}$N.
}\label{Fig:fitting_21}}
\end{figure*}

\begin{figure*}[!htbp]
\centering
\includegraphics[width = 0.45 \textwidth]{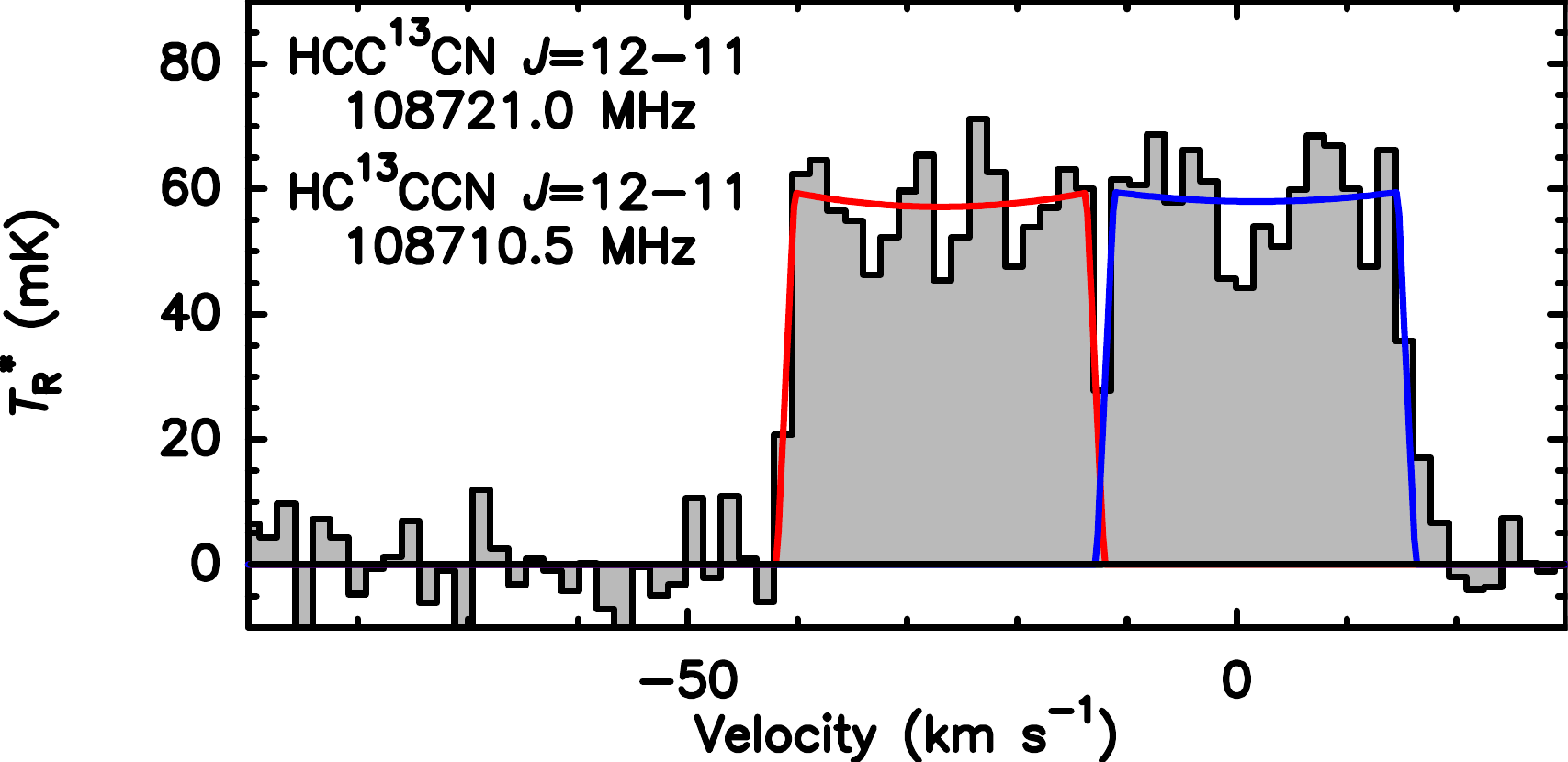}
\hspace{0.05\textwidth}
\includegraphics[width = 0.45 \textwidth]{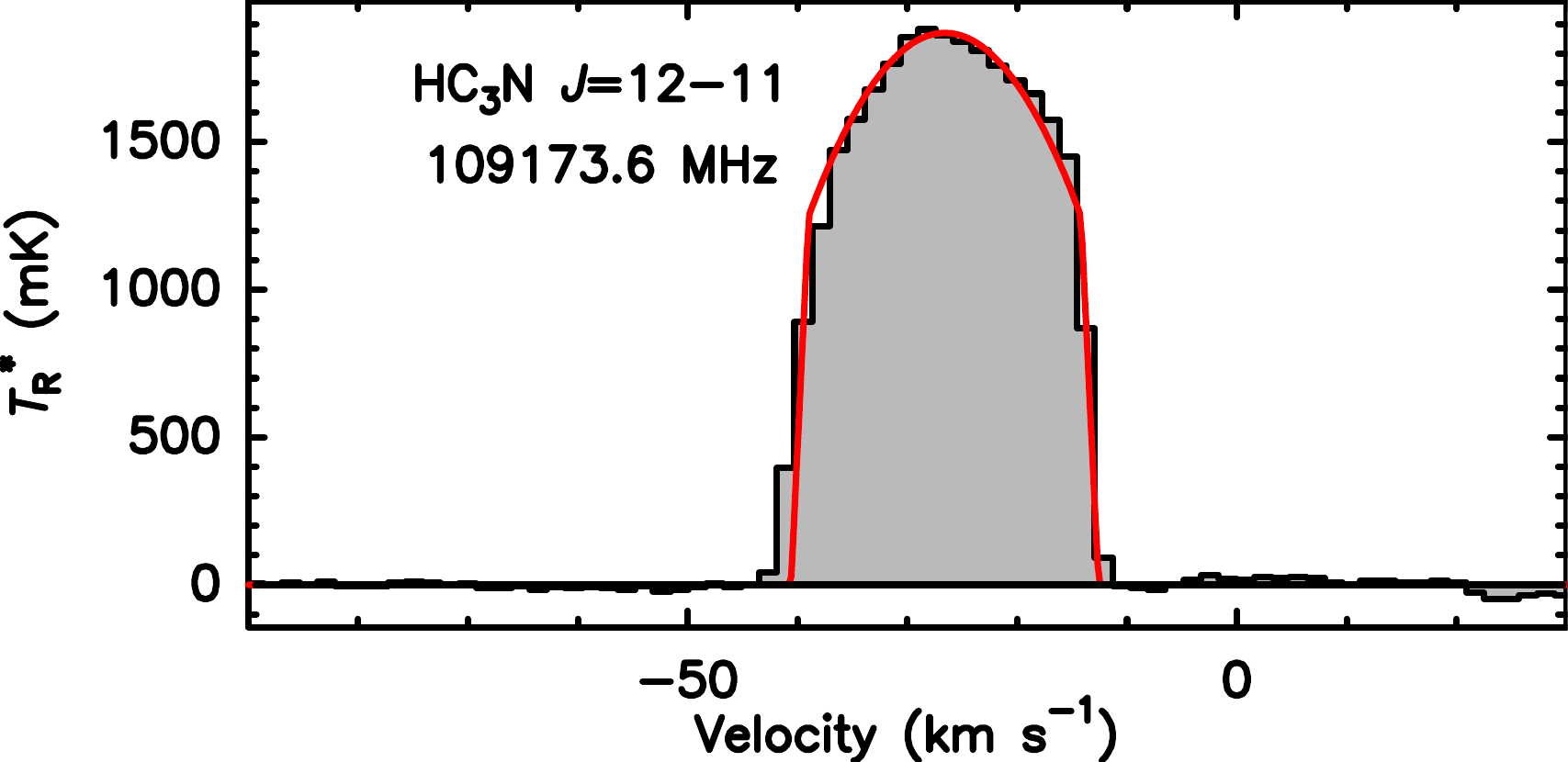}
\vspace{0.1cm}
\includegraphics[width = 0.45 \textwidth]{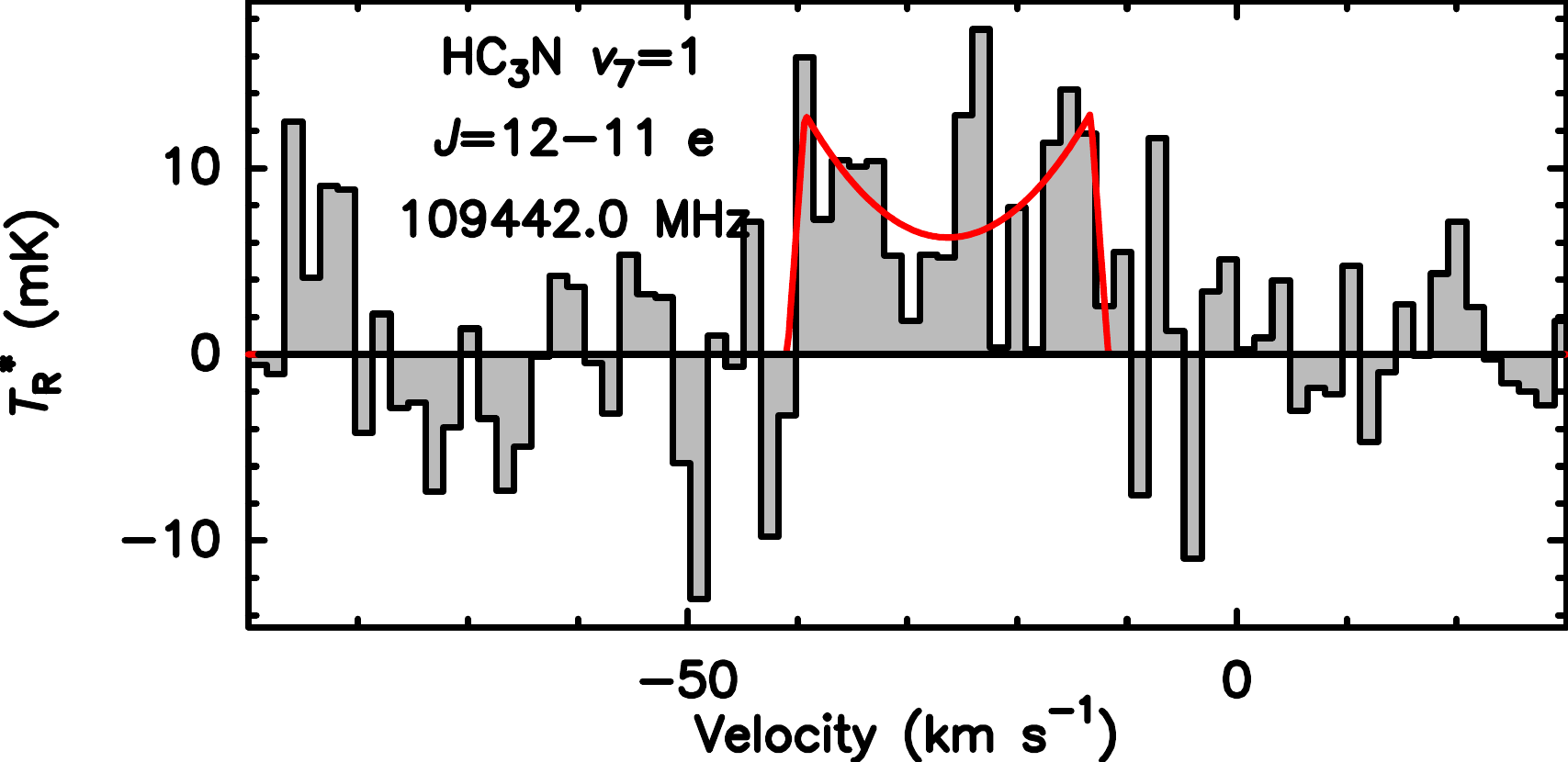}
\hspace{0.05\textwidth}
\includegraphics[width = 0.45 \textwidth]{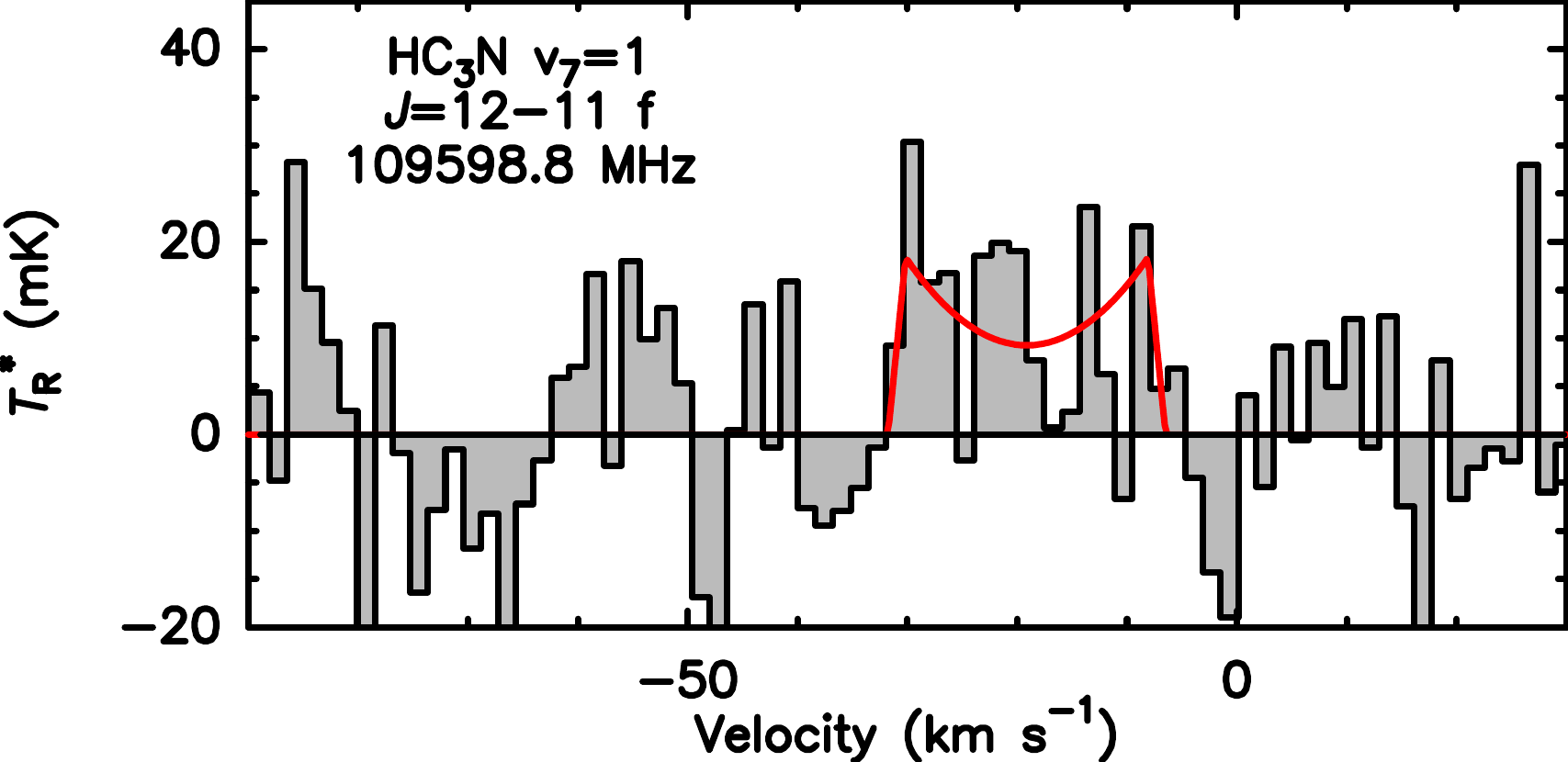}
\vspace{0.1cm}
\includegraphics[width = 0.45 \textwidth]{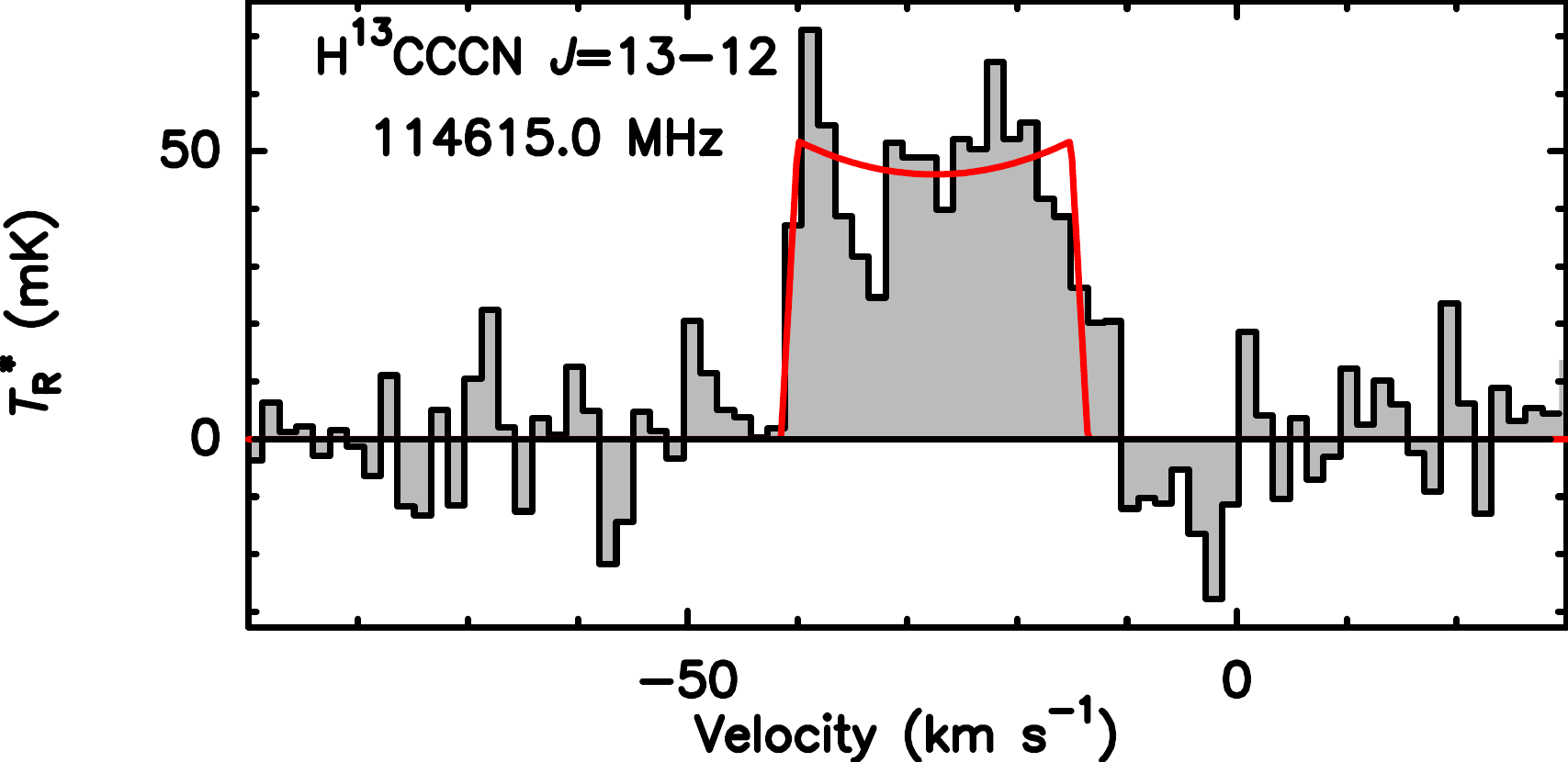}
\hspace{0.05\textwidth}
\includegraphics[width = 0.45 \textwidth]{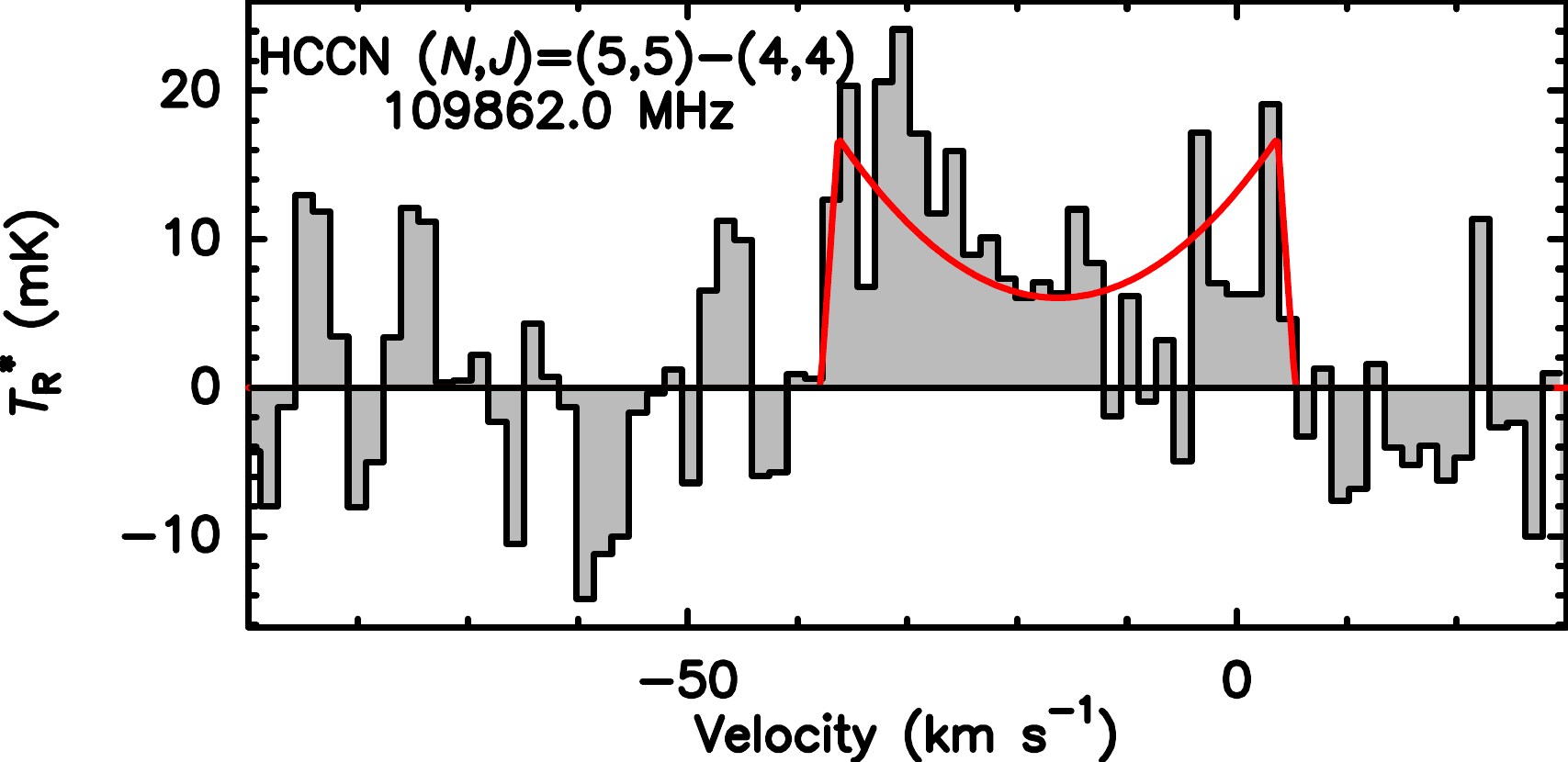}
\centerline{Figure \ref{Fig:fitting_21}. --- continued}
\end{figure*}

\begin{figure*}[!htbp]
\centering
\includegraphics[width = 0.45 \textwidth]{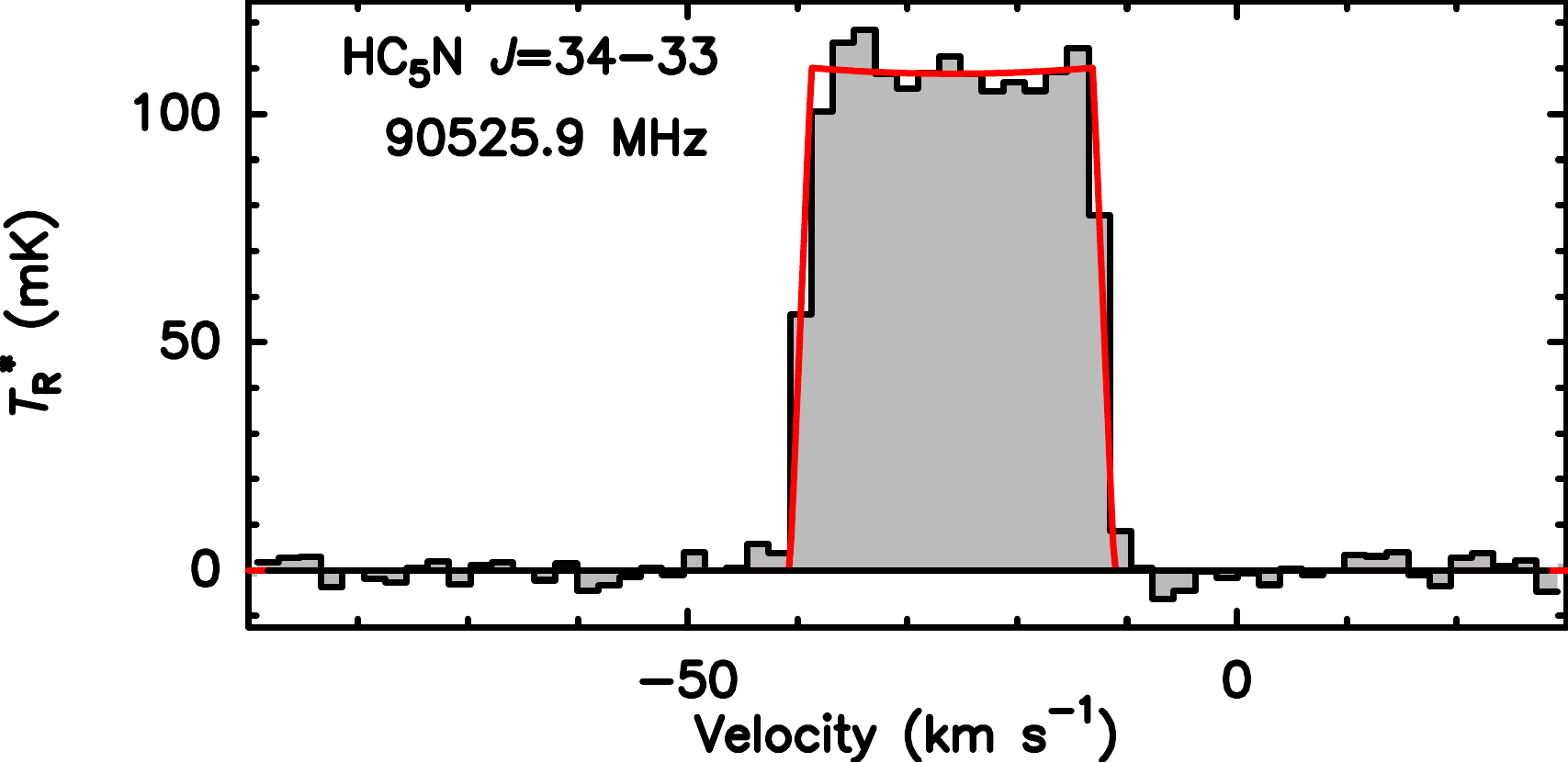}
\hspace{0.05\textwidth}
\includegraphics[width = 0.45 \textwidth]{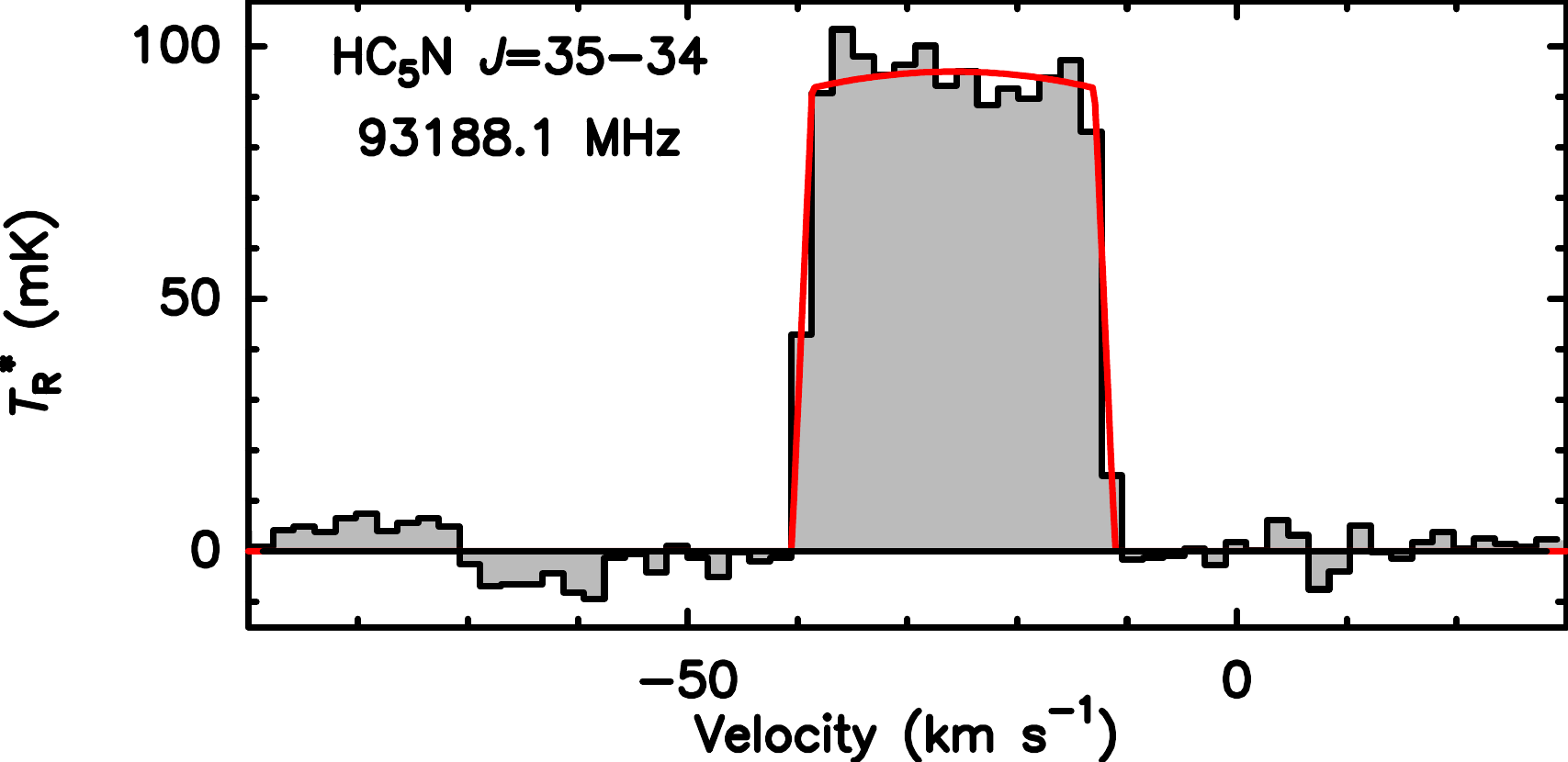}
\vspace{0.1cm}
\includegraphics[width = 0.45 \textwidth]{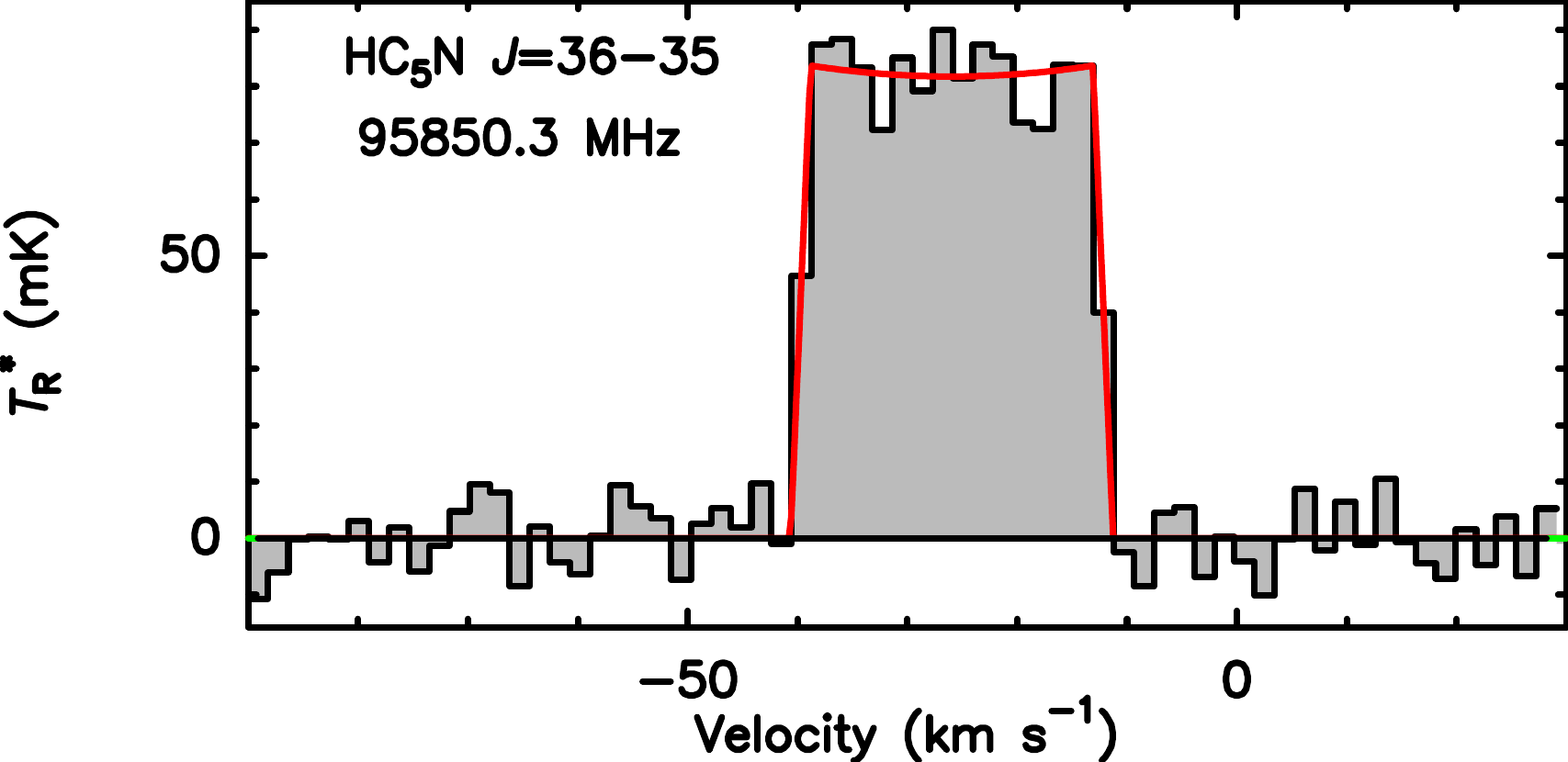}
\hspace{0.05\textwidth}
\includegraphics[width = 0.45 \textwidth]{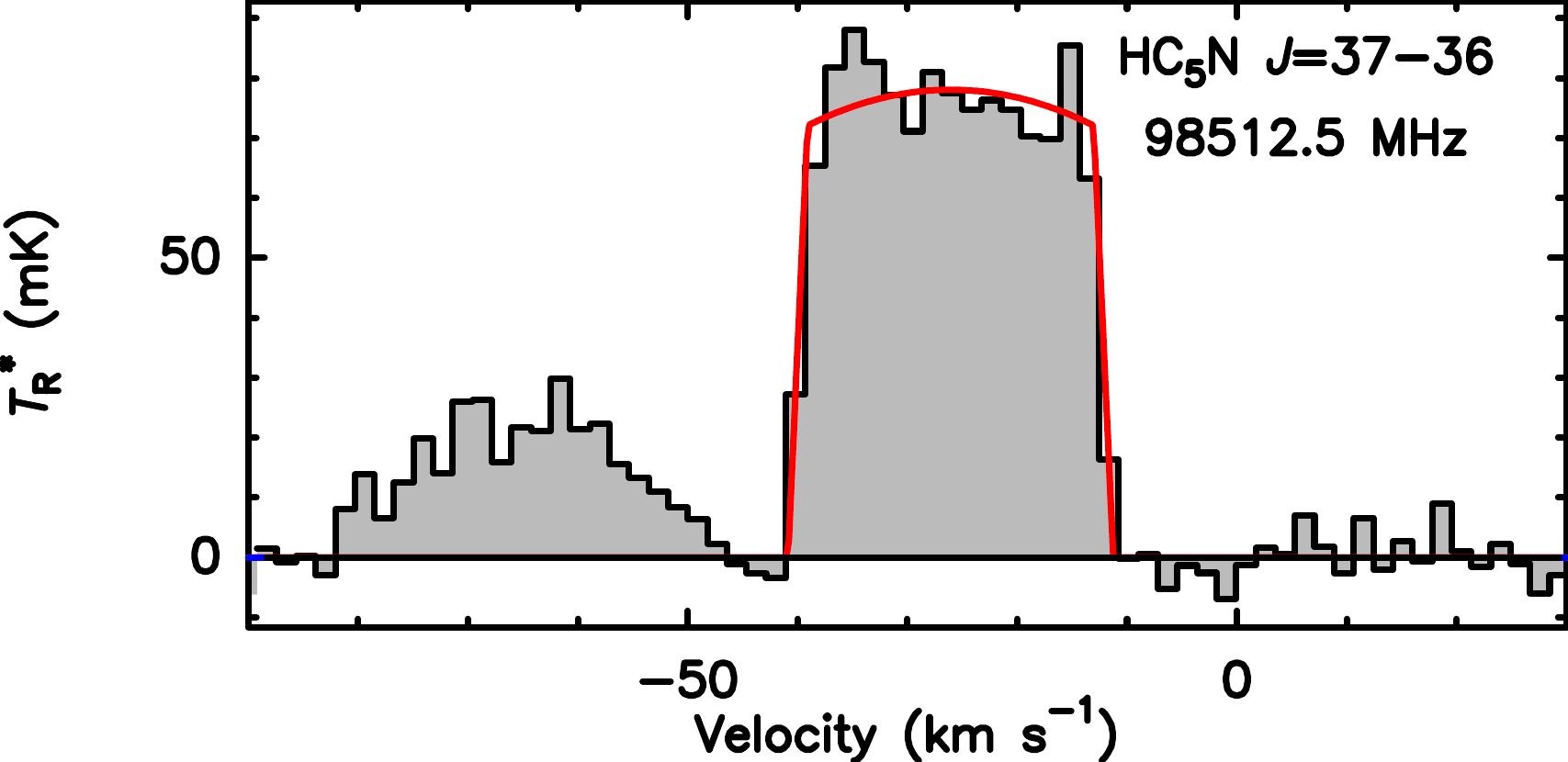}
\vspace{0.1cm}
\includegraphics[width = 0.45 \textwidth]{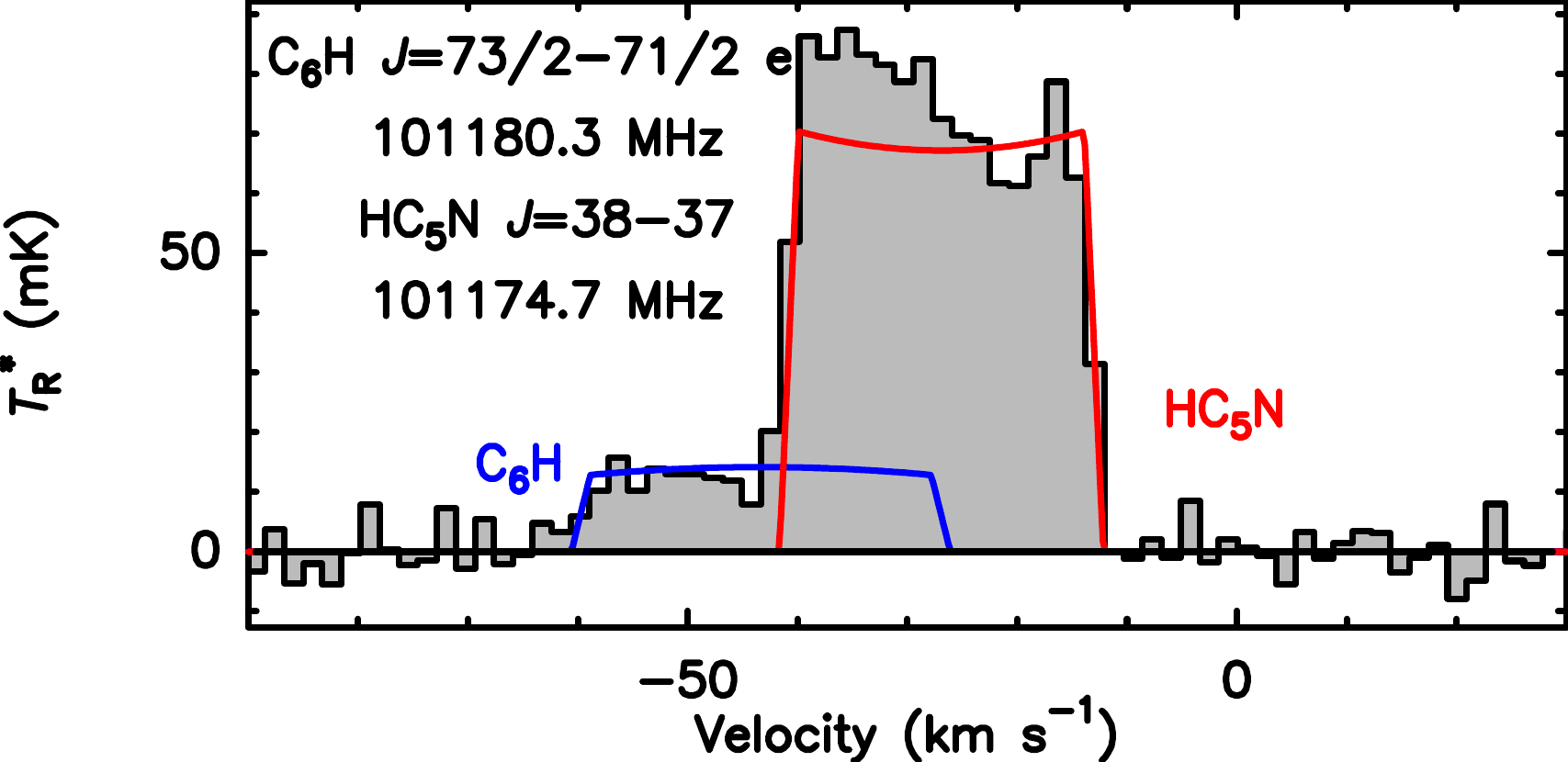}
\hspace{0.05\textwidth}
\includegraphics[width = 0.45 \textwidth]{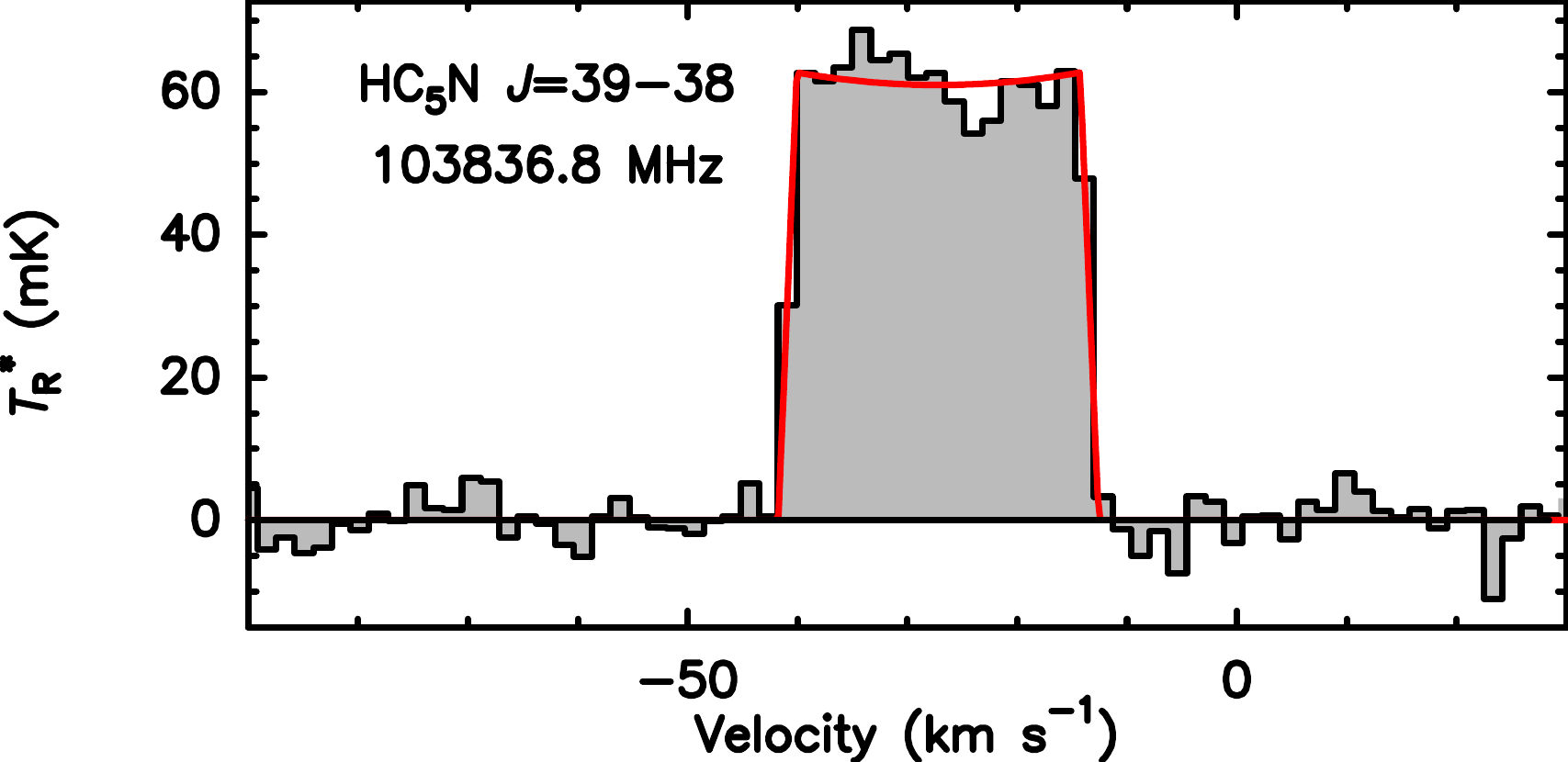}
\vspace{0.1cm}
\includegraphics[width = 0.45 \textwidth]{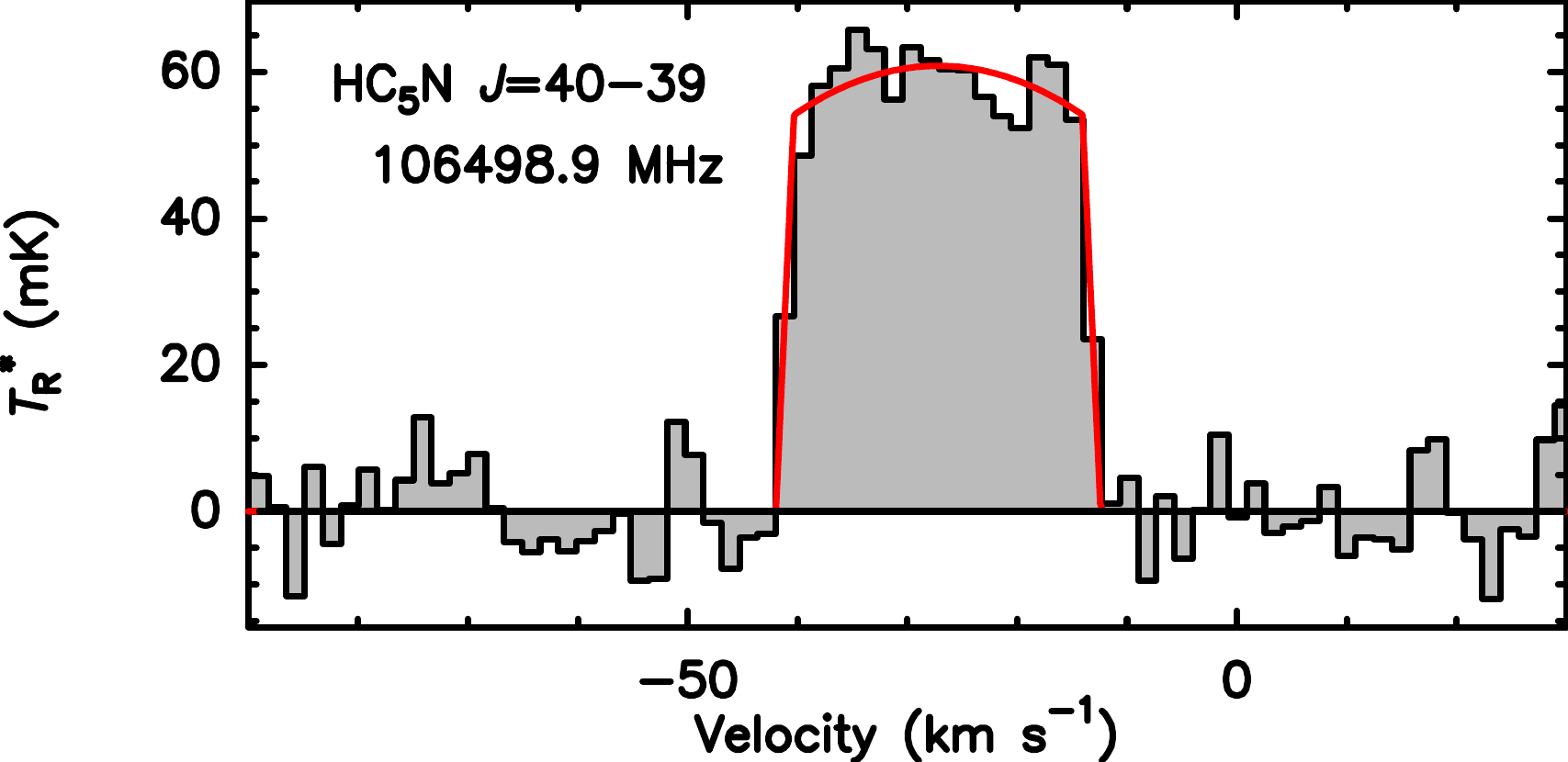}
\hspace{0.05\textwidth}
\includegraphics[width = 0.45 \textwidth]{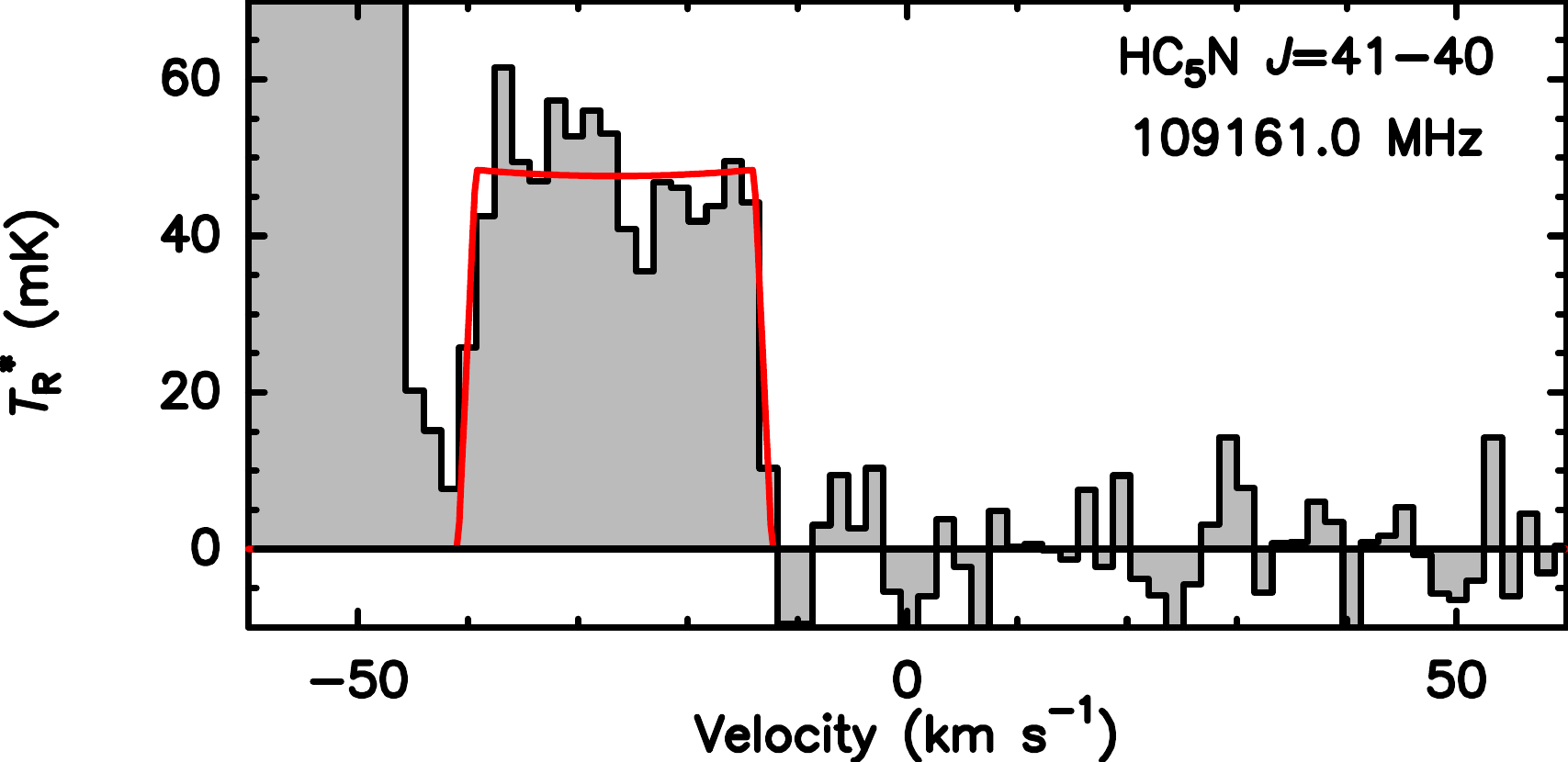}
\vspace{0.1cm}
\includegraphics[width = 0.45 \textwidth]{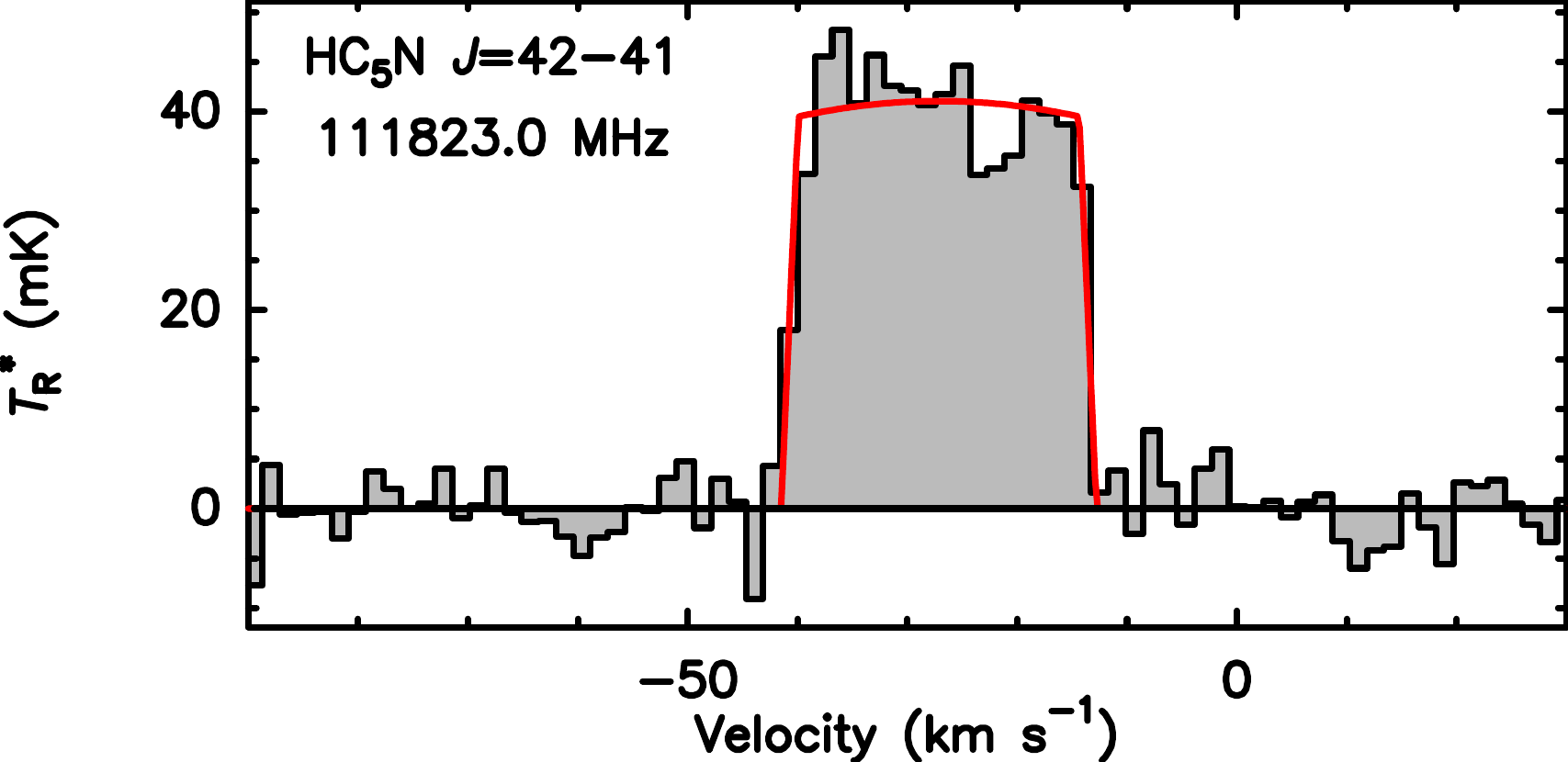}
\hspace{0.05\textwidth}
\includegraphics[width = 0.45 \textwidth]{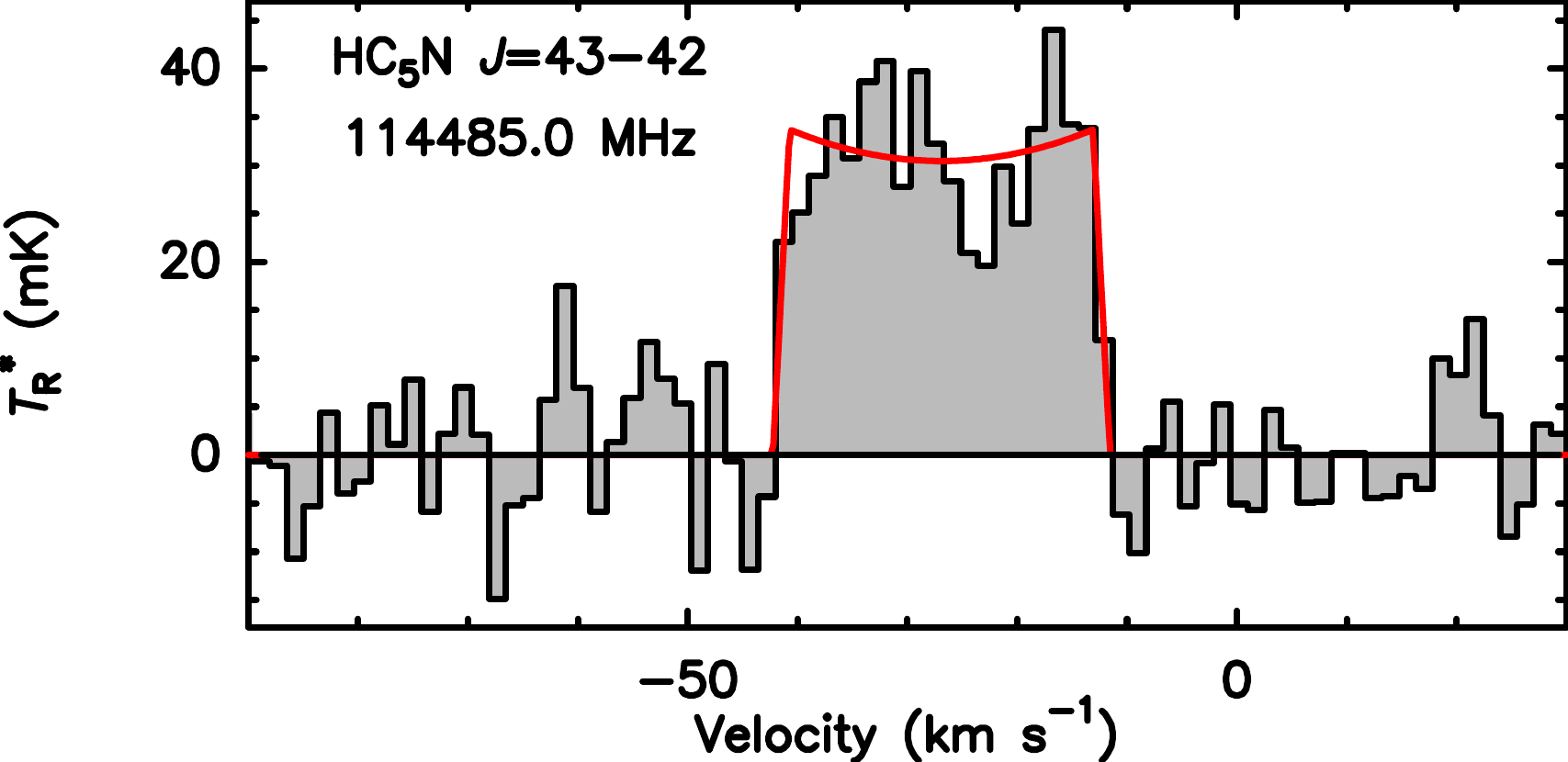}
\caption{{Same as Figure.~\ref{Fig:fitting_1}, but for HC$_{5}$N.}\label{Fig:fitting_22}}
\end{figure*}

\begin{figure*}[!htbp]
\centering
\includegraphics[width = 0.45 \textwidth]{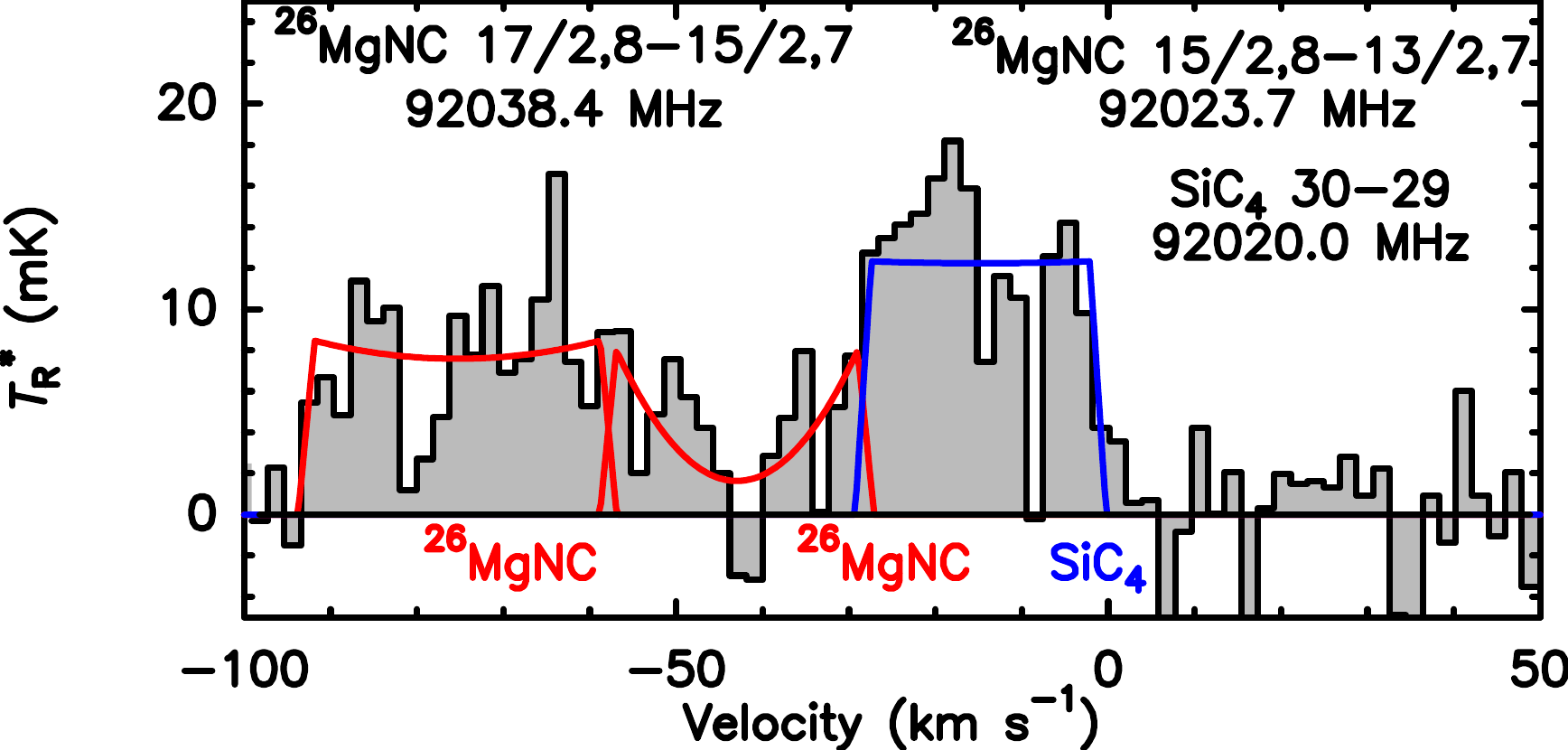}
\hspace{0.05\textwidth}
\includegraphics[width = 0.45 \textwidth]{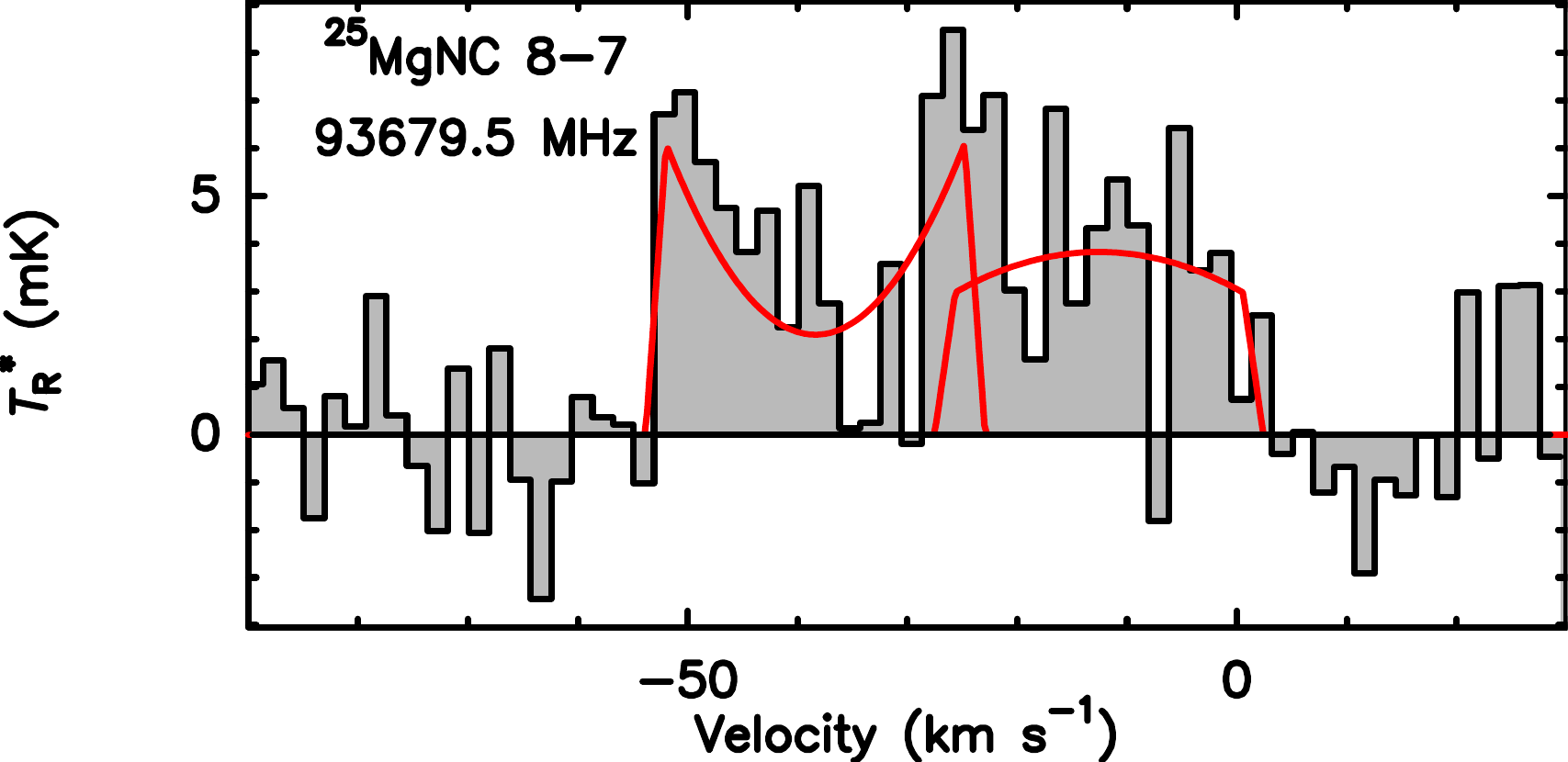}
\vspace{0.1cm}
\includegraphics[width = 0.45 \textwidth]{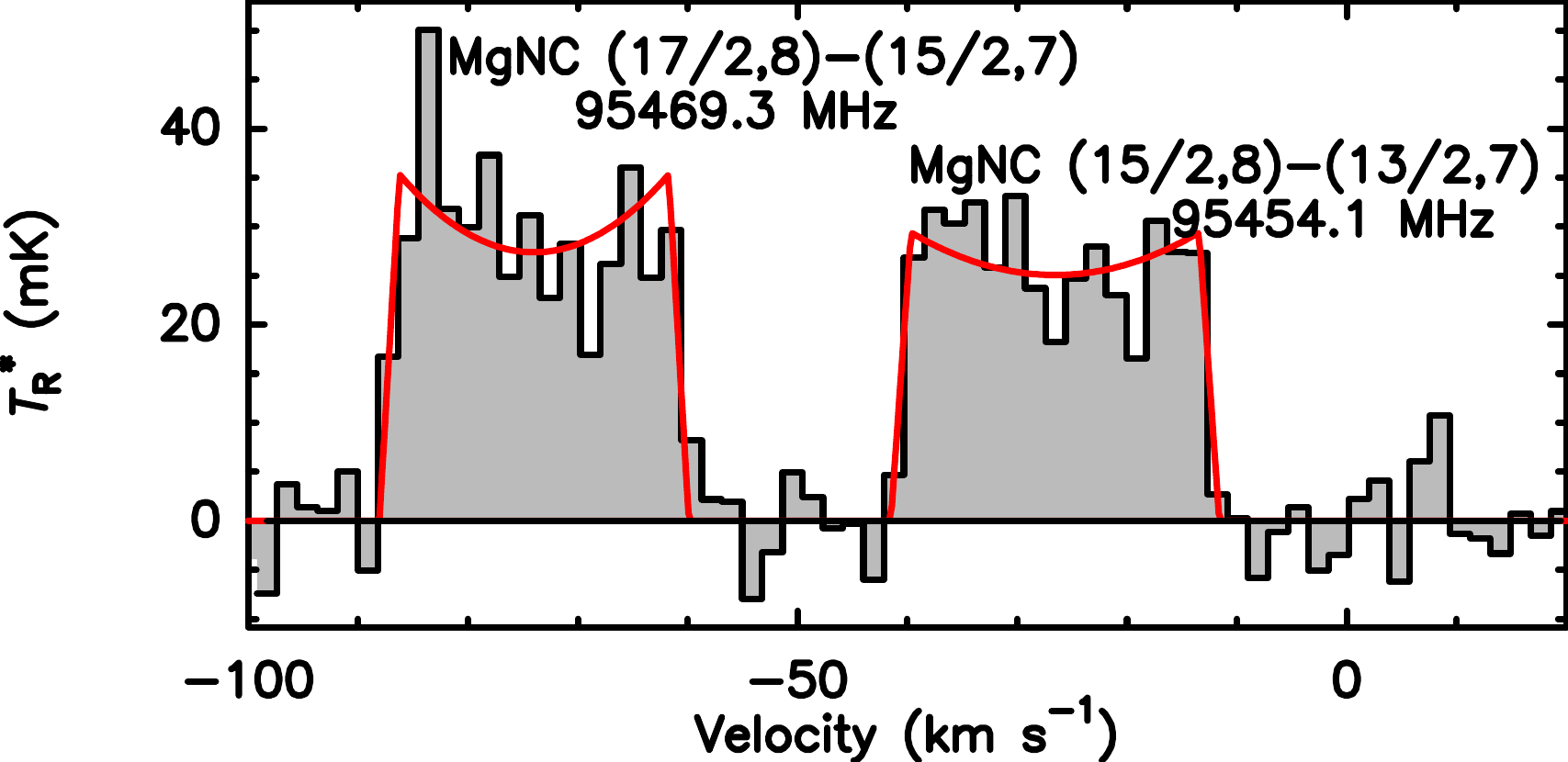}
\hspace{0.05\textwidth}
\includegraphics[width = 0.45 \textwidth]{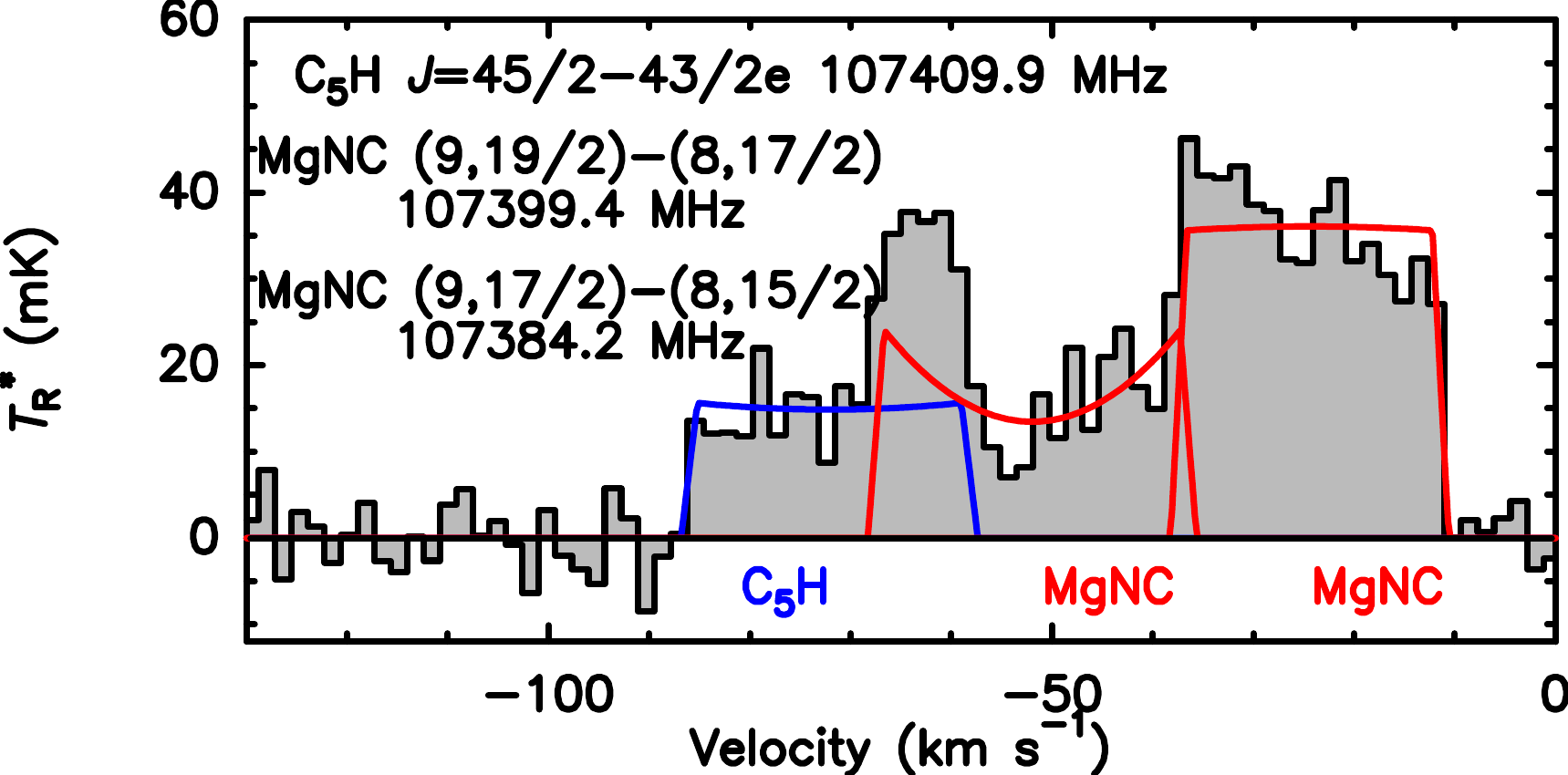}
\caption{{Same as Figure.~\ref{Fig:fitting_1}, but for MgNC and its isotopologues. }\label{Fig:fitting_23}}
\end{figure*}

\begin{figure*}[!htbp]
\centering
\includegraphics[width = 0.45 \textwidth]{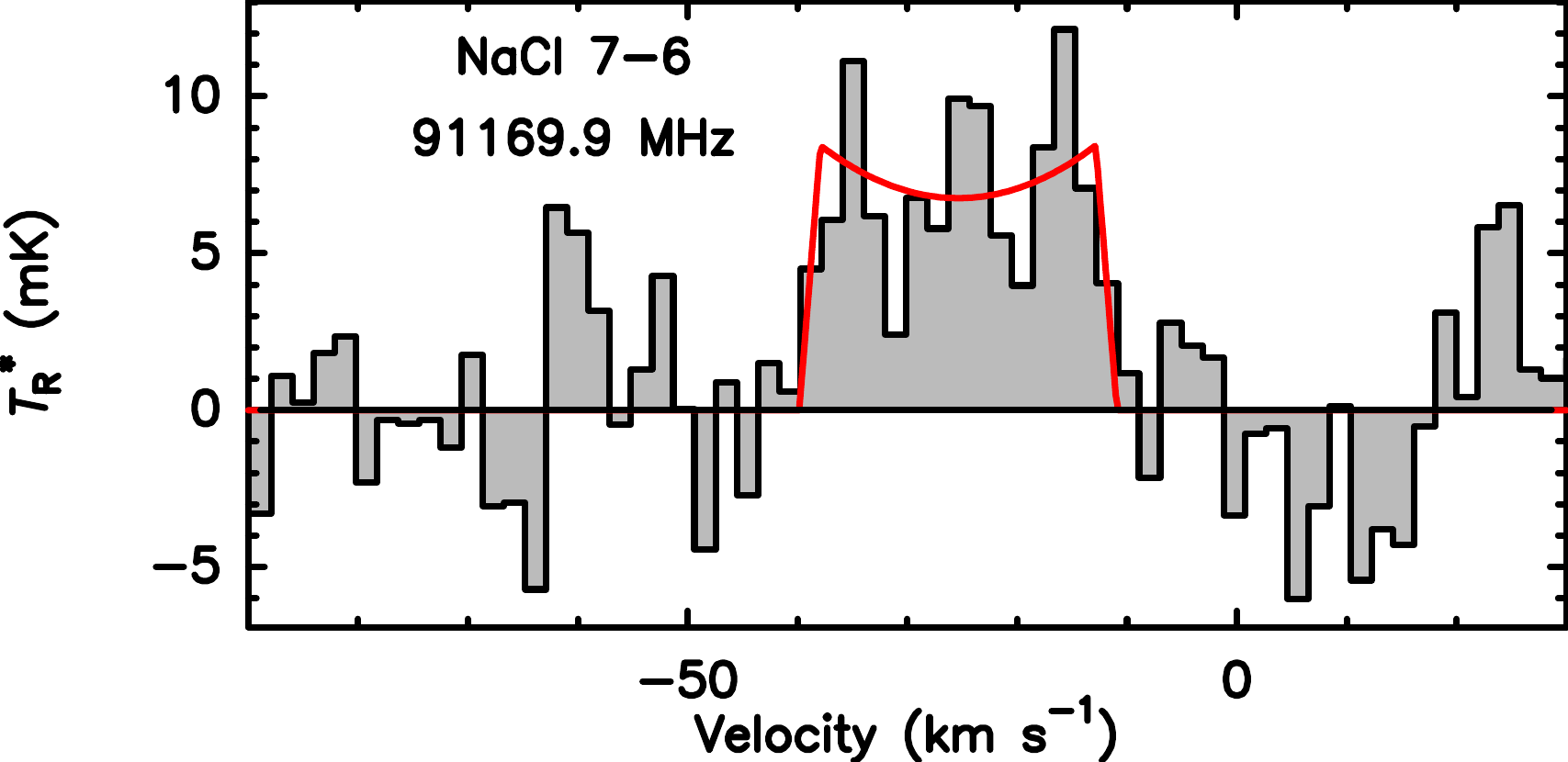}
\hspace{0.05\textwidth}
\includegraphics[width = 0.45 \textwidth]{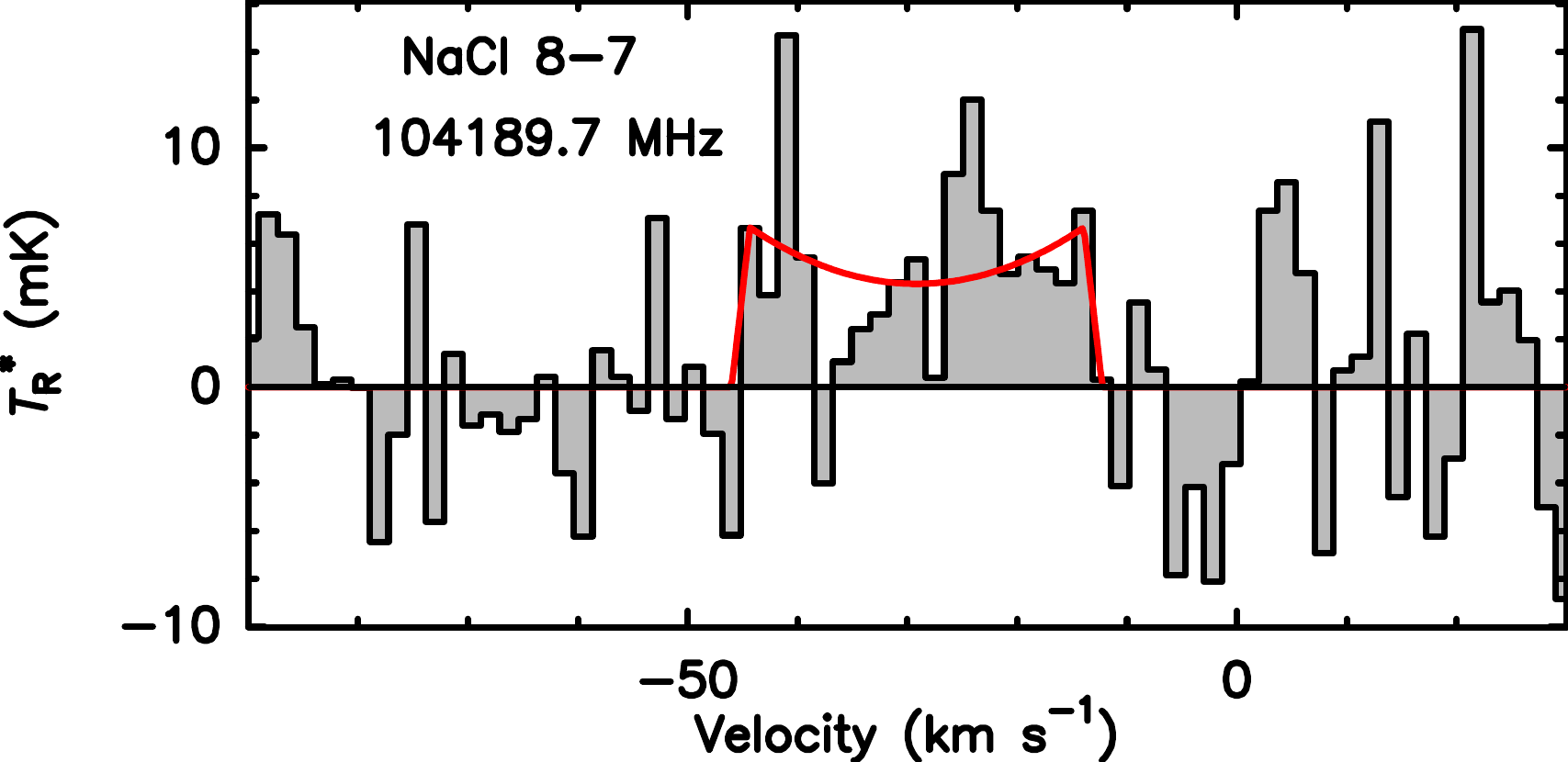}
\caption{{Same as Figure.~\ref{Fig:fitting_1}, but for NaCl. }\label{Fig:fitting_25}}
\end{figure*}

\begin{figure*}[!htbp]
\centering
\includegraphics[width = 0.45 \textwidth]{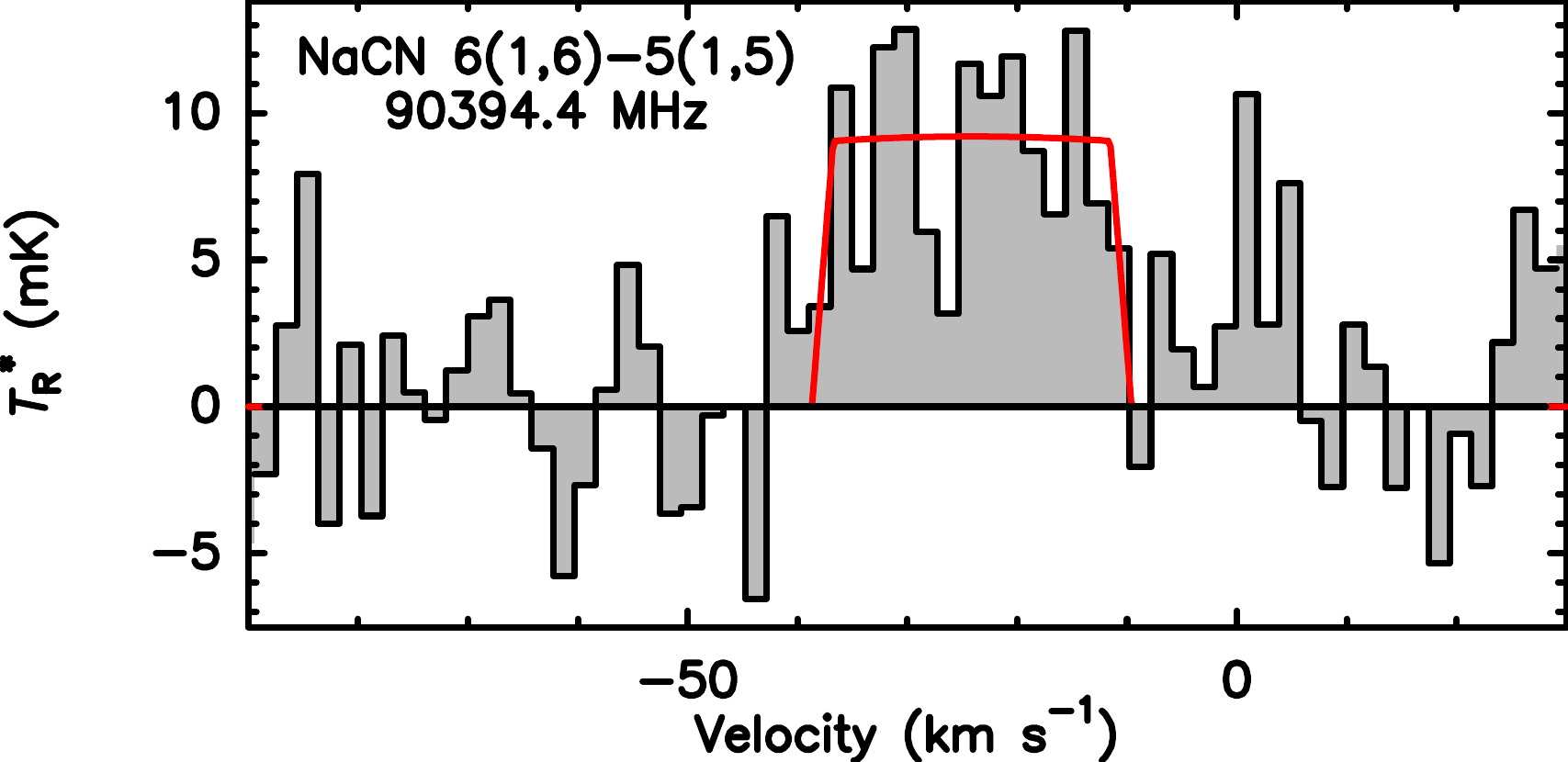}
\hspace{0.05\textwidth}
\includegraphics[width = 0.45 \textwidth]{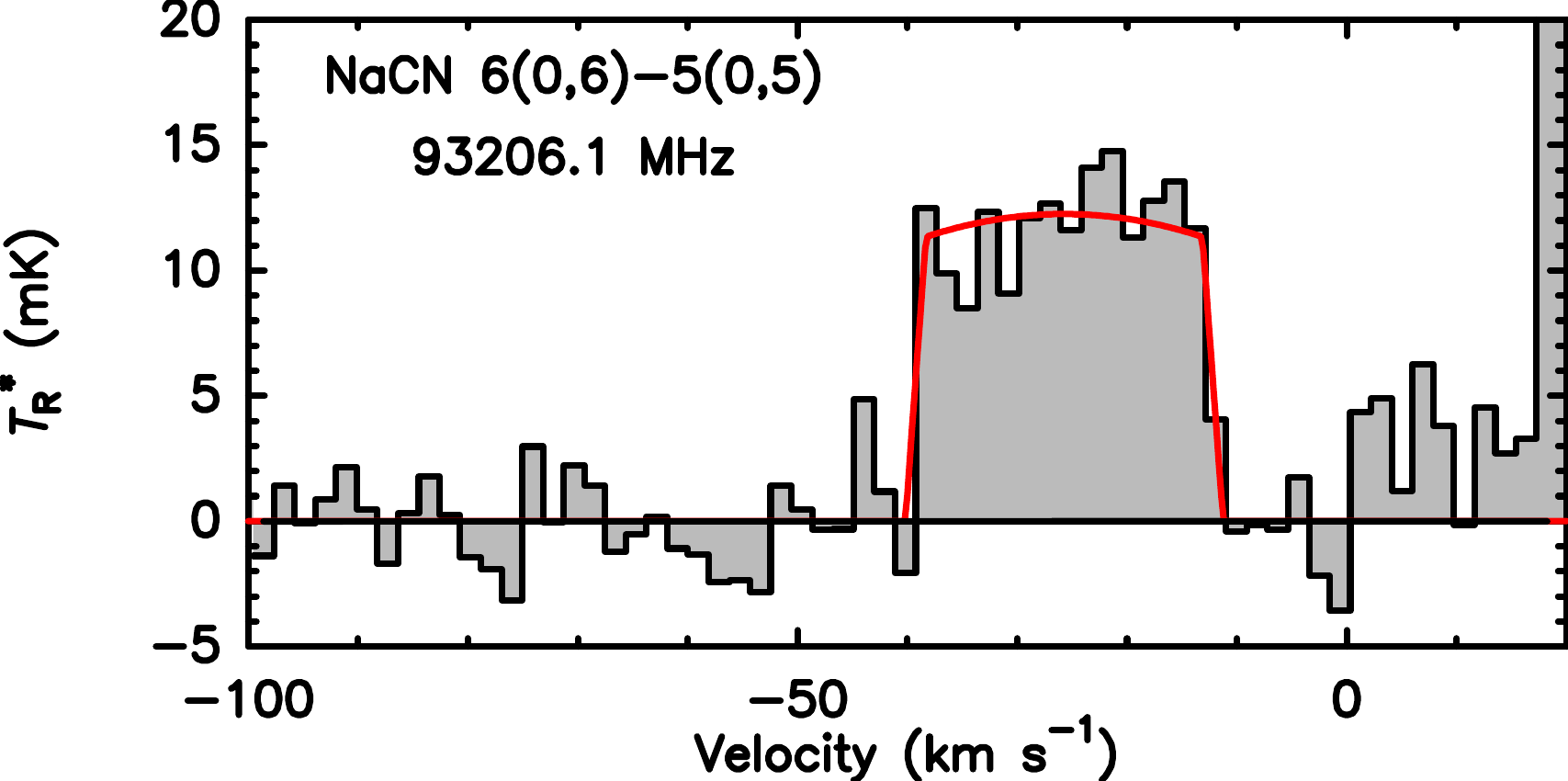}
\vspace{0.1cm}
\includegraphics[width = 0.45 \textwidth]{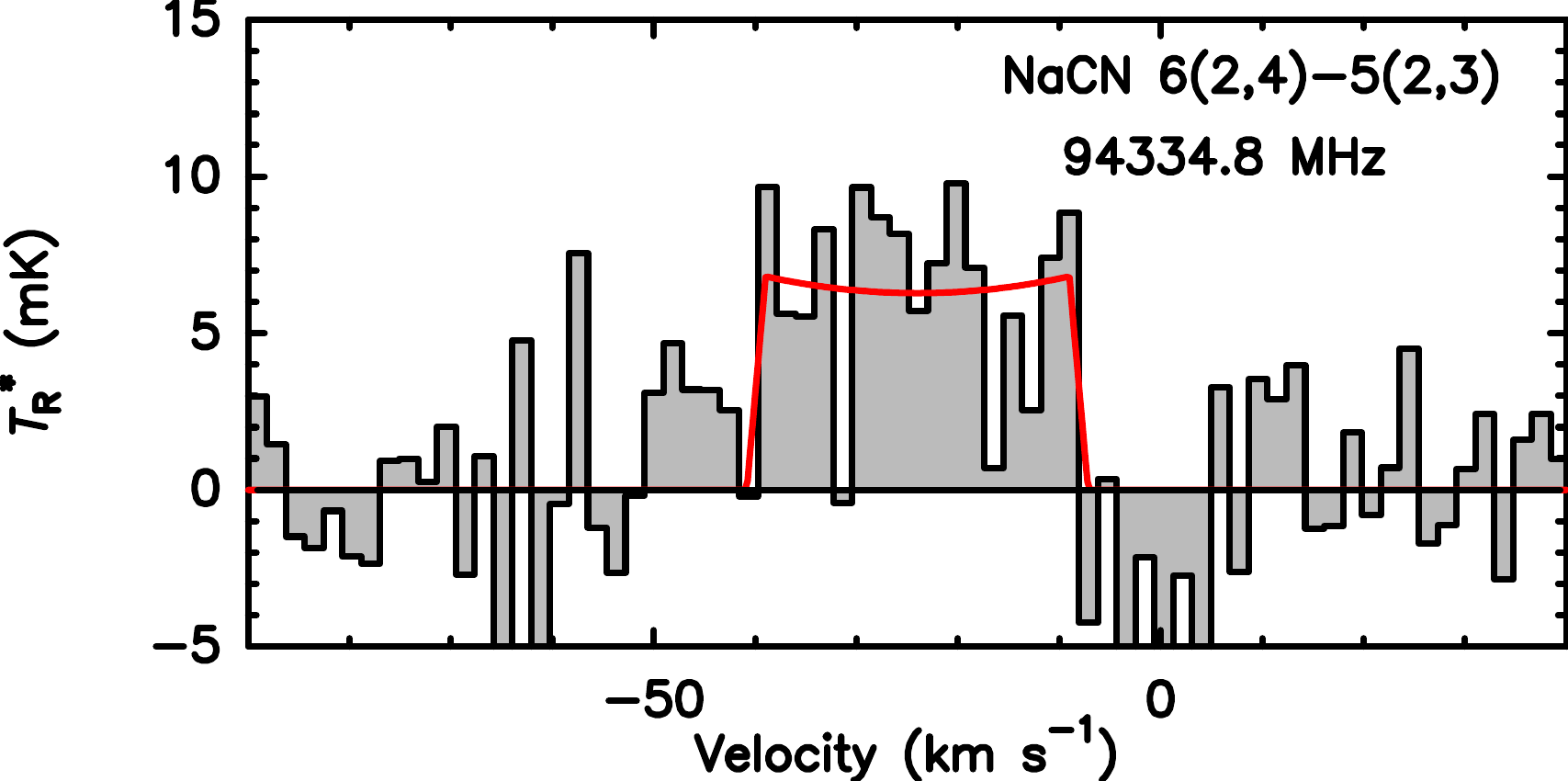}
\hspace{0.05\textwidth}
\includegraphics[width = 0.45 \textwidth]{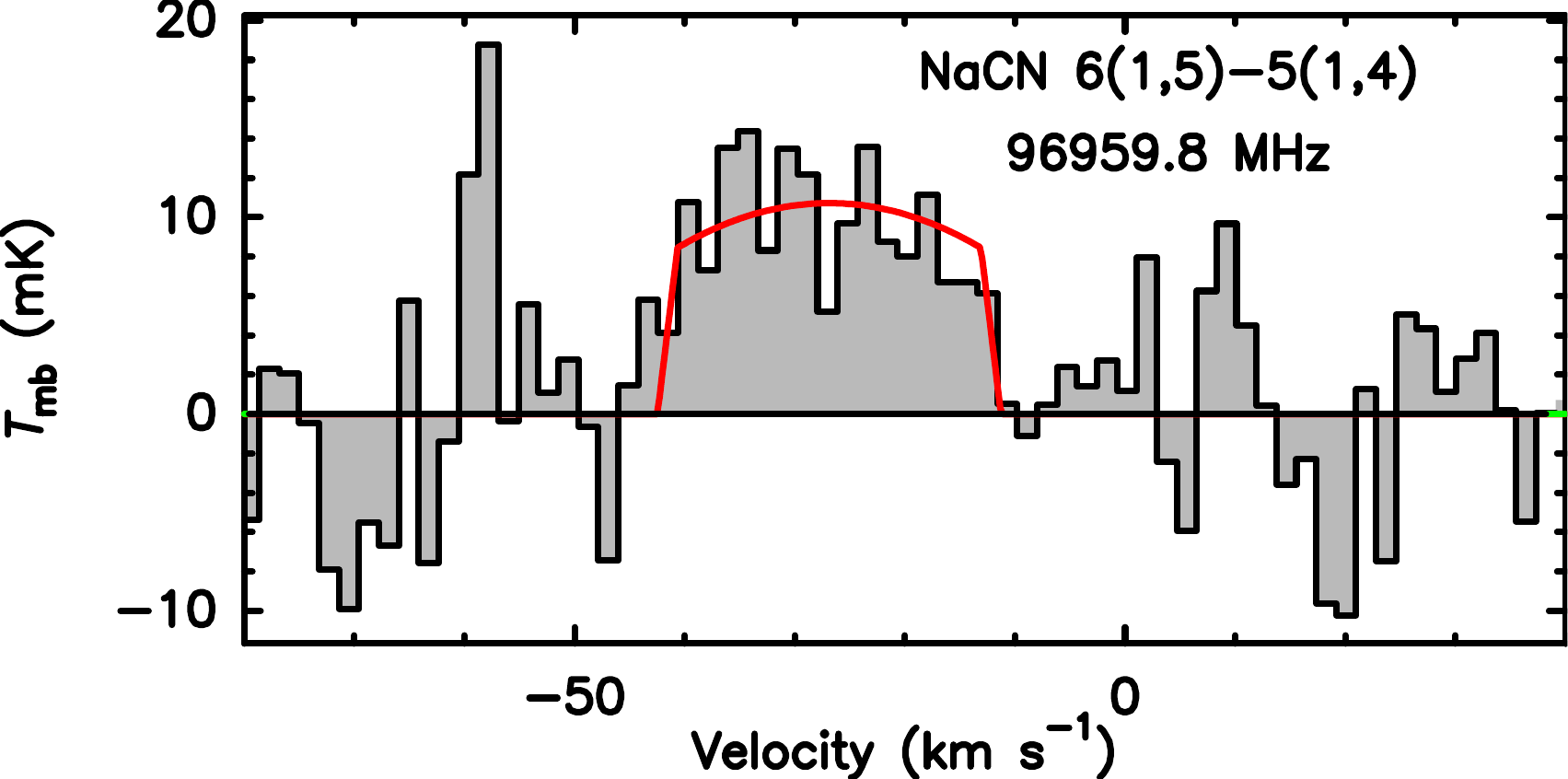}
\vspace{0.1cm}
\includegraphics[width = 0.45 \textwidth]{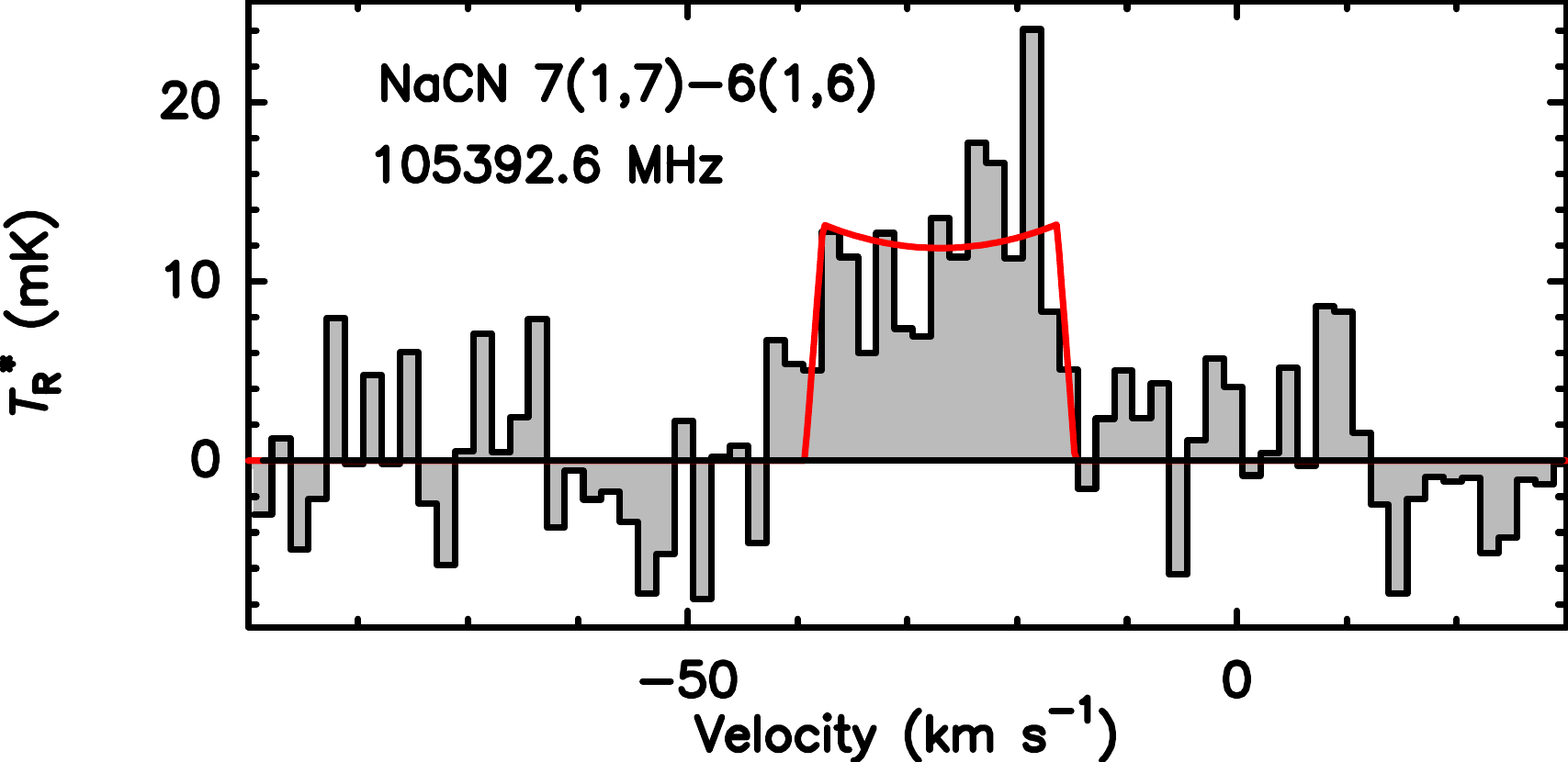}
\hspace{0.05\textwidth}
\includegraphics[width = 0.45 \textwidth]{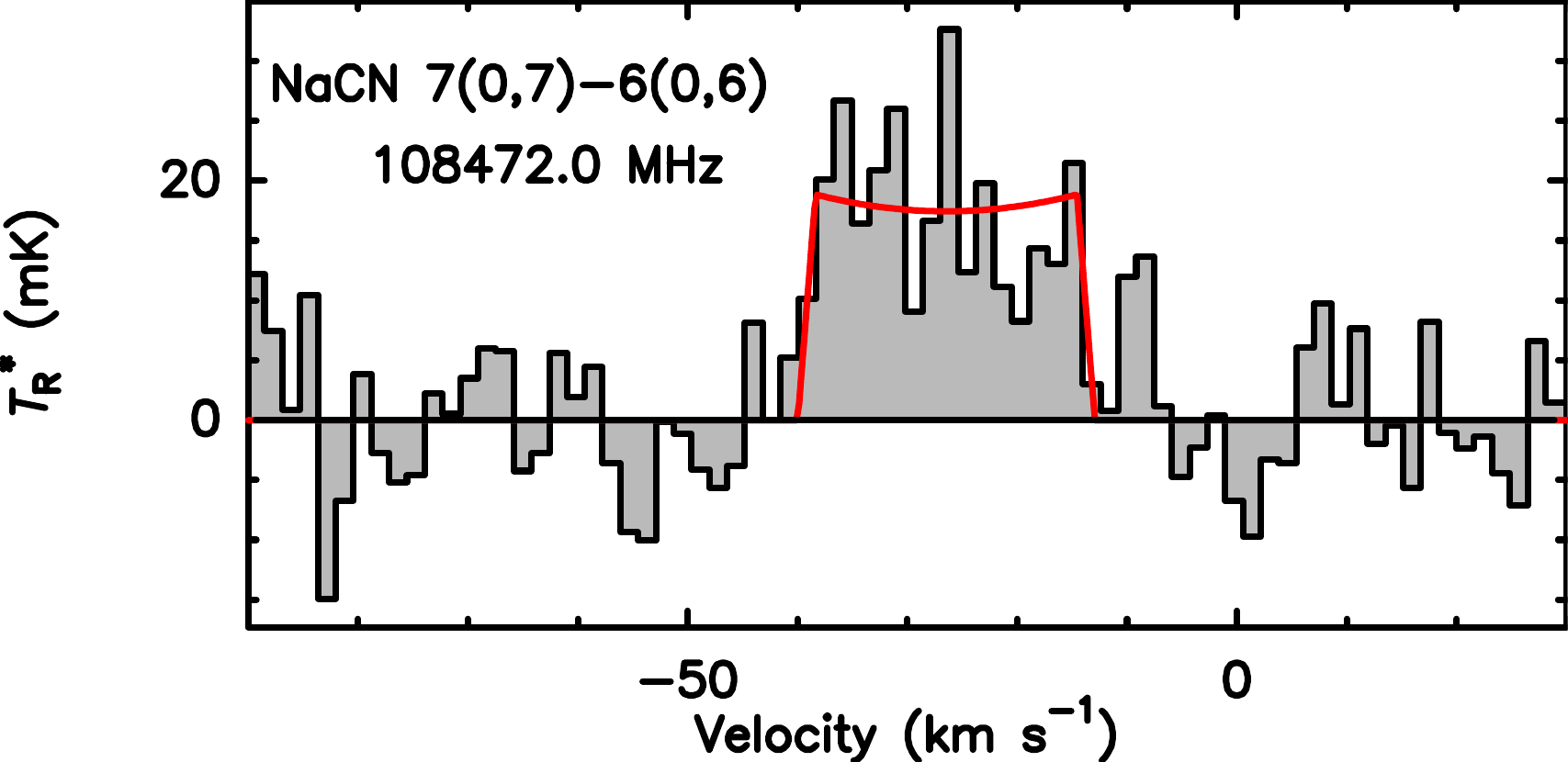}
\vspace{0.1cm}
\includegraphics[width = 0.45 \textwidth]{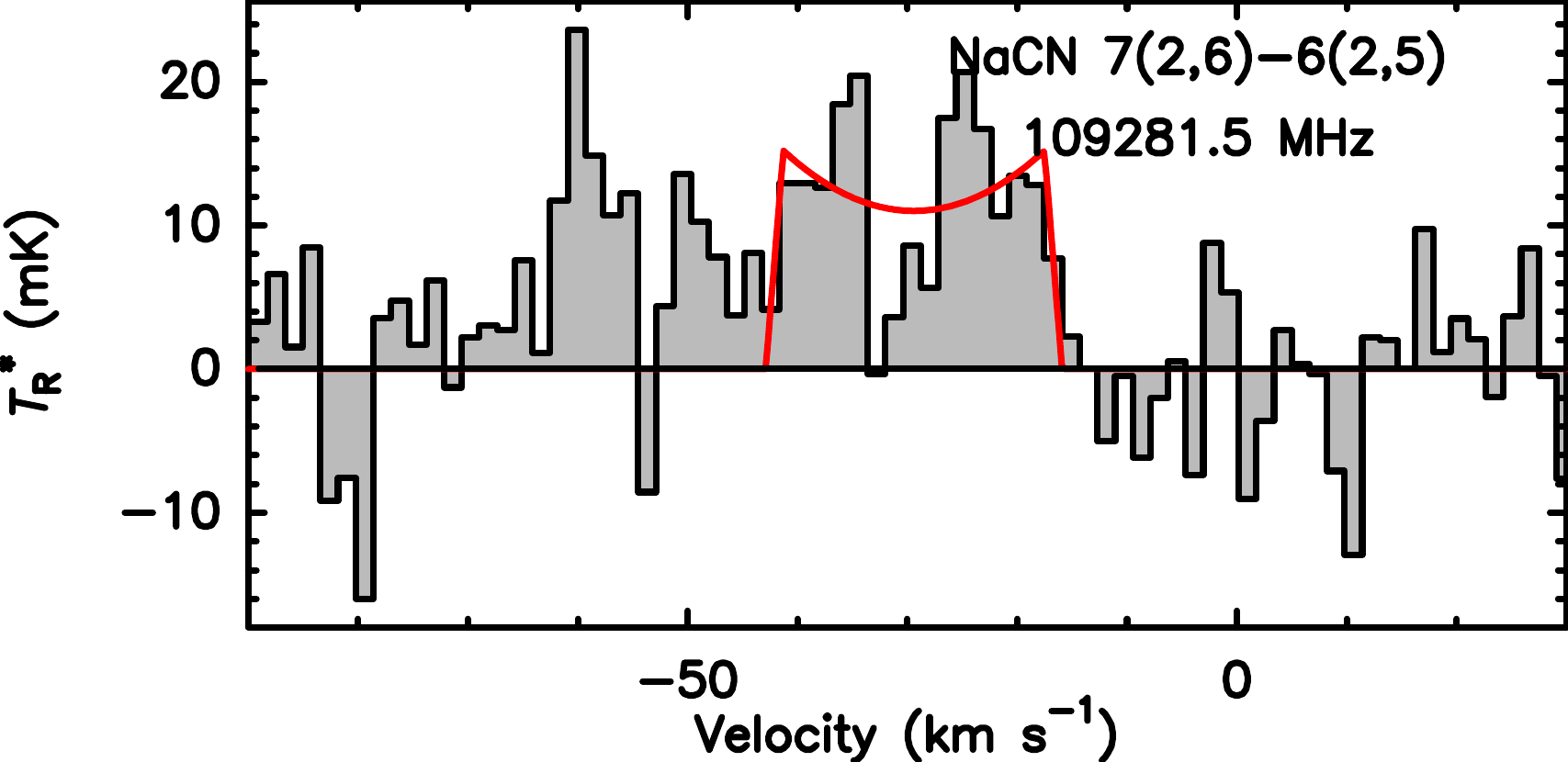}
\hspace{0.05\textwidth}
\includegraphics[width = 0.45 \textwidth]{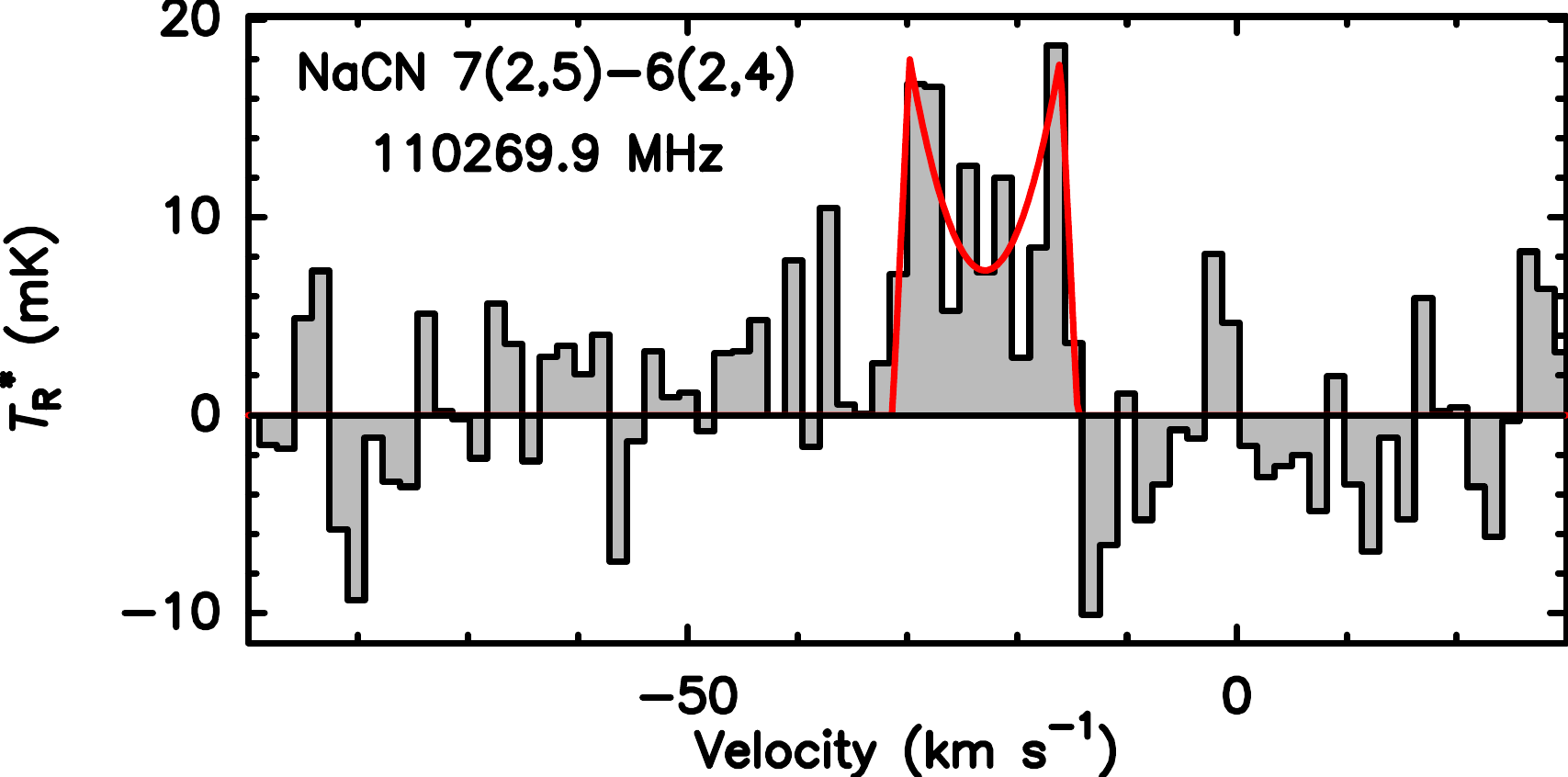}
\caption{{Same as Figure.~\ref{Fig:fitting_1}, but for NaCN.}\label{Fig:fitting_26}}
\end{figure*}

\begin{figure*}[!htbp]
\centering
\includegraphics[width = 0.45 \textwidth]{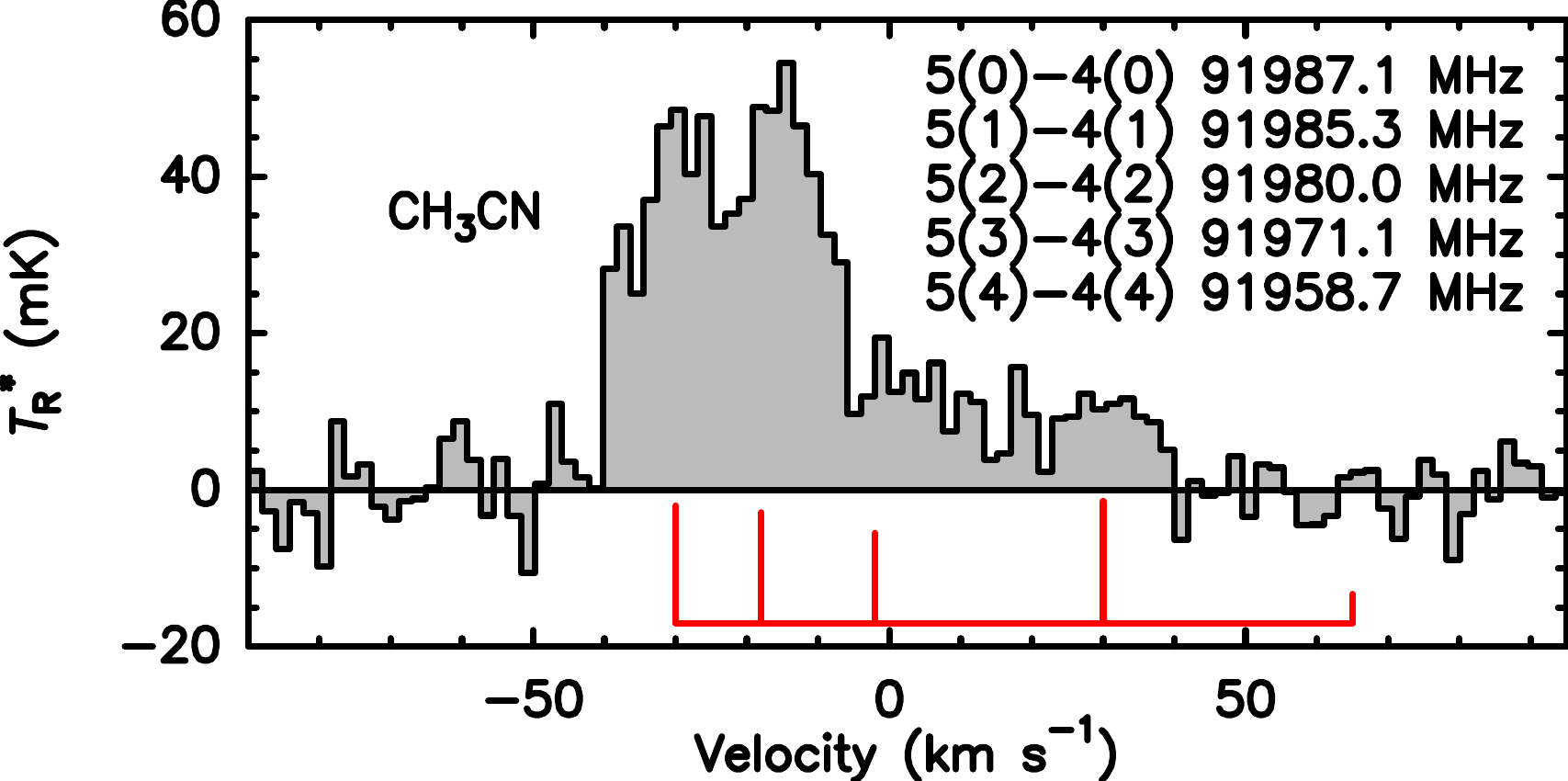}
\hspace{0.05\textwidth}
\includegraphics[width = 0.45 \textwidth]{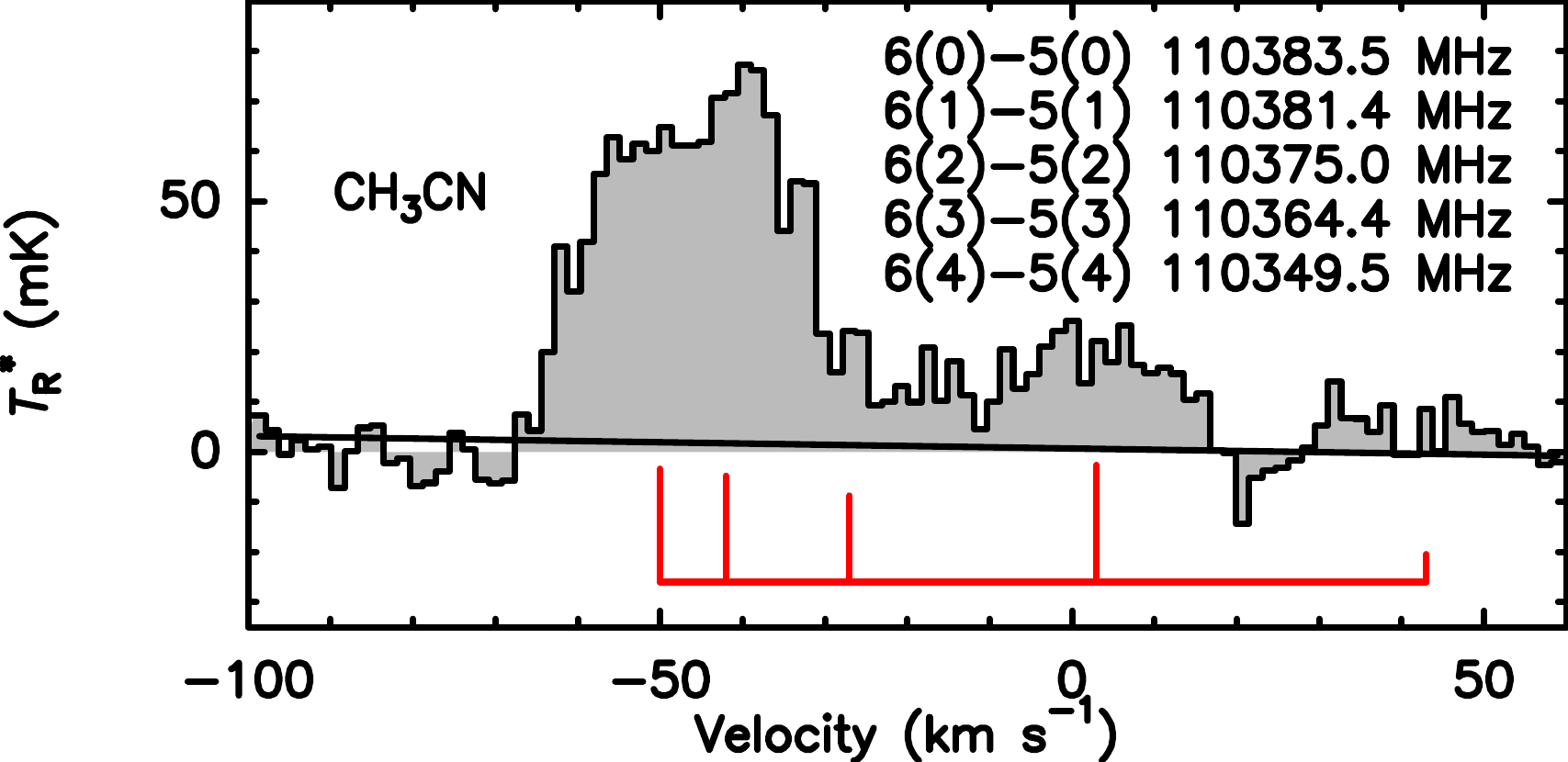}
\caption{{Same as Figure.~\ref{Fig:fitting_1}, but for CH$_{3}$CN.}\label{Fig:fitting_27}}
\end{figure*}

\begin{figure*}[!htbp]
\section{Rotation diagrams}
\centering
\includegraphics[width = 0.47 \textwidth]{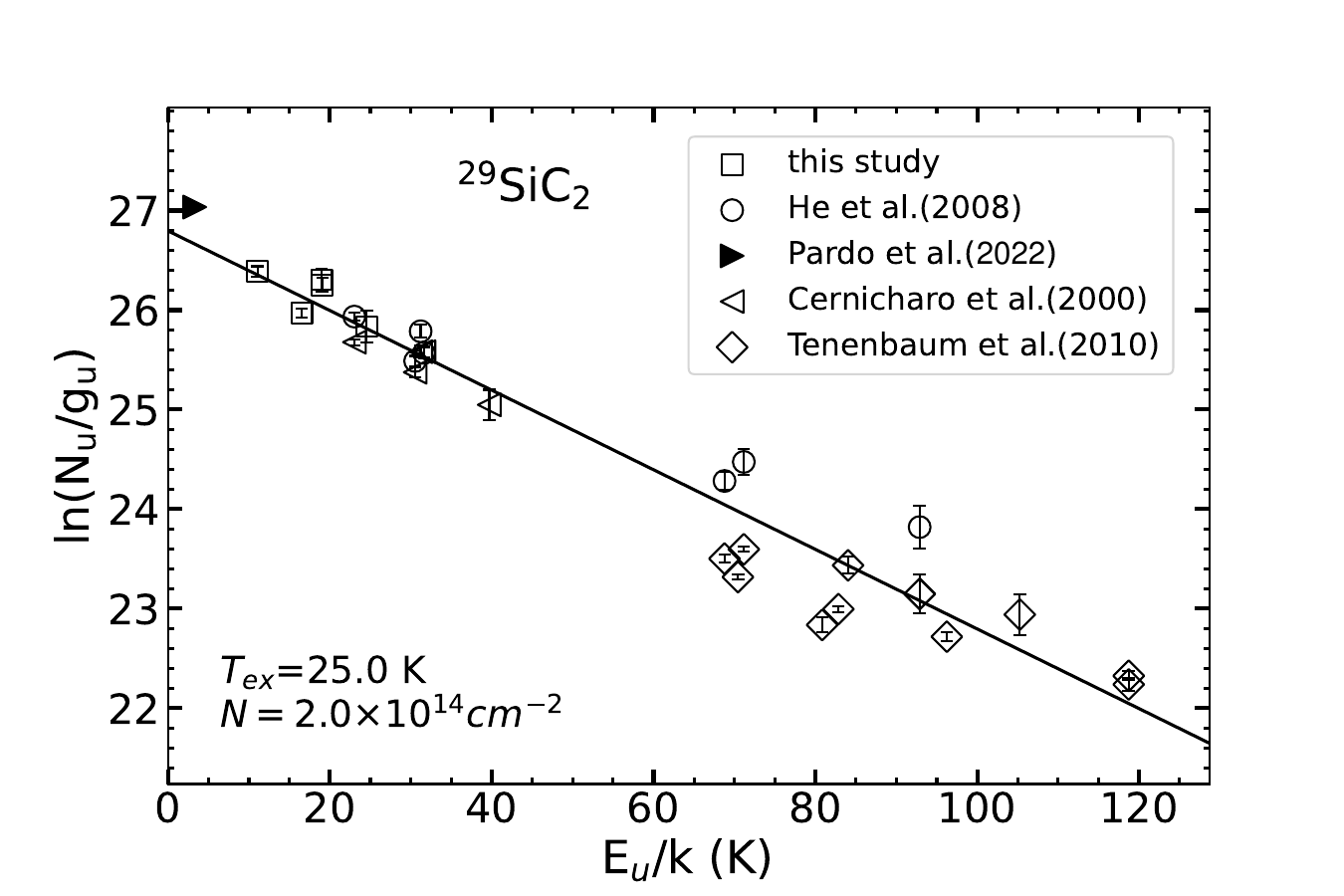}
\includegraphics[width = 0.47 \textwidth]{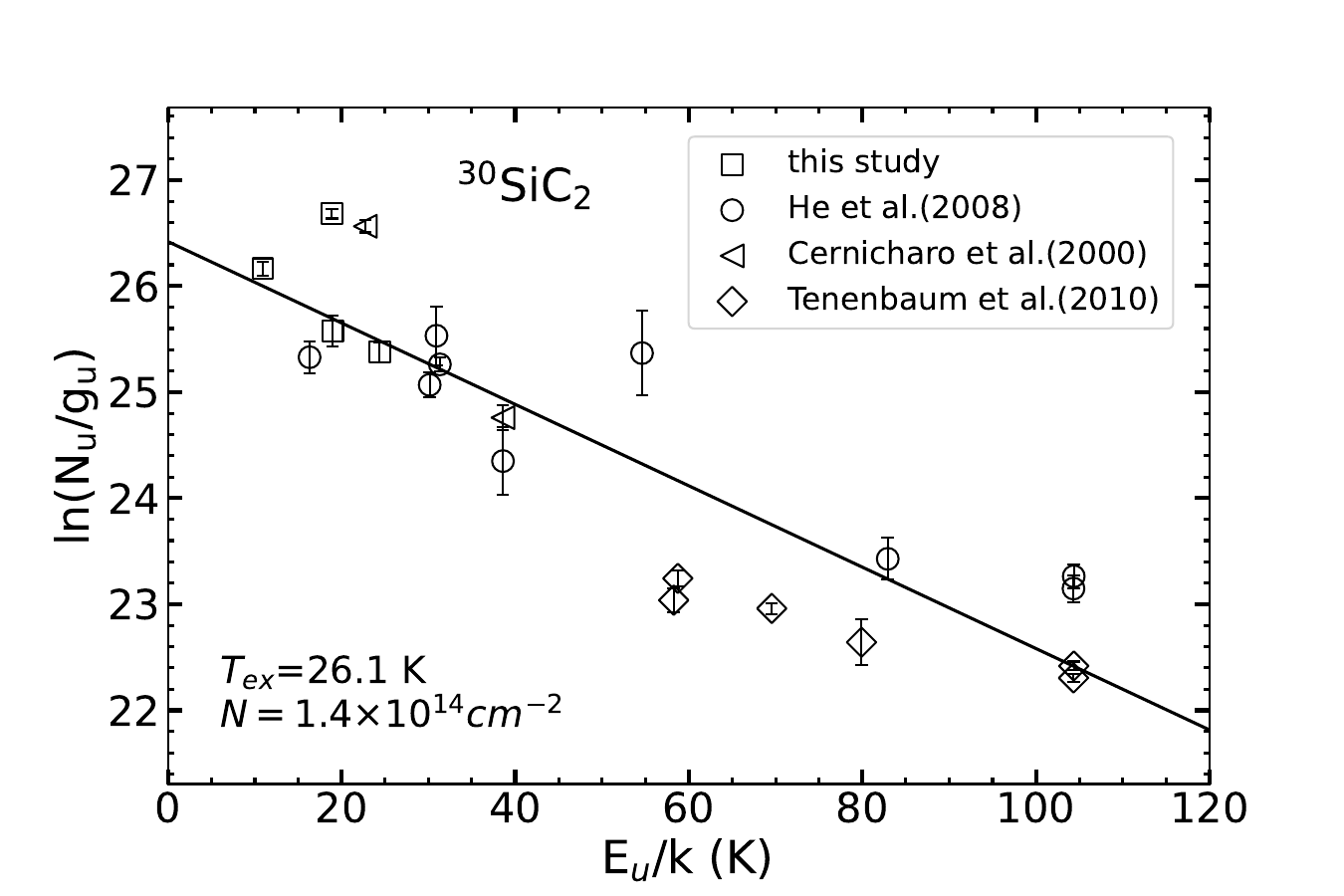}
\includegraphics[width = 0.47 \textwidth]{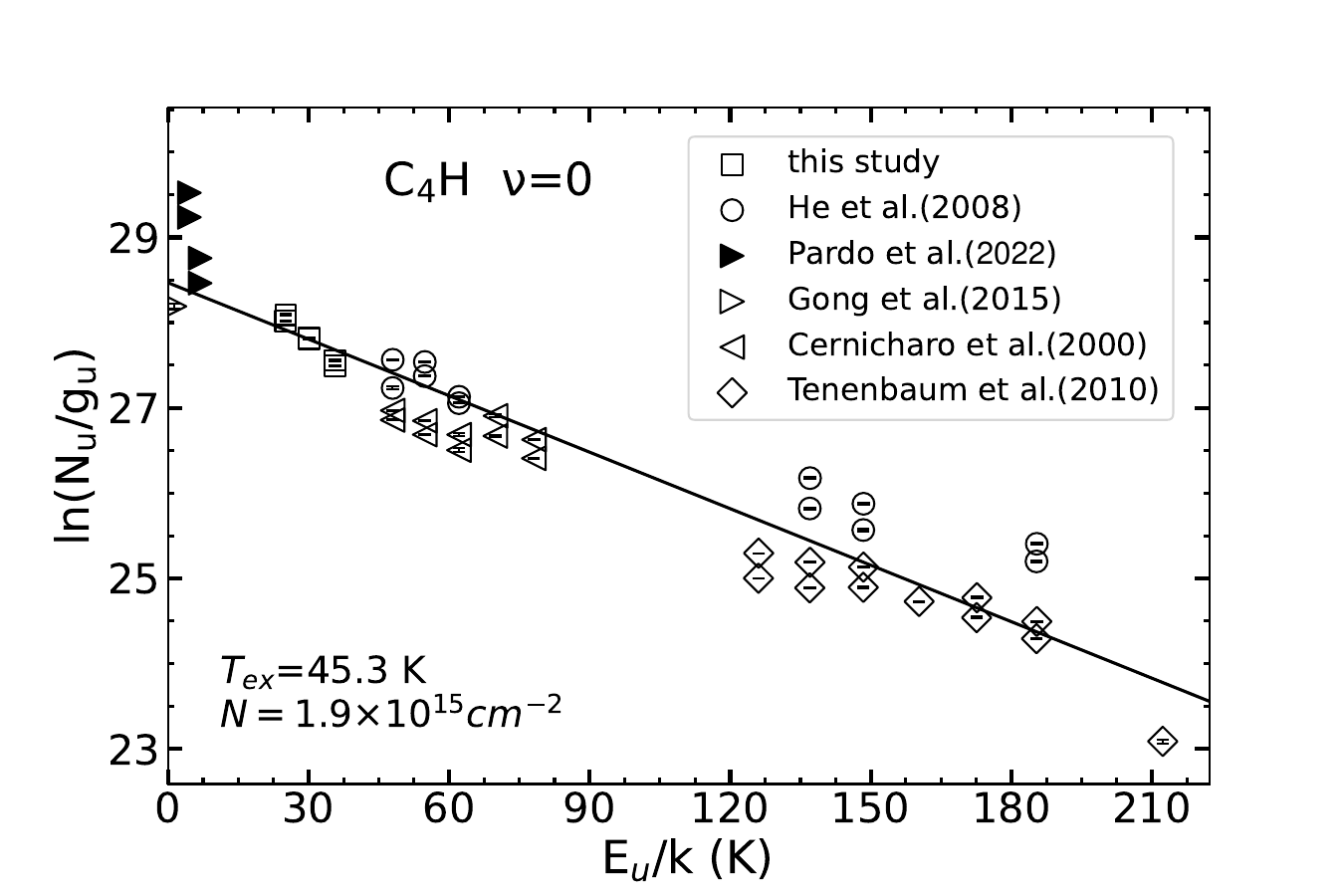}
\includegraphics[width = 0.47 \textwidth]{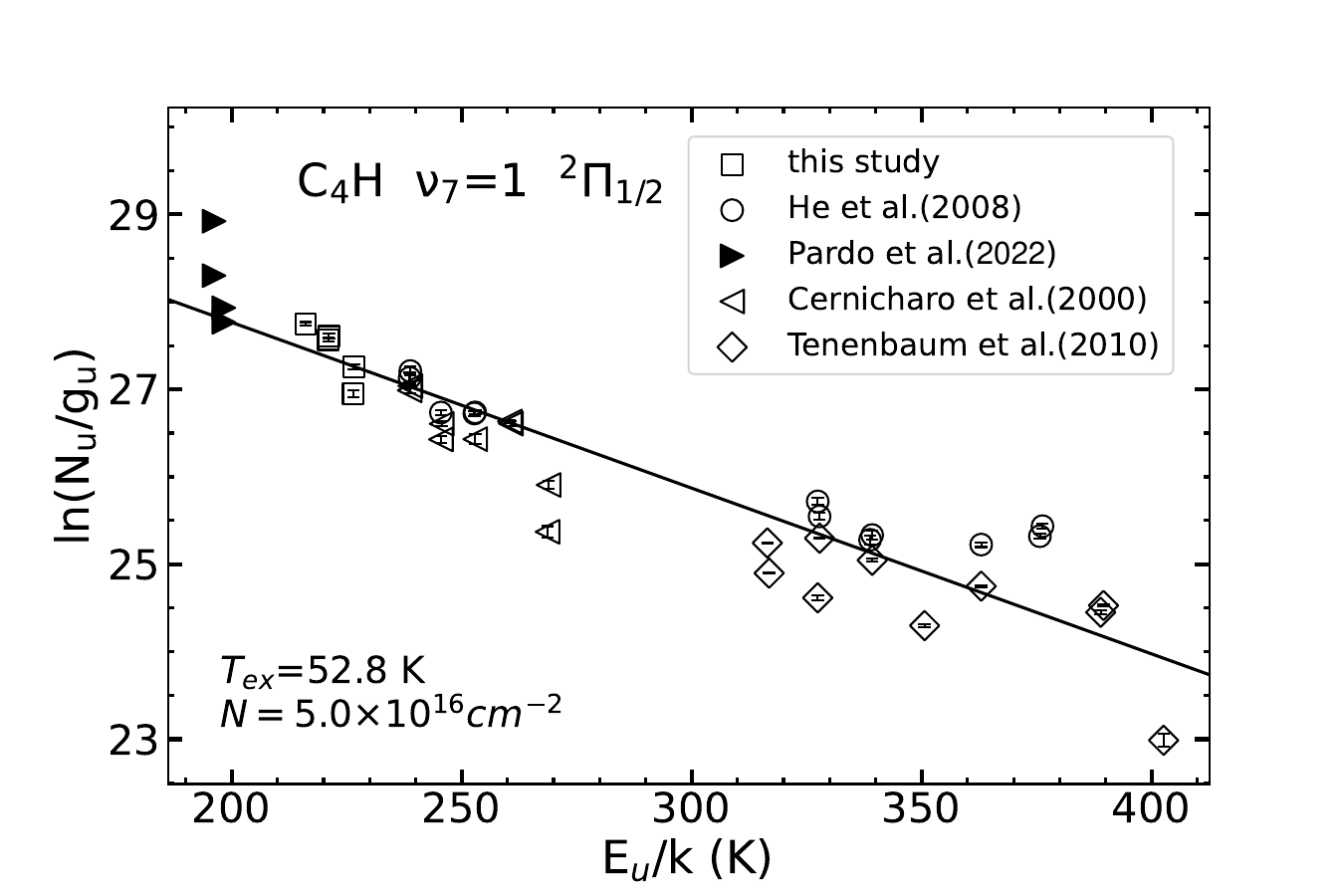}
\includegraphics[width = 0.47 \textwidth]{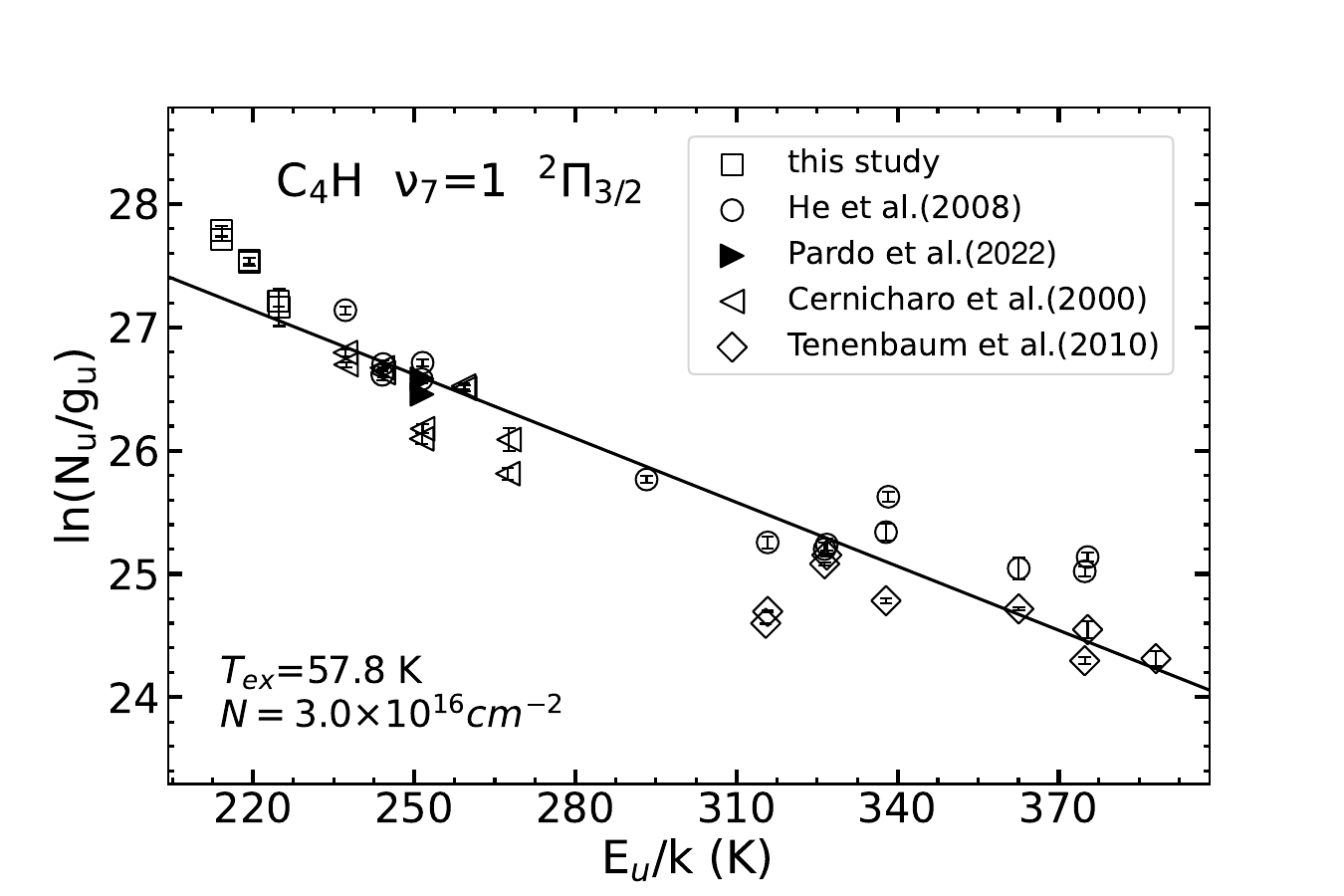}
\includegraphics[width = 0.47 \textwidth]{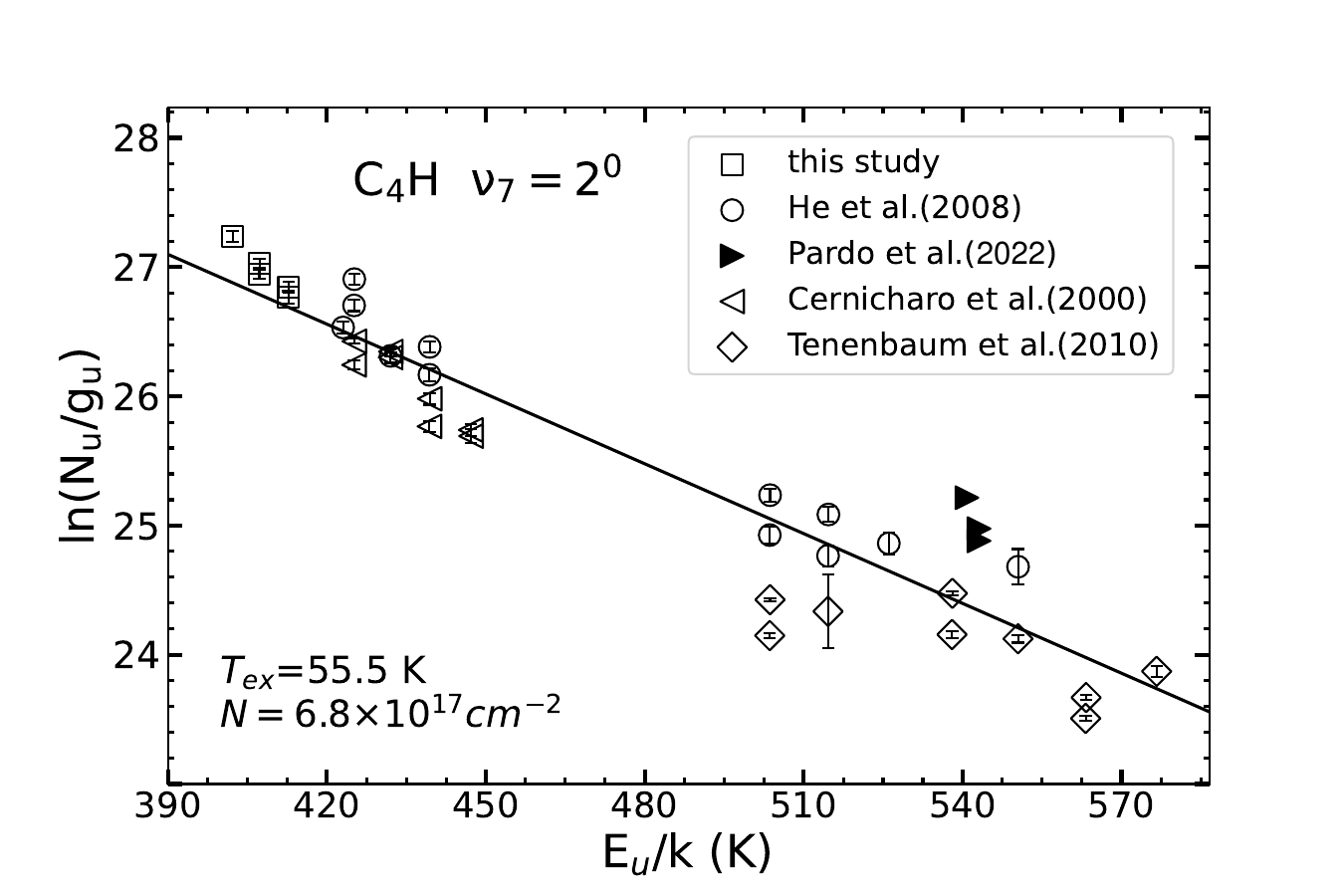}
\includegraphics[width = 0.47 \textwidth]{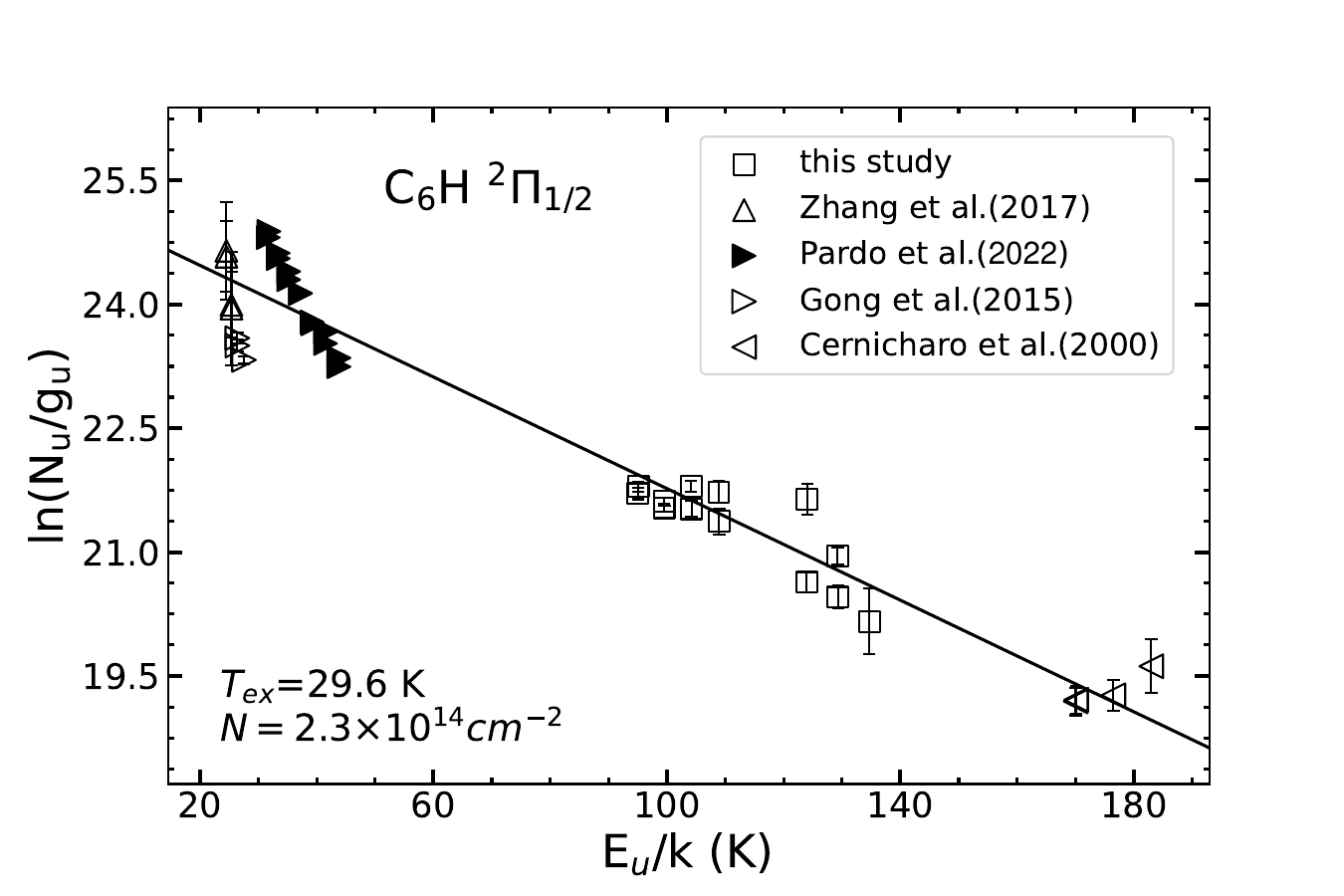}
\includegraphics[width = 0.47 \textwidth]{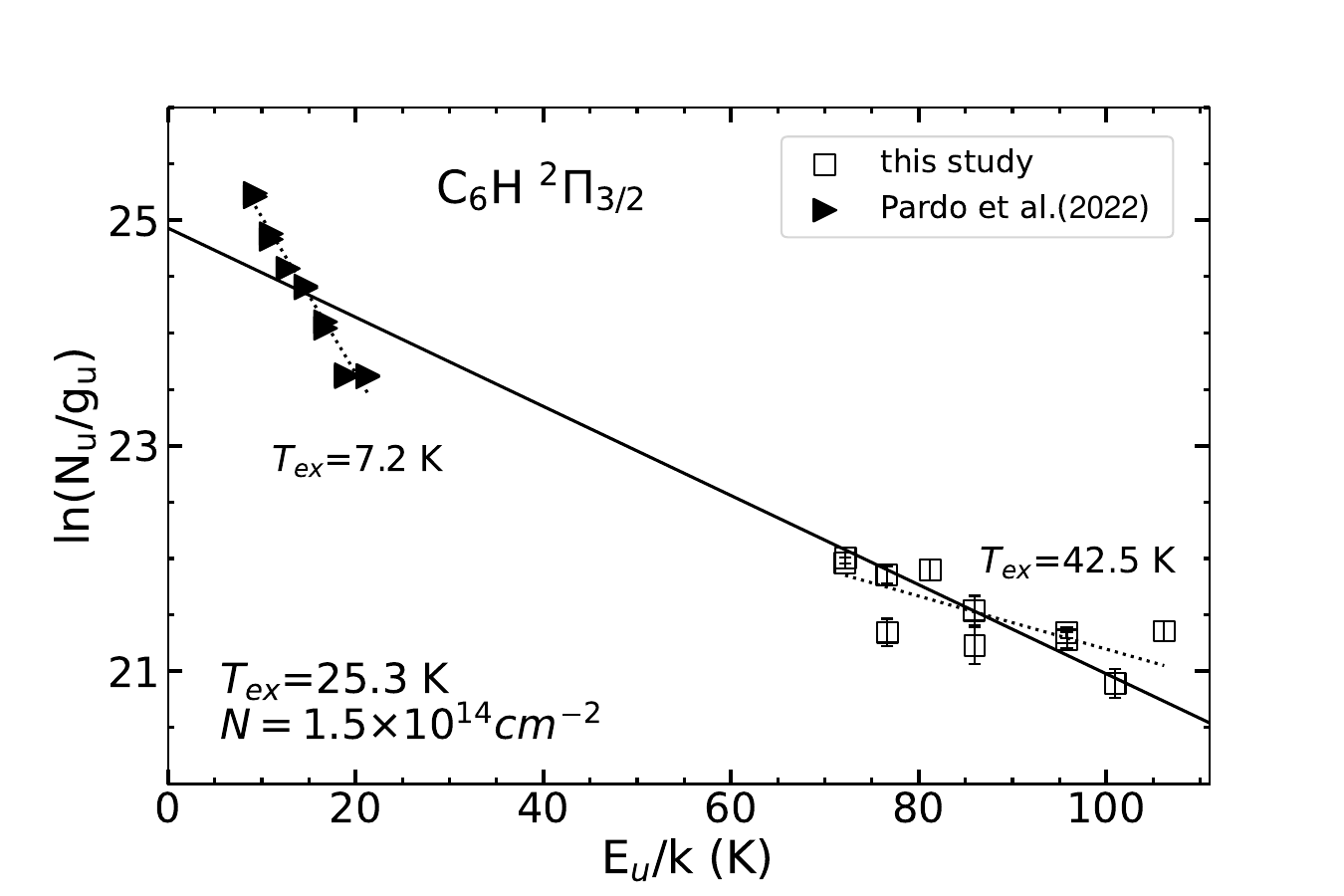}
\caption{{Rotation diagrams for the detected molecules in IRC+10216.}\label{Fig:RD}}
\end{figure*}

\begin{figure*}[!htbp]
\centering
\includegraphics[width = 0.47 \textwidth]{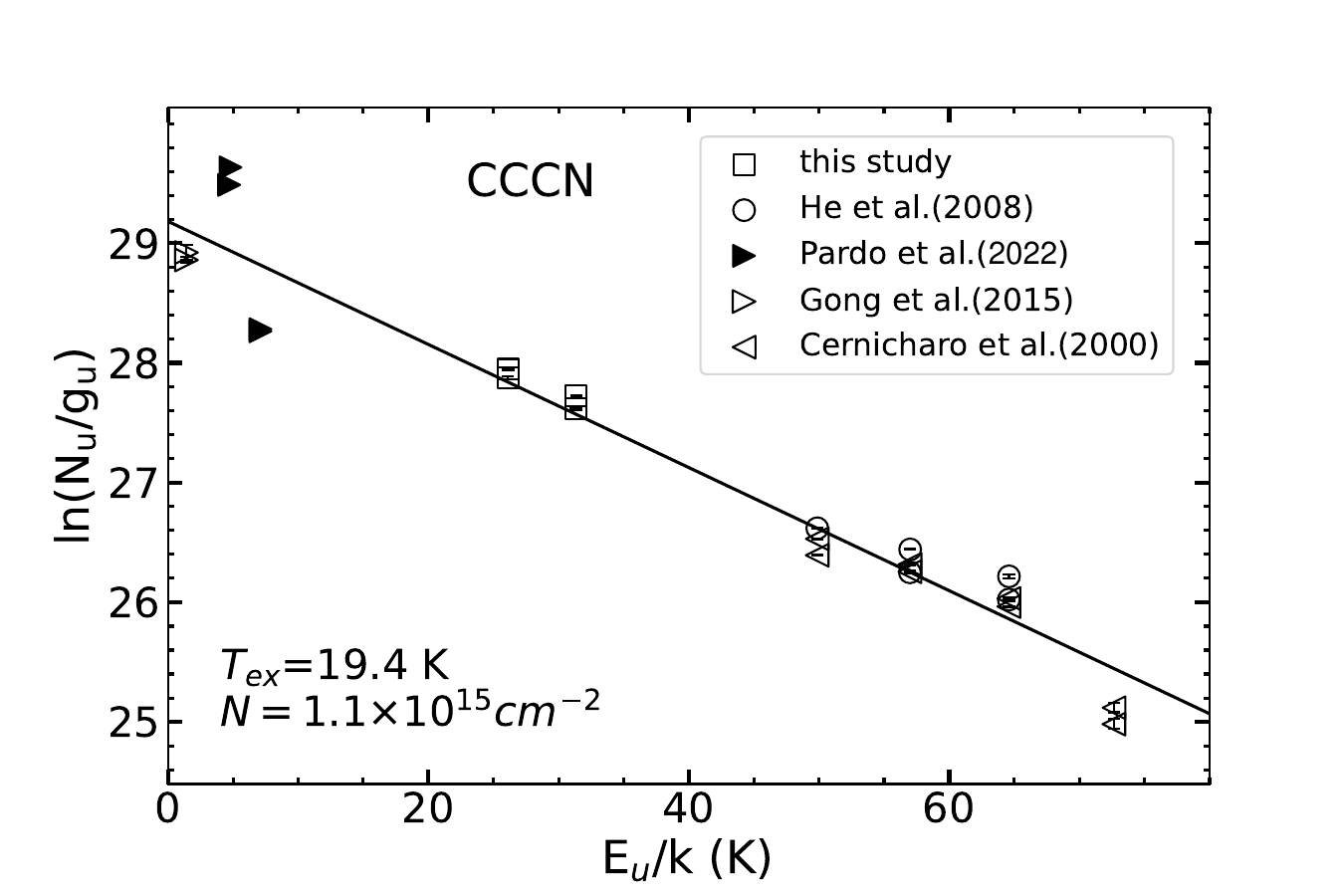}
\includegraphics[width = 0.47 \textwidth]{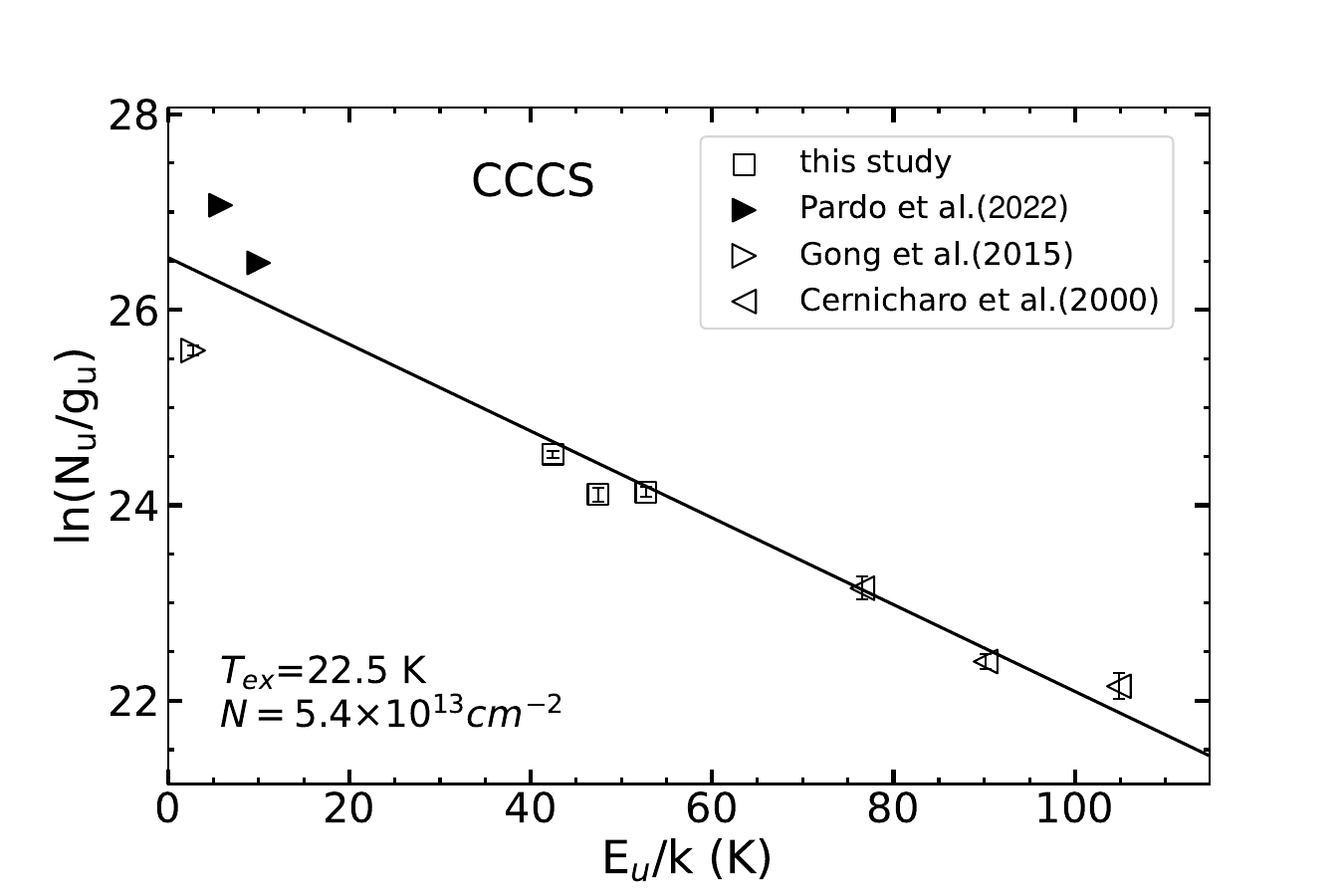}
\includegraphics[width = 0.47 \textwidth]{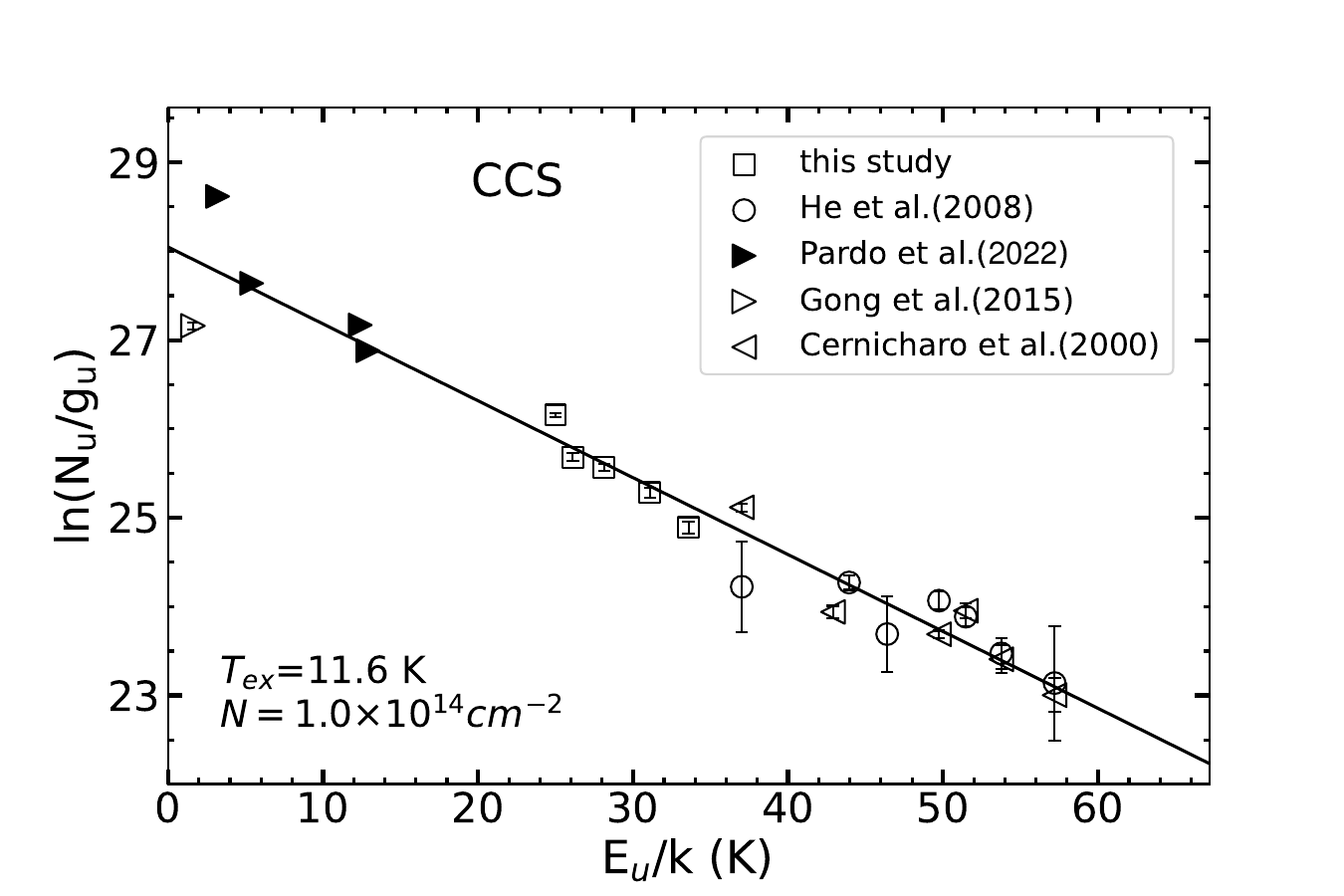}
\includegraphics[width = 0.47 \textwidth]{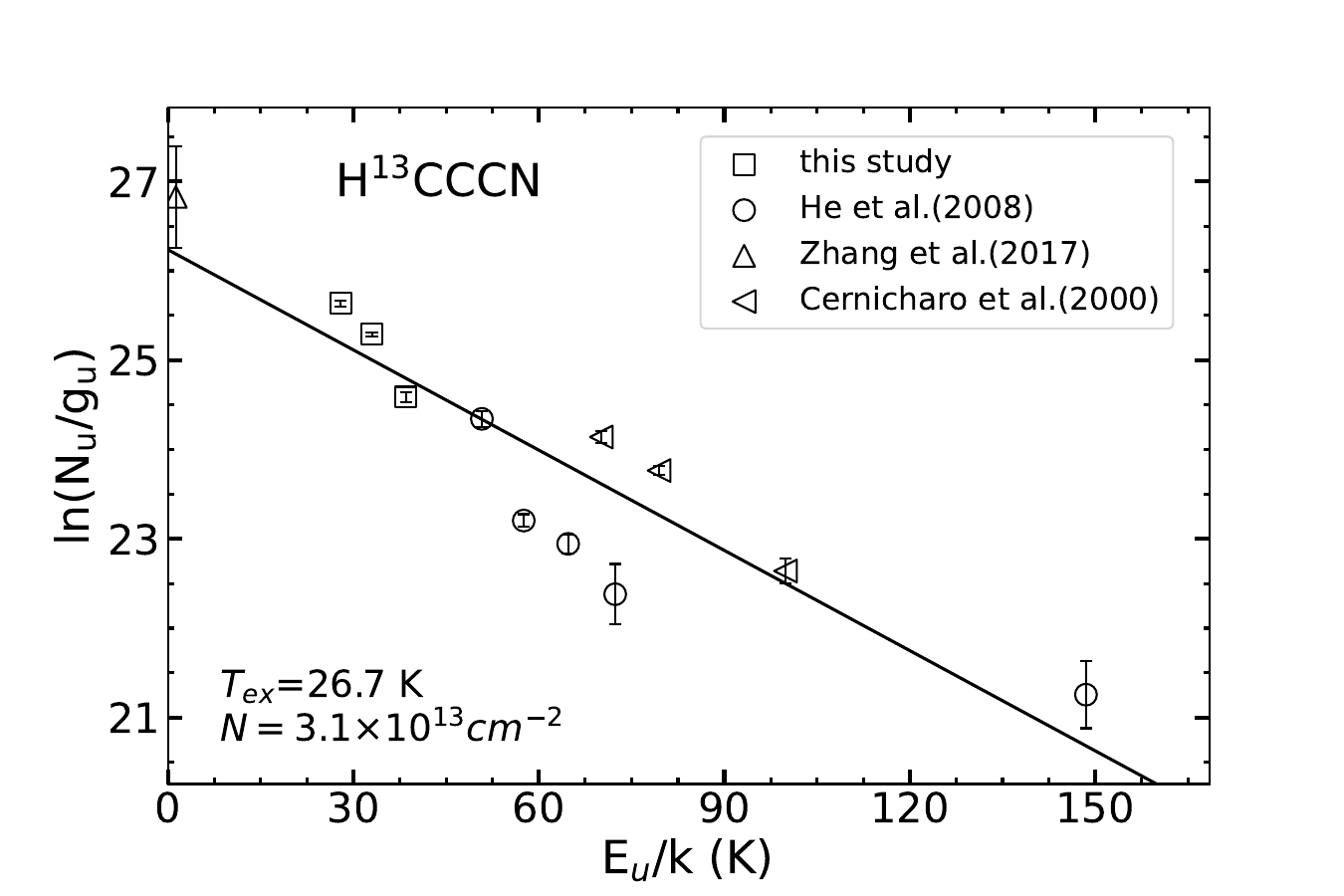}
\includegraphics[width = 0.47 \textwidth]{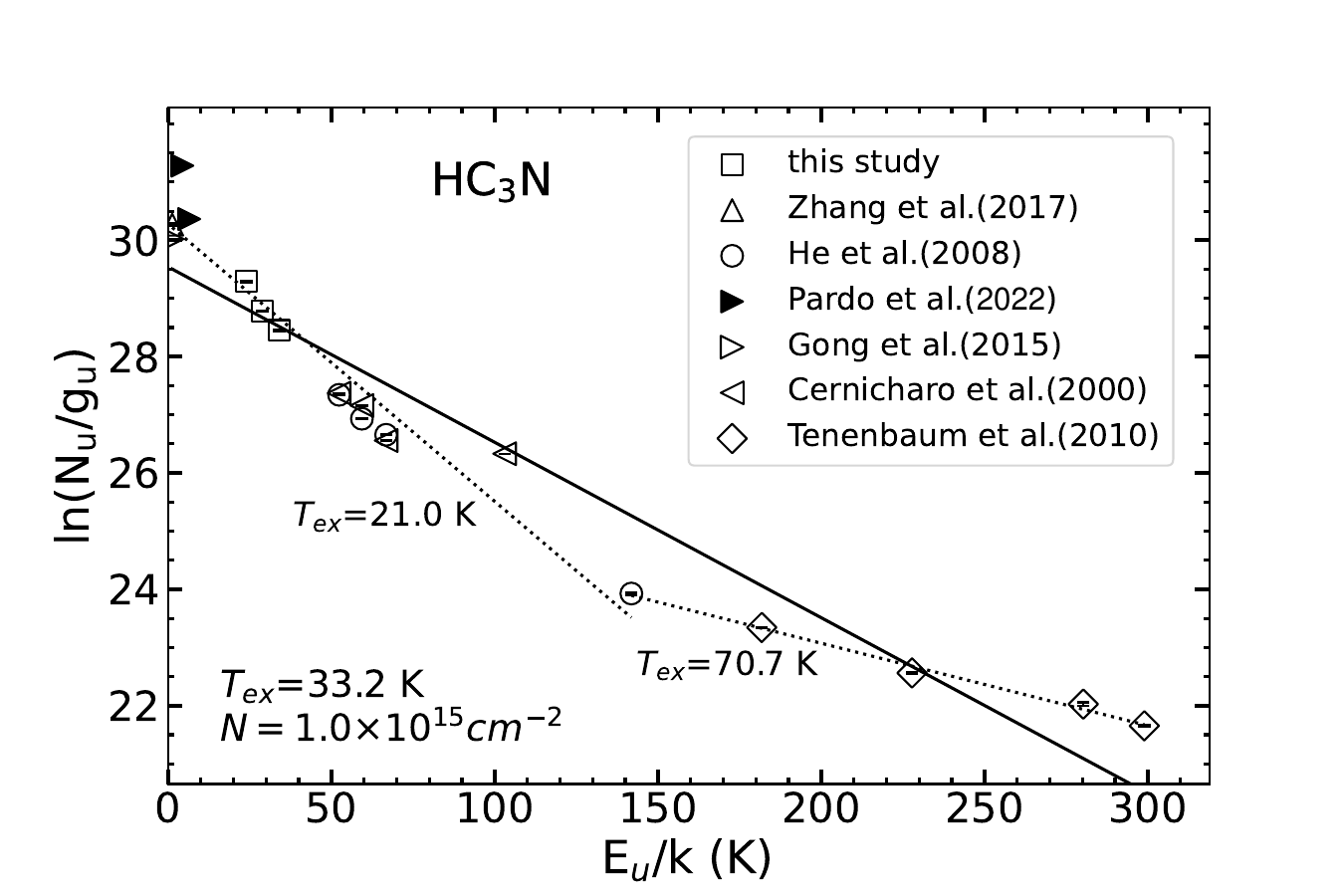}
\includegraphics[width = 0.47 \textwidth]{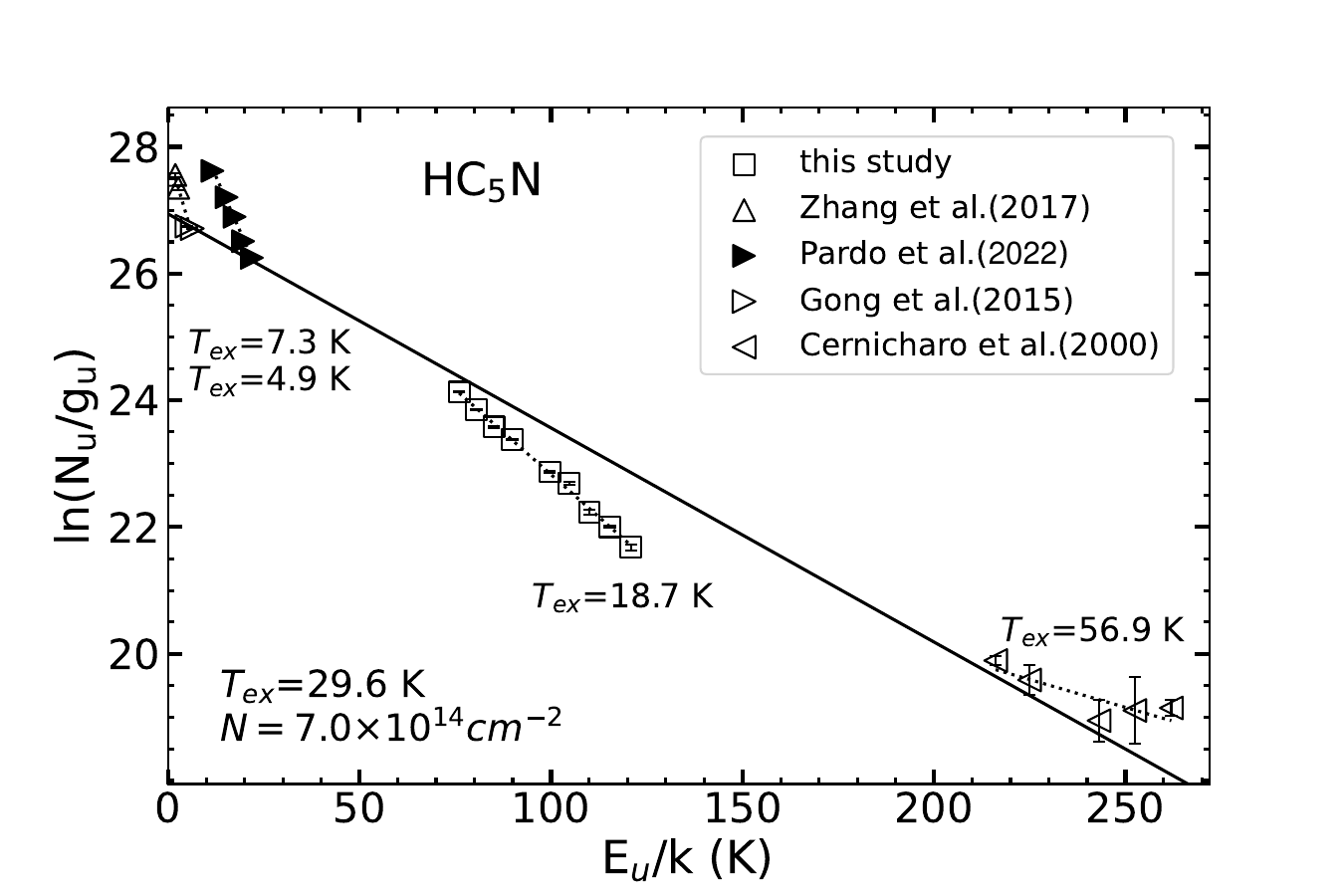}
\includegraphics[width = 0.47 \textwidth]{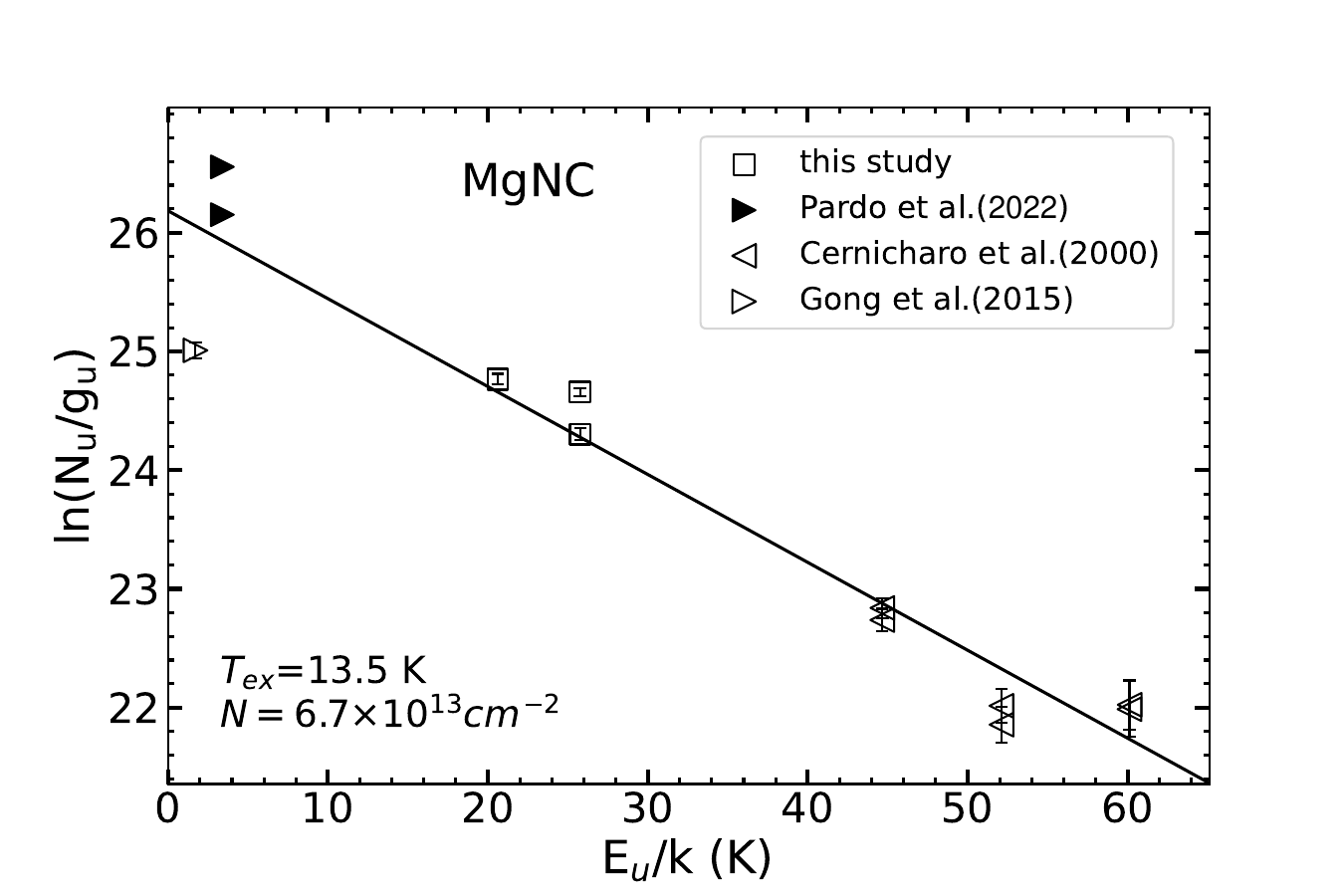}
\includegraphics[width = 0.47 \textwidth]{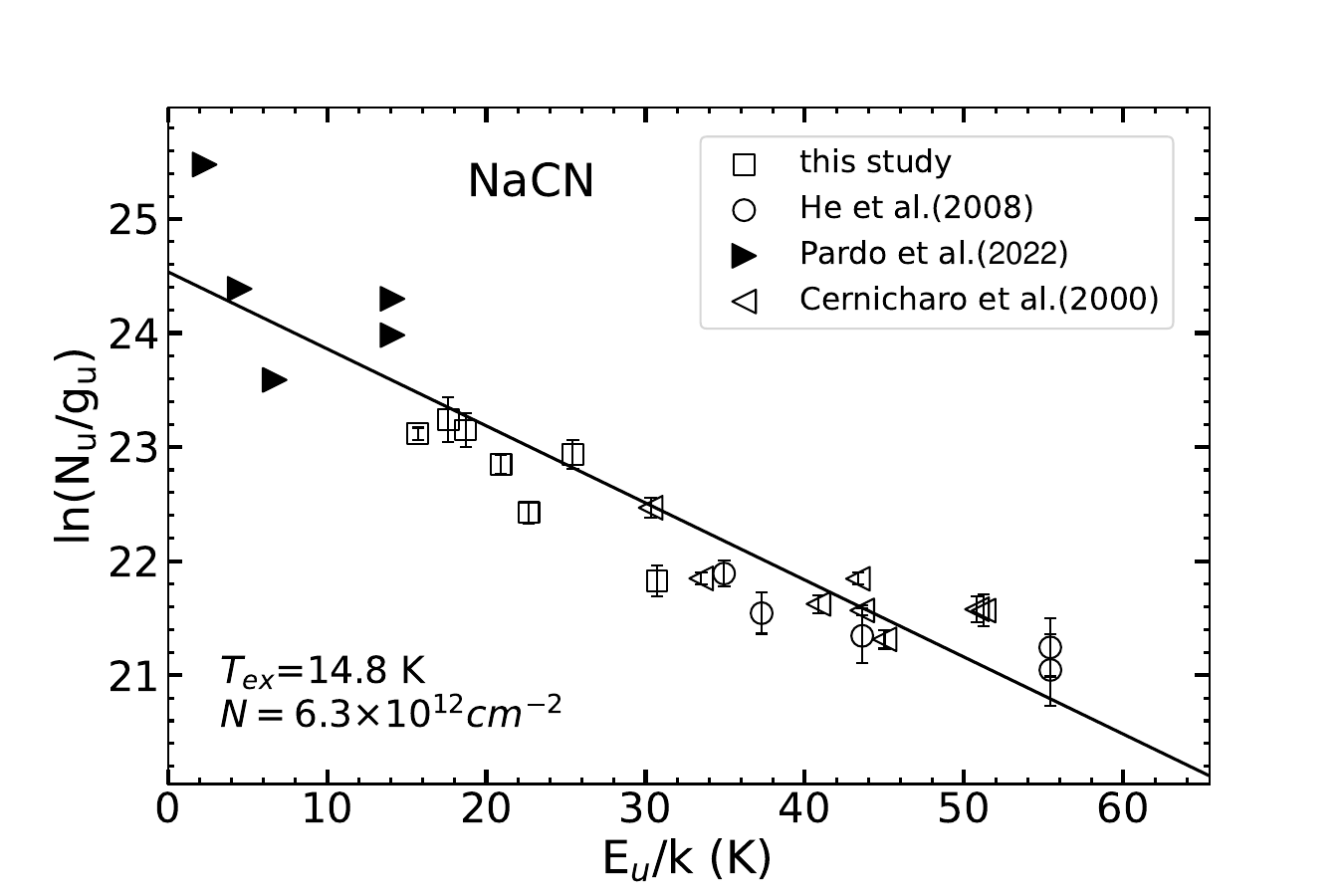}
\centerline{Figure \ref{Fig:RD}. --- continued.}
\end{figure*}

\begin{figure*}[!htbp]
\centering
\includegraphics[width = 0.47 \textwidth]{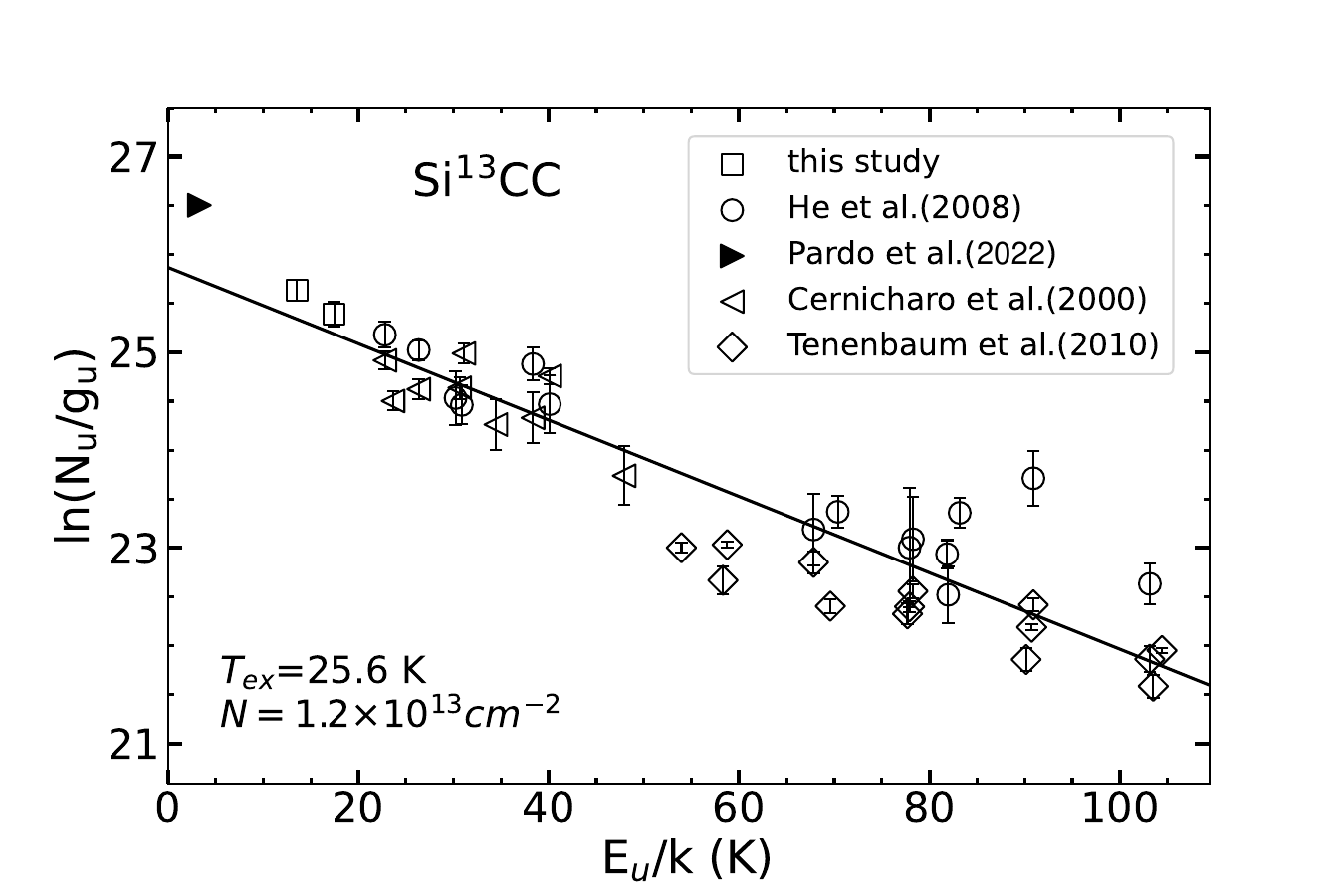}
\includegraphics[width = 0.47 \textwidth]{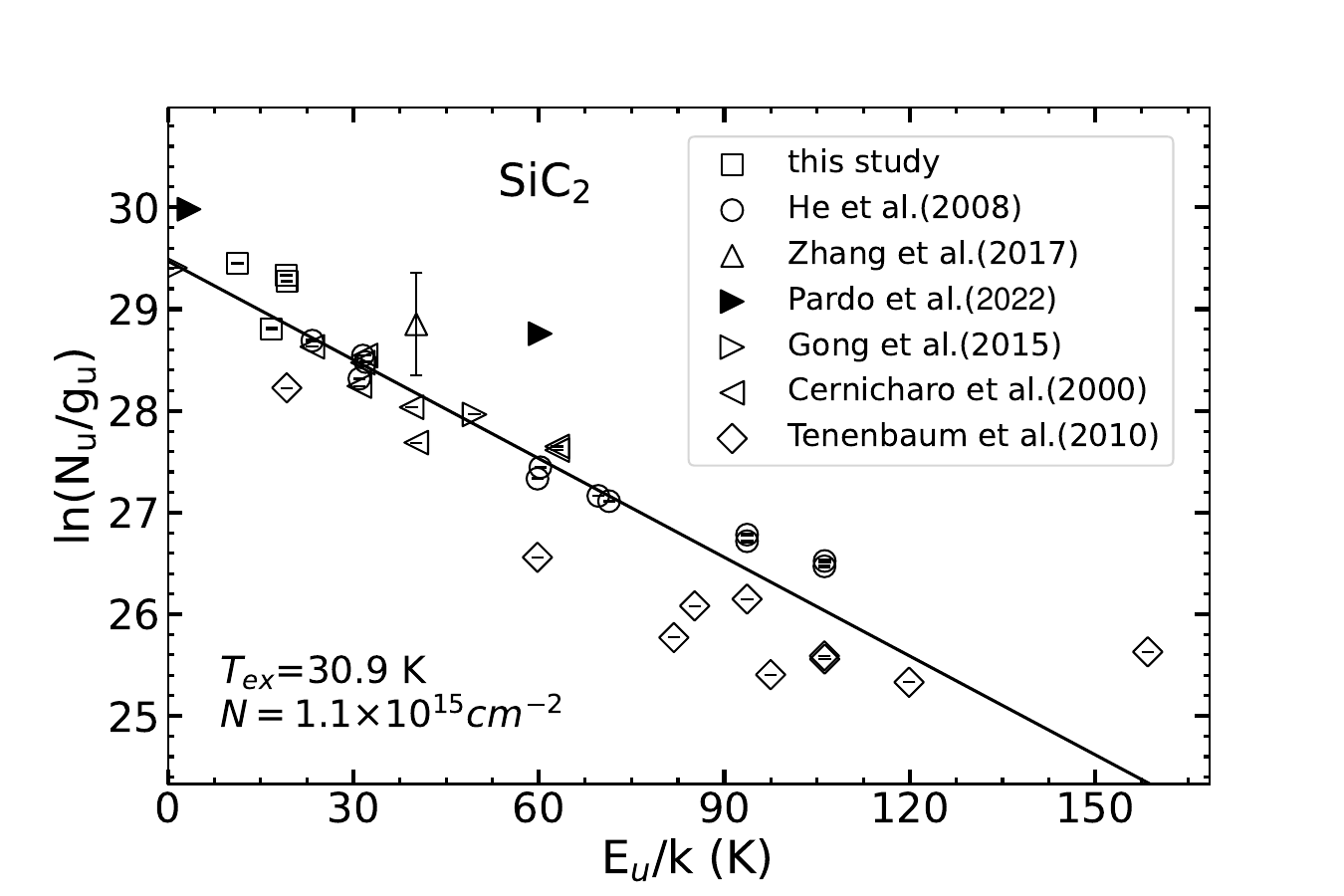}
\includegraphics[width = 0.47 \textwidth]{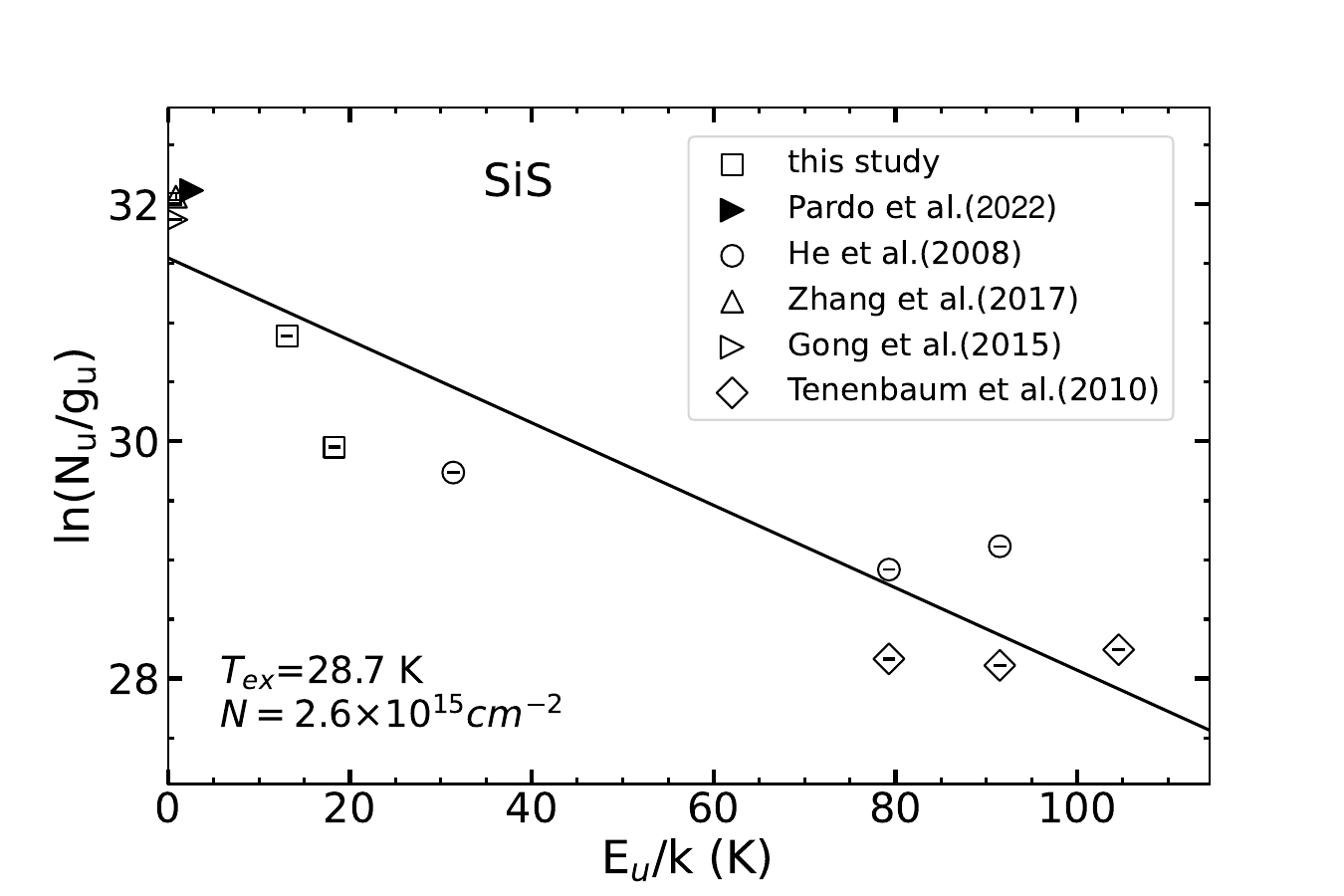}
\includegraphics[width = 0.47 \textwidth]{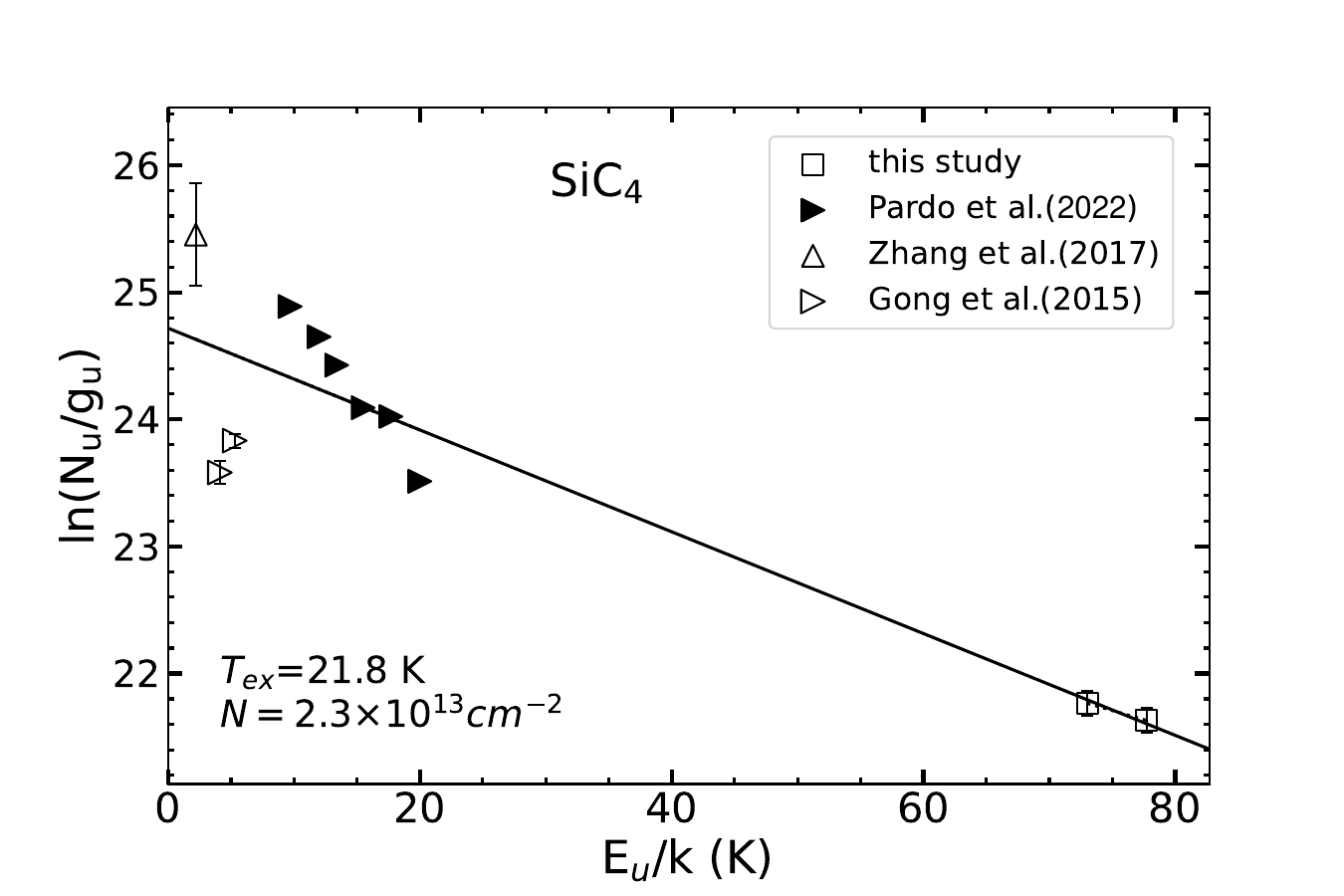}
\centerline{Figure \ref{Fig:RD}. --- continued.}
\end{figure*}

\begin{figure*}[!htbp]
\section{Intensity ratios of molecular lines}
\centering
\includegraphics[width = 0.8 \textwidth]{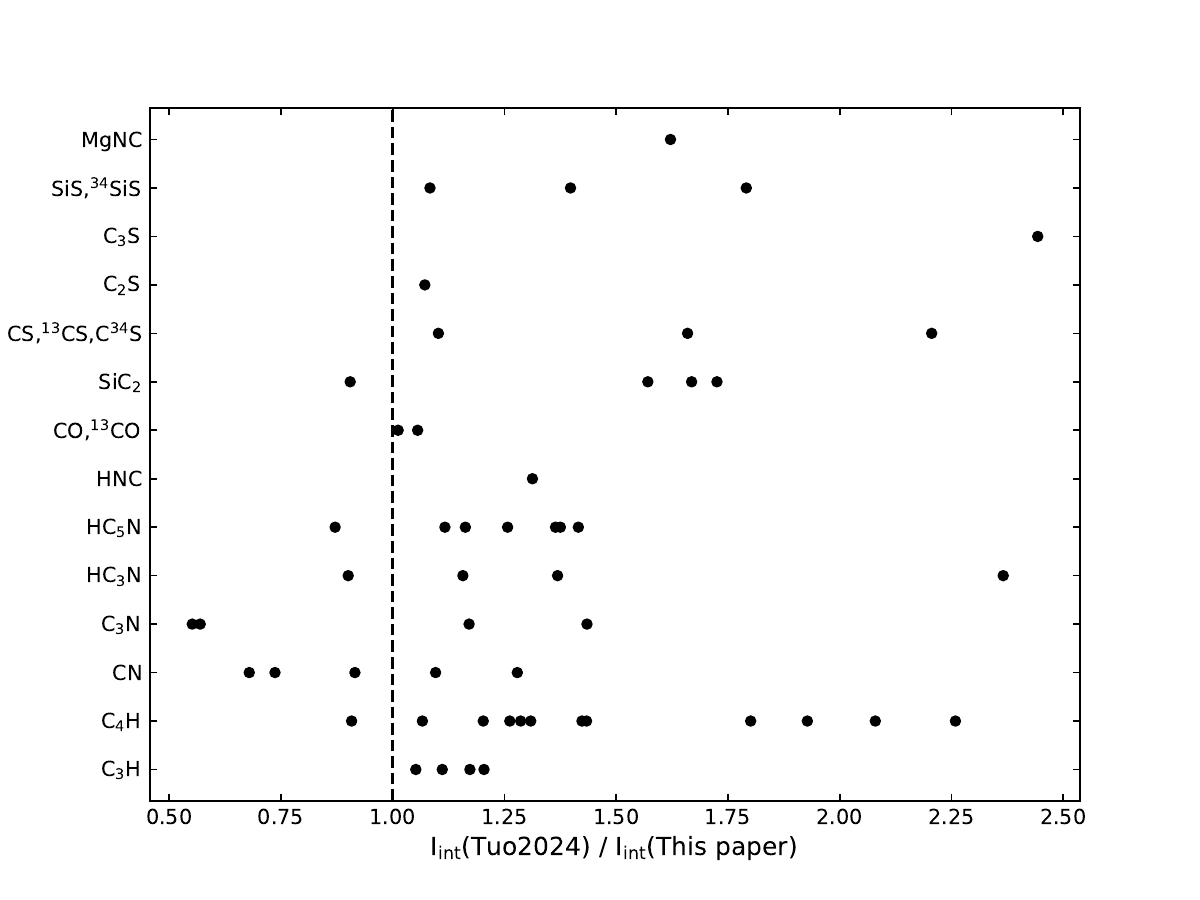}
\caption{{
Intensity ratios of molecular lines between the results of this study and those of \citetalias{2024ApJS..271...45T}.}\label{Fig:compare intensity with Tuo and our}}
\end{figure*}

\begin{figure*}[!htbp]
\section{Isotopic ratios}
\centering
\includegraphics[width = 0.4 \textwidth]{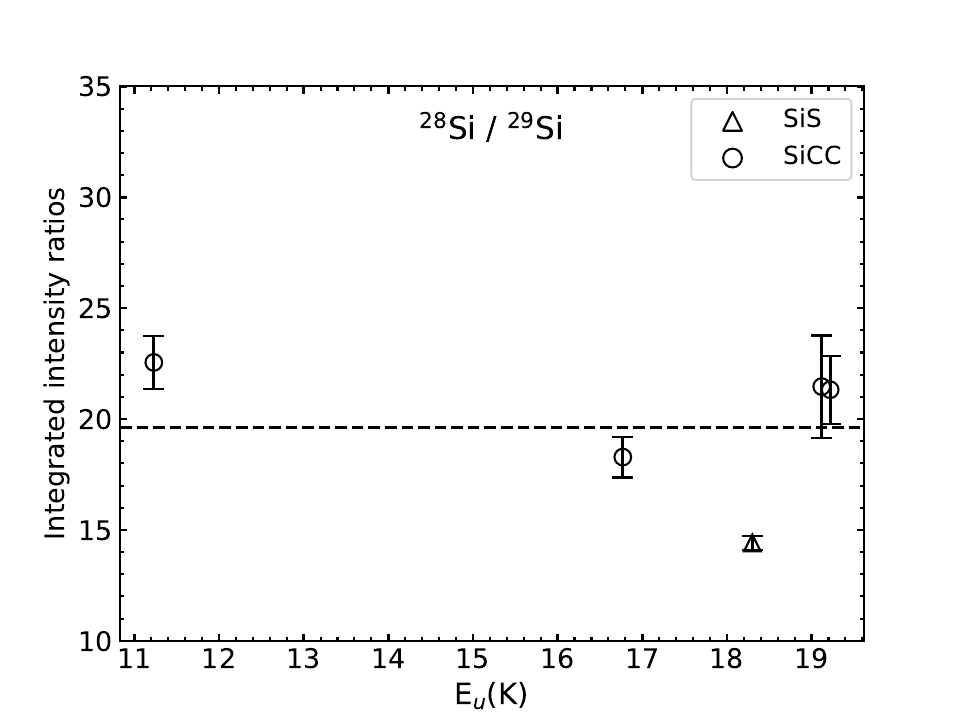}
\includegraphics[width = 0.4 \textwidth]{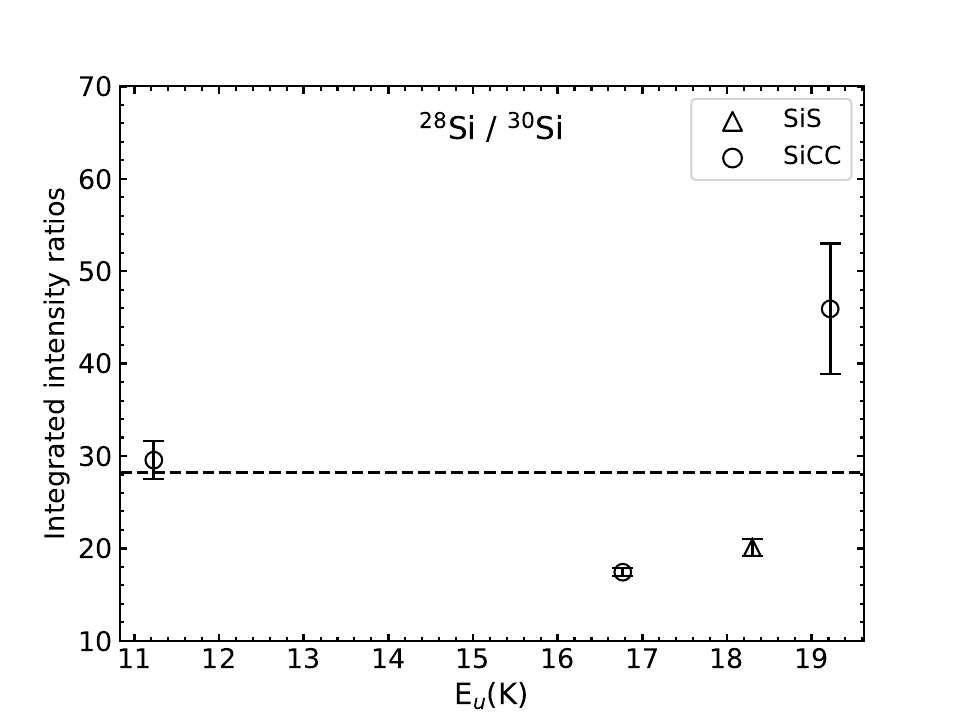}
\includegraphics[width = 0.4 \textwidth]{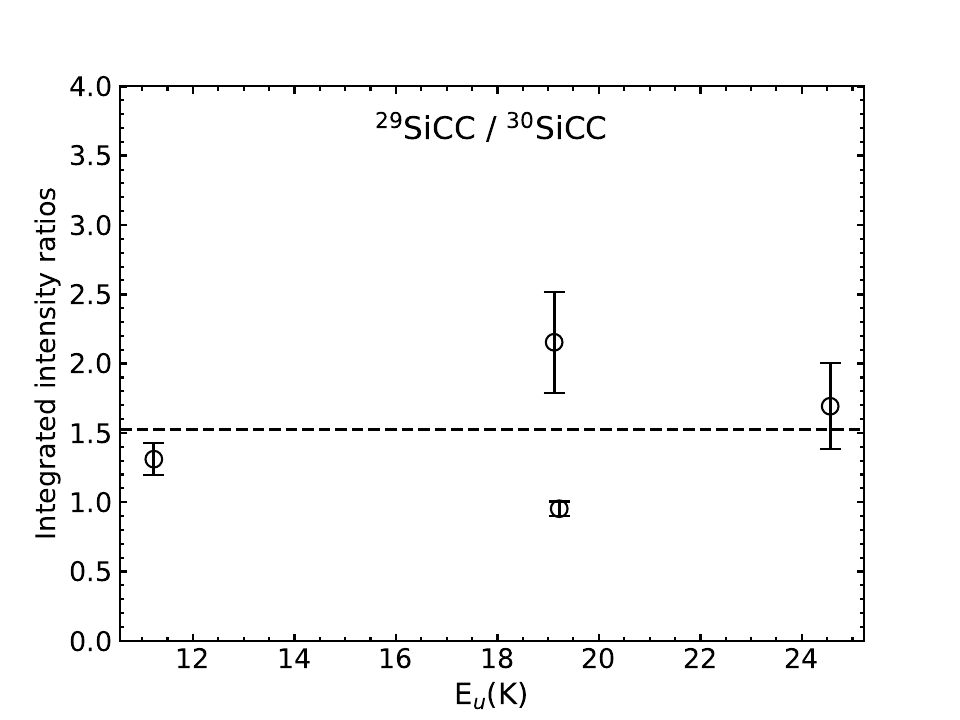}
\caption{{Same as Figure.~\ref{Fig:isotopic ratio CO}, but for Si.}\label{Fig:isotopic ratio 28Si29Si}}
\end{figure*}

\end{appendix}
\end{document}